\documentclass[11pt]{report}
\usepackage[dvips]{graphicx}

\usepackage{cite}

\usepackage{pstricks,pst-node,pst-text,pst-3d,pst-coil}

\usepackage{psfrag}

\newpsobject{showgrid}{psgrid}{subgriddiv=1,griddots=10,gridlabels=6pt}

\def\lsim{\mathrel{\vcenter{\hbox{$<$}\nointerlineskip\hbox{$\sim$}}}}
\def\gsim{\mathrel{\vcenter{\hbox{$>$}\nointerlineskip\hbox{$\sim$}}}}

\renewcommand{\bibname}{References}

\newcommand*{\sinbb}{{\sin2\phi_1}}
\newcommand*{\dM}{\ensuremath{{\Delta m_d}}}

\newcommand*{\bz}{\ensuremath{{B^0}}}
\newcommand*{\bzb}{\ensuremath{{\overline{B}{}^0}}}
\newcommand*{\bp}{\ensuremath{{B^+}}}
\newcommand*{\bminus}{\ensuremath{{B^-}}}
\newcommand*{\piz}{\ensuremath{{\pi^0}}}
\newcommand*{\pip}{\ensuremath{{\pi^+}}}
\newcommand*{\pim}{\ensuremath{{\pi^-}}}
\newcommand*{\kp}{\ensuremath{{K^+}}}
\newcommand*{\km}{\ensuremath{{K^-}}}
\newcommand*{\ks}{\ensuremath{{K_S^0}}}
\newcommand*{\kl}{\ensuremath{{K_L^0}}}
\newcommand*{\kstarz}{\ensuremath{{K^{*0}}}}
\newcommand*{\jpsi}{\ensuremath{{J/\psi}}}

\newcommand*{\Dt}{\ensuremath{{\Delta t}}}
\newcommand*{\Dz}{\ensuremath{{\Delta z}}}

\newcommand*{\fcp}{\ensuremath{{f_{CP}}}}
\newcommand*{\ftag}{\ensuremath{{f_\textrm{tag}}}}

\newcommand*{\ttag}{\ensuremath{{t_\textrm{tag}}}}
\newcommand*{\zcp}{\ensuremath{{z_{CP}}}}
\newcommand*{\ztag}{\ensuremath{{z_\textrm{tag}}}}

\newcommand*{\dE}{\ensuremath{{\Delta E}}}
\newcommand*{\mb}{\ensuremath{{M_\textrm{bc}}}}

\newcommand*{\Nrec}{\ensuremath{{N_\textrm{rec}}}}
\newcommand*{\Nev}{\ensuremath{{N_\textrm{ev}}}}

\newcommand*{\eeff}{\ensuremath{\epsilon_\textrm{eff}}}

\newcommand*{\taubz}{\ensuremath{{\tau_\bz}}}

\newcommand*{\Rsig}{\ensuremath{{R_\textrm{sig}}}}

\newcommand*{\fol}{\ensuremath{{f_\textrm{ol}}}}
\newcommand*{\fsig}{\ensuremath{{f_\textrm{sig}}}}

\newcommand*{\dwl}{\ensuremath{{\Delta w_l}}}
\newcommand*{\fq}{\ensuremath{q}}

\newcommand{\cala}{{\cal A}}
\newcommand{\cals}{{\cal S}}

\newcommand*{\efftot}{0.287 \pm 0.005}

\def\KP{K^+}
\def\KM{K^-}

\def\KS{{K^0_S}}
\def\KL{{K^0_L}}
\def\piZ{{\pi^0}}
\def\piP{{\pi^+}}
\def\piM{{\pi^-}}

\def\KstarZ{{K^{*0}}}
\def\KstarP{{K^{*+}}}

\def\Bbar{\overline{B}{}}

\def\Kbar{\overline{K}{}}

\def\sbar{\overline{s}{}}

\def\bbar{\overline{b}{}}

\def\epem{e^+e^-{}}
\def\mumu{\mu^+\mu^-{}}
\def\elel{\ell^+\ell^-{}}

\def\Bd{B_d^0{}}

\def\btosg{b\to sg}
\def\btosgamma{b\to s\gamma}

\def\btodgamma{b\to d\gamma}
\def\btosll{b\to s\elel}

\def\Btorhogamma{B\to \rho\gamma}

\def\BtoXsgamma{B\to X_s\gamma}
\def\BtoXdgamma{B\to X_d\gamma}
\def\BtoXsll{B\to X_s\elel}

\def\BtoKll{B\to K\elel}
\def\BtoKee{B\to K\epem}
\def\BtoKmumu{B\to K\mumu}
\def\Xsll{X_s\elel}
\def\Xsee{X_s\epem}
\def\Xsmumu{X_s\mumu}

\def\BtoKstarll{B\to K^*\elel}
\def\BtoKstaree{B\to K^*\epem}
\def\BtoKstarmumu{B\to K^*\mumu}
\def\BtoKorKstarll{B\to K^{(*)}\elel}
\def\BtoKstargamma{B\to K^*\gamma}

\def\BtoKstarZG{B^0\to K^{*0}\gamma}
\def\BtoKstarPG{B^+\to K^{*+}\gamma}

\def\fbinv{\mbox{~fb}^{-1}}
\def\abinv{\mbox{~ab}^{-1}}

\def\GeV{\mbox{~GeV}}
\def\GeVc{\mbox{~GeV}/c}
\def\GeVcc{\mbox{~GeV}/c^2}

\def\Vub{V_{ub}}

\def\Vcb{V_{cb}}
\def\Vtd{V_{td}}
\def\Vts{V_{ts}}
\def\Vtb{V_{tb}}

\def\Br{{\cal B}}

\def\Mbc{M_{\rm bc}}
\def\DeltaE{\Delta{E}}

\def\Mll{M(\elel)}
\def\MKpi{M(K\pi)}

\def\shat{\hat{s}}

\def\Egammamin{E_\gamma^{\rm min}}
\def\Acp{A_{CP}}
\def\Deltapz{\Delta_{+0}}
\def\tauBratio{\tau_{B^+}/\tau_{B^0}}

\def\MXs{M(X_s)}
\def\AFB{A_{FB}}

\def\PM#1#2{\,^{+#1}_{-#2}{}}
\def\EM#1{\times10^{-#1}}

\def\etal{\textit{et al.}}
\def\Journal#1#2#3#4{{#1} {\bf #2}, #3 (#4)} 
\def\NCA{\em Nuovo Cimento}                      
\def\NIMA{{\em Nucl. Instrum. Meth.} A}
\def\NPB{{\em Nucl. Phys.} B}
\def\PLB{{\em Phys. Lett.} B}
\def\PRL{\em Phys. Rev. Lett.}
\def\PRD{{\em Phys. Rev.} D}
\def\ZPC{{\em Z. Phys.} C}
\def\EPJD{{\em Eur. Phys. J. direct} C}
\def\EPJC{{\em Eur. Phys. J.} C}

\def\bz{{B^0}}
\def\bzb{{\overline{B}{}^0}}
\def\kl{K_L^0}
\def\dE{{\Delta E}}
\def\mb{{M_{\rm bc}}}
\def\Dt{\Delta t}
\def\Dz{\Delta z}
\def\fol{f_{\rm ol}}
\def\fsig{f_{\rm sig}}
\def\sinbb{{\sin2\phi_1}}
\def\ra{\rightarrow}
\def\myindent{\hspace*{2cm}}  
\def\fCP{f_{CP}}
\def\ftag{f_{\rm tag}}
\def\zCP{z_{CP}}
\def\tCP{t_{CP}}
\def\ttag{t_{\rm tag}}
\def\cala{{\cal A}}
\def\cals{{\cal S}}
\def\dm{\Delta m_d}
\def\dmd{\dm}
\def\taubz{\tau_\bz}
\def\ks{{K_S^0}}
\def\btosqq{b \to s\overline{q}q}
\def\btosss{b \to s\overline{s}s}
\def\dwl{\ensuremath{{\Delta w_l}}}
\def\fq{\ensuremath{q}}

\def\ra{\rightarrow}
\def\myindent{\hspace*{2cm}}  
\def\qbar{\overline{q}}
\def\ubar{\overline{u}}
\def\Bcp{B_{CP}}
\def\UfourS{\Upsilon(4S)}
\def\phiNP{\phi_{\rm NP}}
\def\ACP{{\cal A}_{CP}}
\def\Mbc{M_{\rm bc}}
\def\mbc{\Mbc}
\def\sss{s\bar{s}s}
\def\suu{s\bar{u}u}
\def\uus{u\bar{u}s}
\def\ccs{c\bar{c}s}
\def\sdd{s\bar{d}d}
\def\btosss{b \ra \sss}
\def\btosssss{b \ra \sss\bar{s}s}
\def\btouus{b \ra \uus}
\def\btoccs{b \ra \ccs}
\def\btosuu{b \to \suu}
\def\btosdd{b \to \sdd}
\def\acp{\mathcal{A}_{CP}}
\def\dacp{\delta\acp}
\def\acppar{\mathcal{A}_{CP}^0}
\def\dacppar{\delta\acppar}
\def\nobs{N_{\rm obs}}
\def\bpm{B^{\pm}}
\def\Nb{{\rm N}_B}
\def\etac{\eta_c}
\def\kl{K_L^0}
\def\xs{X_s}
\def\xsp{X_s^+}
\def\xsm{X_s^-}
\def\xspm{X_s^{\pm}}
\def\etack{\etac K}
\def\etackp{\etac K^+}
\def\etackm{\etac K^-}
\def\etackpm{\etac K^{\pm}}
\def\btoetack{B \to \etack}
\def\bptoetackp{\bp \to \etackp}
\def\bmtoetackm{\bminus \to \etackm}
\def\bpmtoetackpm{\bpm \to \etackpm}
\def\bztoetackstarz{\bz \to \etac K^{*0}}
\def\bptoetackstarp{\bp \to \etac K^{*+}}
\def\bpmtoetacxspm{\bpm \to \etac \xspm}
\def\phiphixs{\phi \phi \xs}
\def\phiphixsp{\phi \phi \xsp}
\def\phiphixsm{\phi \phi \xsm}
\def\phiphixspm{\phi \phi \xspm}
\def\phiphik{\phi \phi K}
\def\phiphikp{\phi \phi K^+}
\def\phiphikm{\phi \phi K^-}
\def\phiphikpm{\phi \phi K^{\pm}}
\def\bpmtophiphikpm{\bpm \to \phiphikpm}
\def\btophiphixs{B \to \phiphixs}
\def\bptophiphixsp{\bp \to \phiphixsp}
\def\bmtophiphixsm{\bminus \to \phiphixsm}
\def\bpmtophiphixspm{\bpm \to \phiphixspm}
\def\btophiphik{B \to \phiphik}
\def\tnp{\Theta_{\rm NP}}
\def\sintnp{\sin \tnp}
\def\costnp{\cos \tnp}
\def\ad{a_{\rm D}}
\def\adtwo{a_{\rm D}^2}
\def\ar{a_{\rm R}}
\def\artwo{a_{\rm R}^2}
\def\aNP{a_{\rm NP}}
\def\aNPtwo{a_{\rm NP}^2}
\def\rtwo{r^2}
\def\DSM{D_{\rm SM}}
\def\dpmnp{D^{\pm}_{\rm NP}}
\def\dpm{D^{\pm}}
\def\bnp{\mathcal{B}_{\rm NP}}
\def\bztophiks{\bz \to \phi\ks}
\def\dmd{\Delta m_d}
\def\etap{\eta^\prime}
\def\etapks{\etap\ks}
\def\bztoetapks{\bz \to \etapks}
\def\kkks{K^+ K^- \ks}
\def\bztokkks{\bz \to \kkks}
\def\calsphiks{\cals_{\phi\ks}}
\def\calaphiks{\cala_{\phi\ks}}
\def\calsetapks{\cals_{\etapks}}
\def\calaetapks{\cala_{\etapks}}
\def\calskkks{\cals_{\kkks}}
\def\calakkks{\cala_{\kkks}}
\def\calspizks{\cals_{\pi^0\ks}}
\def\calapizks{\cala_{\pi^0\ks}}
\def\calsksksks{\cals_{\ks\ks\ks}}
\def\calaksksks{\cala_{\ks\ks\ks}}
\def\deltasphiks{\Delta\calsphiks}
\def\deltaaphiks{\Delta\calaphiks}
\def\deltasetapks{\Delta\calsetapks}
\def\deltaaetapks{\Delta\calaetapks}
\def\deltaskkks{\Delta\calskkks}
\def\deltaakkks{\Delta\calakkks}
\def\deltaspizks{\Delta\calspizks}
\def\deltaapizks{\Delta\calapizks}
\def\deltasksksks{\Delta\calsksksks}
\def\deltaaksksks{\Delta\calaksksks}
\def\jpsi{J/\psi}
\def\jpsiks{\jpsi\ks}
\def\bztojpsiks{\bz\to\jpsiks}
\def\calsjpsiks{\cals_{\jpsiks}}
\def\calajpsiks{\cala_{\jpsiks}}

\def\calsksgamma{\cals_{K^{*0}\gamma}}
\def\bztophiks{\bz\to\phi\ks}
\def\bztoksgamma{\bz\to K^{*0}\gamma}

\def\Budker{^{1}~}
\def\Cincinatti{^{2}~}
\def\DESY{^{3}~}
\def\Hawaii{^{4}~}
\def\Hiroshima{^{5}~}
\def\ICRR{^{6}~}
\def\KEK{^{7}~}
\def\Louvain{^{8}~}
\def\Munchen{^{9}~}
\def\Nagoya{^{10}~}
\def\Nara{^{11}~}
\def\Osaka{^{12}~}
\def\TMU{^{13}~}
\def\Tohoku{^{14}~}
\def\Toyama{^{15}~}
\def\Yonsei{^{16}~}
\def\YITP{^{17}~}

\title{Physics at Super $B$ Factory}
\author{
A.~G.~Akeroyd,$\KEK$ 
W.~Bartel,$\DESY$
A.~Bondar,$\Budker$
T.~E.~Browder,$\Hawaii$
A.~Drutskoy,$\Cincinatti$
\\
Y.~Enari,$\Nagoya$ 
T.~Gershon,$\KEK$
T.~Goto,$\YITP$ 
F.~Handa,$\Tohoku$ 
K.~Hara,$\KEK$  
\\
S.~Hashimoto,$\KEK$  
H.~Hayashii,$\Nara$
M.~Hazumi,$\KEK$  
T.~Higuchi,$\KEK$  
J.~Hisano,$\ICRR$
\\
T.~Iijima,$\Nagoya$  
K.~Inami,$\Nagoya$  
R.~Itoh,$\KEK$  
N.~Katayama,$\KEK$
Y.~Y.~Keum,$\Nagoya$  
\\
E.~Kou,$\Louvain$
T.~Kurimoto,$\Toyama$
Y.~Kwon,$\Yonsei$
T.~Matsumoto,$\TMU$
T.~Morozumi,$\Hiroshima$
\\
M.~Nakao,$\KEK$
S.~Nishida,$\KEK$ 
T.~Ohshima,$\Nagoya$  
Y.~Okada,$\KEK$ 
S.~L.~Olsen,$\Hawaii$
\\
T.~Onogi,$\YITP$ 
A.~Poluektov,$\Budker$
S.~Recksiegel,$\Munchen$
H.~Sagawa,$\KEK$ 
M.~Saigo,$\Tohoku$  
\\
Y.~Sakai,$\KEK$ 
A.~I.~Sanda,$\Nagoya$  
K.~Senyo,$\Nagoya$
Y.~Shimizu,$\Nagoya$
T.~Shindou,$\KEK$ 
\\
K.~Sumisawa,$\Osaka$  
M.~Tanaka,$\Osaka$ 
H.~Yamamoto,$\Tohoku$ 
M.~Yamauchi~$\KEK$
\\
\\
\\
(The SuperKEKB Physics Working Group)
\\
\\
\\
$\Budker${\it Budker Institute of Nuclear Physics, Novosibirsk, Russia}\\
$\Cincinatti${\it University of Cincinnati, Cincinnati, Ohio 45221, USA}\\
$\DESY${\it DESY, Hamburg, Germany}\\
$\Hawaii${\it University of Hawaii, Honolulu, Hawaii 96822, USA}\\
$\Hiroshima${\it Graduate School of Science, Hiroshima University,}\\ 
            {\it Higashi-Hiroshima, 739-8526, Japan}\\
$\ICRR${\it ICRR, University of Tokyo, Kashiwa 277-8582, Japan}\\
$\KEK${\it High Energy Accelerator Research Organization (KEK), Tsukuba, Japan}\\
$\Louvain${\it Institut de Physique Th\'{e}orique, Universit\'{e} Catholique de Louvain,}\\{\it B-1348 Louvain-la-Neuve, Belgium}\\
$\Munchen${\it Technische Universit\"{a}t M\"{u}nchen, 
            D-85748 Garching, Germany}\\
$\Nagoya${\it Nagoya University, Nagoya, Japan}\\
$\Nara${\it Nara Women's University, Nara, Japan}\\
$\Osaka${\it Osaka University, Osaka, Japan}\\
$\TMU${\it Tokyo Metropolitan University, Tokyo, Japan}\\
$\Tohoku${\it Tohoku University, Sendai, Japan}\\
$\Toyama${\it Toyama University, Toyama, Japan}\\
$\Yonsei${\it Yonsei University, Seoul, Korea}\\
$\YITP${\it YITP, Kyoto University, Kyoto 606-8502, Japan}
}
\date{\today}

\begin{document}
\maketitle

\begin{abstract}
This report presents the results of studies that
investigate the physics reach at a Super $B$ factory,
an asymmetric-energy $e^+e^-$ collider with a
design luminosity of $5 \times 10^{35}$ cm$^{-2}$s$^{-1}$,
which is around 40 times as large as the peak
luminosity achieved by the KEKB collider.
The studies focus on flavor physics and $CP$ violation measurements
that could be carried out in the LHC era.
The physics motivation, key observables,
measurement methods and expected precisions
are presented.
The sensitivity studies are
a part of the activities associated with the preparation
of a Letter of Intent for SuperKEKB, which has been
submitted recently~\footnote{KEK Report 04-4 ``Letter of Intent for KEK Super $B$ Factory'', available from http://belle.kek.jp/superb/loi/.}.
%
%
%
%
%
%
%
%
%
%
\end{abstract}

\tableofcontents

\chapter{Introduction}
\label{chap:introduction}

\section{Motivation for the Higher Luminosity $B$ Factory} 

What are the most fundamental elements of the Universe?
What is the law which governs their interactions?
These are the questions that theoretical and experimental
particle physicists have been working hard to answer for
more than a century, and it was about thirty years ago that
they arrived at the Standard Model of elementary
particles.
The Standard Model contains three generations of quarks and
leptons, and their interactions are mediated by gauge bosons
according to the $SU(3)_C \times SU(2)_L \times U(1)_Y$ gauge field
theory. 
Over the past thirty years the Standard Model
has been confirmed by many precise experimental measurements.

Nevertheless, there are several reasons why the Standard Model
is not completely satisfactory as \emph{the} theory of
elementary particles.
First of all, it includes many parameters, \textit{i.e.}
the masses and mixing of the quarks and leptons, all of
which are a priori unknown.
The hierarchy of quark and lepton masses and the flavor
mixing matrices suggest that some hidden mechanism occurring
at a  higher energy scale governs their pattern. 
Secondly, due to quadratically divergent radiative
corrections, the Higgs mass is naturally of the same order as its
cutoff scale; this implies that some new physics exists
not far above the electroweak scale.
From a cosmological viewpoint, there is a serious
problem with the matter-antimatter asymmetry in the universe.
This cannot be explained solely by the $CP$ violation 
that occurs in the
Standard Model, which originates from quark-flavor
mixing. These reasons lead us to believe that new physics exists,
and is most likely at the TeV energy scale. 

The most direct way to discover the new physics is to
construct energy frontier machines, such as the Large 
Hadron Collider (LHC) or the Global Linear Collider (GLC),
to realize TeV energy scale collisions in which new heavy
particles may be produced.
The history of particle physics implies, however, that this is
not the only way.
In fact, before its discovery, the existence of the charm quark
was postulated to explain the smallness of
strangeness-changing neutral currents
(the Glashow-Illiopolous-Maiani (GIM) mechanism~\cite{Glashow:gm}).
The third family of quarks and leptons was predicted by
Kobayashi and Maskawa to explain the small $CP$ violation 
seen in
kaon mixing~\cite{Kobayashi:fv}.
These are examples of Flavor Changing Neutral Current (FCNC)
processes, with which one can investigate the effect of
heavier particles appearing only in quantum loop
corrections.

The mechanism to suppress FCNC processes should also be
present in new physics models if the new physics lies at
the TeV energy scale, because otherwise such FCNC processes
would violate current experimental limits. 
Information obtained from flavor physics
experiments are thus essential to uncover the details of the
physics beyond the Standard Model, even after energy
frontier machines discover new particles.

A natural place to investigate a wide range of FCNC processes is
in $B$ meson decays.
This is because the bottom quark belongs to the third
generation and, hence, its decays involve all existing generations
of quarks.
In addition to $B^0-\bar{B}^0$ mixing, which is an
analog of the traditional $K^0-\bar{K}^0$ mixing, there are
many FCNC decay processes induced by so-called penguin
diagrams, 
such as the radiative decay $b\rightarrow s\gamma$, 
the semileptonic decay $b\rightarrow s\ell^+\ell^-$, and 
the hadronic decays $b\rightarrow dq\bar{q}$ and 
$b\rightarrow sq\bar{q}$.
All of these processes are suppressed in the Standard Model
by the GIM mechanism, and, therefore, the effect of new
physics may be relatively enhanced.
The Higher Luminosity B Factory is a machine designed to
explore such interesting $B$ decay processes.

\vspace*{3ex}

In the summer of 2001 the presence of $CP$ violation in the $B$
meson system was established by the Belle collaboration~\cite{Abashian:2001pa,Abe:2001xe,Abe:2002wn,Abe:2002px}
(and simultaneously by the BaBar collaboration~\cite{Aubert:2001sp,Aubert:2002rg,Aubert:2001nu,Aubert:2002ic})
through the measurement of the time dependent asymmetry in the
decay process $B^0(\bar{B}^0)\rightarrow J/\psi \ks$.
This measurement was the main target of the present
asymmetric $e^+e^-$ $B$ Factories, and it was achieved as
originally planned.
The experimental data indicated that the Kobayashi-Maskawa
mechanism, which is now a part of the Standard Model of
elementary particles, is indeed the dominant source of
observed $CP$ violation in Nature.

The Belle experiment also proved its ability to measure 
a number of decay modes of the $B$ meson and to extract
Cabibbo-Kobayashi-Maskawa (CKM) matrix elements and
other interesting observables. 
For instance, the precision of the measurement of the angle
$\phi_1$ of the unitarity triangle through the 
$B^0\rightarrow J/\psi \ks$ time-dependent asymmetry reached
the 10\% level~\cite{Abe:2003yu}; 
a $CP$ asymmetry was also observed in 
$B^0\rightarrow\pi^+\pi^-$ decay, from which one can
extract the angle $\phi_2$~\cite{Abe:2002qq,Abe:2003ja,Abe:2004us}; 
the angle $\phi_3$ could also be measured through the
processes $B\rightarrow DK$ and $D\pi$~\cite{Swain:2003yu,Abe:2003cn,Abe:2003gn};
the semi-leptonic FCNC processes 
$B\rightarrow K\ell^+\ell^-$~\cite{Abe:2001dh},
$B\rightarrow K^*\ell^+\ell^-$~\cite{Ishikawa:2003cp}, and even the corresponding inclusive decay  
$B\rightarrow X_s\ell^+\ell^-$~\cite{Kaneko:2002mr} were
observed. 
Furthermore, the recently observed disagreement between 
the values of the angle $\phi_1$ measured in the penguin process 
$B\rightarrow\phi \ks$ and the precisely measured value
in $B\rightarrow J/\psi \ks$ suggests the existence of a new
$CP$ phase in the penguin process $b\rightarrow sq\bar{q}$~\cite{Abe:2003yt}. 
By collecting many such observations we may probe 
new physics, and, once its existence is established, these
measurements will determine the properties of the new physics.
This is only possible if KEKB's
luminosity is upgraded by a substantial amount.
As we discuss in the following sections, a factor of 50
improvement will greatly enhance the possibility to discover
new physics.

\vspace*{3ex}

In the program of quark flavor physics, one way to explore
physics beyond the Standard Model is to 
improve substantially
the measurement precision of the CKM matrix elements. 
This can be done in many different ways, and
any inconsistency with the Standard Model predictions 
would imply new physics.
In this report we discuss the precision we expect to
achieve at the higher luminosity B factory for various
determinations of the CKM matrix elements.
These consist of measurements of the three angles and three
sides of the Unitarity Triangle. 

Another way to search for the effect of new physics is 
to look at loop-induced rare processes for which the
Standard Model contribution is extremely suppressed. 
Such processes may provide an immediate signature of new  
physics that also contributes through loops but in a
different manner.
We describe several such future measurements including the
mixing-induced $b\rightarrow s\gamma$ asymmetry,
$b\rightarrow s\ell^+\ell^-$ forward-backward asymmetry,
and flavor changing tau decay $\tau\rightarrow\mu\gamma$.

\vspace*{3ex}

$B$ physics programs are also being pursued at hadron machines, 
including the ongoing Tevatron experiments~\cite{Anikeev:2001rk} 
and the $B$ physics programs at the 
Large Hadron Collider (LHC)~\cite{Ball:2000ba}, which is
scheduled to start operation in 2007.
Because of the very large $B\bar{B}$ production cross
section in the hadron environment, some of the quantities we
are planning to measure at the $e^+e^-$ B factory may be
measured with better precision at the hadron colliders.
The study of $B_s$ mesons is probably 
unique to the hadron machines. 
However, an $e^+e^-$ machine provides a much cleaner
environment, which is essential for important
observables that involve $\gamma$'s, $\pi^0$'s, $\kl$'s or
neutrinos in the final states. 
On the $\Upsilon(4S)$ resonance, the $B\bar{B}$ pair is
produced near the threshold and there are no associated particles.
This means that one can reconstruct the full energy-momentum
vector of a $B$ ($\bar{B}$) meson from its daughter
particles (the full reconstruction technique), from this 
one can infer the missing momentum in the decay of the other
$\bar{B}$ ($B$) meson. 
This technique is essential for the measurement of channels
including neutrino(s) in the final state.
The measurement of the CKM element $|V_{ub}|$ through the
semi-leptonic decay $b\rightarrow ul\bar{\nu}$, the search
for a charged Higgs effect in $B\rightarrow D\tau\bar{\nu}$,
and measurements of $B\rightarrow K\nu\bar{\nu}$,
$B\rightarrow\tau\nu$ fall in this class.

\vspace*{3ex}

\section{Belle Status and Prospects}
\label{sec:Belle_status_and_prospect}

By the 2003 summer shutdown,
Belle had accumulated data with an integrated luminosity of
140~fb$^{-1}$ at the $\Upsilon(4S)$ resonance, 
corresponding to 152 million $B\overline{B}$ pairs.
With modest improvements expected for the current KEKB accelerator,
we expect an integrated luminosity of $\sim$ 500~fb$^{-1}$ in several
years.  

In this section, the current physics results are briefly reviewed for
selected topics in order to give an overview of the present status.
The status and future prospects for further topics are presented in the
later sections.

\subsection{Status of $CP$-Violation in $b\to c \overline{c} s$ Processes}

Decays of $B^0$ to the following $b\to c\overline{c} s$ ${CP}$-eigenstates
are reconstructed:
$J/\psi \ks$, $\psi(2S)\ks$, $\chi_{c1}\ks$, $\eta_c \ks$ for
 $\xi_f=-1$  and $J/\psi \kl$ for $\xi_f=+1$
\footnote{
  The inclusion of the charge conjugate decay mode is implied
  unless otherwise stated.
}, where $\xi_f$ denotes $CP$ parity.
The two classes ($\xi_f=\pm 1$)
should have $CP$-asymmetries that are opposite in sign.
$B^0\to J/\psi K^{*0} [\to \ks\pi^0]$ decays are also used,
where the final state is a mixture of even and odd $CP$.
The $CP$ content can, however, be determined from an
angular analysis of other $J\psi K^*$ decays. The $CP$-odd
fraction is found to be small (i.e. ($19\pm 4$)\%). 

\begin{figure}[htbp]
\center
\includegraphics[width=0.45\textwidth,clip]{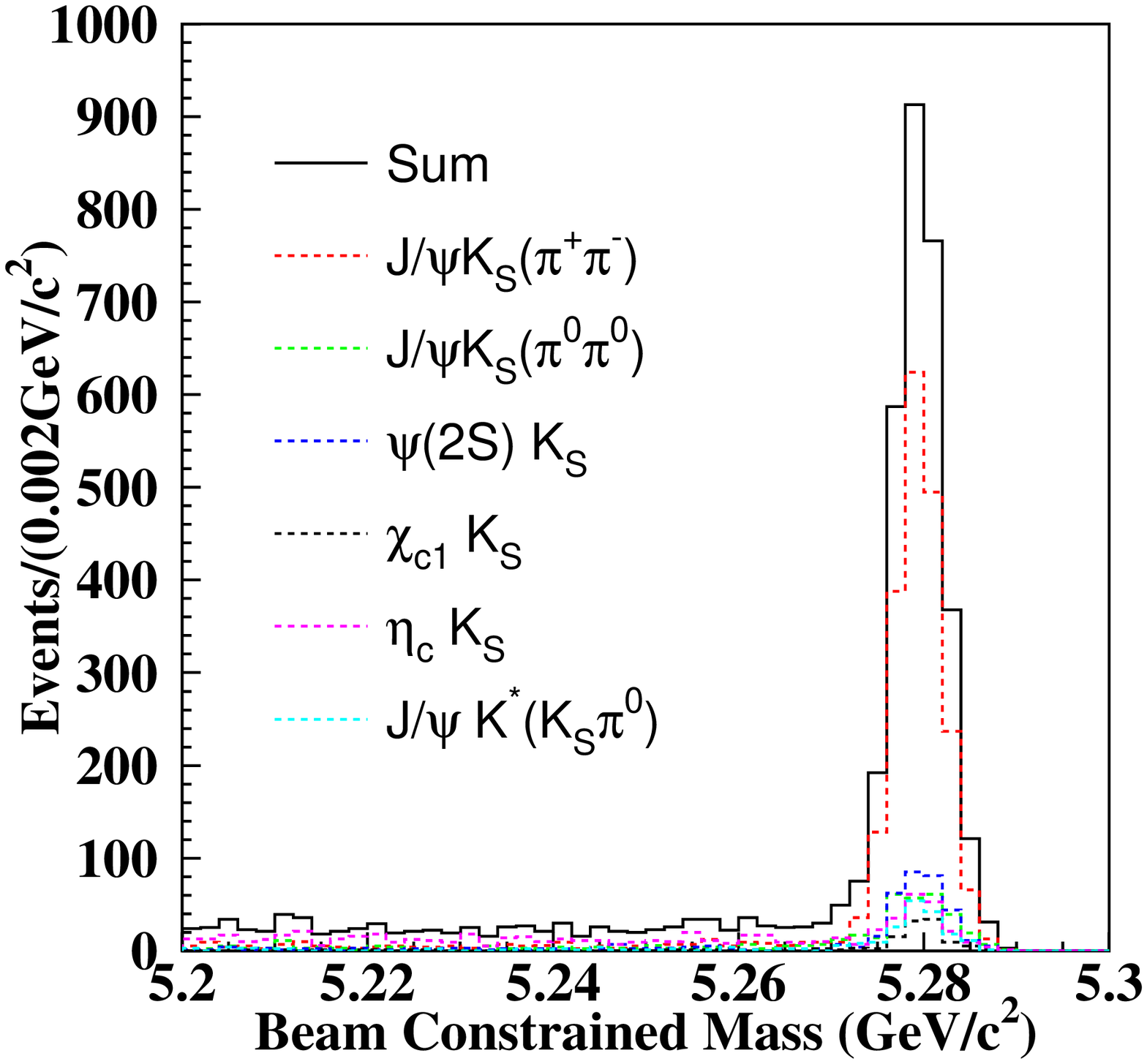}
\includegraphics[width=0.45\textwidth,clip]{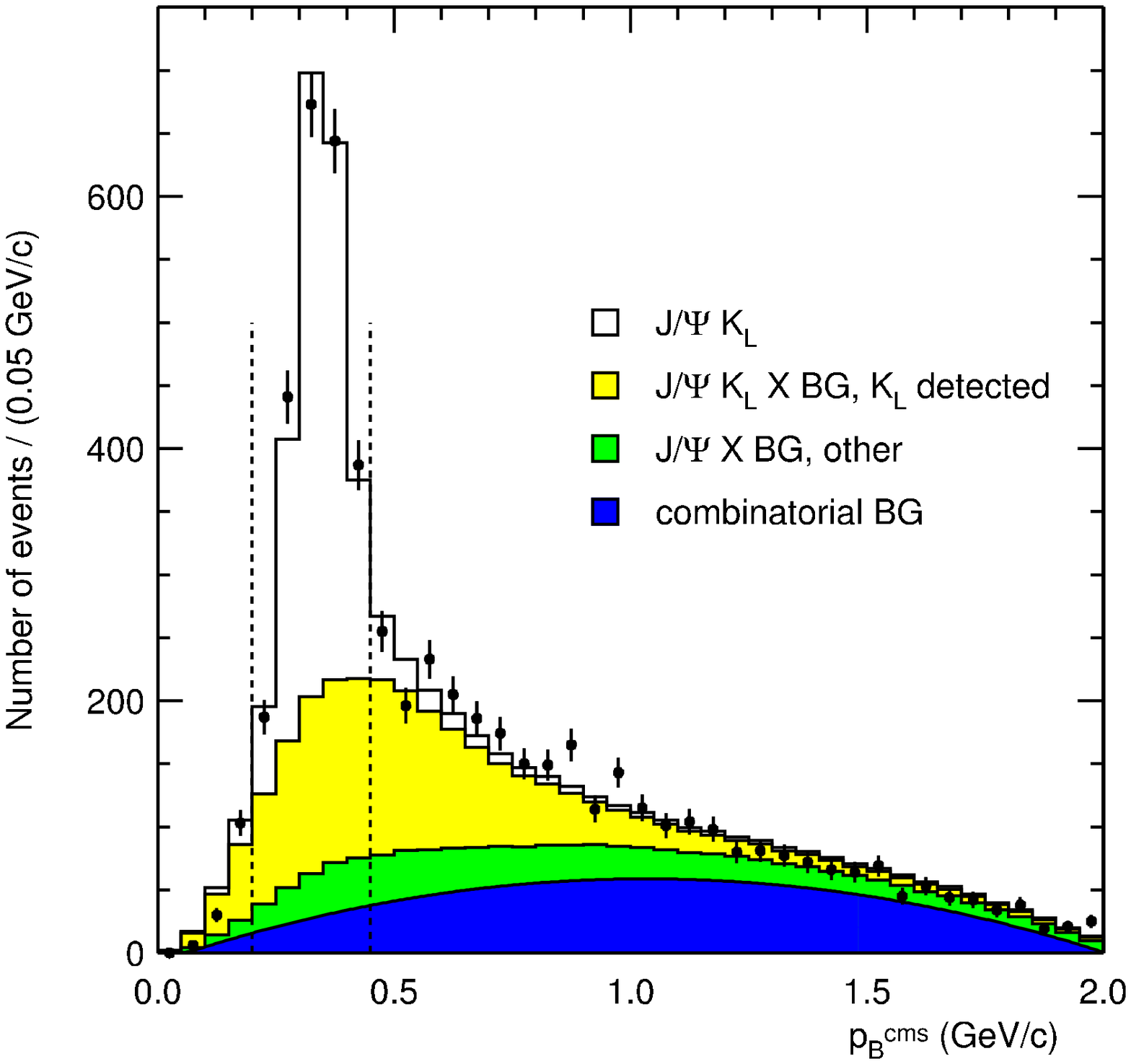}
\caption{ (Left) The fully reconstructed
$CP$-eigenstate sample. 
(Right) The $p_B^*$ ($B$ momentum in the CM frame) distribution for the
$B\to J/\psi \kl$ sample (right).  
The shaded portions show the contributions
of different background components. The vertical dashed lines
indicate the signal region.}
\label{fig:ccs2003}
\end{figure}

The reconstructed samples with 140~fb$^{-1}$ used for the $\sin2\phi_1$ 
measurement~\cite{Abe:2003yu} are shown in Figure~\ref{fig:ccs2003}.
Table~\ref{tab:number} lists the numbers of candidates, $\Nev$,
and the estimated signal purity for each $\fcp$ mode.
It is clear that the $CP$-eigenstate event samples used for the $CP$-violation
measurements in $b\to c \overline{c} s$ are large and clean.

\begin{table}
\center
  \caption{\label{tab:number} 
    The yields for 
    reconstructed $B \to \fcp$
    candidates after flavor tagging and vertex reconstruction, $\Nev$,
    and the estimated signal purity, $p$, in the signal region for 
    each $\fcp$ mode.
    $\jpsi$ mesons are reconstructed in $\jpsi \to \mu^+\mu^-$ or $e^+e^-$
    decays. Candidate $\ks$ mesons are reconstructed in $\ks \to \pi^+\pi^-$
    decays unless explicitly stated otherwise.}
    \begin{tabular}{llrl}
\hline
      \multicolumn{1}{c}{Mode} & $\xi_f$ & $\Nev$ & \multicolumn{1}{c}{$p$} \\
      \hline 
      $J/\psi \ks $                & $-1$ & 1997 & $0.976\pm 0.001$ \\
      $J/\psi \ks(\piz\piz)$       & $-1$ &  288 & $0.82~\pm 0.02$ \\
      $\psi(2S)(\ell^+\ell^-)\ks$  & $-1$ &  145 & $0.93~\pm 0.01$ \\
      $\psi(2S)(\jpsi\pip\pim)\ks$ & $-1$ &  163 & $0.88~\pm 0.01$ \\
      $\chi_{c1}(\jpsi\gamma)\ks$  & $-1$ &  101 & $0.92~\pm 0.01$ \\
      $\eta_c(\ks\km\pip)\ks$      & $-1$ &  123 & $0.72~\pm 0.03$ \\
      $\eta_c(\kp\km\piz)\ks$      & $-1$ &   74 & $0.70~\pm 0.04$ \\
      $\eta_c(p\overline{p})\ks$   & $-1$ &   20 & $0.91~\pm 0.02$ \\
      \cline{3-4}
      All with $\xi_f = -1$        & $-1$ & 2911 & $0.933\pm 0.002$ \\
      \hline
      $J/\psi\kstarz(\ks\piz)$ & +1(81\%)
                                          &  174 & $0.93~\pm 0.01$ \\
      \hline
      $J/\psi\kl$                  & $+1$ & 2332 & $0.60~\pm 0.03$ \\
\hline
    \end{tabular}
\end{table}
%

\begin{figure}[htbp]
\center
\includegraphics[width=0.6\textwidth,clip]{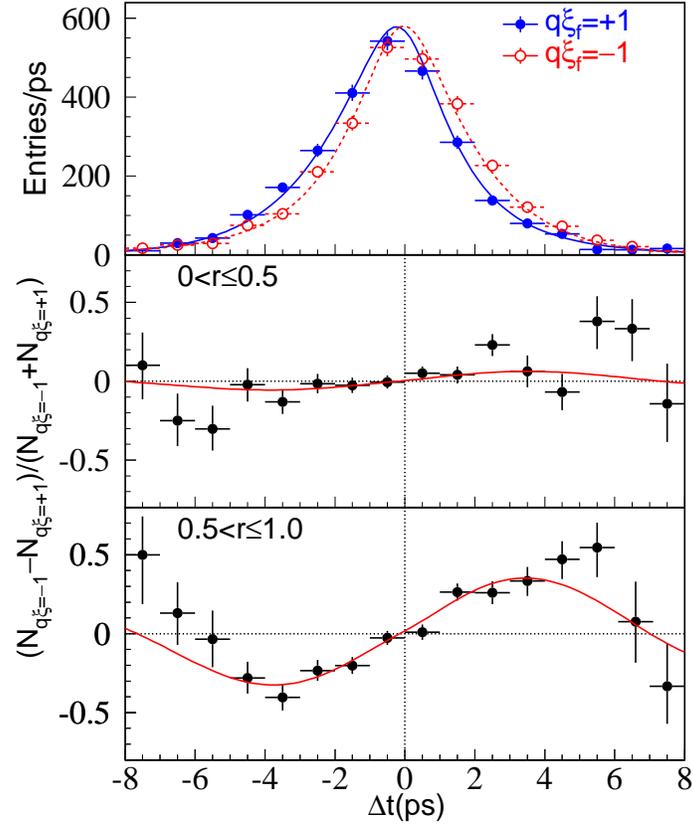}
\caption{ (a) $\Delta t$ distributions for $B^0$
and $\overline{B}^0$ tags (b) raw asymmetry
for low-quality tags and (c) raw asymmetry for high-quality tags.
The smooth curves are projections of the unbinned likelihood fit.}
\label{fig:asym_ccs}
\end{figure}

Figure~\ref{fig:asym_ccs} shows
the $\Delta t$ distributions, where a clear shift between
$B^0$ and $\overline{B}^0$ tags is visible,
and the raw asymmetry plots
for two ranges of the flavor tagging quality variable $r$. For low quality
tags ($0<r<0.5$), which have a large background dilution,
 only a modest asymmetry is visible; in
the high quality tag sub-sample 
($0.5<r<1.0$), a very clear asymmetry with
a sine-like time modulation is present. The final results are extracted
from an unbinned maximum-likelihood fit to the $\Delta t$
distributions that takes into account resolution, mistagging and
background dilution.
The result is 
\begin{equation}
  \sin 2 \phi_1 = 0.733 \pm 0.057 \pm 0.028.
\end{equation} 

This result may be compared to the BaBar result with 82~fb$^{-1}$
of
$\sin 2\phi_1 = 0.741 \pm 0.067 \pm 0.03$ \cite{Aubert:2002ic}.
Both experiments are now in very good agreement; the
average of these results is
\begin{equation}
 \sin 2 \phi_1 = 0.736 \pm 0.049.
\end{equation} 
This average can be interpreted as a constraint
on the CKM angle $\phi_1$. This constraint can be compared 
with the indirect determinations of the unitarity 
triangle \cite{Hocker:2001xe}
and is consistent with the hypothesis that the Kobayashi-Maskawa
phase is the source of $CP$ violation.
The measurement of $\sin 2\phi_1$ in $b\to c\overline{c} s$ modes,
although still statistically limited, is becoming a precision
measurement. 
The systematics are small and well understood.

%

The presence of an asymmetry with a cosine dependence
($|\lambda| \neq 1$) would indicate direct $CP$ violation.
In order to test for this possibility in $b\to c\overline{c} s$ modes,
Belle also performed a fit with
$a_{CP} \equiv -\xi_f {\rm Im}\lambda/|\lambda|$
and $|\lambda|$ as free parameters, keeping everything
thing else the same. They obtain
\begin{eqnarray}
  |\lambda| = 1.007\pm 0.041({\rm stat}) ~~{\rm and}~~
  a_{CP} = 0.733\pm 0.057{\rm (stat)},
\end{eqnarray} 
for all the $b\to c\overline{c} s$ $CP$ modes combined.
This result is consistent with the assumption used in their
primary analysis.

\subsection{Status of $CP$-Violation in $b\to s q \overline{q}$ Penguin Processes}
One of the promising ways to probe additional $CP$-violating
phases from new physics beyond the Standard Model is to
measure the time-dependent $CP$-asymmetry in
penguin-dominated modes such as 
$B^0\to \phi K_S^0$, $B^0\to \eta^{'} K_S^0$,
where heavy new particles may contribute inside the loop, 
and compare it with the asymmetry in $B^0\to J/\psi K_S^0$
and related $b\to c\overline{c} s$ charmonium modes. 
Belle has measured $CP$-violation in 
$B^0\to \phi K_S^0$, $\eta^{'} K_S^0$,
and $K^+K^- K_S^0$ with 140~fb$^{-1}$~\cite{Abe:2003yt}. 

The decay $B^0\to \phi K_S^0$, 
which is dominated by the $b \to s\overline{s} s$
transition, is an especially unambiguous and
sensitive probe of new $CP$-violating phases from 
physics beyond the Standard Model~\cite{Grossman:1996ke}.
The Standard Model predicts that measurements of
$CP$-violation in this mode should yield $\sin 2\phi_1$ to
a very good approximation
\cite{London:1997zk,Grossman:1997gr,Grossman:2003qp}.
A significant deviation in the time-dependent $CP$-asymmetry in
this mode from what is observed
in $b \to c\overline{c}s$ decays would be evidence for a new
$CP$-violating phase.

\begin{figure}[htbp]
\center
\resizebox{0.6\textwidth}{!}{\includegraphics{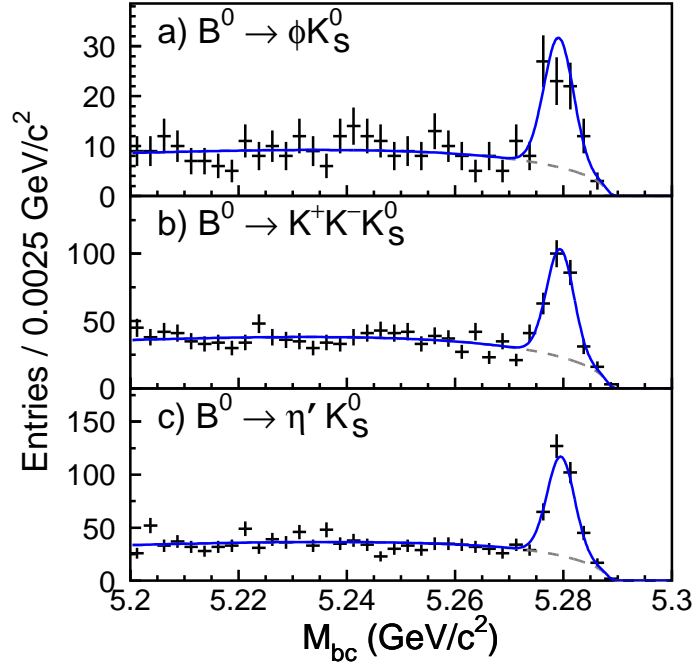}}
\caption{The beam-energy constrained mass distributions for
(a) $\bz\to\phi\ks$, (b) $\bz\to K^+K^-\ks$, and (c) $\bz\to\eta'\ks$
within the $\dE$ signal region.
Solid curves show the fit to signal plus background distributions,
and dashed curves show the background contributions.
The background for $\bz\to\eta'\ks$ decay includes
an MC-estimated $B\overline{B}$ background component.
}\label{fig:sqq_mbc}
\end{figure}
\begin{figure}[htbp]
\resizebox{!}{0.32\textwidth}{\includegraphics{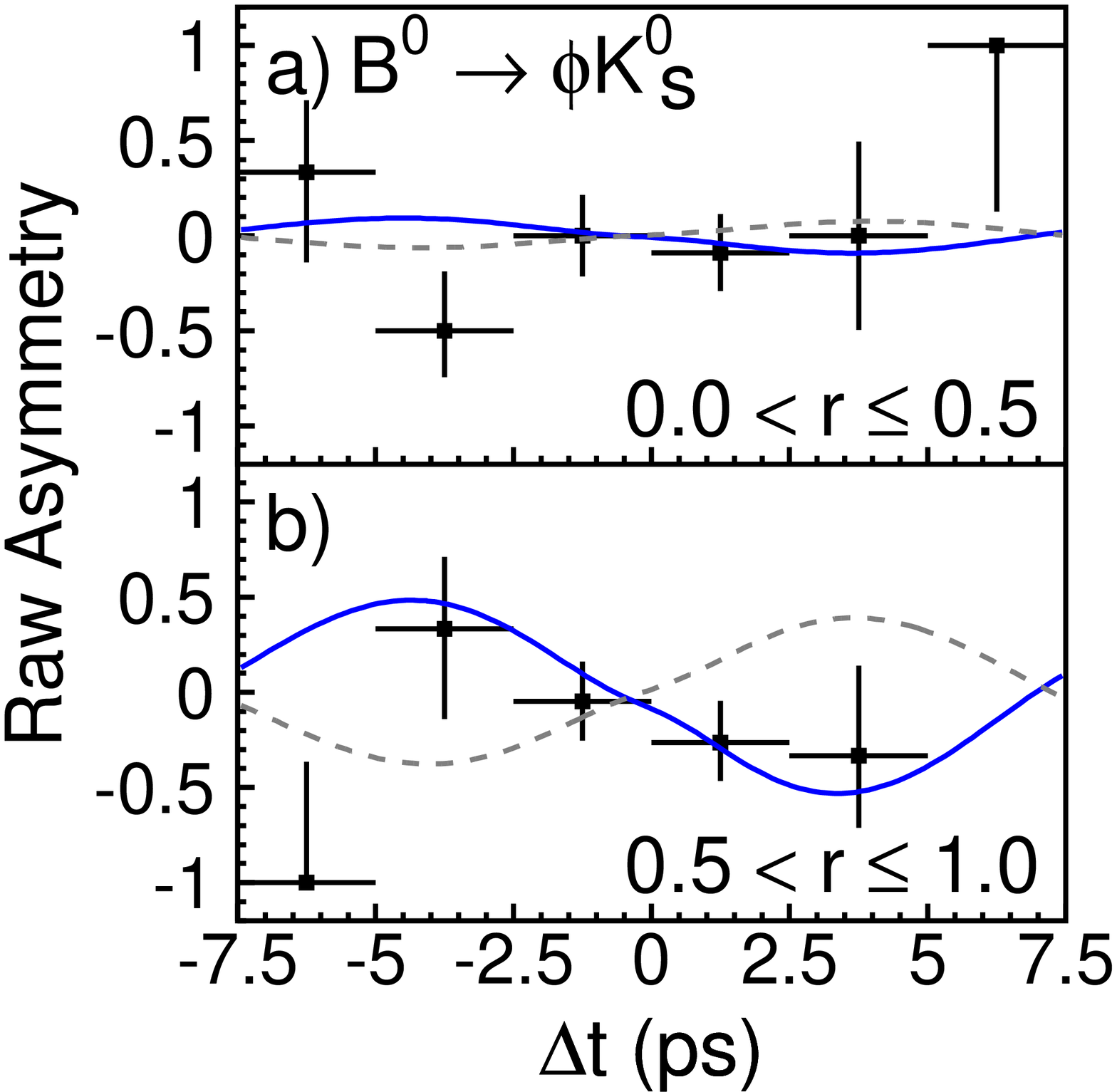}} 
\resizebox{!}{0.32\textwidth}{\includegraphics{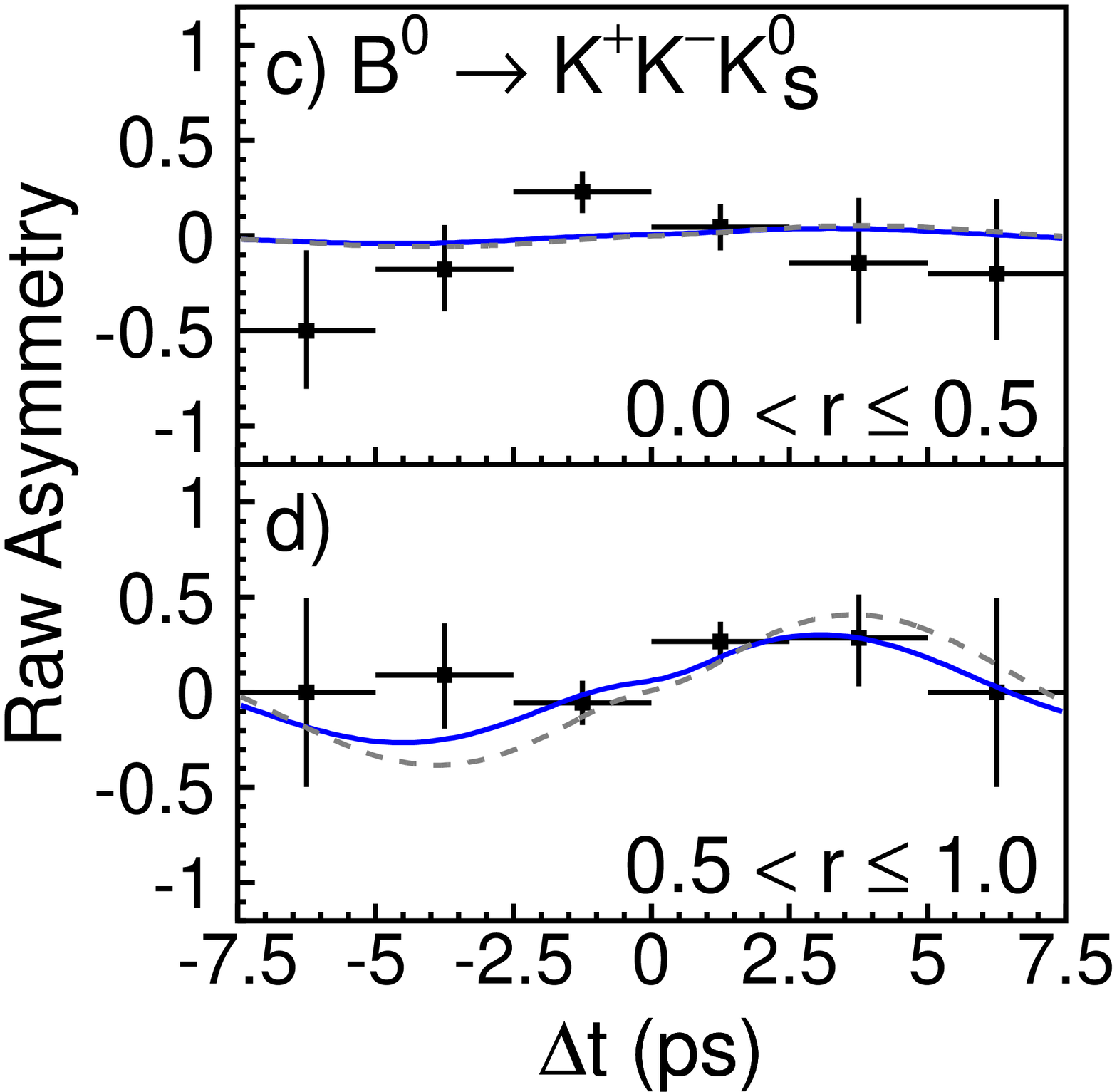}} 
\resizebox{!}{0.32\textwidth}{\includegraphics{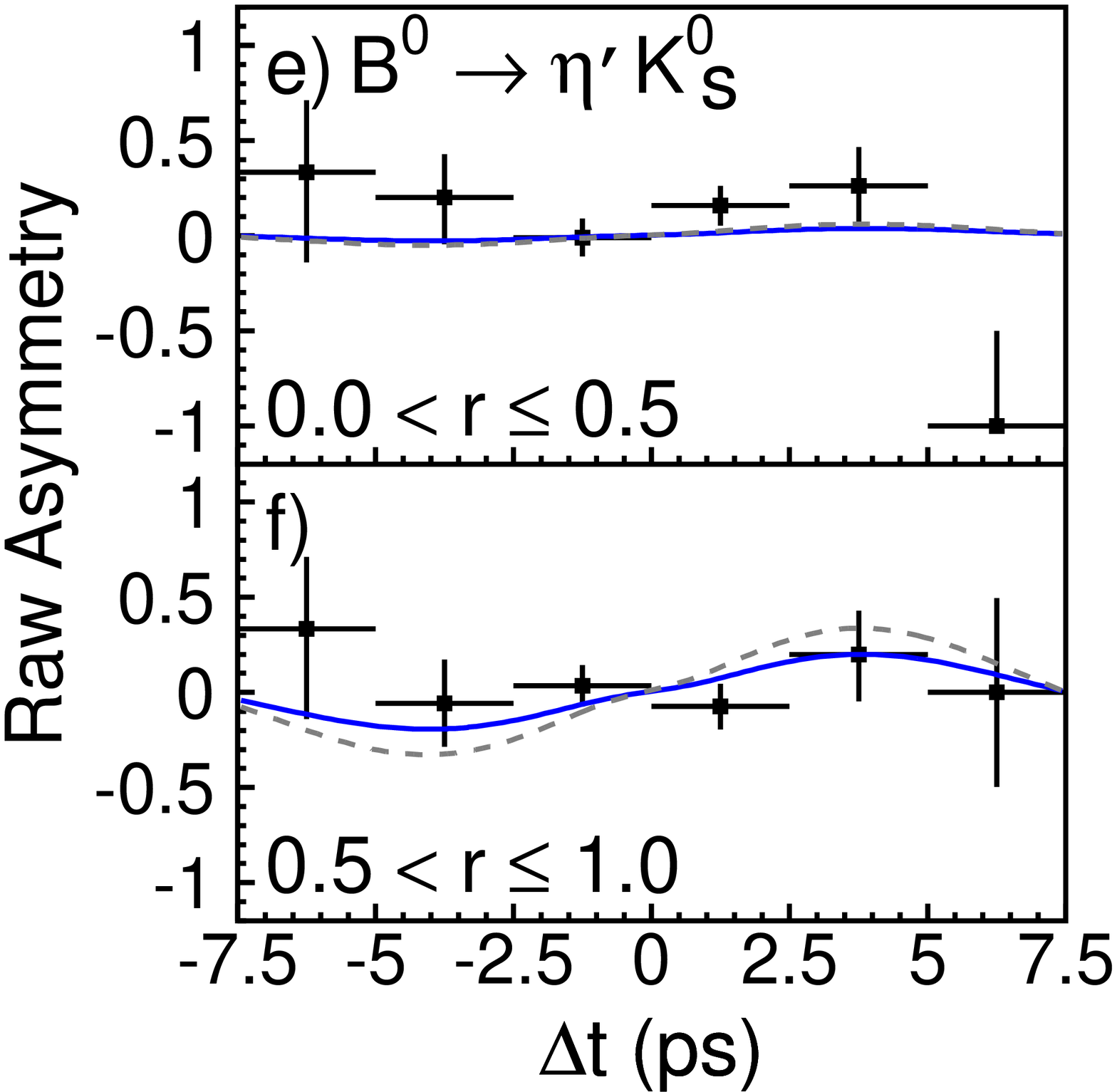}} 
\caption{
(a) The asymmetry, $A$, in each $\Dt$ bin for
$\bz\to\phi\ks$ with $0 < r \le 0.5$, 
(b) with $0.5 < r \le 1.0$,
(c) for $\bz\to K^+K^-\ks$ with $0 < r \le 0.5$, (d) with $0.5 < r \le 1.0$,
(e) for $\bz\to \eta'\ks$ with $0 < r \le 0.5$, and (f) with $0.5 < r \le 1.0$,
respectively. The solid curves show the result of the 
unbinned maximum-likelihood fit.
The dashed curves show the Standard Model expectation with $\sin2\phi_1$ = +0.731 
and $\cala$ = 0.}
\label{fig:sqq_asym}
\end{figure}

The $B\to\phi K_S^0$ sample is shown in
Figure~\ref{fig:sqq_mbc}(a).  
The signal contains $68\pm 11$ events.
Figure~\ref{fig:sqq_asym}(a,b) shows the raw asymmetries 
in two regions of the flavor-tagging parameter $r$. 

The observed $CP$-asymmetry for $B^0 \to \phi K_S^0$
in the region $0.5 < r \le 1.0$ (Figure~\ref{fig:sqq_asym} (b))
indicates a deviation from the Standard Model expectation (dashed curve). 
Note that these projections onto the $\Delta t$ axis do not take into
account event-by-event information (such as the signal fraction, the
wrong tag fraction and the vertex resolution) that is used in the
unbinned maximum likelihood fit.
The contamination of $K^+ K^- K_S^0$ events in the $\phi K_S^0$ sample 
($7.2\pm1.7$\%) is small;
backgrounds from the decay $B^0 \rightarrow f_0(980) K_S^0$, 
which has the opposite 
$CP$-eigenvalue to $\phi K_S^0$, are found to be small
($1.6^{+1.9}_{-1.5}$\%).  The influence of these backgrounds
is treated as a source of systematic uncertainty.

The likelihood fit gives 
\begin{equation}
  \sin 2 \phi_{1{\rm eff}}(B\to \phi K_S^0) = -0.96 \pm 0.5^{+0.09}_{-0.11}.
\end{equation} 
The likelihood function is parabolic and well-behaved.
An evaluation of the significance of the result using the
Feldman-Cousins method and allowing for systematic uncertainties
shows that this result deviates by 3.5$\sigma$ from the Standard
Model expectation \cite{Abe:2003yt}.
%
%
%
BaBar has analyzed a sample with an integrated luminosity of 110~fb$^{-1}$,
containing $70\pm 9$ events, and obtain
$\sin 2 \phi_{1{\rm eff}}(B\to \phi K_S^0) = 0.45 \pm 0.43 \pm 0.07$
\cite{bib:BaBar_sss}. This result differs from the Belle result by
2.1$\sigma$ and the naive average, $-0.14 \pm 0.33$, is still away from
the $\sin2\phi_1$ world average by 2.6$\sigma$.

The decay mode $B\to K^+ K^- K_S^0$, where $K^+ K^-$ combinations
consistent with the $\phi$ have been removed, is found to
be dominantly $CP$-even \cite{Garmash:2003er} and, thus, can
be treated as a $CP$-eigenstate and used for studies of
time-dependent $CP$-violation in $b\to s q \overline{q}$
processes. 
The beam constrained mass distribution for the 
$B\to K^+ K^-K_S^0$ sample used by Belle is shown in 
Figure~\ref{fig:sqq_mbc}(b), where
there are $199\pm 18$ signal events.
The result is
\begin{equation}
  \sin 2 \phi_{1{\rm eff}}(B\to K^+ K^-K_S^0) = 
  0.51\pm 0.26\pm 0.05^{+0.18}_{-0.00},
\end{equation} 
where the third error is due to the uncertainty in the $CP$ content
of this final state \cite{Garmash:2003er}.
The results for $B\to K^+ K^- K_S^0$ are also consistent with
$b\to c\overline{c} s$ decays. However, in this decay there is also
the possibility of ``tree-pollution'', i.e. the contribution of 
a $b\to u \overline{u} s$ tree amplitude that may complicate the
interpretation of the results \cite{Grossman:2003qp}.

The mode $B\to \eta^{\prime} K_S^0$ is expected to include
contributions from $b\to s \overline{u} u$ and $b\to s\overline{d} d$ 
penguin processes. 
The beam constrained mass distribution for the $B\to \eta^{\prime} 
K_S^0$ sample shown in Figure~\ref{fig:sqq_mbc}(c)
contains $244\pm 21$ signal events \cite{Chen:2002af}.
The fit gives (Figure~\ref{fig:sqq_asym}(e,f))
\begin{equation}
  \sin 2 \phi_{1{\rm eff}}(B\to \eta^{\prime} K_S^0) = 0.43\pm 0.27\pm 0.05.
\end{equation} 
The average with the BaBar result 
($0.02\pm 0.34\pm 0.03$ with 82~fb$^{-1}$)\cite{Aubert:2003bq}
is about 2.2$\sigma$ from the $b\to c\overline{c} s$
measurement, which is the Standard Model expectation.

\subsection{Radiative $B$ decays}

\subsubsection{Exclusive $\BtoKstargamma$}
\begin{figure}
  \center
  \includegraphics[width=0.6\textwidth,clip]{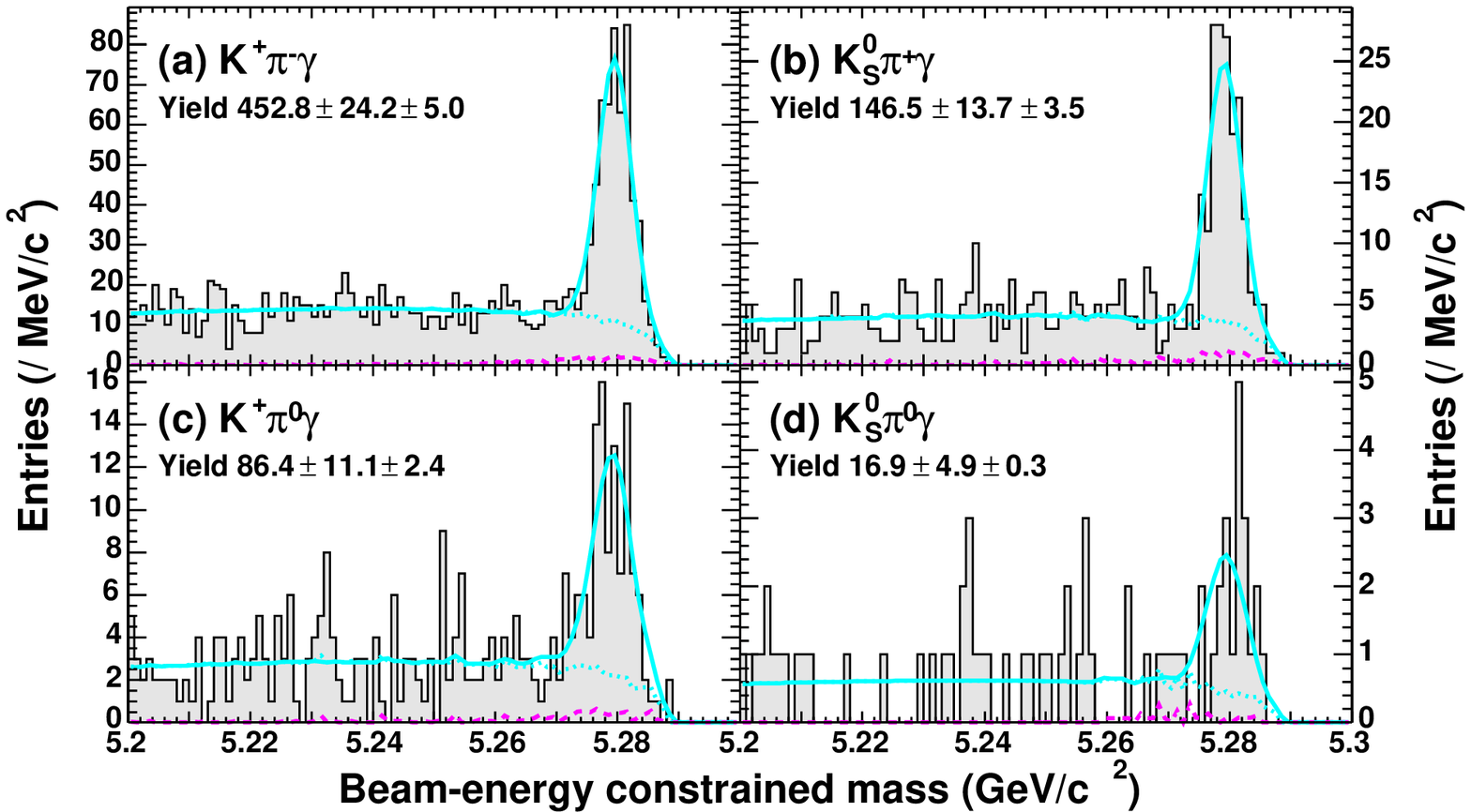}
  \caption{$\BtoKstargamma$ signals from Belle.}
  \label{fig:belle-kstargam}
\end{figure}

Measurement of the $\BtoKstargamma$ exclusive branching fraction is
straightforward, since one can use $\Mbc$, $\DeltaE$ and $K^*$ mass
constraints.  
($K^*$ denotes $K^*(892)$ throughout this section.)  
The latest Belle branching fraction
measurements (Figure~\ref{fig:belle-kstargam}) use
$78\fbinv$ of data, and have a total error that is much less than
10\% for both $B^0$ and $B^+$ decays.  
The results from CLEO~\cite{Coan:1999kh},
BaBar~\cite{Aubert:2001me} and
Belle~\cite{bib:kstargam-belle} are in good agreement and are
listed in Table~\ref{tbl:kstargam}.  
The world averages are
\begin{eqnarray}
  \Br(\BtoKstarZG) &=& (4.17\pm0.23)\times10^{-5},\\
  \Br(\BtoKstarPG) &=&(4.18\pm0.32)\times10^{-5}.
\end{eqnarray}
The corresponding theoretical branching fraction is about 
$(7\pm 2)\times 10^{-5}$, higher than the measurement but with a large
uncertainty~\cite{Ali:2001ez,Bosch:2001gv}.  
A computation based on PQCD has been just completed~\cite{matsumori}:
\begin{eqnarray}
  \Br(\BtoKstarZG) &=& (4.5^{+1.5}_{-1.1})\times10^{-5},\\
  \Br(\BtoKstarPG) &=& (4.3^{+1.6}_{-1.1})\times10^{-5}.
\end{eqnarray}
where the errors quoted come from a parameter in the $B$ meson wave function 
$\omega_B$. The difference in the value of the computed branching ratio 
compared to the previous computation comes from the fact that 
PQCD computation of
the $B\to K^*$ transition form factor is smaller than those of
light-cone QCD sum rule and the lattice QCD simulation.

\begin{table}
\begin{center}
\caption{$B\to K^*\gamma$ branching fractions\label{tbl:kstargam}}
\begin{tabular}{ccc}
\hline
 & $B^0\to K^{*0}\gamma$
 & $B^+\to K^{*+}\gamma$ \\
 & $[\times10^{-5}]$ &  $[\times10^{-5}]$ \\
\hline
CLEO  & $4.55\pm0.70\pm0.34$ & $3.76\pm0.86\pm0.28$ \\
BaBar & $4.23\pm0.40\pm0.22$ & $3.83\pm0.62\pm0.22$ \\
Belle & $4.09\pm0.21\pm0.19$ & $4.40\pm0.33\pm0.24$ \\
\hline
\end{tabular}
\end{center}
\end{table}
A better approach to exploit the $\BtoKstargamma$ branching fraction
measurements is to consider isospin
asymmetry \cite{Kagan:2001zk}.  
A small difference in the branching fractions between
$\BtoKstarZG$ and $\BtoKstarPG$ tells us the sign of the
combination of the Wilson coefficients, $C_6/C_7$.  
Belle has taken into account correlated systematic errors
and obtains 
\begin{equation}
\begin{array}{lll}
\Deltapz&\equiv&{\displaystyle
                {(\tauBratio)\Br(\BtoKstarZG)-\Br(\BtoKstarPG) \over
                 (\tauBratio)\Br(\BtoKstarZG)+\Br(\BtoKstarPG)}}\\[12pt]
&=&(+0.003\pm0.045\pm0.018),
\end{array}
\end{equation}
which is consistent with zero. 
PQCD gives\cite{matsumori}
\begin{equation}
\Deltapz=+0.059^{+0.011}_{-0.012}
\end{equation}
It is extreamly interesting to see if nealy 5
the PQCD computation based on the standard model is varified.

Here, the lifetime ratio $\tauBratio=1.083\pm0.017$ is used,
and the $B^0$ to $B^+$ production ratio is assumed to be
unity.  
The latter is measured to be $f_0/f_+=1.072\pm0.057$ and is
included in the systematic error.

\subsubsection{Other Exclusive Radiative Decays}
The dominant radiative decay channel $\BtoKstargamma$ covers
only 12.5\% of the total $\BtoXsgamma$ branching fraction
(world average $(3.34 \pm 0.38) \times 10^{-4}$).  
The remainder has to be 
accounted for by decays with higher resonances or multi-body decays.
Knowledge of these decay modes will eventually be useful to reduce the
systematic error for the inclusive measurement.  

\begin{figure}
  \center
  \includegraphics[width=0.45\textwidth,clip]{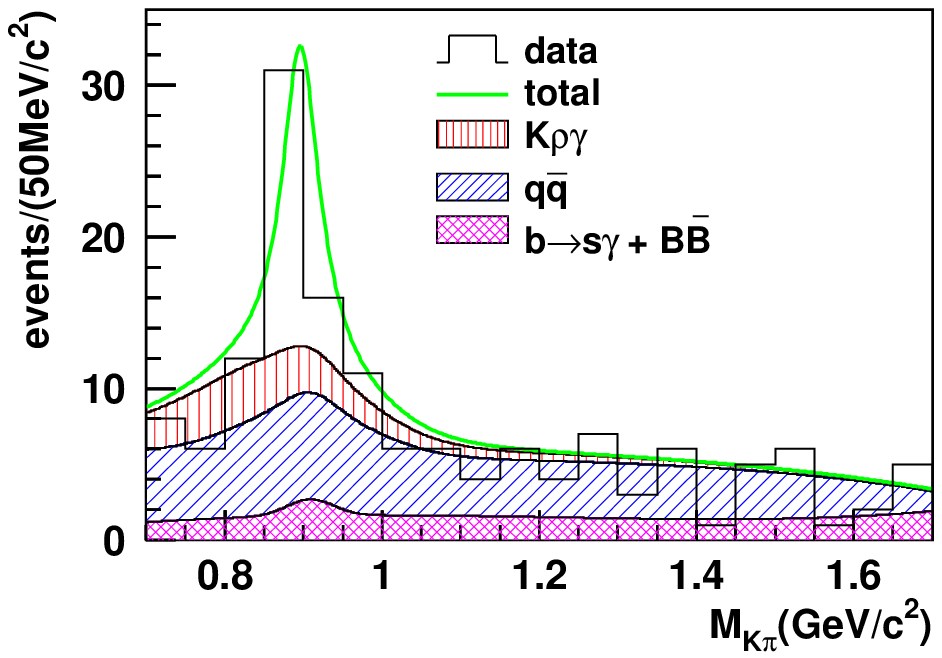}
  \includegraphics[width=0.45\textwidth,clip]{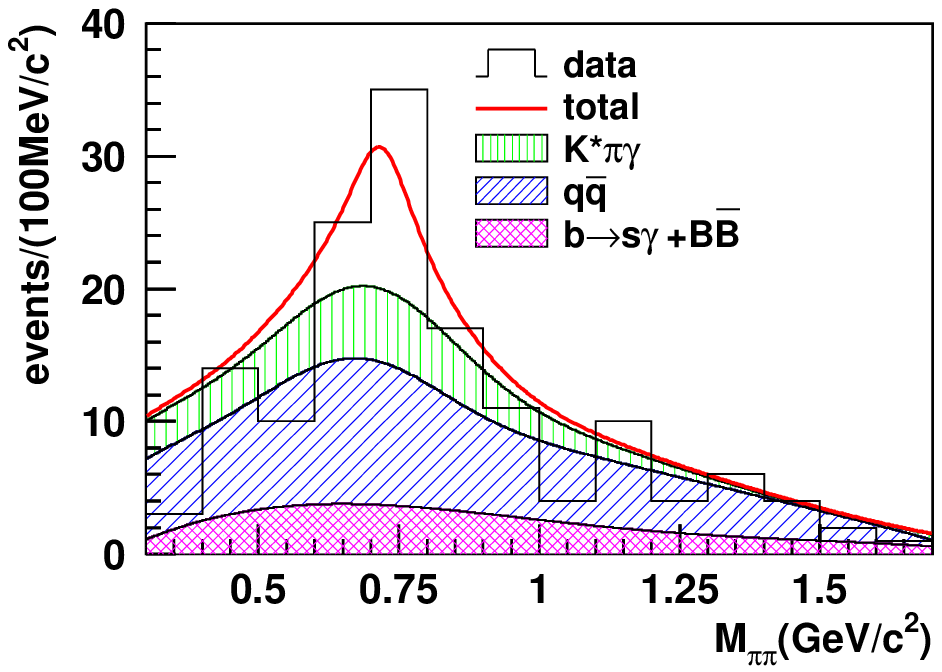}
  \caption{$B^+\to K^{*0}\pi^+\gamma$ and $B\to K^+\rho^0\gamma$ from Belle.}
  \label{fig:belle-kxgam}
\end{figure}

\begin{figure}
  \center
  \includegraphics[width=0.45\textwidth,clip]{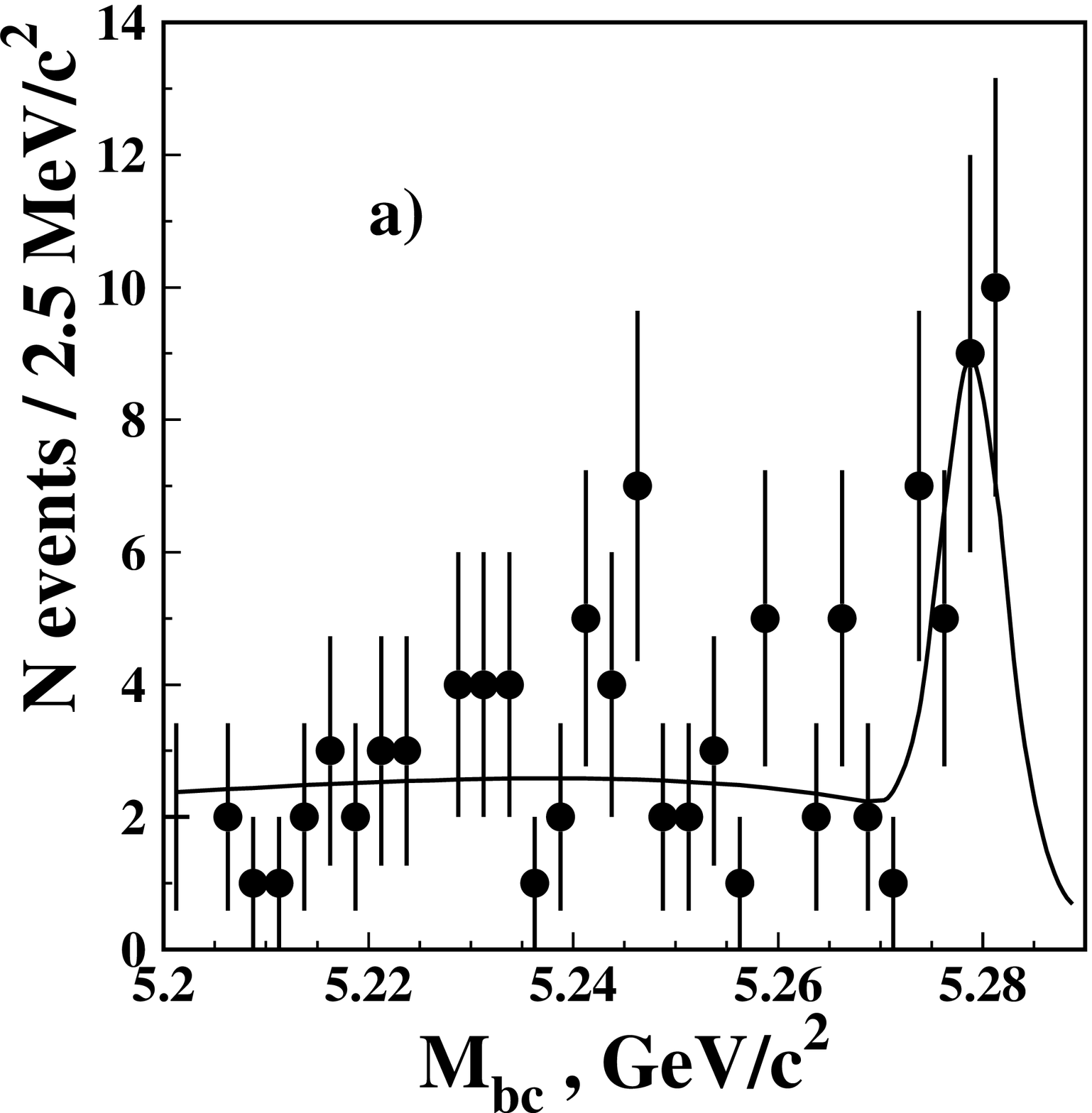}
  \includegraphics[width=0.45\textwidth,clip]{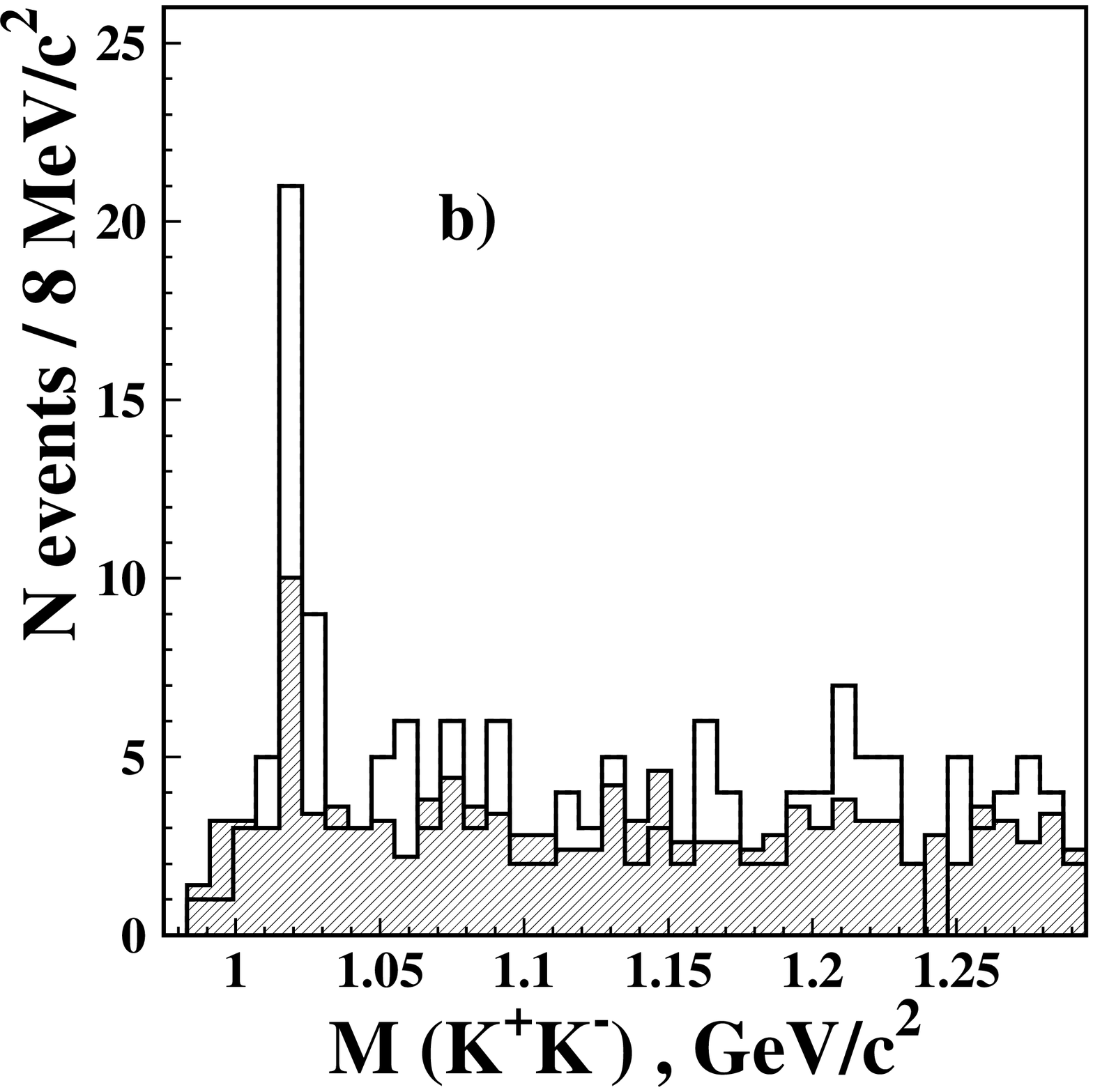}
  \caption{$B^+\to K\phi\gamma$ from Belle.}
  \label{fig:belle-kphigam}
\end{figure}

Belle has extended the analysis into multi-body decay
channels in addition to $B\to K_2^*(1430)[\to K\pi]\gamma$
decay \cite{Nishida:2002me}. 
Using $29\fbinv$ of data, the decay
$B^+\to \KP\piP\piM\gamma$ is measured to have a branching fraction of
$(24\pm5\PM{4}{2})\times10^{-6}$ for $M(K\pi\pi)<2.4\GeV$.  The decay is
dominated by $K^{*0}\piP\gamma$ and $\KP\rho^0\gamma$ final states that
overlap each other as shown in Figure~\ref{fig:belle-kxgam}.  
At this moment, it is not possible to disentangle the
resonant states that decay into $K^*\pi$ or $K\rho$, such as
$K_1(1270)$, $K_1(1400)$, $K^*(1650)$, and so on.
 A clear $B^+\to K^+\phi\gamma$ ($5.5\sigma$)
signal was recently observed by Belle with $90\fbinv$ of data
(Figure~\ref{fig:belle-kphigam}), together with $3.3\sigma$
evidence for $B^0\to\KS\phi\gamma$.  
There is no known $K\phi$ resonant state.  
This is the first example of a $s\sbar s\gamma$ final state.  
The branching fractions for $B\to K\phi\gamma$ are measured
to be \cite{Drutskoy:2003xh} 
\begin{equation}
\begin{array}{rcl}\displaystyle
\Br(B^+\to K^+\phi\gamma)&=&(3.4\pm0.9\pm0.4)\times10^{-6},\\
\Br(B^0\to K^0\phi\gamma)&=&(4.6\pm2.4\pm0.4)\times10^{-6}\\
                         &<&8.3\times10^{-6} \mbox{~~~(90\%~CL)}.
\end{array}
\end{equation}
With more data, one can perform a time-dependent $CP$-asymmetry
measurement with the $\KS\phi\gamma$ decay channel.

At present, $(35\pm6)\%$ of the total $\BtoXsgamma$ rate is measured to be
either $\BtoKstargamma$ (12.5\%), $B\to K_2^*(1430)\gamma$ (4\% after
excluding $K\pi\pi\gamma$), $B\to K^*\pi\gamma$ (9\%), $B\to
K\rho\gamma$ (9\%) or $B\to K\phi\gamma$ (1\%).  The remaining $(65\pm6)\%$
may be accounted for by decays with multi-body final states, baryonic
decays, modes with $\eta$ and $\eta'$, multi-kaon final states other
than $K\phi\gamma$ or in the large $X_s$ mass range.

\subsubsection{Search for Direct $CP$-Asymmetry}
In the Standard Model,
the direct $CP$-asymmetry in $\BtoXsgamma$ is predicted to be 
0.6\% 
with a small error\cite{Soares:1991te,Kagan:1998bh}.
A large $CP$-asymmetry would be a clear sign of new physics. 

\begin{figure}
  \center
  \includegraphics[width=0.45\textwidth,clip]{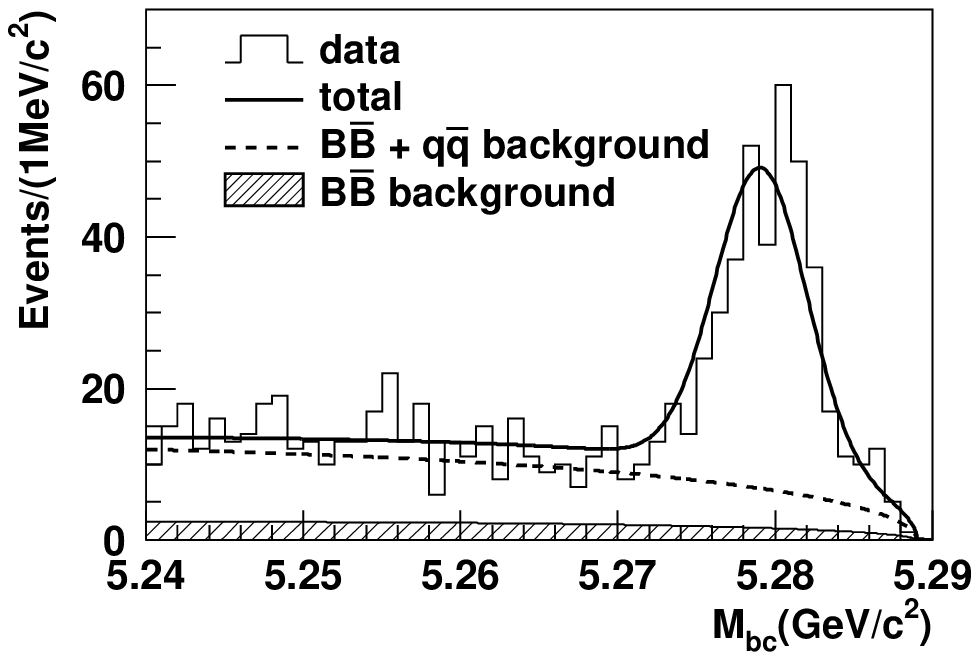}
  \includegraphics[width=0.45\textwidth,clip]{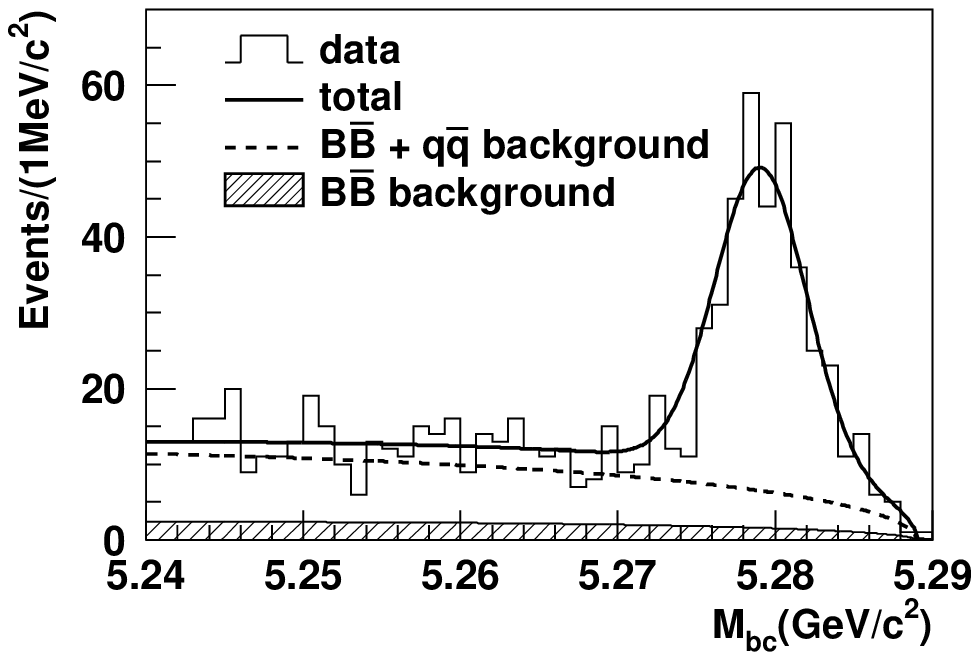}
  \caption{$\Bbar$-tagged (top-left),
           $B$-tagged (top-right) 
           $\BtoXsgamma$ signals from Belle.}
  \label{fig:belle-acpxsgam}
\end{figure}

An $\Acp(\BtoXsgamma)$ measurement was performed by
Belle \cite{Abe:2003yw}, by summing up the exclusive modes
of one kaon plus up to four pions and modes with three kaon
plus up to one pion.  
This result eliminates $\BtoXdgamma$ by exploiting particle
identification devices for the tagged hadronic recoil system.  
$\MXs<2.1\GeV$ is required, which roughly corresponds to
$\Egammamin\sim2.25\GeV$. 
Events are self-tagged as $B$ candidates ($B^0$ or $B^+$) or
$\Bbar$ candidates ($\Bbar^0$ or $B^-$), except for
ambiguous modes with a $\KS$ and zero net charge.  
In order to correct the imperfect knowledge of the hadronic
final state, the signal yields for each exclusive mode are
used to correct the Monte Carlo multiplicity distribution.  
The resulting $\Bbar$-tagged
($342\pm23\PM{7}{14}$ events), $B$-tagged ($349\pm23\PM{7}{14}$ events)
signals are shown in Figure~\ref{fig:belle-acpxsgam}.  
Using the wrong-tag fractions of
$0.019\pm0.014$ between $B$- and $\Bbar$-tagged, $0.240\pm0.192$ from
ambiguous to $B$- or $\Bbar$-tagged, and $0.0075\pm0.0079$ from $B$- or
$\Bbar$-tagged to ambiguous samples, the asymmetry is measured to be
\begin{equation}
\Acp(\BtoXsgamma)=0.004\pm0.051\pm0.038.
\end{equation}
The result corresponds to a 90\% confidence level limit of
$-0.107<\Acp(\BtoXsgamma)<0.099$, and therefore already constrains
part of the new physics parameter space.

For exclusive radiative decays, it is straightforward to extend the
analysis to search for a direct 
$CP$-asymmetry \cite{Coan:1999kh,Aubert:2001me,bib:kstargam-belle}.
Particle identification devices at Belle and BaBar resolve the possible
ambiguity between $K^{*0}\to\KP\piM$ and $\Kbar^{*0}\to\KM\piP$ to an
almost negligible level with a reliable estimation of the wrong-tag
fraction (0.9\% for Belle).  The results of the asymmetry measurements are
listed in Table~\ref{tbl:acpkstargam}, whose average is
\begin{equation}
\Acp(\BtoKstargamma)=(-0.5\pm3.7)\times10^{-2}.
\end{equation}
It is usually thought that large $CP$-violation in $\BtoKstargamma$
is not allowed in the Standard Model and the above
result may be used to constrain new
physics.  However, since the involved strong phase difference may not be
reliably calculated for exclusive decays, the interpretation may be model
dependent.

\begin{table}
\begin{center}
\caption{$B\to K^*\gamma$ direct $CP$-asymmetry}
\label{tbl:acpkstargam}
\begin{tabular}{cc}
\hline
CLEO  $(9.1\fbinv)$  & $(8\pm13\pm3)\times10^{-2}$ \\
BaBar $(20.7\fbinv)$ & $(-4.4\pm7.6\pm1.2)\times10^{-2}$ \\
Belle $(78\fbinv)$   & $(-0.1\pm4.4\pm0.8)\times10^{-2}$ \\
\hline
\end{tabular}
\end{center}
\end{table}

\subsubsection{Search for $\btodgamma$ Final States}
There are various interesting aspects to the $\btodgamma$ transition.
Within the Standard Model, most of the diagrams are the same as those for
$\btosgamma$, except for the replacement of the CKM matrix element
$\Vts$ with $\Vtd$.  A measurement of the $\btodgamma$ process will
therefore provide the ratio $|\Vtd/\Vts|$ without large model-dependent
uncertainties. 
This mode is also one where
a large direct $CP$-asymmetry is predicted 
both within and beyond the Standard Model.

The search for the exclusive decay $B\to\rho\gamma$ is as straightforward as
the measurement of $\BtoKstargamma$, except for its small branching
fraction, the enormous combinatorial background from copious $\rho$
mesons and random pions, and the huge $\BtoKstargamma$ background that
overlaps with the $B\to\rho\gamma$ signal window. 
$B\to\omega\gamma$
is not affected by $\BtoKstargamma$ background, but is still 
unobserved.
The upper limits obtained by BaBar \cite{Aubert:2003me},
Belle \cite{Nakao:2003qt} and CLEO \cite{Coan:1999kh} are
summarized in Table~\ref{tbl:rhogam}.  
The upper limits are still about twice as large as the Standard Model
predictions \cite{Ali:2001ez,Bosch:2001gv} $(9.0\pm3.4)\times10^{-7}$ for
$\rho^+\gamma$, and $(4.9\pm1.8)\times10^{-7}$ for $\rho^0\gamma$ and
$\omega\gamma$.

\begin{table}
\begin{center}
\caption{90\% confidence level upper limits on the $B\to \rho\gamma$ and
$\omega\gamma$ branching fractions.}
\label{tbl:rhogam}
\begin{tabular}{cccc}
\hline
& $\rho^+\gamma$ & $\rho^0\gamma$ & $\omega\gamma$ \\
\hline
CLEO  $(9.1\fbinv)$ & $13\EM6$  & $17\EM6$  & $9.2\EM6$ \\
Belle $(78\fbinv)$  & $2.7\EM6$ & $2.6\EM6$ & $4.4\EM6$ \\
BaBar $(78\fbinv)$  & $2.1\EM6$ & $1.2\EM6$ & $1.0\EM6$ \\ 
\hline
\end{tabular}
\end{center}
\end{table}

From these upper limits the bound $|\Vtd/\Vts|<0.34$ can be obtained, 
which is still weaker than the
corresponding bound derived from $\Delta m_s/\Delta m_d$.

\subsection{Electroweak Rare $B$ Decays}

\subsubsection{Observation of $\BtoKstarll$}
The first signal for $\BtoKll$ was observed by Belle \cite{Abe:2001dh}
using $29\fbinv$ of data and later confirmed by 
BaBar \cite{Aubert:2002pj}; a
$\BtoKstarll$ signal, whose branching fraction is expected to be larger,
was not significant in those data samples.

\begin{figure}
  \center
  \includegraphics[width=0.6\textwidth,clip]{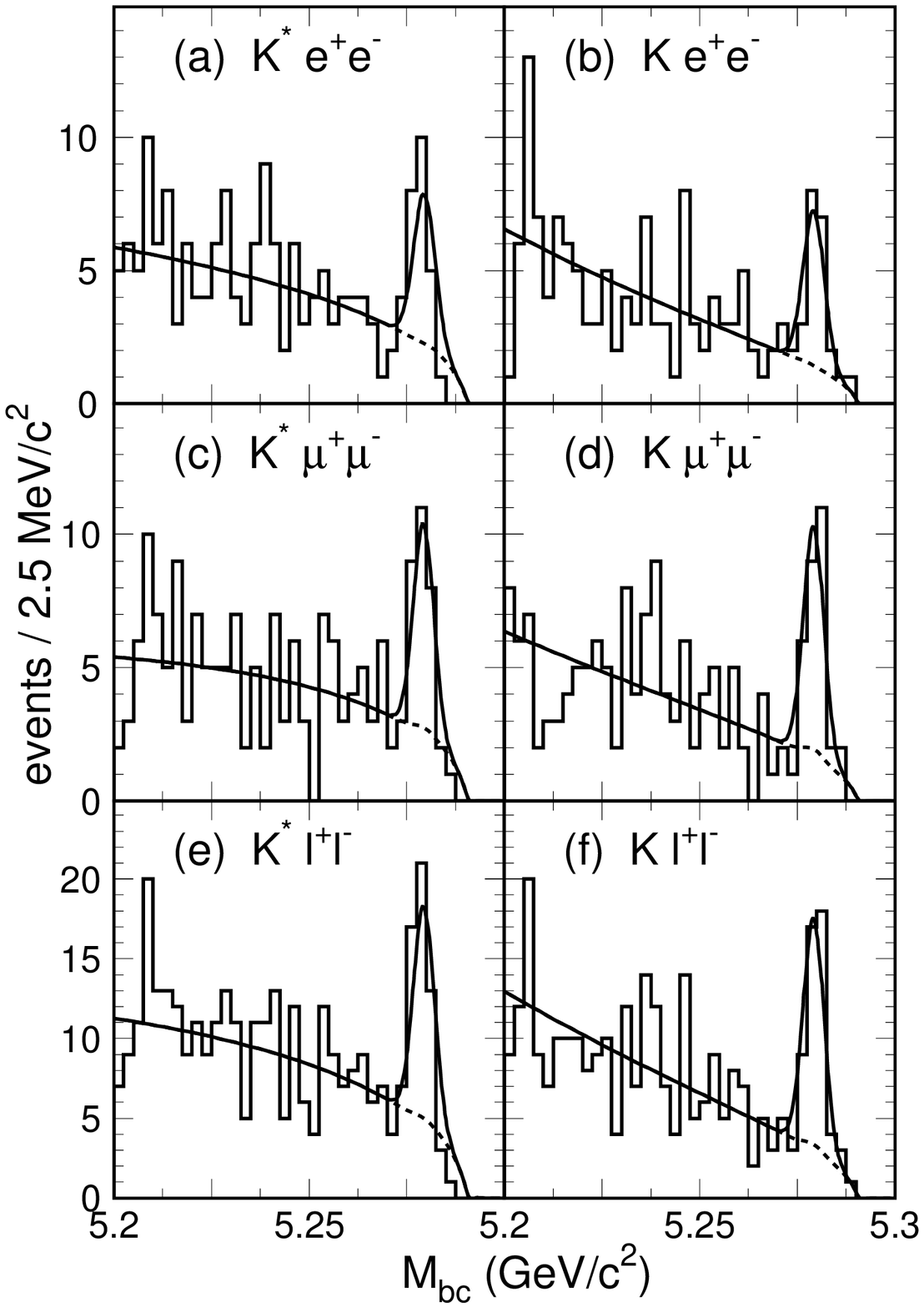}
  \caption{The $\BtoKorKstarll$ signal observed by Belle.}
  \label{fig:kstarll-belle}
\end{figure}

The $\BtoKorKstarll$ signal is identified using $\Mbc$, $\DeltaE$ (and
$\MKpi$ for $K^*\elel$).  
Belle updated the analysis using a $140\fbinv$ data sample, and with a
number of improvements in the analysis
procedure \cite{Ishikawa:2003cp}.  
The most significant improvement was
a lower minimum lepton momentum of 0.7 (0.4) GeV for muons (electrons)
from 1.0 (0.5) GeV that
gained 12\% (7\%) in the total efficiency.  In
addition, a $K^*\elel$ combinations are removed if there can be an
unobserved photon along with one of the leptons that can form a $B\to J/\psi
K\to \elel\gamma K$ decay.  As a result, the first $\BtoKstarll$ signal was
observed with a statistical significance of 5.7$\sigma$ 
from a fit to $\Mbc$, as
shown in Figure~\ref{fig:kstarll-belle}; in addition,
an improved $\BtoKll$ signal, with a significance of 7.4$\sigma$,
was obtained.

The obtained branching fractions are summarized in
Table~\ref{tbl:kstarll}, together with the BaBar results
\cite{Aubert:2003cm}.
For the combined $\BtoKstarll$ results,
$\Br(\BtoKstarll)=\Br(\BtoKstarmumu)=0.75\Br(\BtoKstaree)$ is assumed which
compensates for the enhancement at the $q^2=0$ pole, which is more
significant for $K^*\epem$, using the expected Standard Model
ratio \cite{Ali:2002jg}.  
The measured branching fractions are in
agreement with the Standard Model, for example \cite{Ali:2002jg,Lunghi:2002pm}
$(3.5\pm1.2)\EM7$ for $\BtoKll$ and $(11.9\pm3.9)\EM7$ for
$\BtoKstarll$.  We note that the experimental errors are already much
smaller than both  
the uncertainties in the theoretical predictions of the
Standard Model
and the variations due to different model-dependent assumptions used to
account for the hadronic uncertainties
\cite{Jaus:av,Melikhov:1997zu,Colangelo:1995jv,Aliev:1999gp,Zhong:2002nu}.

\begin{table}
\caption{$\BtoKorKstarll$ branching fractions.}
\label{tbl:kstarll}
\center
\begin{tabular}{lcc}
\hline
Mode & Belle ($140\fbinv$) & BaBar ($113\fbinv$) \\
     & $[\EM7]$ & $[\EM7]$ \\
\hline
  & \\[-0.30cm]
$B\to K\epem$
  & $4.8^{+1.5}_{-1.3}\pm0.3\pm0.1$
  & $7.9^{+1.9}_{-1.7}\pm0.7$ \\[0.05cm]
$B\to K\mumu$
  & $4.8^{+1.3}_{-1.1}\pm0.3\pm0.2$
  & $4.8^{+2.5}_{-2.0}\pm0.4$ \\[0.05cm]
$B\to K\elel$
  & $4.8^{+1.0}_{-0.9}\pm0.3\pm0.1$
  & $6.9^{+1.5}_{-1.3}\pm0.6$ \\[0.05cm]
\hline
  & \\[-0.30cm]
$B\to K^*\epem$
  & $14.9^{+5.2+1.1}_{-4.6-1.3}\pm0.3$
  & $10.0^{+5.0}_{-4.2}\pm1.3$ \\[0.05cm]
$B\to K^*\mumu$
  & $11.7^{+3.6}_{-3.1}\pm0.8\pm0.6$
  & $12.8^{+7.8}_{-6.2}\pm1.7$ \\[0.05cm]
$B\to K^*\elel$
  & $11.5^{+2.6}_{-2.4}\pm0.7\pm0.4$
  & $8.9^{+3.4}_{-2.9}\pm1.1$ \\[0.05cm]
\hline
\end{tabular}
\end{table}

\begin{figure}
  \center
  \includegraphics[width=0.6\textwidth,clip]{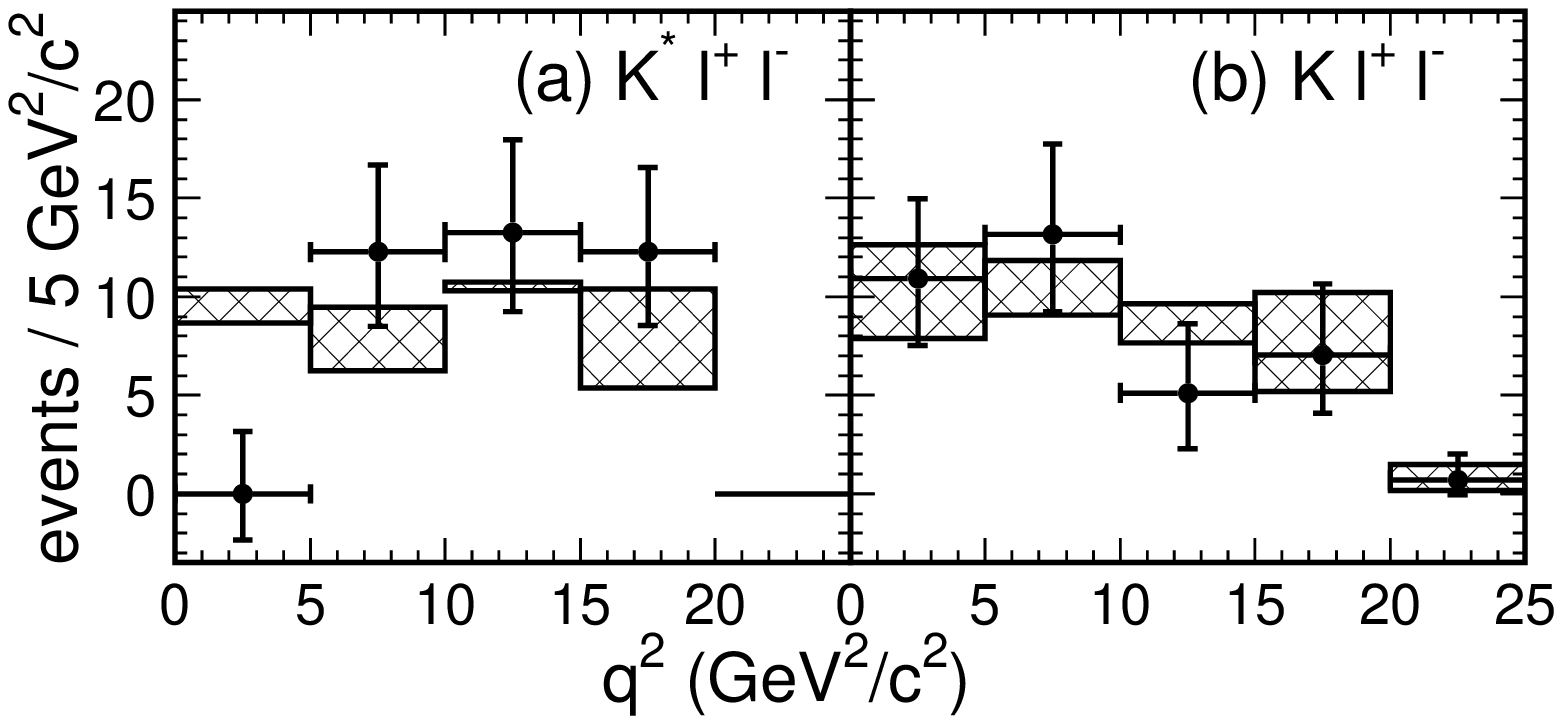}
  \caption{$q^2$ distributions for $\BtoKorKstarll$ from Belle.}
  \label{fig:belle-kll-q2}
\end{figure}

It is still too early to fit the $q^2$ distribution to constrain new
physics.  First attempts by Belle to extract the $q^2$ distribution using the
$\Mbc$ signal yields in individual 
$q^2$ bins are shown in Figure~\ref{fig:belle-kll-q2}.

\subsubsection{Measurement of $\BtoXsll$}
The first measurements of the $\BtoKorKstarll$ branching fractions are
consistent with the Standard Model predictions. However since these predictions have uncertainties that are already
larger than the measurement errors, the inclusive rate for $\BtoXsll$
becomes more important in terms of the search for a deviation from the Standard Model.
In contrast to $\BtoXsgamma$, the lepton pair alone does not provide a
sufficient constraint to suppress the largest background from
semi-leptonic decays.  Therefore, at least for now,
it is only possible to use the
semi-inclusive method to sum up the exclusive modes for now.

\begin{figure}[th]
  \center
  \includegraphics[width=0.6\textwidth,clip]{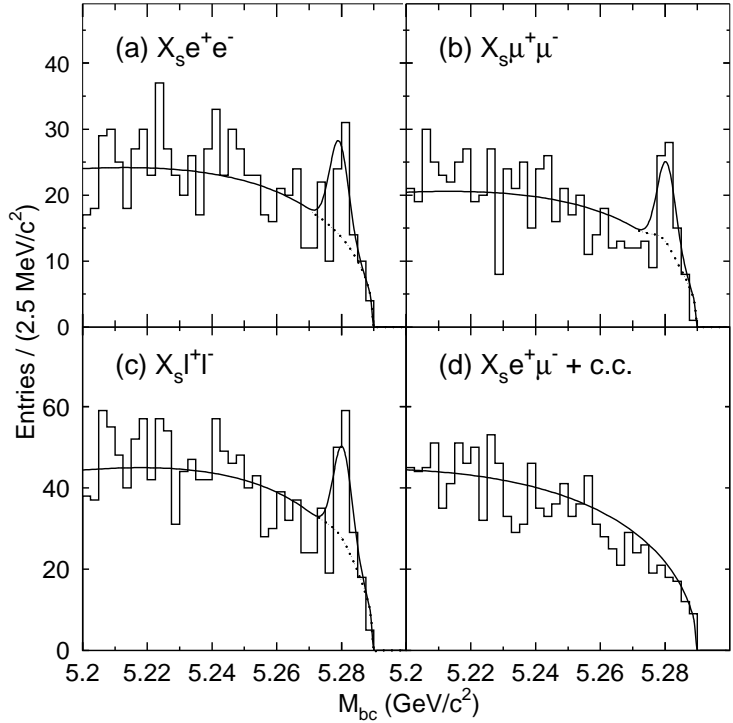}
  \caption{$\BtoXsll$ signal measured by Belle.  The $X_s e^+\mu^-$ sample,
  which is prohibited in the Standard Model, represents the combinatorial backgrounds.}
  \label{fig:xsll-belle}
\end{figure}

Belle has successfully measured the inclusive $\BtoXsll$ branching
fraction \cite{Kaneko:2002mr} with a $60\fbinv$ data sample by
applying a method that reconstructs the $X_s$ final state with one kaon ($K^+$
or $\KS$) and up to four pions, of which one pion is allowed to be a
$\pi^0$.  Assuming the $\KL$ contribution is the same as the $\KS$, this set
of final states covers $82\pm2\%$ of the signal.  In addition, $\MXs$ is
required to be below $2.1\GeV$ in order to reduce backgrounds.  For
leptons, a minimum momentum of $0.5\GeV$ for electrons, $1.0\GeV$ for
muons and $M(\elel)>0.2\GeV$ are required.  Background sources and the
suppression techniques are similar to those for the exclusive decays.
The signal of $60\pm14$ events from Belle with a statistical significance
of 5.4 is shown in Figure~\ref{fig:xsll-belle}.
Corresponding branching fractions are given in
Table~\ref{tbl:xsll}, together with the BaBar results \cite{Aubert:2003rv}.
%
%
The branching fraction results are for the dilepton mass range above
$\Mll>0.2\GeV$/c$^2$ and are interpolated in the $J/\psi$ and $\psi'$ regions
that are removed from the analysis, assuming no interference with these
charmonium states.

\begin{table}
\caption{$\BtoXsll$ branching fractions.}
\label{tbl:xsll}
\begin{center}
\begin{tabular}{lcc}
\hline
Mode & Belle ($60\fbinv$) & BaBar ($78\fbinv$) \\
     & $[\EM6]$ & $[\EM6]$ \\
\hline
$\Xsee$   & $5.0\pm2.3\PM{1.3}{1.1}$ & $6.6\pm1.9\PM{1.9}{1.6}$ \\
$\Xsmumu$ & $7.9\pm2.1\PM{2.1}{1.5}$ & $5.7\pm2.8\PM{1.7}{1.4}$ \\
$\Xsll$   & $6.1\pm1.4\PM{1.4}{1.1}$ & $6.3\pm1.6\PM{1.8}{1.5}$ \\
\hline
\end{tabular}
\end{center}
\end{table}

\begin{figure}
  \center
  \includegraphics[width=0.6\textwidth,clip]{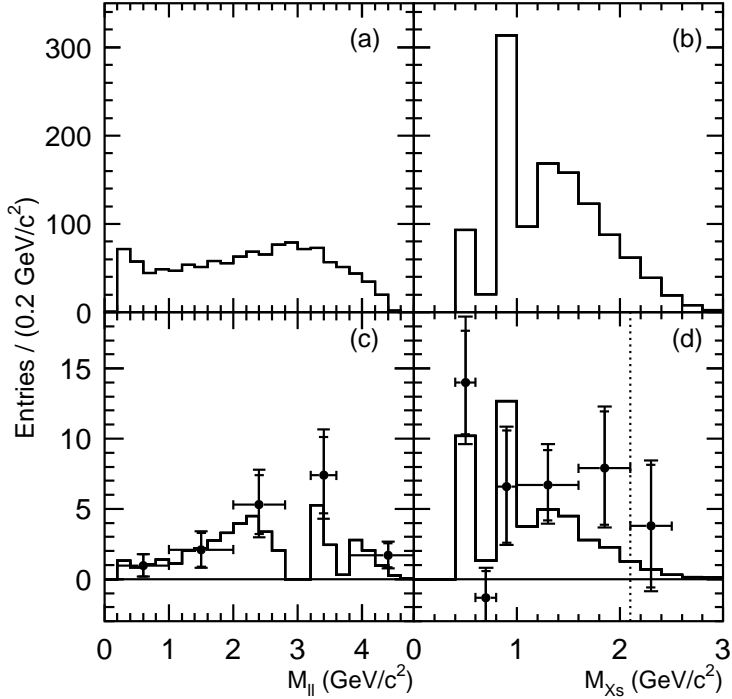}
  \caption{$\Mll$ (left) and $\MXs$ (right) distributions for $\BtoXsll$
  from Belle (points with error bars), compared with the Standard Model predictions
  before (top) and after (bottom) including detector acceptance effects.}
  \label{fig:xsll-dist-belle}
\end{figure}

The results may be compared with the Standard Model
prediction \cite{Ali:2002jg}
of $(4.2\pm0.7)\EM6$ integrated over the same dilepton mass range of
$\Mll>0.2\GeV$/c$^2$.  With this requirement, the effect of the $q^2=0$ pole
becomes insignificant, giving almost equal branching fractions for the
electron and muon modes.  The measured branching fractions are in agreement
with the Standard Model, considering the large error 
in the measurement.  It should be
noted that the large systematic error is dominated by the uncertainty in
the $\MXs$ distribution, in particular the fraction of $\BtoKorKstarll$,
which will be reduced with more statistics.  Distributions for $\MXs$ and
$\Mll$ are shown in Figure~\ref{fig:xsll-dist-belle},
in which no significant deviation from the Standard Model
is observed.

\subsection{Prospects}
With a data sample of $\sim$ 500~fb$^{-1}$, the statistical errors will be
reduced to about a half of the present errors.
The value of $\sin 2\phi_1$ will be measured with an accuracy less than 5\% using
$b \to c\overline{c} s$ processes.  
Though uncertainties will be large, $\phi_2$ and $\phi_3$ will also be 
measured.  Together with some improvement in $|V_{cb}|$ and $|V_{ub}|$ 
measurements, the KM scheme for $CP$-violation will be further
confirmed.  Direct $CP$-violation will be observed in several decay
modes, which also strongly supports the KM scheme.

We expect to obtain
results for observables that are sensitive to new physics but
have not been measured yet, such as
the forward-backward asymmetry of lepton-pairs in 
$B \to K^{(*)}\ell^+\ell^-$, time-dependent $CP$-violation of
radiative decays, {\it etc}.
If the effect of new physics is large, it may be seen;
for example, the current indication of 
the $CP$-violation in $\bz \to \phi\ks$ 
decays if its central value remains as currently measured.
However, the confirmation of any new physics effect and the understanding
its nature will require much larger data samples, which 
will be the primary goal of the proposed SuperKEKB and upgraded Belle
detector.





\chapter{Flavor Structure of the Standard Model}

\section{Flavor Structure of the Standard Model}
In the Standard Model of elementary particles there are
three generations of leptons and quarks
\begin{eqnarray}
  \label{eq:leptons}
  \left(\begin{array}{c}\nu_e   \\ e   \end{array}\right)
  \;\;
  \left(\begin{array}{c}\nu_\mu \\ \mu \end{array}\right)
  \;\;
  \left(\begin{array}{c}\nu_\tau\\ \tau\end{array}\right),
  \\
  \label{eq:quarks_mass-eigenstate}
  \left(\begin{array}{c}u \\ d \end{array}\right)
  \;\;\;\;\;
  \left(\begin{array}{c}c \\ s \end{array}\right)
  \;\;\;\;\;
  \left(\begin{array}{c}t \\ b \end{array}\right),
\end{eqnarray}
and their interactions are described by a gauge field
theory with the gauge group 
$SU(3)_C \times SU(2)_L \times U(1)_Y$.
The group $SU(3)$ denotes Quantum Chromodynamics (QCD), which
governs the strong interaction among quarks.
The transformation property under the electroweak gauge
group $SU(2)_L\times U(1)_Y$ differs for the left and right
chiralities of fermions.
The right-handed components of the leptons and quarks are
singlets under the weak $SU(2)_L$, and the weak hypercharge
$Y$ is 0, $-1$, $2/3$, $-1/3$ for neutrinos, leptons,
up-type quarks and down-type quarks, respectively.
The left-handed components of leptons transform as doublets
under the weak $SU(2)_L$ while the weak doublets of quarks
differ slightly from
(\ref{eq:quarks_mass-eigenstate}) and are given by
\begin{equation}
  \label{eq:quarks_weak-eigenstate}
  \left(\begin{array}{c}u \\ d' \end{array}\right)_L
  \;\;\;
  \left(\begin{array}{c}c \\ s' \end{array}\right)_L
  \;\;\;
  \left(\begin{array}{c}t \\ b' \end{array}\right)_L.
\end{equation}

The weak eigenstates $(d',s',b')$ are a linear combination
of the mass eigenstates $(d,s,b)$, being related 
by a $3\times 3$ unitary matrix, referred to as the 
Cabibbo-Kobayashi-Maskawa (CKM) matrix
$\hat{V}_{\mathrm{CKM}}$ \cite{Cabibbo:yz,Kobayashi:fv}, 
as follows 
\begin{equation}
  \label{eq:CKM_matrix}
  \left(\begin{array}[c]{c} d'\\ s'\\ b'\end{array}\right)
  =
  \hat{V}_{\mathrm{CKM}}
  \left(\begin{array}[c]{c} d\\ s\\ b\end{array}\right)
  \equiv
  \left(
    \begin{array}[c]{ccc}
      V_{ud} & V_{us} & V_{ub}\\
      V_{cd} & V_{cs} & V_{cb}\\
      V_{td} & V_{ts} & V_{tb}
    \end{array}
  \right)
  \left(\begin{array}[c]{c} d\\ s\\ b\end{array}\right).
\end{equation}
The charged current interactions of quarks mediated by the
$W$ boson are described by an interaction Lagrangian
\begin{eqnarray}
  \label{eq:charged_current}
  \mathcal{L}^{\mathrm{CC}}
  & = &
  - \frac{g_2}{\sqrt{2}}
  \left(
    \begin{array}{ccc}
      \overline{u} & \overline{c} & \overline{t}
    \end{array}
  \right)_L
  \gamma^\mu
  \left(\begin{array}[c]{c} d'\\ s'\\ b'\end{array}\right)_L
  W_\mu^\dagger
  + \mathrm{h.c.}
  \nonumber\\
  & = &
  - \frac{g_2}{\sqrt{2}}
  \left(
    \begin{array}{ccc}
      \overline{u} & \overline{c} & \overline{t}
    \end{array}
  \right)_L
  \gamma^\mu
  \hat{V}_{\mathrm{CKM}}
  \left(\begin{array}[c]{c} d\\ s\\ b\end{array}\right)_L
  W_\mu^\dagger
  + \mathrm{h.c.},
\end{eqnarray}
where $W_\mu$ denotes the $W$ boson, and
$g_2$ is the gauge coupling corresponding to the gauge
group $SU(2)_L$.
In the low energy effective Hamiltonian it appears as the
Fermi constant $G_F/\sqrt{2}=g_2^2/8M_W^2$.
Due to the misalignment between the up-type and down-type
quark fields, the charged current induces transitions among
different generations.

In contrast, the neutral current is
flavor-conserving, which is ensured by the unitarity of the
CKM matrix, and, thus, Flavor Changing Neutral Currents
(FCNC) are absent at the tree level in the Standard Model.
This is the Glashow-Iliopoulos-Maiani (GIM) mechanism
\cite{Glashow:gm}.
Even including loop corrections, the FCNC interaction
vanishes in the limit of degenerate (up-type) quark masses,
due to the unitarity of the CKM matrix.

The CKM matrix is a unitary $N\times N$ matrix with $N(=3)$
number of generations, and thus contains $N^2$ parameters in
general. 
However, $2N-1$ phases may be absorbed by rephasing the
$2N$ quark fields (one overall phase is related to the total
baryon number conservation and is irrelevant for the quark
mixing), and $(N-1)^2$ independent parameters remain.
Of these, $\frac{1}{2}(N-1)N$ are real parameters, which
correspond to rotation angles among different generations,
while $\frac{1}{2}(N-2)(N-1)$ are imaginary parameters, which
are sources of $CP$-violation.
In the three-generation Standard Model, there are 3 mixing
angles and 1 $CP$-phase.

The standard parametrization of the CKM matrix is the following:
\cite{Hagiwara:fs}
\begin{equation}
  \label{eq:CKM_standard}
  V_{\mathrm{CKM}} =
  \left(
    \begin{array}[c]{ccc}
      c_{12}c_{13} & s_{12}c_{13} & s_{13}e^{-i\delta}\\
      -s_{12}c_{23}-c_{12}s_{23}s_{13}e^{i\delta} &
      c_{12}c_{23}-s_{12}s_{23}s_{13}e^{i\delta} &
      s_{23}c_{13}\\
      s_{12}s_{23}-c_{12}c_{23}s_{13}e^{i\delta} &
      -s_{23}c_{12}-s_{12}c_{23}s_{13}e^{i\delta} &
      c_{23}c_{13}
    \end{array}
  \right),
\end{equation}
where $c_{ij}=\cos\theta_{ij}$ and $s_{ij}=\sin\theta_{ij}$
with $\theta_{ij}$ ($ij=12$, $13$ and $23$) the mixing
angles, and $\delta$ is the complex phase.
It is known experimentally that the angles are small and
exhibit the hierarchy $1\gg s_{12}\gg s_{23}\gg s_{13}$.
To make this structure manifest, the Wolfenstein
parametrization \cite{Wolfenstein:1983yz} is often used,
in which one sets $\lambda=|V_{us}|\simeq 0.22$ and
\begin{equation}
  \label{eq:CKM_Wolfenstein}
  V_{\mathrm{CKM}} =
  \left(
    \begin{array}[c]{ccc}
      1-\lambda^2/2 & \lambda & A\lambda^3(\rho-i\eta) \\
      -\lambda & 1-\lambda^2/2 & A\lambda^2\\
      A\lambda^3(1-\rho-i\eta) & -A\lambda^2 & 1
    \end{array}
  \right)
  + O(\lambda^4),
\end{equation}
with $A$, $\rho$ and $\eta$ being real parameters of order unity. 
In this parametrization the source of $CP$-violation is
carried by the most off-diagonal elements $V_{ub}$ and
$V_{td}$. 

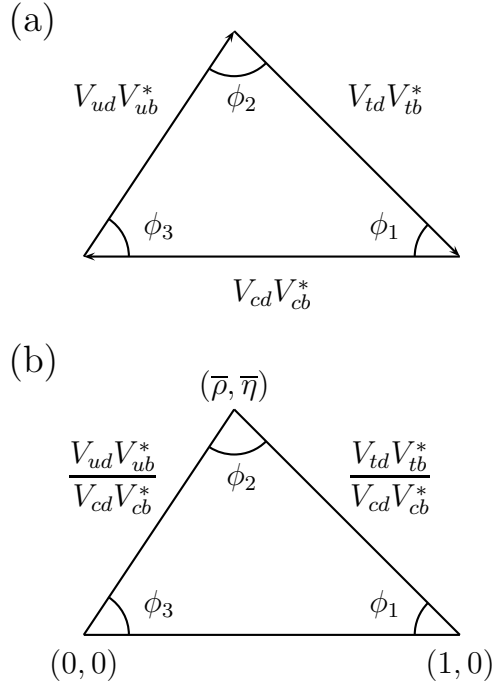
\begin{figure}[tbp]
  \centering
  \begin{pspicture}[1.0](7.0,4.5)
    \rput[Bl](0,4){\Large (a)}
    \psline{->}(1,1)(3,4)
    \psline{->}(3,4)(6,1)
    \psline{->}(6,1)(1,1)
    \rput[B](3.5,0.4){\large $V_{cd}V_{cb}^*$}
    \rput[Bl](4.5,3.0){\large $V_{td}V_{tb}^*$}
    \rput[Br](2.0,3.0){\large $V_{ud}V_{ub}^*$}
    \psarc(6,1){0.6}{135}{180}
    \psarc(1,1){0.6}{0}{57}
    \psarc(3,4){0.6}{237}{315}
    \rput[B](5,1.3){\large $\phi_1$}
    \rput[B](3.1,3){\large $\phi_2$}
    \rput[B](2,1.3){\large $\phi_3$}
  \end{pspicture}
  \\
  \begin{pspicture}[1.0](7.0,5)
    \rput[Bl](0,4.5){\Large (b)}
    \psline(1,1)(3,4)
    \psline(3,4)(6,1)
    \psline(6,1)(1,1)
    \rput[Bl](4.5,3.0){\large 
      $\displaystyle\frac{V_{td}V_{tb}^*}{V_{cd}V_{cb}^*}$}
    \rput[Br](2.0,3.0){\large 
      $\displaystyle\frac{V_{ud}V_{ub}^*}{V_{cd}V_{cb}^*}$}
    \psarc(6,1){0.6}{135}{180}
    \psarc(1,1){0.6}{0}{57}
    \psarc(3,4){0.6}{237}{315}
    \rput[B](5,1.3){\large $\phi_1$}
    \rput[B](3.1,3){\large $\phi_2$}
    \rput[B](2,1.3){\large $\phi_3$}
    \rput[B](1,0.5){\large $(0,0)$}
    \rput[B](6,0.5){\large $(1,0)$}
    \rput[B](3,4.2){\large $(\overline{\rho},\overline{\eta})$}
  \end{pspicture}
  \caption{Unitarity triangle}
  \label{fig:unitarity_triangle}
\end{figure}

Among these four parameters, $\lambda$ and $A$ are
relatively well known from corresponding semi-leptonic
decays: 
$|V_{us}|=0.2196\pm 0.0026$ from $K_{l3}$ decays and
$|V_{cb}|=(41.2\pm 2.0)\times 10^{-3}$ from inclusive and
exclusive $b\rightarrow cl\overline{\nu}_l$ decays
\cite{Hagiwara:fs}.
The determination of the other two parameters $\rho$ and $\eta$
is conveniently depicted as a contour in the plane of
$(\rho,\eta)$. 
It corresponds to the unitarity relation of the CKM matrix 
applied to the first and third columns
\begin{equation}
  \label{eq:unitarity_relation}
  V_{ud}V_{ub}^*+  V_{cd}V_{cb}^*+  V_{td}V_{tb}^*=0.
\end{equation}
This relation may be presented in the complex plane as in 
Fig.~\ref{fig:unitarity_triangle} (a), which is called the
``unitarity triangle''. 
Since $V_{cd}V_{cb}^*$ is real to a good approximation (up to
$O(\lambda^7)$), it is convenient to normalize the triangle
by $|V_{cd}V_{cb}^*|=A\lambda^3$ so that the apex has the
coordinate $(\overline{\rho},\overline{\eta})$ where
\begin{equation}
  \label{eq:rho-eta_bar}
  \overline{\rho} = \rho (1-\lambda^2/2),\;\;\;\;
  \overline{\eta} = \eta (1-\lambda^2/2),
\end{equation}
(Fig.~\ref{fig:unitarity_triangle} (b)).
The three angles of the unitarity triangle represent the 
complex phase of the combinations
\begin{equation}
  \label{eq:angles}
  \phi_1 = \arg\left[-\frac{V_{cd}V_{cb}^*}{V_{td}V_{tb}^*}\right],\;\;
  \phi_2 = \arg\left[-\frac{V_{td}V_{tb}^*}{V_{ud}V_{ub}^*}\right],\;\;
  \phi_3 = \arg\left[-\frac{V_{ud}V_{ub}^*}{V_{cd}V_{cb}^*}\right].
\end{equation}
The notation $\alpha\equiv\phi_2$, $\beta\equiv\phi_1$,
$\gamma\equiv\phi_3$ is also used in the literature.

\begin{figure}[tbp]
  \centering
  \includegraphics*[width=8cm]{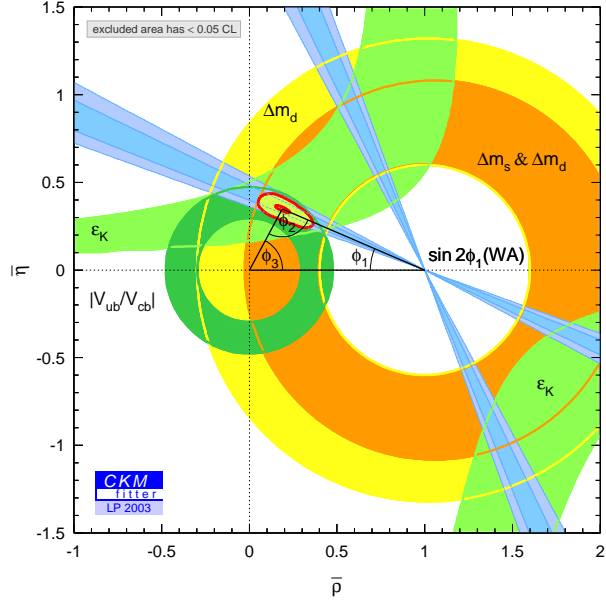}
  \caption{
    A fit of the parameters 
    $(\overline{\rho},\overline{\eta})$ using several
    experimental constraints as of Lepton-Photon 2003.
    The plot is taken from the CKM fitter page
    http://ckmfitter.in2p3.fr/.
  }
  \label{fig:rho_eta_LP2003}
\end{figure}

The present constraints on the parameter 
$(\overline{\rho},\overline{\eta})$ are 
summarized in Fig.~\ref{fig:rho_eta_LP2003}, which is
taken from the CKM fitter group
(http://ckmfitter.in2p3.fr/).
Details of the input parameters are discussed, for instance,
in \cite{Battaglia:2003in}.
Prospects with SuperKEKB will be discussed in the
following sections.

\section{Low Energy Effective Hamiltonians}
In $B$ decays the exchange of a $W$ boson and virtual
loops involving the top quark are effectively point-like
interactions, since the relevant length scale of $B$
meson decays is at least $O(1/m_b)$ while the $W$ exchange 
takes place at a short distance scale $O(1/M_W)$.
It is theoretically inefficient to calculate the physical
amplitudes using the entire $W$ and top quark propagators,
and one may instead introduce a low energy effective
Hamiltonian.
This framework is based on the Operator Product Expansion
(OPE) \cite{Wilson:zs}, which allows one to separate the long
distance physics from the short distance interactions,
occurring at a length scale of $1/M_W$, up to the corrections
of order $m_b/M_W$, which can be safely neglected in many
cases. 

\begin{figure}[tbp]
  \centering
  \includegraphics*{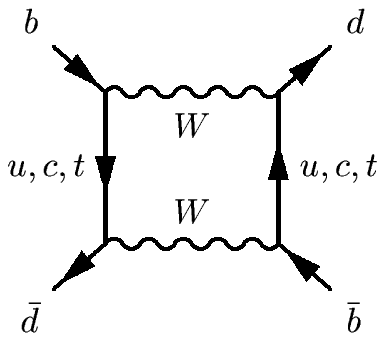}
  \hspace*{4mm}
  \includegraphics*{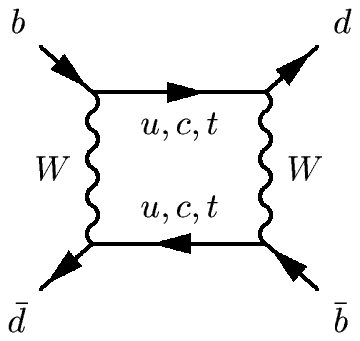}
  \caption{Box diagrams to produce the $\Delta B=2$ four-quark operator}
  \label{fig:box}
\end{figure}

For instance, $B^0-\overline{B}^0$ mixing occurs through the 
box diagrams shown in Fig.~\ref{fig:box}.
The interaction can be described in terms of the $\Delta
B=2$ effective Hamiltonian 
\begin{equation}
  \label{eq:DeltaB=2_Hamiltonian}
  \mathcal{H}_{\mathit{eff}}^{\Delta B=2}
  = \frac{G_F^2}{16\pi^2} (V_{tb}V_{td}^*)^2
  M_W^2 S_0(m_t^2/M_W^2) 
  C^{\Delta B=2}(\mu_b) Q^{\Delta B=2}(\mu_b),
\end{equation}
where the $\Delta B=2$ effective four-quark operator is
\begin{equation}
  \label{eq:DeltaB=2_operator}
  Q^{\Delta B=2} = 
  \overline{d}\gamma_\mu(1-\gamma_5)b\,
  \overline{d}\gamma_\mu(1-\gamma_5)b.
\end{equation}
The operator is defined at the renormalization scale $\mu_b$
and the corresponding Wilson coefficient 
$C^{\Delta B=2}(\mu_b)$ is calculated using the
renormalization group technique as 
$C^{\Delta B=2}(\mu_b)=[\alpha_s^{(5)}(\mu_b)]^{-6/23}$
to leading order.
The function $S_0(m_t^2/M_W^2)$ is called the Inami-Lim
function \cite{Inami:1980fz} and represents the loop
effect through the box diagrams.

The $\Delta B=1$ transitions are described by the following
effective Hamiltonian 
\begin{equation}
  \label{eq:effective_Hamiltonian}
  \mathcal{H}_{\mathit{eff}}^{\Delta B=1} = 
  \frac{4G_F}{\sqrt{2}} V_{\mathrm{CKM}}
  \sum_i C_i(\mu_b) O_i(\mu_b) + \mathrm{h.c.},
\end{equation}
where $V_{\mathrm{CKM}}$ is the corresponding CKM matrix
element.
The operators $O_i(\mu_b)$ defined at the scale $\mu_b$ are
listed below, and the couplings $C_i(\mu_b)$ are the Wilson 
coefficients. 

\begin{figure}[tbp]
  \centering
  \includegraphics*{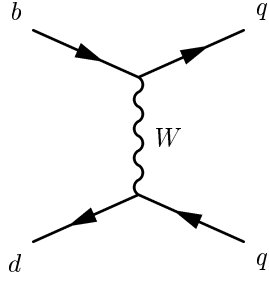}
  \caption{
    Tree-type $W$ boson exchange diagram.
  }
  \label{fig:tree}
\end{figure}

The effective operators representing the tree-level $W$
exchange diagram depicted in Fig.~\ref{fig:tree} are 
\begin{eqnarray}
  O_1^q & = & 
  \overline{d}^\alpha \gamma_\mu(1-\gamma_5) q^\beta\,
  \overline{q}^\beta \gamma_\mu(1-\gamma_5) b^\alpha,
  \;\;\; q = u, c,
  \\
  O_2^q & = & 
  \overline{d}^\alpha \gamma_\mu(1-\gamma_5) q^\alpha\,
  \overline{q}^\beta \gamma_\mu(1-\gamma_5) b^\beta,
  \;\;\; q = u, c.
\end{eqnarray}
Here the superscripts $\alpha$ and $\beta$ specify
the color contraction.
There is another set of operators obtained by replacing
$d$ by $s$.

\begin{figure}[tbp]
  \centering
  \includegraphics*{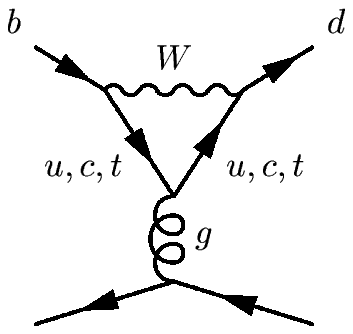}
  \hspace*{6mm}
  \includegraphics*{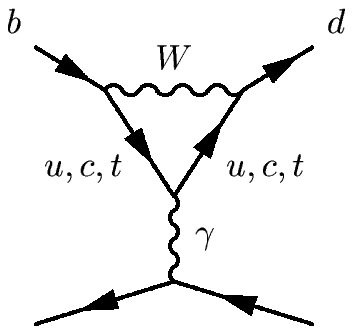}
  \caption{
    Penguin diagrams.
    The fermion line on the bottom are quarks for the gluon
    penguin diagram (left), while either quarks or leptons
    can be involved in the electro-weak penguin (right).
  }
  \label{fig:penguin}
\end{figure}

Many important FCNC decays occur through the so-called
penguin diagrams.
Some examples are shown in Fig.~\ref{fig:penguin}.

The gluon penguin diagram (Fig.~\ref{fig:penguin} (left))
produces the following operators 
\begin{eqnarray}
  O_3 & = &
  \sum_{q=u,d,s,c} 
  \overline{d}^\alpha \gamma_\mu(1-\gamma_5) b^\alpha\,
  \overline{q}^\beta \gamma_\mu(1-\gamma_5) q^\beta,
  \\
  O_4 & = &
  \sum_{q=u,d,s,c} 
  \overline{d}^\alpha \gamma_\mu(1-\gamma_5) b^\beta\,
  \overline{q}^\beta \gamma_\mu(1-\gamma_5) q^\alpha,
  \\
  O_5 & = &
  \sum_{q=u,d,s,c} 
  \overline{d}^\alpha \gamma_\mu(1-\gamma_5) b^\alpha\,
  \overline{q}^\beta \gamma_\mu(1+\gamma_5) q^\beta,
  \\
  O_6 & = &
  \sum_{q=u,d,s,c} 
  \overline{d}^\alpha \gamma_\mu(1-\gamma_5) b^\beta\,
  \overline{q}^\beta \gamma_\mu(1+\gamma_5) q^\alpha.
\end{eqnarray} 
There are also diagrams in which the gluon or photon is not
attached to the fermion line and directly appears in the
final state. 
Such diagrams produce
\begin{eqnarray}
  \label{eq:O_7}
  O_{7\gamma} & = & 
  \frac{e}{8\pi^2}
  m_b \overline{d} \sigma^{\mu\nu}(1+\gamma_5) F_{\mu\nu} b,
  \\
  \label{eq:O_8}
  O_{8g} & = &
  \frac{g}{8\pi^2}
  m_b \overline{d} \sigma^{\mu\nu}(1+\gamma_5) G_{\mu\nu}^a T^a b,
\end{eqnarray}
where $F_{\mu\nu}$ and $G_{\mu\nu}$ are electromagnetic and
QCD field strength tensors, respectively.
These operators are responsible for the 
$b\rightarrow d\gamma$ and $b\rightarrow dg$ transitions.
The operators relevant to the $b\rightarrow s\gamma$ and 
$b\rightarrow sg$ transitions are obtained by replacing $d$
by $s$ in (\ref{eq:O_7}) and (\ref{eq:O_8}).

The electroweak penguin diagram (Fig.~\ref{fig:penguin}
(right)) gives a higher order contribution in the
electromagnetic coupling constant $\alpha$ and is thus very
small in general.
However, since it may become a source of isospin symmetry
breaking, in some cases it could be relevant.
The corresponding operators are
\begin{eqnarray}
  O_7 & = & 
  \frac{3}{2} \sum_{q=u,d,s,c} e_q\,
  \overline{d}^\alpha \gamma_\mu(1-\gamma_5) b^\alpha\,
  \overline{q}^\beta \gamma_\mu(1+\gamma_5) q^\beta,
  \\
  O_8 & = & 
  \frac{3}{2} \sum_{q=u,d,s,c} e_q\,
  \overline{d}^\alpha \gamma_\mu(1-\gamma_5) b^\beta\,
  \overline{q}^\beta \gamma_\mu(1+\gamma_5) q^\alpha,
  \\
  O_9 & = & 
  \frac{3}{2} \sum_{q=u,d,s,c} e_q\,
  \overline{d}^\alpha \gamma_\mu(1-\gamma_5) b^\alpha\,
  \overline{q}^\beta \gamma_\mu(1-\gamma_5) q^\beta,
  \\
  O_{10} & = & 
  \frac{3}{2} \sum_{q=u,d,s,c} e_q\,
  \overline{d}^\alpha \gamma_\mu(1-\gamma_5) b^\beta\,
  \overline{q}^\beta \gamma_\mu(1-\gamma_5) q^\alpha.
\end{eqnarray}
$e_q$ is the electromagnetic charge of quarks;
$2/3$ for up-type quarks and $-1/3$ for down-type quarks.
When the fermion line in the bottom of
Fig.~\ref{fig:penguin} (right) is a lepton ($e$, $\mu$ or
$\tau$), we obtain 
\begin{eqnarray}
  \label{eq:O_9}
  O_{9V} & = & \frac{e^2}{16\pi^2} 
  \overline{d} \gamma_\mu(1-\gamma_5) b\,
  \overline{l} \gamma_\mu l,\\
  O_{10A} & = & \frac{e^2}{16\pi^2} 
  \overline{d} \gamma_\mu(1-\gamma_5) b\,
  \overline{l} \gamma_\mu\gamma_5 l.
\end{eqnarray}
They give rise to the $b\rightarrow d(s)l^+l^-$ transitions.

The Wilson coefficients in Eq.(\ref{eq:effective_Hamiltonian})
are calculated using a perturbation theory to 
next-to-leading order (NLO)
\cite{Altarelli:1980fi,Buras:1989xd,Buras:1991jm,Buras:1992tc}.

\section{$B-\overline{B}$ Mixing}
Neutral $B$ meson mixing is one of the most important
FCNC processes in $B$ physics.
In the Standard Model it involves the CKM matrix element
$V_{td}$ and thus gives a $CP$-violating amplitude, which
induces a variety of $CP$-violating observables through its
quantum mechanical interference with other amplitudes.

A $B^0$ meson produced as an initial state may evolve into
its antiparticle $\overline{B}^0$ through the interaction
given by the $\Delta B=2$ effective Hamiltonian
(\ref{eq:DeltaB=2_Hamiltonian}). 
In quantum mechanics the state $|B^0(t)\rangle$ at time
$t$ is a superposition of two states
$|B^0\rangle$ and $|\overline{B}^0\rangle$.
The time evolution is described by a Shr\"odinger equation
\begin{equation}
  i\frac{d}{dt} |B(t)\rangle = 
  \left( M - i \frac{\Gamma}{2} \right)
  |B(t)\rangle,
\end{equation}
where the two-by-two Hermitian matrices $M$ and $\Gamma$
denote mass and decay matrices, respectively.
The diagonal parts are constrained from $CPT$ invariance
$M_{11}=M_{22}$ and $\Gamma_{11}=\Gamma_{22}$, and the
off-diagonal parts $M_{12}$ ($M_{21}$) and $\Gamma_{12}/2$
($\Gamma_{21}/2$) are dispersive and absorptive parts of the
$\Delta B=2$ transition.
The eigenstates of the matrix $M-i\Gamma/2$ are given by
\begin{eqnarray}
  |B_1\rangle & = & p |B^0\rangle + q |\overline{B}^0\rangle,\\
  |B_2\rangle & = & p |B^0\rangle - q |\overline{B}^0\rangle,
\end{eqnarray}
and their coefficients $p$ and $q$ are obtained by solving 
\begin{equation}
  \frac{q}{p} = +
  \sqrt{\frac{M_{12}^*-i\Gamma_{12}^*/2}{M_{12}-i\Gamma_{12}/2}}
\end{equation}
together with the normalization condition $|p|^2+|q|^2=1$.
The eigenvalues $M_{1,2}-i\Gamma_{1,2}/2$ are related to
observables as follows:
the $B$ meson mass $M=(M_1+M_2)/2$,
the $B$ meson width $\Gamma=(\Gamma_1+\Gamma_2)/2$,
the $B^0-\overline{B}^0$ mixing frequency
$\Delta M \equiv M_2-M_1$, and 
the width difference $\Delta\Gamma \equiv \Gamma_1-\Gamma_2$.

In the $B$ meson system there is a relation 
$\Delta\Gamma\ll\Delta M$, which follows from
$\Gamma_{12}\ll M_{12}$.
We may then approximately obtain 
\begin{eqnarray}
  \Delta M & = & 
  -2|M_{12}| 
  \left[ 1 + O\left(
      \left|\frac{\Gamma_{12}}{M_{12}}\right|^2
    \right)
  \right],
  \\
  \Delta \Gamma & = & 
  2|\Gamma_{12}| \cos\zeta
  \left[ 1 + O\left(
      \left|\frac{\Gamma_{12}}{M_{12}}\right|^2
    \right)
  \right],
\end{eqnarray}
and
\begin{equation}
  \label{eq:q/p}
  \frac{q}{p} = + \sqrt{\frac{M_{12}^*}{M_{12}}}
  \left[
    1 
    - \frac{1}{2}\left|\frac{\Gamma_{12}}{M_{12}}\right|
    \sin\zeta
    +
    O\left(
      \left|\frac{\Gamma_{12}}{M_{12}}\right|^2
    \right)
  \right].
\end{equation}
The angle $\zeta$ is the $CP$ violating phase difference
between $M_{12}$ and $\Gamma_{12}$
\begin{equation}
  \frac{\Gamma_{12}}{M_{12}} =
  \left|\frac{\Gamma_{12}}{M_{12}}\right| e^{i\zeta}.
\end{equation}

The time evolution of the state $|B^0\rangle$
and $|\overline{B}^0\rangle$ produced at time $t=0$ is then given
by 
\begin{eqnarray}
  |B^0(t)\rangle & = &
  g_+(t) |B^0\rangle +
  \frac{q}{p} g_-(t) |\overline{B}^0\rangle,
  \\
  |\overline{B}^0(t)\rangle & = &
  g_+(t) |\overline{B}^0\rangle +
  \frac{p}{q} g_-(t) |B^0\rangle,
\end{eqnarray}
with
\begin{eqnarray}
  g_+(t) & \!\!\!=\!\!\! &
  e^{-iMt-\Gamma t/2}
  \left[
    \cosh\frac{\Delta\Gamma t}{4} \cos\frac{\Delta M t}{2}
    - i
    \sinh\frac{\Delta\Gamma t}{4} \sin\frac{\Delta M t}{2}
  \right],
  \\
  g_-(t) & \!\!\!=\!\!\! &
  e^{-iMt-\Gamma t/2}
  \left[
    -\sinh\frac{\Delta\Gamma t}{4} \cos\frac{\Delta M t}{2}
    + i
    \cosh\frac{\Delta\Gamma t}{4} \sin\frac{\Delta M t}{2}
  \right].
\end{eqnarray}
As the initial $B^0$ (or $\overline{B}^0$) state evolves, it oscillates 
between $B^0$ and $\overline{B}^0$ states with the frequency $\Delta M$.
The $CP$-violating phase arises in the mixing parameter $q/p$,
which carries the phase of $M_{12}$ as shown in
(\ref{eq:q/p}).

\section{Time-dependent Asymmetries}
\label{sec:Time-dependent_asymmetries}
One of the key observables at an asymmetric $B$ factory is
the time-dependent asymmetry between $B^0$ and
$\overline{B}^0$ decays.

Let us consider a decay of the $B$ meson to a final
state $f$.
The decay rate $\Gamma(B^0(t)\rightarrow f)$ is time
dependent, since the decaying state is a time-dependent
superposition of $|B^0\rangle$ and $|\overline{B}^0\rangle$, as
discussed in the previous section.
We write the decay amplitude of flavor eigenstates as
$A_f = \langle f|B^0\rangle$ and
$\overline{A}_f = \langle f|\overline{B}^0\rangle$,
and define a parameter 
\begin{equation}
  \label{eq:lambda_f}
  \lambda_f = \frac{q}{p} \frac{\overline{A}_f}{A_f}.
\end{equation}
If the final state is a $CP$-eigenstate 
$CP |f\rangle=\xi_f |f\rangle$ with an eigenvalue
$\xi_f=\pm 1$, then the time dependent asymmetry
\begin{equation}
  \label{eq:time-dependent_asymmetry}
  a_f(t) = \frac{
    \Gamma(\overline{B}^0(t)\rightarrow f) -
    \Gamma({B}^0(t)\rightarrow f)
  }{
    \Gamma(\overline{B}^0(t)\rightarrow f) +
    \Gamma({B}^0(t)\rightarrow f)
  }
\end{equation}
becomes
\begin{equation}
  a_f(t) = 
  \mathcal{A}_f \cos(\Delta M t) +
  \mathcal{S}_f \sin(\Delta M t),
\end{equation}
neglecting the small width difference of the $B$ meson.
Here, the direct and indirect (or mixing-induced) $CP$
asymmetries are written as
\begin{eqnarray}
  \mathcal{A}_f & = & \
  \frac{|\lambda_f|^2-1}{|\lambda_f|^2+1},
  \\
  \mathcal{S}_f & = & 
  \frac{2\,\mathrm{Im}\lambda_f}{1+|\lambda_f|^2}.
\end{eqnarray}
Since the absolute value of $q/p$ is approximately 1, 
direct $CP$-violation $|\mathcal{A}_f|\not=0$ requires
$|A_f|\not=|\overline{A}_f|$, which 
could happen if $A_f$ is a sum of (more than one) decay
amplitudes having different $CP$-phases.
Indirect $CP$-violation, on the other hand, proves the
quantum mechanical interference between the mixing and decay
amplitudes.

\subsection{Measurement of $\sin 2\phi_1$}
\label{sec:theory_sin2phi1}
The mixing-induced asymmetry provides a variety of methods
to measure the angles of the Unitarity Triangle
(Fig.~\ref{fig:unitarity_triangle}).
It was first proposed by Bigi, Carter, and Sanda in
1980--1981 \cite{Carter:hr,Carter:tk,Bigi:qs}, 
and gave strong motivation to construct the present KEK
B Factory.
The best known example is the case where the final state is
$J/\psi \ks$, whose quark level process is 
$\overline{b}\rightarrow\overline{c}c\overline{s}$ followed by
the $K^0-\overline{K}^0$ mixing. 
In the case where the decay is dominated 
by a single amplitude, the ratio of decay amplitudes is given by
\begin{equation}
  \frac{\overline{A}_{J/\psi \ks}}{A_{J/\psi \ks}} =
  -\left(\frac{V_{cb}V_{cs}^*}{V_{cb}^*V_{cs}}\right)
  \left(\frac{V_{cs}V_{cd}^*}{V_{cs}^*V_{cd}}\right),
\end{equation}
where one minus sign appears
the $CP$ odd final state $J/\psi \ks$.
Together with the phase in the mixing
\begin{equation}
  \frac{q}{p} \simeq
  \frac{V_{tb}^*V_{td}}{V_{tb}V_{td}^*},  
\end{equation}
the entire ratio $\lambda_{J/\psi \ks}$ becomes
\begin{equation}
  \lambda_{J/\psi \ks} = -
  \left(\frac{V_{tb}^*V_{td}}{V_{tb}V_{td}^*}\right)
  \left(\frac{V_{cb}V_{cs}^*}{V_{cb}^*V_{cs}}\right)
  \left(\frac{V_{cs}V_{cd}^*}{V_{cs}^*V_{cd}}\right)
  = - e^{-2i\phi_1}.
\end{equation}
Thus, one can precisely measure the angle $\phi_1$ from the 
time-dependent asymmetry 
\begin{equation}
  \label{eq:J/psi_KS_asymmetry}
  a_{J/\psi \ks}(t)=\sin(2\phi_1)\sin\Delta M t.  
\end{equation}
There exists an additional decay amplitude through the
penguin diagram $\overline{b} \rightarrow \overline{s}c\overline{c}$, which
involves the CKM factor $V_{ts}V_{tb}^*$.
Using the unitarity of the CKM matrix
$V_{ts}V_{tb}^*=-V_{cs}V_{cb}^*-V_{us}V_{ub}^*$,
the weak phase of the penguin contribution is the same as
that of the tree amplitude $V_{cs}V_{cb}^*$ up to a doubly
Cabibbo-suppressed correction.
Therefore, the relation (\ref{eq:J/psi_KS_asymmetry})
holds to an excellent approximation ($\sim$ 1\%), and the
mode $J/\psi \ks$ is called the ``gold-plated'' mode.
The precision expected at SuperKEKB is discussed in
Section~\ref{sec:sin2phi1}.

There are other decay modes which develop the same weak
phase.
Namely, the penguin decays
$\overline{b}\rightarrow\overline{s}s\overline{s}$ are
accompanied by the CKM factor
$V_{ts}V_{tb}^*$ just as the penguin contribution to
$\overline{b}\rightarrow\overline{s}c\overline{c}$.
Since there is no direct tree diagram for 
$\overline{b}\rightarrow\overline{s}s\overline{s}$, measurement of the 
same angle $\sin 2\phi_1$ through its time dependent
asymmetry can be a probe of any new physics phase in the penguin
loop process \cite{London:1997zk}.
At the hadron level, the corresponding modes are
$B^0\rightarrow\phi\ks$ and $B^0\rightarrow\eta'\ks$.
These asymmetries have already been measured as described in
Section~\ref{sec:Belle_status_and_prospect}, and the
expected sensitivity at SuperKEKB is studied in
Section~\ref{sec:sss}.
Possible contaminations from the tree-level process
$\overline{b}\rightarrow\overline{u}u\overline{s}$
with a rescattering of $u\overline{u}$ to $s\overline{s}$ may distort
the measurement.
However, these contributions are expected to be small ($O(\lambda^2)\sim$ 5\%)
\cite{Grossman:1997gr}, and a model independent bound can be
obtained using $SU(3)$ relations provided that the related
modes are observed at higher luminosity $B$ factories
\cite{Grossman:2003qp}.

\subsection{Measurement of $\phi_2$}
\label{sec:measurement_of_sin2phi2}
If a decay can occur through more than one amplitude
with different weak phases, the analysis is more involved.
As an example, we consider $B^0\rightarrow\pi^+\pi^-$,
whose decay amplitude can be parametrized as
\begin{equation}
  \label{eq:B0_to_pipi_amplitude}
  A(B^0\rightarrow\pi^+\pi^-) = 
  T_{\pi\pi} + P_{\pi\pi}.
\end{equation}
The first term represents an amplitude for the tree level $W$
exchange process $\overline{b}\rightarrow\overline{u}u\overline{d}$, which
picks up the CKM matrix elements 
$V_{ud}V_{ub}^*$, while the second term is a penguin diagram
contribution $\overline{b}\rightarrow\overline{d}u\overline{u}$ 
containing the CKM factor $V_{td}V_{tb}^*$.
If the penguin contribution can be neglected, the ratio
$\lambda_{\pi^+\pi^-}$ in (\ref{eq:lambda_f}) reads as
\begin{equation}
  \lambda_{\pi^+\pi^-} = 
  \left(\frac{V_{tb}^*V_{td}}{V_{tb}V_{td}^*}\right)
  \left(\frac{V_{ud}^*V_{ub}}{V_{ud}V_{ub}^*}\right)
  = e^{2i\phi_2},
\end{equation}
and the time-dependent asymmetry could be used to determine 
the angle $\phi_2$.
However, both amplitudes in
(\ref{eq:B0_to_pipi_amplitude}) are the same order in
$\lambda$ ($\sim\lambda^3$), 
and the penguin contribution is not so suppressed compared
to the tree level contribution;
the ratio of amplitudes $|P_{\pi\pi}/T_{\pi\pi}|$
is roughly estimated to be around 0.3 from 
$B\rightarrow K\pi$ decays assuming the flavor $SU(3)$
symmetry. 

One solution to the problem is to consider isospin
symmetry \cite{Gronau:1990ka}. The
$B\rightarrow\pi\pi$ decay amplitudes are written as
\begin{eqnarray}
  A(B^0\rightarrow\pi^+\pi^-) & = &
  \sqrt{2} (A_2 - A_0),\\
  A(B^0\rightarrow\pi^0\pi^0) & = &
  2 A_2 + A_0,\\
  A(B^+\rightarrow\pi^+\pi^0) & = &
  3 A_2.
\end{eqnarray}
$A_0$ and $A_2$ are amplitudes for 
isospin 0 and 2 of two-pion final state, respectively. 
The tree diagram contributes to both $A_0$ and $A_2$, while
the penguin diagram produces only isospin 0.
Then, one obtains a relation
\begin{equation}
  \label{eq:isospin_relation_1}
  A(B^0\rightarrow\pi^+\pi^-) + \sqrt{2}
  A(B^0\rightarrow\pi^0\pi^0) = \sqrt{2}
  A(B^+\rightarrow\pi^+\pi^0).
\end{equation}
and its $CP$ conjugate
\begin{equation}
  \label{eq:isospin_relation_2}
  \overline{A}(B^0\rightarrow\pi^+\pi^-) + \sqrt{2}
  \overline{A}(\overline{B}^0\rightarrow\pi^0\pi^0) = \sqrt{2}
  \overline{A}(B^-\rightarrow\pi^-\pi^0).
\end{equation}
By measuring the branching ratios of the three decay modes and
the time dependent asymmetry of $\pi^+\pi^-$, one can
determine the absolute values $|A_0|$ and $|A_2|$ and their
relative phase difference $\arg(A_0 A_2^*)$, through a
simple geometric reconstruction.
This determines the angle $\phi_2$ up to a four-fold
ambiguity.

Another solution is to consider the isospin relations among
$B\rightarrow\rho\pi$ decays \cite{Snyder:1993mx}.
There are three possible decay chains 
$B^0\rightarrow\{\rho^+\pi^-,\rho^0\pi^0,\rho^-\pi^+\}
\rightarrow\pi^+\pi^-\pi^0$.
Together with their $CP$-conjugate amplitudes, there exist
six different amplitudes, each of which has contributions
from both tree and penguin diagrams.
By combining the time-dependent asymmetry of this process in
the Dalitz plot one may extract the pure tree amplitude, and
thus the angle $\phi_2$.

If there is non-negligible contribution from the electroweak
penguin diagram, the isospin relations are violated, since
up and down quarks have different charges
\cite{Deshpande:1994pw}.
However, such a contribution is suppressed compared to the
gluon penguin by a factor 
$\alpha_{[\mathrm{weak}}(m_t^2/m_Z^2)]/\alpha_s\ln(m_t^2/m_c^2)]
 \sim 0.1$ \cite{Gronau:1995hn}.
At SuperKEKB, the actual size of the electroweak
penguin amplitude can be estimated from the analysis of
$K\pi$ decays \cite{Gronau:1995hn,Neubert:1998re,%
  Yoshikawa:2003hb,Gronau:2003kj,Buras:2003yc}.

The prospects of measuring the angle $\phi_2$ at SuperKEKB
using these methods are discussed in
Section~\ref{sec:phi_2}.


\section{Theoretical Methods}
In order to extract fundamental parameters, such as the
quark masses and CKM matrix elements, from $B$ decay
experiments, one needs model independent calculations of the
decay amplitudes.
However, since $B$ meson decays involve
complicated QCD interactions, which are highly non-perturbative in
general, the theoretical calculation of physical amplitudes
is a non-trivial task. 

\subsection{Heavy Quark Symmetry}
One useful theoretical method for avoiding
hadronic uncertainties is to use symmetries.
Using isospin or $SU(3)$ flavor symmetries, different decay
amplitudes can be related to each other.
This approach is widely used in $B$ decay analyses, \textit{e.g.} the
isospin analysis of $\pi\pi$ decays to extract 
$\sin 2\phi_2$ discussed in
Section~\ref{sec:measurement_of_sin2phi2}.

Another symmetry which is especially important in $B$
physics is the heavy quark symmetry 
\cite{Isgur:vq,Isgur:ed}.
In the limit of an infinitely heavy quark mass, the heavy quark
behaves as a static color source and the QCD interaction
cannot distinguish different flavors, \textit{i.e.} charm or
bottom.
Consequently the decay amplitudes (or form factors) of $b$ and $c$
hadrons are related to each other (heavy quark flavor
symmetry).
Moreover, since the spin-dependent interaction decouples in the
infinitely heavy quark mass limit, some form factors
become redundant (heavy quark spin symmetry).
The most famous example is the heavy-to-heavy semi-leptonic
decay 
$\overline{B}\rightarrow D^{(*)}\ell\overline{\nu}_\ell$ 
form factors.
In general there are 6 independent form factors for these
exclusive decay modes, but in the heavy quark limit they
reduce to one; the Isgur-Wise function, and its
normalization in the zero-recoil limit are determined.

A more general formalism has also been developed 
in the language of effective field theory,
\textit{i.e.} Heavy Quark Effective Theory (HQET)
\cite{Eichten:1989zv,Georgi:1990um,Grinstein:1990mj}.
It provides a systematic expansion in terms of
$\Lambda_{\mathrm{QCD}}/m_Q$.

\subsection{Heavy Quark Expansion}
The inclusive decay rate of a $B$ meson to the
final state $X$ can be written as
\begin{equation}
  \label{eq:inclusive_decay_rate}
  \Gamma(B\rightarrow X) =
  \frac{1}{2m_B}
  \sum_X (2\pi)^4 \delta^4(p_B-p_X)
  |\langle X|\mathcal{H}_{\mathit{eff}}^{\Delta B=1}|B\rangle|^2,
\end{equation}
where the sum runs over all possible final states and
momentum configurations.
The effective Hamiltonian 
$\mathcal{H}_{\mathit{eff}}^{\Delta B=1}$
is proportional to 
$\overline{c}\gamma^\mu P_L b \overline{l}\gamma_\mu P_L \nu_l$ 
when the $b\rightarrow c$ semileptonic decay is considered,
or to 
$\overline{u}\gamma^\mu P_L b \overline{l}\gamma_\mu P_L \nu_l$ 
if we are interested in using the $b\rightarrow u$ semileptonic
decay process to determine $|V_{ub}|$.
It could also describe non-leptonic decay by considering a
four-quark operator
$\overline{c}\gamma^\mu P_L b \overline{q}\gamma_\mu P_L q'$.
Using the optical theorem, Eq.(\ref{eq:inclusive_decay_rate})
can be rewritten in terms of an absorptive part of a $B$
meson matrix element
\begin{equation}
  \label{eq:optical_theorem}
  \Gamma(B\rightarrow X) =
  \frac{1}{m_B}
  \mathrm{Im}\,
  \langle B| \mathcal{T} |B\rangle,
\end{equation}
where the operator $\mathcal{T}$ is
\begin{equation}
  \label{eq:T}
  \mathcal{T} = i\int d^4x 
  T\left(
  \mathcal{H}_{\mathit{eff}}^{\Delta B=1}(x)
  \mathcal{H}_{\mathit{eff}}^{\Delta B=1}(0)
  \right).
\end{equation}
Since the momentum flowing into the final state quark
propagator is large ($\sim m_b$), one can expand the
time-ordered product of operators in terms of local
operators, using the Operator Product Expansion (OPE)
technique \cite{Wilson:zs}.
It gives an expansion in terms of the inverse heavy quark
mass and is called the Heavy Quark Expansion (HQE)
\cite{Chay:1990da,Bigi:1992su,Bigi:1993fe,%
  Manohar:1993qn,Blok:1993va}.

The lowest dimensional operator is $\overline{b}b$, whose matrix
element is unity up to $(\Lambda_{\mathrm{QCD}}/m_b)^2$
corrections.
The first non-trivial higher order correction appears at
$1/m_b^2$ with the chromomagnetic operator
$\overline{b}\sigma_{\mu\nu}g_sG^{\mu\nu}b$.
Therefore, at $O((\Lambda_{\mathrm{QCD}}/m_b)^2)$ the heavy
quark expansion can be expressed in terms of two
non-perturbative parameters
\begin{eqnarray}
  \label{eq:HQE_paramaters}
  \lambda_1 & = & 
  \frac{1}{2m_B} 
  \langle B(v)|\overline{h}_v (i\vec{D})^2 h_v |B(v)\rangle,
  \\
  3\lambda_2 & = &
  \frac{g_s}{2m_B} 
  \langle B(v)|\frac{1}{2} \overline{h}_v
  \sigma_{\mu\nu}G^{\mu\nu} h_v |B(v)\rangle,
\end{eqnarray}
where the operators are defined with the HQET field $h_v$
and the $B$ meson state is also defined in the heavy quark
limit.
$\lambda_2$ is known from the hyper-fine splitting of the $B$
meson ($B$-$B^*$ splitting) to be $\lambda_2\simeq$
0.12~GeV$^2$, while $\lambda_1$ has to be calculated using
non-perturbative methods, such as QCD sum rules
\cite{Ball:1993xv,Neubert:1996wm} or lattice QCD
\cite{Crisafulli:1995pg,Gimenez:1996av,%
  Kronfeld:2000gk,Aoki:2003jf}, or to be fitted with
experimental data of inclusive $B$ decays
\cite{Dikeman:1995ad,Kapustin:1995nr,%
  Falk:1995me,Falk:1995kn,Gremm:1996yn}.

\subsection{Perturbative Methods}
Model independent theoretical calculations of exclusive
non-leptonic decay amplitudes are known to be very challenging, 
as they involve both soft and hard gluon exchanges and clear
separation of the perturbative (hard) and non-perturbative
(soft) parts is intractable.
In the heavy quark limit, however, a formulation to realize
such separation of short and long distance physics has
recently been developed, ameliorating the perturbative
calculation of decay amplitudes.

The intuitive idea is the color transparency
argument due to Bjorken \cite{Bjorken:kk}.
An energetic light meson emitted from $B$ decay resembles
a color dipole, and its soft interaction with the remaining
decay products is suppressed by
$\Lambda_{\mathrm{QCD}}/m_b$.
A systematic formulation of such an idea is provided by QCD
factorization
\cite{Beneke:1999br,Beneke:2000ry,Beneke:2001ev}.
In this formalism, the decay matrix elements of 
$B\rightarrow\pi\pi$ can be factorized in the form
\begin{eqnarray}
  \label{eq:QCD_factorization}
  \langle\pi(p')\pi(q)|Q_i|\overline{B}(p)\rangle & = &
  f^{B\to\pi}(q^2)\int_0^1\! du\, T^I_i(u)\Phi_\pi(u)
  \nonumber\\
  &&
  +\int_0^1\! d\xi du dv\, T^{II}_i(\xi,u,v)
  \Phi_B(\xi)\Phi_\pi(u)\Phi_\pi(v),
\end{eqnarray}
for a four-quark operator $Q_i$.
The first term represents a factorization of the amplitude
into the $B\to\pi$ form factor and an out-going pion wave
function $\Phi_\pi(u)$ convoluted with the hard scattering 
kernel $T^I_i(u)$.
Here, the term ``factorization'' is used for two meanings:
one is the factorization of the diagram to
$\langle\pi|V|\overline{B}\rangle \langle\pi|A|0\rangle$,
while the other is the separation of hard, collinear and
soft interactions.
The kernel $T^I_i(u)$ describes the hard interaction only
and, thus, is calculable using perturbation theory.
The second term in Eq.(\ref{eq:QCD_factorization}) describes a
factorization of the amplitude into three pieces:
$\overline{B}\to 0$, $0\to\pi$, and $0\to\pi$ convoluted with a
hard interaction kernel $T^{II}_i(\xi,u,v)$.
To make the factorization of collinear and soft degrees of
freedom more explicit, an effective theory has also been
developed, which is called the Soft Collinear Effective
Theory (SCET) 
\cite{Bauer:2000ew,Bauer:2000yr,Bauer:2001ct,Bauer:2001yt}.

The form factor $f^{B\to\pi}(q^2)$ and the light-cone
distribution function (or wave function) $\Phi_B(\xi)$ and 
$\Phi_\pi(u)$ contain long-distance dynamics, which has to
be treated with non-perturbative methods.

Another method of factorization,
Perturbative QCD (PQCD)
\cite{Li:1994ck,Li:jr,Li:1994iu,%
  Keum:2000ph,Keum:2000wi,Keum:2000ms},
has also been proposed and is being used for the analysis of various $B$
decay modes. 
%
It relies on the Sudakov suppression of the tail of the wave 
function, which smears the end-point singularity in the QCD-factorization
calculation of the form factor. The form factor $f^{B\to\pi}(q^2)$ then 
becomes factorizable, and the first term in Eq.(\ref{eq:QCD_factorization}) 
can be rewritten into
a formula similar to the second term. The input of the form factor is 
used to determine the wave functions.

\subsection{Lattice QCD}
Lattice QCD provides a method to calculate non-perturbative
hadronic matrix elements from the first principles of QCD
\cite{Wilson:1974sk}.
It is a regularization of QCD on a four-dimensional
hypercubic lattice, which enables numerical simulation on
the computer.
Since the calculation is numerically so demanding, one has
to introduce several approximations in the calculation and
these lead to systematic uncertainties.

\begin{figure}[tbp]
  \centering
  \includegraphics*[width=8cm]{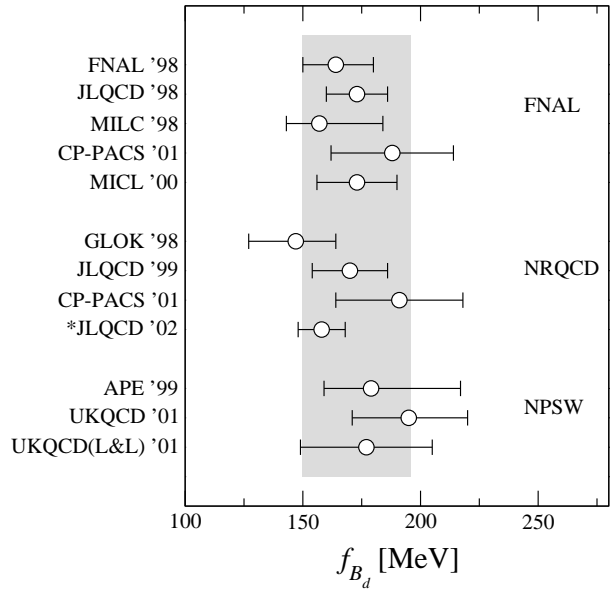}
  \caption{
    Recent quenched lattice calculations of $f_B$.
    The figure is from \cite{Yamada:2002wh}.
  }
  \label{fig:quenched_fB}
\end{figure}

For more than a decade, lattice QCD has been applied to the
calculation of matrix elements relevant to $B$ physics.
The best-known quantity is the $B$ meson leptonic decay
constant $f_B$, for which the systematic uncertainty is now
under control at the level of 10--15\% accuracy.
The important tools to achieve this goal are the following.
\begin{itemize}
\item \textit{Effective theories for heavy quarks}.
  Since the Compton wave-length of the $b$ quark is shorter than
  the lattice spacing $a$, the discretization error is out
  of control with the usual lattice fermion action for
  relativistic particles.
  Instead, Heavy Quark Effective Theory (HQET)
  \cite{Eichten:1989zv} or Non-relativistic QCD (NRQCD)
  \cite{Caswell:1985ui,Thacker:1990bm,Lepage:1992tx} 
  is formulated on the lattice and used to simulate the $b$
  quark.
  Another related effective theory is the so-called Fermilab
  action \cite{El-Khadra:1996mp,Kronfeld:2000ck}, which
  covers the entire (light to heavy) mass regime
  with the same lattice action. 
  For the $B$ meson, the next-to-leading ($1/m_Q$) order
  calculation provides $\lsim$ 5\% accuracy
  \cite{Ishikawa:1997sx}.
\item \textit{Effective theories to describe discretization
    errors}.
  The discretization effect can be expressed in terms of the
  Lagrangian language, \textit{i.e.} Symanzik effective
  theory \cite{Symanzik:1983dc,Symanzik:1983gh}.
  It also provides a method to eliminate the error by adding
  irrelevant operators to the lattice action.
  The $O(a)$ error existing in the Wilson fermion action can
  be removed by adding a dimension-five operator
  \cite{Sheikholeslami:1985ij}. 
  The cancellation of the $O(a)$ error can also be done
  non-perturbatively \cite{Jansen:1995ck,Luscher:1996ug}, so
  that the remaining discretization error is $O(a^2)$ and
  not $O(\alpha_s^na)$. 
\item \textit{Renormalized perturbation theory}.
  To relate the lattice operators to their continuum
  counterparts, one has to rely on perturbation theory.
  For a long time lattice perturbation had bad
  convergence behavior and the perturbative error was too
  large if one calculated only the one-loop terms.
  This problem was cured by Lepage and Mackenzie by taking a
  renormalized coupling constant as an expansion parameter
  \cite{Lepage:1992xa}.
\end{itemize}
A summary of recent lattice calculations of $f_B$ in the
quenched approximation is shown in
Fig.~\ref{fig:quenched_fB}.
Results of many groups obtained with different
discretizations for heavy quarks agree very well within the
error band of $\sim$ 13\%.

\begin{figure}[tbp]
  \centering
  \includegraphics*[width=9cm]{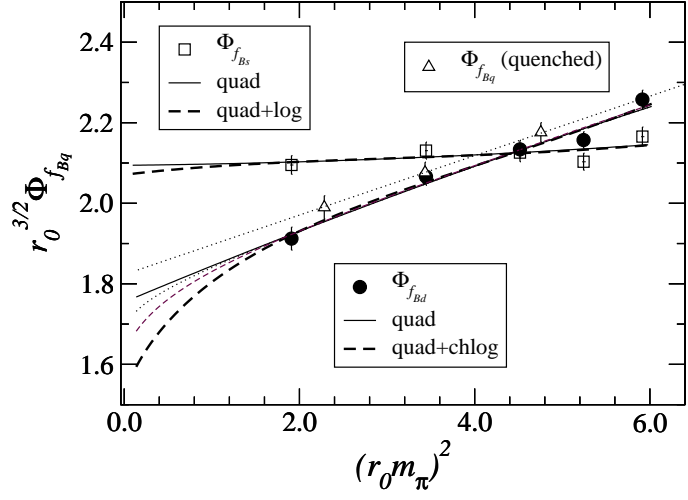}
  \caption{
    Chiral extrapolation of unquenched $f_B$.
    The plot is for $\Phi_{f_B}\equiv f_B\sqrt{m_B}$
    normalized with a Sommer scale $r_0$.
    Figure from \cite{Aoki:2003xb}.
  }
  \label{fig:unquenched_fB}
\end{figure}

The above results are obtained within the approximation of
neglecting the pair creation and annihilation of quarks in the
vacuum, which is called the quenched approximation.
To include such dynamical quark effects requires much more
computer power, and it has only recently become feasible.
A recent calculation by the JLQCD collaboration
\cite{Aoki:2003xb} is shown in
Fig.~\ref{fig:unquenched_fB}, which represents the light
quark mass dependence of $f_B$ in two-flavor QCD.
A major uncertainty in the unquenched simulation comes
from the chiral extrapolation, as the present unquenched
simulation is limited to relatively heavy sea quark masses
($m_q\gsim m_s/2$).
Because chiral perturbation theory predicts the chiral
logarithm $m_q\ln m_q$ \cite{Grinstein:1992qt}, the chiral
extrapolation may bend downwards near the chiral limit as
shown by dashed curves in Fig.~\ref{fig:unquenched_fB}.
To control this extrapolation one needs simulations with
much smaller sea quark masses as adopted in the recent
simulations using staggered fermions for sea quarks
\cite{Davies:2003ik}.

Lattice QCD calculations can be applied to several other
important quantities:
\begin{itemize}
\item $B_B$.
  The $B$ parameter in the $B^0-\overline{B}^0$ mixing has been
  calculated in the unquenched QCD \cite{Aoki:2003xb}, with the
  result $f_B\sqrt{\hat{B}_B}=215(11)(^{+\ 0}_{-23})(15)$~MeV. 
  Here the first error is
  statistical, the second is an uncertainty from the chiral
  extrapolation and the third is from other systematic errors.
\item \textit{Heavy-to-heavy semileptonic decay}.
  The zero recoil form factor of the semileptonic decay
  $B\to D^{(*)}l\nu$ has been calculated rather precisely using a
  double-ratio technique
  \cite{Hashimoto:1999yp,Hashimoto:2001nb}.
\item \textit{Heavy-to-light semileptonic decay}.
  The $B\to\pi \ell\nu$ form factor has been calculated in
  the quenched QCD to an accuracy of order 20\%
  \cite{Bowler:1999xn,Abada:2000ty,El-Khadra:2001rv,Aoki:2001rd}.
\end{itemize}
These and other recent results have been reviewed at recent 
conferences
\cite{Yamada:2002wh,Lellouch:2002nj,Becirevic:2002zp,%
  Kronfeld:2003sd}. 



\chapter{Flavor Structure of the Physics beyond the Standard Model} 

\section{Motivation for New Physics}
The Standard Model of elementary particles has
been very successful in explaining a wide variety of
existing experimental data.
It covers a range of phenomena from low energy (less than a GeV) physics,
such as kaon decays, to high energy (a few hundred
GeV) processes involving real weak gauge bosons ($W$ and
$Z$) and top quarks. 
There is, therefore, little doubt that the present Standard
Model is a theory to describe the physics below the 
energy scale of several hundred GeV, which has been explored so far.

However, the Standard Model is not satisfactory as
\textit{the} theory of elementary particles beyond the TeV
energy scale.
First of all, it does not explain the characteristic pattern
of the mass spectrum of quarks and leptons.
The second generation quarks and leptons are several orders of
magnitude heavier than the corresponding first generation
particles, and the third generation is even heavier by
another order of magnitude.
The quark flavor mixing matrix --- the CKM matrix --- also has
a striking hierarchical structure, \textit{i.e.}
the diagonal terms are close to unity and
$1\gg\theta_{12}\gg\theta_{23}\gg\theta_{13}$, where
$\theta_{ij}$ denotes a mixing angle between the $i$-th and
$j$-th generation.
The recent observation of neutrino oscillations implies 
that there is
also a rich flavor structure in the lepton sector.
All of these masses and mixings are free parameters in the
Standard Model, but ideally they should be explained by
higher scale theories.

The particles in the Standard Model acquire masses from
the Higgs mechanism.
The Higgs potential itself is described by a scalar field
theory, which contains a quadratic mass divergence. 
This means that a Higgs mass of order 100~GeV is realized
only after a huge cancellation between the bare Higgs mass
squared $\mu_0^2$ and the quadratically divergent mass 
renormalization, both of which are quantities of order
$\Lambda^2$ where $\Lambda$ is the cutoff scale.
If $\Lambda$ is of the order of the Planck scale, then
a cancellation of more than 30 orders of magnitude is required.
This is often called the hierarchy problem
\cite{Weinberg:gm,Weinberg:bn,Susskind:1978ms}. 
Therefore it would be highly unnatural if the Standard Model were
\textit{the} theory valid at a very high energy scale, such as
the Planck scale.
Instead, the Standard Model should be considered as an
effective theory of some more fundamental theory, which most
likely lies in the TeV energy region.

$CP$-violation is needed in order to produce the
observed baryon number (or matter-antimatter) asymmetry in
the universe. 
In the Standard Model, the complex phase of the CKM matrix
provides the only source of the $CP$-violation\footnote{
  The $\theta$ parameter in the QCD Lagrangian is another
  possible source of the $CP$-violation, but its value has to
  be unnaturally small $\theta\leq 10^{-10}$ in order to be
  consistent with the neutron electron dipole moment
  experiment. 
  This is another problem --- the strong $CP$ problem.
},
but models of baryogenesis suggest that it is
quantitatively insufficient
(for a review, see \cite{Cohen:1993nk}).
This is another motivation to consider new physics
models.

\section{New physics scenarios}
Several scenarios have been proposed for the physics beyond
the Standard Model.
They introduce new particles, dynamics, symmetries or even
extra-dimensions at the TeV energy scale.
In the supersymmetry (SUSY) scenarios, one introduces a new
symmetry between bosons and fermions, and a number of new
particles that form supersymmetric pairs with the existing
Standard Model particles. 
The quadratic divergence of the Higgs mass term then
cancels out among superpartners
(for reviews, see \cite{Nilles:1983ge,Haber:1984rc}).
Technicolor-type scenarios assume new strong dynamics
(like QCD) at the TeV scale and the Higgs field is realized
as a composite state of more fundamental particles
(for a recent review, see \cite{Hill:2002ap}).
The large extra space-time dimension models
\cite{Arkani-Hamed:1998rs,Randall:1999ee} cure the problem
by extending the number of spacetime dimensions beyond four
(a recent review can be found in \cite{Hewett:2002hv}).
In Little Higgs models the Higgs is a
pseudo-Nambu-Goldstone boson, and thus naturally light
\cite{Arkani-Hamed:2001nc}.

Flavor Changing Neutral Current (FCNC) processes, such
as $B^0-\overline{B}^0$ mixing and the $b\rightarrow s\gamma$
transition, provide strong constraints on new physics
models. 
If there is no suppression mechanism for FCNC processes,
such as the GIM mechanism in the Standard Model, the new
physics contribution can easily become too large to
be consistent with the experimental data.
In fact, if one introduces a FCNC interaction as a higher
dimensional operator to represent some new physics
interaction, the associated energy scale is typically of
order $10^3$~TeV, which is much higher than the expected
scale of the new physics ($\sim$ TeV). 
Therefore, one has to introduce some flavor structure in
new physics models.

\section{Supersymmetric models}
\label{sec:Supersymmetric_models}
Here, let us consider the supersymmetric (SUSY) model as an 
example of new physics models at the TeV scale.
The SUSY model is attractive not only because it solves the
Higgs mass hierarchy problem. It is also consistent with
Grand Unification \cite{Georgi:sy,Pati:1974yy},
\textit{i.e.}
the renormalization group running of the three gauge couplings is
modified by the supersymmetric partners, causing them to
intersect at the same point at $\Lambda_{\rm GUT}\simeq 10^{16}$~GeV.

The general SUSY models have a number of free parameters
corresponding to the masses and mixings of the superpartners for
each Standard Model particle.
Even in the minimal model --- the Minimal Supersymmetric
Standard Model (MSSM) --- the number is more than a
hundred.
These mass and mixing parameters are, at least partly,
governed by the soft supersymmetry breaking mechanism, 
which is necessary to make the superpartners heavy enough
such that they are not detected at existing collider
experiments. 
Therefore, to predict the mass spectrum and flavor mixing of
the SUSY particles one has to specify the details of the
SUSY breaking mechanism, which should be given at
energy scales higher than the TeV scale.

The Minimal Supersymmetric Standard Model (MSSM) is a
minimal supersymmetric extension of the Standard Model, containing
a superpartner for each particle in the Standard Model 
and two Higgs doublets.
Its matter contents are written in terms of the chiral
super-fields as
\begin{equation}
  Q_i(3,2,1/6),\;\; 
  \overline{U}_i(\overline{3},1,-2/3), \;\;
  \overline{D}_i(\overline{3},1,1/3)
\end{equation}
for the left-handed ($Q$) and right-handed ($U$ and $D$)
quark sector,
\begin{equation}
  L_i(1,2,-1/2), \;\; 
  \overline{E}_i(1,1,1)
\end{equation}
for the left-handed ($L$) and right-handed ($E$) lepton
sector, and 
\begin{equation}
  H_1(1,2,-1/2), \;\; H_2(1,2,1/2)
\end{equation}
for the Higgs fields.
The representation (or charge) for the gauge group
$SU(3)_C \times SU(2)_L \times U(1)_Y$ is given in
parentheses, and $i$ (= 1, 2, or 3) is a generation index. 
Under the assumption of $R$-parity conservation, which is
required to avoid an unacceptably large proton decay rate, the
superpotential is written as
\begin{equation}
  \mathcal{W}_{\mathrm{MSSM}} =
  f_D^{ij} \overline{D}_i Q_j H_1 +
  f_U^{ij} \overline{U}_i Q_j H_2 +
  f_E^{ij} \overline{E}_i L_j H_1 +
  \mu H_1 H_2,
\end{equation}
where $f_U$ and $f_D$ are the quark Yukawa couplings.
The soft supersymmetry breaking terms are
\begin{eqnarray}
  -\mathcal{L}_{\mathrm{soft}} & = &
  (m_Q^2)^i_{\ j} \tilde{q}_i \tilde{q}^{\dagger j} +
  (m_D^2)_i^{\ j} \tilde{d}^{\dagger i} \tilde{d}_j +
  (m_U^2)_i^{\ j} \tilde{u}^{\dagger i} \tilde{u}_j +
  (m_E^2)^i_{\ j} \tilde{e}_i \tilde{e}^{\dagger j} +
  (m_L^2)_i^{\ j} \tilde{l}^{\dagger i} \tilde{l}_j
  \nonumber\\
  & &
  +
  \Delta_1^2 h_1^\dagger h_1 +
  \Delta_2^2 h_2^\dagger h_2 
  - (B\mu h_1 h_2 + \mathrm{h.c.})
  \nonumber\\
  & &
  + A_D^{ij} \tilde{d}_i \tilde{q}_j h_1
  + A_U^{ij} \tilde{u}_i \tilde{q}_j h_2
  + A_L^{ij} \tilde{u}_i \tilde{q}_j h_2
  \nonumber\\
  & &
  + \frac{M_1}{2} \overline{\tilde{B}}\tilde{B}
  + \frac{M_2}{2} \overline{\tilde{W}}\tilde{W}
  + \frac{M_3}{2} \overline{\tilde{g}}\tilde{g}.
\end{eqnarray}
These consist of mass terms for scalar fields ($\tilde{q}_i$,
$\tilde{u}_i$, $\tilde{d}_i$, $\tilde{l}_i$, $\tilde{e}_i$,
$h_1$, and $h_2$), Higgs mixing terms, trilinear scalar
couplings, and gaugino ($\tilde{B}$, $\tilde{W}$, and
$\tilde{g}$) mass terms.


Flavor physics already places strong restrictions on the possible 
structure of the SUSY breaking sector, since
arbitrary terms would induce many flavor violating processes which are
easily ruled out by present experimental data.
Therefore, in order to comply with the requirement of
highly suppressed FCNC interactions, one has to introduce some
structure in the soft SUSY breaking terms.
Several scenarios have been proposed.
\begin{itemize}
\item {\it Universality}.
  The SUSY breaking terms have a universal flavor structure
  at a very high energy scale, such as the Planck scale
  ($\sim 10^{18}$~GeV) or the GUT scale ($\sim 10^{16}$~GeV).
  It could also be a lower scale ($\sim 10^{4-6}$~GeV). 
  The universality comes from mediation of the SUSY breaking
  effect by flavor-blind interactions, such as gravity
  (for a review of gravity mediation
  see \cite{Nilles:1983ge}), 
  the Standard Model gauge interaction (gauge mediation
  \cite{Dine:1994vc,Dine:1995ag,Dine:1996xk} the gaugino
  mediation
  \cite{Kaplan:1999ac,Chacko:1999mi,Schmaltz:2000gy}), or 
  the super-Weyl anomaly (anomaly mediation
  \cite{Randall:1998uk,Giudice:1998xp}). 
  Since the soft SUSY breaking terms are flavor-blind, the
  squark masses are degenerate at the high energy scale
  where those terms are generated.
  The GIM mechanism then works as long as the scalar triple
  coupling (squark-squark-Higgs), the $A$ term, is
  proportional to the Yukawa couplings in the Standard
  Model.
  An additional flavor violating effect could appear through 
  the renormalization group running of the squark masses to
  the low energy scale, which depends on the flavor
  \cite{Hall:1985dx}. 
  For the gauge mediation scenario the effect on FCNC
  processes is extremely suppressed, since the SUSY breaking scale is low
  and there is not enough room for the running.
\item {\it Alignment}. 
  Squark and slepton mass matrices could be diagonalized (no 
  flavor changing interaction) in the same basis as quarks
  and leptons, if one assumes some symmetries involving
  different generations
  \cite{Nir:1993mx,Leurer:1993gy}. 
  Flavor violation is then suppressed and
  flavor violating processes are induced by incomplete
  alignment.
\item {\it Decoupling}. 
  The squarks and sleptons of the first and second generations
  are sufficiently heavy, 10--100~TeV, so that flavor violation in
  the first and second generations is suppressed
  \cite{Dine:1990jd,Dine:1993np,Dimopoulos:1995mi,%
    Pomarol:1995xc,Cohen:1996vb,Hisano:1998tm,Hisano:2000wy}.
  In general such models predict large FCNC effects in the third
  generation, \textit{i.e.} the $b$ quark and
  $\tau$ lepton decays.
\end{itemize}
Signals for FCNC processes and $CP$-violation largely
depend on the structure of the soft SUSY breaking terms.  

The Grand Unified Theory (GUT) \cite{Georgi:sy,Pati:1974yy}
is one of the motivations for introducing supersymmetry.
Besides the unification of couplings, GUTs also relate the Yukawa
couplings of the quark and lepton sectors.
Since the particle content and symmetry are modified above
the GUT scale, they could generate different FCNC effects even
if universal soft SUSY breaking is assumed.
GUTs also predict some correlation between quark and lepton
flavor violation processes.
Such studies have been done by several authors
\cite{Moroi:2000tk,Akama:2001em,Baek:2001kh,Goto:2002xt,Goto:2003iu,%
  Chang:2002mq,Ciuchini:2002uv,Hisano:2003bd,Ciuchini:2003rg}

\def\be{\begin{equation}}
\def\ee{\end{equation}}
\def\bea{\begin{eqnarray}}
\def\eea{\end{eqnarray}}
\def\nnb{\nonumber}
\def\lsim{\raise0.3ex\hbox{$\;<$\kern-0.75em\raise-1.1ex\hbox{$\sim\;$}}}
\def\gsim{\raise0.3ex\hbox{$\;>$\kern-0.75em\raise-1.1ex\hbox{$\sim\;$}}}
\def\Frac#1#2{\frac{\displaystyle{#1}}{\displaystyle{#2}}}
\def\no{\nonumber\\}
\def\slash#1{\ooalign{\hfil/\hfil\crcr$#1$}}
\def\ep{\eta^{\prime}}
\def\susy{\mbox{\tiny SUSY}}
\def\sm{\mbox{\tiny SM}}
\def\sphik{S_{\phi \ks}}
\def\dM{\Delta M_s}

\section{SUSY effect on $b\to s$ transitions}
\label{sec:new_physics/btos}
Here we discuss the possible effects of SUSY particles on
$b\to s$ transitions.
The $b\to s$ processes are especially interesting, since the
time-dependent $CP$ asymmetries measured in 
these processes, in particular $B\to\phi \ks$ by Belle,
deviates from its Standard Model 
expectation
$\mathcal{S}_{\phi \ks}=\mathcal{S}_{J/\psi \ks}$
\cite{Abe:2003yt}.
Since the measurements by BaBar are still consistent with the
Standard Model~\cite{bib:BaBar_sss}, much more data are
required to resolve the problem.
In the Standard Model the $B\to\phi \ks$ transition is induced by the
$b\to s$ penguin diagram, in which there is room for
new physics effects to compete with the Standard Model
contribution. 
Supersymmetric GUT models also predict large effects in
$b\to s$ transitions as a large mixing angle is observed
in the neutrino sector between the second and third generations. 

We parametrize the effect of soft SUSY breaking terms
applying the mass insertion approximation (MIA)
\cite{Hall:1985dx}.
In the MIA, one adopts a basis where the fermion and
sfermion mass matrices are rotated in the same way to
diagonalize the fermion mass matrix (the super-CKM basis). 
In this basis, the couplings of fermions and sfermions to
neutral gauginos are flavor diagonal, leaving all the
sources of flavor violation in the off-diagonal terms of
the sfermion mass matrix. 
These terms are denoted by $(\Delta^q_{AB})^{ij}$,
where $A,B=(L, R)$ and $q=(u, d)$. 
The sfermion propagator can then be expanded as 
\begin{equation}
  \langle \tilde q_A^a \tilde q_B^{b*} \rangle 
  = i(k^2 \mathbf{1}-\tilde{m}^2 \mathbf{1}-
  \Delta^q_{AB})^{-1}_{a b}
  \simeq 
  \frac{i\delta_{ab}}{k^2-\tilde{m}^2} +
  \frac{i(\Delta^q_{AB})_{ab}}{(k^2-\tilde{m}^2)^2} +
  \cdots,
\end{equation}
where $\mathbf{1}$ is the unit matrix and $\tilde{m}$ is the
averaged squark mass.
Here we keep only the first term of the expansion. 
In this way, the flavor violation in SUSY models can be
parameterized in a model independent way by the
dimensionless parameters
$(\delta^{q}_{AB})_{ij}=(\Delta^{q}_{AB})_{ij}/\tilde{m}^2$,
where $\tilde{m}$ is an average squark mass.
Constraints on these parameters from presently available
data have been analyzed in
\cite{Gabbiani:1996hi,Ciuchini:1998ix,Causse:2000xv}
(also see \cite{Masiero:xj} for a summary).

\begin{figure}[tbp]
  \begin{center}
    \includegraphics[width=12cm]{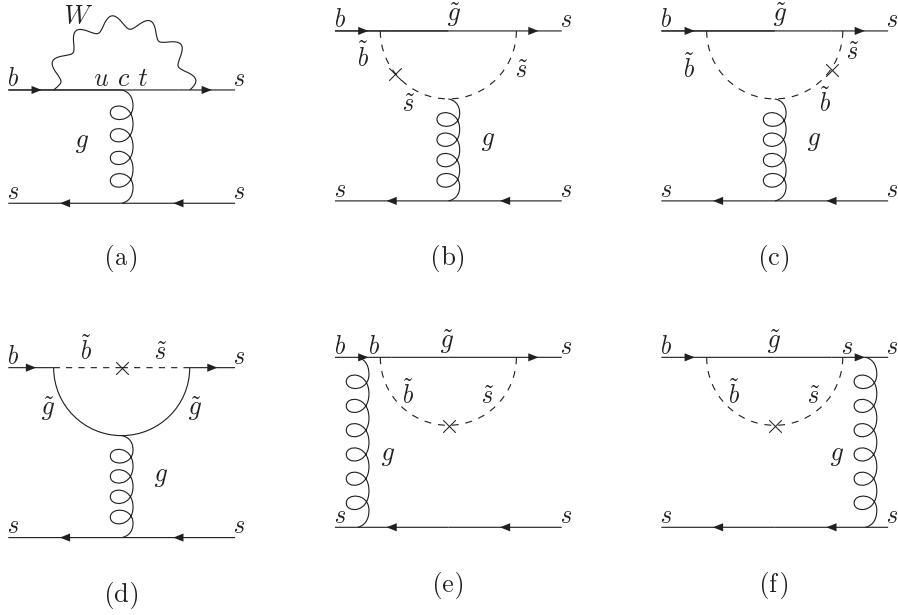}
    \caption{
      The Standard Model contribution (a) and the gluino--down squark 
      contributions (b)--(f) to $B\to\phi \ks$ decay. 
      The cross represents the mass insertions 
      $(\delta^{d}_{AB})_{23}$.
    }
    \label{fig:1}
  \end{center}
\end{figure}

Using the MIA, the $b\to s\overline{s}s$ transition accompanied
by SUSY particles occurs through the diagrams shown in 
Figure~\ref{fig:1}, and the $B\to\phi \ks$ data provide a
constraint on the mass insertions $(\delta^{d}_{AB})_{23}$
where $(A,B)=(L,R)$.
In fact, the absolute value of $(\delta^{d}_{AB})_{23}$ is
constrained by the branching ratio of $B\to X_s\gamma$ as   
\begin{equation}
  \label{eq:present_bound}
  |(\delta^{d}_{LL(RR)})_{23}| < 8.2,
  \;\;\; 
  |(\delta^{d}_{LR(RL)})_{23}| < 1.6\times 10^{-2},
\end{equation}
for $m_{\tilde{g}}=$ 300~GeV and $m_{\tilde{q}}=$ 500~GeV
\cite{Causse:2000xv}
\footnote{
  In the following analysis we use 
  $|(\delta^{d}_{LL(RR)})_{23}| < 1$ instead,
  as the mass insertion approximation does not converge for
  $>1$ and the entire analysis becomes unreliable.
}
Thus, an interesting question is whether the observed
deviation of $\mathcal{S}_{\phi \ks}$ from 
$\mathcal{S}_{J/\psi \ks}$ can be obtained while satisfying
(\ref{eq:present_bound}). 

\begin{figure}[tbp]
\begin{center}
  \psfrag{x}[r][r][2]{$\theta_{\mathrm{SUSY}}$}
  \psfrag{y}[r][r][2]{$\mathcal{S}_{\phi \ks}$}
  \psfrag{2}[r][r][1.8]{$\overline{2}$}
  \psfrag{p}[r][r][1.8]{$\pi$}
  \psfrag{0}[r][r][1.8]{0}
  \includegraphics[width=12cm]{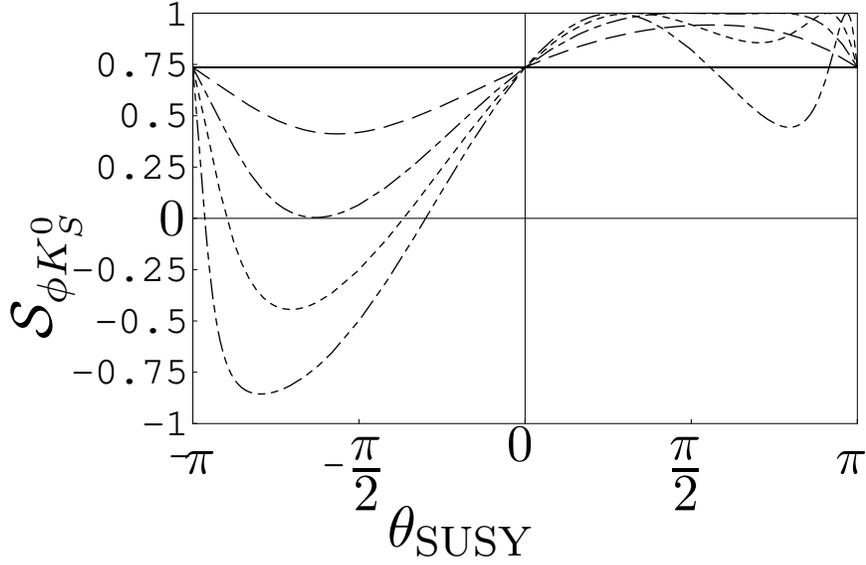}
  \caption{
    $\mathcal{S}_{\phi \ks}$ including SUSY contributions
    (\ref{eq:Sfull}) as a function of $\theta_{\mathrm{SUSY}}$ 
    for different values of 
    $|A^{\mathrm{SUSY}}/A^{\mathrm{SM}}|$:
    0.2 (dashed), 0.4 (dashed-dotted), 0.6 (dotted), 0.8
    (dashed-double-dotted). 
    The solid line represents the central value of
    $\mathcal{S}_{J/\psi \ks}$.
    Here we assume that the strong phase difference between
    SUSY and the Standard Model is negligible $\delta_{12}\simeq0$.  
  }
  \label{fig:2}
\end{center}
\end{figure}

Here we parameterize the Standard Model and SUSY amplitudes as 
\begin{eqnarray}
  A^{\rm SM}(\phi \ks)=|A^{\mathrm{SM}}| e^{i\delta_{\rm SM}}, 
  & \;\;\; &
  A^{\rm SUSY}(\phi \ks) = |A^{\mathrm{SUSY}}| 
  e^{i\theta_{\rm SUSY}}e^{i\delta_{\rm SUSY}},
  \\
  \overline{A}^{\rm SM}(\phi \ks)=|\overline{A}^{\mathrm{SM}}|
  e^{i\delta_{\rm SM}},
  & \;\;\; &
  \overline{A}^{\rm SUSY}(\phi \ks) = |\overline{A}^{\mathrm{SUSY}}|
  e^{-i\theta_{\rm SUSY}}e^{i\delta_{\rm SUSY}},
\end{eqnarray}
where $\delta_{\rm SM(SUSY)}$ is the strong ($CP$-conserving)
phase and $\theta_{\rm SUSY}$ is the weak ($CP$-violating) phase. 
Then, we obtain 
\begin{equation}
  \mathcal{S}_{\phi \ks}
  = 
  \frac{
    \sin 2\phi_1 +
    2\left(\frac{|A^{\mathrm{SUSY}}|}{|A^{\mathrm{SM}}|}\right)
    \cos\delta_{12}\sin(\theta_{\rm SUSY}+2\phi_1) +
    \left(\frac{|A^{\mathrm{SUSY}}|}{|A^{\mathrm{SM}}|}\right)^2
    \sin (2\theta_{\rm SUSY}+2\phi_1 )
  }{
    1+
    2\left(\frac{|A^{\mathrm{SUSY}}|}{|A^{\mathrm{SM}}|}\right)
    \cos\delta_{12}\cos\theta_{\rm SUSY} +
    \left(\frac{|A^{\mathrm{SUSY}}|}{|A^{\mathrm{SM}}|}\right)^2
  },
  \label{eq:Sfull} 
\end{equation}
where $\delta_{12}\equiv \delta_{\rm SM}-\delta_{\rm SUSY}$. 
Let us first see how large $ |A^{\mathrm{SUSY}}/A^{\mathrm{SM}}|$
needs to be in order to have $\mathcal{S}_{\phi \ks}$ as small as the
experimental central value.
Using (\ref{eq:Sfull}), we plot $\mathcal{S}_{\phi \ks}$
in terms of $\theta_{\mathrm{SUSY}}$ for different values of
$|A^{\mathrm{SUSY}}/A^{\mathrm{SM}}|$ in Figure~\ref{fig:2},
by fixing $\delta_{12}=0$ for simplicity and using the
central value of the observed $\sin 2\phi_1$. 
We can see that the deviation of $\mathcal{S}_{\phi \ks}$
from $\mathcal{S}_{J/\psi \ks}$ becomes maximal at around 
$\theta_{\mathrm{SUSY}}=-\pi /2$ and 
$|A^{\mathrm{SUSY}}/A^{\mathrm{SM}}|\gsim 0.4$ is required
in order to have a negative value of 
$\mathcal{S}_{\phi \ks}$.  

The effective Hamiltonian for the penguin process can be
expressed as 
\begin{equation}
  \mathcal{H}^{\Delta B=1}_{\mathrm{eff}} = 
  -\frac{G_F}{\sqrt{2}} V_{tb}V_{ts}^*
  \left[\sum_{i=3}^{6} C_i O_i 
    + C_g O_g +
    \sum_{i=3}^{6} \tilde{C}_i\tilde{O}_i
    + \tilde{C}_g\tilde{O}_g
  \right]
\end{equation}
with
\begin{eqnarray}
  O_3 &=&
  \overline{s}_{\alpha}\gamma^{\mu}Lb_{\alpha} 
  \overline{s}_{\beta}\gamma^{\mu}Ls_{\beta}, \\ 
  O_4 &=&
  \overline{s}_{\alpha}\gamma^{\mu}Lb_{\beta} 
  \overline{s}_{\beta}\gamma^{\mu}Ls_{\alpha}, \\ 
  O_5 &=&
  \overline{s}_{\alpha}\gamma^{\mu}Lb_{\alpha} 
  \overline{s}_{\beta}\gamma^{\mu}Ls_{\beta}, \\ 
  O_6 &=&
  \overline{s}_{\alpha}\gamma^{\mu}Lb_{\beta} 
  \overline{s}_{\beta}\gamma^{\mu}Rs_{\alpha}, \\
  O_{g} &=& 
  \frac{g_s}{8\pi^2}m_b\overline{s}_{\alpha}\sigma^{\mu\nu}
  R\frac{\lambda^A_{\alpha\beta}}{2}b_{\beta}G^A_{\mu\nu}
  \label{og},
\end{eqnarray}
where $L\equiv (1-\gamma_5)/2$ and $R\equiv(1+\gamma_5)/2$.
The terms with a tilde are obtained from $C_{i,g}$ and
$O_{i,g}$ by exchanging $L\leftrightarrow R$. 
The Wilson coefficient $C_{i(g)}$ includes both 
the Standard Model and SUSY
contributions.
In the following analysis, we neglect the effect of the
operator 
$O_{\gamma} 
= \frac{e}{8\pi^2}m_b\overline{s}_{\alpha}\sigma^{\mu\nu}
R b_{\alpha} F_{\mu\nu}$ 
and the electroweak penguin operators, which give very small
contributions. 
The matrix elements are computed in the naive factorization
approximation in the following\footnote{
  The matrix element of $B\to\phi \ks$ can also be
  computed using more recent methods: 
  the QCD factorization and the pQCD approach method. 
  Application of these to analyse the SUSY effect on
  $S_{\phi \ks}$ can be found in 
  \cite{Ciuchini:2002uv,Kane:2003zi} for the
  former and \cite{Mishima:2003ta} for the latter. 
}. 
The SUSY amplitude $A^{\mathrm{SUSY}}(\phi K)$ contains 
gluino and chargino contributions although the latter is
negligible.
The Wilson coefficients for the gluino contributions that
come from box and penguin diagrams are computed in
\cite{Gabbiani:1996hi}.
We note that the dominant SUSY contribution to 
$B\to\phi \ks$ comes from the chromo-magnetic operator,
$O_g$ and $\tilde{O}_g$.

The numerical result for the ratio of the Standard Model to SUSY
amplitude for 
$m_{\tilde{g}}\simeq m_{\tilde{q}}$ = 500~GeV is as follows
\cite{Khalil:2002fm}
\begin{equation}
  \frac{A^{\mathrm{SUSY}}(\phi \ks)}{
    A^{\mathrm{SM}}(\phi \ks)}
  \simeq 
  (0.14+0.02i) [(\delta_{LL}^d)_{23}+(\delta_{RR}^d)_{23}] +
  (65+11i) [(\delta_{LR}^d)_{23}+(\delta_{RL}^d)_{23}].
  \label{eq:res1}
\end{equation}
The tiny imaginary parts in (\ref{eq:res1}) are the strong
phases coming from the QCD correction terms in the effective
Wilson coefficient in \cite{Ali:1997nh}.  
We shall consider SUSY models where one mass insertion
is dominant, which is often the case in
minimal flavour violation models. 
Let us start with the models in which $LL(RR)$ dominates and
the remaining mass insertions, 
$(\delta_{LR}^d)_{23}, (\delta_{RL}^d)_{23}$ and 
$(\delta_{RR(LL)}^d)_{23}$ are all negligible.   
Taking into account the constraints from the $B\to X_s\gamma$ branching
ratio in (\ref{eq:present_bound}), we obtain a maximum
SUSY contribution of 
$(|A^{\mathrm{SUSY}}|/|A^{\mathrm{SM}}|)_{LL(RR)}\simeq 0.14$, 
which can lead to only a small deviation between
$\mathcal{S}_{\phi \ks}$ and $\mathcal{S}_{J/\psi \ks}$ 
as we learned in Figure~\ref{fig:2}. 
On the other hand, even if $LR$ and $RL$ dominated models 
are more severely constrained by $B\to X_s\gamma$, the SUSY
contribution can reach  
$(|A^{\mathrm{SUSY}}/A^{\mathrm{SM}}|)_{LR(RL)}\simeq 1.1$, 
which gives a negative value of 
$\mathcal{S}_{\phi \ks}$ for a large range of 
$\theta_{\mathrm{SUSY}}$.  
Smaller gluino and squark masses,  
$m_{\tilde{g}}\simeq m_{\tilde{q}}=300~ \mathrm{GeV}$, 
result in a larger SUSY contribution: 
\begin{equation}
  \frac{A^{\mathrm{SUSY}}(\phi \ks)}{
    A^{\mathrm{SM}}(\phi \ks)}
  \simeq(0.39+0.07i)[(\delta_{LL}^d)_{23}+(\delta_{RR}^d)_{23}]
  +(110+18i)[(\delta_{LR}^d)_{23}+(\delta_{RL}^d)_{23}].
  \label{aa1}
\end{equation}
In this case, all $LL$, $RR$, $LR$ and $RL$ dominated models
can lead to a negative value of $\mathcal{S}_{\phi \ks}$.
We see that in order to have a significant difference between
$\mathcal{S}_{\phi \ks}$ and $\mathcal{S}_{J/\psi \ks}$,
$LL$ and $RR$ dominated models require small gluino and
squark masses. 
Thus, some $LL$ and $RR$ dominated models with large
$m_{\tilde{g}}$ and $m_{\tilde{q}}$ may be excluded by
measurements of $\mathcal{S}_{\phi \ks}$;   
for instance, an observation of $\mathcal{S}_{\phi \ks}<0$
would exclude 
$m_{\tilde{g}}\simeq m_{\tilde{q}}\gsim$ 300~GeV.

Once more precise experimental data are available and the
existence of a new physics contribution in
$\mathcal{S}_{\phi \ks}$ is confirmed, it will be necessary to find  
further evidence to prove that $\mathcal{S}_{\phi \ks}$
indeed includes a SUSY contribution.
For this purpose, we are able to benefit from other 
$b\to s$ transitions, which can also be described by the
mass insertion $(\delta_{AB}^d)_{23}$. 
In the rest of this section, we shall discuss the process
$B\to\eta'\ks$.

$CP$-violation in $B\to\eta'\ks$ was first reported by 
the Belle collaboration in summer 2002 \cite{Chen:2002af}.
Averaging with the result from the BaBar collaboration
reported in spring 2003 \cite{Aubert:2003ez}, we obtain 
$S_{\ep \ks}=0.33\pm 0.34$.
The $\eta'$ meson is known to be composed of $u\overline{u}$,
$d\overline{d}$ and $s\overline{s}$ accompanied by a small amount of
other particles such as gluonium and $c\overline{c}$ \textit{etc.} 
\footnote{
  We do not consider contributions from exotic
  particles here. 
  However, since an unexpectedly large branching ratio is
  observed in the $B\to\eta' K$ processes, possible large
  contributions from such particles have been considered. 
  The gluonium contributions to $B\to \eta' \ks$ 
  including the possibility that SUSY effects
  also enhance the branching ratio of $B\to\eta' K$ 
  is discussed in \cite{Khalil:2003bi}.
}. 
Apart from the exotic particles, the $B\to\eta' \ks$ process 
comes from the $B_{d}^0-\overline{B}_{d}^0$ mixing box diagram
and the penguin and tree decay diagrams. 
While there are two penguin diagrams $b\to s\overline{s}s$ 
and $b\to d\overline{d}d$, there is only one tree diagram, which
furthermore is Cabibbo suppressed. 
As a result, the tree contribution is very small and
estimated to be less than 1\%.
Thus, $B\to\eta' \ks$ and $B\to\phi \ks$ are approximately 
the same apart from the parity of the final states. 
In the following, we shall investigate whether this parity 
difference can lead to the experimental observation 
$\mathcal{S}_{\phi \ks} < 0 < \mathcal{S}_{\eta'\ks}
< \mathcal{S}_{J/\psi \ks}$. 

Let us first see the consequence of this parity difference
in the SUSY contributions by comparing the computation of
the matrix elements for $B\to\phi \ks$ and $B\to\eta'\ks$ in
the case of the $b\to s\overline{s}s$ transition.
In the naive factorization approximation, the amplitudes are
written as a product of Wilson coefficients, form factors
and decay constants:
\begin{equation}
  A(B \to \phi (\eta') K) \propto 
  C_{\mathrm{Wilson}}\ F^{B\to K} f_{\phi(\eta')}.
\end{equation}
The decay constants appear in the calculation by sandwiching
the $V\pm A$ current, corresponding to $O_i$ and
$\tilde{O}_i$ contributions, respectively, between
$\phi(\eta')$ and the vacuum:
\begin{eqnarray}
  \langle 0|\overline{s} \gamma_\mu (1\pm\gamma_5)s|\phi\rangle 
  &=&
  + m_{\phi}f_{\phi}\epsilon_{\mu}
  \label{eq:decay_const_phi}
  \\
  \langle 0|\overline{s} \gamma_{\mu}(1\pm\gamma_5)s|\eta'\rangle
  &=&
  \pm i f_{\eta'} p_{\mu}
  \label{eq:decay_const_eta}
\end{eqnarray}
As can be seen from (\ref{eq:decay_const_eta}), the overall
sign flips for the $V+A$ and $V-A$ currents in the case of
$\eta'$. 
Thus, the contributions for $\mathcal{S}_{\eta' \ks}$ coming 
from $\tilde{O}_i$ and $O_i$ have opposite signs.
Accordingly, the numerical results are found to be 
\cite{Khalil:2003bi}
\begin{eqnarray}
  \frac{A^{\mathrm{SUSY}}(\eta'\ks)}{A^{\mathrm{SM}}(\eta'\ks)}
  \!\! &\simeq& \!\!
  (0.15+0.03i)[(\delta_{LL}^d)_{23}-(\delta_{RR}^d)_{23}]+
  (69+12i)[(\delta_{LR}^d)_{23}-(\delta_{RL}^d)_{23}],
  \label{eq:res2}\\
  \frac{A^{\mathrm{SUSY}}(\eta'\ks)}{A^{\mathrm{SM}}(\eta'\ks)}
  \!\! &\simeq& \!\!
  (0.41+0.08i)[(\delta_{LL}^d)_{23}-(\delta_{RR}^d)_{23}]+
  (115+20i)[(\delta_{LR}^d)_{23}-(\delta_{RL}^d)_{23}]
  \label{eq:res3}
\end{eqnarray}
for $m_{\tilde{g}}\simeq m_{\tilde{q}}$ = 500~GeV and
300~GeV, respectively.
As expected, the result is very similar to those for 
$B\to\phi \ks$ in (\ref{eq:res1}) and (\ref{aa1}) 
except for the overall signs of the $RR$ and $RL$ mass
insertions.
Thus, if $\mathcal{S}_{\phi \ks}$ and
$\mathcal{S}_{\eta' \ks}$ differ, one would need some contributions
from the $RR$ and/or $RL$ mass insertion.

\begin{figure}[tbp]
\begin{center}
\psfrag{x}[r][r][1.4]{$\theta_{\susy}$}\psfrag{y}[c][c][1.4]{$S_{\phi \ks},\  S_{\ep \ks}$}
\psfrag{p}[r][r][1.25]{$\pi$}
\psfrag{0}[r][r][1.25]{0}
\psfrag{2}[r][r][1.25]{$\overline{2}$}
\scalebox{0.8}{
\includegraphics[width=10cm,height=8cm,keepaspectratio]{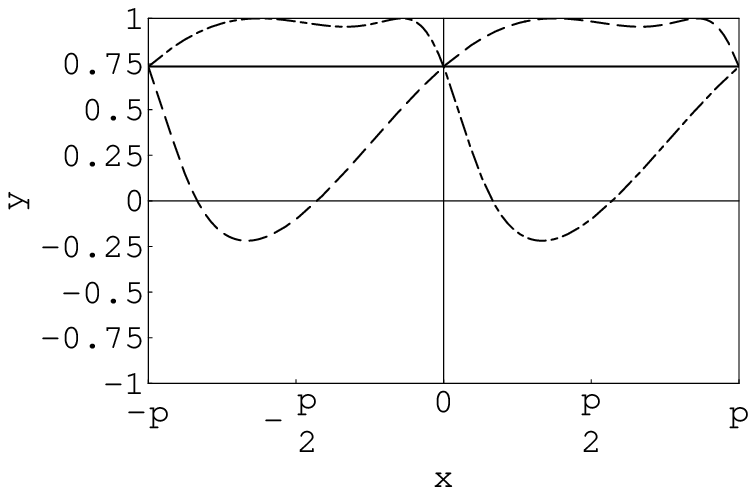}
\includegraphics[width=10cm,height=8cm,keepaspectratio]{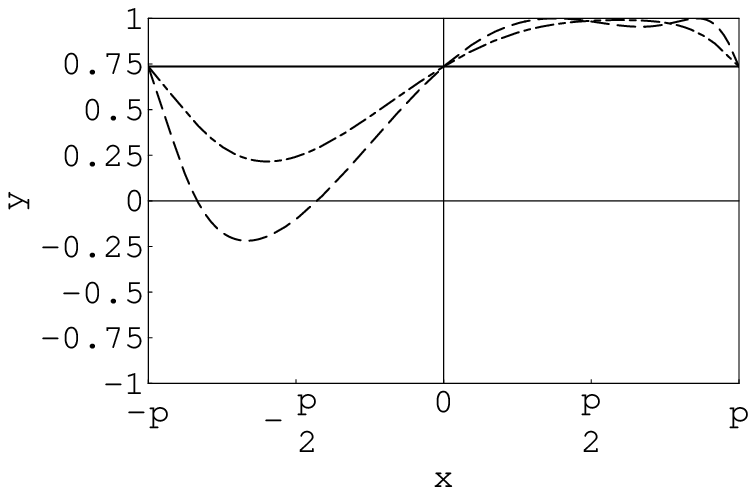}}
\caption{
  Possible solution for the puzzle of 
  $\mathcal{S}_{\phi \ks} < 0 < \mathcal{S}_{\eta'\ks} < 
  \mathcal{S}_{J/\psi \ks}$. 
  Dashed and dashed-dotted lines are results for
  $\mathcal{S}_{\phi \ks}$ and $\mathcal{S}_{\eta'\ks}$,
  respectively. 
  Left figure is for the case that $RR$ and/or $RL$ are
  dominant.
  Right figure is for the case that there are both $LL$
  (and/or $LR$) and $RR$ (and/or $RL$) contributions and the
  $LL$ (and/or $LR$) amplitude is 4 times larger than $RR$
  (and/or $RL$) one whereas the phase of the former is equal
  to the phase of the latter. 
  For both cases, we fix 
  $|A^{\mathrm{SUSY}}(\phi \ks)/A^{\mathrm{SM}}(\phi \ks)|
  \simeq 0.5$. 
}
\label{fig:3}
\end{center}
\end{figure}

Now we shall discuss how we could reproduce the relation
$\mathcal{S}_{\phi \ks} < 0 < \mathcal{S}_{\eta'\ks}
< \mathcal{S}_{J/\psi \ks}$ as indicated by experiment. 
Let us first consider $RR$ and/or $RL$ dominated models with 
$A^{\mathrm{SUSY}}(\phi \ks)/A^{\mathrm{SM}}(\phi \ks)
\simeq 0.5\ e^{i\theta_{\mathrm{SUSY}}}$. 
In this case, the overall sign flips for $\eta' \ks$,
so we obtain 
$A^{\mathrm{SUSY}}(\eta'\ks)/A^{\mathrm{SM}}(\eta'\ks)\simeq$
$-0.5\ e^{i\theta_{\mathrm{SUSY}}}$. 
The results for $S_{\phi \ks}$ and $S_{\eta'\ks}$ are shown
in Figure~\ref{fig:3} (left). 
In this way,   
$\mathcal{S}_{\phi \ks}$ and $\mathcal{S}_{\eta' \ks}$ 
may differ and we can obtain 
$\mathcal{S}_{\phi \ks} < 0 < \mathcal{S}_{\eta' \ks}$. 
However, $\mathcal{S}_{\eta'\ks}$ is rather large compared
to $S_{J/\psi \ks}$.  
To solve this problem, we need to consider models with both
sizable $LL$ (and/or $LR$) and $RR$ (and/or $RL$) mass
insertions. 
Let us give an example: 
the amplitude of the $LL$ (and/or $LR$) mass insertion is 
4 times larger than the one for $RR$ (and/or $RL$)
whereas their phases are equal. 
In this case with the same SUSY contributions to
$\mathcal{S}_{\phi \ks}$, 
$A^{\mathrm{SUSY}}(\phi \ks)/A^{\mathrm{SM}}(\phi \ks)
\simeq 0.5e^{i\theta_{\mathrm{SUSY}}}$, we obtain 
$A^{\mathrm{SUSY}}(\eta'\ks)/A^{\mathrm{SM}}(\eta'\ks)
\simeq 0.3e^{i\theta_{\mathrm{SUSY}}}$. 
In this way, we are able to reproduce the pattern
$\mathcal{S}_{\phi \ks} < 0 < \mathcal{S}_{\eta' \ks} <
 \mathcal{S}_{J/\psi \ks}$ 
(see Figure~\ref{fig:3} (right)).

\section{Model independent study of $b\to s\gamma$ and 
  $b\to s\ell^+\ell^-$ processes}
\label{sec:mode_indep_bsgamma}
\def\bsll{$b \rightarrow s \ell^+ \ell^- $ }

In the previous section we considered the supersymmetric
extension of the Standard Model and discussed its possible
effects on the $b\to s$ transition processes.
In contrast, in this section we explain what we can learn
from the experimental data in a model independent way.
To be specific, we consider the $b\to s\gamma$ and 
$b\to s\ell^+\ell^-$ decays
\cite{Ali:1994bf,Fukae:1998qy,Fukae:2000jv,Hiller:2003js}.

As we discuss below, there are many types of interactions
which may contribute to $b\to s\gamma$ and \bsll. 
The new physics effects originate from the energy scale
higher than the electro-weak scale, and below that new
physics scale, the effects evolve down to the electro-weak
scale.  
Then, at the electro-weak scale, they are matched onto the
low energy effective theory, which is valid below the
electro-weak scale.  
The new physics effects can thus be expressed in terms of
higher dimensional operators such as four-Fermi interactions 
and dimension-five interactions. 
For \bsll and  $b\to s\gamma$ processes, 
$\bar{s} b \ell^+ \ell^-$ ($O_9$ and $O_{10}$), 
$\bar{s} \sigma_{\mu \nu} b F^{\mu \nu}$ ($O_7$) and 
$\bar{s} \sigma_{\mu \nu}T^a b G^a_{\mu \nu}$ ($O_8$)
are such operators.
(There are several these operators with different chiral 
structures.)
For each operator there is the Wilson coeffeicient
$C_i^{\mathrm{eff}}$, in which the new physics effects are
encoded. 
The four-quark operators can contribute to \bsll through the
one-loop matrix elements and the operator mixing
\cite{Hiller:2003js}.
We assume that such effects are small compared to the
contributions from the tree level matrix elements of 
$\bar{s} b \ell^+ \ell^-$ operators.

Within the leading logarithmic approximation of QCD
corrections, we obtain the following amplitude for 
$b\to s\gamma$.
\begin{equation}
  M(b\to s\gamma)= 
  \frac{4 G_F}{\sqrt{2}} \frac{e}{16 \pi^2}
  V_{tb} V_{ts}^* m_b 
  \left[
    C_{7L}^{\mathrm{eff}}  
    (\bar{s}  \sigma_{\mu \nu} b_R)
    + C_{7R}^{\mathrm{eff}}
    (\bar{s} \sigma_{\mu \nu} b_L )
  \right] 
  F^{\mu \nu},
\end{equation}
where $F_{\mu \nu}$ stands for 
$-i (q_{\mu} \epsilon_{\nu}^{\ast}-q_{\nu} \epsilon_{\mu}^{\ast})$. 
In the Standard Model 
$C_{7R}^{\mathrm{eff}}=\frac{m_s}{m_b} C_{7L}^{\mathrm{eff}}$.
In the left-right symmetric models \cite{Senjanovic:1975rk},
$C_{7R}^{\mathrm{eff}}$ can be as large as
$C_{7L}^{\mathrm{eff}}$
\cite{Fujikawa:1993zu,Babu:1993hx,Cho:zb}.
Using the branching ratio of $B \to X_s \gamma$,
we can constrain 
$|C_{7L}^{\mathrm{eff}}|^2+|C_{7R}^{\mathrm{eff}}|^2$. 

Since the branching fraction does not tell us about
the ratio $C_{7R}^{\mathrm{eff}}/C_{7L}^{\mathrm{eff}}$,
the observables which are sensitive to the ratio are
needed. 
Three methods using $B$ meson decays have been proposed. 
One can extract the ratio from the time dependent CP asymmetry
of $b\to s\gamma$ \cite{Atwood:1997zr}.
Another measurement which is sensitive to the ratio is the
transverse polarization of $B\to K^*\gamma$
\cite{Melikhov:1998cd}.
This can be measured by using the decay chains
$B\to K^* \gamma^* \to  K\pi l^+ l^-$
\cite{Kruger:1999xa,Kim:2000dq}.
The azimuthal angle distribution is sensitive to the ratio.
The distribution at low invariant dilepton mass region must
be measured so that the decay amplitude from a $Z$ exchange
and box diagram contribution is suppressed. 
The other method uses 
$B \to K_{res} \gamma \to K \pi \pi \gamma$ and 
triple momentum correlation
$p_{\gamma} \cdot (p_K \times p_{\pi})$
\cite{Gronau:2001ng,Gronau:2002rz}.

We next consider the possible new physics effects on 
$b \to s \ell^+ \ell^-$.
In addition to $C_7^{\mathrm{eff}}$, 
there are ten local four Fermi interactions which contribute
to \bsll. 
One can write down the amplitude including all the
contributions \cite{Fukae:1998qy,Fukae:2000jv}:
\begin{eqnarray}
  \lefteqn{
    \mathcal{M}(b \to s \ell^+ \ell^-)
    = \frac{G_F \alpha }{\sqrt{2}\pi }V_{ts}^*V_{tb}
  }
  \nonumber\\
  &&
  \left[
    (C_{LL}+C_9^{\mathrm{eff}}-C_{10})
    ~\bar{s}_L \gamma_\mu b_L 
    ~\bar{l}_L \gamma^\mu l_L
    +
    (C_{LR}+C_9^{\mathrm{eff}}+C_{10})
    ~\bar{s}_L \gamma_\mu b_L  
    ~\bar{l}_R \gamma^\mu l_R
  \right.
  \nonumber\\
  &&
  + C_{RL}
  ~\bar{s}_R \gamma_\mu b_R 
  ~\bar{l}_L \gamma^\mu l_L
  + C_{RR}
  ~\bar{s}_R \gamma_\mu b_R  
  ~\bar{l}_R \gamma^\mu l_R
  \nonumber \\
  &&
  + C_{LRLR}
  ~\bar{s}_L b_R ~\bar{l}_L l_R
  + C_{RLLR}~\bar{s}_R b_L ~\bar{l}_L l_R
  \nonumber \\
  &&
  + C_{LRRL} ~\bar{s}_L b_R ~\bar{l}_R l_L  
  + C_{RLRL} ~\bar{s}_R b_L ~\bar{l}_R l_L
  \nonumber \\
  &&
  + C_T
  ~\bar{s} \sigma_{\mu \nu } b 
  ~\bar{l} \sigma^{\mu \nu } l 
  + i C_{TE}
  ~\bar{s} \sigma_{\mu \nu } b 
  ~\bar{l} \sigma_{\alpha \beta } l 
  ~\epsilon^{\mu \nu \alpha \beta }
  \nonumber \\
  && 
  \left.
    - 2 m_b C_{7R}^{\mathrm{eff}}  
    (\bar{s} 
    i \sigma_{\mu \nu } b_L) 
    (\bar{l} \gamma^\mu l) \frac{q^\nu}{q^2}
    - 2 m_b C_{7L}^{\mathrm{eff}} (\bar{s} i \sigma_{\mu \nu} b_R
    )(\bar{l} \gamma^\mu l) \frac{q^\nu}{q^2}
  \right].
\end{eqnarray}
The Standard Model predicts to
\begin{eqnarray}
  &&
  (C_{LL},C_{LR},C_{RR},C_{RL},C_{RLRL},C_{LRLR},C_{LRRL},C_{RLLR})=0, 
  \\
  && 
  (C_{7R}^{\mathrm{eff}},C_{7L}^{\mathrm{eff}}) = 
  (\frac{m_s}{m_b}, 1) C_{7SM}^{\mathrm{eff}},
  \\
  &&
  (C_T,C_{TE}) = 0,
\end{eqnarray}
where $(C_9^{\mathrm{eff}},C_{10}, C_{7SM}^{\mathrm{eff}})$
are the Standard Model coefficients.  
The numerical values of the corresponding Wilson coefficients are 
$C_9^{NDR}=4.153$, $C_{10}=-4.546$,
$C_{7SM}^{\mathrm{eff}}=-0.311$.  
$C_9^{\mathrm{eff}}$ is close to $-C_{10}$.
Two-loop calculation has also been completed recently
\cite{Bobeth:1999mk,Asatrian:2001de}. 
Beyond the Standard Model, one-loop calculation
of these Wilson coefficients is available for the Minimal
Supersymmetric Standard Model (see, for example,
\cite{Xiong:up} for a recent paper).

The branching ratio of \bsll is most sensitive to the
coefficient $C_{LL}$, since the interference of the $C_{LL}$
and $C_{9}^{\mathrm{eff}}-C_{10}$ is large.
Depending on the sign of $C_{LL}$, the interference with the
Standard Model contribution can be constructive or destructive.

\begin{figure}[tbp]
  \centering
  \includegraphics[viewport=50 300 780 760,clip,scale=0.5]{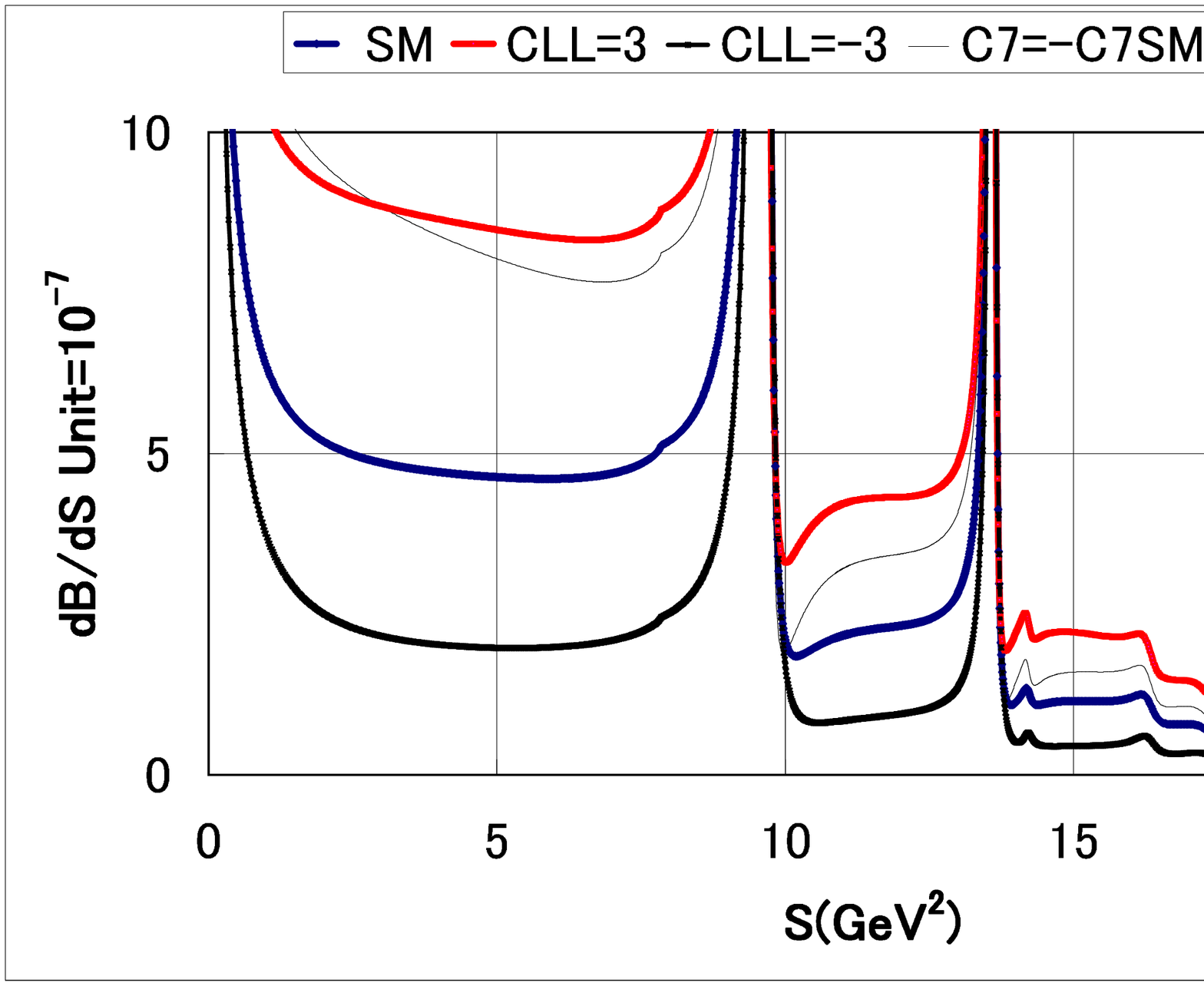}
  \caption{
    Differential branching ratio $ \frac{dB}{dS}$ for
    \bsll.
  }
  \label{dbds}
\end{figure}

\begin{figure}[tbp]
  \centering
  \includegraphics[viewport=50 300 780 760,clip,scale=0.5]{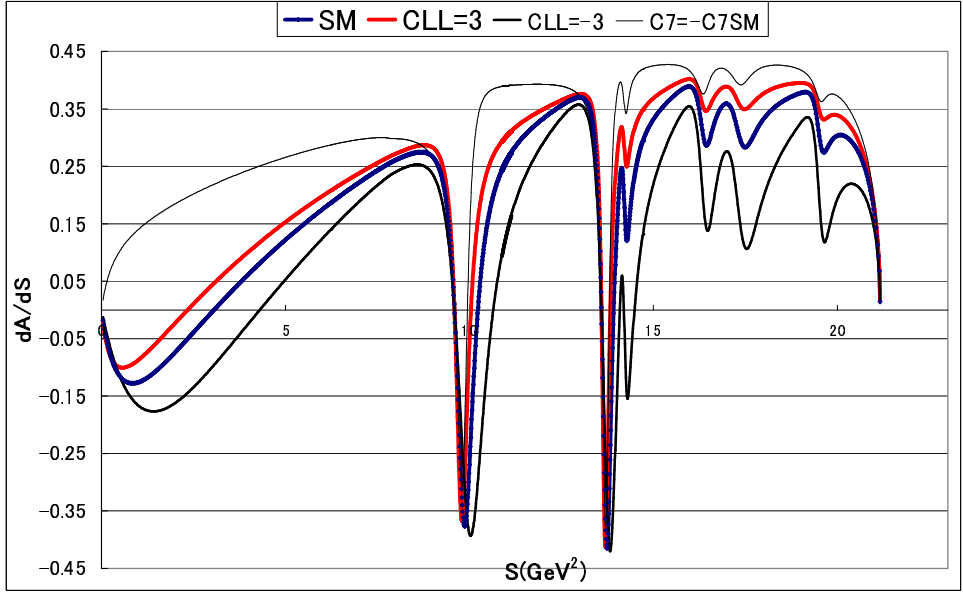}
  \caption{
    Forward-backward asymmetry $\frac{dA}{ds}$ for \bsll.
  }
\label{dads}
\end{figure}

In Figures \ref{dbds} and \ref{dads}, we show the
differential branching ratio $\frac{dB}{ds}$ and the
forward-backward asymmetry $\frac{dA}{ds}$, respectively. 
We choose three sets of the coefficients and show dilepton
mass squared ($s$) distribution. 
The three cases correspond to:
(1) $C_{LL}=3$ , (2) $C_{LL}=-3$, and 
(3) $(C_{7R}^{\mathrm{eff}},C_{7L}^{\mathrm{eff}})$ = 
$-(\frac{m_s}{m_b},1) \ C_{7 SM}$.
When $C_{LL}$ is positive, the branching ratio is larger
than the Standard Model value. 
If $C_{LL}$  is negative, it decreases. 
If we change the sign of $C_7$ compared with the Standard Model,
the branching ratio also increases as in the case of $C_{LL}>0$.
The case (3) can be clearly distinguished from the Standard Model
by studying the forward and backward asymmetry.

As shown in Figure~\ref{dads}, in the Standard Model there
is a zero crossing point of the forward-backward asymmetry
in the low invariant mass squared region. 
If the sign of $C_{7L}$ is different from the Standard Model, 
the zero crossing point may disappear from the
forward-backward asymmetry. 
Such a scenario is suggested in a supergravity model
\cite{Goto:1996dh}. 
In another new physics model scenario including the down
type SU(2) singlet quark, $Z$ FCNC current may contribute to
$b \to sl^+l^-$ at the tree level. 
The tree level $Z$ FCNC contribution may give a large
contribution to $C_{10}$ and can change the sign of the
forward-backward asymmetry. 
Besides the observables mentioned above, the
forward-backward CP asymmetry of \bsll is useful under the
presence of new source of CP violation
\cite{Buchalla:2000sk}. 
To conclude, with various observables and combination of them,
we can test the new physics scenarios if they contribute to
\bsll interactions.

\section{Lepton Flavor Violation}
\label{sec:Lepton_flavor_violation}
Strictly speaking, the Standard Model already has to be
modified by introducing tiny neutrino masses,
in order to incorporate the Lepton Flavor Violating
(LFV) phenomena observed in the neutrino sector.
Neutrino mixing $\nu_\mu$-$\nu_\tau$ was first discovered in 
atmospheric neutrino measurements
at Super-Kamiokande~\cite{Fukuda:1998mi}, and it is
being further confirmed by the K2K experiment \cite{Ahn:2002up}.
The neutrino oscillation was also confirmed in solar
neutrinos, which come from $\nu_e$-$\nu_\mu$ mixing, 
by both the Super-Kamiokande \cite{Fukuda:2001nj} and SNO
\cite{Ahmad:2002jz,Ahmad:2002ka} experiments.
More recently, the Kamland experiment pinned down the 
explanation of the 
solar neutrino problem to the large mixing angle (LMA) solution
\cite{Eguchi:2002dm}. 
Thus, neutrino oscillation experiments are providing
high precision measurements in the neutrino sector. 
It is very interesting that the $\nu_e$-$\nu_\mu$ and
$\nu_\mu$-$\nu_\tau$ mixing angles are found to be almost
maximal and the neutrino mass structure is quite different
from that of the quark sector.

Now it is known that lepton-flavor symmetries are not exact in
Nature. 
However, the magnitude of LFV processes in the charged lepton
sector is not obvious.
The tiny neutrino masses do not lead to sizable LFV
processes in the charged lepton sector, since the event rates
are suppressed by the fourth power of $(m_\nu/m_W)$.
Thus, searches for LFV in the charged lepton sector will
probe physics beyond the Standard Model and the origin of
the neutrino masses. 

The $\tau$ lepton is a member of the third generation and is the
heaviest charged lepton. 
It can decay into quarks and leptons in the first and second
generations. 
This may imply that $\tau$ lepton physics could provide some
clues to puzzles in the family structure.  
In fact, one naively expects the heavier quarks and
leptons to be more sensitive to the dynamics responsible for
fermion mass generation. 

In the following we discuss LFV $\tau$ lepton decay in SUSY
models, especially in the supersymmetric seesaw mechanism and SUSY GUTs,
and other models such as extra-dimension models and R-parity
violating SUSY models.

\subsection{LFV in the Supersymmetric Models}
The LFV in the SUSY extension of the Standard Model comes from the soft
SUSY breaking terms, since the supersymmetric interactions
have the same flavor structure as in the Standard Model. 
The soft SUSY breaking terms in the lepton sector are as
follows, 
\begin{equation}
  -\mathcal{L}
  = (m_{E}^2)_{ij} \tilde{e}_{Ri}^\dagger \tilde{e}_{Rj}
  + (m_{L}^2)_{ij} \tilde{l}_{Li}^\dagger \tilde{l}_{Lj}
  + \left\{
    (A_{e})_{ij}H_1 \tilde{e}_{Ri}^\dagger \tilde{l}_{Lj}+h.c.
  \right\}
  \label{dim5}
\end{equation}
where $(m_{E}^2)_{ij}$ and $(m_{L}^2)_{ij}$ are mass
matrices for the right-handed sleptons $\tilde{e}_{Rj}$ and
the left-handed sleptons 
$\tilde{l}_{Lj}(\equiv(\tilde{\nu}_{Lj}, \tilde{e}_{Lj}))$,
respectively.
$(A_{e})_{ij}$ is the trilinear scalar coupling matrix. 

\begin{figure}[tbp]
  \begin{center}
    \includegraphics[width=8cm]{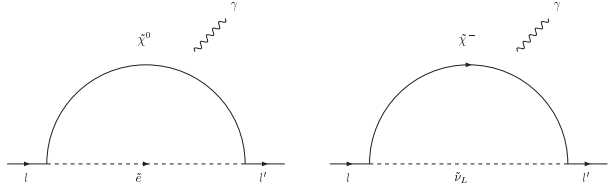}
  \end{center}
  \caption{
    Feynman diagrams to generate 
    $l^-\rightarrow l^{\prime-} \gamma$ in the 
    SUSY models with conserved R parity. 
    $\tilde{e}$, $\tilde{\nu}$, $\tilde{\chi}^0$ and
    $\tilde{\chi}^-$ represent charged slepton, sneutrino, 
    chargino, and neutralino, respectively.
  }
  \label{fig:mueg1}
\end{figure}

LFV processes for charged leptons are radiative if R
parity is conserved, since the SUSY interactions must be
bilinears of the SUSY fields. 
Thus, $\tau^-\rightarrow\mu^-(e^-)\gamma$ and 
$\mu^- \rightarrow e^-\gamma$ are the most sensitive to the
flavor structure of the soft SUSY breaking terms except for
some exceptional cases. 
These processes are generated by diagrams in
Figure~\ref{fig:mueg1}.
The effective operators relevant to 
$l^-\rightarrow l^{\prime-}\gamma$ 
are flavor-violating magnetic moment operators,
\begin{equation}
  \mathcal{H}=
  \sum_{l > l^\prime}\frac{4G_F}{\sqrt{2}}
  \left[
    m_l A^{l l^\prime}_R \overline{l}\sigma^{\mu\nu} P_R l^\prime
    +
    m_l A^{l l^\prime}_L \overline{l}\sigma^{\mu\nu} P_L l^\prime
  \right]F_{\mu\nu} +h.c.,
  \label{eff_op_lfv}
\end{equation}
where $P_{L/R}=(1\mp\gamma_5)/2$, and the branching ratios
are given as 
\begin{equation}
  Br(l^-\rightarrow l^{\prime-} \gamma) = 
  384 \pi^2 (|A^{l l^\prime}_R|^2+|A^{l l^\prime}_L|^2)~
  Br(l^-\rightarrow l^{\prime-} \nu_l \overline{\nu}_{l^\prime}).
\end{equation}
Here, 
$Br(\tau^-\rightarrow\mu^-(e^-)\nu_\tau\overline{\nu}_{\mu(e)})
\simeq 0.17$ and 
$Br(\mu^-\rightarrow e^-\nu_\mu\overline{\nu}_e)= 1$.

The coefficients in Eq.~(\ref{eff_op_lfv}) are approximately given
as 
\begin{eqnarray}
  \label{eq:AtaulR}
  A^{\tau l^\prime}_R
  &=&
  \frac{\sqrt{2}e}{4 G_F} 
  \frac{\alpha_Y}{4\pi} \frac{\tan\beta}{m_{SUSY}^2}
  \left[
    -\frac{1}{120} \delta^R_{\tau l^\prime}
  \right],
  \\
  \label{eq:AtaulL}
  A^{\tau l^\prime}_L
  &=&
  \frac{\sqrt{2}e}{4 G_F} 
  \frac{\alpha_2}{4\pi} \frac{\tan\beta}{m_{SUSY}^2}
  \left[
    (\frac{1}{30}+\frac{t_W^2}{24}) \delta^L_{\tau l^\prime}
  \right],
  \\
  \label{eq:AmueR}
  A^{\mu e}_R
  &=&
  \frac{\sqrt{2}e}{4 G_F} 
  \frac{\alpha_Y}{4\pi} \frac{\tan\beta}{m_{SUSY}^2}
  \left[
    -\frac{1}{120} \delta^R_{\mu e} 
    +\frac{1}{120} \delta^R_{\mu \tau} \delta^R_{\tau e}
    -\frac{1}{60} \frac{m_\tau}{m_\mu} \delta^L_{\mu \tau}\delta^R_{\tau e}
  \right],
  \\
  \label{eq:AmueL}
  A^{\mu e}_L
  &=&
  \frac{\sqrt{2}e}{4 G_F} 
  \frac{\alpha_2}{4\pi} \frac{\tan\beta}{m_{SUSY}^2}
  \left[
    (\frac{1}{30}+\frac{t_W^2}{24}) \delta^L_{\mu e}
    -(\frac{1}{80}+\frac{7t_W^2}{240}) \delta^L_{\mu \tau} \delta^L_{\tau e}
  \right.
  \nonumber\\
  &&
  \;\;\;\;\;\;\;\;\;\;\;\;\;\;
  \left.
    -\frac{t^2_W}{60} \frac{m_\tau}{m_\mu} \delta^R_{\mu \tau}\delta^L_{\tau e}
  \right],
\end{eqnarray}
assuming for simplicity that all SUSY particle masses are equal
to $m_{SUSY}$ and $\tan\beta\gg 1$. 
Here, $t_W\equiv \tan\theta_W$, where $\theta_W$ is the
Weinberg angle, and the mass insertion parameters are given
as 
\begin{eqnarray}
  \delta^R_{l l^\prime}
  = \left(\frac{(m_{E}^2)_{l l^\prime}}{m_{SUSY}^2}\right),
  &&
  \delta^L_{l l^\prime}
  = \left(\frac{(m_{L}^2)_{l l^\prime}}{m_{SUSY}^2}\right).
\end{eqnarray}
When both the 13 and 23 generation components of
the slepton mass matrices are non-vanishing, 
$\mu^-\rightarrow e^-\gamma$ is generated via a scalar tau
lepton exchange. 
In particular, if both the left-handed and right-handed mixings
are sizable, the branching ratio is enhanced by 
$(m_\tau/m_\mu)^2$ compared to the case where 
only left-handed or right-handed mixing angles are non-vanishing. 
The off-diagonal components in $(A_{e})_{ij}$ are
sub-dominant in these processes since the contribution is
not proportional to $\tan\beta$.

We list constraints on $\delta^R_{ll^\prime}$ and
$\delta^L_{ll^\prime}$ from current experimental bounds on
$Br(\tau^-\rightarrow\mu^-(e^-)\gamma)$, which are derived by the
Belle experiment, and $Br(\mu^-\rightarrow e^-\gamma)$ in
Table~\ref{delta_LFV1}.  In this table, we take $\tan\beta=10$ and
$m_{\rm SUSY}=100$~GeV and $300$~GeV.  The constraints from $\mu^-\rightarrow
e^-\gamma$ on the slepton mixings are quite stringent.  On the other
hand, the current bounds on the LFV $\tau$ lepton decay modes 
independently give sizable constraints on $|\delta^L_{\tau\mu}|$ 
and $|\delta^L_{\tau
e}|$. Furthermore, while the current constraint on
$|\delta^L_{\mu\tau}\delta^L_{\tau e}|$ from the LFV $\tau$ lepton decay
is weaker than that from the LFV muon decay, improvement of the
LFV $\tau$ lepton decay modes by about an order of magnitude will give a
bound on $|\delta^L_{\mu\tau}\delta^L_{\tau e}|$ competitive with
that from the LFV muon decay. 

\begin{table}[tbp]
  \begin{center}
    \begin{tabular}{|c|c|c|c|c|c|}
      \hline
      $m_{SUSY}$ &
      $|\delta^L_{\tau \mu}|$ & 
      $|\delta^L_{\tau e}|$ & 
      $|\delta^L_{\mu e}|$ & 
      $|\delta^L_{\mu\tau }\delta^L_{\tau e}|$ & 
      $|\delta^R_{\mu\tau}\delta^L_{\tau e}|$
      \\
      \hline
      $100$~GeV& 
      $2\times 10^{-2}$& 
      $2\times 10^{-2}$& 
      $4\times 10^{-5}$&
      $1\times 10^{-4}$&
      $2\times 10^{-5}$
      \\
      & & & &
      $(3\times 10^{-4})$ &
      $(7\times 10^{-3})$ 
      \\
      \hline
      $300$~GeV& 
      $2\times 10^{-1}$& 
      $2\times 10^{-1}$& 
      $4\times 10^{-4}$&
      $9\times 10^{-4}$&
      $2\times 10^{-4}$
      \\
      & & & &
      $(3\times 10^{-2})$ &
      $(6\times 10^{-1})$ 
      \\
      \hline\hline
      ${m_{SUSY}}$&     
      $|\delta^R_{\tau \mu}|$ &
      $|\delta^R_{\tau e}|$ &
      $|\delta^R_{\mu e}|$ &
      $|\delta^R_{\mu\tau }\delta^R_{\tau e}|$ &
      $|\delta^L_{\mu\tau }\delta^R_{\tau e}|$ 
      \\
      \hline
      100~GeV&
      $3\times 10^{-1}$ & 
      $3\times 10^{-1}$ &
      $9\times 10^{-4}$ &
      $9\times 10^{-4}$ &
      $2\times 10^{-5}$ \\
      & & & & 
      $(1\times10^{-1})$ & 
      $(7\times 10^{-3})$ \\
      \hline
      300~GeV&
      $3$ & 
      $3$ &
      $8\times 10^{-3}$ &
      $8\times 10^{-3}$ &
      $2\times 10^{-4}$ \\
      & & & & 
      $(9)$ & 
      $(6\times 10^{-1})$ \\
      \hline
    \end{tabular}
  \end{center}
  \caption{
    Constraints on $\delta^R_{ll^\prime}$ and 
    $\delta^L_{ll^\prime}$ from current experimental bounds
    on $Br(l^-\rightarrow l^{\prime-}\gamma)$.  Here, we use
    the results of the LFV tau decay search from the Belle experiment.
    We take $\tan\beta=10$ and $m_{\rm SUSY}=100$~GeV and  $300$~GeV. 
    The numbers in parentheses are derived from constraints
    on $|\delta^{(L/R)}_{\tau\mu}|$ and 
    $|\delta^{(L/R)}_{\tau e}|$.
  }
  \label{delta_LFV1}
\end{table}

The flavor structure of the soft SUSY breaking terms depends on the origin
of the SUSY breaking and the physics beyond the MSSM, as discussed in
Section~\ref{sec:Supersymmetric_models}. Even in the Universal scalar
mass scenario, the LFV Yukawa interaction may induce LFV slepton
mass terms radiatively. If the heavier leptons have larger LFV Yukawa
interactions, the $\tau$ lepton is the most sensitive to them. 
In the decoupling 
scenario, scalar $\tau$ leptons may be much lighter than other sleptons and 
have LFV interactions. Thus, the search for LFV $\tau$ lepton decay 
is important
to probe such new physics. In the next section we present predictions for 
the LFV $\tau$ lepton decay in the SUSY seesaw model and SUSY GUTs.

Finally, we discuss other tau LFV processes in SUSY models.  In a
broader parameter space, $\tau^-\rightarrow \mu^-(e^-)
\gamma$ are the largest tau LFV processes, unless they are
suppressed by some accidental cancellation or much heavier
SUSY particle masses. 
The LFV $\tau$ lepton decay modes to three leptons are
dominantly induced by the photon-penguin contributions, and
are correlated with $\tau^-\rightarrow\mu^-(e^-)\gamma$
as follows
\begin{eqnarray}
  Br(\tau^- \rightarrow \mu^- e^+e^-)/
  Br(\tau^- \rightarrow \mu^-\gamma) 
  &\simeq&
  1/94,
  \\ 
  Br(\tau^- \rightarrow \mu^-\mu^+\mu^-)/
  Br(\tau^- \rightarrow \mu^- \gamma) 
  &\simeq&
  1/440,
  \\
  Br(\tau^- \rightarrow e^-e^+e^-)/
  Br(\tau^- \rightarrow e^- \gamma)
  &\simeq&
  1/94,
  \\ 
  Br(\tau^- \rightarrow e^-\mu^+\mu^-)/
  Br(\tau^- \rightarrow e^- \gamma) 
  &\simeq&
  1/440.
\end{eqnarray}
The LFV $\tau$ decay modes into pseudoscalar mesons tend to be 
smaller than those to three leptons since the branching
ratios are not proportional to $\tan^2\beta$.
 
\begin{figure}[tbp]
  \begin{center}
    \includegraphics[width=10cm]{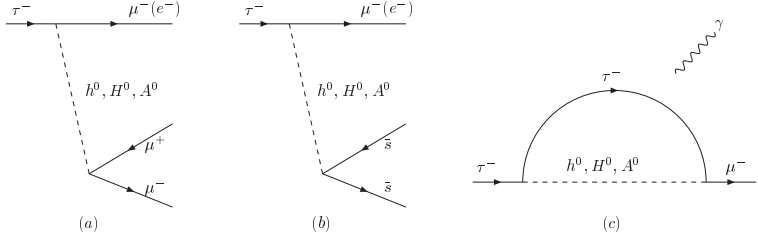}
  \end{center}
  \caption{
Feynman diagrams to generate 
(a) $\tau^-\rightarrow \mu^-(e^-) \mu^+\mu^-$,
(b) $\tau^-\rightarrow \mu^-(e^-) \eta$,
(c) $\tau^-\rightarrow \mu^-(e^-) \mu^-\gamma$,
induced by anomalous Higgs boson couplings.
  }
  \label{fig:lfv_higgsmediation}
\end{figure}

When sleptons are much heavier than the weak scale,
$Br(\tau^-\rightarrow\mu^-(e^-)\gamma)$ is suppressed. In this case,
the modes
$\tau^-\rightarrow \mu^-(e^-)\mu^+\mu^-$ and $\tau^-\rightarrow \mu^-
(e^-) \eta$ induced by Higgs boson exchange become relatively
important \cite{Babu:2002et}. The LFV anomalous Yukawa coupling for the Higgs
bosons is generated by radiative corrections, and it is not
suppressed by powers of the slepton masses. While these processes are
suppressed by a small Yukawa coupling constant for muons or strange
quarks, they may acquire sizable branching ratios when $\tan\beta$ is
large since the branching ratios are proportional to
$\tan^6\beta$. When $\delta_{\tau\mu}^{L}$ is non-vanishing, the
approximate formula for $Br(\tau^-\rightarrow \mu^-\mu^+\mu^-)$ is
given as \cite{Babu:2002et,Dedes:2002rh}
\begin{eqnarray}
Br(\tau^-\rightarrow \mu^-\mu^+\mu^-)
&=&
\frac{m_\mu^2 m_\tau^2 \epsilon_2^2 |\delta_{\tau\mu}^{L}|^2}
     {8\cos^6\beta}
Br(\tau^-\rightarrow\mu^-\nu_\tau\overline{\nu}_{\mu})
\nonumber\\
&&\times
\left[
\left(
\frac{\sin(\alpha-\beta) \cos\alpha}{M_{H^0}^2}
-\frac{\cos(\alpha-\beta) \sin\alpha}{M_{h^0}^2}
\right)^2
+
\frac{\sin^2\beta}{M_{A^0}^4}
\right],\nonumber\\
&\simeq& 3.8\times 10^{-7} \times |\delta_{\tau\mu}^{L}|^2
\left(\frac{\tan\beta}{60}\right)^6
\left(\frac{M_{A^0}}{100~{\rm GeV}}\right)^{-4},
\label{eq:tauetamu}
\end{eqnarray}
and $Br(\tau^-\rightarrow\mu^-\eta)$ is 8.4 times larger than
$Br(\tau^-\rightarrow \mu^-\mu^+\mu^-)$ \cite{Sher:2002ew}.  
Here, $\epsilon_2$ is a function of the SUSY particle
masses. 
We take limits of large $\tan\beta$ and equal SUSY breaking
mass parameters in the last step (\ref{eq:tauetamu}). 
Notice that
$\tau^-\rightarrow\mu^-\gamma$ also has a comparable branching ratio
to them since the Higgs loop diagram 
[Figure~\ref{fig:lfv_higgsmediation}(c)] is enhanced by the
$\tau$ lepton Yukawa coupling constant \cite{hs}. 
As a result, the ratio of the branching ratios for these LFV
$\tau$ lepton decay modes, which are induced by the Higgs
boson exchange, is 
$Br(\tau^- \rightarrow \mu^-\eta):
 Br(\tau^- \rightarrow \mu^-\gamma):
 Br(\tau^- \rightarrow \mu^-\mu^+\mu^-)
 = 8.4:1.5:1$.

\subsection{SUSY Seesaw Mechanism and SUSY GUTs}
In general, in seesaw and GUT models, LFV Yukawa
interactions are introduced. 
If the SUSY breaking mediation scale is higher than the GUT
\cite{Barbieri:1994pv,Barbieri:1995tw,Hisano:1996qq} or the
right-handed neutrino mass scale
\cite{Borzumati:1986qx,Hisano:1995nq,Hisano:1995cp,Casas:2001sr},
sizable LFV processes are predicted as mentioned in the previous section.

The most economical way to generate the tiny neutrino masses
is the seesaw mechanism. 
In the seesaw model, a neutrino Yukawa coupling $Y_\nu$ is
introduced, which is lepton-flavor violating.
In the supersymmetric extension, the off-diagonal components
of the left-handed slepton mass matrix are radiatively
induced, and they are approximately given by 
\begin{equation}
  (\delta m_{\tilde{L}}^2)_{ij}
  \simeq
  -\frac{1}{8\pi^2}(3m_0^2+A_0^2) 
  \sum_k (Y_\nu^\dagger)_{ki} (Y_\nu)_{kj}
  \log\frac{M_G}{M_{N_k}}, 
  \label{eq:delmseesaw}
\end{equation}
where ${M}_{N_i}$ are the $i$-th right-handed neutrino masses,
and $M_G$ is the Planck scale.
Here we assume the gravity mediation scenario, and the
parameters $m_0$ and $A_0$ are the universal scalar mass and
the universal trilinear scalar coupling. 
The predicted small neutrino mass matrix is
\begin{equation}
  ({m_\nu})_{ij} =
  \sum_k 
  \frac{(Y_\nu)_{ki}(Y_\nu)_{kj}\langle H_2\rangle^2}{{M}_{N_k}}.
  \label{eq:seesawmass}
\end{equation}
(\ref{eq:delmseesaw}) has a different structure from
(\ref{eq:seesawmass}). 
Thus, we may obtain independent information about the seesaw
mechanism from the charged LFV searches and the neutrino
oscillation experiments.

\begin{figure}[tbp]
  \begin{center}
    \includegraphics[width=10cm]{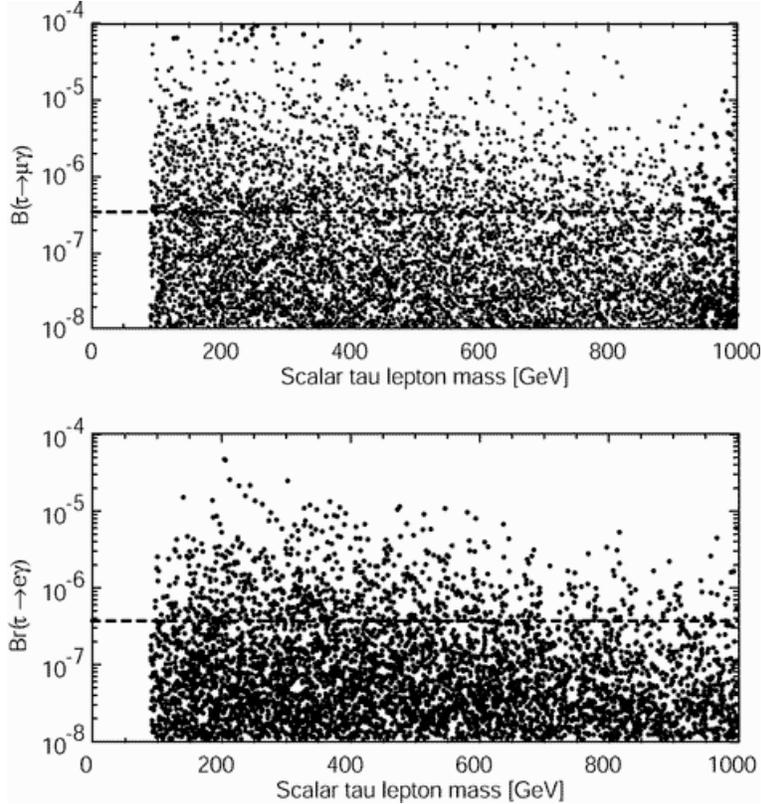}
  \end{center}
  \caption{
    $Br(\tau^-\rightarrow\mu^- \gamma)$ and 
    $Br(\tau^-\rightarrow e^- \gamma)$
    in the SUSY seesaw mechanism, assuming the gravity
    mediation scenario for the SUSY breaking. 
    Dashed lines show the current experimental bounds from the Belle 
    experiment.
    Here, we take $\tan\beta=30$, the $SU(2)_L$ gaugino mass
    200~GeV, $A_0=0$, and positive Higgsino mass.  
    We fix the neutrino Yukawa coupling by using the
    neutrino oscillation data under assumptions of the
    neutrino Yukawa coupling.
  } 
  \label{fig:seesaw}
\end{figure}

In Figure~\ref{fig:seesaw} we show 
$Br(\tau^-\rightarrow\mu^- \gamma)$ and
$Br(\tau^-\rightarrow e^- \gamma)$ in the SUSY seesaw
mechanism, assuming the gravity mediation scenario for the
SUSY breaking.
We fix the neutrino Yukawa coupling using the neutrino
oscillation data under assumptions for the neutrino Yukawa
coupling $Y_\nu$, which suppresses 
$Br(\mu^-\rightarrow e^-\gamma)$. 
The experimental bounds on these $\tau$ processes have already
excluded some of the parameter space.  
While a natural candidate for the largest LFV $\tau$ lepton
decay mode is $\tau^-\rightarrow\mu^-\gamma$ from the
atmospheric neutrino result, some model-parameters in the
seesaw model predict larger 
$Br(\tau^-\rightarrow e^- \gamma)$ 
\cite{Ellis:2002fe}. This is
because (\ref{eq:delmseesaw}) and (\ref{eq:seesawmass}) have
different dependences on $Y_\nu$ and $M_N$ as mentioned above.

In GUT models, even if the neutrino Yukawa contribution is
negligible, LFV processes are predicted. In the $SU(5)$ SUSY GUT,
the right-handed charged leptons are embedded in the
\textbf{10}-dimensional multiplets with the left-handed
quarks and the right-handed up-type quarks.
The LFV SUSY breaking terms for the right-handed sleptons
are generated by the top-quark Yukawa coupling $Y_t$ above
the GUT scale. 
The off-diagonal components in the right-handed slepton mass
matrix are given as
\begin{equation}
  (\delta m_{\tilde{E}}^2)_{ij}
  \simeq
  -\frac{3}{8\pi^2}(3m_0^2+A_0^2) 
  V_{i3}V_{j3}^\star |Y_t|^2 \log\frac{M_G}{M_{\rm GUT}},
\end{equation}
where $V_{ij}$ is the CKM matrix in the SUSY $SU(5)$ GUT and
$M_{\rm GUT}$ is the GUT scale.  

In the minimal $SU(5)$ SUSY GUT, in which the neutrino Yukawa coupling
is negligible, $Br(\tau^-\rightarrow \mu^-\gamma)$ is smaller than
$10^{-(9-10)}$. While the process is enhanced by the top-quark Yukawa
coupling, it is suppressed by the CKM matrix element and the $U(1)_Y$
gauge coupling constant. Furthermore, when the right-handed sleptons
have LFV mass terms, an accidental cancellation among the diagrams
tends to suppress the branching ratio.  However, we notice that the
CKM matrix elements at the GUT scale may not be the same as ones
extrapolated from the low energy data.  This is because the quark and
lepton mass ratios are not well explained in this model.  If $V_{32}$
is larger, the processes are enhanced
\cite{Arkani-Hamed:1995fs,Hisano:1998cx}.


\subsection{Other Theoretical Models}
In the previous subsections we discussed LFV $\tau$ lepton
decays in SUSY models assuming that R parity is conserved.
In this subsection we review some examples of other models
which predict LFV $\tau$ lepton decays.

In extra-dimension models
\cite{Arkani-Hamed:1998rs,Randall:1999ee} 
the ``fundamental'' scale is expected to be comparable to
the weak scale, and the classical seesaw mechanism does not
explain the tiny neutrino masses. 
Instead, singlet neutrinos in
the bulk space are introduced
\cite{Arkani-Hamed:1998vp,Dienes:1998sb,Grossman:1999ra}.
The Yukawa couplings of the left-handed neutrinos with the
bulk neutrinos are suppressed by an overlap of the wave
functions, and small Dirac neutrino masses are predicted.

In this model the loop diagrams from the Kaluza-Klein states of
the bulk neutrinos generate the LFV $\tau$ lepton and $\mu$
decay \cite{Faraggi:1999bm,DeGouvea:2001mz}.  
On the other hand, this model is constrained by short
baseline experiments and charged current universality,
since the kinetic mixing term for neutrinos is not negligible.
In the minimal model, in which the unique source of LFV is the
Yukawa coupling of the left-handed neutrinos with the bulk 
neutrinos, 
$Br(\tau^-\rightarrow\mu^-\gamma)\lsim 10^{-9}$ 
from the existing constraints, while 
$Br(\mu^-\rightarrow e^-\gamma)\lsim 10^{-(11-13)}$
depending on $U_{e3}$ \cite{DeGouvea:2001mz}. 
In the non-minimal case, the constraints may be looser.

In R-parity violating SUSY models, there exist lepton
flavor and baryon number non-conserving interactions at the
tree level. 
Proton stability gives stringent bounds on such models,
as well as other processes, such as the decays of charged
leptons and $B$ and $D$ mesons, neutral current processes,
and FCNC processes.
Since the $\tau$ lepton is the heaviest lepton, various
$\tau$ LFV modes can be induced.
These include
$\tau^-\rightarrow \mu^-\mu^+\mu^-$, 
$\mu^-e^+ e^-$, $\mu^+ e^-e^-$ and 
$\tau^-\rightarrow \mu^- M^0$, $\mu^- V^0$.
Here, $M^0$ and $V^0$ are pseudoscalar and vector mesons,
respectively. 
Comprehensive studies of LFV $\tau$ lepton decay
have been performed in \cite{Saha:2002kt}. 

If the masses of the right-handed neutrinos are $O(1-10)$ TeV in the
seesaw mechanism, sizable LFV $\tau$ lepton decay might be
possible \cite{Ilakovac:1994kj}. 
It is found in \cite{Cvetic:2002jy} that 
$Br(\tau^-\rightarrow\mu^-\gamma)$ and 
$Br(\tau^-\rightarrow \mu^-\mu^+\mu^-)$ are smaller than
$10^{-9}$ and $10^{-10}$, respectively.
It is also argued that 
$Br(\tau^-\rightarrow e^-\gamma)$ and 
$Br(\tau^-\rightarrow e^-e^+e^-)$  can reach $10^{-8}$ and
$10^{-9}$, respectively, in various models
\cite{Cvetic:2002jy}. 



\chapter{Sensitivity at SuperKEKB}

\def\bz{{B^0}}
\def\bzb{{\overline{B}{}^0}}
\def\kl{K_L^0}
\def\dE{{\Delta E}}
\def\mb{{M_{\rm bc}}}
\def\Dt{\Delta t}
\def\Dz{\Delta z}
\def\fol{f_{\rm ol}}
\def\fsig{f_{\rm sig}}
\def\sinbb{{\sin2\phi_1}}

\def\ra{\rightarrow}
\def\myindent{\hspace*{2cm}}  
\def\fCP{f_{CP}}
\def\ftag{f_{\rm tag}}
\def\zCP{z_{CP}}
\def\tCP{t_{CP}}
\def\ttag{t_{\rm tag}}
\def\cala{{\cal A}}
\def\cals{{\cal S}}
\def\dm{\Delta m_d}
\def\dmd{\dm}
\def\taubz{\tau_\bz}
\def\ks{{K_S^0}}
\def\btosqq{b \to s\overline{q}q}
\def\btosss{b \to s\overline{s}s}
\def\dwl{\ensuremath{{\Delta w_l}}}
\def\fq{\ensuremath{q}}

\section{Overview}

\subsection{Goals of sensitivity studies}

As described in Chapter~\ref{chap:introduction},
the primary purpose of SuperKEKB is
to perform a comprehensive study of $B$ decays
governed by quark transitions
induced by quantum loops, where
sizable effects from physics beyond the Standard
Model are expected.
If the study leads to a conclusion that
an observed pattern 
is inconsistent with the Standard Model expectation,
we further proceed to
identify underlying flavor structure.

One of major goals of sensitivity studies is
to clarify the meaning of the statement above
quantitatively. To this end, we 
estimate statistical, systematic and
theoretical errors on key observables at SuperKEKB.
The target luminosity of SuperKEKB is
5 $\times$ 10$^{35}$ cm$^{-2}$s$^{-1}$, which
corresponds to an annual integrated luminosity of 5 ab$^{-1}$
assuming 100 days of operation (i.e. the Snowmass year definition).
Therefore we estimate expected errors at 5 ab$^{-1}$
to describe what will be achieved at an early stage of the SuperKEKB
experiment.
We also provide errors at 50 ab$^{-1}$ as ultimate
measurements that can be performed at SuperKEKB. 
Although this looks rather ambitious,
our experience at the Belle experiment tells that
an integrated luminosity that is 10 times as large
as a target annual luminosity is not a mere dream.

There are a huge number of observables that can be measured
at SuperKEKB. It is not the main purpose of our sensitivity
studies to cover all of them. Instead, our strategy is
to concentrate on observables that are indispensable
to reach the primary goal mentioned above.
The following points are considered to select such
observables:
\begin{itemize}
\item A sizable deviation from a prediction of the Standard Model is expected.
\item A hadronic uncertainty is negligible or very small.
\item A measurement at SuperKEKB is (much) superior to others, in particular to LHCb, whose expected physics performance is regarded as the benchmark of next-generation $B$ physics programs at hadron colliders.
\end{itemize}
Selected topics do not necessarily satisfy all of these
criteria, but do have clear advantages over others
not covered in this chapter.

At SuperKEKB we expect a background level that is about 20 times
larger than what we currently observe at Belle. Our investigation leads to
a conclusion that under such conditions
there is a feasible detector design that
guarantees performance equivalent to 
the present Belle detector.
Therefore, throughout our studies,
we assume a detector
that has the same performance as the present Belle detector
unless otherwise noticed.
Any possible gain from an improved detector performance
is regarded as a bonus, and our aim is to demonstrate that
most of physics goals at SuperKEKB are achieved 
even without any improvement in the detector performance.
One important exception, however, is $B$ meson vertex
reconstruction using the $\ks$. Here we clearly need to
introduce a larger vertex detector to increase
vertex reconstruction efficiencies. 
Therefore we incorporate the proposed detector design
in this case.

One of the biggest advantages of the sensitivity studies for
Super-KEKB compared with those for hadron collider experiments
is that 
we can fully utilize information obtained by analyzing
Belle data. 
In particular, 
many of studies described in this chapter
rely on Monte Carlo
pseudo-experiments (also called ``toy Monte Carlo experiments'')
in which PDFs are constructed from data.
A clear advantage of this approach over genuine Monte Carlo
simulations (e.g. GEANT simulation)
is that background fractions and detector resolutions
are more reliable.
For some topics for which pseudo-experiments are not available,
however, we also use GEANT simulation
and/or FSIM, a parametric Monte Carlo simulator that
requires much less CPU power than GEANT.

In the rest of this section we overview
two important analysis techniques that are
used repeatedly in our studies. One is
the procedure to fit a
proper-time difference distribution for
a time-dependent $CP$ asymmetry measurement.
The other is
to reconstruct one $B$ meson exclusively (or semi-inclusively)
so as to study decays of an accompanying $B$ meson
in the cleanest environment. This is called
``full reconstruction $B$ tagging''.

\subsection{Time-dependent $CP$ asymmetries}
\label{sec:overview:icpv}

In the decay chain $\Upsilon(4S)\to \bz\bzb \to f_{CP}f_{\rm tag}$,
where one of the $B$ mesons decays at time $t_{CP}$ to a final state $f_{CP}$ 
and the other decays at time $t_{\rm tag}$ to a final state  
$f_{\rm tag}$ that distinguishes between $B^0$ and $\bzb$, 
the decay rate has a time dependence
given by~\cite{Carter:hr,Carter:tk,Bigi:qs}
\begin{equation}
\label{eq:psig}
{\cal P}(\Delta{t}) = 
\frac{e^{-|\Delta{t}|/{\taubz}}}{4{\taubz}}
\biggl\{1 + \fq\cdot 
\Bigl[ \cals\sin(\dmd\Delta{t})
   + \cala\cos(\dmd\Delta{t})
\Bigr]
\biggr\},
\end{equation}
where $\taubz$ is the $B^0$ lifetime, $\dmd$ is the mass difference 
between the two $B^0$ mass
eigenstates, $\Delta{t}$ = $t_{CP}$ $-$ $t_{\rm tag}$, and
the $b$-flavor charge $\fq$ = +1 ($-1$) when the tagging $B$ meson
is a $B^0$ 
($\bzb$).
$\cals$ and $\cala$ are $CP$-violation parameters.
For example, to a good approximation,
the Standard Model 
predicts $\cals = -\xi_f\sin 2\phi_1$, where $\xi_f = +1 (-1)$ 
corresponds to  $CP$-even (-odd) final states, and $\cala =0$
for both $b \to c\overline{c}s$ and 
$b \to s\overline{s}s$ transitions. 

We determine $q$ and $\Delta t$ for each event.
Charged leptons, pions, kaons, and $\Lambda$ baryons
that are not associated with a reconstructed $CP$ eigenstate decay
are used to identify the $b$-flavor of the accompanying $B$ meson.
The tagging algorithm is described in detail 
elsewhere~\cite{Abe:2002px}.
We use two parameters, $\fq$ and $r$, to represent the tagging information.
The first, $q$, is already defined above.
The parameter $r$ is an event-by-event Monte Carlo-determined
flavor-tagging dilution parameter that ranges
from $r=0$ for no flavor discrimination
to $r=1$ for an unambiguous flavor assignment.
It is used only to sort data into six intervals of $r$,
according to estimated flavor purity.
We determine directly from data
the average wrong-tag probabilities, 
$w_l \equiv (w_l^+ + w_l^-)/2~(l=1,6)$,
and differences between $\bz$ and $\bzb$ decays, 
$\dwl \equiv w_l^+ - w_l^-$,
where $w_l^{+(-)}$ is the wrong-tag probability
for the $\bz(\bzb)$ decay in each $r$ interval. 
The event fractions and wrong-tag fractions
are summarized in Table~\ref{tab:wtag}.
The total effective tagging efficiency is determined to be
$\eeff \equiv \sum_{l=1}^6 \epsilon_l(1-2w_l)^2 = \efftot$,
where $\epsilon_l$ is the event fraction for each $r$ interval.
The error includes both statistical and systematic uncertainties.

\begin{table}[tbp]
\begin{center}
    \begin{tabular}{ccclll}
\hline
      $l$ & $r$ interval & $\epsilon_l$ &\multicolumn{1}{c}{$w_l$} 
          & \multicolumn{1}{c}{$\dwl$}  &\multicolumn{1}{c}{$\eeff^l$} \\
      \hline
 1 & 0.000 -- 0.250 & 0.398 & $0.464\pm0.006$ &$-0.011\pm0.006$ &$0.002\pm0.001$ \\
 2 & 0.250 -- 0.500 & 0.146 & $0.331\pm0.008$ &$+0.004\pm0.010$ &$0.017\pm0.002$ \\
 3 & 0.500 -- 0.625 & 0.104 & $0.231\pm0.009$ &$-0.011\pm0.010$ &$0.030\pm0.002$ \\
 4 & 0.625 -- 0.750 & 0.122 & $0.163\pm0.008$ &$-0.007\pm0.009$ &$0.055\pm0.003$ \\
 5 & 0.750 -- 0.875 & 0.094 & $0.109\pm0.007$ &$+0.016\pm0.009$ &$0.057\pm0.002$ \\
 6 & 0.875 -- 1.000 & 0.136 & $0.020\pm0.005$ &$+0.003\pm0.006$ &$0.126\pm0.003$ \\
\hline
    \end{tabular}
  \end{center}
  \caption{The event fractions $\epsilon_l$,
    wrong-tag fractions $w_l$, wrong-tag fraction differences $\dwl$,
    and average effective tagging efficiencies
    $\eeff^l = \epsilon_l(1-2w_l)^2$ for each $r$ interval.
    The errors include both statistical and systematic uncertainties.
    The event fractions are obtained from the $\jpsi\ks$ simulation.}
  \label{tab:wtag}
\end{table}

The vertex position for the $\fcp$ decay is reconstructed
using leptons from $\jpsi$ decays or charged hadrons from $\eta_c$ decays,
and that for $\ftag$ is obtained with well reconstructed tracks
that are not assigned to $\fcp$.
Tracks that are consistent with coming from a $\ks\to\pip\pim$ decay
are not used.
Each vertex position is required to be consistent with
the interaction region profile, determined run-by-run,
smeared in the $r$-$\phi$ plane to account for the $B$ meson decay length.
With these requirements, we are able to determine a vertex
even with a single track;
the fraction of single-track vertices is about 10\% for $\zcp$
and 22\% for $\ztag$.
The proper-time interval resolution function $\Rsig(\Dt)$
is formed by convolving four components:
the detector resolutions for $\zcp$ and $\ztag$,
the shift in the $\ztag$ vertex position
due to secondary tracks originating from charmed particle decays,
and the kinematic approximation that the $B$ mesons are
at rest in the cms~\cite{Tajima:2003bu}.
A small component of broad outliers in the $\Dz$ distribution,
caused by mis-reconstruction, is represented by a Gaussian function.
We determine fourteen resolution parameters from the
aforementioned fit to the control samples.
We find that the average $\Dt$ resolution is $\sim 1.43$~ps (rms).
The width of the outlier component
is determined to be $(39\pm 2)$~ps;
the fractions of the outlier components are $(2.1 \pm 0.6) \times 10^{-4}$
for events with both vertices reconstructed with more than one track,
and $(3.1 \pm 0.1) \times 10^{-2}$ for events with at least
one single-track vertex.

We determine $\cals$ and $\cala$ for each mode by performing an unbinned
maximum-likelihood fit to the observed $\Dt$ distribution.
The probability density function (PDF) expected for the signal
distribution, ${\cal P}_{\rm sig}(\Dt;\cals,\cala,\fq,w_l,\dwl)$, 
is given by Eq.~(\ref{eq:psig}) incorporating
the effect of incorrect flavor assignment. 
The distribution is also convolved with the
proper-time interval resolution function $R_{\rm sig}(\Dt)$,
which takes into account the finite vertex resolution. 
We determine the following likelihood value for each
event:
\begin{eqnarray}
P_i
&=& (1-\fol)\int_{-\infty}^{\infty}\biggl[
\fsig{\cal P}_{\rm sig}(\Dt')R_{\rm sig}(\Dt_i-\Dt') \nonumber \\
&+&(1-\fsig){\cal P}_{\rm bkg}(\Dt')R_{\rm bkg}(\Dt_i-\Dt')\biggr]
d(\Dt')  \nonumber \\
&+&\fol P_{\rm ol}(\Dt_i) 
\end{eqnarray}
where $P_{\rm ol}(\Dt)$ is a broad Gaussian function that represents
an outlier component with a small fraction $\fol$.
The signal probability $\fsig$ depends on the $r$ region and
is calculated on an event-by-event basis
as a function of $\dE$ and $\mb$.
${\cal P}_{\rm bkg}(\Dt)$ is a PDF for background events,
which is modeled as a sum of exponential and prompt components, and
is convolved with a sum of two Gaussians $R_{\rm bkg}$.
All parameters in ${\cal P}_{\rm bkg} (\Dt)$
and $R_{\rm bkg}$ are determined by the fit to the $\Dt$ distribution of a 
background-enhanced control sample; i.e. events away from the $\dE$-$\mb$ signal region.
We fix $\tau_\bz$ and $\dmd$ at
their world-average values.
The only free parameters in the final fit
are $\cals$ and $\cala$, which are determined by maximizing the
likelihood function
$L = \prod_iP_i(\Dt_i;\cals,\cala)$
where the product is over all events.

\subsection{$B$ tagging with full reconstruction}
\label{sec:full_recon}
At SuperKEKB, $B\bar{B}$ meson pairs will be
produced from $\Upsilon(4S)$ decays.  To study
$B$ meson decays that include neutrinos, photons, $\pi^0$ mesons
in the final states,
it is useful 
to tag one of the $B$ mesons through the full reconstruction. 
This method has the following attractive features: 
\begin{itemize}
\item The momentum vector and flavor of the other $B$ meson
  can be identified,
  \textit{i.e.}
  single $B$ meson beams can practically be obtained offline.
\item Continuum and combinatoric $B\bar{B}$ backgrounds can
  be significantly reduced. 
\end{itemize}
If we take advantage of these features, it will be possible to measure
the $B$ decays listed in Table \ref{frec:physics_items}. 
Some of these decays have more than one neutrino in the
final states. Therefore, it is very difficult to perform
such studies even in the clean environment of $e^{+}e^{-}$  
collider unless the full reconstruction $B$ tagging is applied. 
 
Because of the modest full reconstruction efficiency
[${\cal O}(0.1\%)$], this method has not  
been extensively applied at the current $B$ factory experiments. However, a
very large $B$ meson sample at SuperKEKB will make it possible
to extract useful results. 
 
\begin{table}[tbp]
  \begin{center}
  \begin{tabular}{|c|c|}
    \hline
 Decay mode        &  Motivation  \\
    \hline
 $B \to X_u l \nu$  &  Precise measurement of $|V_{ub}|$  \\
    \hline
 $B \to \tau \nu$   &  Measurement of $f_B$  \\
    \hline
 $B \to K \nu \bar{\nu}$, $D \tau \nu$ &  Search for new physics \\
    \hline
 Inclusive $B$ decays     &  Detailed study, model independent analysis etc. \\
    \hline
  \end{tabular}
 \end{center}
\caption{Physics topics that will be studied
         with the fully-reconstructed $B$ sample.}
\label{frec:physics_items}
\end{table}

\subsubsection{Hadronic $B$ Tagging}
\begin{figure}[tbp]
  \begin{center}
    \includegraphics[width=7cm]{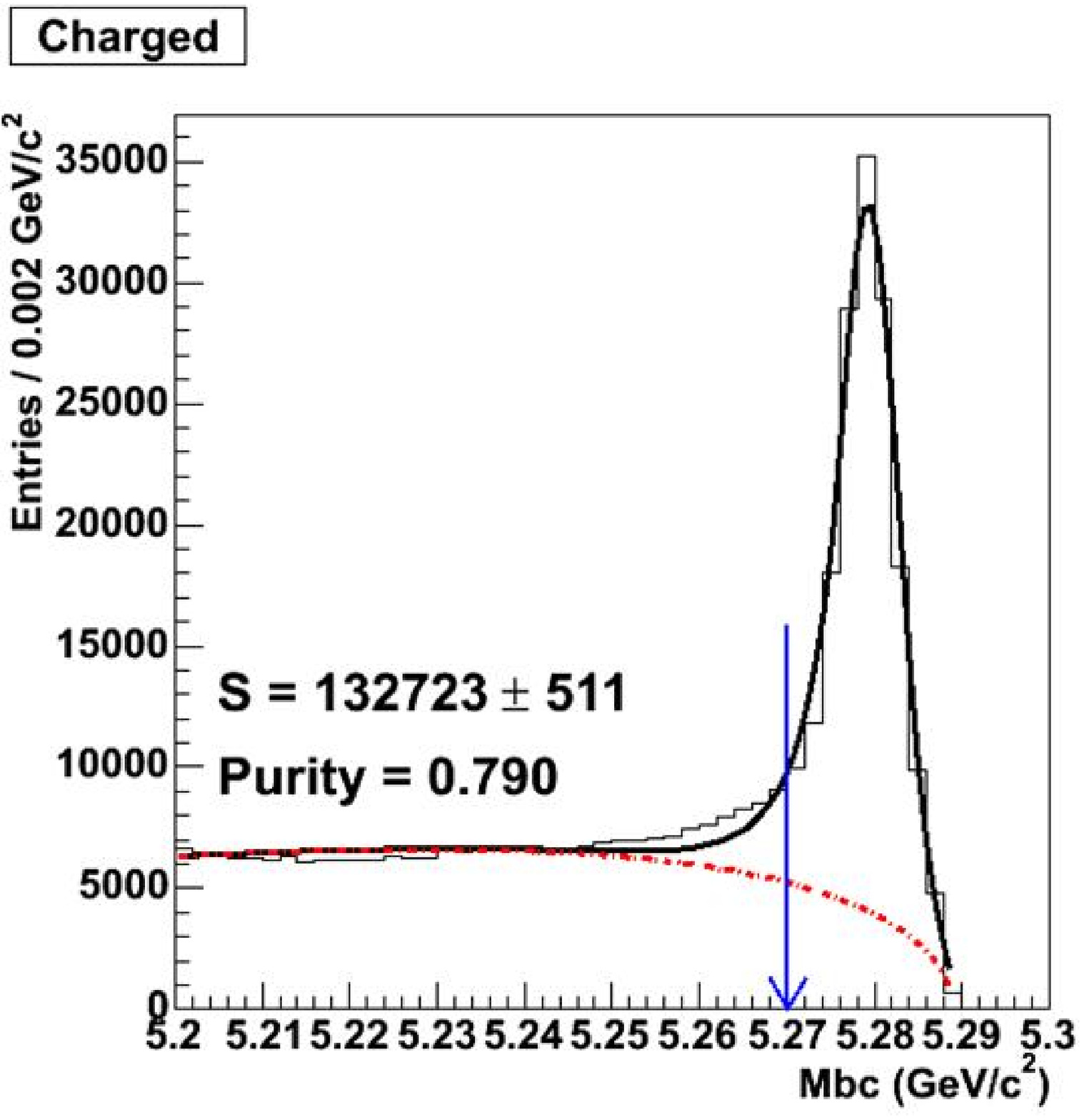}
    \includegraphics[width=7cm]{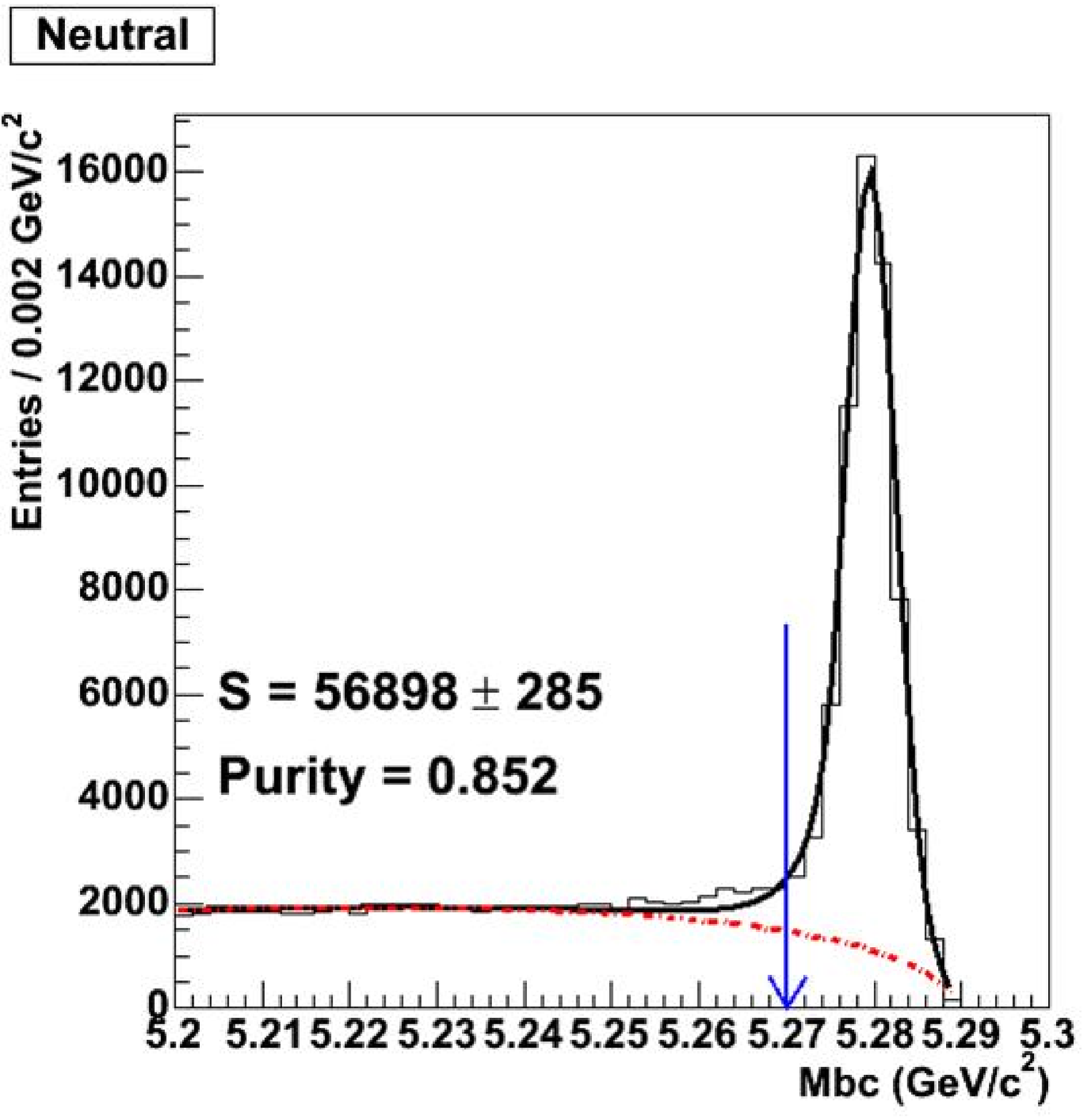}
  \end{center}
  \caption{
    Beam-constrained mass distribution for $B \to D^{(*)}(\pi, \rho, a_1)^{-}$
  with the 152 million $B\bar{B}$ sample recorded by Belle.}
  \label{fig:b2dnpi}
\end{figure}

Most $B$ meson decays are hadronic decays. 
However, the branching fraction of each 
mode is less than ${\cal O}(1 \%)$.
Therefore, we need to collect as many modes as possible to 
achieve a high efficiency.
In Fig. \ref{fig:b2dnpi}, the beam-constrained mass distribution for 
the main decay modes, $B \to D^{(*)} (\pi, \rho, a_1)^{-}$, are shown.
The yields including other decay modes are also shown in Table \ref{frec:hadronic_tag}.
With a 152 million $B\bar{B}$ sample, we already have 
a sample of more than $10^5$
fully-reconstructed $B$ meson tags.
The tagging efficiency is 0.20\% (0.09\%) 
for charged (neutral) $B$ mesons with a purity of 78\% (83\%) 
\cite{Matsumoto:ACAT03}.

The tagging efficiency can be improved largely by loosening selection criteria.
In this case, however, the purity becomes lower.
Therefore the best selection criteria should be searched for
in each analysis. 
In the case of hadronic $B$ tagging mentioned above, we achieve
a tagging efficiency of 0.33\% (0.20\%) for
charged (neutral) $B$ mesons with a purity of 58\% (52\%).

\begin{table}[htbp]
  \begin{center}
  \begin{tabular}{|l|l|c|c|c|}
    \hline
            & decay mode   &  yield   &  eff. (\%)  & purity(\%)    \\
    \hline
 Charged B  &  $B^- \to D^{(*)0}(\pi, \rho, a_1)^{-}$ 
            &  132723  &  0.17   &   79 \\
    \cline{2-5}
            &  $B^- \to D^{(*)0}D_s^{(*)-}$
            &  8700   &   0.01  &  60  \\
    \cline{2-5}
            &  $B^- \to J/\psi K^-$
            & 9373    &   0.01  &  96  \\
    \cline{2-5}
            &  total  
            & 150796   &  0.20   & 78   \\
    \hline
 Neutral B  &  $\bar{B^0} \to D^{(*)+}(\pi, \rho, a_1)^{-}$   
            & 56898   &  0.07  &  85  \\
    \cline{2-5}
            &  $\bar{B^0} \to D^{(*)+}D_s^{(*)-}$   
            & 4390   &  0.006  &  60  \\
    \cline{2-5}
            &  $\bar{B^0} \to (J/\psi, \psi(2s), \chi_{c1} )K_S$, $J/\psi K^{*0}$
            & 7275   &  0.01  &  94  \\
    \cline{2-5}
            &  total  
            & 68563   & 0.09  &  83  \\
    \hline
   \end{tabular}
 \end{center}
 \caption{Full reconstructed $B$ events for hadronic modes with 
 152 million $B\bar{B}$ sample.}
 \label{frec:hadronic_tag}
\end{table}

 If a $B$ meson is 
reconstructed semi-inclusively, we can further improve the efficiency.
At BaBar, the $B \to D^{(*)}( n_1 \pi^{\pm} n_2 K^{\pm} n_3 K_S n_4 \pi^0)^{-}$ 
process is reconstructed, where $n_1 + n_2 \le 5$, $n_3 \le 2$ and $n_4 \le 2$.
As a result,
the higher efficiency, 0.5 (0.3) \% for charged (neutral) $B$ mesons,
is obtained. However, due to combinatoric backgrounds, the purity is only 
around 25 \%.  These samples are used for the $|V_{ub}|$ measurement using 
$B \to X_u l \nu$ decays, 
where the background can be reduced by requiring 
a lepton in the recoil side 
\cite{Aubert:2003zw,Daniele:Vub}. 

\subsubsection{Semileptonic $B$ tagging}

If we use semileptonic $B$ decays to tag one $B$ meson, 
we cannot use information about the $B$ momentum 
 vector due to the missing neutrino. 
However, we still have relatively clean tagged 
 samples. Semileptonic $B$ decays are dominated by $B \to D l \nu$ 
 and $D^{*} l \nu$, with a total branching fraction of around 15\%. 
At Belle, these samples were used for a $|V_{ub}|$ measurement with 
inclusive $B \to X_u l \nu$ decays. 
If we require a semileptonic decay on the recoil 
 side, we can apply a kinematical constraint, and the 
$B$ momentum vector can be 
 determined with a two-fold ambiguity.
The background contribution is 
 significantly suppressed in this case. 
This method yields a $B \to D^{*} l \nu$
 efficiency 
of 0.3 (0.2)\% for charged (neutral) $B$ decays \cite{Sugiyama:Vub}.
 
\begin{figure}[tbp]
  \begin{center}
    \includegraphics[width=12cm]{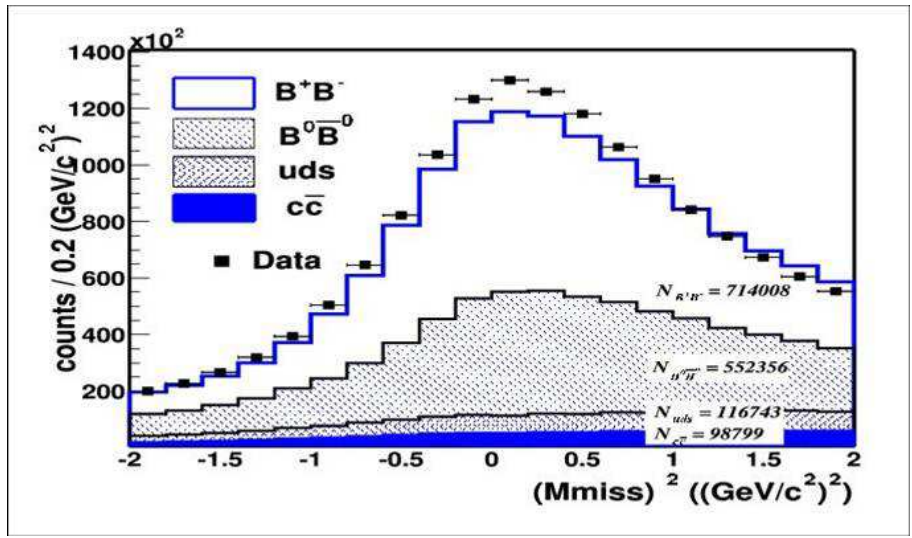}
  \end{center}
  \caption{
    Missing mass squared distribution for $B^{-} \to D^0 l \nu$
  with a sample of $85\times 10^6$ $B\bar{B}$ pairs.}
  \label{fig:b2semileptonic}
\end{figure}

It is also notable that we include
other $B \to D X l \nu$ decays as $B$ tags in the $D l \nu$ mode, because 
the signal in the missing mass spectrum is too broad for separation.
The missing mass distribution for $B^- \to D^0 l \nu$ is shown in 
Fig. \ref{fig:b2semileptonic}. 
The full reconstruction tagging efficiency is estimated to be 1.7\%
including the
contribution from the $B^- \to D^{(*)0} l \nu$ decays
of around $15\%$ \cite{Matsumoto:ACAT03}. 
Other background is mostly combinatoric.
Although the tagging quality
is not good in this case, it is still useful for the study 
of rare $B$ decays with low track multiplicities in the final states.
In BaBar, this method was used to search for 
the $B \to K \nu \bar{\nu}$ and $\tau \nu$ decays
\cite{Aubert:2002nw,Aubert:2003wu}
to provide improved upper limits.

\subsubsection{Conclusions}

  In summary, the tagging efficiency 
will be around 0.2\% (0.1\%) for the clean hadronic tagging for 
charged (neutral) $B$ mesons, and
around 1\% for the semileptonic $B$ tagging 
that has a lower purity.
The efficiency for the hadronic tagging can be
around 0.3\% (0.2\%) if we loosen the selection criteria,
with a tolerable decrease in the purity of the tagging. 

With a $1~{\rm ab}^{-1}$ of data, 
there will be at least
2 million clean tags and around 10 million semileptonic tags. 
With these samples, 
the physics topics listed in Table \ref{frec:physics_items} 
will be studied.

\clearpage \newpage
\def\ra{\rightarrow}
\def\myindent{\hspace*{2cm}}  
\def\qbar{\overline{q}}
\def\ubar{\overline{u}}
\def\Bcp{B_{CP}}
\def\UfourS{\Upsilon(4S)}
\def\phiNP{\phi_{\rm NP}}
\def\ACP{{\cal A}_{CP}}
\def\Mbc{M_{\rm bc}}
\def\mbc{\Mbc}
\def\sss{s\bar{s}s}
\def\suu{s\bar{u}u}
\def\uus{u\bar{u}s}
\def\ccs{c\bar{c}s}
\def\sdd{s\bar{d}d}
\def\btosss{b \ra \sss}
\def\btosssss{b \ra \sss\bar{s}s}
\def\btouus{b \ra \uus}
\def\btoccs{b \ra \ccs}
\def\btosuu{b \to \suu}
\def\btosdd{b \to \sdd}
\def\acp{\mathcal{A}_{CP}}
\def\dacp{\delta\acp}
\def\acppar{\mathcal{A}_{CP}^0}
\def\dacppar{\delta\acppar}
\def\nobs{N_{\rm obs}}
\def\bpm{B^{\pm}}
\def\Nb{{\rm N}_B}
\def\etac{\eta_c}
\def\kl{K_L^0}
\def\xs{X_s}
\def\xsp{X_s^+}
\def\xsm{X_s^-}
\def\xspm{X_s^{\pm}}
\def\etack{\etac K}
\def\etackp{\etac K^+}
\def\etackm{\etac K^-}
\def\etackpm{\etac K^{\pm}}
\def\btoetack{B \to \etack}
\def\bptoetackp{\bp \to \etackp}
\def\bmtoetackm{\bminus \to \etackm}
\def\bpmtoetackpm{\bpm \to \etackpm}
\def\bztoetackstarz{\bz \to \etac K^{*0}}
\def\bptoetackstarp{\bp \to \etac K^{*+}}
\def\bpmtoetacxspm{\bpm \to \etac \xspm}
\def\phiphixs{\phi \phi \xs}
\def\phiphixsp{\phi \phi \xsp}
\def\phiphixsm{\phi \phi \xsm}
\def\phiphixspm{\phi \phi \xspm}
\def\phiphik{\phi \phi K}
\def\phiphikp{\phi \phi K^+}
\def\phiphikm{\phi \phi K^-}
\def\phiphikpm{\phi \phi K^{\pm}}
\def\bpmtophiphikpm{\bpm \to \phiphikpm}
\def\btophiphixs{B \to \phiphixs}
\def\bptophiphixsp{\bp \to \phiphixsp}
\def\bmtophiphixsm{\bminus \to \phiphixsm}
\def\bpmtophiphixspm{\bpm \to \phiphixspm}
\def\btophiphik{B \to \phiphik}
\def\tnp{\Theta_{\rm NP}}
\def\sintnp{\sin \tnp}
\def\costnp{\cos \tnp}
\def\ad{a_{\rm D}}
\def\adtwo{a_{\rm D}^2}
\def\ar{a_{\rm R}}
\def\artwo{a_{\rm R}^2}
\def\aNP{a_{\rm NP}}
\def\aNPtwo{a_{\rm NP}^2}
\def\rtwo{r^2}
\def\DSM{D_{\rm SM}}
\def\dpmnp{D^{\pm}_{\rm NP}}
\def\dpm{D^{\pm}}
\def\bnp{\mathcal{B}_{\rm NP}}
\def\bztophiks{\bz \to \phi\ks}
\def\dmd{\Delta m_d}
\def\etap{\eta^\prime}
\def\etapks{\etap\ks}
\def\bztoetapks{\bz \to \etapks}
\def\kkks{K^+ K^- \ks}
\def\bztokkks{\bz \to \kkks}
\def\calsphiks{\cals_{\phi\ks}}
\def\calaphiks{\cala_{\phi\ks}}
\def\calsetapks{\cals_{\etapks}}
\def\calaetapks{\cala_{\etapks}}
\def\calskkks{\cals_{\kkks}}
\def\calakkks{\cala_{\kkks}}
\def\calspizks{\cals_{\pi^0\ks}}
\def\calapizks{\cala_{\pi^0\ks}}
\def\calsksksks{\cals_{\ks\ks\ks}}
\def\calaksksks{\cala_{\ks\ks\ks}}
\def\deltasphiks{\Delta\calsphiks}
\def\deltaaphiks{\Delta\calaphiks}
\def\deltasetapks{\Delta\calsetapks}
\def\deltaaetapks{\Delta\calaetapks}
\def\deltaskkks{\Delta\calskkks}
\def\deltaakkks{\Delta\calakkks}
\def\deltaspizks{\Delta\calspizks}
\def\deltaapizks{\Delta\calapizks}
\def\deltasksksks{\Delta\calsksksks}
\def\deltaaksksks{\Delta\calaksksks}
\def\jpsi{J/\psi}
\def\jpsiks{\jpsi\ks}
\def\bztojpsiks{\bz\to\jpsiks}
\def\calsjpsiks{\cals_{\jpsiks}}
\def\calajpsiks{\cala_{\jpsiks}}

\section{New $CP$-violating phase in $b\rightarrow s \overline{q}q$}
\label{sec:sss}

\subsection{Introduction}
Despite the great success of the KM mechanism, 
additional $CP$-violating phases
are inevitable in most theories involving new physics (NP)
beyond the SM~\cite{Nir:2002gu}.
Some of them allow large deviations from the SM predictions 
for $B$ meson decays. 
Examples include supersymmetric
grand-unified theories with the see-saw mechanism that can
accommodate large neutrino
mixing~\cite{Moroi:2000tk,Baek:2000sj,Chang:2002mq}. 
Therefore it is of fundamental importance to 
measure $CP$ asymmetries that are sensitive to the
difference between the SM and NP.
Additional sources of $CP$ violation are also highly
desirable to understand the origin of the matter-antimatter
asymmetry of the universe;
detailed studies have found no way that $CP$ violation
in the SM alone could explain
baryogenesis~\cite{Bernreuther:2002uj}. 
Many methods to search for a new source of $CP$ violation 
in $B$ meson decays have been proposed up to now. 
One of the most promising ways is to compare the
mixing-induced $CP$ asymmetries in the $B \to \phi \ks$
decay~\cite{Grossman:1996ke}, which is dominated by  
the $\btosss$ transition that is known to be sensitive
to possible NP effects, with those in the 
$\bz \to J/\psi\ks$ decay~\cite{Carter:tk,Bigi:qs}. 
Ignoring a strong phase difference between 
the amplitude of NP ($A_{\rm NP}$) and SM 
($A_{\rm SM}$)\footnote{The formula with the strong phase is given 
in Section~\ref{sec:new_physics/btos} [Eq.~(\ref{eq:Sfull})].},
we obtain
\begin{equation}
\calsphiks = \frac{\sin2\phi_1+2\rho\sin(2\phi_1+\tnp)+\rho^2\sin(2\phi_1+2\tnp)}
              {1+\rho^2+2\rho\cos\tnp},
\end{equation}
where $\rho \equiv A_{\rm NP}/A_{\rm SM}$ is an amplitude
ratio of NP to the SM. 
Since $\calsjpsiks \simeq \sinbb$ is expected in many
extensions of the SM, the difference 
$\deltasphiks \equiv  (-\xi_f)\calsphiks - \calsjpsiks$
is a gold-plated observable to search for a new
$CP$-violating phase. 

\begin{figure}[tbp]
  \begin{center}
    \includegraphics[width=0.9\textwidth,clip]{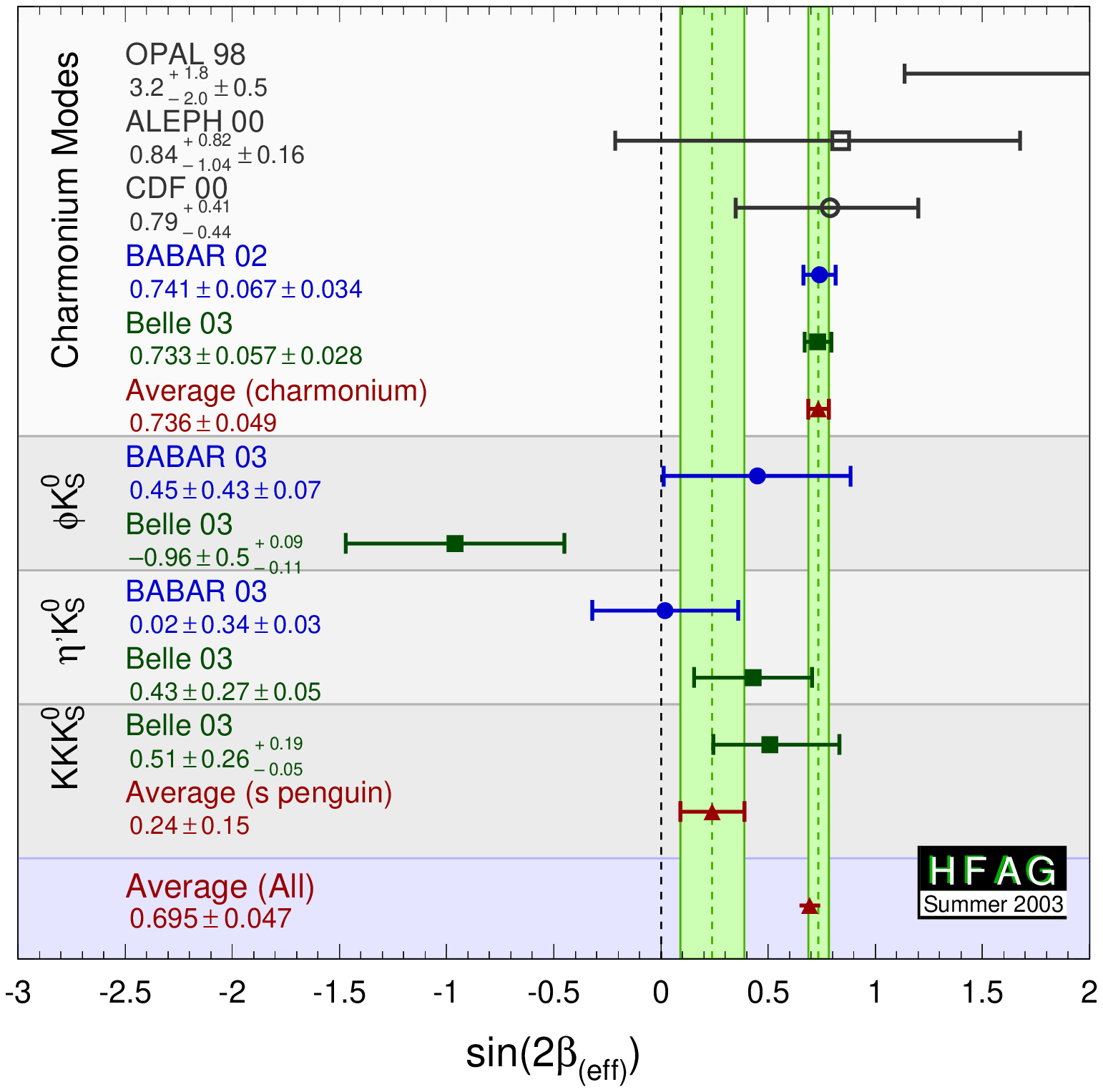}
  \end{center}
  \caption{
    $\cals$ terms measured with decay modes governed by
    $\btoccs$ or $\btosss$ transitions.
  }
  \label{fig:hfagbtos}
\end{figure}

Recent measurements by Belle~\cite{Abe:2003yt} and
BaBar~\cite{bib:BaBar_sss} collaborations
yield values smaller than the SM expectation;
a difference by 2.6 standard deviations 
is obtained when two results are combined.
The other charmless decays $\bz \to \eta' \ks$ and
$\bz \to K^+K^-\ks$, which are
mediated by $\btosss$, $\suu$ and $\sdd$ transitions,
also provide additional
information~\cite{Abe:2003yt,bib:BaBar_sss}. 
The present world average (as of August 2003)
with $\bztophiks$, $\eta' \ks$
and $K^+K^-\ks$ combined is different from the average
with $\bz \to J/\psi \ks$ and related modes
by 3.1 standard deviations~\cite{HFAG} as shown in
Figure~\ref{fig:hfagbtos}.
Possible theoretical implications of these
measurements are already discussed in
Section~\ref{sec:new_physics/btos}. 

In this section we describe the expected sensitivities for
$\cals(\phi\ks)$, $\cals(\eta'\ks)$ and $\cals(\kkks)$ based
on the measurements performed with the present Belle detector.
It will be crucial to measure as many observables
as possible and to check correlations among them. Therefore
we also describe several other new methods to access
a new $CP$-violating phase in the $b\to s$ transition.

\subsection{$\bztophiks$, $\etap\ks$ and $\kkks$}
\label{sec:phiks}
As mentioned in the previous section, the $\bztophiks$ decay
is one of the most promising decays in which to search for
a new $CP$-violating phase in $b \to \sss$ transitions.
The other charmless decays $\bz \to \eta' \ks$ and
$\bz \to K^+K^-\ks$ provide additional information.
As most of the experimental procedure for these
three modes is common, we discuss them here together.
Note that the theoretical uncertainties for $\Delta\cals$
within the SM depend on the decay mode. We discuss this issue
in Section~\ref{sec:sss/discussion}.
Note also that the three-body decay $\bztokkks$ is in
general a mixture of $CP$-even and $CP$-odd eigenstates.
The Belle collaboration finds that
the $K^+K^-\ks$ state is primarily $\xi_f=+1$;
a measurement of the $\xi_f=+1$ fraction with a 140~fb$^{-1}$ data set 
yields $1.03 \pm
0.15\mbox{(stat)}\pm0.05\mbox{(syst)}$~\cite{Abe:2003yt}. 
The uncertainty in the $CP$-asymmetry parameters, which arises
from the $\xi_f=-1$ component, is included in the systematic error.

We estimate the expected sensitivities at SuperKEKB
by extrapolating the present experimental results.
Therefore we first explain Belle's analysis
with a data sample of 140 fb$^{-1}$~\cite{Abe:2003yt}.

We reconstruct $\bz$ decays to
$\phi\ks$ and $\eta'\ks$ final states for $\xi_f=-1$, and
$\bz\to K^+K^-\ks$ decays
that are a mixture of $\xi_f=+1$ and $-1$.
$K^+K^-$ pairs that are consistent with $\phi \to K^+K^-$ decay are excluded
from the $\bz \to K^+K^-\ks$ sample.
We find that the $K^+K^-\ks$ state is primarily $\xi_f=+1$;
a measurement of the $\xi_f=+1$ fraction with a 140~fb$^{-1}$ data set 
yields $1.03 \pm 0.15\mbox{(stat)}\pm0.05\mbox{(syst)}$.
In the following determination of $\cals$ and $\cala$,
we fix $\xi_f=+1$ for this mode.
The intermediate meson states are reconstructed from the
following decay chains: 
$\eta'\to\rho^0 (\to \pi^+\pi^-) \gamma$ or 
$\eta'\to\pi^+\pi^-\eta (\to \gamma\gamma)$,
$\ks \to \pi^+\pi^-$, and $\phi\to K^+K^-$.

Pairs of oppositely charged tracks are used to reconstruct
$K^0_S \to \pi^+\pi^-$ decays. 
The $\pi^+\pi^-$ vertex is required to be displaced from 
the IP by a minimum transverse distance of 0.22~cm for
high momentum ($>1.5$ GeV/$c$) candidates and 0.08~cm for
those with momentum less than 1.5~GeV/$c$.
The direction of the pion pair momentum must agree with
the direction defined by the IP and the vertex displacement
within 0.03 rad for high-momentum candidates, and within 0.1
rad for the remaining candidates. 

Candidate $\phi \to K^+K^-$ decays are found by selecting
pairs of oppositely charged tracks that are not pion-like 
($P(K/\pi)>0.1$), where a kaon likelihood ratio, 
$P(K/\pi) = \mathcal{L}_K /(\mathcal{L}_K + \mathcal{L}_\pi)$, has values 
between 0 (likely to be a pion) and 1 (likely to be a kaon).
The likelihood $\mathcal{L}_{K(\pi)}$ is derived 
from $dE/dx$, ACC and TOF measurements.
The vertex of the candidate charged tracks is required to be consistent with
the interaction point (IP) to suppress poorly measured tracks.
In addition, candidates are required to have a $K^+K^-$
invariant mass that is less than 10 MeV/$c^2$ from the nominal
$\phi$ meson mass.

Since the $\phi$ meson selection is effective in reducing background events,
we impose only minimal kaon-identification requirements.
We use more stringent kaon-identification requirements
to select non-resonant $K^+K^-$ candidates
for the $\bz \to K^+K^-\ks$ decay.
We reject $K^+K^-$ pairs that are consistent with
$D^0 \to K^+K^-$, $\chi_{c0} \to K^+K^-$, or $J/\psi \to K^+K^-$ decay.
$D^+ \to \ks K^+$ candidates are also removed.

To reconstruct $\eta'$ candidates,
we first require that all of the
tracks have associated SVD hits and radial impact parameters $|dr|
< 0.1$~cm projected on the $r$-$\phi$ plane. 
Particle identification information
from the ACC, TOF and CDC $dE/dx$ measurements are used to form a
likelihood ratio in order to distinguish pions from kaons with at
least 2.5$\sigma$ separation for laboratory momenta up to 3.5
GeV/$c$. Candidate photons from
$\eta_{\gamma\gamma}~(\eta^\prime_{\rho\gamma})$ decays are
required to be isolated and have $E_\gamma > 50~(100)$~MeV from
the ECL measurement. The invariant mass of $\eta_{\gamma \gamma}$
candidates is required to be between 500~MeV/$c^2$ and
570~MeV/$c^2$. A kinematic fit with an $\eta$ mass constraint is
performed using the fitted vertex of the $\pi^+\pi^-$ tracks from
the $\eta^\prime$ as the decay point. For
$\eta^\prime_{\rho\gamma}$ decays, the candidate $\rho^0$ mesons
are reconstructed from pairs of vertex-constrained $\pi^+\pi^-$
tracks with an invariant mass between 550 and 920~MeV/$c^2$. The
$\eta^\prime$ candidates are required to have a reconstructed mass
from 940 to 970~MeV/$c^2$ for the $\eta^\prime_{\eta\pi\pi}$ mode
and 935 to 975~MeV/$c^2$ for $\eta^\prime_{\rho\gamma}$ mode.
Charged $K^{\pm}$ candidates are selected for the decay $B^{\pm}
\to \eta^\prime K^{\pm}$ based on the particle identification
information.

We also reconstruct events where
only one of the charged pions has associated SVD hits.
In this case, the requirement on the impact parameter
is relaxed for the track without SVD hits, while 
a higher threshold is imposed on the likelihood ratio.

For reconstructed $B\to\fcp$ candidates, we identify $B$ meson decays using the
energy difference $\dE\equiv E_B^{\rm cms}-E_{\rm beam}^{\rm cms}$ and
the beam-energy constrained mass $\mb\equiv\sqrt{(E_{\rm beam}^{\rm cms})^2-
(p_B^{\rm cms})^2}$, where $E_{\rm beam}^{\rm cms}$ is
the beam energy in the cms, and
$E_B^{\rm cms}$ and $p_B^{\rm cms}$ are the cms energy and momentum of the 
reconstructed $B$ candidate, respectively.
The $B$ meson signal region is defined as 
$|\dE|<0.06$ GeV for $\bz \to \phi \ks$,
$|\dE|<0.04$ GeV for $\bz \to K^+K^-\ks$,
$|\dE|<0.06$ GeV for $\bz \to \eta'(\to \rho\gamma) \ks$, or
$-0.10$ GeV $< \dE <0.08$ GeV for $\bz \to \eta'(\to \pi^+\pi^-\eta) \ks$,
and $5.27~{\rm GeV}/c^2 <\mb<5.29~{\rm GeV}/c^2$ for all decays.
In order to suppress background from the $e^+e^- \rightarrow 
u\overline{u},~d\overline{d},~s\overline{s}$, or $c\overline{c}$
continuum, we form signal and background
likelihood functions, ${\cal L}_{\rm S}$ and ${\cal L}_{\rm BG}$, 
from a set of variables that characterize the event topology,
and impose thresholds on the likelihood ratio 
${\cal L}_{\rm S}/({\cal L}_{\rm S}+{\cal L}_{\rm BG})$.
The threshold value depends both on the decay mode and 
on the flavor-tagging quality.

The vertex reconstruction, flavor tagging and the
unbinned maximum likelihood fit to the $\Dt$ distributions
are the same as those described for the $\sin2\phi_1$
measurement (Section~\ref{sec:sin2phi1}) 
with $\bz \to J/\psi \ks$ decays. 
By using the identical procedure for both cases, we can
reduce the systematic uncertainties in the differences
$\deltasphiks \equiv \calsphiks -\calsjpsiks$ etc.

After flavor tagging and vertex reconstruction,
we expect the signal yields and the purities listed in
Table~\ref{tbl:num}.

\begin{table}
\begin{center}
\begin{tabular}{llrlll}
\hline
\multicolumn{1}{c}{Mode} & \multicolumn{1}{c}{$\xi_f$} 
                 & $N_{\rm sig}$ & \multicolumn{1}{c}{Purity} 
                 & \multicolumn{2}{c}{Statistical error} \\
 & & & & \multicolumn{1}{c}{$\cals$} 
       & \multicolumn{1}{c}{$\cala$}\\
\hline
$\phi\ks$   & $-1$        & 2400 & 0.64 & 0.074 & 0.051\\
$K^+K^-\ks$ & $+1(100\%)$ & 7100 & 0.55 & 0.040 & 0.028\\
$\eta'\ks$  & $-1$        & 8700 & 0.58 & 0.042 & 0.028\\
\hline
\end{tabular}
\end{center}
\caption{
Expected numbers of $\bz\to\fcp$ signal events, $N_{\rm sig}$, 
and estimated signal purities in the $\dE$-$\mbc$ signal region 
for each $\fcp$ mode at 5 ab$^{-1}$.}
\label{tbl:num}
\end{table}

\begin{figure}[tbp]
  \begin{center}
    \includegraphics[width=0.32\textwidth,clip]{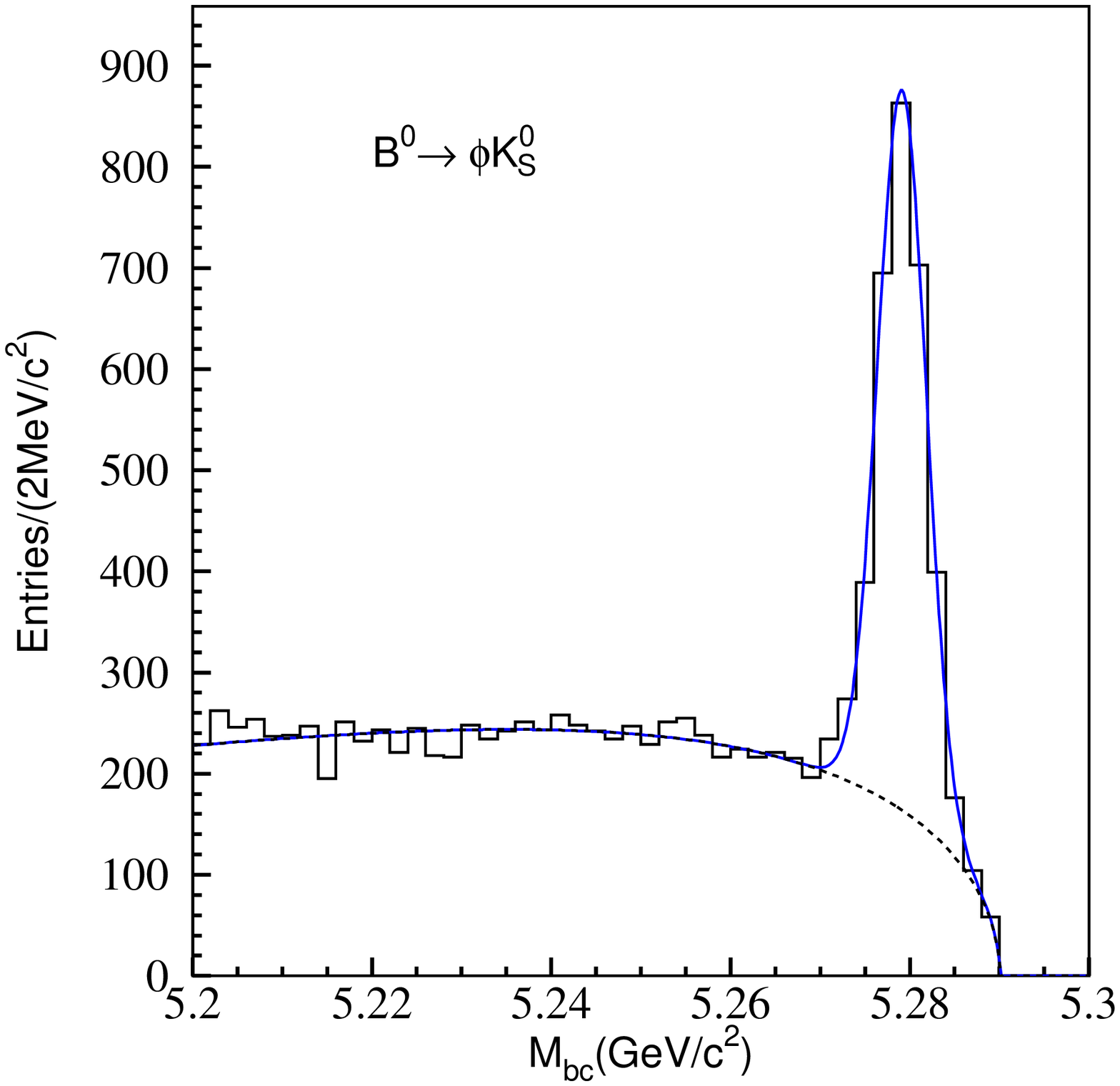}
    \includegraphics[width=0.32\textwidth,clip]{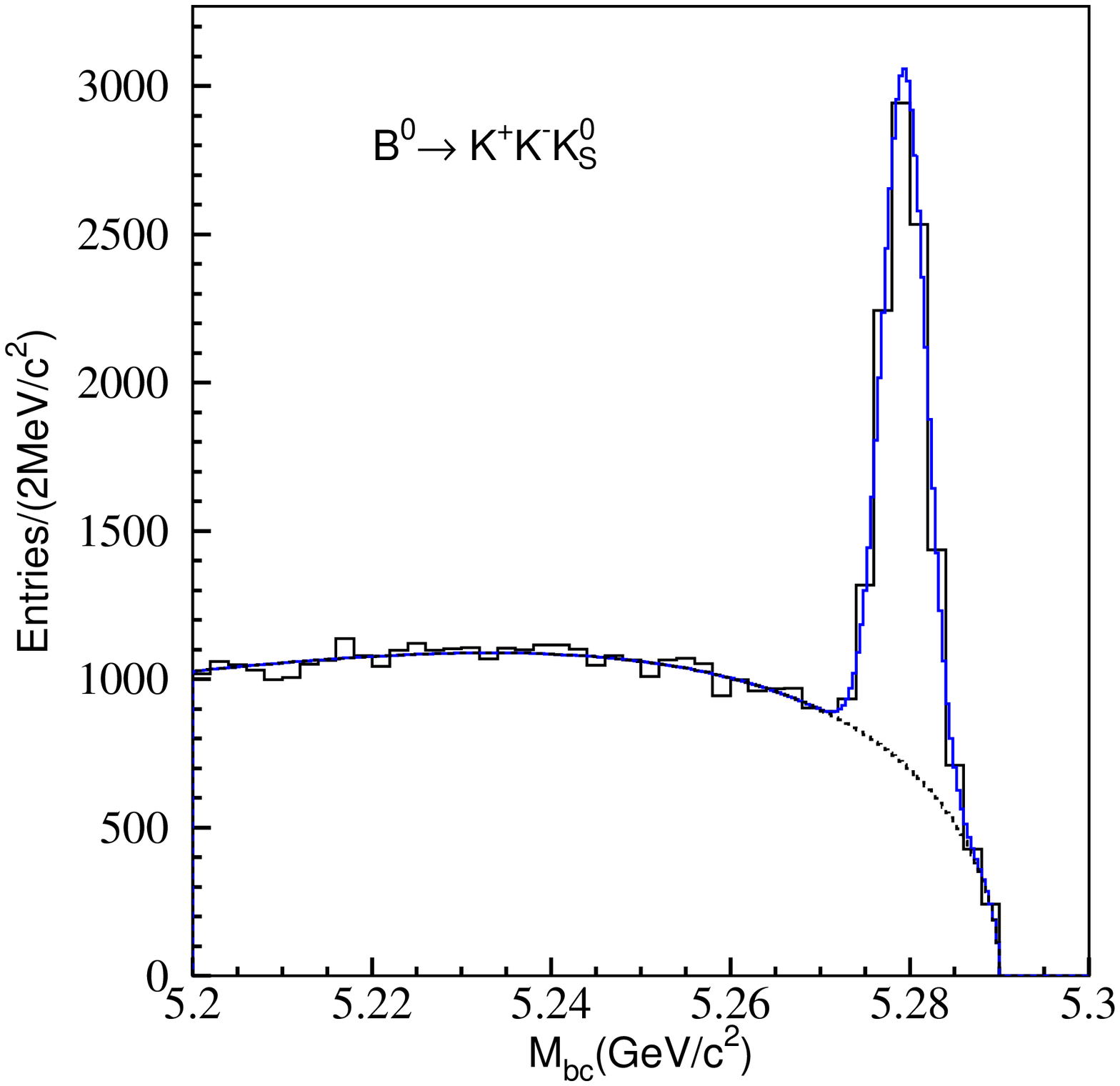}
    \includegraphics[width=0.32\textwidth,clip]{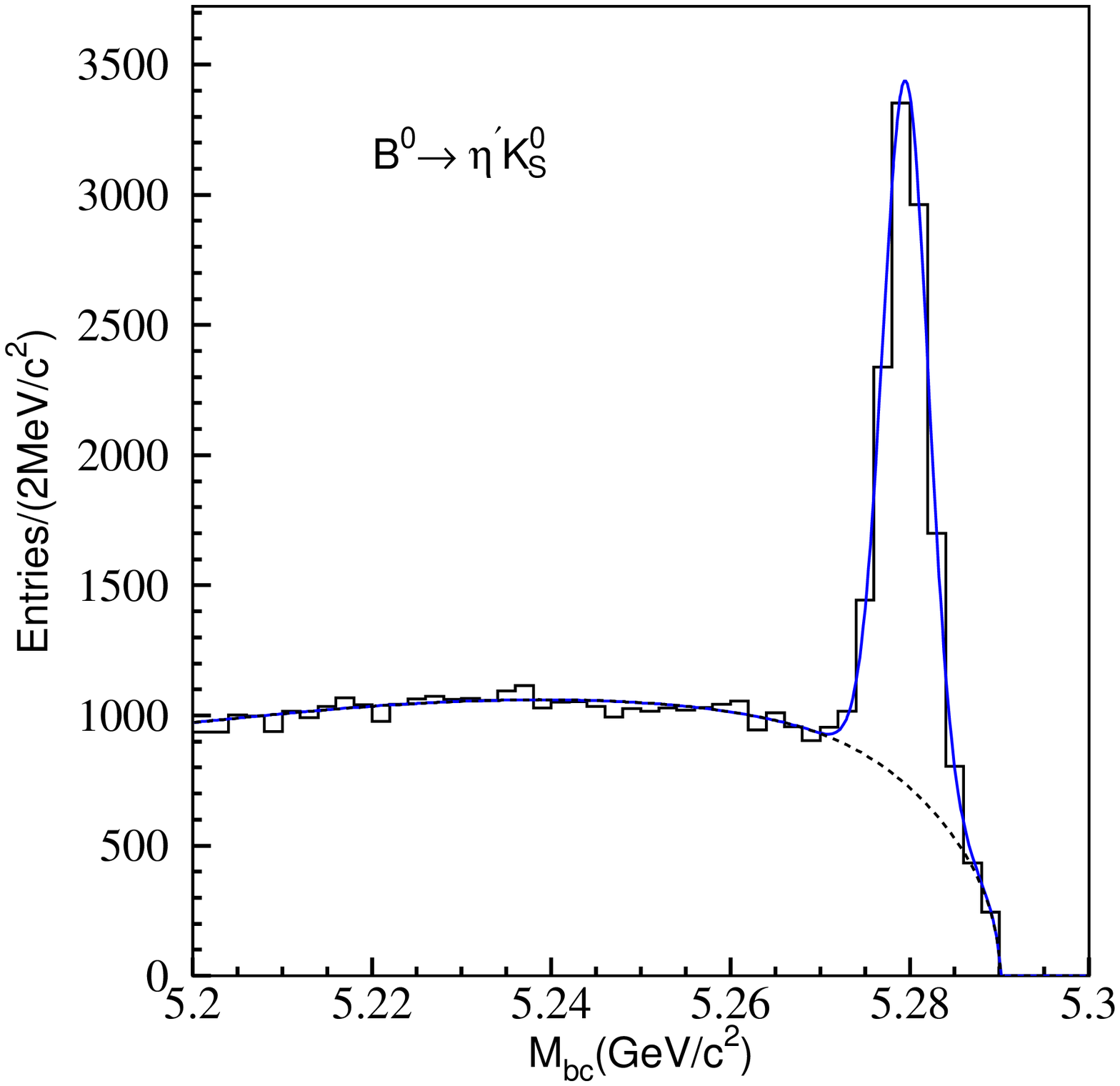}
  \end{center}
\caption{Beam-constrained mass ($\mb$) distributions for
(left) $\bztophiks$, (middle) $\bztokkks$, and (right) $\bztoetapks$
decays within the $\dE$ signal region expected at 5 ab$^{-1}$.
Solid curves show results of fits to signal plus background distributions,
and dashed curves show the background contributions.
}\label{fig:sss_mbc}
\end{figure}

Figure~\ref{fig:sss_mbc} shows the $\mb$ distributions for the 
reconstructed $B$ candidates that have $\dE$ values within
the signal region. 

Unbinned maximum likelihood fits yield expected statistical
errors on $\cals$ and $\cala$ shown also in Table~\ref{tbl:num}.
It is seen that errors are rather small even for these
rare $B$ decays. Hence systematic uncertainties become crucial.

Major sources of the systematic uncertainties 
on $\cals$ and $\cala$ are common
to those for $\bz \to \jpsi\ks$, which will be
described in Section~\ref{sec:sin2phi1}.
For the $\kkks$ mode, the $CP$-even fraction is expected
to be determined more precisely as the integrated luminosity
increases. Therefore we assume that the systematic uncertainty
due to the uncertainty in the $CP$-even fraction 
is {\it reducible}; \textit{e.g.} it is proportional to
1/$\sqrt{{\cal L}_{\rm tot}}$.
An additional systematic error for $\bz \to \phi\ks$ 
arises from the $\bz \to f_0 \ks$ and
$\kkks$ backgrounds. These background fractions are estimated
at 140 fb$^{-1}$ from the $K^+K^-$ invariant mass distribution
to be $1.6^{+1.9}_{-1.5}$\% and $7.2\pm 1.7$\%,
respectively. 
We assume that the errors on these fractions will be reduced
as the integrated luminosity increases.
We also assume that the $CP$-violating parameters for
these decays are also determined experimentally
with errors that will also decrease as the 
integrated luminosity increases.
Therefore, the systematic errors on $\cals$ and $\cala$
due to these backgrounds 
are also assumed to be {\it reducible}.

Some of the systematic errors cancel
when we calculate $\Delta\cala$ or $\Delta\cals$.
For example, the effect of the tag-side interference 
cancels in $\deltaaphiks$ and
$\deltasphiks$
since it causes a bias in the same direction for
$\calsphiks$ and $\calsjpsiks$ measurements.
On the other hand,
a special care on the systematic bias from the
tag-side interference needs to be taken for $\deltaakkks$.
In this case, the effect does not cancel
because the bias has the opposite sign to each other.
We use information from $\bz\to J/\psi\kl$ decays
to reduce this uncertainty on $\deltaakkks$.

\begin{figure}[tbp]
  \begin{center}
    \includegraphics[width=0.8\textwidth,clip]{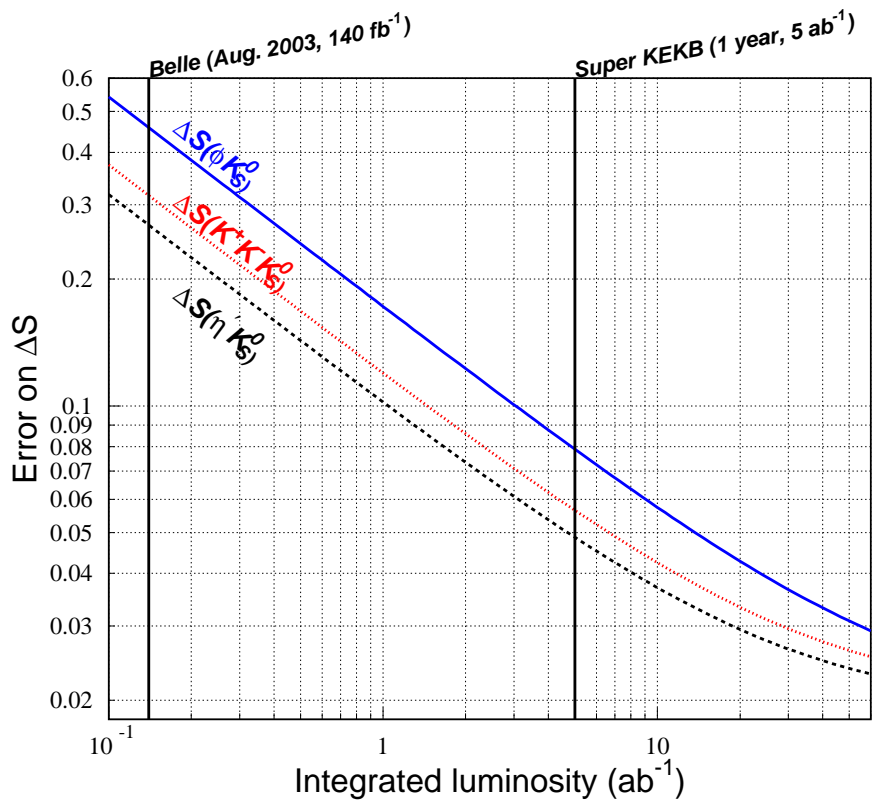}
    \includegraphics[width=0.8\textwidth,clip]{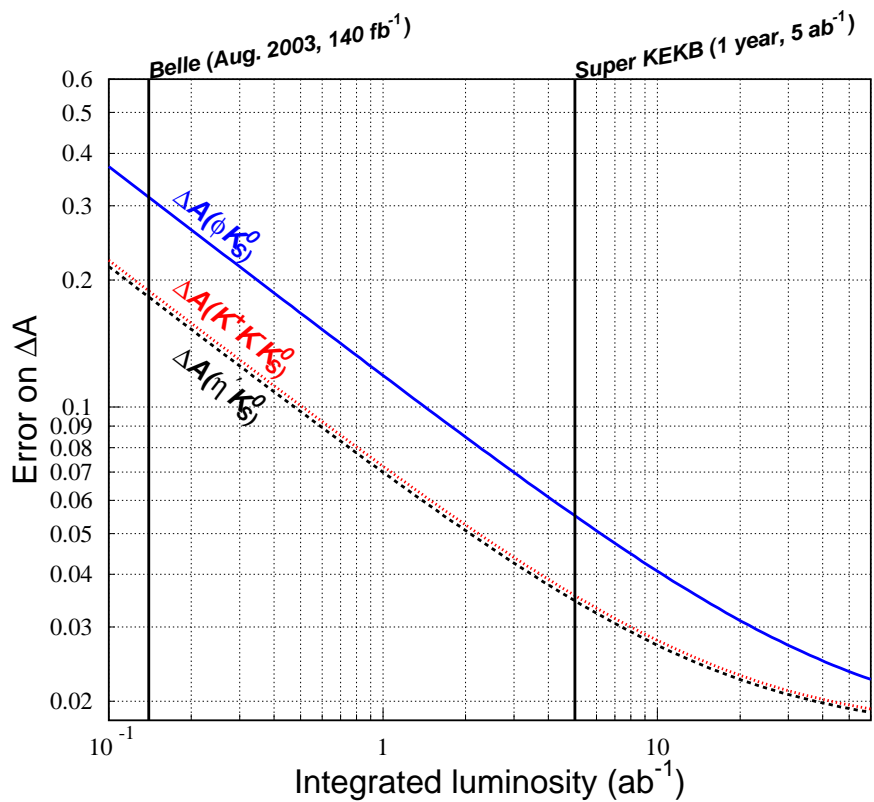}
  \end{center}
 \caption{Expected total errors on $\Delta\cals$ (top) and
          $\Delta\cala$ (bottom) as a function of
          integrated luminosity.}
\label{fig:ds_da_sss_loi}
\end{figure}

Figure~\ref{fig:ds_da_sss_loi}
shows the resulting total errors
on $\Delta\cals$ and $\Delta\cala$
as a function of integrated luminosity.
Table~\ref{tbl:sss_summary} also shows 
the corresponding values at 5 ab$^{-1}$ and 50 ab$^{-1}$.

\begin{table}[tbp]
\begin{center}
\begin{tabular}{|l|cc|cc|}
\hline
\multicolumn{1}{|c|}{Mode} & \multicolumn{2}{|c|}{5 ab$^{-1}$}
             & \multicolumn{2}{|c|}{50 ab$^{-1}$} \\
  & $\Delta\cals$ & $\Delta\cala$ & $\Delta\cals$ & $\Delta\cala$ \\
\hline 
$\phi\ks$   & 0.079 & 0.055 & 0.031 & 0.024 \\
$K^+K^-\ks$ & 0.056 & 0.036 & 0.026 & 0.020 \\
$\eta'\ks$  & 0.049 & 0.035 & 0.024 & 0.019 \\
\hline
\end{tabular}
\end{center}
\caption{Expected total errors on $\Delta\cals$
and $\Delta\cala$ at 5 ab$^{-1}$ and 50 ab$^{-1}$.} 
\label{tbl:sss_summary}
\end{table}

\begin{figure}[tbp]
  \begin{center}
    \includegraphics[width=0.32\textwidth,clip]{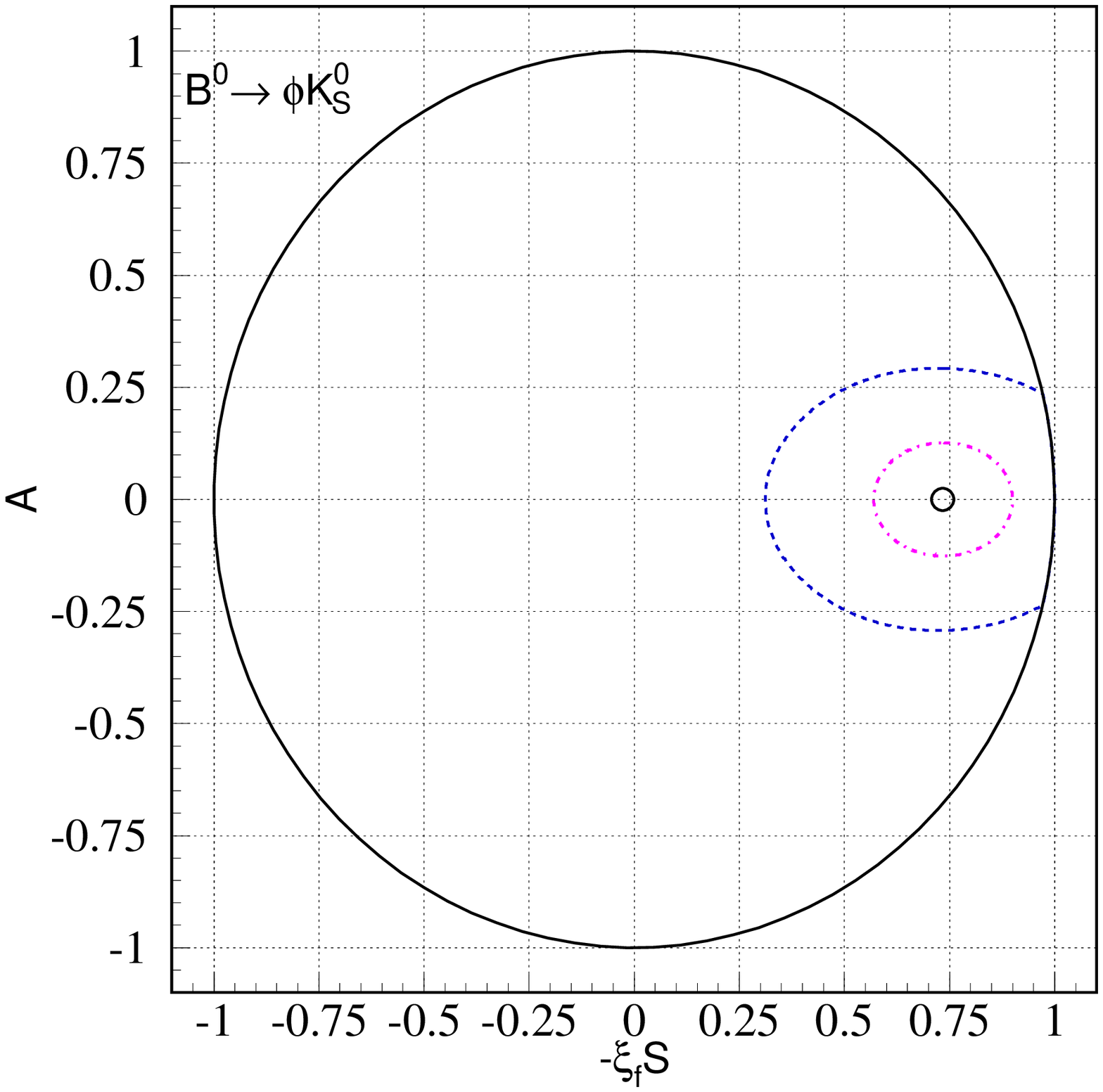}
    \includegraphics[width=0.32\textwidth,clip]{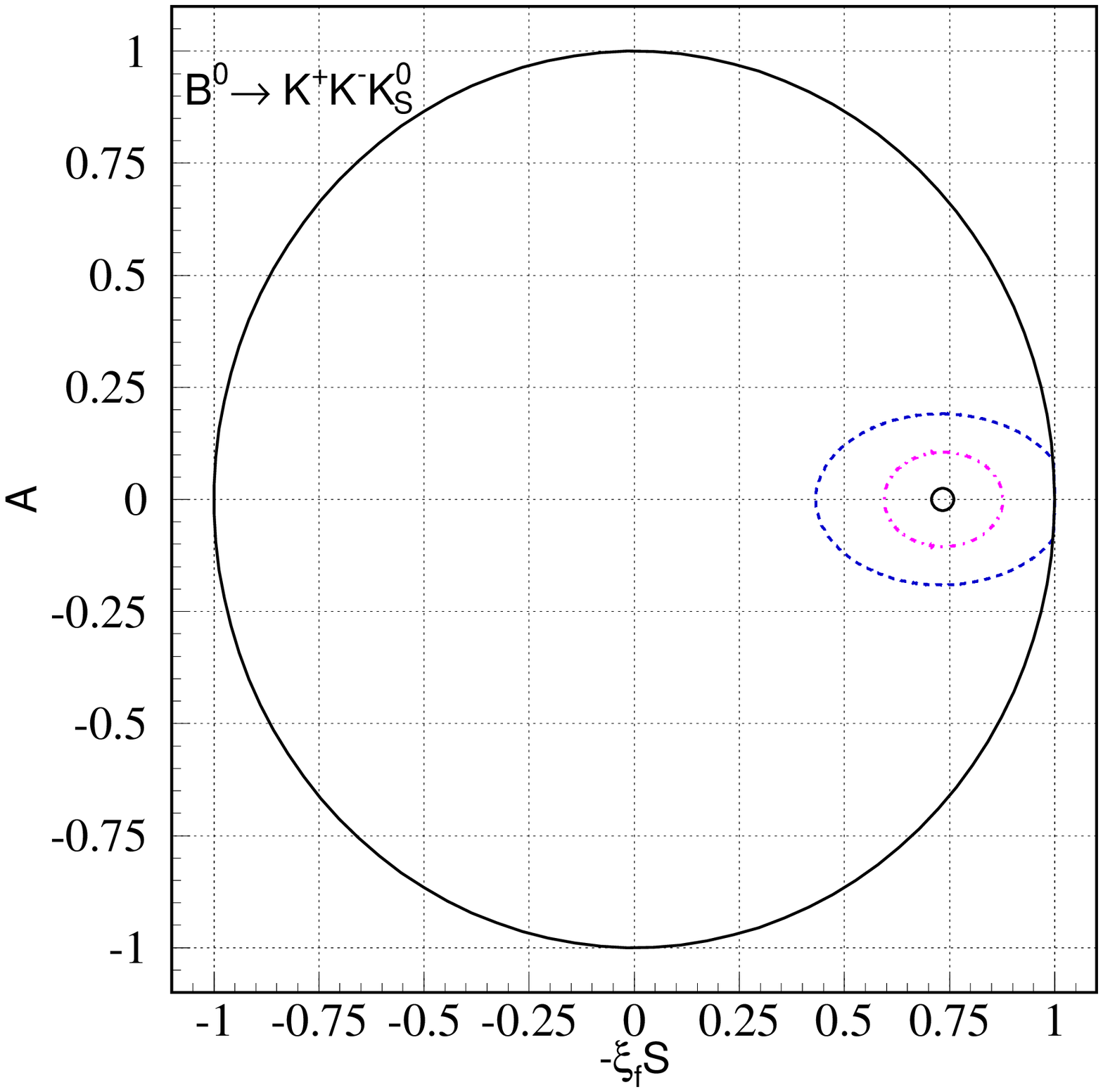}
    \includegraphics[width=0.32\textwidth,clip]{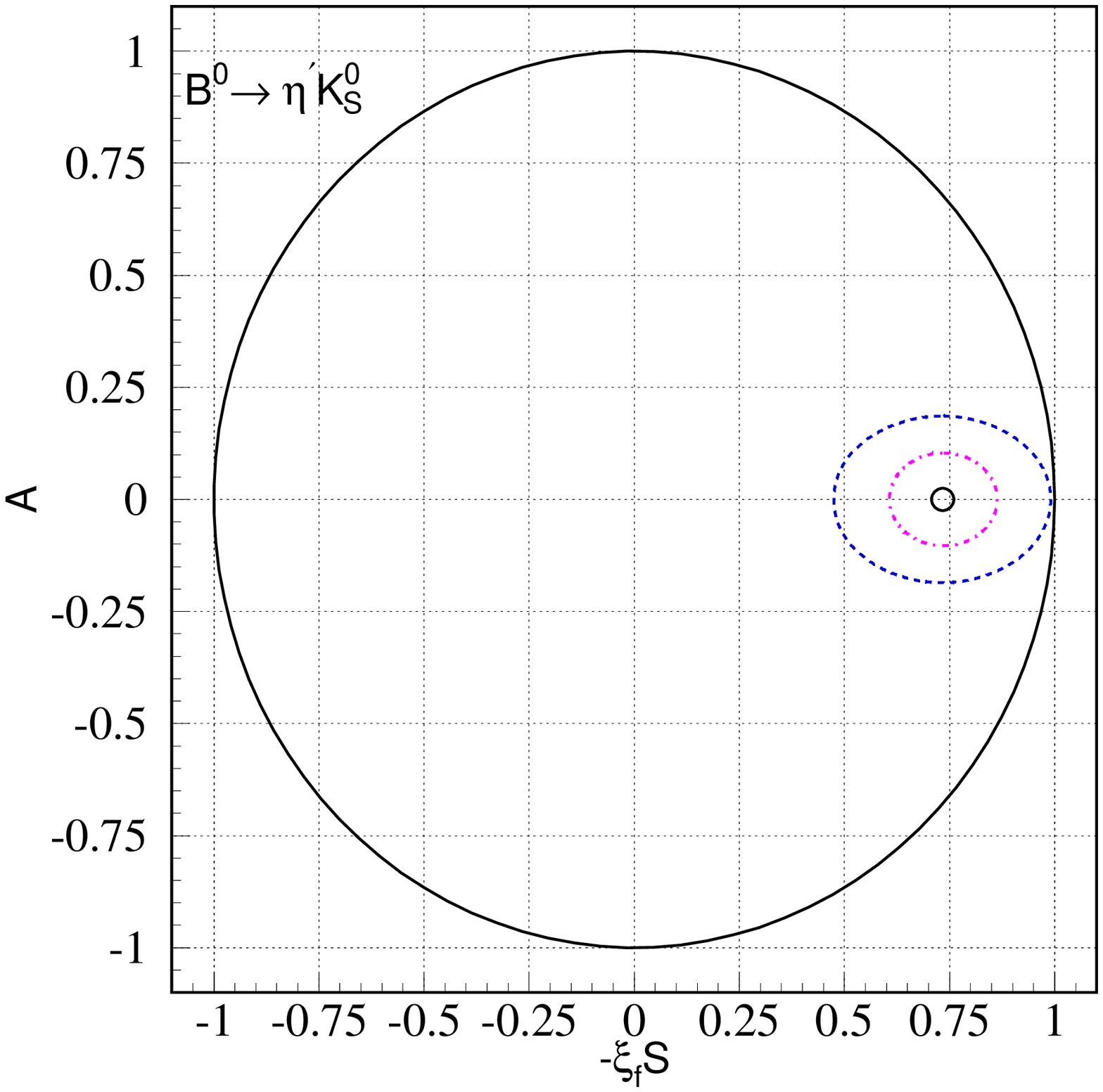}
  \end{center}
 \caption{5$\sigma$ confidence regions for $\cala$ and $\cals$ in
          (left) $\bztophiks$, (middle) $\bztokkks$ 
          and (right)$\bztoetapks$ decays at 5 ab$^{-1}$ and 50 ab$^{-1}$.
          Input values are $\cals = 0.73$ and $\cala = 0$.
          }
\label{fig:sss_2d_5sigma}
\end{figure}

Based on the above estimates, we perform Feldman-Cousins
analyses to obtain 5$\sigma$ discovery regions
at 5 ab$^{-1}$ and at 50 ab$^{-1}$
in the 2-dimensional plane of $\cala$ and $\cals$.
Results are shown in Figure~\ref{fig:sss_2d_5sigma}.
At SuperKEKB, even a small deviation of
$\Delta\cals \sim 0.1$ can be
established with a 5$\sigma$ significance
as far as the statistical and systematic errors are concerned.
Therefore it is important to understand levels of 
theoretical uncertainties within the SM very well.
This issue will be discussed in 
Section~\ref{sec:sss/discussion}.

\subsection{$\bz \to \ks\ks\ks$ and $\pi^0\ks$}
In this section we discuss the feasibility of
time-dependent $CP$ asymmetry measurements using
only a $\ks$ and a constraint from the interaction point
to determine the $B$ decay vertices.
In particular, we consider $\bz \to \ks\ks\ks$ and
$\pi^0\ks$ decays as the most promising modes in this
class of decays to
study a new $CP$-violating phase in $b \to s$
transitions
(another important mode $\bz \to K^{*0}(\to \ks\piz)\gamma$
is discussed in Sec.~\ref{sec:sg_sll:s_kstargamma}).
Recently, the BaBar collaboration has succeeded in measuring
the $B$ decay vertex in $B^0 \to \ks \pi^0$~\cite{Browder:2003ii}.
The possibility of obtaining $B$ vertex information from $\ks$ mesons 
makes time-dependent analyses of these decays feasible.

Recently it was pointed out~\cite{Gershon:2004tk} that
in decays of the type $\bz \to P^0Q^0X^0$, where
$P^0$, $Q^0$ and $X^0$ represent any $CP$ eigenstate spin-0 neutral particles,
the final state is a $CP$ eigenstate.
First we give a brief proof of this statement.
In what follows, $L$ denotes the angular momentum of the $P^0Q^0$ system,
and $L^{\prime}$ denotes the angular momentum of 
$X^0$ relative to the $P^0Q^0$ system.
By conservation of angular momentum in the decay $B^0 \to P^0Q^0X^0$, 
we obtain
\begin{eqnarray}
  \label{eq:consJ} 
  {\bf J}_{B^0} & = &
  {\bf L} + {\bf L}^{\prime} + 
  {\bf S}_{P^0} + {\bf S}_{Q^0} + {\bf S}_{X^0} 
  \\
  {\bf 0} & = & {\bf L} + {\bf L}^{\prime},
\end{eqnarray}
since the neutral $B$ meson is a spin-0 particle,
as are $P^0$, $Q^0$ and $X^0$.
In the above equations, ${\bf J}$, ${\bf L}$ and ${\bf S}$ represent
the total, orbital and intrinsic angular momentum, respectively
(and elsewhere $J$, $L$ and $S$ represent their magnitudes).
Therefore, the angular momentum between the $P^0Q^0$ system and $X^0$ 
($L^{\prime}$) must be equal to $L$, 
and we can write down the $CP$ of the $P^0P^0X^0$ system:
\begin{eqnarray}
  \label{eq:cp_p0q0x0}
  CP\left(P^0Q^0X^0\right) & = & 
  CP\left(P^0\right) \times CP\left(Q^0\right) \times CP\left(X^0\right) 
  \times \left(-1\right)^{L} \times \left(-1\right)^{L^{\prime}} 
  \\
  & = & 
  CP\left(P^0\right) \times CP\left(Q^0\right) \times CP\left(X^0\right).
\end{eqnarray}
Thus there are many three-body $B$ decays that can be used
in time-dependent $CP$ asymmetry measurements
{\it without any angular analysis}.
In particular, the decay $\bz \to \ks\ks\ks$ is promising.
In terms of phenomenology, 
it has an advantage over $B^0 \to K^+K^- \ks$, 
since the latter includes a contribution from the $b \to u$ tree diagram, 
which has a different weak phase. Although this contribution 
(``tree pollution'') is expected
to be small as will be discussed in Section~\ref{sec:sss/discussion},
it might be an issue at SuperKEKB.
In the case of $\bz \to \ks\ks\ks$, however,
there is no $u$ quark in the final state.
The $\btosuu$ tree diagram
followed by rescattering into $\sdd$ or $\sss$ 
is OZI-suppressed.
Therefore these are almost pure penguin decays.
There can be contributions from both 
$\btosss$ with additional $d \bar{d}$ production, and 
$\btosdd$ with additional $s \bar{s}$ production, 
but these diagrams have the same weak phase.
Any new physics contribution expected in the $B^0 \to \phi \ks$ decay
can also affect
$B^0 \to \ks\ks\ks$,
and in the absence of new physics it should exhibit the same $CP$ violating
effects as $J/\psi \ks$.

\begin{figure}[tbp]
  \begin{center}
    \includegraphics[width=0.49\textwidth,clip]{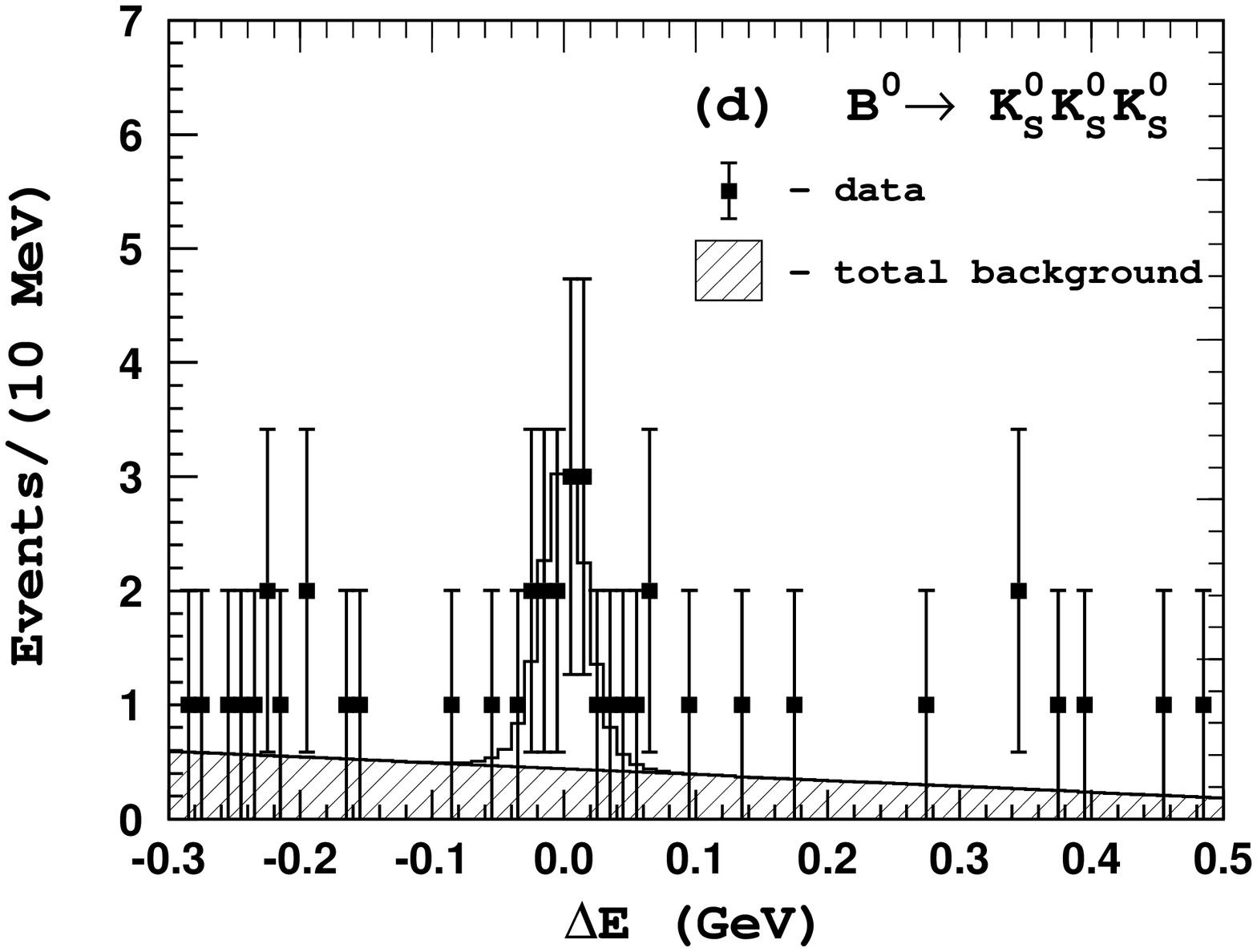}
    \includegraphics[width=0.49\textwidth,clip]{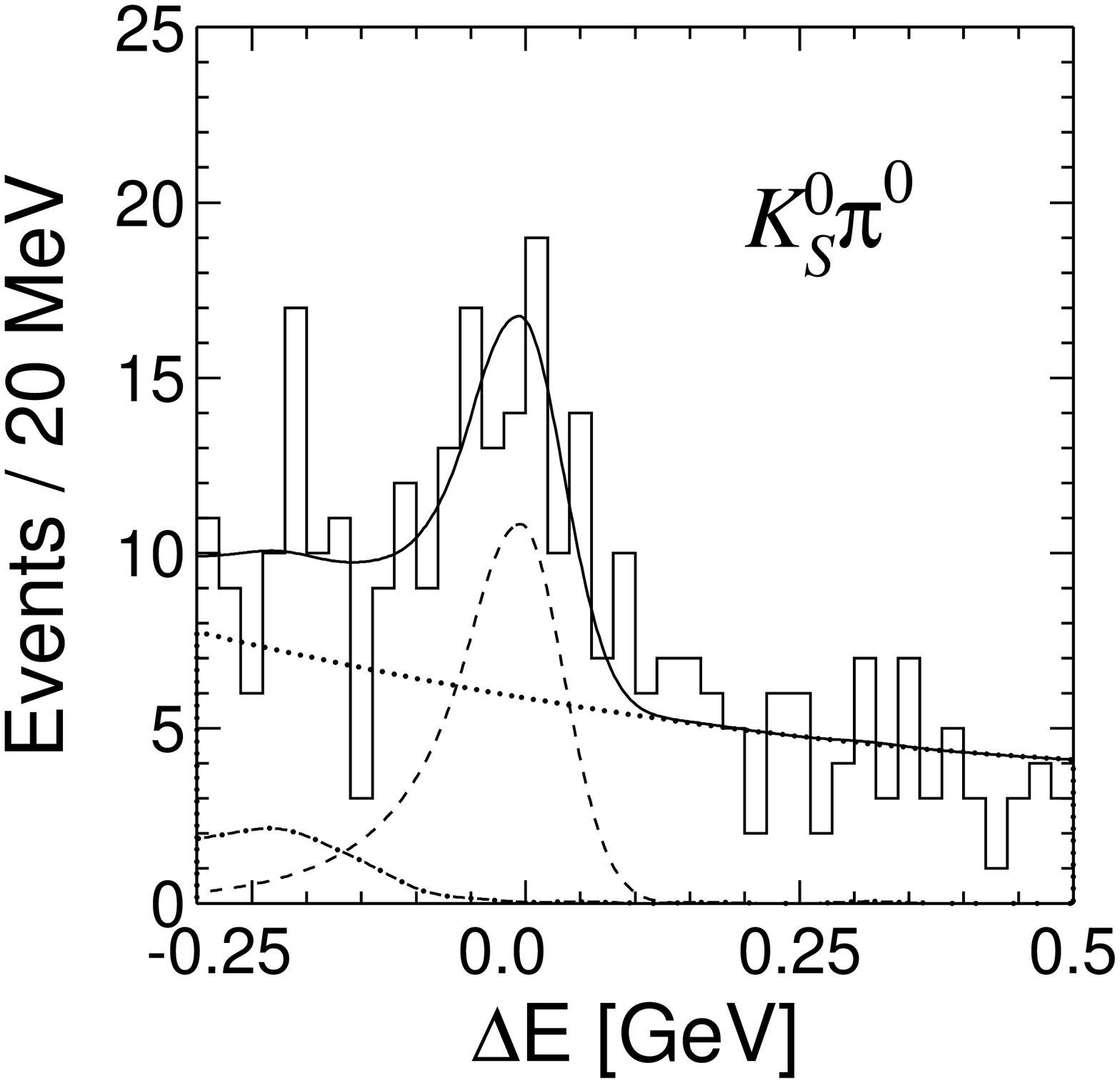}
  \end{center}
 \caption{(left) 
    Observation of $B^0 \to \ks \ks \ks$~\cite{Garmash:2003er}.
    From $78 \ {\rm fb}^{-1}$ of data recorded on the $\Upsilon(4S)$ resonance,
    a signal yield of $12.2^{+4.5}_{-3.8}$ is obtained, 
    leading to a branching fraction of 
    ${\cal B}\left(B^0 \to \ks \ks \ks\right) = 
    \left( 4.2^{+1.6}_{-1.3} \pm 0.8 \right) \times 10^{-6}$.
    (right) Observation of $B^0 \to \ks \pi^0$~\cite{Chao:2003ue}.
    From $78 \ {\rm fb}^{-1}$ of data recorded on the $\Upsilon(4S)$ resonance,
    a signal yield of $72.6\pm 14.0$ is obtained, 
    leading to a branching fraction of 
    ${\cal B}\left(B^0 \to \ks \ks \ks\right) = 
    (11.7 \pm 2.3^{+1.2}_{-1.3}) \times 10^{-6}$.
  }
\label{fig:ksksks_kspi0}
\end{figure}

Turning to experimental considerations, 
we note that this mode has been observed at Belle~\cite{Garmash:2003er}.
From $78 \ {\rm fb}^{-1}$ of data recorded on the $\Upsilon(4S)$ resonance, 
$12.2^{+4.5}_{-3.8}$ signal events are found.
From Figure~\ref{fig:ksksks_kspi0} (left), we can count the
number of candidates 
in the region $-0.1 \ { \rm GeV} < \Delta E < 0.1 \ { \rm GeV}$, 
to estimate the signal purity.
There are 21 candidates in this region giving a purity
of $\sim$0.6, which is approximately the same purity as $\etap\ks$.
The efficiency to reconstruct a $\ks$ vertex reflects the
probability for the $\ks$ to decay inside the vertex detector,
and so depends on the $\ks$ momentum and on the size of the vertex detector.
In a three body $B^0 \to \ks\ks\ks$ decay, 
at least one $\ks$ must have fairly low momentum in the $B^0$ rest frame.
Therefore, we expect a high vertex efficiency for $B^0 \to \ks\ks\ks$,
if the vertex efficiency for $B^0 \to \ks \pi^0$ 
(where the $\ks$ has high momentum) is moderate.
In the time-dependent analysis of $B^0 \to \ks \pi^0$ 
by the BaBar collaboration, 
a vertex efficiency of 65\% is obtained~\cite{Browder:2003ii}.
Since the size of SuperKEKB vertex detector is comparable
to the BaBar SVT,
we expect that vertex efficiencies for $B^0 \to \ks\ks\ks$ 
of close to 100\% may be obtained.
Using these assumptions, we estimate the statistical error on
$\calsksksks)$ to be
\begin{equation}
\begin{array}{c}
\delta \calsksksks = 0.14 \mbox{~(at $5\abinv$)}\\
\delta \calsksksks = 0.04 \mbox{~(at $50\abinv$)}.
\end{array}
\end{equation}

The authors of \cite{Ciuchini:1997zp} claim that
$K_S \pi^0$ is one of the gold-plated modes to study
a new $CP$-violating phase in the $b \to s$ transition.
As mentioned before, the BaBar group
recently showed a preliminary result on
the time-dependent $CP$ asymmetries in this decay,
demonstrating the feasibility of measuring
the $B$ meson decay point using only a $\ks$
and the interaction point constraint.

Belle's result based on 78 fb$^{-1}$ of data 
is shown in Figure~\ref{fig:ksksks_kspi0} (right).
The signal yield is $72.6 \pm 14.0$.
Since this estimation includes the tail region in the $\Delta E$
distribution where rare $B$ decay backgrounds are not negligible,
a more stringent selection is necessary for the time-dependent
$CP$ asymmetry measurements. 
Using $|\Delta E| < 0.1$ GeV
and assuming a vertex reconstruction efficiency of 50\%,
the signal yield with reconstructed vertices
at 5 ab$^{-1}$ is $\sim$1900 for a purity of 0.46.
Based on these numbers, we obtain
\begin{equation}
\begin{array}{c}
\delta \calspizks = 0.10 \mbox{~(at $5\abinv$)}\\
\delta \calspizks = 0.03 \mbox{~(at $50\abinv$)}.
\end{array}
\end{equation}

\subsection{$\bpmtophiphixspm$}
In this section we discuss a new method to study
direct $CP$ violation that arises from a new $CP$-violating phase
in $\bpmtophiphixspm$ decays~\cite{Hazumi:2003it}.
Here $\xspm$ represents a final state with a specific strange flavor
such as $K^{\pm}$ or $K^{*\pm}$.
These non-resonant direct decay amplitudes are dominated by
the $\btosssss$ transition.
A contribution from the $\btouus$ transition followed by rescattering
into $s\bar{s}s$ is expected to be below 1\% because of
CKM suppression and
the OZI rule~\cite{Hazumi:2003it}.
In these decays, when the invariant mass of the $\phi\phi$ system
is within the $\etac$ resonance region, they interfere with
the $\bpm \to \etac (\to \phi \phi) \xspm$ decay
that is dominated by the $\btoccs$ transition.
The decay width of
$\etac$ is sufficiently large~\cite{Hagiwara:fs,Fang:2002gi}
to provide a sizable interference.
Within the SM, this interference does not cause sizable direct $CP$ violation
because there is no weak phase difference between the $\btosssss$ 
and the $\btoccs$ transitions.
On the other hand, a NP contribution with a new $CP$-violating phase
can create a large weak phase difference.
Thus large $CP$ asymmetries can appear only from NP amplitudes, and
an observation of direct $CP$ violation in these decays is an unambiguous
manifestation of physics beyond the SM.

Although the same argument so far is applicable 
to the $\bpm \to \phi \xspm$ decays,
there is no guaranteed strong phase difference that is calculable
reliably for these decays.
In contrast,
the Breit-Wigner resonance provides 
the maximal strong phase difference in the case of
$\bpm \to (\phi \phi)_{m \sim m_{\etac}} \xspm$ decays.
Since present experimental knowledge of the decay rate for $\btosss$ 
is still limited, a large $CP$ asymmetry up to 0.4 is allowed.

The Belle Collaboration recently announced evidence for
$\btophiphik$ decays~\cite{Huang:gc}.
The signal purity is close to 100\% 
when the $\phi\phi$ invariant mass is within the $\etac$ mass region.
Belle~\cite{Fang:2002gi} has also reported 
the first observation of the $\bztoetackstarz$ decay.
This implies that other modes such as 
$\bptoetackstarp$ will also be seen with a similar branching
fraction, so that we will be able to study
semi-inclusive $\bpmtoetacxspm$ transitions experimentally.
The semi-inclusive branching fraction of $\bpmtoetacxspm$ is not yet measured,
but is expected
to be comparable to the branching fraction of the semi-inclusive
decay $\bpm \to J/\psi\xspm$
\cite{Deshpande:1994mk,Gourdin:1994xx,Ahmady:1997ph}. 

\begin{figure}[tbp]
  \begin{center}
    \includegraphics[width=0.9\textwidth,clip]{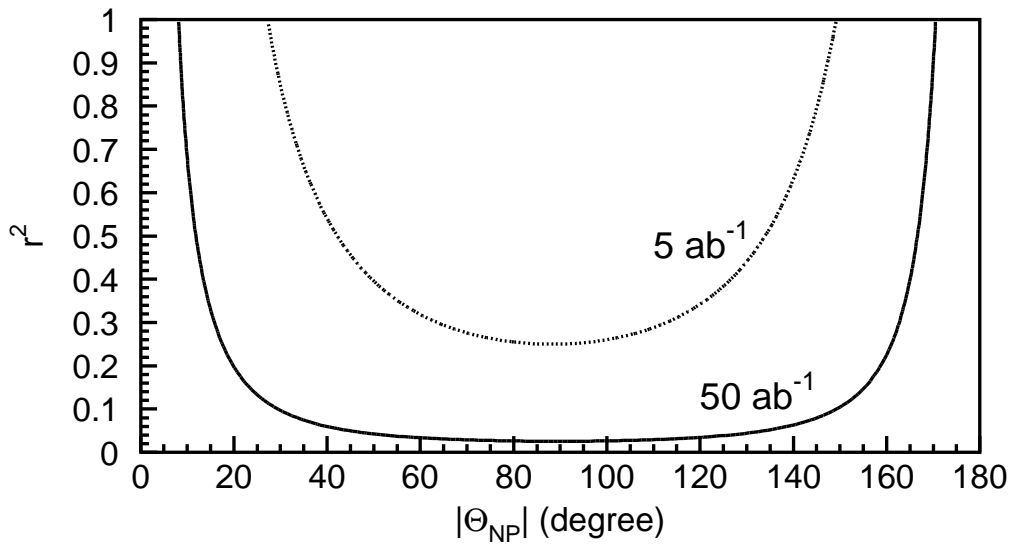}
  \end{center}
 \caption{Expected sensitivities on direct $CP$ violation in 
          the $\bpmtophiphikpm$ decay
          at 5 ab$^{-1}$ (dotted line) 
          and at 50 ab$^{-1}$ (solid line).
          In the regions above the curves, direct $CP$ violation
          can be measured with a 5$\sigma$ significance or larger.}
\label{fig:tnpvsr2}
\end{figure}

We have performed Monte Carlo simulation for
the $\bpmtophiphikpm$ decay and estimated statistical errors
on the $CP$ asymmetry parameter.
The procedure and the fit parameters are the same as those described
in \cite{Hazumi:2003it}.
The reconstruction efficiency and the $\phi\phi$ mass resolution
are estimated using a GEANT-based detector simulator for
the present Belle detector~\cite{:2000cg}.
We perform an unbinned maximum-likelihood fit
to the differential decay rate distribution.
Figure~\ref{fig:tnpvsr2} shows the 5$\sigma$ search regions
at 5 ab$^{-1}$ (dotted line) and
at 50 ab$^{-1}$ (solid line),
where $r^2$ is the ratio between the NP amplitude and the SM amplitude,
and $\tnp$ is the $CP$-violating phase from NP.
Direct $CP$ violation can be observed in a large parameter space
with significance above 5$\sigma$.

\begin{figure}[tbp]  
  \begin{center}
    \includegraphics[width=0.9\textwidth,clip]{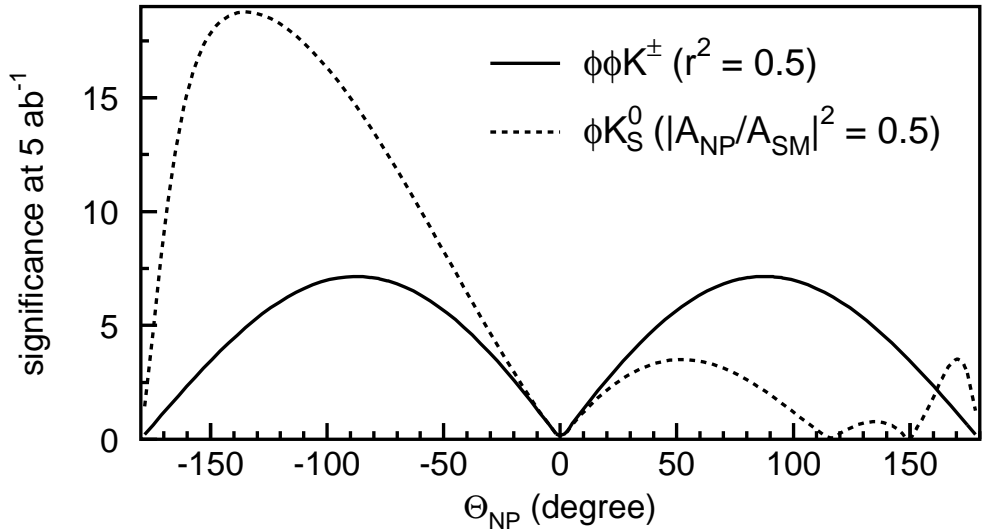}
  \end{center}
 \caption{Expected statistical significance of deviations from
          the SM for direct $CP$ violation in the $\bpmtophiphikpm$ decay
          with $\rtwo = 0.5$ (solid line)
          and for time-dependent $CP$ violation in
          the $\bztophiks$ decay with $|A_{\rm NP}/A_{\rm SM}|^2 = 0.5$
          (dashed line). For each case, significance is calculated
          at 5 ab$^{-1}$.}
\label{fig:signp1010}
\end{figure}

Figure~\ref{fig:signp1010} shows the expected
significance of the new phase $\tnp$ at 5 ab$^{-1}$ for
$\bpmtophiphikpm$ decay ($\rtwo = 0.5$) and for
time-dependent $CP$ violation in
the $\bztophiks$ decay ($|A_{\rm NP}/A_{\rm SM}|^2 = 0.5$).

The significance for $\deltasphiks$
depends on the sign of $\tnp$, which
is not the case for the $\bpmtophiphikpm$ decay.
The sign dependence arises from an asymmetric
range for $\deltasphiks$; to a good approximation,
we have
$-1-\sin 2\phi_1 \leq \Delta\cals \leq 1-\sin 2\phi_1$
where $\sin 2\phi_1 = +0.736\pm 0.049$~\cite{HFAG}.
Therefore the $\bpmtophiphikpm$ decay plays a unique
role in searching for a new $CP$-violating phase.

Experimental sensitivities can be improved by
adding more final states. The
technique to reconstruct $\xs$,
which has been successfully adopted for the measurements
of semi-inclusive
$B \to \xs \ell\ell$ transitions~\cite{Kaneko:2002mr},
can be used for this purpose.
Flavor-specific neutral $B$ meson decays,
such as $\bz \to \phi\phi K^{*0}(\to K^+\pi^-)$,
and other charmonia such as
the $\chi_{c0} \to \phi\phi$ decay can also
be included.

\subsection{Discussion}
\label{sec:sss/discussion}
\begin{figure}[tbp]
 \begin{center}
 \includegraphics[width=0.9\textwidth,clip]{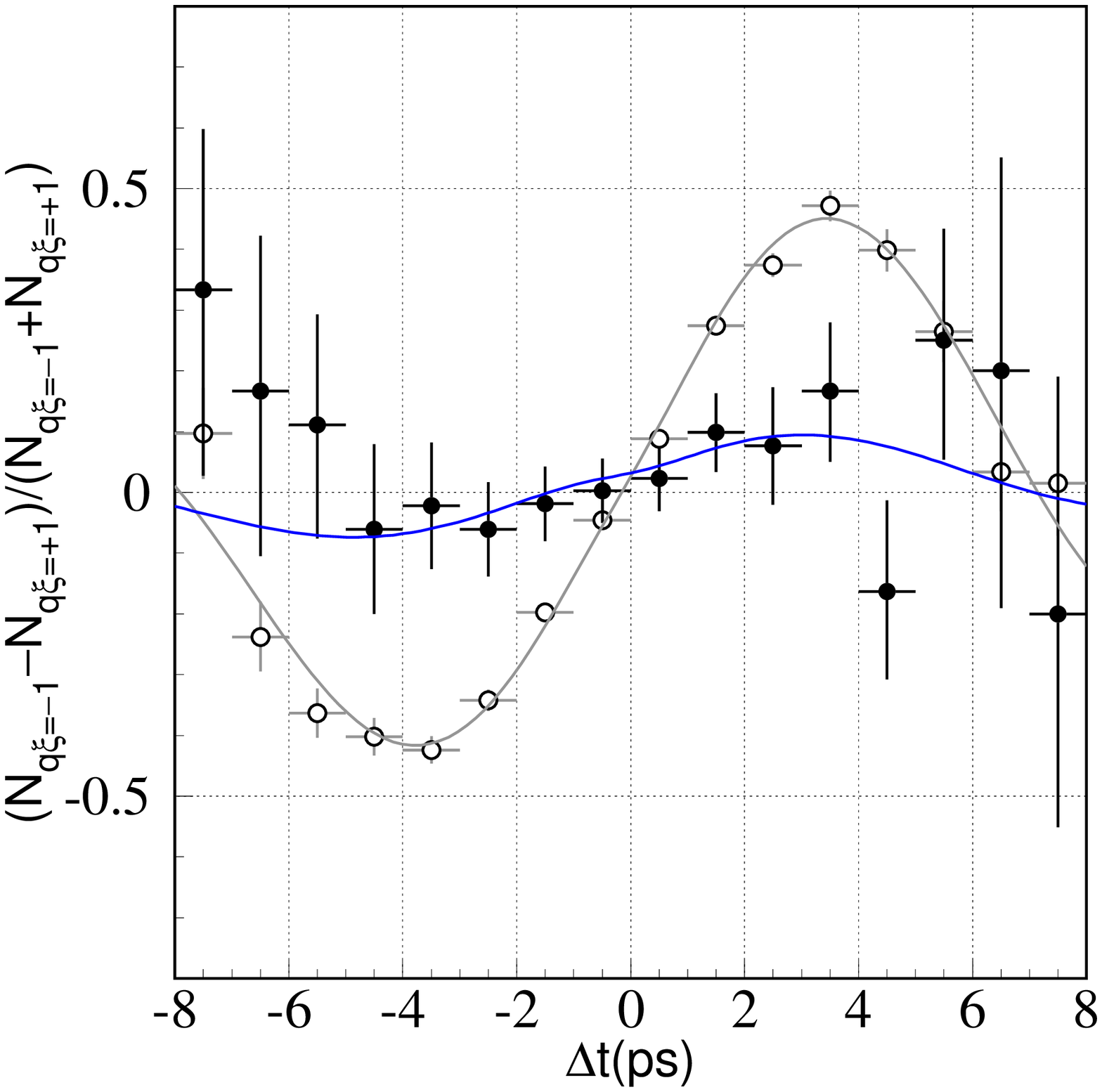}
 \end{center}
 \caption{Raw asymmetries for $\bztophiks$ and $\bz\to\jpsi\ks$ at 5 ab$^{-1}$.
          Input values are $\calsphiks = +0.24$ and $\calaphiks = +0.07$,
          which are the HFAG average for the $b \to s$ transition
          as of LP03~\cite{HFAG}.}
\label{fig:phiks_vs_jpsiks}
\end{figure}

As is discussed in the previous sections,
statistical errors in new phase measurements
can be at a few percent level at SuperKEKB.
Figure~\ref{fig:phiks_vs_jpsiks} shows an example
of a fit to events in a MC pseudo-experiment for
the $\bztophiks$ and $\jpsi\ks$ decays at 5 ab$^{-1}$,
where the input value of $\cals(\phi\ks) = +0.24$ is chosen to be
the world average value of the $\cals$ term
for $\bztophiks$, $\etapks$ and $\kkks$.

This level of large deviation can easily be observed 
only with a single decay channel $\bztophiks$ at SuperKEKB.
Combining all the available modes described in
the previous sections allows us to measure a deviation
of $\sim 0.1$.
At this level, even the SM may be able to create
non-zero values of $\Delta\cals$. Therefore
it is important to evaluate $\Delta\cals$ within
the SM.

Grossman, Isidori and Worah \cite{Grossman:1997gr} 
analyzed the possible
pollution in the $\bz \to \phi\ks$ decay, which
comes from the $b \to u\overline{u} s$ transition that contains $V_{ub}$.
They estimate that the pollution is at most 
$O(\lambda^2) \sim$ 5\%.
In addition, they claim that the upper limit of
the pollution will be obtained experimentally from
the ratios of branching fractions
${\cal B}(B^+ \ra \phi \pi^+)/{\cal B}(B^0 \ra \phi K_S)$ and
${\cal B}(B^+ \ra K^*K^+)/{\cal B}(B^0 \ra \phi K_S)$. This
is due to the fact that 
enhancement of $b \ra u\ubar s$ should also be detected in these
modes. Therefore they conclude that
new physics is guaranteed
if $|\Delta\cals(\phi\ks)| > 0.05$ is established.

For the $\bztoetapks$ decay, London and Soni \cite{London:1997zk} 
discussed the tree ($b \ra u\ubar s$)
pollution by evaluating
$T(\eta '  K_S)/P(\eta ' K_S) = 
T(\eta' K_S)/T(\pi^+ \pi^-) \times T(\pi^+ \pi^-)/P(\eta' K_S)$,
and concluded that manifestation of new physics is established
if $|\Delta\cals(\etapks)| > 0.1$ is observed.

The above studies aim at providing estimates of
the possible deviation within the SM based on some models
of QCD. A different approach has recently been 
proposed by Y.~Grossman, Z.~Ligeti,
Y.~Nir and H.~Quinn~\cite{Grossman:2003qp}.
They do not rely on specific QCD models but instead
use SU(3) relations to estimate or bound the contributions
to these amplitudes proportional to $V^*_{ub}V_{us}$,
which induce a non-zero $\cals$ value within the SM.
At present, the power of the method is limited by the
uncertainties on branching fractions of charmless two-body decays.
As a result, they conclude that 
$\Delta\cals(\phi\ks) < 0.25$,
$\Delta\cals(\kkks) \sim 0.13$ and
$\Delta\cals(\etapks) < 0.36$.
As data improve, these bounds could become significantly stronger.
Taking these theoretical considerations into account,
we conclude that SuperKEKB can provide precision
measurements of $\Delta\cals$ up to the limit of
hadronic uncertainties, which will be a few percent level.

%
%
%


\clearpage \newpage




\def\SuperB{SuperKEKB}

\def\alphaem{\alpha_{\rm em}}

\def\BtophiKS{B^0\to\phi\KS}
\def\btosss{b\to s\bar{s}s}

\def\Bd{B_d^0}
\def\Bdbar{\overline{B}{}_d^0}
\def\BdBdbar{\Bd\mbox{-}\Bdbar}

\def\piZ{\pi^0{}}
\def\piP{\pi^+{}}
\def\piM{\pi^-{}}
\def\KL{K_L^0{}}
\def\KS{K_S^0{}}
\def\KstarZ{K^{*0}}
\def\KstarP{K^{*+}}

\def\elel{\ell^+\ell^-{}}
\def\mumu{\mu^+\mu^-{}}
\def\epem{e^+e^-{}}

\def\btos{b\to s}
\def\btod{b\to d}
\def\btosg{b\to sg}
\def\btosgamma{b\to s\gamma}
\def\btodgamma{b\to d\gamma}
\def\btosdgamma{b\to s(d)\gamma}
\def\btosll{b\to s\elel}
\def\btodll{b\to d\elel}
\def\btosdll{b\to s(d)\elel}

\def\BtoXsgamma{B\to X_s\gamma}
\def\BtoKstargamma{B\to K^*\gamma}
\def\BtoKstarZG{B^0\to K^{*0}\gamma}

\def\BtoXsll{B\to X_s\elel}
\def\BtoKll{B\to K\elel}
\def\BtoKmumu{B\to K\mumu}
\def\BtoKee{B\to K\epem}
\def\BtoKstarll{B\to K^*\elel}
\def\BtoKstaree{B\to K^*\epem}
\def\BtoKstarmumu{B\to K^*\mumu}
\def\BtoKorKstarll{B\to K^{(*)}\elel}
\def\BtoKorKstaree{B\to K^{(*)}\epem}
\def\BtoKorKstarmumu{B\to K^{(*)}\mumu}

\def\BtoXdgamma{B\to X_d\gamma}
\def\Btorhogamma{B\to\rho\gamma}
\def\Btopill{B\to\pi\elel}

\def\GeV{\mbox{~GeV}}
\def\GeVc{\GeV/c}
\def\GeVcc{\GeVc^2}

\def\abinv{\mbox{~ab}^{-1}}
\def\fbinv{\mbox{~fb}^{-1}}

\def\Br{{\cal B}}
\def\Acp{A_{CP}}
\def\AFB{\overline{A}{}_{\rm FB}}

\def\shat{\hat{s}}

\def\Vtd{V_{td}}
\def\Vts{V_{ts}}
\def\Vtb{V_{tb}}
\def\Vtsstar{{\Vts}^{\!\!*}}
\def\Vtbstar{{\Vtb}^{\!\!*}}

\def\PM#1#2{\,^{+#1}_{-#2}}

\def\etal{\textit{et al.}}
\def\Journal#1#2#3#4{{#1} {\bf #2}, #3 (#4)} 
\def\NCA{Nuovo Cimento}                      
\def\NIMA{{Nucl.} {Instrum.} {Meth.} A}
\def\NPB{{Nucl.} {Phys.} B}
\def\PLB{Phys.\ Lett. B}
\def\PRL{{Phys.} {Rev.} {Lett.}}
\def\PRD{{Phys.} {Rev.} D}
\def\ZPC{{Z.} {Phys.} C}
\def\EPJD{{Eur.} {Phys.} {J.} direct C}
\def\EPJC{{Eur.} {Phys.} {J.} C}
\def\MPLA{{Mod.} {Phys.} {Lett.} A}

\section{$\btosgamma$ and $\btosll$}

\subsection{Introduction}
In this section, we discuss the radiative and electroweak processes
$\btosdgamma$ and $\btosdll$.  The corresponding exclusive decays are
$B\to K^*(\rho)\gamma$ and $\BtoKorKstarll$.  
Due to the GIM mechanism, the radiative process $\btosgamma$
starts at one-loop order, but still has a large
branching fraction because of the non-decoupling effect of
the top quark loop and the large CKM factor $\Vtb\Vtsstar$.  
The other processes, $\btosll$ and $\btodgamma$, are
suppressed with respect to $\btosgamma$ by two orders of
magnitude mainly due to additional $\alphaem$ and
$|\Vtd/\Vts|^2$ factors, respectively; $\btodll$ is
suppressed by four orders of magnitude due to both of them.  
These decay processes are sensitive to new physics effects
that are predicted in extensions to the Standard Model.  
Moreover, new physics effects from flavor changing
neutral interactions contribute to $\btosdgamma$ and to
$\btosdll$ in a different way.  
Typically in the former case new physics effects always
appear at a one-loop or higher orders, while in the latter
process they may arise at the tree-level, \textit{i.e.},
violating the GIM mechanism.   
Therefore, even if no new physics effect is found in
$\btosgamma$, there could be a significant effect in
$\btosdll$. 


The $\btosgamma$ process was first observed by CLEO a decade
ago and has been extensively studied since then.  
The $\btosll$ process has recently been measured 
first by Belle and later by BaBar.
The measured branching fractions are consistent with 
Standard Model predictions. 
No $\btodgamma$ process has been measured yet, but it is
expected to be observed sooner or later.  
With a much larger data sample expected at SuperKEKB,
it will become possible to measure various distributions and
asymmetries accurately enough to observe possible deviations
from the Standard Model, in addition to significant
improvements in branching fraction measurements.  
Various properties of $\btod$ transitions can also be
measured.  
These measurements will be essential in order to understand
the parity, chirality and Lorentz structures that may differ
from the Standard Model, before, and especially after, the
discovery of new physics beyond the Standard Model,
elsewhere if not in these decays. 

The major targets of SuperKEKB for these decays are as
follows. 
\begin{enumerate}
\item Precision test of the Standard Model with improved
  accuracy, 
\item search for a deviation from the Standard Model in
  various distributions, \textit{e.g.} in the
  forward-backward asymmetry of $\BtoKstarll$,
\item search for a lepton flavor dependence of $\btosll$,
\item measurement of mixing induced $CP$ violation in
  $\btosgamma$, 
\item search for direct $CP$ violation in $\BtoKstargamma$,
  and 
\item study of flavor changing transitions in the $b \to d$
  sector, \textit{i.e.} $B\to\rho\gamma$, and hopefully an
  exclusive $\btodll$ processes such as $\Btopill$.
\end{enumerate}

Both exclusive and inclusive modes are useful to test the
strong interaction in weak decays.  
In the past decade, perturbative QCD corrections to the
radiative and electroweak $B$ decays were computed beyond the
leading order which leads to the predictions with higher
accuracy \cite{Asatrian:2002va}. 
The theoretical errors from this part are reduced to a few
percent level coming from the renormalization point dependence
and a charm quark mass uncertainty.  
To include  non-perturbative effects for inclusive
decays, we may apply the heavy quark expansion.  
The photon energy spectrum $d\Gamma/dE_\gamma$ in
$\btosgamma$ is sensitive to the non-perturbative effects,
\textit{e.g.} Fermi-motion effects that are encoded in the
shape function of the $B$ meson.  
Such non-perturbative effects can be determined by measuring 
the moments of the photon energy spectrum.  
Because the structure function is process independent, we
can apply the one determined in $\btosgamma$ to other
inclusive modes such as $\BtoXsll$ \cite{Ali:1996bm}.  
As for the exclusive processes, there remain large
uncertainties from the form factors of the hadronic matrix
elements; these form factors have usually been calculated
using QCD sum rules, for which no clear idea on how to reduce their
errors is available.  
The situation might improve by using a new approach with lattice
QCD.  
It is usually thought that a large fraction of
uncertainty is canceled by measuring a ratio or an asymmetry of
two decay rates.  
Experimental information is essential to check the various
theoretical models, to evaluate the size of the errors
especially for cases where the uncertainties are
expected to cancel.

Predictions for exclusive rare $B$ decays 
by factorization, perturbative QCD (PQCD), light-front QCD,
lattice QCD and other models 
can be compared with the experimental data.
This certainly improves understanding of the weak decays of
$B$ mesons.  
Interestingly, a complete understanding of the
long-distance effects in $\BtoXsll$ is not yet available.
This effect comes from the the decay chains 
$B\to J/\psi (\psi') X_s \to \elel X_s$ 
\cite{Lim:1988yu}.  
By measuring the dilepton mass squared distribution near the
charmonium resonances, we may study the long-distance
effects experimentally. 

In order to search for or constrain so many predictions from
extensions of the Standard Model described above, a model
independent study is a useful approach.  
New physics effects may appear as a modification to
the short-distance couplings, which can be expressed as
modifications to the Wilson coefficients.  
The $\btosgamma$ transition is sensitive to the coefficient
$C_7$ for the $bs\gamma$-coupling and to a lesser extent to 
$C_8$ for the $bsg$-coupling through higher order
corrections; the $\btosll$ transition is sensitive to $C_9$
and $C_{10}$ for the vector and axial-vector
$bs\elel$-coupling in addition. 

One can further generalize this approach 
\cite{Fukae:1998qy,Hiller:2003js}. 
New physics effects that may affect $\btosgamma$ can be
parametrized by four types of interactions, which include two
types of $\btosg$ interactions and two types of $\btosgamma$
transitions.  
In addition to these, there are four Fermi-interactions
with the form of the bilinear products of $\bar{b} s$ and $\elel$.
They can be parametrized by 12 types of interactions in $\BtoXsll$
\cite{Fukae:1998qy}. 
With the data at SuperKEKB, we can improve the
situation and may pinpoint the new physics effect in a
model independent way, which covers a wider class of models than
the present analysis. 

Among the various interesting aspects of these decay channels,
we select the following observables that will be
tested with SuperKEKB:
the $\BtoXsgamma$ inclusive branching fraction,
the forward-backward asymmetry of $\BtoKstarll$, the ratio
of $\BtoKmumu$ to $\BtoKee$, the mixing induced CP asymmetry in
$\BtoKstargamma$, and the direct CP asymmetry in $\BtoXsgamma$.

\subsection{$\BtoXsgamma$ branching fraction}
The excellent agreement between the measured $\BtoXsgamma$
branching fraction and the theoretical prediction, within
about 10\% accuracy both for the measurement and theory, has
constrained various new physics scenarios; 
for example, the charged Higgs constructively interferes with
the SM amplitude, and its mass must be above 3.5 TeV if no other
new physics contribution cancels the enhancement.  
Neither the
measurement nor theory errors can be reduced substantially
due to systematic uncertainties.
Therefore one can
not expect to measure a significant deviation in the near
future.
Nevertheless, improved measurements of $\BtoXsgamma$ are
extremely important to fix the magnitude of the Wilson
coefficient $C_7$ and the photon energy spectrum for other
measurements. 

The theoretical uncertainty, which is now determined by the
accuracy of the next-to-leading-order corrections, is
expected to be reduced down to 5\% when all
next-to-next-to-leading-order corrections are included. 
The work is on-going, and is expected to be completed within
a few years. 

There have been two techniques to measure $\BtoXsgamma$: a fully inclusive
method just to tag the photon in which the background subtraction is the
main issue, and a method of summing up exclusive modes, in which the
extrapolation to the un-measured modes is the key issue.  The former
method has several options; to eliminate backgrounds we can either require a
lepton-tag or a full-reconstruction tag for the other side $B$, if it is not
statistically limited.  The latter method gives more stringent 
constraints, but is
seriously affected by the hadronization model uncertainty that makes it
unsuitable for a precision measurement.

\begin{figure}[tbp]
  \centering
  \resizebox{0.8\textwidth}{!}{\includegraphics{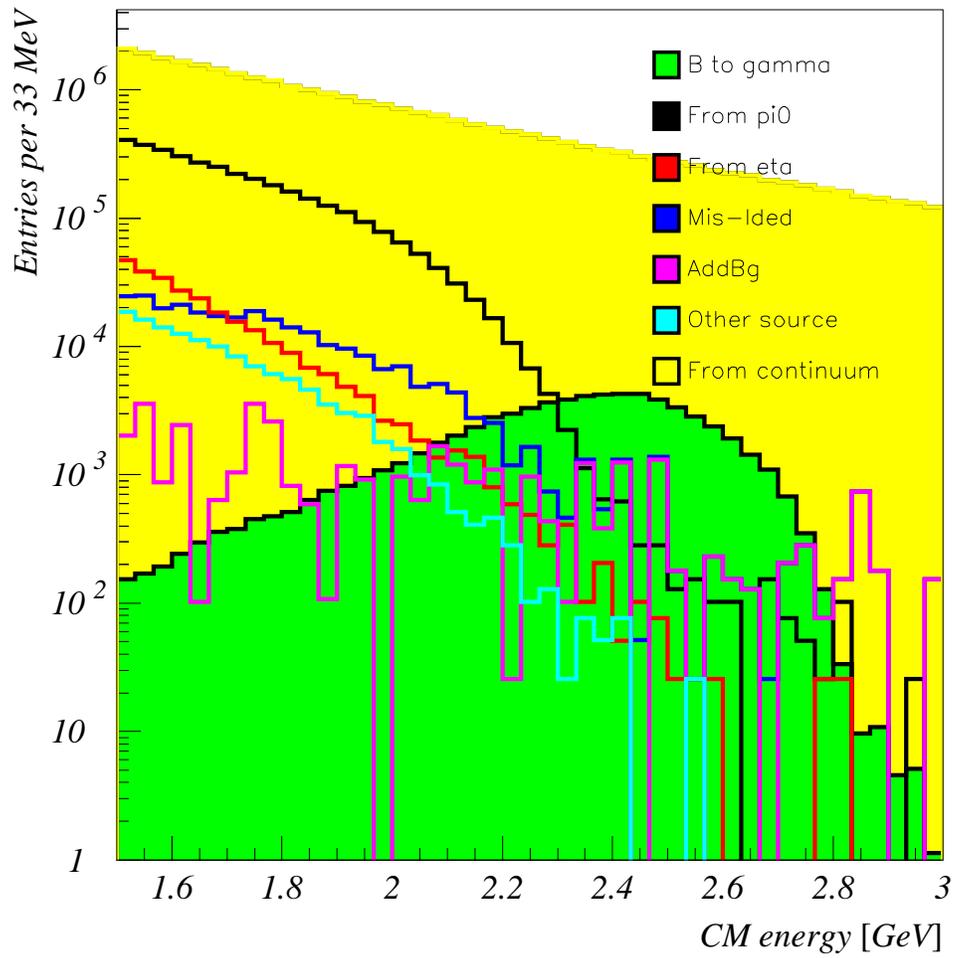}}
  \caption{Photon energy spectrum (MC) for $140\fbinv$.}
  \label{fig:xsgam}
\end{figure}

The measurements are currently limited by a systematic error 
that arises from a
minimum photon energy requirement typically at $2.0\GeV$.  
One can in principle reduce the systematic error to 5\%
by reducing the minimum photon energy requirement close to $1.5\GeV$.  
This is not a straightforward task, however, since the background
increases rapidly in the lower photon energy range while the
signal decreases, as shown in Figure~\ref{fig:xsgam}.
Typically, twice as much data is required to decrease the
minimum photon energy by 0.1~GeV while keeping the same
statistical error.  
Therefore, $\BtoXsgamma$ is a suitable measurement at
\SuperB; scaling from the currently available results, about
$5\abinv$ on-resonance plus $0.5\abinv$ off-resonance data will be
needed to decrease the photon energy requirement down to
1.5~GeV and to reduce the statistical error to be a half at the
same time, so that the total error becomes around 5\%.  
The dominant background is from continuum, which
can be reliably subtracted using the off-resonance sample.  
The second largest background is from $B\to\piZ X$ (and
$\eta X$), which can be also subtracted using a photon
energy spectrum inferred from a measured $\piZ$ momentum
distribution.  
The problematic part will be the background from neutral
hadrons ($\KL$, neutrons and anti-neutrons), for which control samples of
$\epem\to\phi\gamma$, $\phi\to\KS\KL$ and inclusive $\Lambda$
($\overline{\Lambda}$) events have to be studied.

\subsection{$\BtoKstarll$ forward-backward asymmetry}
The forward-backward asymmetry in $\BtoKstarll$, defined as
\begin{equation}
\AFB(q^2) = {N(q^2;\,\theta_{B\ell^+}>\theta_{B\ell^-})
           - N(q^2;\,\theta_{B\ell^+}<\theta_{B\ell^-}) \over
             N(q^2;\,\theta_{B\ell^+}>\theta_{B\ell^-})
           + N(q^2;\,\theta_{B\ell^+}<\theta_{B\ell^-})},
\end{equation}
is an ideal quantity to disentangle the Wilson coefficients $C_9$
and $C_{10}$ together with the sign of $C_7$.

As discussed in Section~\ref{sec:mode_indep_bsgamma}, in a
SUSY scenario the sign of the $\btosgamma$ amplitude ($C_7$) can 
be opposite to the Standard Model prediction, while the
transition rate may be the same as in the Standard Model.
This case can be discriminated by measuring the
forward-backward asymmetry of $\BtoXsll$ or $\BtoKstarll$.  
Within the Standard Model, there is a zero crossing
point of the forward-backward asymmetry in the low dilepton invariant
mass region, while the crossing point may disappear in some SUSY
scenarios.  Another important new physics effect can be searched for by
using the $\BtoKstarll$ forward-backward asymmetry.  In a model with
$SU(2)$ singlet down-type quarks, tree-level $Z$
flavor-changing-neutral-currents are induced.  In this case, the
larger effect is expected on the
axial-vector coupling ($C_{10}$) to the dilepton than on
the vector coupling ($C_9$).  Because the forward-backward asymmetry is
proportional to the axial-vector coupling, the sign of the asymmetry can
be opposite to the Standard Model.  The same new physics effect is also
effective for $\BtophiKS$ where anomalous mixing-induced $CP$
violation can occur.
A correlation is expected between $\btosll$
and $\btosss$.

\begin{figure}[tbp]
  \centering
  \resizebox{0.70\textwidth}{!}{\includegraphics{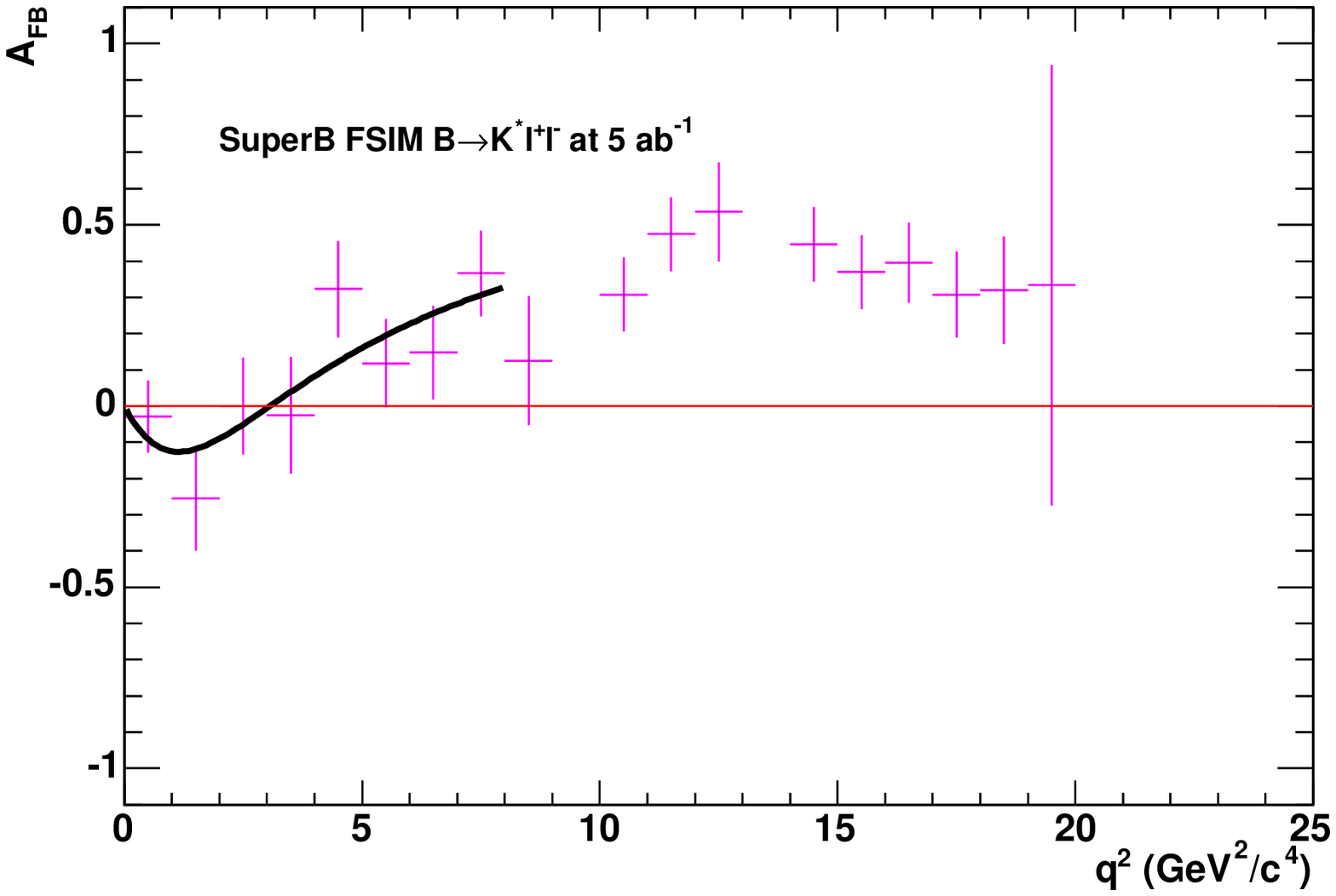}}\\
  \resizebox{0.70\textwidth}{!}{\includegraphics{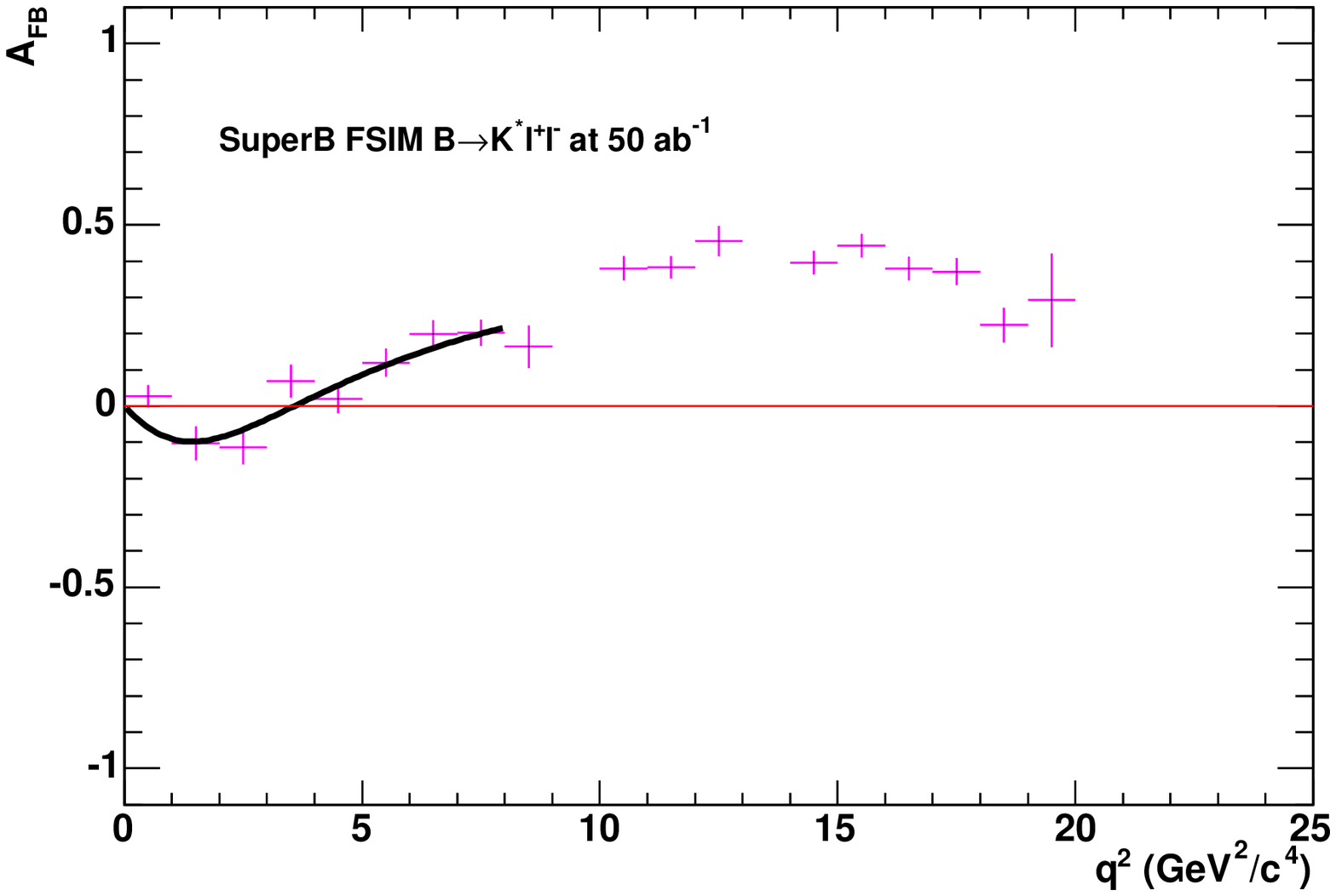}}
  \caption{
    Forward-backward asymmetry in $\BtoKstarll$ at $5\abinv$ (top)
    and $50\abinv$ (bottom).
  }
  \label{fig:afb}
\end{figure}

The forward-backward asymmetry is roughly proportional to
$C_{10}(2C_7+C_9\hat{s})$.  In the SM, $C_7$ is about $-0.3$
and $C_9$ is about 4, so this function crosses zero around 
$\shat \sim 0.15$, or $q^2 \sim (3.5\GeVcc)^2$.  
Figure~\ref{fig:afb} shows the expected $\AFB$ at 5
and $50\abinv$.  It can be seen that this crossing pattern of the
forward-backward asymmetry will be already visible at $5\abinv$, and 
will be
clearly observed at $50\abinv$.  The fitted curves in Fig.~\ref{fig:afb}
have empirical functional forms that are proportional to
$C_{10}(2C_7+C_9\hat{s})$.  From these fits, it is possible to
disentangle $C_{10}$ and $C_9$, if the size of $C_7$ is fixed using
$\btosgamma$.  From a fit to the $50\abinv$ sample, we obtain
$\delta C_9=0.43$ and $\delta C_{10}=0.59$, or roughly 10\% and 14\%
errors, respectively.  The uncertainty due to $C_7$ is small. 
We note that a more realistic fitting function that includes all
next-to-next-to-leading order corrections and form factors is needed.
The result will be improved by combining it with a $\BtoXsll$ branching
fraction measurement and $q^2$ fit results, which will only be possible at
$e^+e^-$ $B$-factories.

\subsection{$\BtoKmumu$ versus $\BtoKee$}
Branching fractions for the exclusive decays $\BtoKorKstarll$ have 
already been measured to be consistent with SM predictions. The
measurement error is already smaller than that for the theory.
Theory predictions suffer from large model dependent and irreducible
uncertainties in the form-factors of at least $\pm30\%$, and the
currently available predictions vary by a factor of two or more.
However, one can still utilize the measurements in such a way that the
theory uncertainties cancel.

In new physics models with a different Higgs sector from that of the Standard
Model, scalar and pseudo-scalar types of interactions may arise in
$\btosll$.  Depending on the lepton flavor $\ell=e$ and $\ell=\mu$, the new
physics effects can differ.  By measuring
$R_{K^{(*)}}=\Br(\BtoKorKstarmumu)/\Br(\BtoKorKstaree)$, such new
physics effects can be searched for.  
A particular example can be found in a minimal supergravity
model \cite{Hiller:2003js,Wang:2003je}.

In the SM, the branching fractions for $\BtoKee$ and $\BtoKmumu$ are
predicted to be equal except for a tiny phase space difference due to
the lepton masses.  For  $\BtoKstarll$ modes, the branching fraction
for $\BtoKstaree$ is larger than $\BtoKstarmumu$ for small dilepton
masses, due to a larger interference contribution from $\BtoKstargamma$
in $\BtoKstaree$.  However, this situation may be modified in the models
mentioned above, in which a neutral SUSY Higgs contribution can
significantly enhance only the $\BtoKorKstarmumu$ channel
if $\tan\beta$ is large.  Therefore, the ratio $R_K=\Br(\BtoKmumu)/
\Br(\BtoKee)$ is an observable that is sensitive to new physics if it is
larger than unity.

From current Belle results on the branching fractions, we obtain
$R_K=1.0\pm0.4$ or $R_K<1.7$ (90\% CL).  The error on $R_K$ is currently
dominated by the statistical error, and the error will simply scale with
the luminosity even at $50\abinv$.  The expected error is
\begin{equation}
\begin{array}{c}
\delta R_K=0.07\;\;\; \mbox{~(at $5\abinv$)},\\
\delta R_K=0.02\;\;\; \mbox{~(at $50\abinv$)}.
\end{array}
\end{equation}
The systematic error in the ratio is dominated by the error on the
lepton identification efficiency, which will still be much smaller than
the statistical error even at $50\abinv$.


Since the $\btosll$ transition diagram is equivalent to $B_s\to\mumu$,
there already exists a bound from the CDF's limit, which corresponds to
$R_K<1.6$.  However, this limit is model dependent unless $B_s\to\epem$
is observed, which is very unlikely.

\subsection{Mixing induced CP asymmetry in $\BtoKstargamma$}
\label{sec:sg_sll:s_kstargamma}
Mixing-induced $CP$ asymmetry in $\btosgamma$ is an excellent window to study
a new phase that may also be observed in the hadronic transition $\btosss$.
The $CP$ asymmetry parameter $\cals_{K^*\gamma}$ is expected to be
${\cal O}(m_s/m_b)$ in the SM
within an error of a few percent, and any deviation would be a sign of new
physics.

For example, the existence of the neutrino mass suggests that 
the left-right symmetry is restored at a higher energy, while parity is
spontaneously broken at a low energy.  In left-right symmetric models,
the helicity of the photon from $\btosgamma$ can be a mixed state of two
possible photon helicities, while the left-handed photon is dominant in
the Standard Model.  This case can be tested by using the mixing induced
$CP$ asymmetry in $\btosgamma$.  It can also be checked by measuring the
azimuthal angle distribution of $\BtoKstarll$ 
\cite{Kim:2000dq}.

In order to measure the mixing-induced $CP$ asymmetry, a suitable $CP$ final
state is needed.  The final state  $\BtoKstarZG$, $\KstarZ\to\KS\piZ$
is such a $CP$ eigenstate, and the technique for the $\sin2\phi_1$
measurement is applicable.  This mode was previously considered to be
technically challenging due to the detached $\KS$ decay vertex; however,
it is now found to be feasible from recent studies of $B\to\KS\piZ$, if
one requires hits for the $\KS\to\piP\piM$ tracks in the silicon vertex
detector.

\begin{figure}
\resizebox{0.49\textwidth}{!}{\includegraphics{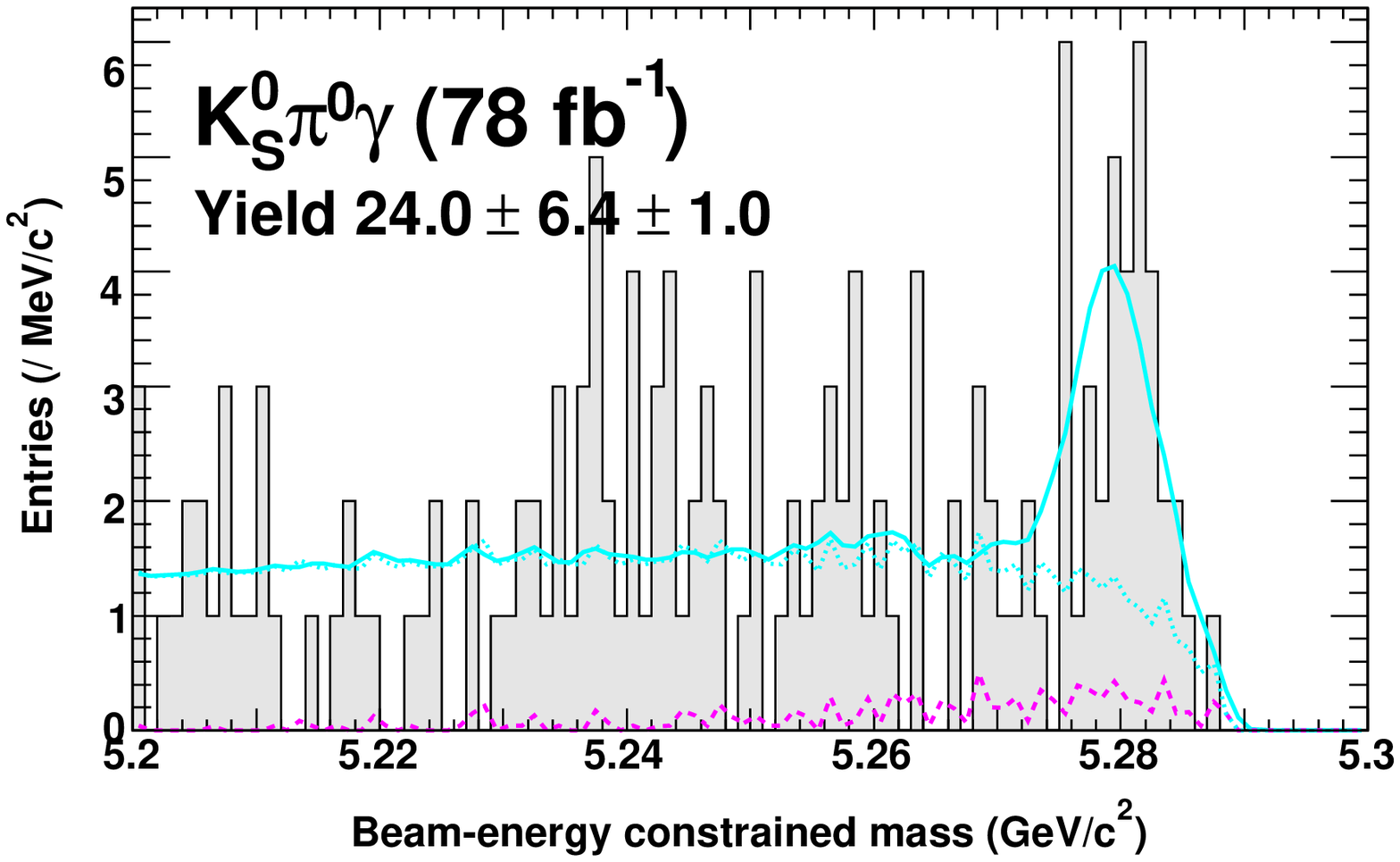}}
\resizebox{0.49\textwidth}{!}{\includegraphics{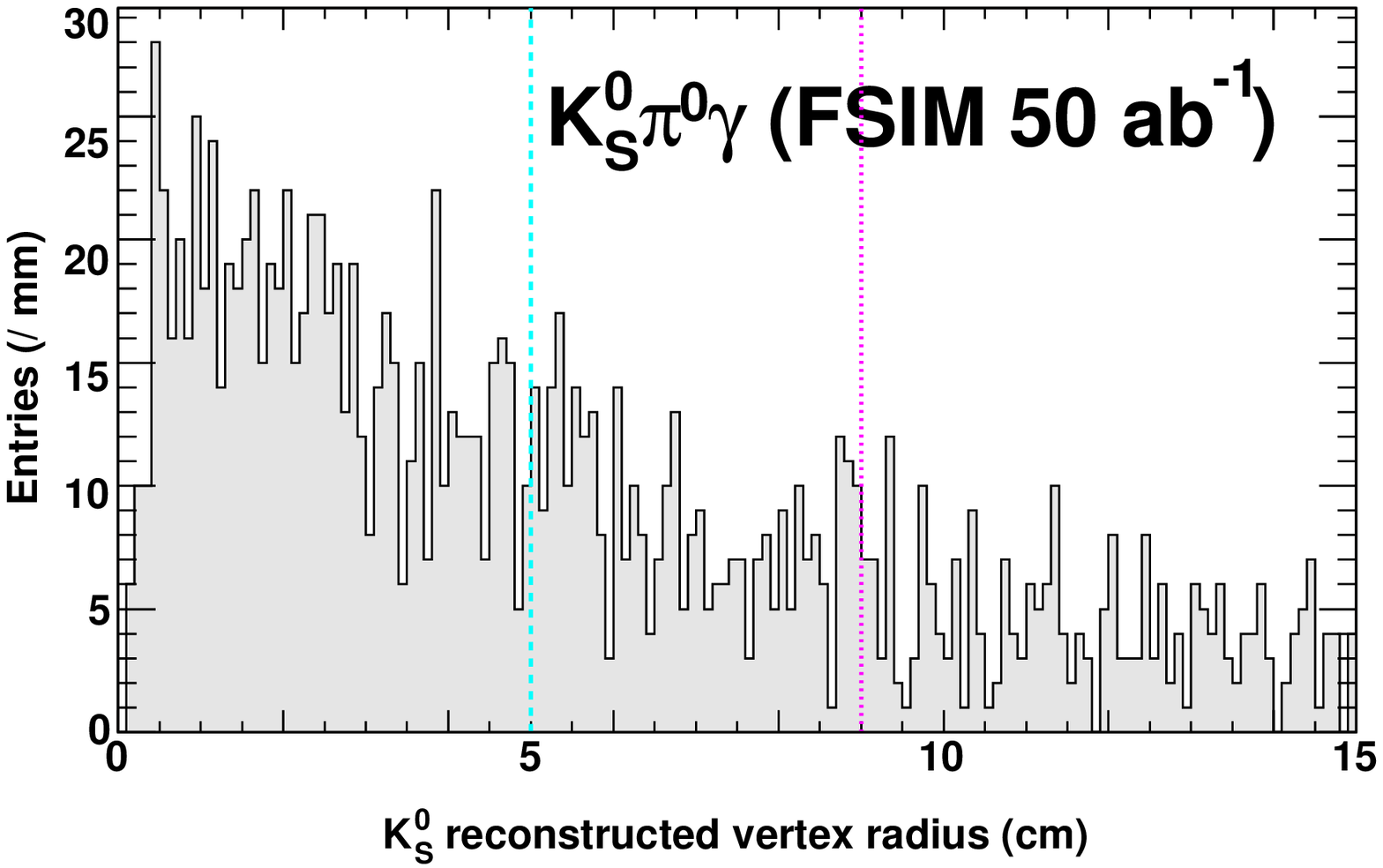}}
\caption{$\BtoKstarZG$, $\KstarZ\to\KS\piZ$ signal with $78\fbinv$
(left) and the radial $\KS$ decay vertex distribution (right).}
\label{fig:k0spi0gamma}
\end{figure}

Although the branching fraction for $\BtoKstarZG$ is sizable
($\sim4\times10^{-5}$), the number of events in the $\KS\piZ\gamma$
final state is rather limited due to its small sub-decay branching
fraction and a low efficiency. However, 
the signal is clearly observed already 
at $78\fbinv$ as shown in Fig.~\ref{fig:k0spi0gamma}-a.

The efficiency for the current data is about 1\%.  Assuming that the
efficiency is unchanged for \SuperB, one expects about $2\times10^3$
($2\times10^4$) events at 5 (50)$\abinv$.  The useful events with at
least two hits in the present Silicon Vertex Detector (SVD)
must have the decay vertex within
about 5 cm in radius, which corresponds to a half of the total, as shown in
Fig.~\ref{fig:k0spi0gamma}-b.  This will increase to about 70\% for a
radius of 9 cm.  Using the result of $B\to\eta'\KS$, which has a similar
signal-to-noise ratio, and assuming that a worse
$\Delta t$ resolution in $\ks\pi^0\gamma$ than in $\eta^\prime\ks$
would be
compensated by the expected better SVD performance
for \SuperB, the anticipated statistical
error of the $S$ term is
\begin{equation}
\begin{array}{c}
\delta S_{K^*\gamma} = 0.14\mbox{~(at $5\abinv$)},\\
\delta S_{K^*\gamma} = 0.04\mbox{~(at $50\abinv$)}.
\end{array}
\end{equation}

There are other exclusive decay channels for the mixing-induced $CP$
asymmetry study in which no difficulty exists in the vertex reconstruction.
One possible channel is $B^0\to\KS\phi\gamma$ for which a $3\sigma$
signal is already observed ($5\sigma$ signal is observed for the
corresponding charged decay $B^+\to K^+\phi\gamma$).  Since
$\KS\phi\gamma$ is not a $CP$ eigenstate, one has to perform an angular
analysis.  From the past experience with $B\to J/\psi\KstarZ$, we 
obtain an error of 0.63 with 89 events.  At 5 (50)$\abinv$, we expect
150 (1500) $\KS\phi\gamma$ events that will lead to an error of 0.5
(0.15) in the $S$ term, and it therefore becomes interesting with
$50\abinv$ or more.  The other possibility is to use higher resonances,
namely $B\to K_1(1270)^0\gamma$, $K_1(1270)\to\KS\rho^0$.  However, this
is now considered to be difficult, because these modes have not been 
observed yet and
one has to disentangle them from $K_1(1270)\to\KstarP\piM$ that has the
same $\KS\piP\piM$ final state.

\subsection{$\BtoXsgamma$ direct $CP$ asymmetry}
The direct $CP$ asymmetry for $\BtoXsgamma$ is one of the 
quantities with theoretical uncertainties smaller than
experimental errors.
The predicted SM $CP$ asymmetry is
$\Acp=0.0042\PM{0.0017}{0.0012}$ \cite{Hurth:2003dk}, while its
magnitude could be above 10\% in many extensions of the SM.  Therefore,
a large space remains to be explored.

The sensitivity at SuperKEKB can be obtained by extrapolating the latest
Belle results,
$\Acp(\BtoXsgamma)=-0.004\pm0.051\rm{(stat)}\pm0.038\rm{(syst)}$.  This
measurement was performed by summing up exclusive modes, $X_s\to
Kn(\pi)$ $(n=1\mbox{~to~}4)$ and $X_s\to KK^+K^-(\pi)$, where $K$ is a
$K^\pm$ or $\KS$ and most one $\piZ$ is allowed.  This method has an
advantage of suppressing the $B\to X_d\gamma$ contribution to a
negligible level.  Another method that uses the inclusive photon and
tags the charge by an additional lepton, cannot distinguish the $B\to
X_d\gamma$ contribution; however it is an another interesting subject
because of the partial sensitivity to the $B\to X_d\gamma$ channel.

The SM prediction for the combined asymmetry $A^{s\gamma+d\gamma}_{CP}$
is essentially zero, in contrast to $A_{CP}$ for $b\to s\gamma$ alone.  
This is because $b\to s\gamma$ and $b\to d\gamma$ contribute to 
$A^{s\gamma+d\gamma}_{CP}$ with an opposite sign and practically 
with an equal magnitude, which is a consequence of the unitarity of
the CKM matrix, the small mass difference $m_s-m_d$ and the real Wilson
coefficient $c_7$ \cite{Soares:1991te}. 
In models beyond the SM, $A^{s\gamma+d\gamma}_{CP}$ can be 
non-zero, and is usually dominated by the
$b\to s\gamma$ component. In addition, the 
contributions to $A^{s\gamma+d\gamma}_{CP}$ 
from $b\to s\gamma$ and $b\to d\gamma$ can have the same or opposite sign
\cite{Akeroyd:2001cy,Akeroyd:2001gf,Hurth:2003dk}.

Although the systematic error is not much smaller than the statistical
error, most of the systematic errors are limited by the statistics of
the control samples and hence can be reduced with a larger statistics.
Note that systematic errors in the tracking efficiency and particle
identification cancel in the asymmetry.  The breakdown of the errors at
$140\fbinv$ are, from the signal shape ($\sim0.008$) partly due to the
uncertainty in the $M(X_s)$ spectrum and partly due to the multiplicity
distribution, from the possible $\Acp$ in the charmless $B$ decay
background ($0.02$), and from a charge asymmetry in the background
suppression requirements ($0.029$).  The $M(X_s)$ shape and charmless
contributions will be known better with more data, and other errors just
scale with the statistics.  The irreducible model error of the first
error is about 0.003, giving the expected errors of
\begin{equation}
\begin{array}{lcl}
\delta\Acp(\mbox{at~}5\abinv)
   &{=}& \pm 0.009\rm{~(stat)}\pm0.006\rm{~(syst)},
\\
\delta\Acp(\mbox{at~}50\abinv)
   &{=}& \pm 0.003\rm{~(stat)}\pm0.002\rm{~(syst)}\pm0.003\rm{~(model)}.
\end{array}
\end{equation}
Therefore, at $50\abinv$, the measurement is more or less
limited by the systematic error. 
While it is still insufficient to measure the predicted SM
asymmetry beyond a $1\sigma$ significance at $50\abinv$,
this precision is sufficient to observe a $CP$ asymmetry
of $\sim 0.03$ with a 5$\sigma$ significance.
\begin{figure}[tbp]
\centering
\resizebox{0.32\textwidth}{!}{\includegraphics{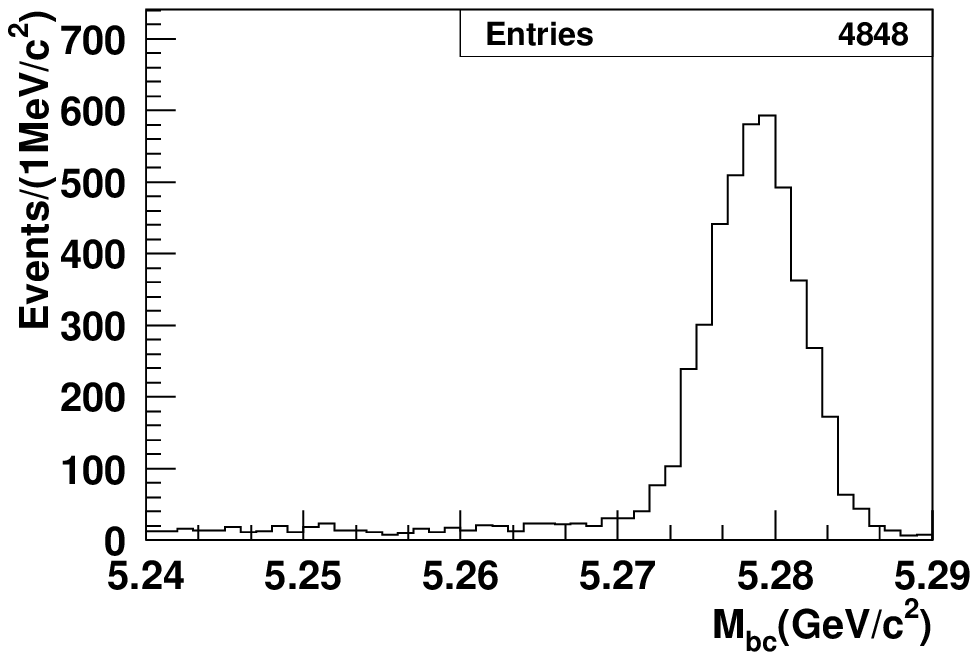}}
\resizebox{0.32\textwidth}{!}{\includegraphics{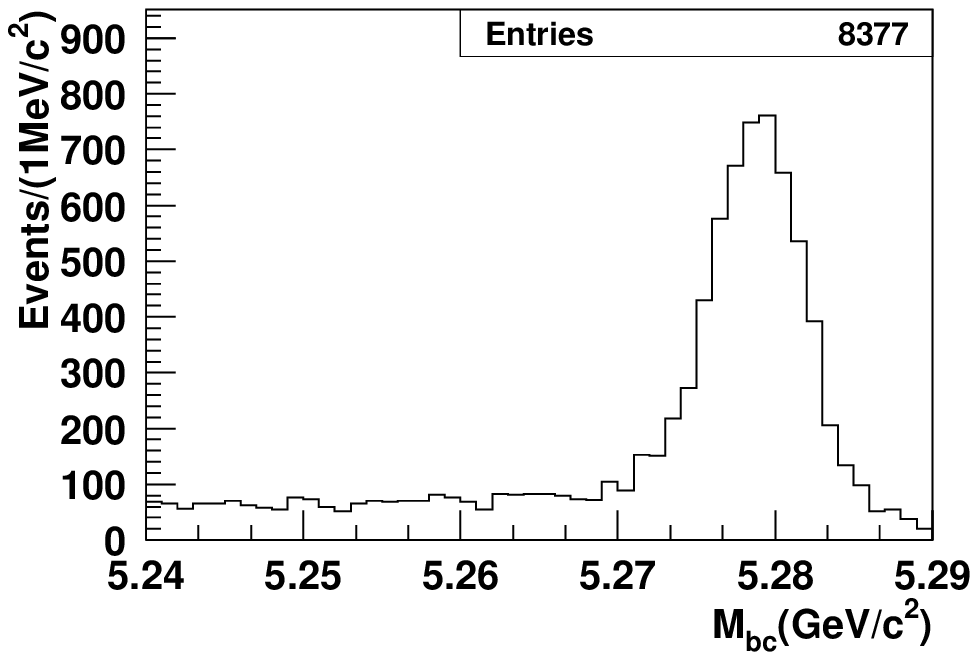}}
\resizebox{0.32\textwidth}{!}{\includegraphics{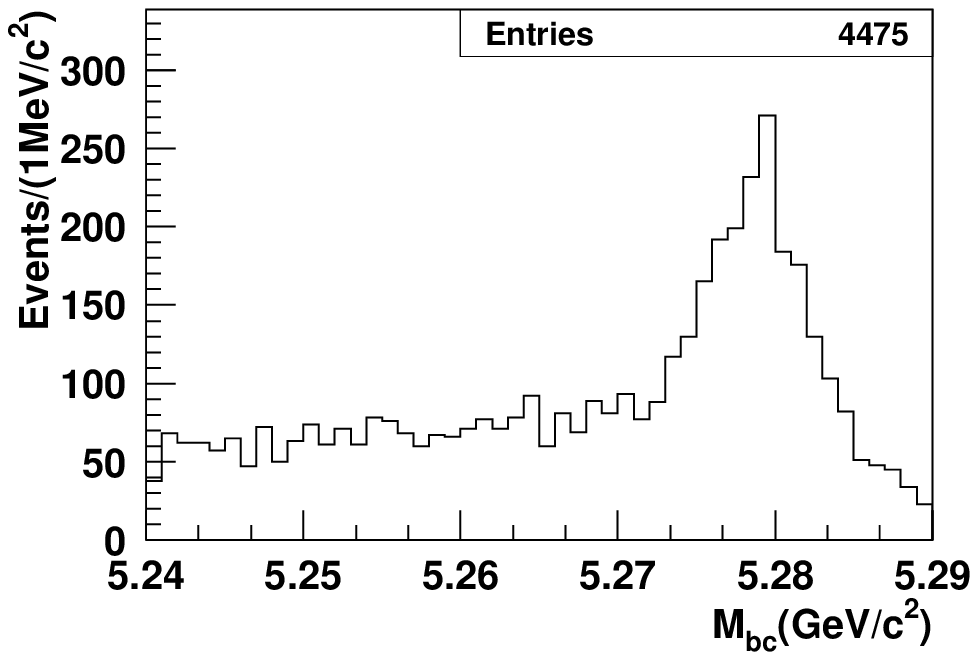}}\\
\resizebox{0.32\textwidth}{!}{\includegraphics{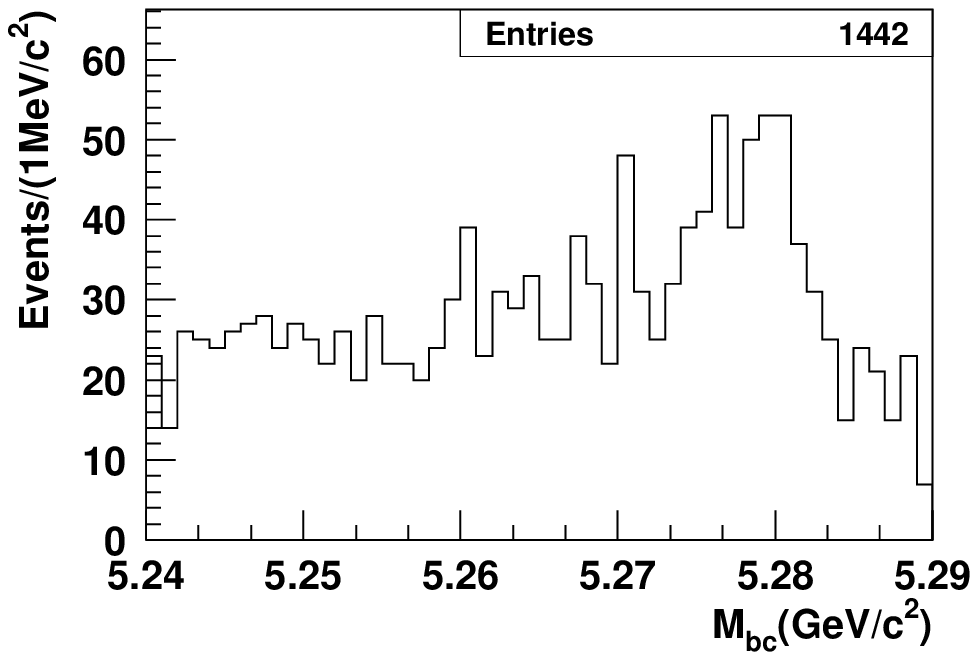}}
\resizebox{0.32\textwidth}{!}{\includegraphics{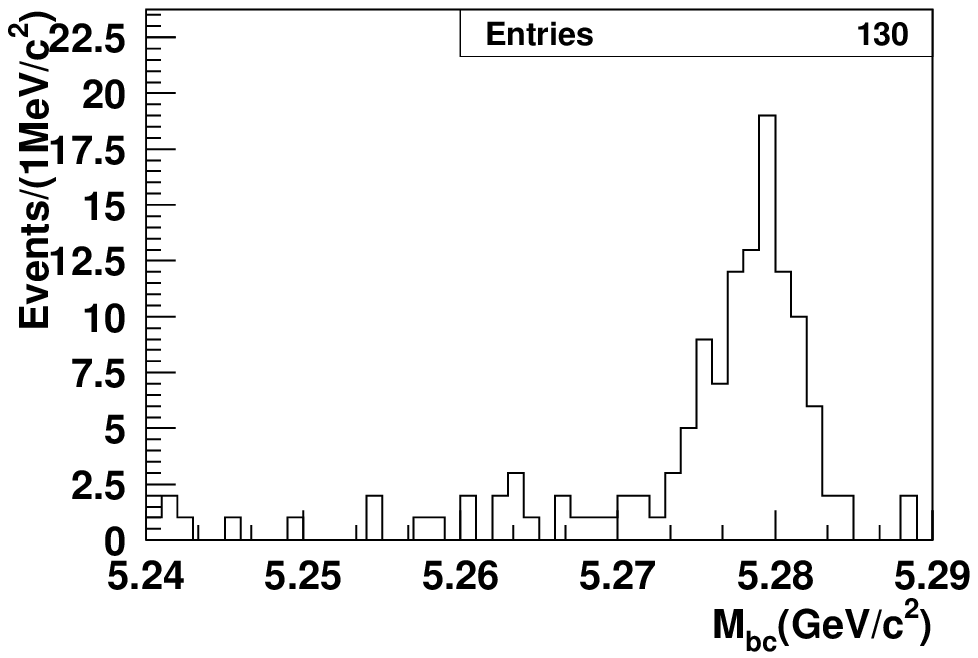}}\\
\caption{Breakdown of the expected $\BtoXsgamma$ exclusive modes: 
$K\pi\gamma$ other than $\BtoKstargamma$,
$K\pi\pi\gamma$,
$K\pi\pi\pi\gamma$,
$K\pi\pi\pi\pi\gamma$, and
$KK^+K^-\gamma$ from left-top to right-bottom.}
\label{fig:xsgamacp}
\end{figure}

\subsection{$\btodgamma$ and $\btodll$}
Finding new physics effects in a $b\to d$ transition may be
easier than in a $\btos$ transition because the Standard
Model amplitude is suppressed in the $\btod$ transition.  
Therefore, measurements of the decay rates of $\Btorhogamma$
will be a good test of the Standard Model.  
By combining $|\Vtd\Vtbstar|$ measured in the $\BdBdbar$ mixing,
we can search for new physics effects in the $\btodgamma$
transition \cite{Handoko:1994xw}. 
In the SM, $b\to s\gamma$ and $b\to d\gamma$ are both
mediated by a common Wilson coefficient, $c_7$. 
This is also true in any model with a minimal 
flavor-violating framework \textit{i.e.} where flavor changing 
interactions are determined by the CKM angles. 
However, in models with tree-level FCNCs,  $C_7$ for $b\to
d\gamma$ can differ from $C_7$ for $b\to s\gamma$. 
Examples include SUSY models with gluino mediated FCNCs
\cite{Arhrib:2001gr} and models with a non-unitary CKM matrix
\cite{Akeroyd:2001gf}.
Since the Standard Model prediction for the $\Btorhogamma$
branching fraction suffers from a large model-dependent
uncertainty, ideally it is necessary to measure the
inclusive rate for $\BtoXdgamma$. 

Although it has yet to be measured, the $\btodgamma$ process will
provide various interesting new physics probes.

\subsection{Summary}
We have discussed various decay channels of $\btosdgamma$ and
$\btosll$ that are good probes of new physics effects.  The sensitivity
results are summarized in Table~\ref{tbl:summary-radiative}.

\begin{table}[tbp]
\begin{center}
\begin{tabular}{lcc}
\hline 
mode & $5\abinv$ & $50\abinv$ \\
\hline 
$\Br(\BtoXsgamma)$ & 5\% & 5\% \\
$\Acp(\BtoXsgamma)$
 & $0.009\oplus0.006$
 & $0.003\oplus0.002\oplus0.003$ \\
Mixing induced $S_{K^*\gamma}$ & 0.14 & 0.04 \\
\hline 
$R_K(\BtoKll)$ & 0.07 & 0.02 \\
$\AFB(\BtoKll)$ & & \\
~~$C_9$    from $\AFB(\BtoKstarll)$ & 32\% & 10\% \\
~~$C_{10}$ from $\AFB(\BtoKstarll)$ & 44\% & 14\% \\
\hline 
\end{tabular}
\end{center}
\caption{Summary of the expected errors for $\btosdgamma$ and $\btosll$ at \SuperB.}
\label{tbl:summary-radiative}
\end{table}

\clearpage \newpage
\section{More than one neutrino I: 
 $B \to K^{(*)}\nu\bar{\nu}$ and $B \to \tau \nu$}

\subsection{Introduction}
In the Standard Model the rare FCNC decay $b\to s\nu\bar{\nu}$
proceeds at the one-loop level through penguin and box
diagrams.
Additional new physics heavy particles may therefore
contribute to this decay mode, leading to significant
enhancements of the branching fraction.
Since the final state leptons do not have electric charge,
this mode is not affected by long distance effects from vector
resonances ($\rho$, $J/\psi$, $\psi'$ \textit{etc.}) and its
theoretical predictions are cleaner than 
those for $b\to sl^+l^-$.
The inclusive branching fraction is estimated to be
$4\times10^{-5}$ \cite{Buchalla:1993bv,Grossman:1995gt} for
the sum of three neutrino flavors, whereas the exclusive
branching fractions are predicted to be 
$Br(B^-\rightarrow K^-\nu\bar{\nu}) \approx 4\times
10^{-6}$ \cite{Colangelo:1996ay}.

However, the experimental measurement of 
$b\to s\nu\bar{\nu}$ is quite challenging due to
two missing neutrinos. 
The best inclusive limit to date is from ALEPH
$Br(b\to s\nu\bar{\nu})<6.4\times10^{-4}$ \cite{Barate:2000rc}, 
and a limit of $<2.4\times 10^{-4}$ at $90\%$ confidence level 
on the exclusive branching fraction of
$B\to K\nu\bar{\nu}$ was set by CLEO\cite{Browder:2000qr}.
BABAR has recently reported a preliminary
upper limit $Br(B\to K\nu\bar{\nu})< 7.0\times10^{-5}$
\cite{Aubert:2003yh}.
At SuperKEKB, measurements of these decay branching
fractions will become possible, as millions of fully
reconstructed $B$ samples will be accumulated.

The purely leptonic decays $B^\pm\to l^\pm\nu$ 
in the SM proceed via annihilation
to a $W^\pm$ and are proportional to the square of the $B$ meson decay constant
($f_B$). In models beyond the SM there can be additional tree-level 
contributions such as a $H^\pm$ ($s$-channel) \cite{Hou:1992sy},
\cite{Akeroyd:2003zr} or  sfermions ($s,t$-channels) in
$R$ parity violating SUSY models \cite{Guetta:1997fw}.
The SM predictions and the current experimental
upper limits are shown in Table~\ref{explimits}, where we take $f_B=200$ MeV.
\begin{table}\begin{center}
\begin{tabular} {|c|c|c|c|} \hline
Decay & SM Prediction   & {\sc Belle} (60 fb$^{-1}$)& {\sc BABAR}
(81 fb$^{-1}$)  \\ \hline
 $B^\pm\to e^\pm\nu$ & $9.2\times 10^{-12}$ & $\le 5.4\times 10^{-6}$ & 
\\ \hline
 $B^\pm\to \mu^\pm\nu$ & $3.9\times 10^{-7}$ 
 & $\le 6.8\times 10^{-6}$  & $\le 6.6\times 10^{-6}$   \\ \hline
  $B^\pm\to \tau^\pm\nu$ & $8.7\times 10^{-5}$  & & 
$\le 4.1\times 10^{-4}$   \\ \hline
\end{tabular}\end{center}
\caption{SM predictions and current experimental limits from BELLE/BABAR.}
\label{explimits}
\end{table}
Observation of such decays would provide the first direct measurement of 
$f_B$ or even evidence for new physics. In particular, the sensitivity
of the $\tau^\pm\nu$ and $\mu^\pm\nu$ channels
to any $H^\pm$ contribution is complementary and
competitive with that of the exclusive semi--leptonic decay 
$B\to {\overline D}\tau^\pm\nu$ that is described in 
Sec.~\ref{sec:sensitivity_dtaunu}.
The tree--level partial width (including only $W^\pm$ and
$H^\pm$ contributions) is as follows \cite{Hou:1992sy}:
\begin{equation}
\Gamma(B^\pm\to \ell^\pm\nu)={G_F^2 m_{B} m_l^2 f_{B}^2\over 8\pi}
|V_{ub}|^2 \left(1-{m_l^2\over m^2_{B}}\right)^2 \,\times\, r_H
\end{equation}
where $r_H$ is independent of the lepton flavour and is given by
\begin{equation}
r_H=\left(1-{\tan^2\beta\over 1+\tilde\epsilon_0\,\tan\beta}\,{m^2_B\over 
m^2_{H^\pm}}\right)^2
\end{equation}

\begin{figure}
\begin{center}
\includegraphics[width=0.8\textwidth,clip]{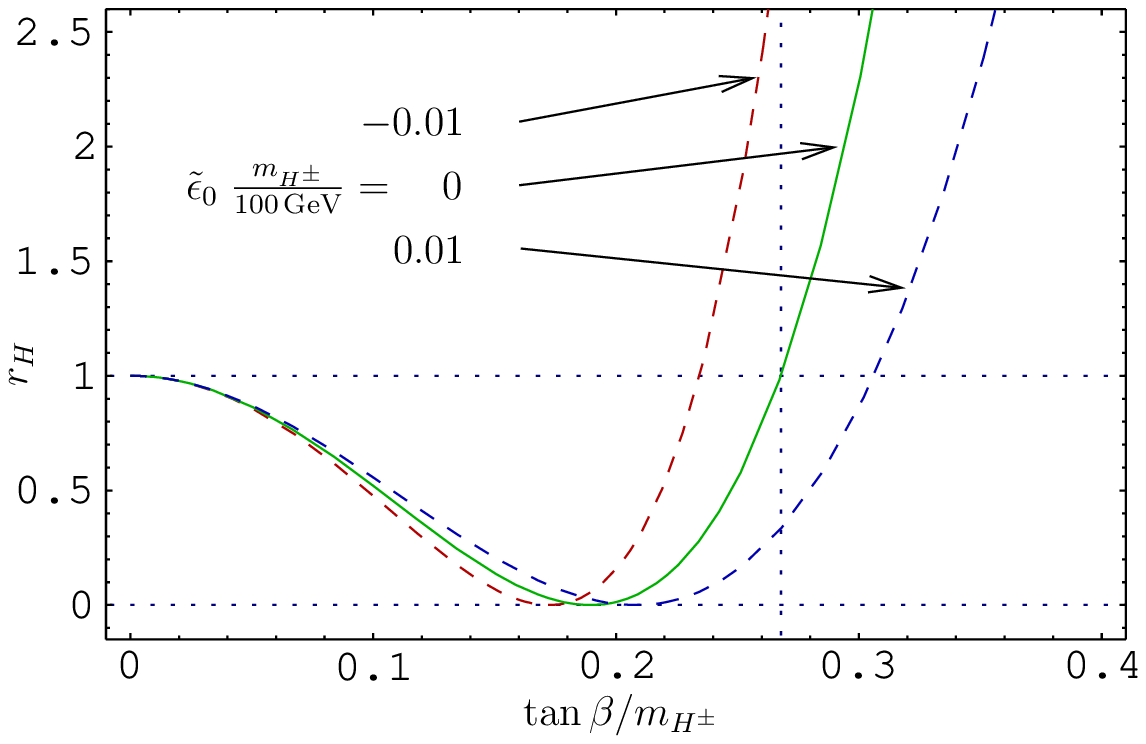}
\end{center}
\caption{$r_H$ as a function of $\tan\beta/m_{H^\pm}$. We show curves
for $\tilde\epsilon_0\; {m_{H^\pm}\over 100\,{\rm GeV}}= 0,\pm 0.01$.
\label{rH}}
\end{figure}

The overall factor of $m_l^2$ arises from helicity suppression
for $W^\pm$ and Yukawa suppression for $H^\pm$. The parameter
$\tilde\epsilon_0$ encodes the effects of large SUSY
corrections to the $b$ quark Yukawa coupling, 
and is typically constrained $|\tilde\epsilon_0|\le 0.01$. 
In Fig.~\ref{rH} we plot $r_H$ as a function of $\tan\beta/m_{H^\pm}$ 
for several values of $\tilde\epsilon_0 (m_{H^\pm}/100\, GeV)$. 
The current experimental limits from Table \ref{explimits} constrain 
$r_H< 4.7$ and $r_H< 17.4$ from the $\tau^\pm\nu$ and $\mu^\pm\nu$
channels, respectively. Sensitivity to the SM rate ($r_H=1$) 
is expected in both channels with data samples of a few $ab^{-1}$,
so that Super-KEKB might actually be able to measure $r_H$.
One can see that $r_H=1$ is obtained for
$\tan\beta/m_{H^\pm}= 0.27\pm 0.03\, {\rm GeV}^{-1}$, while
complete cancellation can occur around $\tan\beta/m_{H^\pm}=0.2\,
{\rm GeV}^{-1}$. Importantly, any signal of
a $H^\pm$ in the decay $B\to {\overline D}\tau^\pm\nu$ (which is sensitive to
$\tan\beta/m_{H^\pm} > 0.14$ with 5 $ab^{-1}$) would
also manifest itself in both $B^\pm\to \tau^\pm\nu$ and
$B^\pm\to \mu^\pm\nu$.

An additional observable in which
$f_B$ cancels out is the ratio $R_{\tau\mu}$ defined by: 
\begin{equation}
R_{\tau\mu}={BR(B^\pm\to \tau^\pm\nu)
\over BR(B^\pm\to \mu^\pm\nu)}
\end{equation}
Assuming only $W^\pm$ and $H^\pm$ contributions, $r_H$ would
also cancel out, and thus $R_{\tau\mu}\sim m^2_\tau/m^2_\mu$.  
However, sizeable deviations of 
$R_{\tau\mu}$ from this value are possible in $R$ parity 
violating SUSY models, since $r_H$ is in general no longer independent
of the lepton flavour. In addition, in such models the decay 
$B^\pm\to e^\pm\nu$ may be enhanced to experimental observability.

\subsection{Estimation of signal and background}

To estimate the sensitivity for
these decays, we start with the decay mode,
$B^- \to K^-\nu\bar{\nu}$,
taking a conservative full reconstruction efficiency of 0.2\% for
$B^+$ using high quality hadronic tags, with an
integrated luminosity of $50~\mathrm{ab}^{-1}$. 
Although the
physics is different, the B decay to $\tau\nu$ has the same topology
when one prong $\tau$ decay modes are used to observe the
signal. Thus, in the following the $B\rightarrow\tau\nu$ decay will be
analyzed as well.

\begin{figure}
 \begin{center}
  \begin{minipage}{0.32\textwidth}
   \begin{center}
    (a) $B^- \to K^-\nu\nu$ \\
    \includegraphics[scale=0.45]{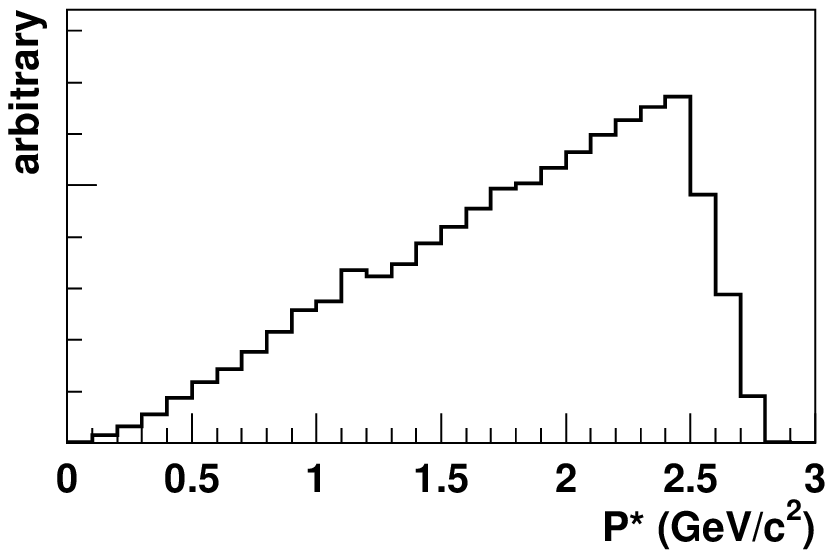} \\
    (d) $B^- \to D^0\rho^-$ \\
    \includegraphics[scale=0.45]{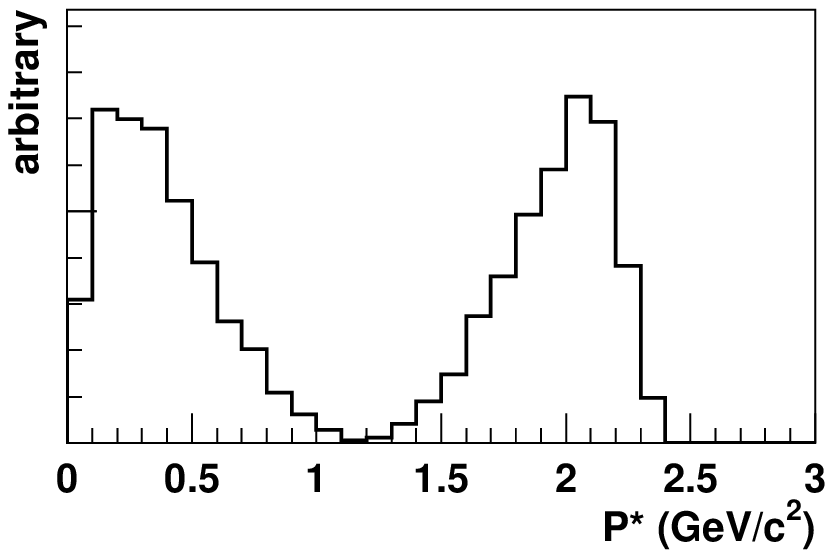}
   \end{center}  
  \end{minipage}
  \hfil
  \begin{minipage}{0.32\textwidth}
   \begin{center}
    (b) $B^- \to \tau^-\nu$ \\
    \includegraphics[scale=0.45]{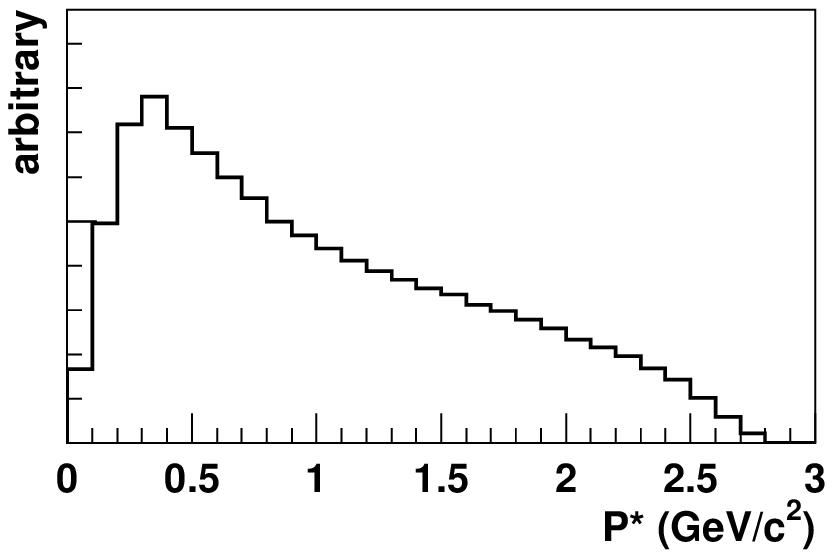} \\
    (e) $B^- \to D^0\pi^-$ \\
    \includegraphics[scale=0.45]{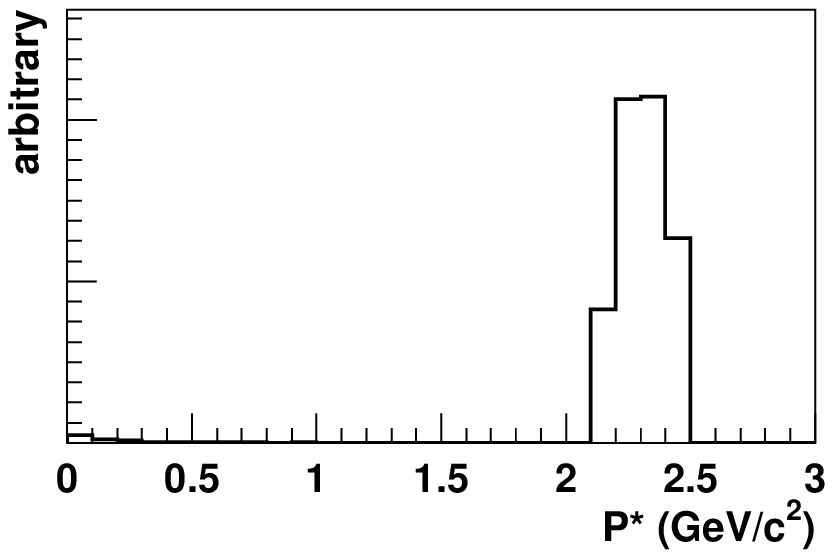}
   \end{center}  
  \end{minipage}
  \hfil
  \begin{minipage}{0.32\textwidth}
   \begin{center}
    (c) $B^- \to D^0\ell^-\nu$ \\
    \includegraphics[scale=0.45]{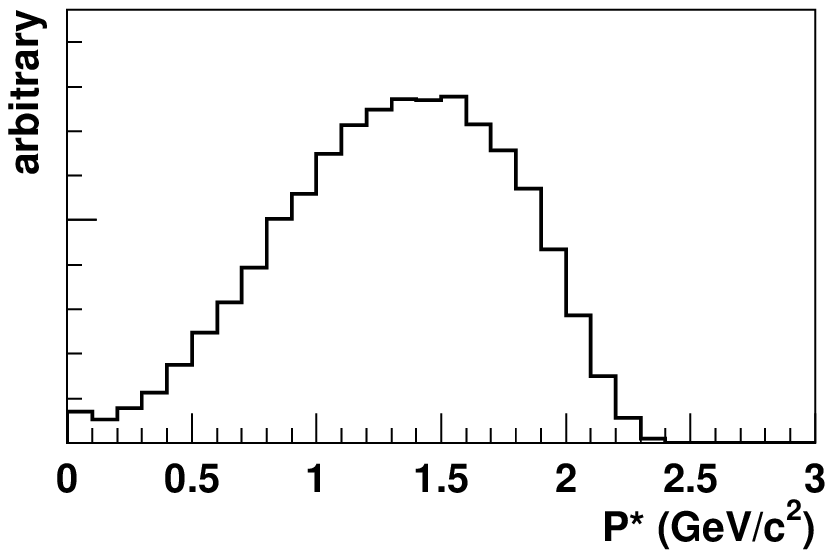} \\
    (f) $B^- \to \rho^+\pi^0$ \\
    \includegraphics[scale=0.45]{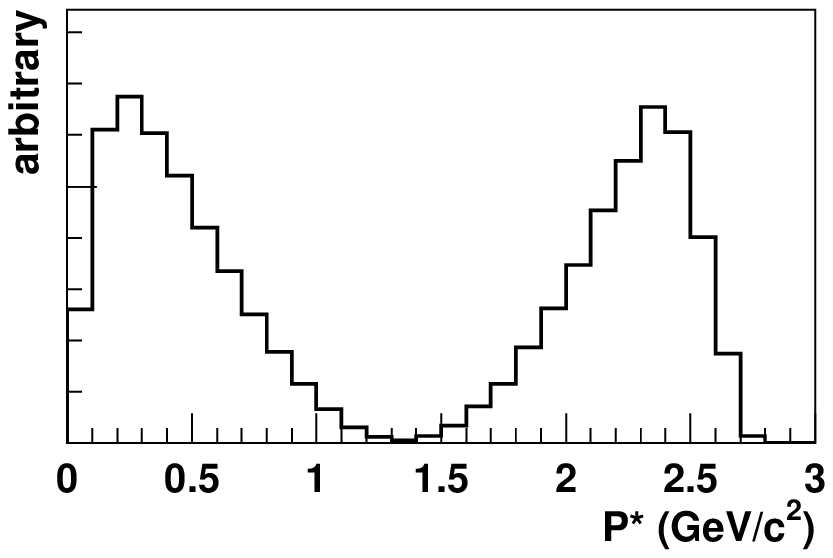}
   \end{center}
  \end{minipage}
  \caption{\label{fig:knunu-qqmc}%
  Momentum distributions of the only charged particle
  for various $B$ decays.
  }
 \end{center}
\end{figure}

The signature of a $B^- \to K^- \nu\bar{\nu}$ decay is
a single kaon and nothing else recoiling against
the reconstructed $B$ meson. In order to estimate backgrounds for
this decay, we first identify other $B$ decay modes that have only
one charged particle in the final state,
using the QQ generator. We find that
examples of such decays are:
semi-leptonic $B$ decays, $B \to D^{(*)}\ell\nu$ in which
the $D$ decays to $K_L\pi^0$, 
$B \to D\rho$, $D\pi$ with $D \to K_L\pi^0$;
charmless $B$ decays such as $B^+ \to \rho^+\pi^0$.
Figure~\ref{fig:knunu-qqmc} shows the momentum distributions
of the charged particle in these decays as well as in the signal
modes. A three body phase space decay for $B \to K\nu\bar{\nu}$ is assumed.


\begin{figure}
 \begin{center}

  \includegraphics[width=0.8\textwidth,clip]{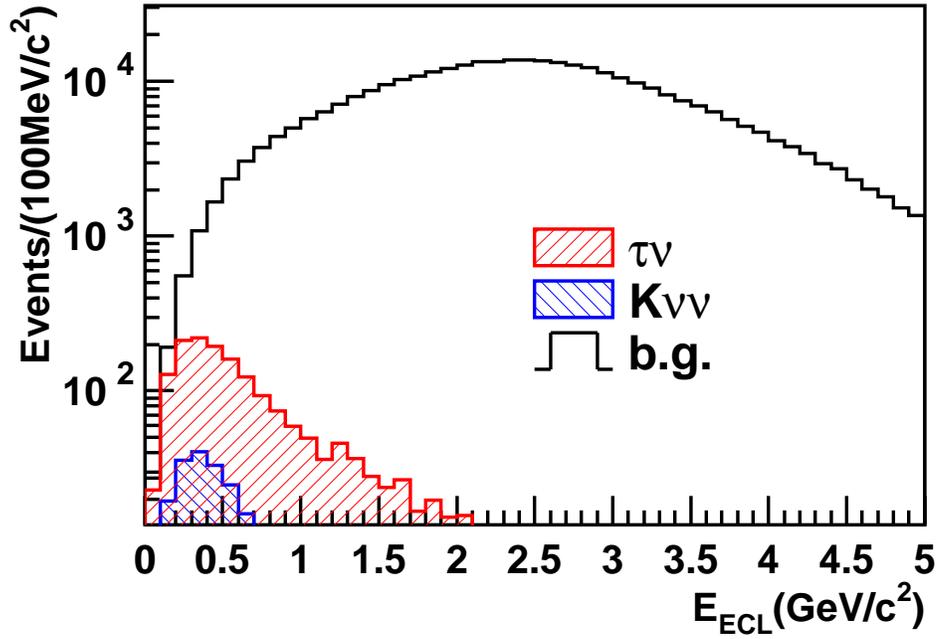}
  \caption{\label{fig:knunu-ecl-nocut}%
  ECL energy distribution for $B \to K\nu\bar{\nu}$, $B \to \tau\nu$
  and background MC (log scale).
The only requirement applied to the charged particle
  is$|\vec{p}^{\,*}| > 0.7~\mathrm{GeV}/c$.
  }
 \end{center}
\end{figure}

We then estimate the amount of signal and background
assuming $50~\mathrm{ab}^{-1}$
using a full simulation program based on GEANT3 for the Belle
experiment. 
We assume that the full reconstruction is perfect and
its efficiency is $0.2\%$.
Figure~\ref{fig:knunu-ecl-nocut} shows the total energy deposited
in the electro-magnetic calorimeter by the signal side ``B''
meson decay, unmatched hadronic interactions and beam related
backgrounds.
We select events in which only one charged track is observed with a
momentum cut; $|\vec{p}^{\,*}| > 0.7~\mathrm{GeV}/c$.
Particle identification cuts are not applied on the charged particle.
These criteria are rather similar to the recent analyses.
It is obvious that we need tighter selection criteria to
observe signals. 

\begin{figure}
 \begin{center}
  \includegraphics[width=0.8\textwidth,clip]{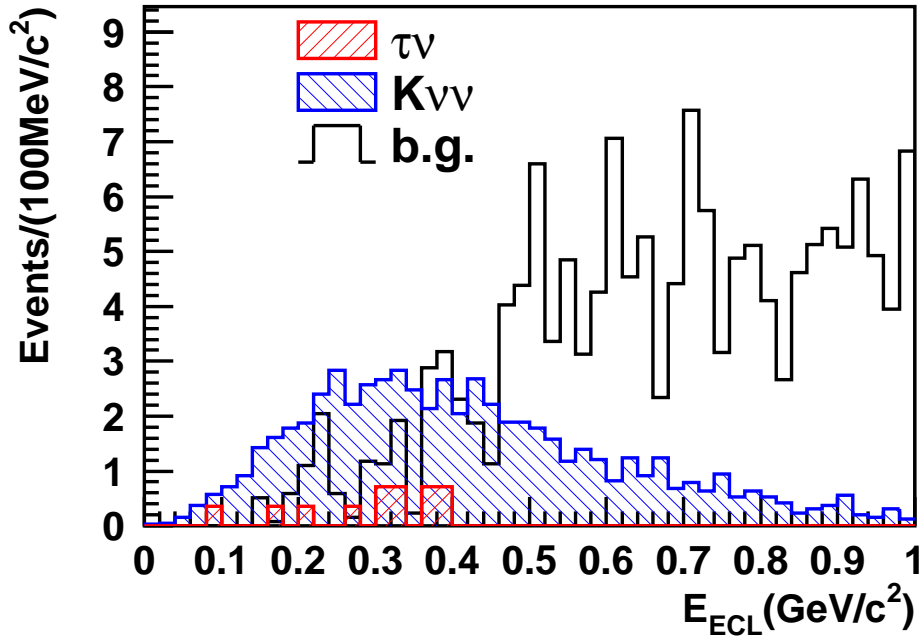}
  \caption{\label{fig:knunu-ecl-kaon-23s}%
  ECL energy distribution for $B \to K\nu\bar{\nu}$, $B \to \tau\nu$
  and background MC, with
  $2.0~\mathrm{GeV}/c < |\vec{p}^{\,*}| < 2.7~\mathrm{GeV}/c$
  and a tight KID requirement applied to
  the charged particle.
  }
 \end{center}
\end{figure}

We optimize the momentum selection to
$2.0~\mathrm{GeV}/c < |\vec{p}^{\,*}| < 2.7~\mathrm{GeV}/c$
using Fig.~\ref{fig:knunu-qqmc}.
We also apply tight KID requirements.
Figure~\ref{fig:knunu-ecl-kaon-23s} shows the total energy
distribution after applying these selection criteria.
If we define the signal region to be $E < 0.5~\mathrm{GeV}$,
the number of $B \to K\nu\bar{\nu}$ events is estimated to be
$43.0 \pm 1.1$, while the contribution from the background is $29.3 \pm
3.4$ events. This corresponds to a significance of $5.1\sigma$.
Table~\ref{table:knunu-bg-knunu} summarizes
the background contribution for each decay mode.
Large contributions from semi-leptonic decays
are reduced by applying a tight kaon identification selection.

\begin{table}
 \begin{center}
  \caption{\label{table:knunu-bg-knunu} Backgrounds in the
  $B \rightarrow K \nu \bar{\nu}$}
  \begin{math}
 \begin{array}{cccc}
  \hline\hline
   \mbox{Mode} & E<1.0~\mathrm{GeV} & \mbox{$E<1.0~\mathrm{GeV}$ with KID}
   & \mbox{$E<0.5~\mathrm{GeV}$ with KID} \\ \hline
  B^- \to D^{*0}\mu^-\bar{\nu}  &  265.7 \pm 11.7 &  8.2 \pm 2.1 &  2.1 \pm 1.0 \\
  B^- \to D^0\pi^-              &  199.0 \pm 10.1 &  7.7 \pm 2.0 &  3.1 \pm 1.3 \\
  B^- \to D^{*0}e^-\bar{\nu}    &  181.1 \pm  9.6 &  8.7 \pm 2.1 &  2.1 \pm 1.0 \\
  B^- \to D^0\mu^-\bar{\nu}     &  131.3 \pm  8.2 & 25.1 \pm 3.6 &  5.1 \pm 1.6 \\
  B^- \to D^{*0}\pi^-           &  110.3 \pm  7.5 &  3.6 \pm 1.4 &  0.0 \pm 0.0 \\
  B^- \to D^0\rho^-             &   95.9 \pm  7.0 &  6.2 \pm 1.8 &  1.0 \pm 0.7 \\
  B^- \to D^0e^-\bar{\nu}       &   85.7 \pm  6.6 & 19.5 \pm 3.2 &  4.1 \pm 1.5 \\
  B^- \to D^{*0}\rho^-          &   43.6 \pm  4.7 &  3.6 \pm 1.4 &  0.5 \pm 0.5 \\
  B^- \to D^0K^-                &   13.3 \pm  2.6 &  8.7 \pm 2.1 &  2.1 \pm 1.0 \\
  B^- \to D^{*0}K^-             &    5.6 \pm  1.7 &  4.6 \pm 1.5 &  0.5 \pm 0.5 \\
  \mbox{Other $b \to c$ decays} &   29.8 \pm  3.9 &  3.1 \pm 1.3 &  1.0 \pm 0.7 \\ \hline
  \mbox{$b \to c$ decays total} & 1161.4 \pm 24.4 & 99.0 \pm 7.1 & 21.5 \pm 3.3\\ \hline
  B^- \to K^{*-}\pi^-\pi^+         &  28.0 \pm 1.5 &  0.8 \pm 0.2 & 0.2 \pm 0.1 \\
  B^- \to \pi^-\bar{K}^0           &  22.8 \pm 1.3 &  0.5 \pm 0.2 & 0.0 \pm 0.0 \\
  B^- \to \rho^+\pi^0              &  21.1 \pm 1.3 &  0.4 \pm 0.2 & 0.0 \pm 0.0 \\
  B^- \to K^-f_2^{\prime}(1525)    &  19.1 \pm 1.2 & 12.5 \pm 1.0 & 1.1 \pm 0.3 \\
  B^- \to K^-\pi^0                 &  18.5 \pm 1.2 & 12.7 \pm 1.0 & 3.3 \pm 0.5 \\
  B^- \to K^{*-}\rho^0             &  13.6 \pm 1.0 &  0.2 \pm 0.1 & 0.1 \pm 0.1 \\
  B^- \to K^{*-}f_2(1270)          &  10.9 \pm 0.9 &  0.4 \pm 0.2 & 0.0 \pm 0.0 \\
  B^- \to K^{*-}f_2^{\prime}(1525) &  10.7 \pm 0.9 &  8.2 \pm 0.8 & 1.3 \pm 0.3 \\
  B^- \to \pi^-\pi^0               &   9.3 \pm 0.8 &  0.4 \pm 0.2 & 0.2 \pm 0.1 \\
  B^- \to K^{*-}\gamma             &   7.9 \pm 0.8 &  0.0 \pm 0.0 & 0.0 \pm 0.0 \\
  B^- \to \rho^+K^0                &   7.1 \pm 0.7 &  0.1 \pm 0.1 & 0.0 \pm 0.0 \\
  B^- \to \pi^-K^{*0}              &   6.0 \pm 0.7 &  0.1 \pm 0.1 & 0.0 \pm 0.0 \\
  B^- \to K^-\eta^{\prime}         &   4.5 \pm 0.6 &  3.2 \pm 0.5 & 0.2 \pm 0.1 \\
  B^- \to K^-f_0(1370)             &   4.0 \pm 0.6 &  2.2 \pm 0.4 & 0.2 \pm 0.1 \\
  B^- \to K^-\eta                  &   3.9 \pm 0.5 &  2.7 \pm 0.5 & 0.5 \pm 0.2 \\
  B^- \to K^-f_2(1270)             &   3.5 \pm 0.5 &  2.4 \pm 0.4 & 0.2 \pm 0.1 \\
  B^- \to \bar{K}_0^*(1430)^0\pi^- &   2.8 \pm 0.5 &  0.2 \pm 0.1 & 0.1 \pm 0.1 \\
  B^- \to \rho^-\eta               &   2.8 \pm 0.5 &  0.0 \pm 0.0 & 0.0 \pm 0.0 \\
  B^- \to K^-K^0                   &   2.5 \pm 0.4 &  1.7 \pm 0.4 & 0.4 \pm 0.2 \\
  \mbox{Other rare $B$ decays}     &  16.7 \pm 1.1 &  3.2 \pm 0.5 & 0.2 \pm 0.1 \\ \hline
  \mbox{Rare $B$ decays total}     & 215.6 \pm 4.1 & 51.6 \pm 2.0 & 7.8 \pm 0.8 \\ \hline
  \mbox{Total}                 & 1377.0 \pm 24.7 & 150.6 \pm 7.3 & 29.3 \pm 3.4 \\ \hline\hline
 \end{array}
\end{math}

 \end{center}
\end{table}




\subsection{Discussion}

Further improvements for this analysis are expected. The efficiency
for the full reconstruction tagging technique can be
improved. Optimisation of the tag side efficiency and background can be 
studied separately for the discovery of the decay and setting upper
limit. Likelihood methods can be used for the tag side as
well as information on the signal single charged track and energy
deposits in the calorimeter. Better solid angle coverage will
improve both the tagging efficiency and the signal reconstruction efficiency. 
Much better
kaon identification and rejection of other spieces will reduce the
backgrounds. More effective rejection of the events that contain $K_L$'s
can be studied. On the other hand, the environment may become
much harsher as luminosity increases and beam related backgrounds will
certainly increase if the detector performance remains the same.
As a result we may suffer from a smaller
tagging efficiency and more energy deposits in the
calorimeter. So far we are not able to estimate such effects
quantitatively, but will continue to study this important
decay mode.

\clearpage \newpage


\newcommand{\btodtaunu}{B \to \bar{D}\tau^+\nu_\tau}

\section{More than one neutrino II: $\btodtaunu$}
\label{sec:sensitivity_dtaunu}

The decay $\btodtaunu$ is sensitive to the exchange of
charged Higgs boson as introduced in the Minimal
Supersymmetric Standard Model (MSSM), since the amplitude is
roughly proportional to $m_\tau m_b\tan\beta$.
The branching fraction of $\btodtaunu$ is expected to be
large ($\sim 8\times10^{-3}$) in the SM, but much more data are
required to measure this process because of the presence of
two or more neutrinos in the decay final state.
This mode consequently requires the tagging of the other side
$B$, for which a reconstruction efficiency is estimated to be
$0.2\%$,
which has been discussed in Section~\ref{sec:full_recon}.
Therefore, the study of the MSSM Higgs through this decay
mode is only possible with the statistical power of
SuperKEKB. 

\subsection{Introduction}
In the MSSM, the coupling of the charged Higgs bosons,
$H^\pm$, to quarks and leptons is given by
\begin{eqnarray}
  {\cal L}_H &=& (2\sqrt{2}G_F)^{1/2} 
  \left[
    \tan\beta 
    \left( \bar{u}_L V_{KM} M_d d_R + \bar{\nu}_L M_\ell \ell_R \right)
    + \frac{1}{\tan\beta}
    \bar{u}_R V_{CKM} M_u d_L
  \right] H^\pm \nonumber
  \\&&+\; h.c. ,
\end{eqnarray}
where $M_u$ and $M_d$ are the quark mass matrices, $M_\ell$
is the lepton ($\ell=e$, $\mu$, or $\tau$) mass matrix, 
$V_{CKM}$ is the Cabibbo-Kobayashi-Maskawa matrix, and 
$\tan\beta = v_2/v_1$ is the ratio of the vacuum expectation
values of the Higgs fields. 
Therefore, the decay amplitude of $\btodtaunu$ that is
mediated by $\bar{b}\to c\tau^+\nu$ has a term
proportional to $M_bM_\tau\tan^2\beta$
\cite{Tanaka:1994ay,Miki:2002nz}.
The large $\tau$ mass is an advantage of this decay in
measuring the charged Higgs mass compared to other
semi-leptonic decays. 

\begin{figure}[tbp]
\begin{center}
  \includegraphics*[width=12cm]{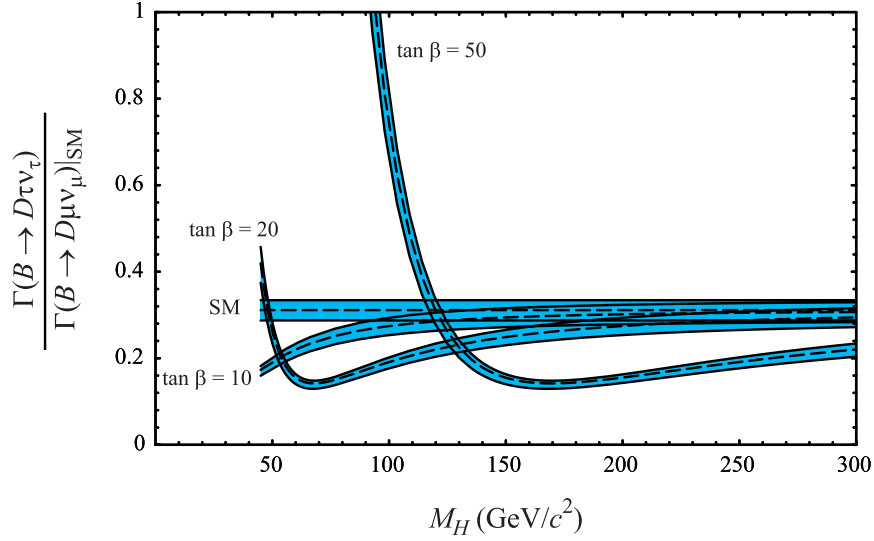}
  \\[10mm]
  \includegraphics*[width=10cm]{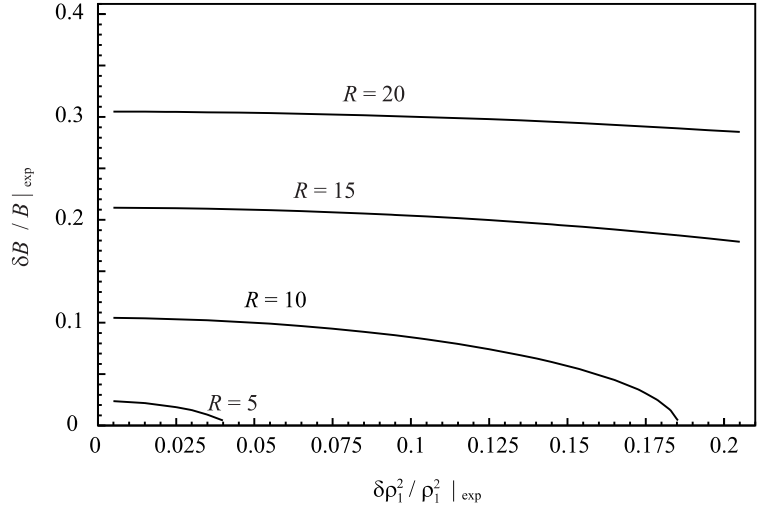}
\end{center}
\caption{
  The top figure shows the ratio of $\Gamma(\btodtaunu)$ to
  $\Gamma(B\to\bar{D}\mu^+\nu_\mu)$. 
  The flat band is the prediction of the Standard Model;
  also shown is the charged Higgs contribution, which is a function of 
  the charged Higgs mass with several $\tan\beta$ values.
  The width of each band is due to uncertainty in the form factor.
  The bottom figure shows the $\delta B/B|_{\rm exp}$
  distribution as a function of 
  $\delta  \rho_1^2/\rho_1^2|_{\rm exp}$ with varying 
  $R \equiv M_W\tan\beta/M_H^\pm$. 
}
\label{fig:tan2beta-BR}
\end{figure}

Figure~\ref{fig:tan2beta-BR} (top) shows 
the ratio $B$:
\begin{equation}
  B = \frac{\Gamma(\btodtaunu)}{\Gamma(B\to \bar{D}\mu^+\nu_\mu)_{\rm SM}}
\end{equation}
as a function of the charged Higgs mass with several
$\tan\beta$ values. 
The width of each band represents uncertainty in the 
$B\to D$ semi-leptonic form factor. 
The form factor is modeled as a function of the momentum
transfer $q^2$ using a slope parameter
$\rho^2_1$ \cite{Caprini:1997mu}, and the uncertainty is
from the error of $\rho^2_1$.
The Belle collaboration has measured $\rho^2_1$ as
$\rho^2_1 = 1.33\pm0.22$
\cite{Abe:2001nn}.
Figure~\ref{fig:tan2beta-BR} (bottom) shows the 
$\delta B/B|_{\rm exp}$ distribution as a function of 
$\delta \rho_1^2/\rho_1^2|_{\rm exp}$.
In the figure we show several curves with varying 
$R \equiv M_W\tan\beta/M_H^\pm$.

\subsection{$\btodtaunu$ Reconstruction}
Final states for the $\btodtaunu$ decay include two $\tau$-neutrinos 
after taking into account the sub-decay of the $\tau$;
one additional neutrino exists when $\tau$ decays into a leptonic mode.
Because of the presence of two or more invisible particles, 
the decay of $\btodtaunu$ has few kinematic constraints.
To reduce the number of combinations of the reconstructed $D$ and $\tau$,
we first remove particles originating from the other $B$ ($B_{\rm ful}$),
which does not decay into $\btodtaunu$ ($B_{\rm sig}$).
We then reconstruct $D$ and $\tau$ candidates using the remaining particles.
Finally, we apply kinematic selection requirements on the reconstructed 
$D$ and $\tau$ combination to reduce background.

In the following paragraphs we describe the reconstruction of $\btodtaunu$.
A detailed description of the full reconstruction of $B_{\rm ful}$ mesons 
is given in subsection~4.1.2.

\paragraph{Light Meson Reconstructions}
The $\ks$ mesons are reconstructed from a pair of oppositely charged $\pi$ tracks.
The invariant mass of the $\pi$ pair should be within $|M_{\pi\pi} - M_{\ks}|< 15 {\rm GeV}/c^2$.
The position of closest approach of the  $\pi$ tracks should be displaced 
from the interaction point in the plane perpendicular to the beam axis.

Neutral pions are reconstructed from a pair of two photons.
The $\piz$ invariant mass is required to be within $|M_{\gamma\gamma}-M_{\piz}| < 16~{\rm MeV}/c^2$.
We require the photon energy deposit in the calorimeter to be greater than 50~MeV.

\paragraph{Charmed Meson Reconstructions}
We reconstruct $D$ mesons from $\bar{D}^0\to K^+\pi^-(\piz)$,
$K^+\pi^+\pi^-\pi^-(\piz)$, $\ks\pi^+\pi^-(\piz)$, and
$\ks\pi^0$, 
where $(\piz)$ indicates zero or one neutral pion.
Charge conjugate modes are implicitly included throughout
this section.
These channels cover $35\%$ of the total $\bar{D}^0$ decay width.
The charged $D$ meson is reconstructed from $D^+\to K^-\pi^+\pi^+(\piz)$ and $\ks\pi^+$,
covering $16\%$ of the total $D^+$ decay width.

We construct kaon (${\cal L}(K)$) and pion (${\cal L}(\pi)$) likelihoods 
to identify the particle species of each charged particle
by combining the $dE/dx$ in the drift chamber, the hit in the time of flight 
counter and the  ring imaging Cherenkov counter.
The charged particle is identified as a kaon 
if ${\cal L}(K)/[{\cal L}(\pi) + {\cal L}(K)]$ exceeds 0.4;
otherwise it is a pion.

The reconstructed $D^0$ or $D^+$ should have an
invariant mass within $|M_{Kn\pi} - M_D| < 30~{\rm MeV}/c^2$.

If more than one $D$ meson is reconstructed, the $D$ meson that has the closest invariant mass
to the world average value is taken \cite{Hagiwara:fs}.

\paragraph{$B$ Meson Reconstructions}
The $\btodtaunu$ decay is reconstructed from $\bz\to D^-\tau^+\nu_\tau$ and $\bp\to \bar{D}^0\tau^+\nu_\tau$,
where the $\tau^+$ is identified in one of four following sub-decays:
$\tau^+\to\pip\nu_\tau$, $\rho^+(\pip\piz)\nu_\tau$, $e^+\nu_\tau\bar{\nu}_e$, and $\mu^+\nu_\tau\bar{\nu}_\mu$.

The $\btodtaunu$ candidate is reconstructed by adding one charged particle,
which is assumed to originate from the $\tau^+$ decay, to the reconstructed $D$ meson.
Positively identified protons are rejected.

If the additional charged particle is consistent with the electron or 
muon hypothesis,
the $\tau^+$ decay is treated as a leptonic mode 
($\tau^+\to \ell^+\nu_\tau\bar{\nu}_\ell$)\footnote{
  Throughout this section, the symbol $\ell$ indicates
  leptons except $\tau$ unless otherwise specified.
};
otherwise the $\tau^+$ decay is considered as a hadronic
mode ($\tau^+ \to \pi^+\nu_\tau$). 

In the case of $\tau^+\to\pi^+\nu_\tau$ decay, one neutral pion (if it exists) 
that is associated
to neither $B_{\rm ful}$ nor $D$ decay is added to the $\pi^+\nu_\tau$ final state
to reconstruct $\tau^+ \to \rho^+(\pi^+\piz)\nu_\tau$ mode.
In this case, the $\pi^+\piz$ invariant mass should be within $|M_{\pi^+\piz}-M_{\rho^+}|< 300~{\rm MeV}/c^2$.

To reject $B\to \bar{D}^*\tau\nu_\tau$ events, we reconstruct $D^*$'s by adding
a $\pi^0$ to the reconstructed $D$.
When $|(M_{Kn\pi\piz} - M_{Kn\pi}) - \Delta M_{D^*-D}| < 10~{\rm MeV}/c^2$, we discard the event.

If any charged particle and/or $\kl$ candidate remains 
after the reconstructions of $B_{\rm ful}$ and $B_{\rm sig}$, the event is rejected.

\subsection{Kinematic Event Selection}
We use three kinematic parameters to select $\btodtaunu$ signal events from the reconstructed candidates.

The first is the residual cluster energy in the calorimeter ($E_{\rm res}$).
We require $E_{\rm res} < 100~{\rm MeV}$.

The second is the missing-mass squared defined by
\begin{equation}
|MM|^2 \equiv |p_{B_{\rm sig}}- p_D - p_{X^+}|^2,
\end{equation}
where $p_{X^+}$ is the charged particle momentum originating from the $\tau^+$
decay.  The momentum $p_{B_{\rm sig}}$ is given by 
$p_{B_{\rm sig}} = p_{\Upsilon(4S)}-p_{B_{\rm ful}}$.

The last is the cosine of the angle between the momenta of the 
two $\tau$-neutrinos ($\cos\theta$) in the frame where ${\vec p}_{B_{\rm sig}} = {\vec p}_D$.
This parameter can only be defined for the 
$\tau^+\to h^+\bar{\nu}_\tau$ sub-decay.
Energy-momentum conservation for 
the $B_{\rm sig}\to D\tau^+(h^+\bar{\nu}_\tau)\nu_\tau$ decay is expressed by
\begin{equation}
p_{B_{\rm sig}} = p_D + p_{h^+} + p_{\bar{\nu}_\tau} + p_{\nu_\tau}. \label{eq:emcon}
\end{equation}
The $\tau^+$ and neutrino masses are given by
\begin{equation}
\left(p_{h^+} + p_{\bar{\nu}_\tau}\right)^2 = m_\tau^2,
\end{equation}
\begin{equation}
p_{\bar{\nu}_\tau}^2 = p_{\nu_\tau}^2 = 0.\label{eq:mneutrino}
\end{equation}
We then boost the system to the frame where
\begin{equation}
{\vec p}_{B_{\rm sig}} = {\vec p}_D. \label{eq:frame}
\end{equation}
Using Eq.~(\ref{eq:emcon})-(\ref{eq:frame}), we have
\begin{equation}
(E_{B_{\rm sig}} - E_D)^2 - 2E_{\nu_\tau}(E_{B_{\rm sig}} - E_D) = m_\tau^2.
\end{equation}
Then, the energies of the two neutrinos can be expressed in terms of
measurable parameters as
\begin{equation}
E_{\nu_\tau} = \frac{(E_{B_{\rm sig}} - E_D)^2 - m_\tau^2}{2(E_{B_{\rm sig}} - E_D)}, \quad\quad
E_{\bar{\nu}_\tau} = E_{B_{\rm sig}} - E_D - E_{h^+} - E_{\nu_\tau}.
\end{equation}
Finally, we can measure $\cos\theta$ using the following equation:
\begin{eqnarray}
(\vec{p}_{B_{\rm sig}} - \vec{p}_D - \vec{p}_{h^+})^2 &=& (\vec{p}_{\bar{\nu}_\tau}+\vec{p}_{\nu_\tau})^2 \nonumber \\ 
& = & 2 \vec{p}_{\bar{\nu}_\tau}\cdot\vec{p}_{\nu_\tau} \nonumber\\
& = & 2 E_{\bar{\nu}_\tau} E_{\nu_\tau} \cos\theta.
\end{eqnarray}
The $\cos\theta$ of the signal events is limited from $-1$ to $+1$, while that of the background events is unrestricted.

The background contamination is studied by using generic $B$ decay MC samples.
We generate samples of $3.2\times10^6$ MC events
with the fast simulator for generic
charged and neutral $B$ decays.
In this case, $8\times10^{-3}$ is taken as the branching fraction of $\btodtaunu$ decay for both charged and neutral $B$.
To increase statistics,
the $B_{\rm ful}$ is pseudo-reconstructed using generator information,
i.e.: the $B_{\rm ful}$ is fully reconstructed with perfect efficiency and purity.

\begin{figure}
\begin{center}
\resizebox{0.90\textwidth}{!}{\includegraphics{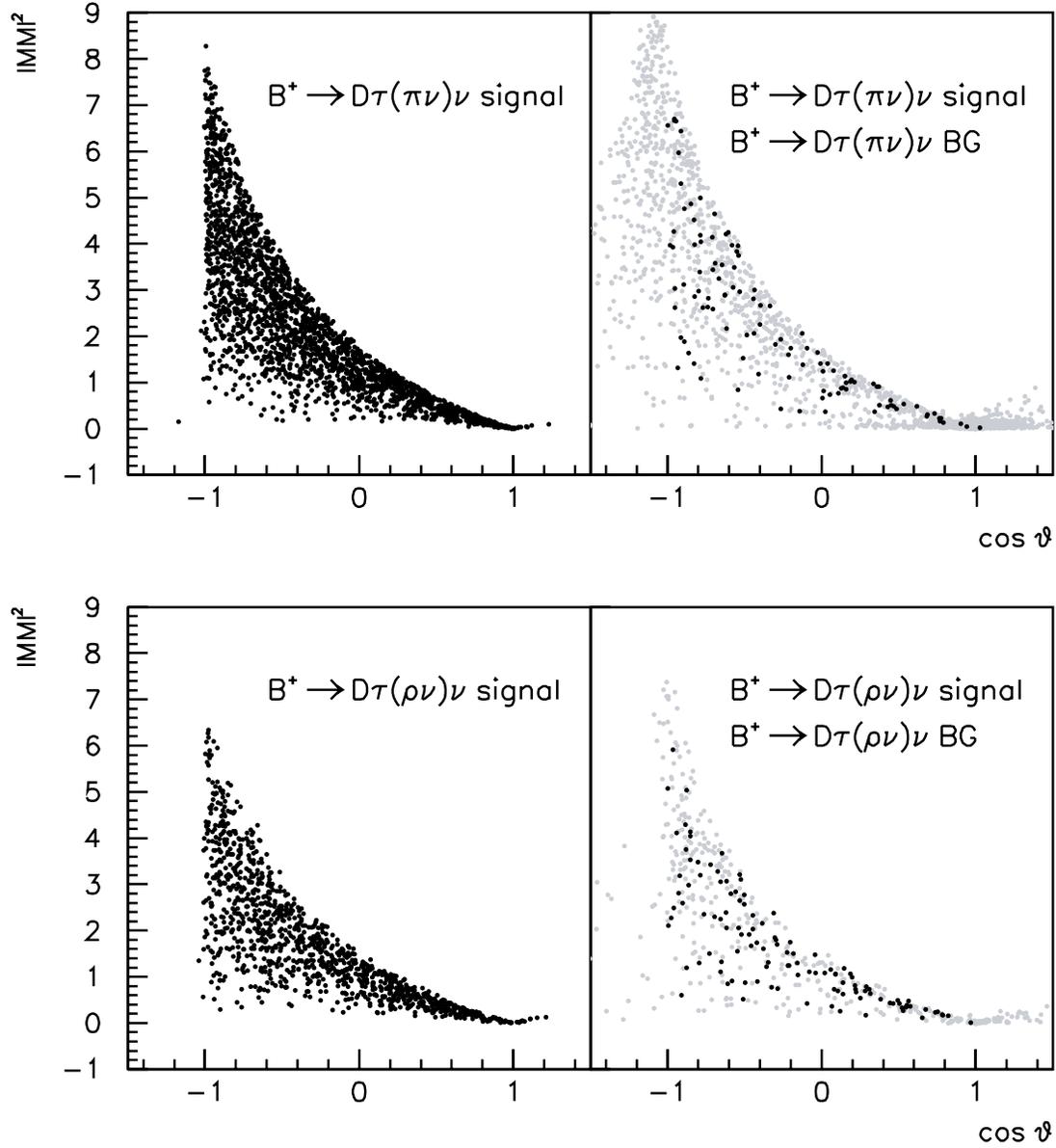}}
\end{center}
\caption{
A scatter plot of the reconstructed $B^+\to \bar{D}^0\tau^+(h^+\bar{\nu}_\tau)\nu_\tau$ candidates
in the $|MM|^2$-$\cos\theta$ plane.
The upper two figures are obtained in the $\tau^+\to\pi^+\bar{\nu}_\tau$ decay, and
the lower two figures are obtained in the $\tau^+\to\rho^+(\pi^+\piz)\bar{\nu}_\tau$ decay.
The left figures show the distributions obtained from signal MC.
In the right figures, the background distributions are shown by gray points
and the signal distributions are shown by black points.
}
\label{fig:mm2costh12-had}
\end{figure}

Figure~\ref{fig:mm2costh12-had} shows the distribution of the reconstructed
$B^+\to \bar{D}^0\tau^+(h^+\bar{\nu}_\tau)\nu_\tau$ candidates in the
$|MM|^2$-$\cos\theta$ plane.  The left column figures show the signal events only
and the right ones show the background distribution (gray points)
with the signal distributions (black points) superimposed .
We determine the optimal signal region to be $|MM|^2>0.1~({\rm GeV}/c^2)^2$ 
and $-1.0<\cos\theta<0.8$,
irrespectively of the $\tau^+\to\pi^+\bar{\nu}_\tau$ or $\tau^+\to\rho^+\bar{\nu}_\tau$
by maximizing $S/\sqrt{S+B}$ (FOM), where $S$ and $B$ are the numbers of reconstructed signal and background events, respectively.

\begin{figure}
\begin{center}
\resizebox{0.90\textwidth}{!}{\includegraphics{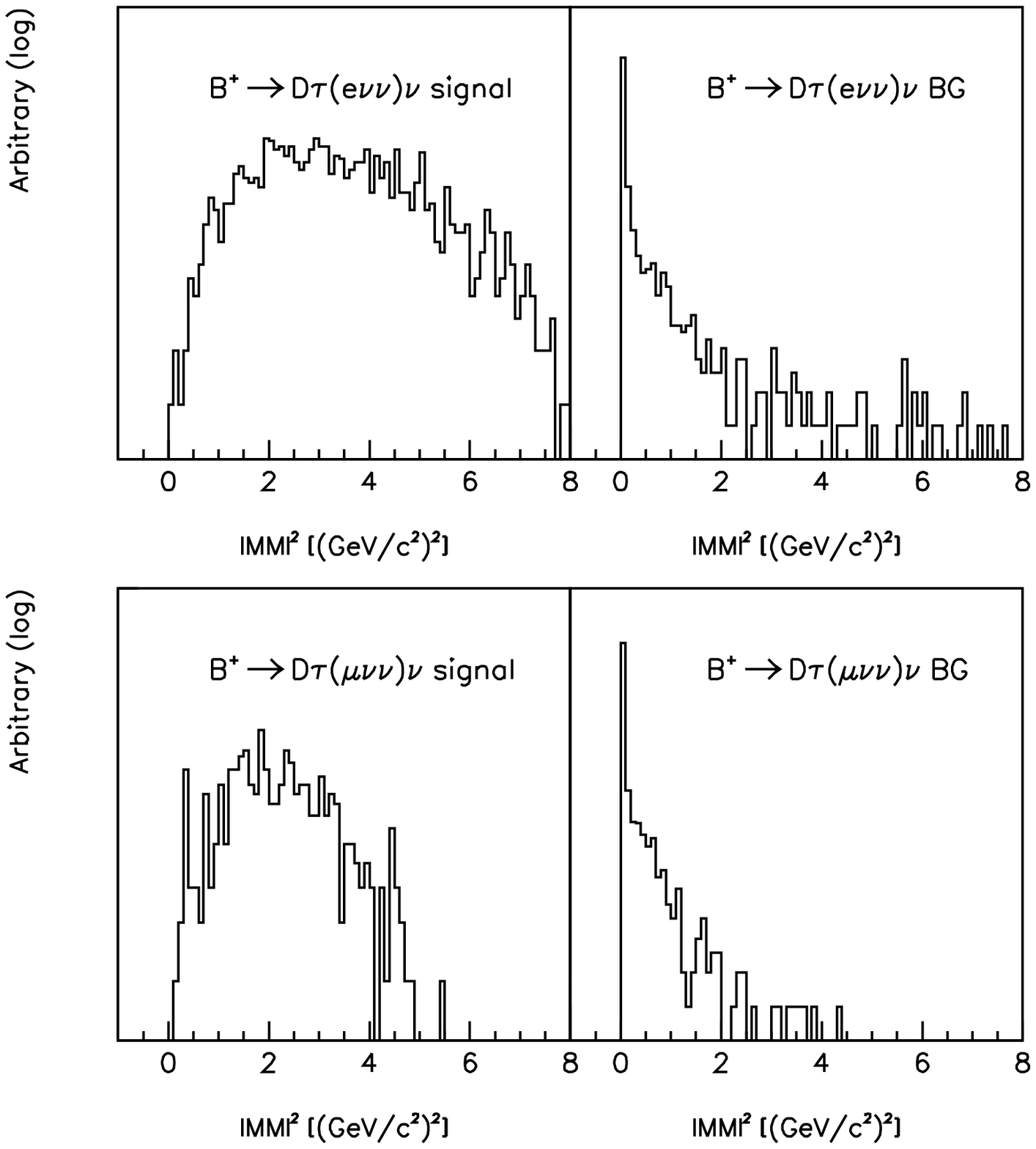}}
\end{center}
\caption{
The $|MM|^2$ distribution for reconstructed $B^+\to \bar{D}^0\tau^+(e^+\bar{\nu}_\tau\nu_e)\nu_\tau$ candidates (upper)
and $B^+\to \bar{D}^0\tau^+(\mu^+\bar{\nu}_\tau\nu_\mu)\nu_\tau$ candidates (lower).
The left figures show the distributions obtained from signal MC, 
and the right figures show the background.
}
\label{fig:mm2-lep}
\end{figure}

Figure~\ref{fig:mm2-lep} shows the $|MM|^2$ distribution of the reconstructed candidates of
$B^+\to \bar{D}^0\tau^+(\ell^+\bar{\nu}_\tau\nu_\ell)\nu_\tau$.
For that decay irrespectively of the lepton flavor from the $\tau$ decay, 
we determine the selection criteria 
$|MM|^2>1.2~({\rm GeV}/c^2)^2$ by maximizing the FOM.

We find that $|MM|^2$ and $\cos\theta$ for the neutral $B$ decay have 
similar distributions
in both signal and background as the charged $B$ decay.
Therefore, the selection criteria determined by charged $B$ decays
are also applied to neutral $B$ decays.

Table~\ref{tab:eff-SN} summarizes the reconstruction efficiency.
The numbers of reconstructed signal and the background events in the generic $B$ decay MC samples
are also shown.

\begin{table}[tbp]
\begin{center}
\begin{tabular}{|l|c|c|c|c|}

\hline
Decay mode                                                         & efficiency (\%) & $Br$ & $N_{\rm sig}$ & $N_{\rm bkg}$ \\
\hline
\hline
$\bar{D}^0\tau^+(\ell^+\bar{\nu}_\tau\nu_\ell)\nu_\tau$     & $ 4.9 \pm 0.3$  & $13.5\times10^{-4}$ & 213   & 293     \\
\hline
$\bar{D}^0\tau^+(h^+\bar{\nu}_\tau)\nu_\tau$                & $10.9 \pm 0.5$  & $13.6\times10^{-4}$ & 476   &2085     \\
\hline
\hline
$D^-\tau^+(\ell^+\bar{\nu}_\tau\nu_\ell)\nu_\tau$           &  $0.7 \pm 0.2$  &  $6.2\times10^{-4}$ &  15   &  22     \\
\hline
$D^-\tau^+(h^+\bar{\nu}_\tau)\nu_\tau$                      &  $3.3 \pm 0.4$  &  $6.4\times10^{-4}$ &  68   & 194     \\
\hline
\end{tabular}
\end{center}
\caption{
Summary of the reconstruction efficiencies.
The numbers of reconstructed signal ($N_{\rm sig}$) and the background events ($N_{\rm bkg}$)
in generic $B$ decay MC samples are also shown,
where $Br(\btodtaunu)=8\times10^{-3}$ is assumed.}
\label{tab:eff-SN}
\end{table}

\subsection{Background Components}
The dominant background source in the $B^+\to \bar{D}^0\tau^+(h^+\bar{\nu}_\tau)\nu_\tau$ signal region
is the decay $B^+\to D^{*-}\ell^+\nu_\ell\pi^+$, due mainly to a missing 
$\ell^+$ and a slow pion from the $D^{*-}$.
The next largest background modes are
$B^+\to \bar{D}^{*0}\ell^+\nu_\ell$ with a missing $\gamma$ or $\pi^0$ 
from the $\bar{D}^{*0}$ and misidentification of $\ell^+$ as $\pi^+$, and
the $B^+\to \bar{D}^{*0}\tau^+\nu_\tau$ with a missing slow pion.
In this study the $B^+\to \bar{D}^{0}\tau^+(\ell^+\bar{\nu}_\tau\nu_\ell)\nu_\tau$ decay is considered as a background for the $B^+\to D^-\tau^+(h^+\bar{\nu}_\tau)\nu_\tau$ analysis; they
contribute to the signal region due mostly to misidentification of 
the $\ell^+$  from the $\tau^+$ decay as $\pi^+$.
The sum of the background modes listed above are $\sim45\%$ of the total background.
The contribution from  $D_s$ inclusive decays is $\sim8\%$ of the total.

The largest contribution to the $B^+\to \bar{D}^-\tau^+(\ell^+\bar{\nu}_\tau\nu_\ell)\nu_\tau$ signal region comes from
$B^+\to \bar{D}^{*-}\ell^+\nu_\ell\pi^+$ where both pions from $B^+$ 
and $\bar{D}^{*-}$ are missed.
The next largest background components are $B^+\to\bar{D}^{*0}\ell^+\nu_\ell$ 
and $B^+\to\bar{D}^{*0}\tau^+\nu_\tau$ 
when the $\gamma$ or $\piz$ from the $\bar{D}^{*0}$ are missed.

The dominant background mode in the 
$B^0\to D^-\tau^+(h^+\bar{\nu}_\tau)\nu_\tau$ signal region is
$B^0\to D^-\tau^+\nu_\tau$ with a mis-reconstructed $\tau^+$; 
the next largest is $B^0\to D^{*-}\mu^+\nu_\mu$,
although it is only $\sim20\%$ of the total.
No other background modes make significant contributions
in the signal region.

For the $B^0\to D^-\tau^+(\ell^+\bar{\nu}_\tau\nu_\ell)\nu_\tau$ mode,
because of the small MC statistics, we cannot yet evaluate the background.

\subsection{Statistical Significance}
As described in subsection~4.1.2, the full reconstruction efficiencies
for charged (neutral) $B$ mesons
are estimated to be around $0.2\%$ ($0.1\%$)
for a purity of about $80\%$.

Table~\ref{tab:5-50ab} lists the expected signal yields and backgrounds at 
integrated luminosities of $5$ and $50~{\rm ab}^{-1}$.
The values include a correction for the purity of $B_{\rm ful}$ reconstruction.
We assume $Br(\btodtaunu) = 8\times10^{-3}$ in the table.
The values listed are obtained by scaling the results in 
Table~\ref{tab:eff-SN} according to the integrated luminosity.
The expected uncertainties in the measured branching fraction ($\delta(Br)/Br$) are also shown.

\begin{table}[tbp]
\begin{center}
\begin{tabular}{|c|c|c|c|c|c|c|}
\hline
Decay mode                                                   & \multicolumn{3}{|c|}{$5~{\rm ab}^{-1}$}  & \multicolumn{3}{|c|}{$50~{\rm ab}^{-1}$} \\
\cline{2-7}
                                                    & $N_{\rm sig}$ & $N_{\rm bkg}$ & $\delta(Br)/Br$ & $N_{\rm sig}$ & $N_{\rm bkg}$ & $\delta(Br)/Br$ \\
\hline
\hline
$\bar{D}^0\tau^+(\ell^+\bar{\nu}_\tau\nu_\ell)\nu_\tau$ & $280  \pm 20 $  & $550  \pm20  $  & 7.9\% & $2800 \pm 50 $  & $5500 \pm 70 $  & 2.5\% \\
\cline{1-3} \cline{5-6}
$\bar{D}^0\tau^+(h^+\bar{\nu}_\tau)\nu_\tau$            & $620  \pm 20 $  & $3600 \pm60  $  &       & $6200 \pm 80 $  & $36000\pm200 $  &       \\
\hline
\hline
$D^-\tau^+(\ell^+\bar{\nu}_\tau\nu_\ell)\nu_\tau$       & $10   \pm 3  $  & $21   \pm5   $  &28.5\% & $98   \pm 10 $  & $210  \pm 10 $  & 9.0\% \\
\cline{1-3} \cline{5-6}
$D^-\tau^+(h^+\bar{\nu}_\tau)\nu_\tau$                  & $45   \pm 7  $  & $170  \pm10  $  &       & $450  \pm 20 $  & $1700 \pm 40 $  &       \\
\hline
\end{tabular}
\end{center}
\caption{The expected signal yields and backgrounds at integrated 
 luminosities of $5$ and $50~{\rm ab}^{-1}$,
 assuming $Br(\btodtaunu) = 8\times10^{-3}$.
 The expected uncertainties in the measured branching fraction ($\delta(Br)/Br$) are also shown.}
\label{tab:5-50ab}
\end{table}

The branching fraction for $B^+\to\bar{D}^0\tau^+\nu_\tau$ is expected 
to be determined with $12\sigma$ statistical significance at an integrated 
luminosity of 5~ab$^{-1}$.
The branching fractions of the neutral $B$ modes can also be measured 
at 50~ab$^{-1}$ with $11\sigma$ significance.

\subsection{Systematic Uncertainty}
Major sources of systematic uncertainty in the branching fraction measurement in $\btodtaunu$ decay are expected to be
$B_{\rm ful}$ reconstruction efficiency and purity,
particle identification efficiency and purity, and the slow pion
detection efficiency.

\subsection{Constraints on the Charged Higgs Mass}
\begin{figure}
\begin{center}
\resizebox{0.90\textwidth}{!}{\includegraphics{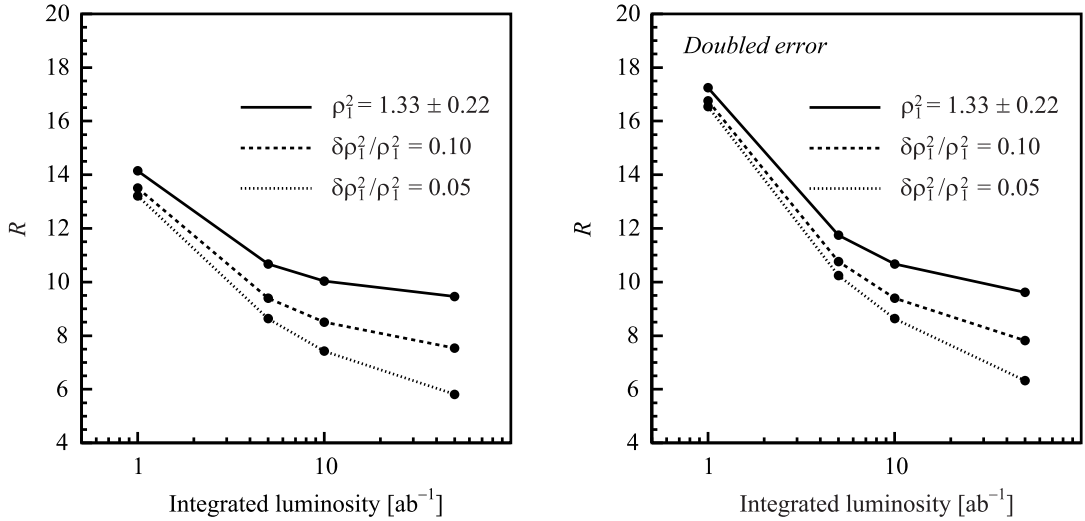}}
\end{center}
\caption{
The left figure shows the maximum $R$ constrained by the $B^+\to\bar{D}^0\tau^+\nu_\tau$ branching fraction measurement
as a function of the integrated luminosity for several values of 
$\delta\rho_1^2/\rho_1^2|_{\rm exp}$.
The right figure shows the same plot for the case with double the 
uncertainty in the measured branching fraction
}
\label{fig:R-lum}
\end{figure}

Figure~\ref{fig:R-lum} (left) shows the maximum $R$ value constrained by the branching fraction measurement of $B^+\to\bar{D}^0\tau^+\nu_\tau$
as a function of the integrated luminosity for several
$\delta\rho_1^2/\rho_1^2|_{\rm exp}$.
The right figure shows the case assuming a doubled uncertainty of the measured branching fraction.
The systematic uncertainty in the branching fraction measurement is not considered.
We ignore the uncertainty in $\delta B/B|_{\rm exp}$ that comes from the uncertainty in
the $B^+\to \bar{D}^0\mu^+\nu_\mu$ branching fraction because
it can be determined much precisely than that of $\bar{D}^0\tau^+\nu_\tau$.
We expect
\[
M_H>\frac{M_W\tan\beta}{11}
\]
at an integrated luminosity of $5~{\rm ab}^{-1}$, assuming that 
the current $\rho_1^2$ precision is unchanged.



\subsection{Summary}
The $\btodtaunu$ decay is a sensitive mode to probe the charged Higgs in the MSSM.
Using this decay mode we expect that the charged Higgs mass will be constrained
by $M_H >M_W\tan\beta/11$ at an integrated luminosity of $5~{\rm ab}^{-1}$.


\clearpage \newpage
\section{$\sinbb$}
\label{sec:sin2phi1}
\def\errtot    {\sigma_{\rm tot}}
\def\errtottwo {\sigma^2_{\rm tot}}
\def\lint {{\cal L}_{\rm int}}
\def\sjpsiks {\cals_{\jpsi\ks}}
\def\sphiks {\cals_{\phi\ks}}
\def\ajpsiks {\cala_{\jpsi\ks}}
\def\setapks {\cals_{\eta'\ks}}
\def\dM {\Delta m_d}
%
Very precise measurements of $\sinbb$,
or $\sjpsiks$, will remain
important at SuperKEKB.
There are two major reasons. 
One is to search for a new $CP$-violating phase
from the SM in $CP$ violation in $b \to s$ transitions
by testing a SM prediction $\sphiks = \sjpsiks$.
The other is to check the consistency of the Unitarity
Triangle.
As explained in Section~\ref{sec:theory_sin2phi1},
$\sinbb$ is determined using the $\bz\to\jpsi\ks$ mode 
with very small hadronic uncertainties.
It is also insensitive to effects beyond the SM.
Thus it serves as a reliable reference point for the SM.

\begin{figure}[tbp]
\begin{center}
\includegraphics[width=0.9\textwidth,clip]{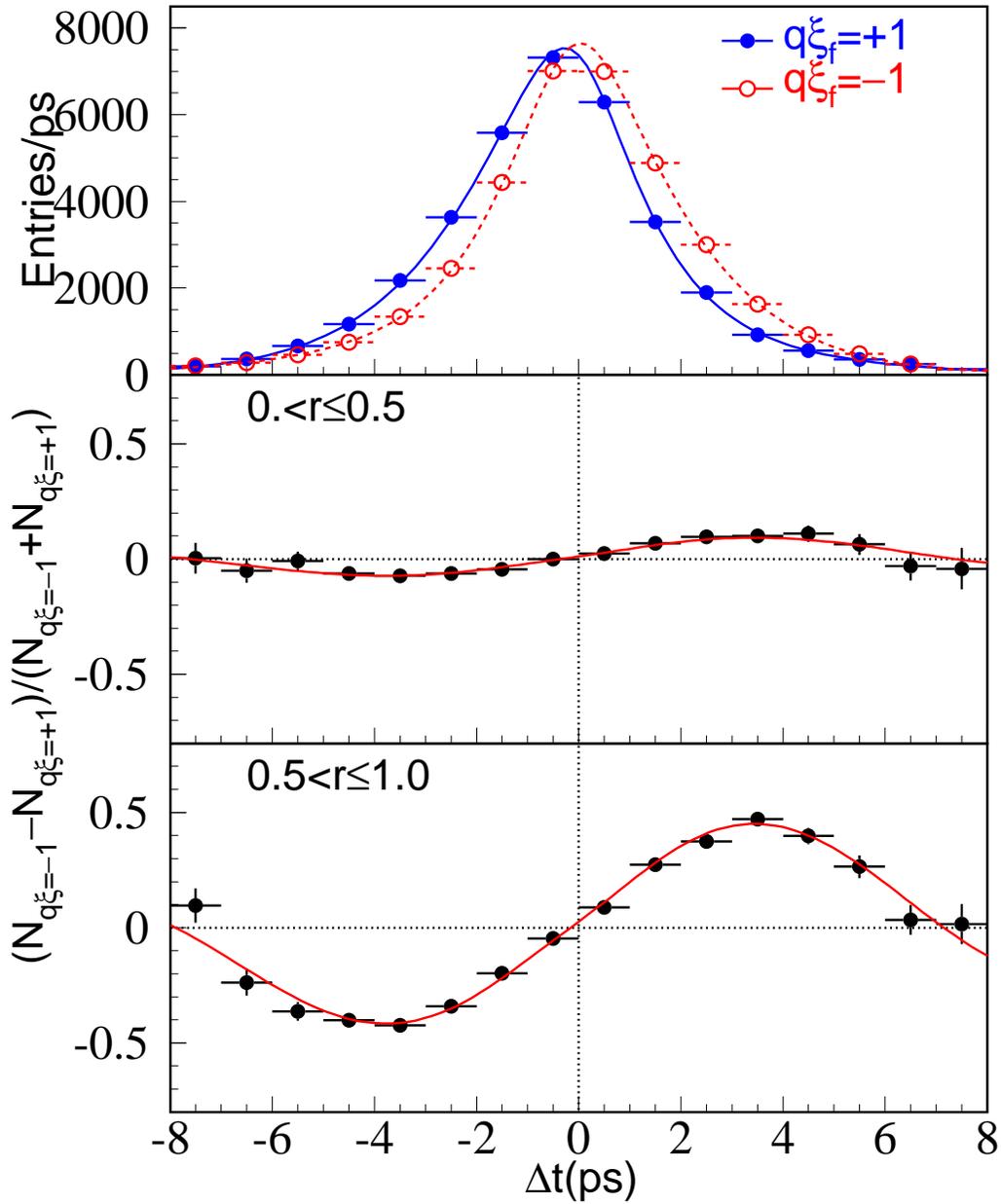}
\end{center}
\caption{An example of a fit to a MC pseudo-experiment at 
5 ab$^{-1}$.}
\label{fig:sin2phi1fit}
\end{figure}

The present world average value for $\sinbb$ is obtained
with the modes $\bz\to\jpsi \kl$, $\jpsi K^{*0}$,
$\psi(2S)\ks$, $\chi_{c1}\ks$ and $\eta_c \ks$
in addition to the $\bz\to\jpsi\ks$ decay.
This is to reduce statistical uncertainties.
With an integrated luminosity of 5 ab$^{-1}$, however,
the systematic uncertainties will be dominant.
Therefore, in this study we use only the gold-plated mode
$\bz\to\jpsi(\to \ell^+\ell^-)\ks(\pi^+\pi^-)$
to minimize systematic uncertainties.
We perform MC pseudo-experiments assuming
that the performance of the detector at SuperKEKB is
identical to that of the present Belle detector.
The analysis procedure for the measurements of time-dependent $CP$
asymmetries at Belle is described in Section~\ref{sec:overview:icpv}.
In addition to the standard 2-parameter fit,
we also test a 1-parameter fit with $\sinbb$ as the free parameter
assuming $\ajpsiks = 0$.
Because of the additional assumption, the error
on $\sinbb$ is slightly smaller than that for
$\sjpsiks$. In the following, we treat these two cases separately.
An example of a fit to a $\Delta t$ distribution for
$\bz\to\jpsi\ks$ candidates in a MC pseudo-experiment 
at 5 ab$^{-1}$ is shown in Fig.~\ref{fig:sin2phi1fit}.

Sources of systematic errors include uncertainties
in the flavor tagging,
in the vertex reconstruction,
in the background fractions and $\Dt$ distributions,
in the resolution function,
in $\dM$ and $\taubz$,
a possible bias in the fit,
and the effect of interference~\cite{Long:2003wq} 
in the $f_{\rm tag}$ final state.
Some of these uncertainties are evaluated from
control samples, which have large but finite statistics.
As the integrated luminosity increases, this part of the
systematic error will decrease.
In order to estimate the expected systematic error at
5 ab$^{-1}$, we therefore need to separate such {\it reducible}
systematic errors from the other part, which is {\it irreducible}.
In this study, we conservatively assume that
uncertainties that do not arise from statistics of control samples
are irreducible, and use estimates obtained for the 140 fb$^{-1}$
data for any integrated luminosity.
Further studies on these ``{\it irreducible}'' errors 
will probably find a way to reduce them. 
Table~\ref{tbl:sin2phi1_syserr} and \ref{tbl:s_a_jpsiks_syserr}
summarize the sources of systematic errors for $\sinbb$
(1-parameter fit) and $\cals$ and $\cala$ (2-parameter fit),
respectively. All the values are evaluated at 140 fb$^{-1}$.

\begin{table}[tbp]
\begin{center}
\begin{tabular}{lcc}\hline
Source & Irreducible & Error of $\sin2\phi_1$ \\
\hline
Wrong tag &  & 0.007 \\
Physics parameters & & 0.002 \\
Vertexing & $\surd$ & 0.012 \\
Background fraction & & 0.006 \\
Background $|\Delta t|$ shape & & 0.001 \\
Resolution function & & 0.005 \\
Resolution parameterization & $\surd$ & 0.006 \\
Tag-side interference & $\surd$ & 0.001 \\
Possible fit bias & & 0.008 \\ \hline
Total & & 0.019 \\ \hline
\end{tabular}
\end{center}
\caption{Systematic errors for $\sin2\phi_1$ measured with the $J/\psi K_S$ mode at 140 fb$^{-1}$.}
\label{tbl:sin2phi1_syserr}
\end{table}

\begin{table}[tbp]
\begin{center}
\begin{tabular}{lccc}\hline \hline
Source & Irreducible & Error of $\mathcal{S}$ & Error of $\mathcal{A}$ \\ \hline
Wrong tag & &0.007 & 0.008 \\
Physics parameters & &0.002 & 0.001 \\
Vertexing & $\surd$ &0.012 & 0.026 \\
Background fraction & &0.006 & 0.012 \\
Background $|\Delta t|$ shape & &0.001 & 0.000 \\
Resolution function & &0.005 & 0.007 \\
Resolution parameterization & $\surd$ &0.006 & 0.002 \\
Tag-side interference & $\surd$ &0.001 & 0.027 \\
Possible fit bias & &0.008 & 0.006 \\ \hline
Total & &0.019 & 0.041 \\ \hline \hline
\end{tabular}
\end{center}
\caption{Systematic errors on $\mathcal{A}$ and $\mathcal{S}$ measured with the $J/\psi K_S$ mode at 140 fb$^{-1}$.}
\label{tbl:s_a_jpsiks_syserr}
\end{table}

The total irreducible systematic error for $\cals$ is estimated to be 0.014 (0.013) for 
$\sjpsiks$ ($\sinbb$). The dominant sources of the irreducible systematic
error are the effect of detector misalignment (0.008)
and uncertainties in the resolution function determination (0.007).
The systematic error for $\cala$ is dominated by the tag-side
interference. As mentioned above, there will be a possibility to
reduce these ``{\it irreducible}'' errors from dedicated studies.
Table~\ref{tbl:sin2phi1err} lists the
expected errors at 140 fb$^{-1}$, 5 ab$^{-1}$ and 50 ab$^{-1}$.

\begin{table}[tbp]
\begin{center}
\begin{tabular}{|ll|c|c|c|c|}
\hline
           &  & Statistical & \multicolumn{2}{|c|}{Systematic}& Total\\
           &  &      & reducible & irreducible   &      \\
\hline
$\sinbb$   & (140 fb$^{-1}$) & 0.080 & 0.014 &       & 0.082\\
           & (5 ab$^{-1})$   & 0.013 & 0.002 & 0.013 & 0.019\\
           & (50 ab$^{-1})$  & 0.004 & 0.001 &       & 0.014\\
\hline
$\sjpsiks$ & (140 fb$^{-1}$) & 0.080 & 0.014 &       & 0.082\\
           & (5 ab$^{-1})$   & 0.013 & 0.002 & 0.014 & 0.019\\
           & (50 ab$^{-1})$  & 0.004 & 0.001 &       & 0.015\\
\hline
$\ajpsiks$ & (140 fb$^{-1}$) & 0.056 & 0.017 &       & 0.070\\
           & (5 ab$^{-1})$   & 0.009 & 0.003 & 0.038 & 0.039\\
           & (50 ab$^{-1})$  & 0.003 & 0.001 &       & 0.038\\
\hline
\end{tabular}
\end{center}
\caption{Expected errors at 140 fb$^{-1}$, 5 ab$^{-1}$ and 50 ab$^{-1}$.}
\label{tbl:sin2phi1err}
\end{table}

The total error for $\sjpsiks$, $\errtot(\sjpsiks)$, is obtained from
\begin{equation}
\errtot(\sjpsiks) = \sqrt{0.080^2 \times 0.14/\lint + 0.014^2 \times 0.14/\lint + 0.014^2}, 
\end{equation}
where $\lint$ is the integrated luminosity in the unit of ab$^{-1}$.
We obtain $\errtot(\sjpsiks) = 0.019$ at $\lint = 5$ ab$^{-1}$, which
is much smaller than the statistical uncertainties for
$\sphiks$ and $\setapks$.



\clearpage \newpage
\section{$\phi_2$}
\label{sec:phi_2}

\newcommand{\spipi}{{\cal S}_{\pi\pi}}
\newcommand{\apipi}{{\cal A}_{\pi\pi}}
\newcommand{\DCRP}{\Delta{C}_{\rp}}
\newcommand{\DSRP}{\Delta{S}_{\rp}}
\newcommand{\Bzbarrp}{{\Bzbar_{\rp}}}
\newcommand{\Bzrp}{\Bz_{\rp}}
\newcommand{\rto}{\rightarrow}
\newcommand{\ArpCP}{A^{\rp}_{CP}}
\newcommand{\Crp}{C_{\rp}}
\newcommand{\Srp}{S_{\rp}}
\newcommand{\Crh}{C_{\rh}}
\newcommand{\Srh}{S_{\rh}}
\newcommand{\ArhCP}{A^{\rh}_{CP}}
\newcommand{\ArKCP}{A^{\rho K}_{CP}}
\def\dm{\Delta m_d}
\newcommand{\dt}{\Delta t}
\newcommand{\Bz}{B^0}
\newcommand{\Bzbar}{\bar{B}^0}
\newcommand{\rp}{\rho\pi}
\newcommand{\rh}{\rho h}
\newcommand{\rppm}{\rho^+\pi^-}
\newcommand{\rmpp}{\rho^-\pi^+}
\newcommand{\rzpz}{\rho^0\pi^0}
\newcommand{\Tbz}{\tau_{B^0}}
\newcommand{\BBmix}{\Bz{-}\Bzbar}
\newcommand{\lampipi}{\lambda_{\pi\pi}}
\newcommand{\Apm}{A_{+-}}
\newcommand{\Amp}{A_{-+}}
\newcommand{\CrK}{C_{\rho{K}}}
\newcommand{\dCrK}{\Delta\CrK}
\newcommand{\SrK}{S_{\rho{K}}}
\newcommand{\dSrK}{\Delta\SrK}
\newcommand{\sTb}{\sin{2\beta}}
\newcommand{\sTfO}{\sin{2\phi_1}}
\newcommand{\sTa}{\sin{2\alpha}}
\newcommand{\sTfT}{\sin{2\phi_2}}
\newcommand{\Btopippim}{\Bz\rto\pi^+\pi^-}
\newcommand{\Btorp}{\Bz\rto\rho\pi}
\newcommand{\cpipi}{{\cal C}_{\pi\pi}}
\newcommand{\qq}{q\bar{q}}
\newcommand{\EBcms}{E^{cms}_B}
\newcommand{\Ebeamcms}{E^{cms}_{beam}}
\newcommand{\pBcms}{p^{cms}_B}
\newcommand{\rpipi}{{\cal R}_{\pi\pi}}
\newcommand{\ifb}{~fb^{-1}}
\newcommand{\iab}{~ab^{-1}}
\newcommand{\apizpiz}{{\cal A}_{\pi^0\pi^0}}


The CKM angle $\phi_2$ can be measured using multi-pion
final states coming from the quark-level decay 
$b\rightarrow u\bar{u}d$.
As discussed in Section~\ref{sec:Time-dependent_asymmetries}
the determination through the time-dependent asymmetry of
$B^0\rightarrow\pi^+\pi^-$ suffers from large penguin
contributions.
Here we consider two methods to eliminate an effect of the penguin
amplitude:
an isospin analysis of $B\rightarrow\pi\pi$ 
\cite{Gronau:1990ka},
and  Dalitz analysis of $B\rightarrow\rho\pi$ 
\cite{Snyder:1993mx}.

\subsection{Status of the $B^0\rightarrow\pi^+\pi^-$ analysis}
Time-dependent $CP$ asymmetries defined in
(\ref{eq:time-dependent_asymmetry}) for
$B^0\rightarrow\pi^+\pi^-$ decays have been measured by the
Belle~\cite{Abe:2003ja,Abe:2004ja} and BaBar~\cite{Aubert:2002jb}
collaborations. 
Based on $152\times 10^6$ \cite{Abe:2003ja} and 
$88\times 10^6$ \cite{Aubert:2002jb} $B\bar{B}$ pairs, 
they obtain
\begin{eqnarray}
  \mathcal{A}_{\pi\pi}
  = +0.58\pm{0.15}\pm{0.07}, &\!\!\!\! & \!\!\!\!
  \mathcal{S}_{\pi\pi}
  = -1.00\pm{0.21}\pm{0.07}
  \;\;\;\; \mbox{\cite{Abe:2004ja}},
  \\
  \mathcal{A}_{\pi\pi}
  = +0.30\pm{0.25}\pm{0.04}, &\!\!\!\! & \!\!\!\!
  \mathcal{S}_{\pi\pi}
  = -0.02\pm{0.34}\pm{0.05}
  \;\;\;\; \mbox{\cite{Aubert:2002jb}}.
\end{eqnarray}
The first and the second errors are statistical and
systematic errors, respectively.
BaBar also showed a preliminary result at Lepton Photon 2003
based on 123 million $B\bar{B}$ pairs; they obtain
$\apipi = +0.19\pm{0.19}\pm{0.05}$ and
$\spipi = -0.40\pm{0.22}\pm{0.03}$.
The world average values using the latest results are
$\apipi = +0.46 \pm{0.13}$ and $\spipi = -0.74 \pm{0.16}$ 
\cite{HFAG}.

From the latest Belle result~\cite{Abe:2004ja}, 
the case that $CP$ symmetry is conserved, 
$\mathcal{A}_{\pi\pi} = \mathcal{S}_{\pi\pi} = 0$, is ruled
out at a level of 5.2 standard deviations. 
Thus this is the first observation of $CP$-violating 
asymmetries in $B^0 \to \pi^+\pi^-$ decays.
A 95.5\% CL region of $\mathcal{A}_{\pi\pi}$ and
$\mathcal{S}_{\pi\pi}$ gives 
$90^\circ\leq\phi_2\leq 146^\circ$ with a modest assumption
of $|P_{\pi\pi}/T_{\pi\pi}| < 0.45$.
From the theoretical side, QCD factorization 
gives $-6\pm 12\%$ for the direct $CP$ asymmetry
$\mathcal{A}_{\pi\pi}$ \cite{Beneke:2001ev}, 
while perturbative QCD (pQCD) suggests a larger
direct asymmetry in the range (16$-$30\%) \cite{Keum:2002vi}.

\subsection{Isospin analysis for $B\rightarrow\pi\pi$}
Here we describe the expected sensitivity 
for the determination of $\phi_2$ at
SuperKEKB  using the $\pi\pi$ isospin analysis.

Estimated errors for the branching ratios and asymmetries 
at the target luminosities are shown in
Table~\ref{tbl:input_param}.
We assume the current world average \cite{HFAG} for
the central values, and also assume that the direct $CP$
violation is absent for the $B^0\rightarrow\pi^0\pi^0$ 
decay \textit{i.e.}, $\mathcal{A}_{\pi^0\pi^0} = 0.0$.
In the following analysis we use the CKMfitter program
\cite{CKMfitter}. 

\begin{table}[tbp]
  \begin{center}
    \begin{tabular}{c|c|c|c|c|c}
      \hline
      & central value & error at & error at & error at & error at \\
      & WA from \cite{HFAG} & 140~fb$^{-1}$ & 500~fb$^{-1}$ 
      & 5~ab$^{-1}$ & 50~ab$^{-1}$ \\
      \hline
      $BR(\pi^+\pi^-)$ & 4.55$\pm$0.44 & 0.50 & 0.25 & 0.082 & 0.026 \\
      $BR(\pi^+\pi^0)$ & 5.27$\pm$0.79 & 0.88 & 0.47 & 0.15  & 0.47  \\
      $BR(\pi^0\pi^0)$ & 1.90$\pm$0.47 & 0.65 & 0.34 & 0.11  & 0.034 \\
      $\mathcal{S}_{\pi\pi}$ & $-0.58\pm0.20$ & 0.23 & 0.12 & 0.039 & 0.012 \\
      $\mathcal{A}_{\pi\pi}$ & $+0.38\pm0.16$ & 0.19 & 0.10 & 0.031 & 0.010 \\
      $\mathcal{A}_{\pi^0\pi^0}$ & 0.0 & 1.00 & 0.53 & 0.17  & 0.053 \\
      \hline
    \end{tabular}
  \end{center}
  \caption{
    Estimated errors for the $\pi\pi$ branching ratios and
    asymmetries.
  }
  \label{tbl:input_param}
\end{table}

From the $\pi^+\pi^-$ time-dependent asymmetry we may
extract the parameters $\mathcal{S}_{\pi\pi}$ and 
$\mathcal{A}_{\pi\pi}$.
The phase of the parameter 
$\lambda_{\pi\pi}=(q/p)(\bar{A}_{\pi\pi}/A_{\pi\pi})$
is extracted as
\begin{equation}
  \sin 2\phi_{2\mathrm{eff}} = 
  \frac{\mathcal{S}_{\pi\pi}}{\sqrt{1-\mathcal{A}_{\pi\pi}^2}},
\end{equation}
which is equal to $\sin 2\phi_2$ if the decay is dominated
by the tree amplitude.
In the presence of a penguin contribution, we have to subtract
the penguin amplitude using isospin relations 
(\ref{eq:isospin_relation_1}) and
(\ref{eq:isospin_relation_2}).

\begin{figure}[tbp]
  \begin{center}
    \includegraphics[width=0.8\textwidth,height=0.4\textheight,clip]{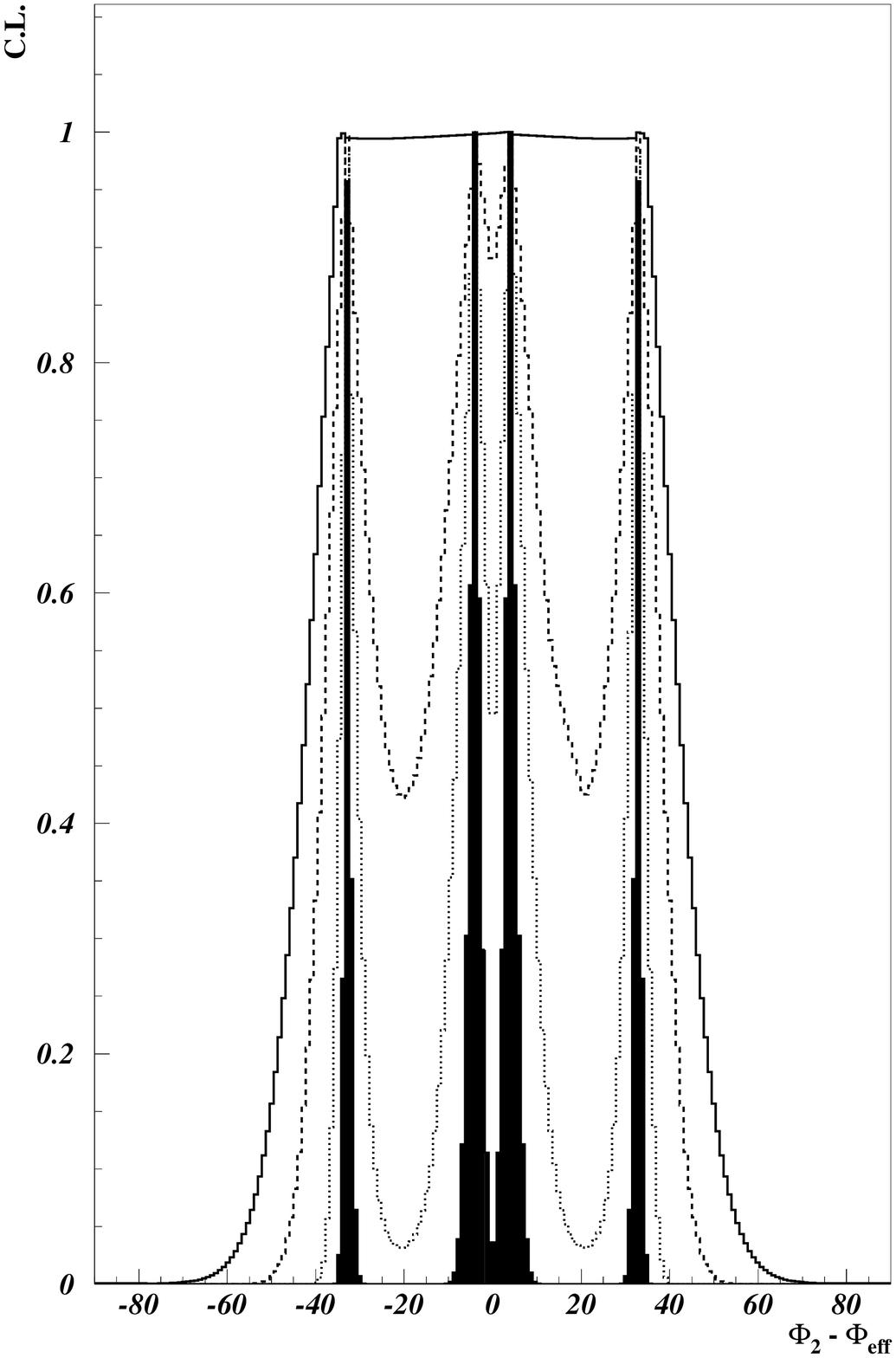}
    \caption{
      CL versus $\phi_2-\phi_{2\mathrm{eff}}$~(deg) for the
      inputs given in Table~\ref{tbl:input_param} at 
      140~fb$^{-1}$ (solid curve), 500~fb$^{-1}$ (dashed curve),
      5~ab$^{-1}$ (dotted curve), and 50~ab$^{-1}$ (shaded area).
    }
    \label{fig:phi2-phi2eff-vs-lum}
  \end{center}
\end{figure}
\begin{figure}[tbp]
  \begin{center}
    \includegraphics[width=0.8\textwidth,height=0.4\textheight,clip]{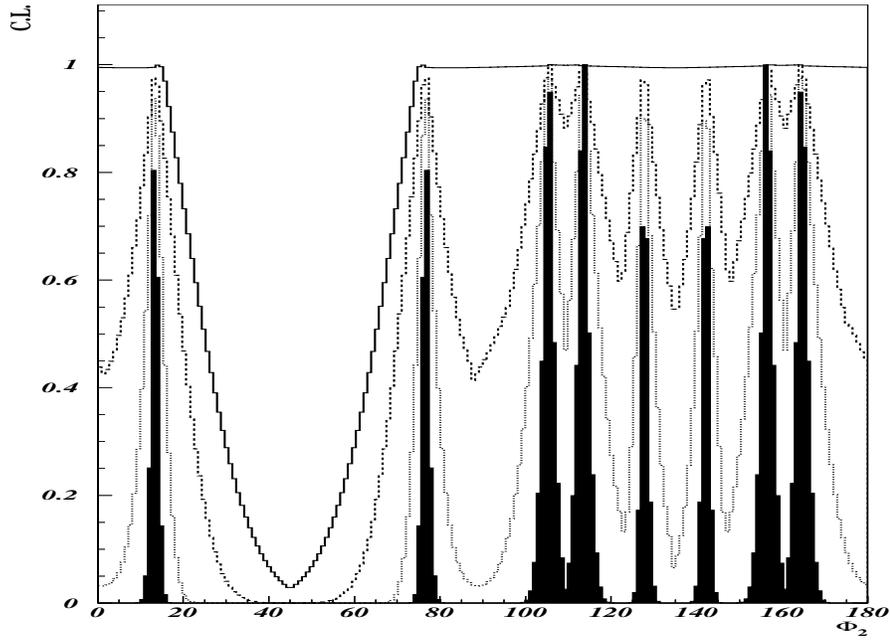}

    \caption{
      CL versus $\phi_2$ (deg) for the inputs given in
      Table~\ref{tbl:input_param} at 
      140~fb$^{-1}$ (solid curve), 500~fb$^{-1}$ (dashed curve),
      5~ab$^{-1}$ (dotted curve), and 50~ab$^{-1}$ (shaded area).
    }
    \label{fig:phi2-vs-lum}
  \end{center}
\end{figure}

The results for $\phi_2-\phi_{2\mathrm{eff}}$ 
and $\phi_2$ are plotted in 
Figures~\ref{fig:phi2-phi2eff-vs-lum}
and \ref{fig:phi2-vs-lum}, respectively.
In those figures the confidence level is plotted at target
luminosities up to 50~ab$^{-1}$.
The absence of direct $CP$ violation for
the $B^0\rightarrow\pi^0\pi^0$ decay leads to the symmetric
solutions shown in the figures.

When  flavor tagging for $B^0\rightarrow\pi^0\pi^0$ is
missing as in the 140~fb$^{-1}$ case, only the outer borders
of the curves can be obtained. 
By knowing the flavor of $B^0\rightarrow\pi^0\pi^0$
the inner structure shows up, and 
$\phi_2-\phi_{2\mathrm{eff}}$ is determined up to a
four-fold ambiguity
and $\phi_2$ is determined up to an eight-fold ambiguity
in the range ($0^\circ,180^\circ$).

\begin{figure}[p]
  \begin{center}
\includegraphics[width=0.8\textwidth,height=0.4\textheight,clip]{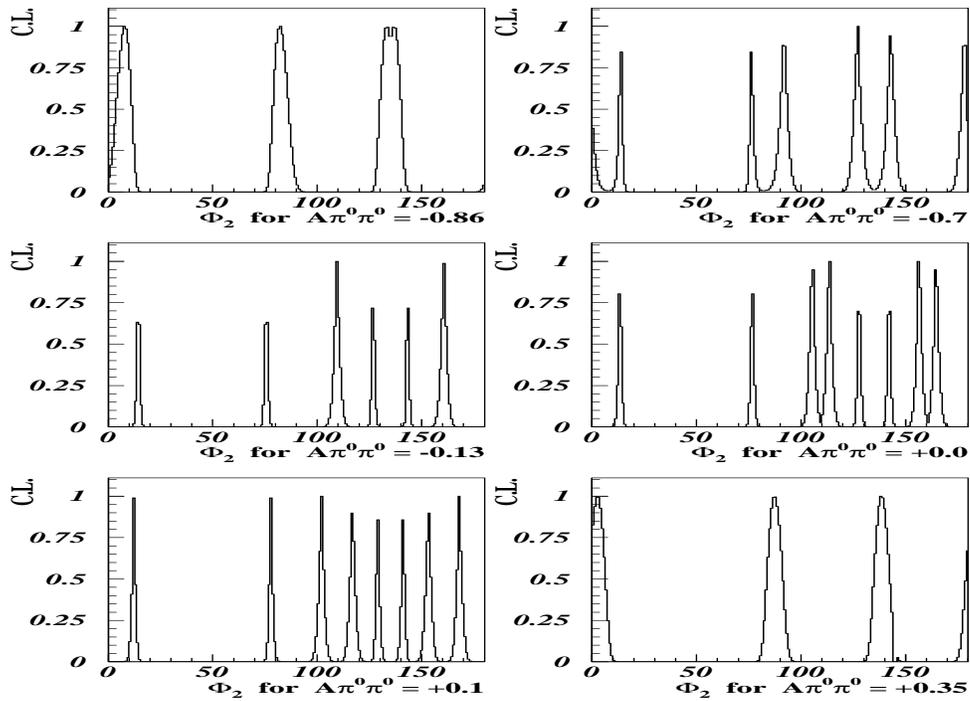}
    \caption{
      CL versus $\phi_2$~(deg) at 50~ab$^{-1}$  
      for $\mathcal{A}_{\pi^0\pi^0}$ = $-0.86$ (top left), 
      $-0.7$ (top right), $-0.13$ (middle left), 
      $0.0$ (middle right), $+0.1$ (bottom left), 
      and $+0.35$ (bottom right).
    }
    \label{fig:phi2-vs-lum-apz}
  \end{center}
\end{figure}

Although we have assumed $\mathcal{A}_{\pi^0\pi^0} = 0$ 
in the estimation above,
theoretical predictions for $\mathcal{A}_{\pi^0\pi^0}$ 
allow large values~\cite{Lu:2001}.
Figure~\ref{fig:phi2-vs-lum-apz} shows CL as a function of
$\phi_2$ at 50~ab$^{-1}$ for several values of
$\mathcal{A}_{\pi^0\pi^0}$.
The best $\phi_2$ resolution ($0.8^\circ \sim 1.4^\circ$) is
expected at $\mathcal{A}_{\pi^0\pi^0}=-0.13$, for which the
angle between $A^{+0}$ and $A^{+-}$ is the same as the angle
between $\bar{A}^{+0}$ and $\bar{A}^{+-}$. 
At $\mathcal{A}_{\pi^0\pi^0}=-0.86$ and +0.35, on the other
hand, one isospin triangle is squashed, and the expected
$\phi_2$ resolution is about $2.6^\circ$ except for the region
where two solutions overlap.

Since the decay $B^+\rightarrow\pi^+\pi^0$ has no
strong penguin contribution, the occurrence of the direct
$CP$ violation in this decay mode would imply an electroweak
penguin contribution. 
The effect of the electroweak penguin to $\pi\pi$ amplitudes 
is estimated in Table~\ref{tbl:ewp-pipi} using the
perturbative QCD calculation with and without the electroweak
penguin (EWP) amplitude.
This calculation suggests that the effect on $\pi^+\pi^-$
and $\pi^{\pm}\pi^0$ is negligible, while the effect 
on $\pi^0\pi^0$ is at a level of
several percent.

\begin{table}
  \begin{center}
    \begin{tabular}{cccc} \hline
      amplitudes & 
      $|A^{+0}|~(\times 10^{-4})$ & 
      $|A^{+-}|~(\times 10^{-4})$ & 
      $|A^{00}|~(\times 10^{-4})$ \\
      \hline
      with EWP   & 1.68 & 2.30 & 0.82 \\
      w/o EWP    & 1.69 & 2.31 & 0.87 \\
      \hline
      difference & 0.5\% & 0.5\% & 6.5\%\\
      \hline\hline
      amplitudes & 
      $|\bar{A}^{+0}|~(\times 10^{-4})$ & 
      $|\bar{A}^{+-}|~(\times 10^{-4})$ &
      $|\bar{A}^{00}|~(\times 10^{-4})$ \\
      \hline
      with EWP   & 1.68 & 2.90 & 0.96 \\
      w/o EWP    & 1.69 & 2.90 & 1.01 \\
      \hline
      difference & 0.5\% & 0.0\% & 5.5\%\\
      \hline
    \end{tabular}
  \end{center}
  \caption{A perturbative QCD calculation of the
      $B\rightarrow\pi\pi$ amplitudes with and without
      electroweak penguin contributions.
    }
  \label{tbl:ewp-pipi}
\end{table}

\subsection{Status of the $B^0\rightarrow\rho\pi$ analysis}
In principle, the CKM angle $\phi_2$ can be measured even in
the presence of penguin contributions, using a full Dalitz
plot analysis. 
However, there are difficulties from combinatorics 
and from the low
efficiency in a three-body topology with a $\pi^0$ as well as large
backgrounds from mis-reconstructed signal events and other 
decays. 
In order to extract $\phi_2$ cleanly, data with large
statistics are thus required.

Unlike the $B^0\rightarrow\pi^+\pi^-$ decay, 
$B^0\rightarrow\rho^\pm\pi^\mp$ is not a $CP$
eigenstate.
In $B^0\rightarrow\rho^\pm\pi^\mp$ decay, four different
flavor decays ($B^0(\bar{B}^0)\rightarrow\rho^\pm\pi^\mp$)
must be considered.
Following a quasi-two-body approach, 
the current analysis by the BaBar collaboration
\cite{Aubert:2003wr} is restricted to the two regions of the 
$\pi^\pm\pi^0{h^\pm}$ Dalitz plot ($h$ = $\pi$ or $K$) that
are dominated by $\rho^\pm{h^\mp}$. 
The decay rate is given by
\begin{eqnarray}
  \lefteqn{
    f^{\rho^{\pm}h^{\mp}}_{q}(\dt)
    =
    (1\pm\ArhCP)
    \frac{e^{-|\Delta t|/{\Tbz}}}{4{\Tbz}} 
  } 
  \nonumber \\
  &&
  \times
  [1+q{\cdot}\{(\Srh\pm\Delta\Srh) \sin (\dm\dt)
  -(\Crh\pm\Delta\Crh) \cos (\dm\dt)\}],
  \nonumber\\
\end{eqnarray}
where $\Delta t=t_{\rho h}-t_{\mathrm{tag}}$ is the time
interval between the decay of $B^0_{\rho h}$ and that of the
other $B^0$ meson. 
One finds a relation 
\begin{equation}
  \Srp\pm\Delta\Srp = 
  \sqrt{1-(\Crp\pm\Delta\Crp)^2}\sin(2\phi^\pm_{\rm 2eff}\pm\delta),
\end{equation}
where 
$2\phi^\pm_{2\mathrm{eff}}=
\arg[(q/p)(\bar{A}^\pm_{\rho\pi}/A^\mp_{\rho\pi})]$ and
$\delta$ = arg[$A^-_{\rho\pi}/A^+_{\rho\pi}$].
arg[$q/p$] is the $B^0-\bar{B}^0$ mixing phase, and
$A^+_{\rho\pi}(\bar{A}^+_{\rho\pi})$ and
$A^-_{\rho\pi}(\bar{A}^-_{\rho\pi})$ are the transition
amplitudes for the processes 
$B^0(\bar{B}^0)\rightarrow\rho^+\pi^-$ and
$B^0(\bar{B}^0)\rightarrow\rho^-\pi^+$, respectively.
The angles $\phi^\pm_{2\mathrm{eff}}$ are equal to $\phi_2$ 
if contributions from penguin amplitudes are absent.

With a data sample of $89\times 10^6$ $B\bar{B}$ pairs
the BaBar Collaboration obtained
\cite{Aubert:2003wr}
\begin{eqnarray}
  A_{CP}^{\rho\pi} &=& -0.18 \pm 0.08 \pm 0.03,
  \\
  C_{\rho\pi} &=& +0.36 \pm 0.18 \pm 0.04,
  \quad
  S_{\rho\pi} \;=\; +0.19 \pm 0.24 \pm 0.03,
  \\
  \Delta C_{\rho\pi} &=& +0.28 \pm 0.19 \pm 0.04,
  \quad
  \Delta S_{\rho\pi} \;=\; +0.15 \pm 0.25 \pm 0.03.
\end{eqnarray}

\subsection{Dalitz plot analysis of $B^0\rightarrow\rho\pi$}
A measurement of $\phi_2$ using isospin relations among
$B\rightarrow\pi\pi$ decays has a four-fold ambiguity as
we mentioned in
Section~\ref{sec:Time-dependent_asymmetries}.
This ambiguity can be avoided with a full Dalitz plot
analysis of the 
$B^0\rightarrow\rho\pi\rightarrow\pi^+\pi^-\pi^0$ decay
\cite{Snyder:1993mx}.

Using the notation of \cite{Lipkin:1991st}, the decay
amplitudes of $B\rightarrow\rho\pi$ are expressed as 
\begin{eqnarray}
  \sqrt{2}A(B^+\rightarrow\rho^+ \pi^0)
  (\equiv S_1) & = & T^{+0}+2P_1,
  \\
  \sqrt{2}A(B^+\rightarrow\rho^0 \pi^+)
  (\equiv S_2) & = & T^{0+}-2P_1,
  \\
  A(B^0\rightarrow\rho^+ \pi^-)
  (\equiv S_3) & = & T^{+-}+P_1+P_0,
  \\
  A(B^0\rightarrow\rho^- \pi^+)
  (\equiv S_4) & = & T^{-+}-P_1+P_0,
  \\
  A(B^0\to \rho^0 \pi^0)
  (\equiv S_5) & = & T^{+0}+T^{0+}-T^{+-}-T^{-+}-2P_0.
\end{eqnarray}
Here $T^{ij}$ ($i$, $j$ = $+$, $-$, or 0) are the tree
amplitudes, and $P_0$ and $P_1$ are the penguin amplitudes
for $I = 0$ and 1 final states, respectively.
Similarly, for the $CP$-conjugate channels, we define the
amplitudes $\bar{S}$, $\bar{T}^{ij}$, and $\bar{P}_i$ which
differ from the original amplitudes only in the sign of the
weak phase of each term. 
From isospin constraints, the following relation is held:
\begin{eqnarray}
 S_1 + S_2 &=& S_3 + S_4 + S_5,\\
 \bar{S}_1 + \bar{S}_2 &=& \bar{S}_3 + \bar{S}_4 + \bar{S}_5.
\end{eqnarray}

In the full Dalitz plot analysis for the $\pi^+\pi^-\pi^0$
final states, we do not specify which intermediate state the
$\pi^+\pi^-\pi^0$ final state comes from.
Thus, we can see the quantum interference.
The amplitude of the $B^0\rightarrow\pi^+\pi^-\pi^0$ decay is
expressed as 
\begin{equation}
  A(f) = f^+S_{3}+f^-S_{4}+f^0S_{5}/2,
\end{equation}
while the amplitude of the $CP$-conjugate channel is given
by 
\begin{equation}
  \bar{A}(f) = f^-\bar{S}_{3}+f^+\bar{S}_{4}+f^0\bar{S}_{5}/2.
\end{equation}
$f^i$ is the Breit-Wigner kinematical distribution function
for $\rho^{\pm,0}$ 
\begin{equation}
  f(m, \theta) = 
  \frac{\cos\theta\,\Gamma_\rho/2}{m_\rho-m-i\Gamma_\rho/2},
\end{equation}
where $m_\rho$ and $\Gamma_\rho$ are the mass and width of
$\rho$ meson, respectively.
The decay rate is given as a function of $\Delta t$,
the invariant mass $m_i$ and helicity angle $\theta_i$ of $\rho$
by 
\begin{eqnarray}
  \lefteqn{\mathcal{P}_{sig}(
    \Delta t, m_+, m_-, m_0, \theta_+, \theta_-, \theta_0 )
    = \frac{e^{-|\Delta t|/\tau_{B^0}}}{4\tau_{B^0}} 
    \Biggl[
    (|\bar{A}|^2+|A|^2)
  } \nonumber \\
  && 
  + q\left(
    (\bar{|A|}^2-|A|^2) \cos(\Delta m \Delta t)
    +2\mathrm{Im}(\frac{\bar{A}}{A}) \sin(\Delta m \Delta t)
  \right)
  \Biggr]
\end{eqnarray}
where $q=1$ for $B_{tag}=B^0$ and $q=-1$ for
$B_{tag}=\bar{B}^0$.
To clarify fit parameters, we denote $T^{ij}$ and $P_0$ as
follows: 
\begin{eqnarray}
  &&
  T=T^{+0}+T^{0+}=|T|e^{i\phi_3}e^{i\delta_{T}},\\
  &&
  \bar{T}=\bar{T}^{+0}+\bar{T}^{0+}=|T|e^{-i\phi_3}e^{i\delta_{T}},\\
  &&
  T^{+-}=|T^{+-}|e^{i\phi_3}e^{i\delta_{+-}},\;
  \bar{T}^{+-}=|T^{+-}|e^{-i\phi_3}e^{i\delta_{+-}},\\
  &&
  T^{-+}=|T^{-+}|e^{i\phi_3}e^{i\delta_{-+}},\;
  \bar{T}^{-+}=|T^{-+}|e^{-i\phi_3}e^{i\delta_{-+}},\\
  &&
  P_{0}=|P_{0}|e^{-i\phi_1}e^{i\delta_{0}},\;
  \bar{P}^{0}=|P_{0}|e^{i\phi_1}e^{i\delta_{0}},\\
  &&P_{1}=|P_{1}|e^{-i\phi_1}e^{i\delta_{1}},\;
  \bar{P}^{1}=|P_{1}|e^{i\phi_1}e^{i\delta_{1}}.
\end{eqnarray}
We have 9 fit parameters
\begin{equation}
  \phi_2,\; |T|,\; \delta_{T},\; |T^{-+}|,\; \delta_{-+},\;
  |P_0|,\; \delta_0,\; |P_1|,\; \delta_1
\end{equation}
and set $|T^{+-}|=1$ and $\delta_{+-}=0$.

We use  ensembles of the Monte Carlo (MC)
pseudo-experiments for this study. 
Each pseudo-experiment consists of events generated with the 
nominal probability density functions (PDFs).
At present, we consider only continuum events as a source of
backgrounds.
Assuming $BR(B^0\rightarrow\rho\pi) = 25\times10^{-6}$ and
the detection efficiency of 15\%, we generate 10080
candidate events equivalent to a 300fb$^{-1}$ data sample. 
There are about 1240 signal event in this sample.
We parametrize the continuum background as
\begin{eqnarray}
  \mathcal{P}_{q\bar{q}}
  =
  \frac{1}{2} {\cal M}(m_{+}, m_{-}, m_{0},
  \theta_{+}, \theta_{-}, \theta_{0})
  \Bigl\{f_\tau \frac{ e^{-|\Delta t|/\tau_{\rm bkg}} }{2\tau_{\rm bkg}}
  + ( 1 - f_\tau )\delta(\Delta t)\Bigl\},
  \label{phi2:eq:cont}
\end{eqnarray} 
where $f_\tau$ is the fraction of the background with
an effective lifetime $\tau_{\mathrm{bkg}}$, and $\mathcal{M}$
is a PDF in the Dalitz plane for the continuum background.
We define a likelihood value for each event as a function
of the 9 parameters:
\begin{eqnarray}
  \lefteqn{
    P_i(\Delta t_i,  m_{+}, m_{-}, m_{0},
    \theta_{+}, \theta_{-}, \theta_{0}) = 
    (1-f_{ol}) 
    \int^{+\infty}_{-\infty}\!\! d\Delta t^\prime 
  }
  \nonumber\\
  \!\!\!\!\!\!\!\! & \!\!\! & \!\!\!\!\!\!
  \{
  f_{sig} \mathcal{P}_{sig}(\Delta t^\prime, q, w_l)
  + f_{q\bar{q}} \mathcal{P}_{q\bar{q}}(\Delta t^\prime)
  \cdot R_{q\bar{q}}(\Delta t_i-\Delta t^\prime)
  \} 
  + f_{ol}{\cal P}_{ol}(\Delta{t_i}),
  \nonumber\\
  \label{eq:Pevent}
\end{eqnarray}
where $f_{sig}$ and $f_{q\bar{q}}$ are the probability
functions for signal events and the continuum background,
respectively.
They are determined on an event-by-event basis.
The signal fraction is estimated using the data 
taken by the Belle detector by the summer of 2002.
For signal events, we use the same values of resolution
parameters as those used for the $\sin2\phi_1$ analysis.
We estimate the parameters for the continuum background using
sideband data. 

\begin{table}
 \begin{center}
  \begin{tabular}{c|r||c|r}\hline \hline
   $\phi_2$ & \multicolumn{3}{l}{$1.57$  $(90 \deg)$} \\ \hline
   $|T|$ & $1.59$ & $\delta_T$ & $-0.65$ \\ \hline
   $|T^{+-}|$ & $1$ & $\delta_{T_{+-}}$ & $0$ \\ \hline
   $|T^{-+}|$ & $0.78$ & $\delta_{T_{-+}}$ & $-2.91$ \\ \hline
   $|P_{0}|$ & $0.19$ & $\delta_{P_{0}}$ & $-0.61$ \\ \hline
   $|P_{1}|$ & $0.19$ & $\delta_{P_{1}}$ & $1.04$ \\ \hline
   \hline
  \end{tabular}
  \caption{
    Input parameters for a full Dalitz plot
    analysis of $B^0\rightarrow\pi^+\pi^-\pi^0$ decay.
  }
  \label{tab:rptoymc_input}
 \end{center}
\end{table}

\begin{figure}
  \begin{center}
    \includegraphics[width=0.8\textwidth,clip]{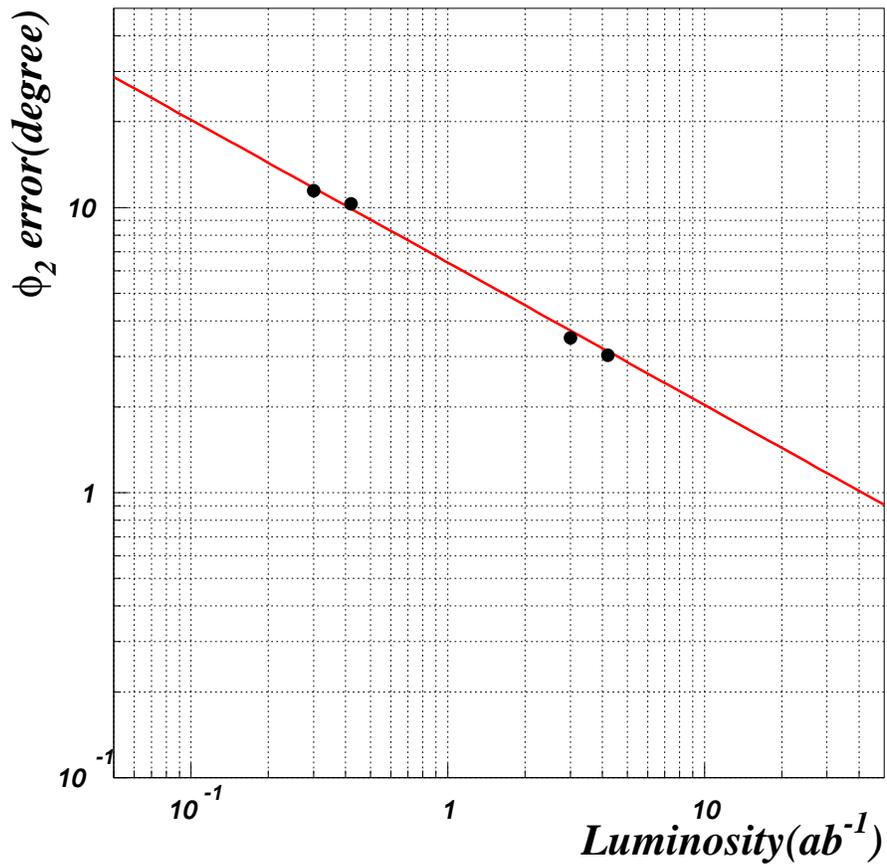}
  \end{center}
  \caption{
    $\phi_2$ error as a function of luminosity. 
    Dots are the results from the MC  pseudo-experiments and
    the line is a linear fit to these dots.
  } 
  \label{fig:phi2err_rp}
\end{figure}

The input values of the fit parameters used here are listed
in Table \ref{tab:rptoymc_input} \cite{Stark:2003nq}.
The result is shown in Figure \ref{fig:phi2err_rp}.
We obtain an error for $\phi_2$ of 
$\delta\phi_2 = 11.4^\circ$ at 300~fb$^{-1}$ and 
$\delta\phi_2 = 3.5^\circ$ at 3~ab$^{-1}$.
From these results, the error of $\phi_2$ is expected to be
2.9$^\circ$ at 5~ab$^{-1}$ and 0.9$^\circ$ at 50ab$^{-1}$.

\clearpage \newpage


\def\ovl{\overline}
\def\bra{\langle}
\def\ket{\rangle}
\def\beq{\begin{equation}}
\def\eeq{\end{equation}}
\def\beqa{\begin{eqnarray}}
\def\eeqa{\end{eqnarray}}
\def\dspl{\displaystyle}
\def\para{{\scriptstyle\|}}
\def\tbst{\vrule width0pt height15pt depth10pt}


\section{$\phi_3$}

\subsection{Introduction}

The angle $\phi_3$ is defined as
\begin{equation}
  \phi_3 \equiv - \arg \left[\frac{V_{ud}V_{ub}^*}{V_{cd}V_{cb}^*}\right],
\end{equation}
which is independent of the quark phase convention. In the standard
phase convention, all elements of the CKM matrix except for 
$V_{ub}$ and $V_{td}$ are nearly real, and in particular
both $V_{ud}$ and $-V_{cd}V_{cb}^*$ are (nearly) real and positive,
and we have
\beq
      \phi_3 \equiv \arg V_{ub}^*\,.\quad\hbox{(standard phase convention)}
\eeq
There are several decay channels that can be used to extract information
on $\phi_3$, and each has distinct merits and drawbacks. Here, we will
investigate the following channels:
\begin{description}
\item[$D^{(*)-}\pi^+$ modes]
  The flavor-tagged time-dependent measurement of $D^{(*)-}\pi^+$
  and its charge-conjugate mode.
  This mode measures $\sin(2\phi_1 + \phi_3)$ 
  and is affected by a strong
  phase. There is, however, no penguin contribution. One could
  fully reconstruct the $D^-\pi^+$ and $D^{*-}\pi^+$ final states, or
  one could use a partial-reconstruction technique where $\bar D^0$
  of the decay $D^{*-}\to \bar D^0 \pi^+$ is not explicitly detected. 
  The latter has more statistics
  but with more background. In both cases, the value of $r$, the ratio
  of Cabibbo-favored amplitude to the Cabibbo-suppressed amplitude, needs
  to be input externally.
\item[$B^+ \to D K^{(*)+}$ (ADS method)]
  $B^+ \to D K^{(*)+}$, $D\to PP$, where $D$ is $D^0$ or $\bar D^0$, and
  their charge-conjugate modes. Even though there are strong phases involved,
  $\phi_3$ can be extracted in a theoretically-clean manner 
  if one uses more than one kind of $D$ decays. 
  The required statistics, however, is quite large.
  The value of the relevant amplitude ratio $r$ can be obtained by the
  fit. If it is known, one can improve the statistical power
  significantly.
\item[$B^\pm\to D K^\pm$ (Dalitz analysis)]
  $B^+ \to D K^+$, $D\to PPP$, where $D$ is $D^0$ or $\bar D^0$, and
  their charge-conjugate modes. This mode takes advantage of the
  interferences that occur in the Dalitz plot of the $D\to PPP$
  decay to extract $\phi_3$ as well as the strong phase.
  The value of the amplitude ratio $r$ is also obtained in the fit.
  This analysis has a good statistical power; it requires, however,
  a detailed understanding of the structure of the Dalitz plot.
\end{description}
More on the theoretical frameworks for each mode as well as
the sensitivities are covered in later subsections.

\subsection{$D^{(*)-}\pi^+$ modes}

\begin{figure}[tbp]
  \begin{center}
\includegraphics[width=0.9\textwidth,clip]{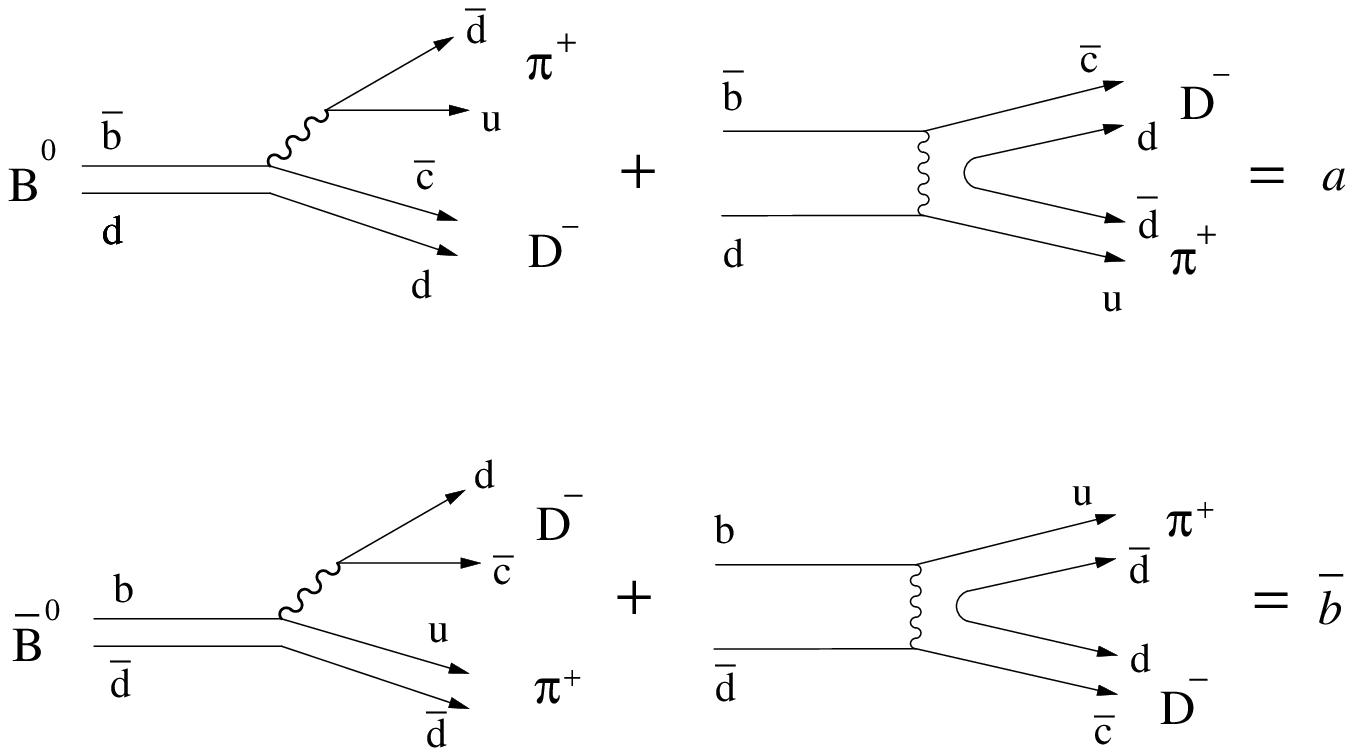}
   \caption{Diagrams for $B^0,\bar B^0\to D^-\pi^+$.}
  \end{center}
   \label{fg:bdpiamps}
\end{figure}

For the final state $D^{(*)-}\pi^+$,
the amplitudes of diagrams shown in Figure~\ref{fg:bdpiamps} interfere.
The two diagrams contributing to the amplitude $a$ (or $\bar b$) have
the same CKM factors, and information on $\phi_3$ is contained in
\beq
    \rho \equiv {q \bar b\over p a} = -r \exp(\delta-\phi_w)\,,
   \quad\phi_w\equiv 2\phi_1 - \phi_3\,,
\eeq
where $B_H = pB^0 -q\bar B^0$, $r\equiv |\rho|$ and $\delta$ is the 
relative strong phase between $\bar b$ and $a$. The time-dependent
distributions are then given by
\beq
 \begin{array}{rc@{\tbst}l}
    \Gamma_{\ell^-,D^-\pi^+}(\Delta t) &=& N
      e^{-\gamma |\Delta t|} \left[ (1+r^2) + (1-r^2)\cos\delta m\Delta t 
                         - 2r\sin(\phi_w-\delta)\,\sin\delta m\Delta t \right] \\
    \Gamma_{\ell^+, D^+\pi^-}(\Delta t) &=& N
      e^{-\gamma |\Delta t|}\left[ (1+r^2) + (1-r^2)\cos\delta m\Delta t 
                         +  2r\sin(\phi_w+\delta) \,\sin\delta m\Delta t \right] \\
    \Gamma_{\ell^-, D^+\pi^-}(\Delta t) &=& N
      e^{-\gamma |\Delta t|} \left[ (1+r^2) - (1-r^2)\cos\delta m\Delta t 
                         -  2r\sin(\phi_w+\delta) \,\sin\delta m\Delta t \right] \\
    \Gamma_{\ell^+, D^-\pi^+}(\Delta t) &=& N
      e^{-\gamma |\Delta t|} \left[ (1+r^2) - (1-r^2)\cos\delta m\Delta t 
                         +  2r\sin(\phi_w-\delta)\,\sin\delta m\Delta t \right]
 \end{array}\,.
  \label{eq:ratesdpi}
\eeq
where $\Gamma_{\ell^-,D^-\pi^+}(\Delta t)$ is the distribution of 
$\Delta t \equiv t_{D\pi} - t_{\rm tag}$ when the tag-side is $\bar B^0$, etc.,
and $D\pi$ can also be $D^*\pi$ in which case $r$ and $\delta$ will be
replaced by $r^*$ and $\delta^*$, respectively. The two relevant
observables are $r\sin(\phi_w+\delta)$ and 
$r\sin(\phi_w-\delta)$. The value of $r$
cannot be obtained by the fit itself, while the expected value of $r^{(*)}$ is
roughly 0.02. One way to obtain $r^{*}$ experimentally is to
use the $SU(3)$-related modes $B^0 \to D^{(*)-}_s\pi^+$. However,
there will be uncertainty associated with the validity of $SU(3)$ and
the size of the exchange diagram which is missing for $D^{(*)-}_s\pi^+$.

\begin{figure}[tbp]
  \begin{center}
\includegraphics[width=0.9\textwidth,clip]{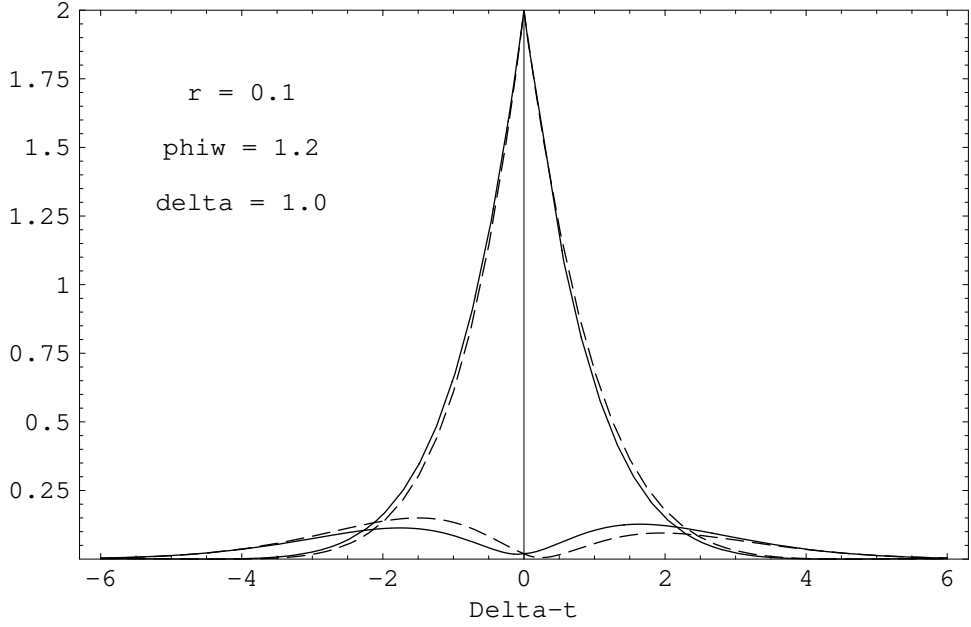}
  \caption{The $\Delta t$ distributions of the flavor-tagged $D\pi$ modes
   for $r=0.1$, $\phi_w = 1.2$ radian, and $\delta = 1.0$ radian. The solid lines are
    for $D^-\pi^+$ final state and dashed lines are for $D^+\pi^-$
    final states. The mixing parameter $x$ is taken to be 0.71.}
  \label{fg:dstpi}
  \end{center}
\end{figure}

The distributions are plotted in Figure~\ref{fg:dstpi} where the 
value of $r$ is artificially enhanced to 0.1 in order to show the $CP$ violating
effects clearly.
The top two of (\ref{eq:ratesdpi}) can be called unmixed modes and
the bottom two mixed modes. 
As may be noticeable in the figure, most of information on CP violation
is in the  mixed modes where CP violation appears as the height asymmetry
of $\Delta t>0$ vs $\Delta t<0$ and the shift of the minimum from 
$\Delta t=0$. Note that $r\sin(\phi_w+\delta)$ can be obtained from
the third distribution alone and $r\sin(\phi_w-\delta)$ from
the fourth alone. The final state $D^{*-}\pi^+$ can be detected
by full reconstruction using the standard technique, it or can also
be reconstructed by a partial reconstruction method where 
the $\bar D^0$ meson in $D^{*-}\to \bar D^0 \pi^-$ is not
explicitly reconstructed. 
Note that the $D^{(*)-}\pi^+$ methods work
even when there are sizable exchange diagrams while the value of $r^{(*)}$
needs to be supplied externally.

\begin{figure}[tbp]
  \begin{center}
\includegraphics[width=0.49\textwidth,clip]{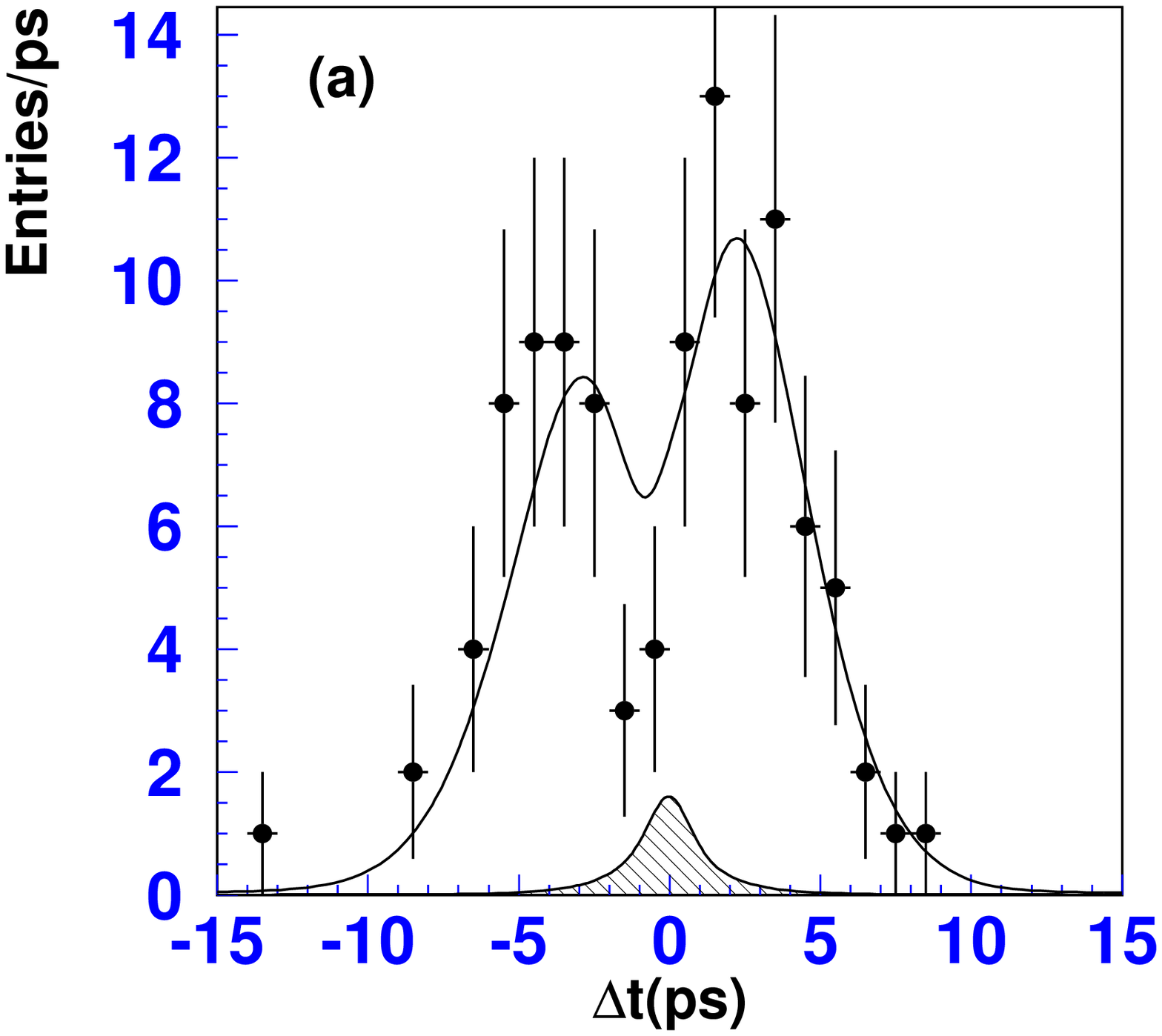}
\includegraphics[width=0.49\textwidth,clip]{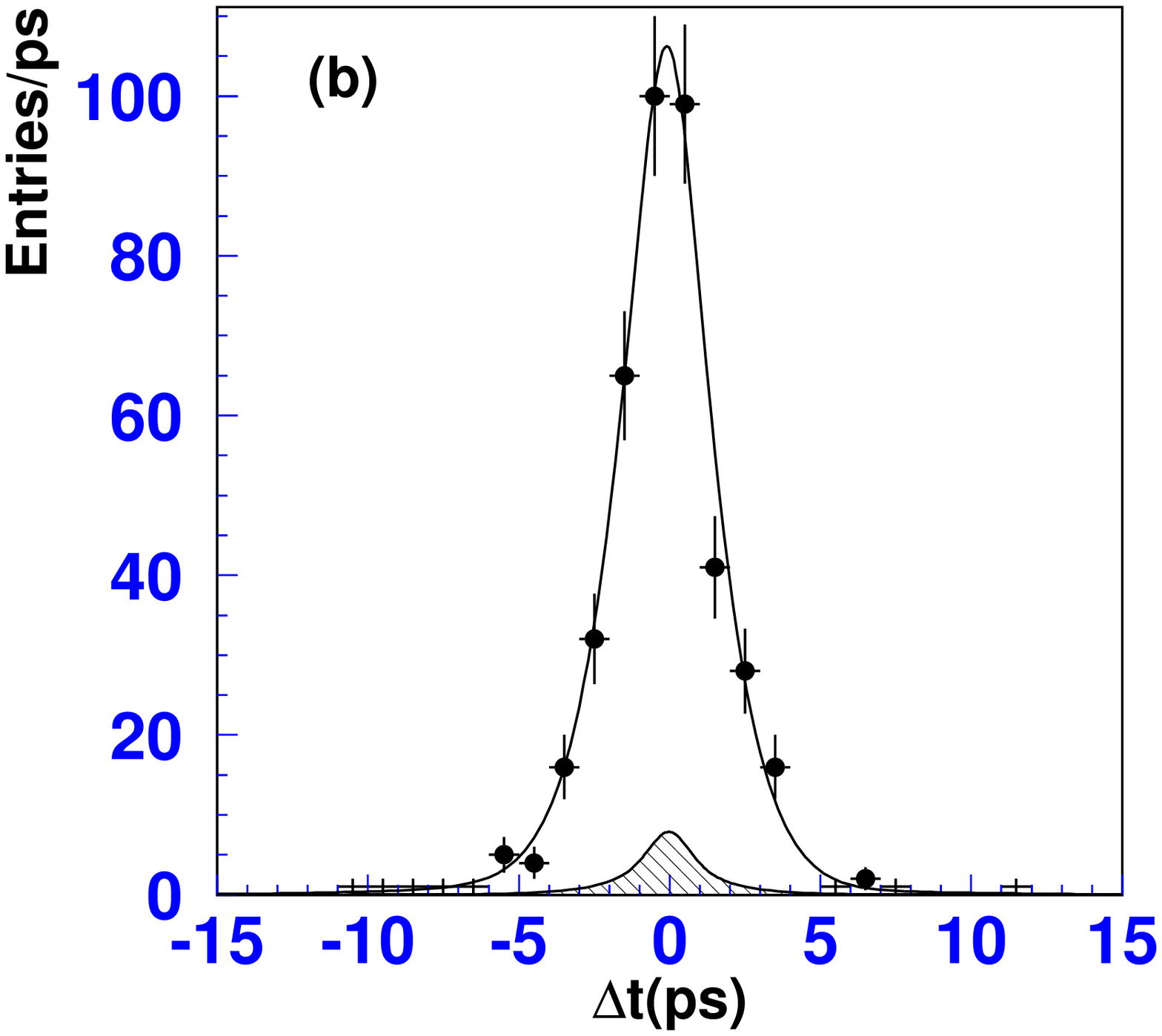}
   \caption{The distributions of $\Delta t$ for $\bar B^0$-tagged
        $D^{*+}\pi^-$ (a) and $D^{*+}\pi^-$ (b). 
        For fully-reconstructed events
        with good-quality tags. (140$fb^{-1}$)}
   \label{fg:dpifull}
  \end{center}
\end{figure}

The distributions of $\Delta t$ for $\bar B^0$-tagged
$D^{*+}\pi^-$ and $D^{*+}\pi^-$ are shown
in Figure~\ref{fg:dpifull} for events with good tagging quality.
The analysis was performed on 140$fb^{-1}$ 
of data~\cite{Sarangi:2004}.
The full-reconstruction analysis gave the following preliminary results:
\beqa
   2r^*\sin(\phi_w + \delta^*) &=&   0.109 \pm 0.057 \pm 0.019 \nonumber\\
   2r^*\sin(\phi_w - \delta^*) &=&   0.011 \pm 0.057 \pm 0.019 \nonumber\\
   2r\sin(\phi_w + \delta) &=&  0.087 \pm 0.054 \pm 0.018  \nonumber\\
   2r\sin(\phi_w - \delta) &=&  0.037 \pm 0.052 \pm 0.018 \,.
\eeqa

For 5 ab$^{-1}$ and 50 ab$^{-1}$ of data, the statistical errors will be
\beq
    \delta\, 2r^*\sin(\phi_w + \delta^*) \sim
    \delta\, 2r\sin(\phi_w + \delta)     \sim
     \left\{ \begin{array}{cc}
                  0.009 & (\hbox{5 ab$^{-1}$ }) \\
                  0.003 & (\hbox{50 ab$^{-1}$ })
             \end{array} \right. \,.
\eeq
We note that the value of $r^{(*)}$ is approximately 0.02; thus,
the above errors correspond to the errors on 
$\sin(\phi_w \pm \delta^{(*)})$ of 0.23 and 0.07, respectively.
If the value of $r^{(*)}$ is known, $\phi_w = 2\phi_1 + \phi_3$
(and $\delta^{(*)}$)
can be extracted from $\sin(\phi_w \pm \delta^{(*)})$.
At this time, determining the value of $r^{(*)}$ to 23\%\ seems
reasonably possible while 7\%\ seems quite challenging. 
On the other hand, our knowledge on $r^{(*)}$  will improve significantly
by the time 50 ab$^{-1}$ of data is taken. It is possible that 
the systematic error on $\phi_3$ due to $r^{(*)}$ is not
overwhelming even with 50 ab$^{-1}$ of data.

\begin{figure}[tbp]
  \begin{center}
\includegraphics[width=0.9\textwidth,clip]{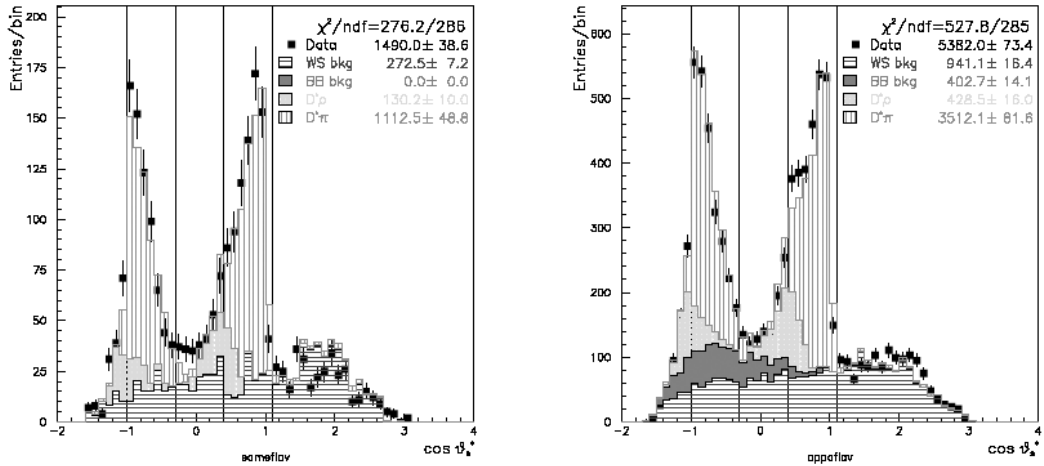}
  \caption{The partial reconstruction of $D^*\pi$.
    The distributions of $\cos\theta^*$ for two
    lepton tagged samples, same-flavor and opposite-flavor tags.
    Based on 78 fb$^{-1}$ of data.}
  \label{fg:dstpi-par}
  \end{center}
\end{figure}

The statistical situation with the partially-reconstructed
$D~^{*+}\pi^-$ is substantially better. 
The helicity of $D^{*+}$ in the $B$ rest frame is $0$ due to
conservation of angular momentum. Thus,
the decay angle $\theta^*$ of the decay $D^{*+}\to D^0 \pi^+$
should have the $\cos^2\theta^*$ shape. One can 
reconstruct $\cos\theta^*$ without explicitly
detecting $D^0$, and the distributions based on 78 fb$^{-1}$ of 
data are shown
in Figure~\ref{fg:dstpi-par} separately for
the two lepton-tagged datasets. 
There are $1110\pm50$ signal events in the same-flavor sample and
$3510\pm80$ signal events in the opposite-flavor sample.
As a result of a $\Delta t$ fit, 
the error on $2r^*\sin(\phi_w \pm \delta^*)$ was found to be
0.029. The statistical errors extrapolated to 5 ab$^{-1}$ and 50 ab$^{-1}$ of data
are
\beq
    \delta\, 2r^*\sin(\phi_w + \delta^*) \sim
     \left\{ \begin{array}{cc}
                  0.005  & (\hbox{ 5 ab$^{-1}$ }) \\
                  0.0015 & (\hbox{50 ab$^{-1}$ })
             \end{array} \right.\quad (\hbox{partial rec.}) \,.
\eeq
The background is about 30\%\ and 
comes mainly from $B$ decays such as $D^*\rho$ and
$D^{**}\pi$. The accuracy of these background estimations is
likely to improve with statistics. The uncertainty of
$r^*$, however, may become a limiting systematics.

\subsection{$B^\pm \to D K^{*\pm}$ (ADS method)}

\begin{figure}[tbp]
  \begin{center}
\includegraphics[width=0.9\textwidth,clip]{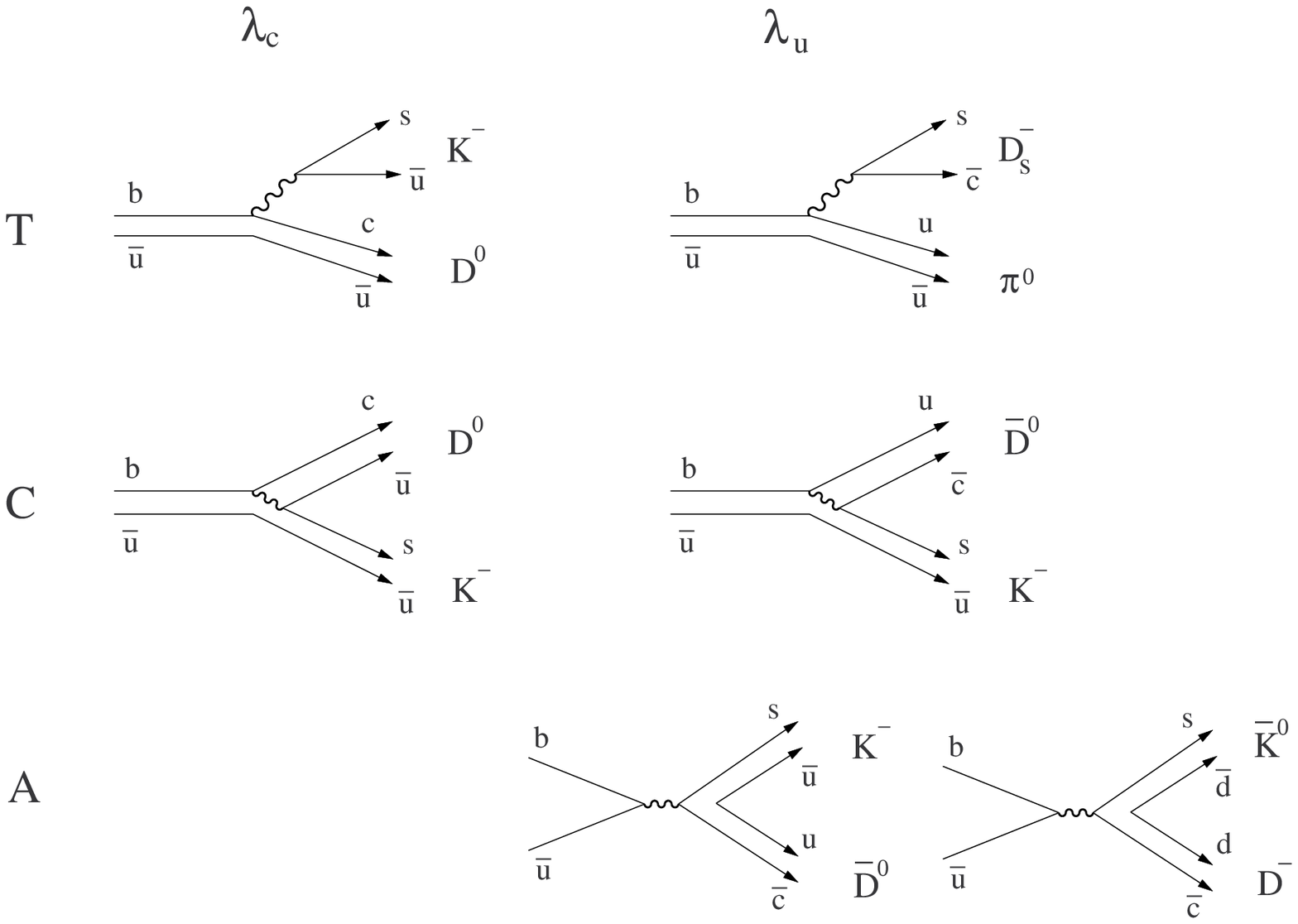}
  \caption{Diagrams contributing to $B^\pm \to D K^\pm$ and related modes. }
  \label{fg:DKdiagB-}
  \end{center}
\end{figure}

In $B^\pm \to D K^\pm$, the number of
$c$ or $\bar c$ quark in the final state is one, and as a result the
penguin processes $b\to s/d$ cannot contribute. When the neutral $D$
meson is detected in a final state that can come from $D^0$ and $\bar D^0$,
the two processes $B^- \to D^0 K^-$ and $B^- \to \bar D^0 K^-$ interfere.
Diagrams contributing to $B^\pm \to D K^\pm$ and related modes are shown in
Figure~\ref{fg:DKdiagB-}. They are categorized in terms of the CKM factors
involved ($\lambda_c = V_{cb}V_{us}^*$ and $\lambda_u = V_{ub}V_{cs}^*$) and
type of processes (T: color-favored tree, C: color-suppressed tree, A:
annihilation).
$B^- \to D^0 K^-$ receives contributions from T and C both with the
CKM factor $\lambda_c$, while $B^- \to \bar D^0 K^-$ can proceed by
C and A both with the CKM factor $\lambda_u$. Thus, 
\beq
   \arg{Amp(B^- \to \bar D^0 K^-) \over Amp(B^- \to D^0 K^-)} = 
   \delta_B - \phi_3\,,\quad
   \arg{Amp(B^+ \to D^0 K^-) \over Amp(B^+ \to \bar D^0 K^-)} = 
   \delta_B + \phi_3\,,
\eeq
where $\delta_B$ is the strong phase. We have noted
\beq
   \arg{\lambda_u\over\lambda_c} = \arg{V_{ub}V_{cs}^*\over V_{cb}V_{us}^*}
    = -\phi_3\,,\quad\hbox{(standard phase convention)}
\eeq
and that when charge conjugate decay is taken, the weak phase changes
sign while the strong phase does not. On the other hand, the absolute size
of $Amp(B^- \to \bar D^0 K^-)$ or $Amp(B^- \to D^0 K^-)$ is the same for
charge conjugate decays:
\beqa
   |Amp(B^- \to \bar D^0 K^-)| &=& |Amp(B^+ \to D^0 K^+)| \equiv B, \\
   |Amp(B^- \to  D^0 K^-)| &=& |Amp(B^+ \to \bar D^0 K^+)| \equiv A.
\eeqa
If $D^0$ and $\bar D^0$ are
detected in a $CP$ eigenstate such as $K^-K^+$ (which is $CP+$; i.e.  
$D_1=(D^0 + \bar D^0)/\sqrt2$), the decay rates of $B^\pm$ are, 
up to an overall constants of $1/2$,
\beqa
   \Gamma(B^- \to D_1 K^-) &=& \left| 
         A + B e^{i(\delta_B - \phi_3)}\right|^2 
          = A^2 + B^2 + 2AB \cos(\delta_B - \phi_3),\\
   \Gamma(B^+ \to D_1 K^+) &=& \left| 
         A + B e^{i(\delta_B + \phi_3)}\right|^2 
          = A^2 + B^2 + 2AB \cos(\delta_B + \phi_3).
\eeqa
Then, there can be a decay rate asymmetry \cite{Gronau:1990ra,Gronau:1991dp}
between $B^- \to D_1 K^-$ and $B^+ \to D_1 K^+$:
\beq
     A_1 \equiv {\Gamma_{D_1 K^-} - \Gamma_{D_1D^+} \over
          \Gamma_{D_1 K^-} + \Gamma_{D_1 K^+}}
  = { - 2r \sin\delta_B\sin\phi_3 \over
      1 + r^2 + 2r \cos\delta_B \cos\phi_3} \,,
\eeq
with
\beq
     r \equiv {B\over A}\,.
\eeq
The value of this $r$ is expected to be around 0.1 to 0.2.
For a $CP-$ final states (such as $K_S \pi^0$; i.e.
$D_2=(D^0 - \bar D^0)/\sqrt2$),  the asymmetry becomes
\beq
     A_2 \equiv {\Gamma_{D_2 K^-} - \Gamma_{D_2 K^+} \over
          \Gamma_{D_2 K^-} + \Gamma_{D_2 K^+}}
  = { 2r \sin\delta_B\sin\phi_3 \over
      1 + r^2 + 2r \cos\delta_B \cos\phi_3} \,.
\eeq
Thus, once the value of $r$ is given, the measurements of $A_1$ and $A_2$ 
can give $\delta_B$ and $\phi_3$ (2 equations and 2 unknowns). In practice,
however, it is difficult to measure the value of $r$ because of the
doubly-Cabibbo-suppressed decays of $D^0$
\cite{Atwood:1996ci,Atwood:2000ck}. It was pointed out, 
however, that by taking more than one final states to which 
both $D^0$ and $\bar D^0$
can decay, one can solve for $r$ and $\delta_B$ 
(the ADS method \cite{Atwood:1996ci,Atwood:2000ck}).

\begin{figure}[tbp]
  \begin{center}
\includegraphics[width=0.7\textwidth,clip]{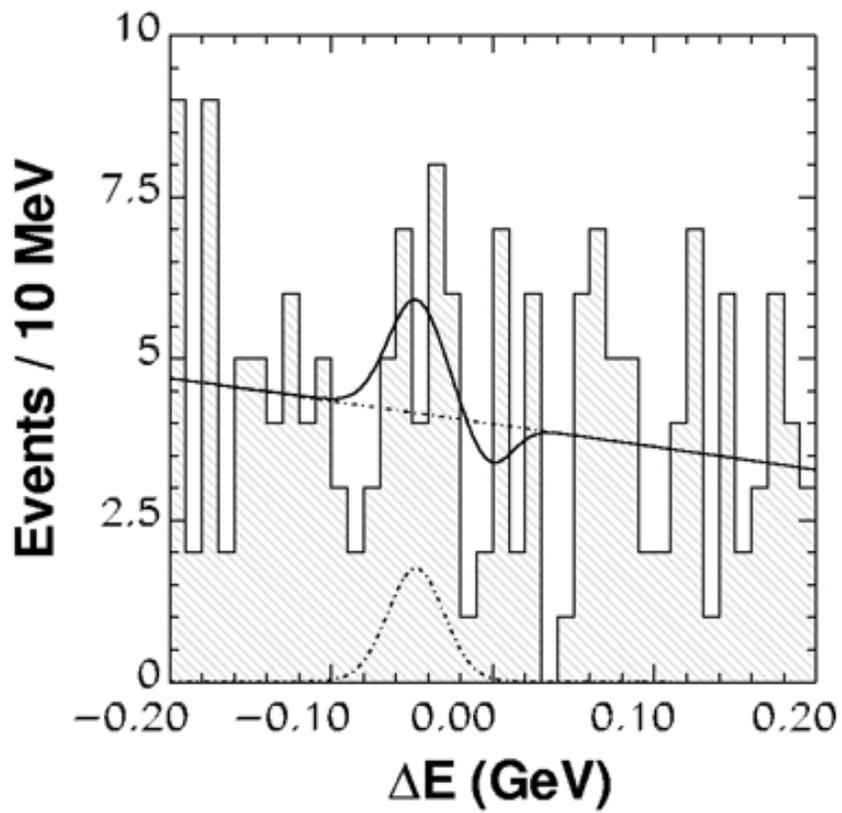}
  \caption{Search for $B^-\to DK^-$, $D\to K^+\pi^-$. 
     The data is 78 fb$^{-1}$.}
  \label{fg:DKdcsd}
  \end{center}
\end{figure}

The authors of \cite{Atwood:1996ci,Atwood:2000ck} used $B^- \to D K^{*-}$ and 
assumed the $D$ decay modes
listed in Table~\ref{tb:ADS}. 
The strong phase $\delta_i$ includes that for the $B$ decay
as well as that for the $D$ decay.
The table also shows the number of events expected for 
$10^8$ $B^\pm$ or roughly 0.1 ab$^{-1}$,
where the detection efficiencies for each $DK^{*-}$
are taken to be the same as the
branching fraction of $D\to i$ where $i$ is the $D$ decay mode
shown. Namely, the detection efficiency of $K^{*-}$ and those
of the $D$ decay final states shown are assumed to be unity.
Under these assumptions, the estimated sensitivity 
on $\phi_3$ is $9^\circ$.
The scale factor includes 2/3 per track, 1/2 per $\pi^0$, and
relevant branching fractions where $K^{*+}$ is assumed to be
detected as $K^-\pi^0$ and $K_S \pi^-$. The value of
$\phi_3$ is taken to be $60^\circ$ and $r$ to be 0.1.
Figure~\ref{fg:DKdcsd} shows the $\Delta E$ distribution 
for $B^-\to DK^-$, $D\to K^+\pi^-$ using 78 fb$^{-1}$ of data.
The detection efficiency was estimated by MC to be 0.27.
With the measured background and the expected signal yield,
S/N is estimated to be 1/7. 
Combining all the above, the required integrated luminosity
for $\delta \phi_3$ of $9^\circ$ becomes 31 ab$^{-1}$, or
equivalently
\beq
    \delta \phi_3 =
     \left\{ \begin{array}{cc}  
               22^\circ & (\hbox{5 ab$^{-1}$}) \\
               7^\circ & (\hbox{50 ab$^{-1}$}) 
             \end{array} \right.\quad (DK^*)\,.
\eeq
If the value of $r$ is larger than 0.1, the sensitivities will be
better. This analysis is theoretically clean and 
will probably be limited by statistics even for 50 ab$^{-1}$ of data. 
Approximately
the same sensitivity is expected from $DK^-$ mode and the combined
sensitivities will be $1/\sqrt2$ times those shown above; namely,
\beq
    \delta \phi_3 =
     \left\{ \begin{array}{cc}  
               16^\circ & (\hbox{5 ab$^{-1}$}) \\
               5^\circ & (\hbox{50 ab$^{-1}$}) 
             \end{array} \right.\quad (DK^* + DK)\,.
\eeq

If the value of $r$ is known, then one can use
$CP$ eigenstates of $D$ decays such as $K^+K^-$ ($CP+$) and
$K_s\pi^0$ ($CP-$) to extract $\phi_3$ as well as 
the strong phase $\delta_B$. This method is statistically
more powerful than the ADS method described above; the
value of $r$, however, is not known well at this time
and seems to be more difficult than the value of $r$
for the $D^{(*)}\pi$ modes. It should be noted, however,
that the value of $r$ measured in the Dalitz analysis
of the following section may be used for this analysis.

\begin{table}[tbp]
  \begin{center}
    \begin{tabular}{|c|ccc|}
      \hline
       mode & \#event & scale factor & strong phase $\delta_i$ \\
      \hline
       $K^+\pi^-$  &  83 &   1  & $10^\circ$ \\
       $K_s\pi^0$  & 791 &   0.34  & $20^\circ$ \\
       $K^+\rho^-$ & 224 &   0.5  & $30^\circ$ \\
       $K^+a_1-$   & 791 &   0.22  & $49^\circ$ \\
       $K_s\rho^0$  & 362 &  0.30   & $200^\circ$ \\
       $K^{*+}\pi^-$  &  65 &  0.33   & $50^\circ$ \\
      \hline
    \end{tabular}
    \caption{The $D$ decay modes used for the sensitivity study
      of $B^-\to DK^{*-}$ mode and the strong phases assumed.
      The number of events are for $10^8$ $B^\pm$
      when the $DK^{*+}$ detection efficiency is assumed to be $Br(D\to i)$.
      The scale factors are relative to the $K^+\pi^-$ mode
      for realistic detection efficiencies described
      in the text.} \label{tb:ADS}
  \end{center}
\end{table}

\subsection{$B^\pm \to D K^\pm$ (Dalitz analysis)}
Use of the Dalitz plots in $B^\pm \to DK^\pm$ has been
suggested in \cite{Atwood:1996ci,Atwood:2000ck} and recently by 
\cite{Giri:2003ty}.
In this analysis, the $D$ meson is detected in the final state
$K_S\pi^+\pi^-$. If the amplitude of $D^0$ decaying to a point
in the Dalitz plot,
$m^2(K_S\pi^+) = m^2_+$ and $m^2(K_S\pi^-) = m^2_-$, is given by
$f(m^2_+,m^2_-)$, then that of $\bar D^0$ decaying to the same
final state can be written as $f(m^2_-,m^2_+)$ where
the two arguments are simply exchanged. This is due to the
approximate $CP$ conservation of the $D$ decay. Explicitly,
\beqa
   Amp(D^0)(m_{K_S\pi^+} = m_+,m_{K_S\pi^-} =m_-) \equiv f(m^2_+,m^2_-) 
         \nonumber \\
   Amp(\bar D^0)(m_{K_S\pi^+} = m_+,m_{K_S\pi^-} =m_-) \equiv f(m^2_-,m^2_+) 
\eeqa
The total decay amplitude for
$B^- \to DK^-$, $D\to K_S\pi^+\pi^-$ is then
\beqa
     M_- &=& Amp(D^0_{\to K_S\pi^+\pi^-} K^-) +
             Amp(\bar D^0_{\to K_S\pi^+\pi^-}K^-) \nonumber\\
    &=& A [f(m^2_+,m^2_-) + r e^{i(\delta_B-\phi_3)} f(m^2_-,m^2_+)]\,,
   \label{eq:dal1}
\eeqa
where $A, r, \delta_B$ are defined in the previous section.
The total decay amplitude for
$B^+ \to DK^+$, $D\to K_S\pi^+\pi^-$, where
$m^2(K_S\pi^+) = m^2_+$ and $m^2(K_S\pi^-) = m^2_-$, is similarly
\beqa
     M_+ &=& Amp(\bar D^0_{\to K_S\pi^+\pi^-} K^+) +
             Amp(D^0_{\to K_S\pi^+\pi^-}K^+) \nonumber\\
    &=& A [f(m^2_-,m^2_+) + r e^{i(\delta_B+\phi_3)} f(m^2_+,m^2_-)]\,,
   \label{eq:dal2}
\eeqa
where the sign of $\phi_3$ has flipped while the strong phase
$\delta_B$ remains the same. 

\begin{figure}
  \begin{center}
\includegraphics[width=0.9\textwidth,clip]{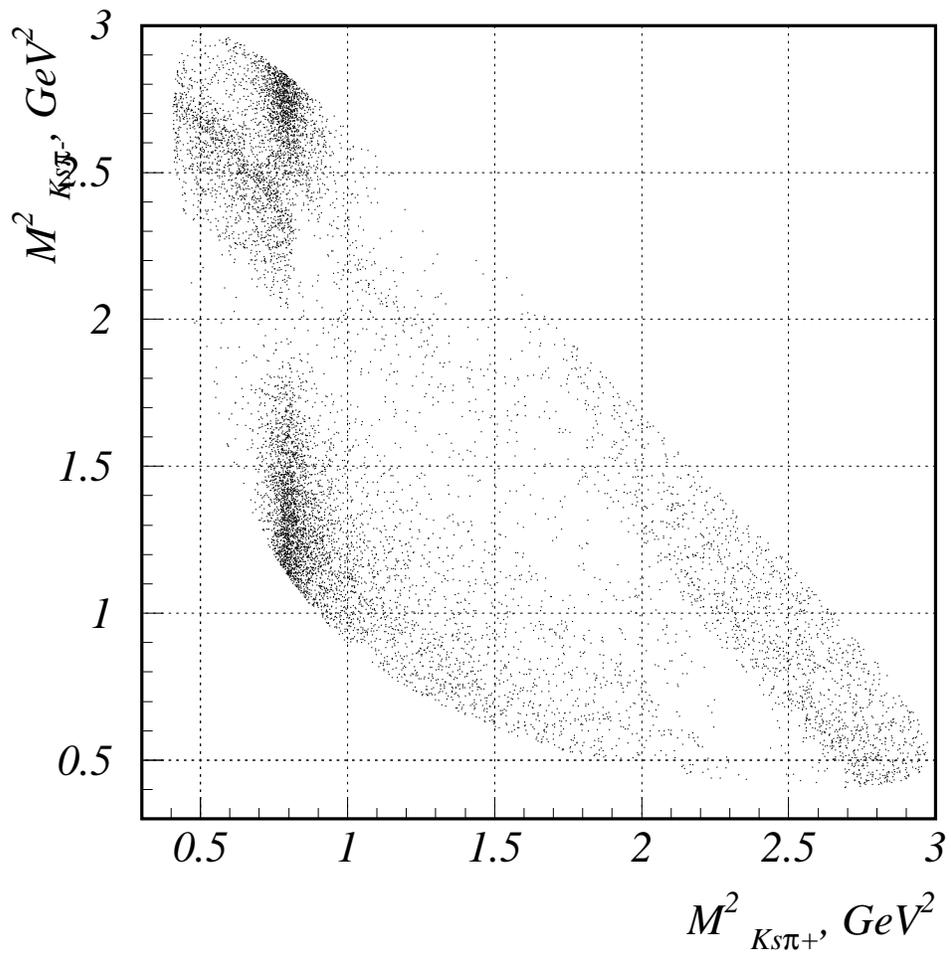}
  \caption{The Dalitz plot for $D^0 \to K_S\pi^+\pi^-$ based on the CLEO 
   measurement.}  \label{fg:dalitz}
  \end{center}
\end{figure}

The Dalitz plot for $D^0 \to K_S\pi^+\pi^-$ based on the CLEO 
measurement~\cite{Muramatsu:2002jp} is shown in Figure~\ref{fg:dalitz}.
As can be seen from the Dalitz plot, the distribution is highly
asymmetric under the exchange of $m_{K_S\pi^+}$ and $m_{K_S\pi^-}$
which indicates that the interference in the Dalitz plot has
a good sensitivity on the phase between the two terms of
(\ref{eq:dal1}) and (\ref{eq:dal2}). The phase angles
$\delta_B - \phi_3$ and $\delta_B + \phi_3$ are obtained
from the separate fits of $B^-$ and $B^+$ decays and
$\phi_3$ can be extracted therefrom.

\begin{figure}
  \begin{center}
\includegraphics[width=0.49\textwidth,clip]{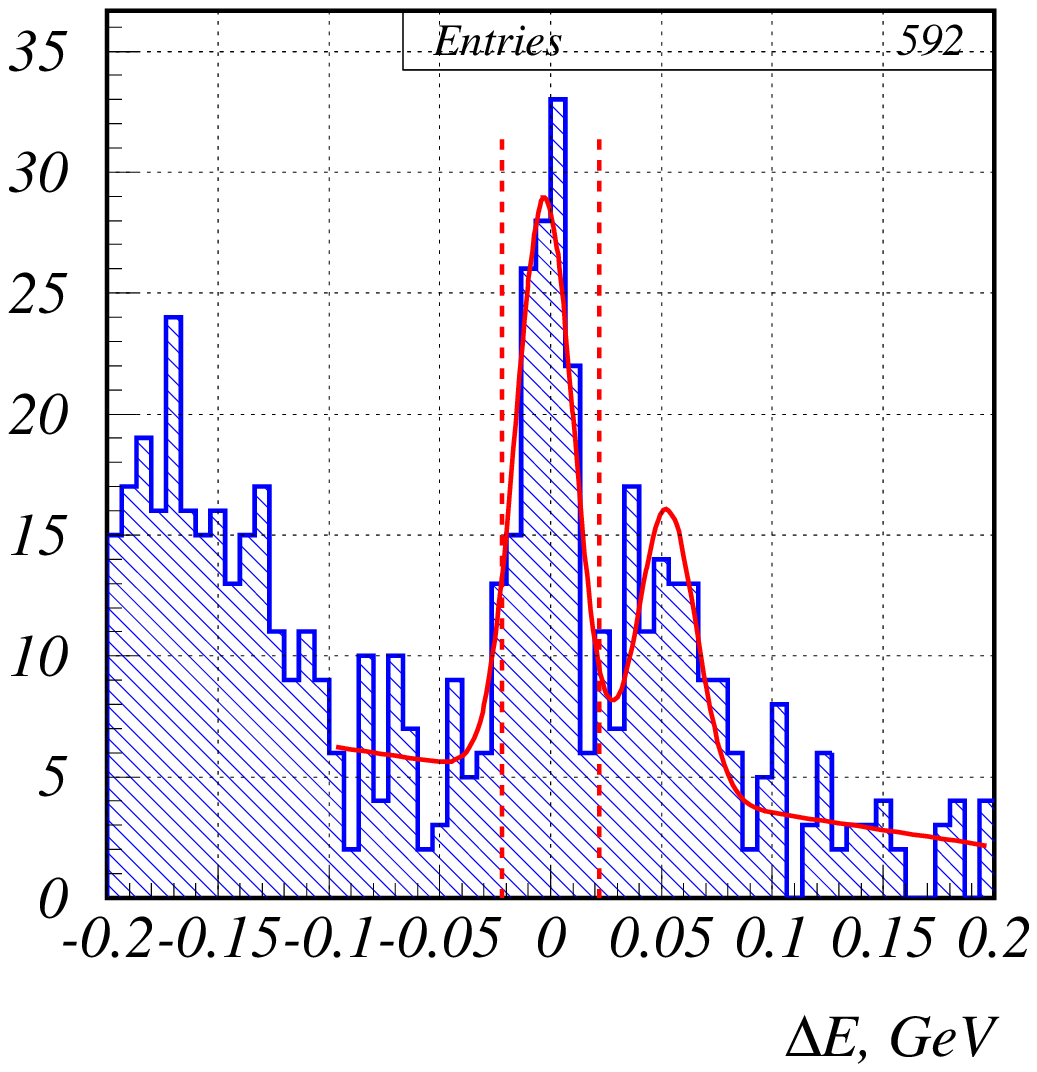}
\includegraphics[width=0.49\textwidth,clip]{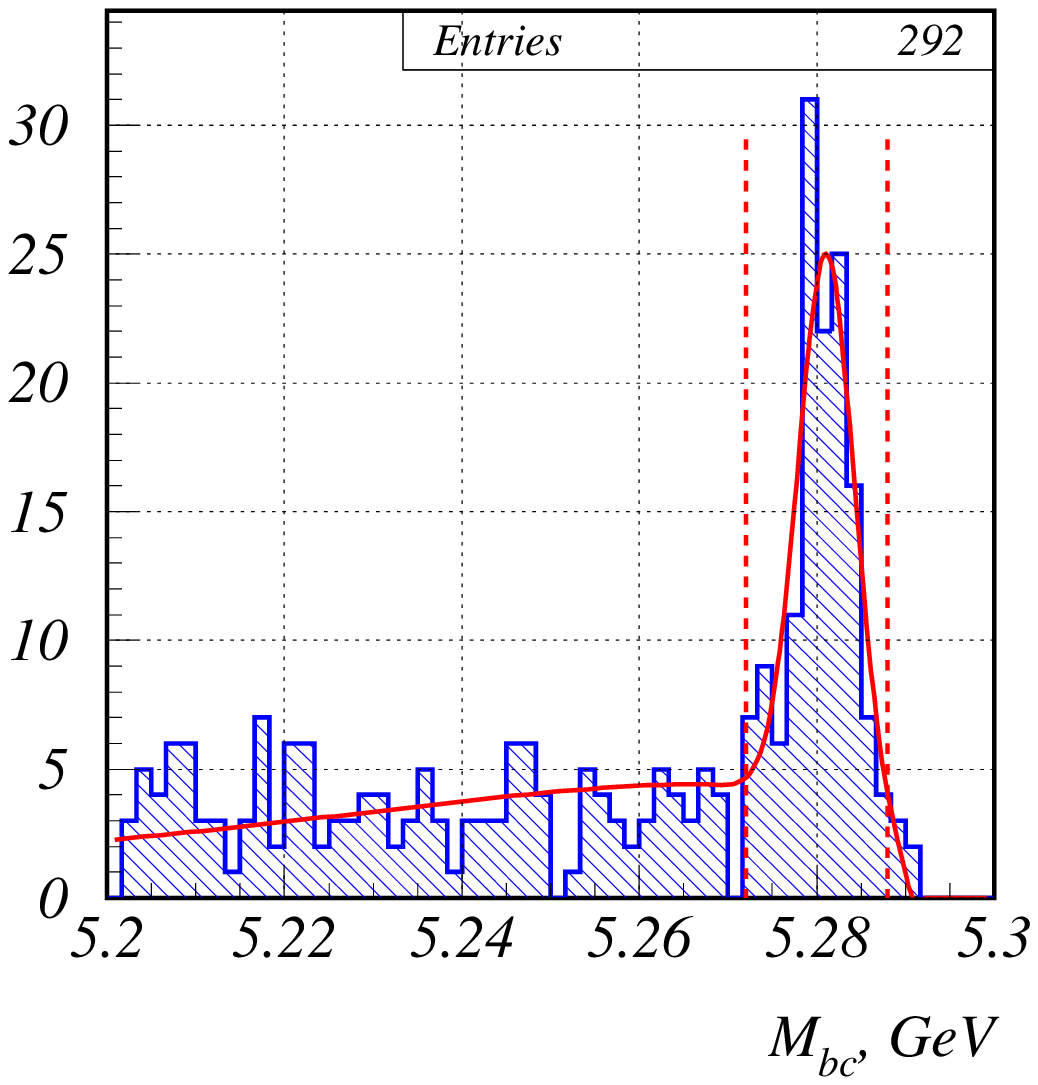}
   \caption{The $\Delta E$ and $M_B$ distributions for
      $B^\pm\to DK^\pm$, $D\to K_S\pi^+\pi^-$, based on
      140 fb$^{^1}$ of data.}
   \label{fg:dal-demb}
  \end{center}
\end{figure}

\begin{figure}
  \begin{center}
\includegraphics[width=0.49\textwidth,clip]{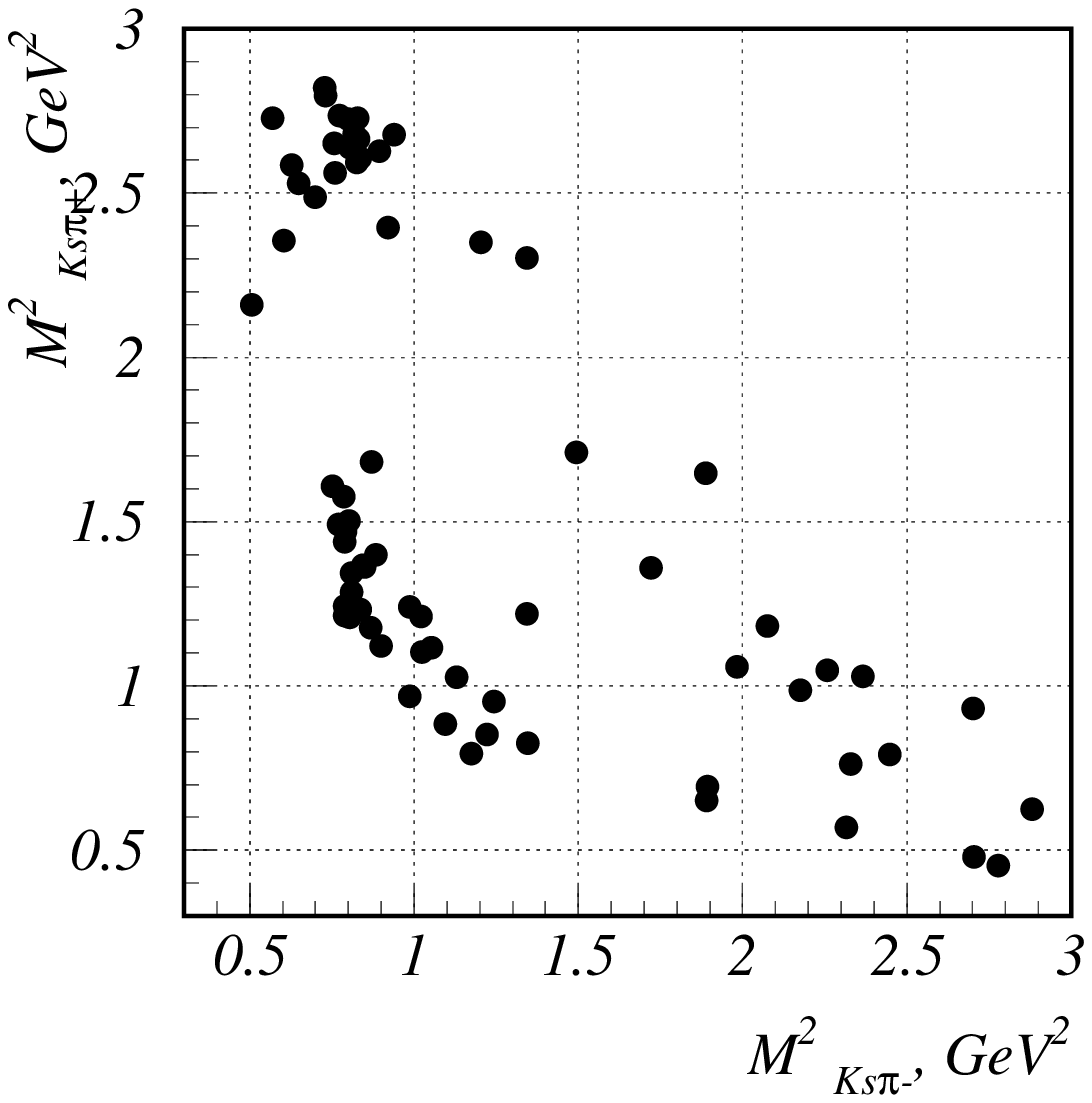}
\includegraphics[width=0.49\textwidth,clip]{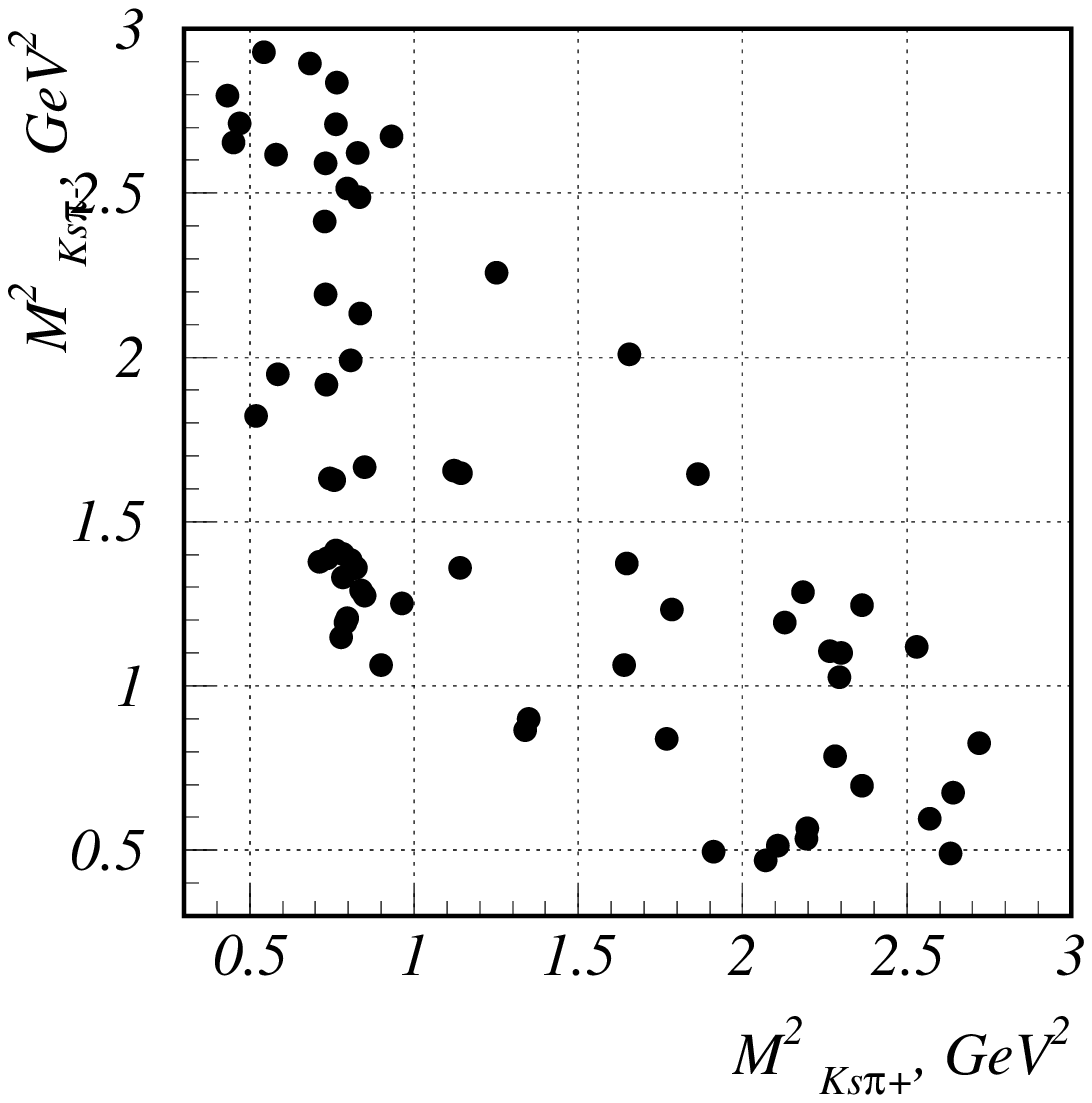}
   \caption{The Dalitz distributions for
      $B^+\to DK^+$ (left) and $B^-\to DK^-$ (right), where
      $D\to K_S\pi^+\pi^-$ based on
      140 fb$^{^1}$ of data.}
   \label{fg:dal-b+b-}
  \end{center}
\end{figure}

\begin{figure}
  \begin{center}
\includegraphics[width=0.49\textwidth,clip]{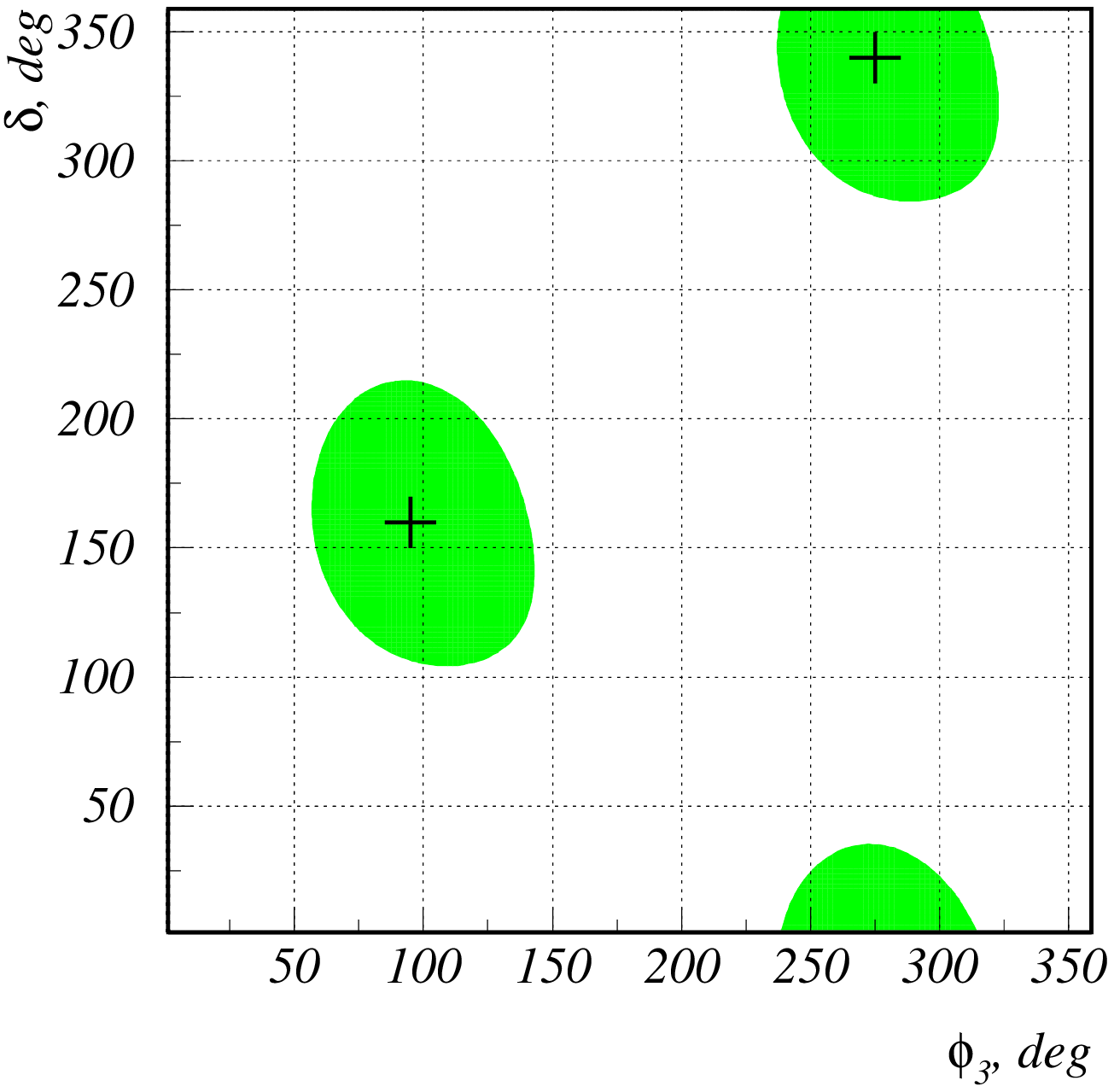}
\includegraphics[width=0.49\textwidth,clip]{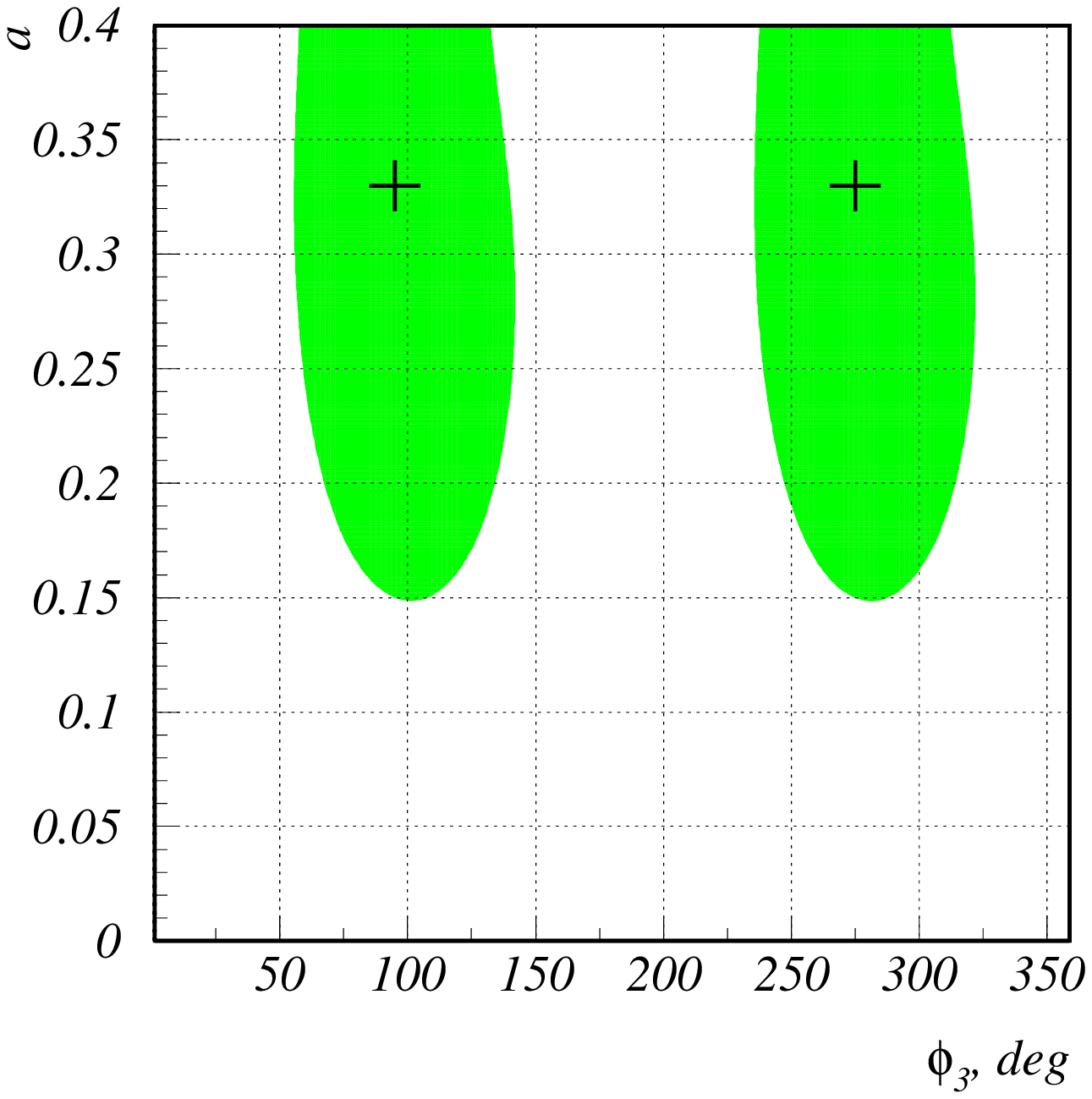}
   \caption{The results of the fit for 
      $r(=a)$, $\delta_B (= \delta)$, and $\phi_3$.}
   \label{fg:dal-results}
  \end{center}
\end{figure}

Figure~\ref{fg:dal-demb} shows the  $\Delta E$ and $M_B$ distributions for
$B^\pm\to DK^\pm$, $D\to K_S\pi^+\pi^-$, based on 140 fb$^{-1}$ of data,
and the Dalitz plots of the $D$ decays are shown in 
Figure~\ref{fg:dal-b+b-} separately for $B^+$ and $B^-$~\cite{Abe:2003cn}.
The results of an unbinned maximum likelihood fit 
for the parameters $r(=a)$, $\delta_B (= \delta)$, and $\phi_3$ are shown in
Figure~\ref{fg:dal-results}.
For $\phi_3$ and $\delta_B$, the fit yields
\beqa
     \phi_3 &=& 95^{+25}_{-20}\pm13\pm10\, ({}^\circ) \nonumber\\
     \delta_B &=& 162^{+20}_{-25}\pm12\pm24\, ({}^\circ) 
\eeqa
where the first errors are statistical, the second
are experimental systematic errors, such as
from background shapes and efficiency shapes, and the third
errors are additional
systematic errors due to the $D$ decay model dependence.
With 5 ab$^{-1}$ and 50 ab$^{-1}$ of data, the statistical
errors will be
\beq
   \delta \phi_3 = \left\{
      \begin{array}{cc}
         4^\circ & (\hbox{5 ab$^{-1}$}) \\
         1.2^\circ & (\hbox{50 ab$^{-1}$}) \\
      \end{array}
      \right. \,.
\eeq
With more statistics, the Dalitz distribution of $D^0 \to K_S\pi^+\pi^-$
will be measured more accurately using $D^{*+}\to D^0\pi^+$. 
Even though it is not clear at present if the uncertainty
due to the $D$ decay modeling can be reduced to a level of one degree,
it is quite possible that the measurement is not overwhelmed by
systematics at 5 ab$^{-1}$. Furthermore, the value of $r$ measured
in this mode can be used in the $B^\pm \to DK^\pm$ modes.

%

\clearpage \newpage
\def\Journal#1#2#3#4{{#1} {\bf #2}, #3 (#4)}
\def\NCA{\em Nuovo Cimento}
\def\NIM{\em Nucl. Instrum. Methods}
\def\NIMA{{\em Nucl. Instrum. Methods} A}
\def\NPB{{\em Nucl. Phys.} B}
\def\PLB{{\em Phys. Lett.}  B}
\def\PRL{\em Phys. Rev. Lett.}
\def\PRD{{\em Phys. Rev.} D}
\def\ZPC{{\em Z. Phys.} C}

\def\gev{{\rm GeV}}
\def\mev{{\rm MeV}}
\def\lqcd{\Lambda_{\rm QCD}}
\def\qcut{q_{\rm cut}}
\def\mxcut{m_{\rm cut}}
\def\gcut{G(\qcut^2,\mxcut)}
\def\mbups{m_b^{1S}}

\newcommand{\boldvec}[1]{\mbox{\boldmath$#1$}}
\newcommand{\backvec}[1]{\stackrel{\leftarrow}{#1}}

\newcommand{\Btopi}{B\rightarrow\pi l\nu}
\newcommand{\Dtopi}{D\rightarrow\pi l\nu}
\newcommand{\Btorho}{B\rightarrow\rho l\nu}
\newcommand{\BtoD}{B\rightarrow D^{(*)} l\nu}
\newcommand{\BtoK}{B\rightarrow K^{(*)} l^+ l^-}
\newcommand{\vdotk}{v \cdot k_{\pi}}

\def\epem{e^+ e^-}
\def\pip{\pi^+}
\def\pim{\pi^-}
\def\piz{\pi^0}
\def\rhop{\rho^+}
\def\rhom{\rho^-}
\def\rhoz{\rho^0}
\def\lnu{\ell^+ \nu}
\def\Vcb{V_{cb}}
\def\Vub{V_{ub}}
\def\magVcb{|V_{cb}|}
\def\magVub{|V_{ub}|}
\def\ifb{\rm fb^{-1}}
\def\iab{\rm ab^{-1}}


\section{$|V_{ub}|$}
\subsection{Introduction}

Precise determination of the magnitude of the Cabibbo-Kobayashi-Maskawa 
matrix element $V_{ub}$ is of fundamental importance in over-constraining
the unitarity triangle, and thereby finding effects of physics beyond
the Standard Model.
Such a serious examination would require precision of a few percent.
A high luminosity SuperKEKB/Belle experiment will provide such a unique
opportunity.

In principle, $|V_{ub}|$ can be determined using data for semileptonic
$B \rightarrow X_u \ell \nu$ decays, where $X_u$ denotes a hadronic 
system containing a $u$-quark.
Both inclusive and exclusive measurements are useful.
However, the present error in $|V_{ub}|$ determination is about 
$\pm 20$\%~\cite{Hagiwara:fs}, and limited by both experimental and
theoretical systematic errors.
The high luminosity at the SuperKEKB will enable us to perform 
high statistics measurements of $B \rightarrow X_u \ell \nu$ decays 
with ``$B$ tagging''.
This will lead to rate measurements with significantly reduced 
experimental systematic errors and also to $|V_{ub}|$ extraction 
much less biased by theoretical ambiguities.
These features are unique at a high luminosity $e^+e^-$ $B$-factory,
and cannot be achieved at $e^+e^-$ machines with the luminosity 
available so far nor at present or future hadron machines.

In this section, we discuss strategy and prospects for determination of 
$|V_{ub}|$ at the SuperKEKB/Belle experiment using both inclusive and 
exclusive semileptonic $B \rightarrow X_u \ell \nu$ decays.

\subsection{Theoretical formalisms for the semileptonic $B$ decays}
\label{sec:vub_theory}

\subsubsection{Inclusive decays}
\label{sec:vub_theory_incl}
The amplitude for $B\rightarrow X_u l\nu$ inclusive decay
can be computed in perturbative QCD using the Operator
Product Expansion (OPE).
Since the $b$ quark inside the $B$ meson has momentum 
$(m_b v + k )^{\mu}$, where $k$ is the residual momentum 
of $O(\Lambda_{QCD})$, 
the OPE is carried out by expanding the quark propagator as 
\begin{equation}
  \frac{1}{(m_b v + k -q)^2} =
  \frac{1}{(m_b v -q)^2} 
  \left[ 
    1 - \frac{(m_b v -q)\cdot k}{(m_b v -q)^2}
    - \frac{k^2}{(m_b v -q)^2} + \cdots 
  \right].
\end{equation}
Denoting the invariant mass and the energy of the hadron state $X_u$ 
as $m_X$ and $E_X$, the first term is of order 
$E_X \Lambda_{QCD}/m_X^2$ while the second term is of order  
$\Lambda_{QCD}^2/m_X^2$.

\begin{figure}[tbp]
\begin{center}
\includegraphics[width=0.9\textwidth,clip]{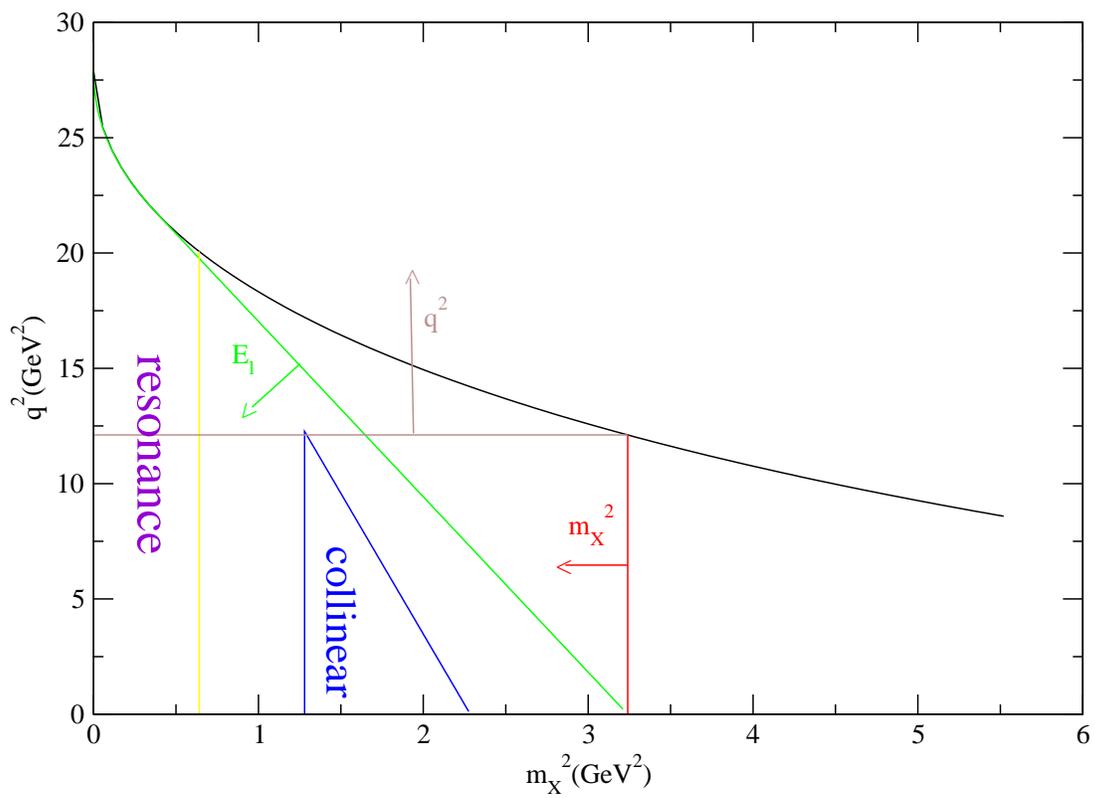}
\end{center}
\caption{Phase space for $B\rightarrow X_u l\nu$ decay.}
\label{fig:bu_kin}
\end{figure}

The phase space can be divided into the following three regions
(see Figure~\ref{fig:bu_kin}): 
(i) a generic region where $\Lambda_{QCD}/m_X$,
$(E_X \Lambda_{QCD})/m_X^2 \ll 1$. 
In this region, the differential decay rate can be
successfully expanded by the OPE;
(ii) a shape function region (or collinear region) 
where $\Lambda_{QCD}/m_X \ll 1$
and  $(E_X \Lambda_{QCD})/m_X^2 \sim 1$. 
In this region, a class of $\Lambda_{QCD}E/m_X^2$ terms must
be resummed, which can be described by the shape function of
the $B$ meson 
\cite{Neubert:1993ch,Mannel:1994pm,Bigi:1993ex}
and its higher twist corrections;
(iii) a resonance region where $\Lambda_{QCD}/m_X \sim 1$.
In this region, the differential decay rate is dominated by 
a few exclusive states so that neither the OPE nor the twist
expansion work. 

Since the $B\rightarrow X_u l\nu$ decay suffers from 
$B\rightarrow X_c l\nu$ decay background one has to 
introduce the cut of the following kinds:
\begin{itemize}
\item lepton energy cut : $E_l > (m_B^2-m_D^2)/(2 m_B)$.
\item hadron invariant mass cut :  $m_X^2 < m_D^2$.
\item lepton mass cut :  $q^2 < (m_B-m_D)^2$.
\item combined $(q^2,m_X^2)$ cut.
\end{itemize}

For the $E_l$ cut, only 10\% of the rate is available, 
which is dominated by the shape function region. 
In this region, the lepton energy spectrum at the leading order 
is given by 
\begin{equation}
  \frac{d\Gamma}{dE_l}
  =
  \frac{G_F^2 |V_{ub}|^2 m_b^4}{96\pi^3}
  \int d\omega \theta(m_b-2 E_l -\omega) f(\omega),
\end{equation}
where $f(\omega)$ is the shape function define by 
\begin{equation}
  f(\omega)
  =
  \frac{1}{2m_B} \langle B| \bar{h} \delta(in\cdot D+\omega) | B \rangle.
\end{equation}
Since this function is a universal quantity of the $B$ meson,
it can be measured experimentally through other processes.
For example, the photon spectrum of $B\to X_s \gamma$ decay
is given by \cite{Neubert:1993um}
\begin{equation}
  \frac{d\Gamma}{dE_\gamma}
  = 
  \frac{G_F^2 |V_{tb}V_{ts}^{\ast}|^2 \alpha_{QED} m_b^5}{32\pi^4}
  f(\omega).
\end{equation}
Therefore, the extraction of $|V_{ub}|$ can be done without
the theoretical uncertainty of the shape function, if one
considers a ratio of weighted integrals 
over the endpoint regions of $B\rightarrow X_u l \nu$ decays
and the photon spectrum in $B\rightarrow X_s \gamma$ decays as 
\cite{Leibovich:1999xf,Neubert:2001sk}
\begin{equation}
  \left|\frac{V_{ub}}{V_{tb}V_{ts}^{\ast}}\right|^2 
  = \frac{3\alpha_{QED}}{\pi} K_{pert}(E_0) 
  \frac{\Gamma_u(E_0)}{\Gamma_s(E_0)}
  + \mathcal{O}(\Lambda/M_B),
\end{equation}
where
\begin{eqnarray}
  \Gamma_u(E_0) &=& 
  \int_{E_0}^{M_B/2} dE_l 
  \frac{d\Gamma(B\rightarrow X_u l\nu )}{dE_l},
  \\
  \Gamma_s(E_0)
  &=& 
  \frac{2}{M_B}\int_{E_0}^{M_B/2} dE_\gamma
  (E_\gamma-E_0)\frac{d\Gamma(B\rightarrow X_s \gamma)}{dE_\gamma}.
\end{eqnarray}
The coefficient $K_{pert}(E_0)$ is a factor from
short-distance effect which can be calculated in
perturbative QCD. 
There are two major sources of uncertainties. 
The first is the unknown higher twist corrections to the
shape function $1/m_b$
\cite{Bauer:2001mh,Bauer:2002yu,Leibovich:2002ys,Neubert:2002yx}.  
Based on model calculations this higher twist correction 
is expected to be of order 15\% for $E_l^{cut}=2.3$ GeV.
The second is the weak annihilation contribution of
$O(1/m_b^2)$, which is estimated as 10\% for
$E_l^{cut}=2.3$~GeV
\cite{Bigi:1993bh,Voloshin:2001xi}. 
These two uncertainties can be reduced below 10\% by
lowering the lepton energy cut by combining with other cuts. 

For the $m_X^2$ cut
\cite{Falk:1997gj,Dikeman:1997es,Bigi:1997dn,%
  Leibovich:2000ig,Leibovich:2000ey},
80\% of the kinematic range is 
available but still the results are sensitive to the 
shape function.
This cut also suffers from the singularity due to the
bremsstrahlung diagram when the partonic invariant mass
$s=(m_b-v)^2$ is zero.

For the pure $q^2$ cut, 20\% of the rate is available and
the decay rate is not sensitive to the shape function
\cite{Bauer:2000xf}. 
However, the kinematic range is sensitive to the resonance
region and the convergence of the OPE as well as the
convergence in the perturbative expansion in $\alpha_s$ are
slower.
The largest error comes from the weak annihilation
contribution of $O(1/m_b^3)$. 
The rate with $q^2$ is  also sensitive to the uncertainty of
$m_b$ \cite{Neubert:2000ch} as can be seen from the $m_b$
dependence of the partial decay rate parametrized as 
$\Gamma(q^2>q^2_{cut}) \propto  m_b^{\Delta(q^2_{cut})}$, 
where $\Delta(q^2_{cut}) \sim 10 + 
\displaystyle{\frac{q^2_{cut}-(m_B-m_D)^2}{1 \mathrm{GeV}^2}}$.

To summarize, possible sources of errors are 
(1) perturbative error from unknown two-loop corrections,
(2) shape function contributions and bremsstrahlung,
(3) uncertainties in $m_b$, and
(4) $O(1/m_b^3)$ power corrections.
The optimized method would be obtained by combining the
$q^2$ and $m_X^2$ cuts \cite{Bauer:2001rc}.
The kinematical constraints $m_X<m_D$ 
and $q_{cut}^2 > m_B m_b - (m_X^{cut})^2$
reduce the charm background.
If we raise $q^2_{cut}$ the errors (3) and (4) gets larger 
while if we lower $q^2_{cut}$ the errors (2)  gets larger
and (1) is small in the intermediate $q^2$ region. 
Thus it is important to find the best cut that minimizes 
the sum of these errors.

In Table~\ref{finaltable}, we give results of
\cite{Bauer:2001rc} for the errors of the partial decay rate
normalized by the total tree level parton decay rate defined as 
\begin{equation}
  \frac{G_F^2 |V_{ub}|^2 m_b^5} {192\pi^3}
  \gcut \equiv 
  \int_{\hat\qcut^2}^1 {\rm d}\hat q^2 \int_0^{\hat s_0}
  {\rm d}\hat s\, \frac{{\rm d}\Gamma}{{\rm d}\hat q^2 {\rm d}\hat s}.
  \label{defineg}
\end{equation}
In order to achieve the $|V_{ub}|$ determination with a few
percent accuracy, $q^2_{cut}$=6 GeV$^2$ (and $m_{X cut}^2=m_D^2$) 
is the optimal choice. 
The dominant error in this case is the uncertainty in $m_b$. 
It is therefore important to determine the bottom quark mass
to 30 MeV/c$^2$ accuracy. 

\begin{table}[tbp]
\begin{tabular}{ccccccc}
  \hline\hline
  Cuts on $(q^2,\,m_X^2)$  &  $\gcut$
  & $\Delta_{\rm struct}G$ & $\Delta_{\rm pert}G$ &  
  $\matrix{\Delta_{m_b}G \cr {\pm 80/30\,\mev}}$  &
  $\Delta_{1/m^3} G$ & $\Delta G$ \\
  \hline\hline
  \multicolumn{1}{c}{Combined cuts} & \multicolumn{6}{c}{} \\ 
  $6\,\gev^2, 1.86\,\gev$ &  0.38  &  $ -4\%$  &4\%  &  13\%/5\%  &  6\%  &
  15\%/9\%  \\
  $8\,\gev^2, 1.7\,\gev$  &  0.27 &  $-6\%$&  6\%  &  15\%/6\%  &  8\%  &
  18\%/12\% \\ 
  $ 11\,\gev^2, 1.5\,\gev$  & 0.15 &  $-7\%$ &13\% & 18\%/7\% & 16\% &
  27\%/22\% \\ 
  \hline
  \multicolumn{1}{c}{Pure $q^2$ cuts} & \multicolumn{6}{c}{} \\ 
  ~$(m_B-m_D)^2, m_D$~ & 0.14 & --\,--&15\% &19\%/7\% & 18\% & 
  ~30\%/24\%~ \\
  $(m_B-m_{D^*})^2, m_{D^*}$& 0.17 & --\,--&13\% &17\%/7\% & 14\%
  &26\%/20\% \\
  \hline\hline
\end{tabular}
\caption{
  $\gcut$ and its errors for different choices of
  $(\qcut^2,\mxcut)$.  
  $\Delta_{\rm struct}G$ gives the fractional  effect of the
  structure function $f(k_+)$ in the simple model which is
  not included in the error estimate.
  The total error is obtained by adding each error in quadrature.   
  The two values correspond to $\Delta \mbups=\pm 80\,\mev$
  and $\pm 30\,\mev$. 
  Table is from \cite{Bauer:2001rc}.
}
\label{finaltable}
\end{table}

\subsubsection{Exclusive decays}
\label{sec:vub_theory_excl}
The exclusive semileptonic decay $\Btopi$ determines the CKM
matrix element $|V_{ub}|$ through the following formula,
\begin{eqnarray}
\frac{d\Gamma}{dq^2} &=& 
\frac{G_F^2}{24\pi^3}|(v\cdot k_{\pi})^2 - m_{\pi}^2|^{3/2} 
|V_{ub}|^2 |f^+(q^2)|^2, 
\end{eqnarray}
where the form factor $f^+$ is defined as
\begin{eqnarray}
\langle\pi(k)|\bar{q}\gamma^{\mu}b|B(p)\rangle
&= &
f^+(q^2)  \left[ 
    (p+k)^{\mu} 
    - \frac{m_B^2-m_{\pi}^2}{q^2} q^{\mu}
  \right] 
 + 
f^0(q^2) \frac{m_B^2-m_{\pi}^2}{q^2} q^{\mu},\
\end{eqnarray}
with  $p$ and $k$ the $B$ and $\pi$ meson momenta.
$q = p-k$ is the momentum transfer and 
$q^2=m_B^2+m_{\pi}^2 - 2 m_B v\cdot k$, 
where $v$ is the velocity of the B meson.
Since the most promising approach in which systematic
improvement based on the first principle calculations is
possible is lattice QCD, we focus on the lattice computation
of the form factors.

Lattice calculation suffers from three major limitations.
One is the discretization error from the large energy of the
initial and final hadrons. 
In order to avoid such error, spatial momenta must be much
smaller than the cutoff, \textit{i.e.} 
$|\vec{p}_B|$, $|\vec{k}_{\pi}| < $ 1~GeV.
This means that the form factors can be computed reliably
only in the range of 
$v \cdot k \equiv E_{\pi} < $ 1 GeV or equivalently 
$q^2 > $ 18 GeV$^2$.
Another limitation is the fact that due to the limited
computer power the light quark mass range for practical
simulations is 
$m_s/3 \leq m_q \leq m_s $ or $m_{\pi} = 0.4 \sim 0.8 $ GeV. 
In order to obtain physical results chiral extrapolation
in the light quark masses is necessary.
The last limitation is the large discretization error from
the $b$-quark mass. 
In the present simulations, the lattice cutoff is limited to
$a^{-1}=2-3$ GeV, so that the $b$-quark mass in lattice unit
is larger than unity. 
This makes the discretization error of $O(am_b)$ completely
out of control. 
In order to avoid this error, one either carries out
simulations around a charm quark mass region and extrapolate
the result in the inverse heavy quark mass $1/m_Q$
(extrapolation method), 
or use heavy quark effective theories, such as NRQCD action or
Fermilab action (HQET). 
In both cases, extrapolation or interpolation of the form
factors in $1/m_Q$ may be performed using the HQET-motivated
form factors $f_1(v\cdot k)$ and $f_2(v\cdot k)$ 
\cite{Burdman:1993es} 
\begin{equation}
  \langle\pi(k)|\bar{q}\gamma^{\mu}b|B(p)\rangle
  =
  2\left[
    f_1(v\cdot k) v^{\mu} + f_2(v\cdot k)\frac{k^{\mu}}{v\cdot k}
  \right].
\end{equation}
With this choice the heavy quark scaling law is explicit, and
the form factors are simply expanded in terms of $1/m_Q$.

So far all lattice calculations of the form factors have
been performed only in the quenched approximation, in which
the sea quark effects are neglected. 
The systematic error due to quenching is hard to estimate
but typical errors in many quantities such as light hadron
masses and decay constants are expected to be at 10-15\%
level. 

Recently five lattice collaborations have carried out 
quenched QCD calculations of $\Btopi$ form factors 
using extrapolation methods~\cite{Bowler:1999xn,Abada:2000ty}, 
the Fermilab action \cite{El-Khadra:2001rv}, and the NRQCD action 
\cite{Aoki:2001rd,Shigemitsu:2002wh} for the heavy quark.
Figure~\ref{fig:fpf0_2} shows the results from different
lattice groups.
$f^+(q^2)$ agrees within systematic errors, while 
$f^0(q^2)$ shows deviations among different methods.
The reason for the discrepancies in $f^0(q^2)$ can be attributed 
to the systematic error in the chiral extrapolation and
heavy quark mass extrapolation (interpolation) error.
The error of the form factors in the present calculations is 
around 20\%. 
In addition to the quenching error and the chiral 
extrapolation error, the major errors are the statistical
error, the discretization error and the $1/M$ extrapolation
error.  

\begin{figure}[tbp]
  \begin{center}
    \includegraphics[width=0.9\textwidth,clip=true]{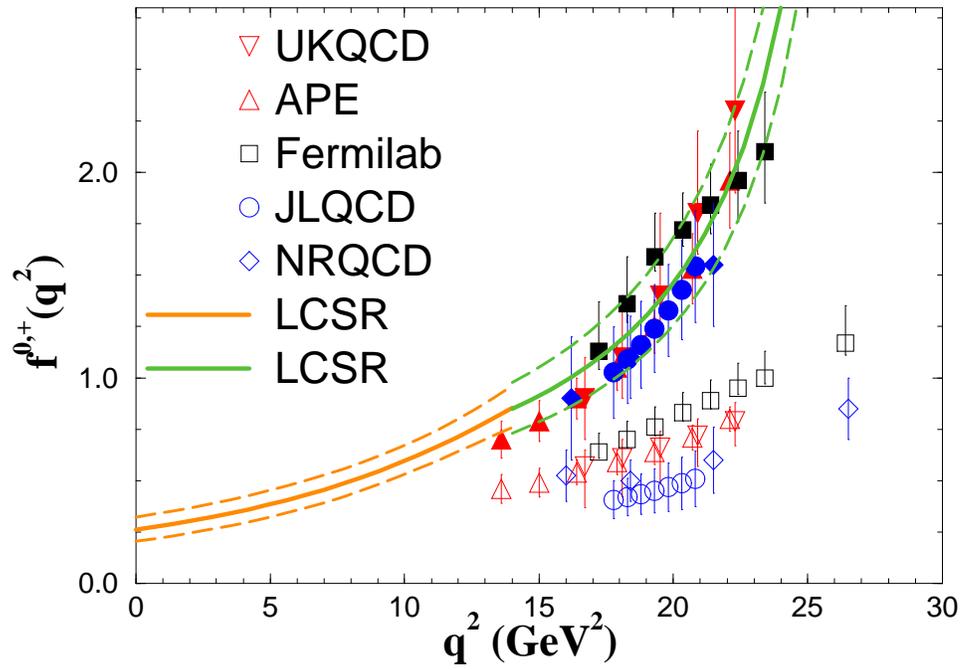}
  \end{center}
  \caption{
    $\Btopi$ form factors by different lattice groups.
    Filled symbols represent $f^+(q^2)$ and open symbols are
    $f^0(q^2)$.
  }
  \label{fig:fpf0_2}
\end{figure}

\begin{table}[tbp]
\begin{center}
\begin{tabular}{llllll}
\hline
method                & stat.   & disc.   &  $1/M$ extrap. & pert.\\
\hline
NRQCD (JLQCD)          & 10\%    &  16\%   &  --            & 4\% \\
extrap. (APE)          & 10\%    &  5\%    &  15\%          & -- \\
\hline
NRQCD (future)        & $<$ 5\% &  4-10\% &  --            & 4\% \\
extrap.+ HQET (future) & $<$ 5\% &  5\%    &  5\%           & -- \\
\hline
\end{tabular}
\end{center}
\caption{Typical errors of the form factor $f^+(q^2)$ for 
$q^2> 16$ GeV$^2$ at present and future prospects. 
``stat.'', ``disc.'', ``$1/M$ extrap'', and ``pert.'' stand
for statistical, discretizations, $1/M$ extrapolation, and
perturbative errors.}
\label{error_fp}
\end{table}

Table~\ref{error_fp} shows the errors by the JLQCD
collaboration (NRQCD) and the APE collaboration
(the extrapolation method). 
We also list expected errors in unquenched lattice
calculations with $a^{-1}=2^3$ GeV in the near future.
The present NRQCD method (JLQCD) has a large discretization error 
since $a^{-1}=1.6$ GeV is used. 
It would be possible to carry out simulations with $a^{-1} =
2-3$ GeV, so that the error of $O((ak)^2)$ are reduced to
4-10\%. 
The discretization error in the extrapolation method appears
in the lattice results for the charm quark region themselves,
which is propagated through the final result by
extrapolation in $1/M$. 
Since the present quenched calculation is carried out with
$a^{-1} = 2.7$ GeV, they will remain to be of the same
order. 

There are several proposals to improve the form factor
determination. 
The quenching error can be resolved only by performing the
unquenched calculations. 
Recently, the JLQCD collaboration has accumulated $n_f=2$
unquenched lattice configurations with $O(a)$-improved
Wilson fermions \cite{Aoki:2002uc}, and $n_f=2+1$ unquenched
configurations with improved staggered fermions have been
produced by the MILC collaboration \cite{Davies:2003ik}. 
These unquenched QCD data should be applied to form factor
calculations. 


Using the heavy quark symmetry is another way of improvement. 
Since the CLEO-c experiment can measure form factors for
$\Dtopi$ to a few percent accuracy, their results will be a
good approximation for the $\Btopi$ form factors. 
Then the task for lattice QCD is to provide the $1/m_Q$
dependence of the form factors. 
The $B$ to $D$ ratio 
$\displaystyle
 \frac{d\Gamma(\Btopi)/d(v\cdot k_\pi)}{d\Gamma(\Dtopi)/d(v\cdot k_\pi)}
$ 
with the same recoil energy $\vdotk$ would be a nice
quantity to measure on the lattice, since a large part of
the statistical error, the perturbative error and  the
chiral extrapolation errors are expected to cancel in this
ratio. 

In the large recoil momentum region the light-cone QCD sum
rule (LCSR) may be used to calculate the form factors
\cite{Khodjamirian:1998ji,Ball:1998tj,Khodjamirian:2000ds,Ball:2001fp}.
In Figure~\ref{fig:fpf0_2} the latest result \cite{Ball:2001fp}
is also shown on top of the lattice calculations.
The curve below $q^2$ = 14~GeV$^2$ is the LCSR result.
At $q^2$ = 14~GeV$^2$, it is connected to a pole dominance
model $f^+(q^2)=c/(1-q^2/m_{B^*}^2)$, which is expected to
be a good approximation when $q^2$ is close to $m_{B^*}^2$.

Model independent bounds for the whole $q^2$ range
can be obtained with dispersion relation, perturbative QCD,
and lattice QCD data \cite{Lellouch:1995yv}.
Reducing the lattice errors or having other inputs would
significantly improve the results. 
More elaborate studies along this line would be important. 

Recently, the UKQCD collaboration \cite{Bowler:2004zb} and 
the SPQcdR collaboration \cite{Abada:2002ie} performed
studies of $\Btorho$ form factors. 
Both collaborations use $O(a)$-improved Wilson action for
the heavy quark and extrapolate the numerical results of
$m_Q \sim m_c$  toward the physical b quark mass. 
UKQCD obtained the partially integrated decay rate in the
region 
$12.7\,\mathrm{GeV}^2<q^2<18.2\,\mathrm{GeV}^2$
as
$\Gamma = (4.9^{+12+\ 0}_{-10-14})\times 10^{12} s^{-1} |V_{ub}|^2$.

\subsection{Measurement of inclusive $b \rightarrow u$ semileptonic decays}
Measurement of the inclusive rate for 
$B \rightarrow X_u\ell \nu$ decays  
is the most straightforward approach to determine $|V_{ub}|$.
OPE provides a firm theoretical basis to convert the measured rate 
to $|V_{ub}|$.
However, this is true only when the total rate can be measured
with a small enough error.
Experimentally, however, we have to introduce cuts on kinematical
variables, such as $E_{\ell}$, $m_{X}$ and $q^2$, to reduce the
huge $B \rightarrow X_c \ell \nu$ background, and only a limited phase 
space is available in a practical measurement.
This complicates the situation, and the cut has to be chosen 
carefully to minimize the theoretical uncertainties.
As discussed in the above section, one of the best strategies is 
to apply a combined cut on $(q^2, m_X)$.

In $\Upsilon(4S)$ experiments, a correct measurement of $(q^2, m_X)$ 
is possible when the accompanying $B$ decays are fully reconstructed.
This technique, referred to as ``full reconstruction tagging'', 
allows us to isolate tracks from the signal $B$ decays for correct 
reconstruction of $m_X$, and also to determine the momentum vector
of the signal-side $B$ meson.
The latter helps to improve reconstruction of the missing neutrino,
leading to correct reconstruction of $q^2$ and better discrimination 
of $B \rightarrow X_c \ell \nu$ background leaking into the signal 
phase space.
Full reconstruction tagging also allows us to determine the
flavor and charge of the signal-side $B$.
It helps to identify the signal lepton using the correlation between the
$B$ flavor and lepton charge, and also to measure the decay rate 
separately for neutral and charged $B$ mesons.
However, this ultimate measurement requires a large accumulation of 
$B\overline{B}$ data because of the relatively small efficiency in the 
full reconstruction of the accompanying $B$'s (a few times 0.1\%).
Therefore, the high luminosity of the SuperKEKB provides an unique
opportunity to perform this measurement with high statistics.

Such analysis with the full reconstruction tagging is being performed
with the present Belle data, although with limited statistics.
Figure~\ref{fig:mx_present} shows the distribution of reconstructed
$m_X$ for the $q^2$ region above 6 GeV$^2$, based on  140 fb$^{-1}$ 
of data accumulated by Summer 2003. 
In this data sample, the number of fully reconstructed events is
about $1.4 \times 10^{5}$ ($9.1 \times 10^{4}$ of $B^+$ and 
$4.7 \times 10^{4}$ of $B^0$ events), and it corresponds to the
full reconstruction efficiency ($\epsilon_{frecon}$) of 0.1\%.
In the low $m_X$ region, one can see a clear enhancement of the 
$B \rightarrow X_u \ell \nu$ signal over the expected
$B \rightarrow X_c \ell \nu$ background.
In the region  $m_X < 1.5$ GeV/$c^2$, for example, the observed 
signal ($S$) is 68 with a predicted background ($B$) of 56, 
resulting in a signal-to-background ratio ($S/B$) of 1.2.
Table~\ref{tbl:eff_and_sn} summarizes the efficiency to detect the
$B \rightarrow X_u \ell \nu$ signal once the accompanying $B$ is
fully reconstructed ($\epsilon_{b \rightarrow u}$), and the 
signal-to-background ratio ($S/B$).
Here, values are shown for three $m_X$ cuts, 
$m_X < 1.5, 1.7$ and 
1.86\,GeV/$c^2$, with the $q^2$ cut fixed at 6 GeV$^2$.
We note that a similar analysis by the BaBar collaboration 
has better $S/B$ by a factor of 2~\cite{Aubert:2003zw}.

\begin{figure}[tbp]
\begin{center}
\includegraphics[width=0.9\textwidth,clip]{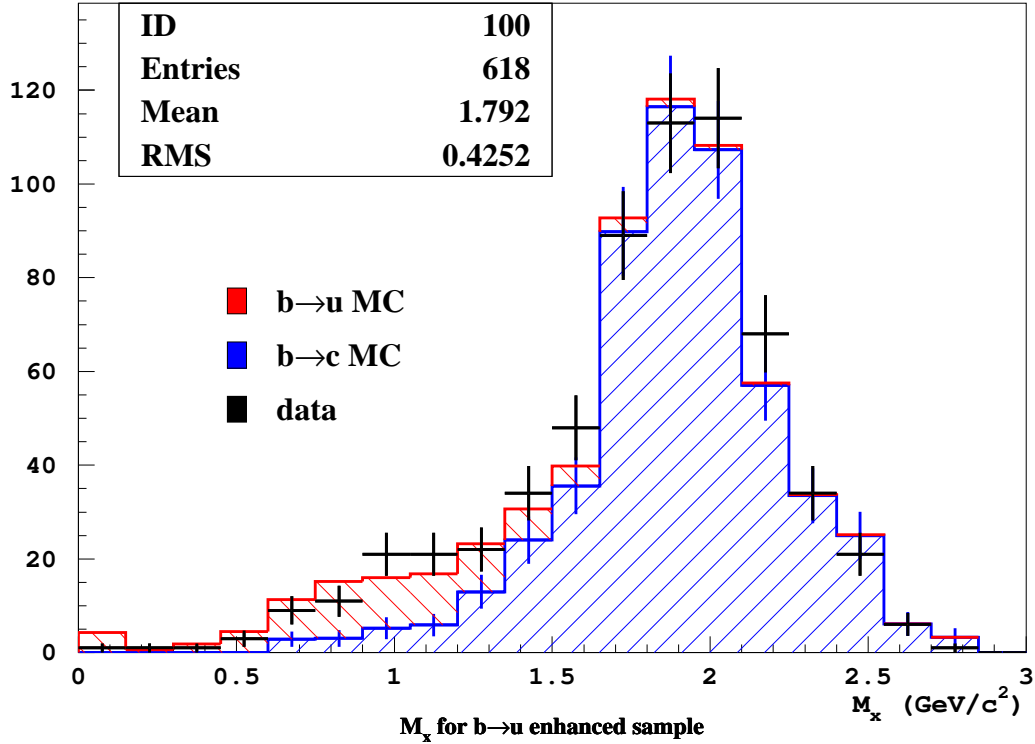}
\end{center}
\caption{Distribution of reconstructed $m_X$ with a full reconstruction
analysis using 141 fb$^{-1}$ of data.}
\label{fig:mx_present}
\end{figure}

\begin{table}[tbp]
\begin{center}
\begin{tabular}{cccc}
\hline\hline
Quantity & \multicolumn{3}{c}{$m_{X cut}$}                \\
         & 1.5 GeV/$c^2$ & 1.7 GeV/$c^2$ & 1.86 GeV/$c^2$ \\
\hline
$\epsilon_{b \rightarrow u}$ & 0.14 & 0.17 & 0.17         \\
$S/B$                        & 1.21 & 0.61 & 0.32         \\
\hline\hline
\end{tabular}
\end{center}
\caption{Efficiencies to detect the 
         $B \rightarrow X_{u} \ell \nu$ signal for the fully 
         reconstructed events ($\epsilon_{b \rightarrow u}$) and
         the signal-to-background ratio ($S/B$) for three 
         different cut values.}
\label{tbl:eff_and_sn}
\end{table}

Based on extrapolation of the present Belle results and realistic
assumptions for the improvement in $\epsilon_{frec}$ and $S/B$, 
we have estimated the $|V_{ub}|$ precision at a SuperKEKB/Belle
experiment. 
We consider here the statistical, experimental systematic and 
theoretical errors, and estimate them as follows.

\begin{description}
\item[Statistical error:]
The statistical error ($\Delta_{stat}$) is simply scaled with the 
integrated luminosity ($L$),
\begin{equation}
\Delta_{stat} = \frac{1}{2} \times 
\sqrt{\frac{1+(S/B \times f_{S/B})^{-1}}
{n_0 \times f_{frec} \times \epsilon_{b \rightarrow u} \times L}}
~~~~,
\end{equation}
where $n_0$ is the rate of $B \rightarrow X_u \ell \nu$ decays after  
full reconstruction, estimated as $68/(0.14 \times 141) = 3.2/$fb$^{-1}$.
The factor $f_{frec}$ accounts for the improvement in $\epsilon_{frec}$,
that is $2.2\times 10^{5}/1.4 \times 10^{5} = 1.6$ 
(see Table~\ref{frec:hadronic_tag}).  
The factor $f_{S/B}$ accounts for a possible improvement in $S/B$ from
the results in Table~\ref{tbl:eff_and_sn}, and is assumed
to be $f_{S/B} = 2$ based on the above mentioned BaBar result.

\item[Experimental systematic error:]
The major source of experimental systematic error ($\Delta_{syst}$) 
will be associated with the background subtraction, and largely
depend on the signal-to-background ratio.
For instance, the $(q^2, m_X)$ measurement by Belle using an advanced 
neutrino reconstruction technique~\cite{Kakuno:2003fk} has
$S/B = 0.18$ and the total experimental systematic error 
(in $Br(B \rightarrow X_u \ell \nu)$) of 18.1\%, 
dominated by the $B\overline{B}$ background subtraction error of 16.8\%.
Relying on this result and assuming a naive scaling of the background 
subtraction error with $(S/B)^{-1}$, $\Delta_{syst}$ is estimated as,
\begin{equation}
\Delta_{syst} = \frac{1}{2} \times
[0.168 \times \frac{0.18}{S/B \times f_{S/B}} \oplus 0.03]~~~~. 
\end{equation}
Here, we add in quadrature a 3\% error associated with the signal detection 
efficiency. 
The estimation then gives $\Delta_{syst} = 2.8$\% in $|V_{ub}|$.
 
\item[Theoretical error:]
As discussed in Section~\ref{sec:vub_theory_incl}, the theoretical 
error ($\Delta_{theo}$) is minimum at a choice of 
$(q^2_{cut}, m_{X cut}) = (6 \mbox{GeV}^2, 1.86 \mbox{GeV}/c^2)$.
The dominant error in this case is the uncertainty in the $b$-quark mass.
Based on a combined fit to recent experimental data of $B$ semileptonic 
decays, Bauer, Ligeti, Luke and Manohar have deduced 
$m_b^{1S} = 4.74 \pm 0.10$\,GeV, where the error is dominated by 
experimental uncertainties~\cite{Bauer:2002sh}.
If the experimental uncertainties are eliminated in the future, the present
100\,MeV error shrinks to 30\,MeV.
Then $\Delta_{theo} = 1/2 \times 9 = 4.5$\% in $|V_{ub}|$, 
according to Table~\ref{finaltable}.
\end{description}

Adding the above three error sources in quadrature, 
Figure~\ref{fig:vub_error_incl} demonstrates the expected improvement 
of $|V_{ub}|$ error as a function of the integrated luminosity $L$, 
with the optimal choice of 
$(q^2_{cut}, m_{X cut}) = (6 \mbox{GeV}^2, 1.86 \mbox{GeV}/c^2)$.
In conclusion, with a precise determination of the $b$-quark mass, 
a $|V_{ub}|$ error of less than 5\% is achievable at $L = 5$ ab$^{-1}$.

\begin{figure}[htb]
\begin{center}
\includegraphics[width=0.9\textwidth,clip]{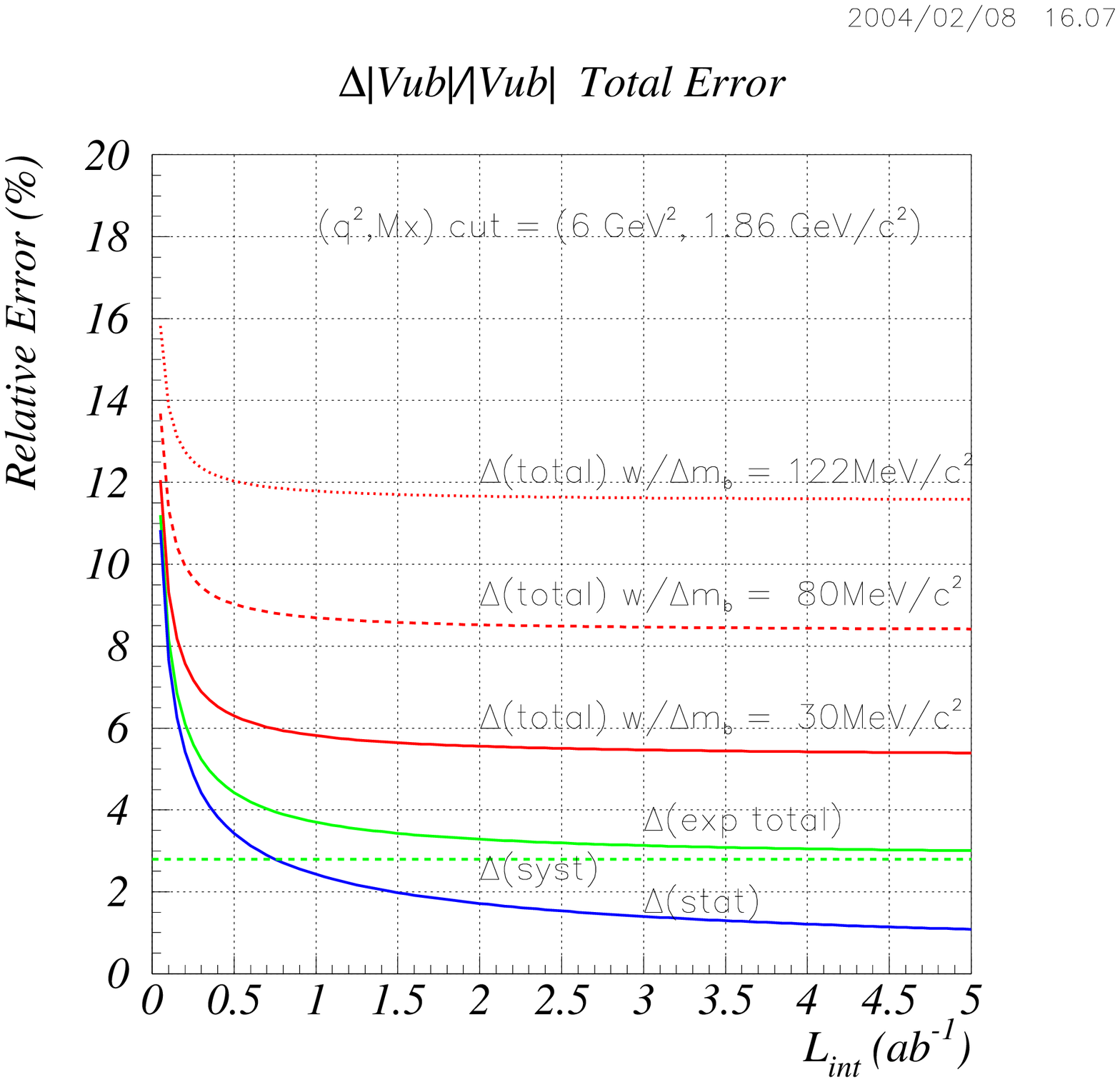}
\end{center}
\caption{Expected improvement of $|V_{ub}|$ error as a function
of $L$.}
\label{fig:vub_error_incl}
\end{figure}

\subsection{Measurement of exclusive $b\rightarrow u$ semileptonic decays}
The measurement of the inclusive $B \rightarrow X_u \ell \nu$ decay
is insensitive to theoretical ambiguities, but experimentally 
challenging because of the large background from 
$B \rightarrow X_c \ell \nu$ decays.
Complementary to this, the measurement of exclusive decays, 
such as $B \rightarrow \pi \ell \nu$ and $B \rightarrow \rho \ell \nu$,
provides experimentally cleaner information, but is subject to 
large theoretical uncertainties in the form factors.
On the experimental side, it is essential to provide precise data
for the  differential rates $d\Gamma/dq^2$ of each exclusive channel.
This is because $d\Gamma/dq^2$ varies depending on theory models,
and such data helps to test the model.
Precise data in the high $q^2$ region is especially important,
since lattice-QCD calculations, the most promising tool for
reliable model-independent determination of $|V_{ub}|$, are
possible only in the region $q^2 > 18$GeV$^2$, as discussed
in Section~\ref{sec:vub_theory_excl}.
While there have been several exclusive measurements in the literature
~\cite{Alexander:1996qu,Behrens:1999vv,Aubert:2003zd,Athar:2003yg,%
  Schwanda:2003bj,Abe:2003rh},
these data lack information on the $q^2$ distribution or, even if they
include such information, suffer from poor statistics and from 
relatively large systematic errors.
A high luminosity SuperKEKB/Belle experiment will enable us to 
measure the $q^2$ distributions with high statistics and 
less systematic uncertainties.
Combined with improvements in lattice QCD with unquenched calculations
in future, this will lead to useful determinations of $|V_{ub}|$.

One of key experimental issues in measuring the exclusive 
$B \to \pi\lnu/\rho\lnu$ decays is the reconstruction of the
undetected neutrino in the final state.
Information on missing energy and momentum of the event have been 
used to infer information about the missing
neutrino (``neutrino reconstruction'').  
This method, originally developed by CLEO, has been applied in
existing measurements and exploits the known kinematics of the 
$\epem \to \Upsilon(4S)$ reaction and near $4\pi$ coverage (``{\it
hermeticity}'') of the detector.
In reality, however, hermeticity of a detector is never complete.
This lack of hermeticity allows background from
 both $B\overline{B}$  and 
cross-feed ($e.g.$, $\pi \ell \nu \leftrightarrow \rho \ell \nu$) 
to contribute,
where the latter is serious especially in the high $q^2$ region.
For instance, the recent CLEO measurement~\cite{Athar:2003yg}
provides $Br(B \rightarrow \pi \ell \nu)$ with a statistical error 
of $13.5(36.0)$\% and an experimental systematic error of $8.6(18.3)$\% 
for the whole $q^2$ ($q^2$($>16$GeV$^2$) region.
The experimental systematic errors are mainly associated with the 
neutrino reconstruction; 6.8\% in the whole $q^2$ region and 17.2\% 
in the $q^2 > 16$GeV$^2$ region.
The CLEO result has been obtained with event sample of only 
$\sim 10~\ifb$, and we can quickly improve the statistical error
to a few \%.
However, the systematic error arising mainly from the neutrino 
reconstruction will soon limit the experimental uncertainties.
  
As in the case of the inclusive measurement, which is 
discussed in the
previous section, analyses with full reconstruction tagging
improve the situation substantially, and make best use of the
high luminosity at  SuperKEKB/Belle.
Figure~\ref{fig:mm2comp} compares the missing mass resolution for 
(a) full reconstruction tagging (for $B\to D^0 \lnu$ with $78~\ifb$), 
and (b) classical $\nu$ reconstruction (for $B\to \omega \lnu$ 
with $78~\ifb$) in the present Belle analyses.  
An improvement in the FWHM by almost a factor of 50 is seen for the 
full reconstruction analysis.  
We can also consider semileptonic tagging, where one tags more
abundant $B\to D^{(*)}\lnu$ decays in the accompanying $B$ decays.
This technique provides about 4 times more statistics by sacrificing
purity and $q^2$ resolution.
Belle has presented a preliminary inclusive analysis result
using this 
method~\cite{Sugiyama:2003rh}. 
An exclusive analysis using this method is also in progress.

\begin{figure}[tbp]
\begin{center}
\includegraphics[width=0.49\textwidth,clip]{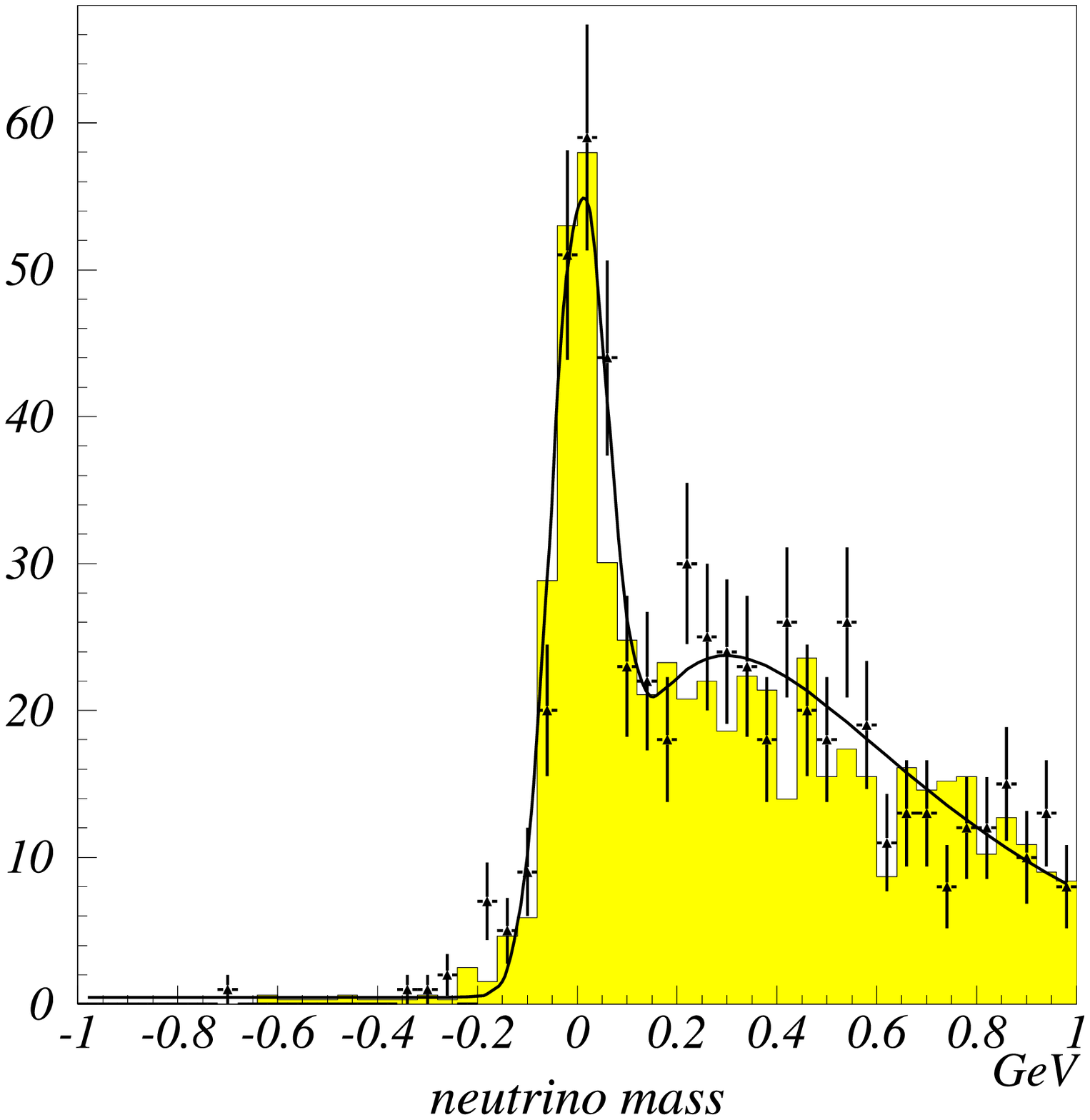}
\includegraphics[width=0.49\textwidth,clip]{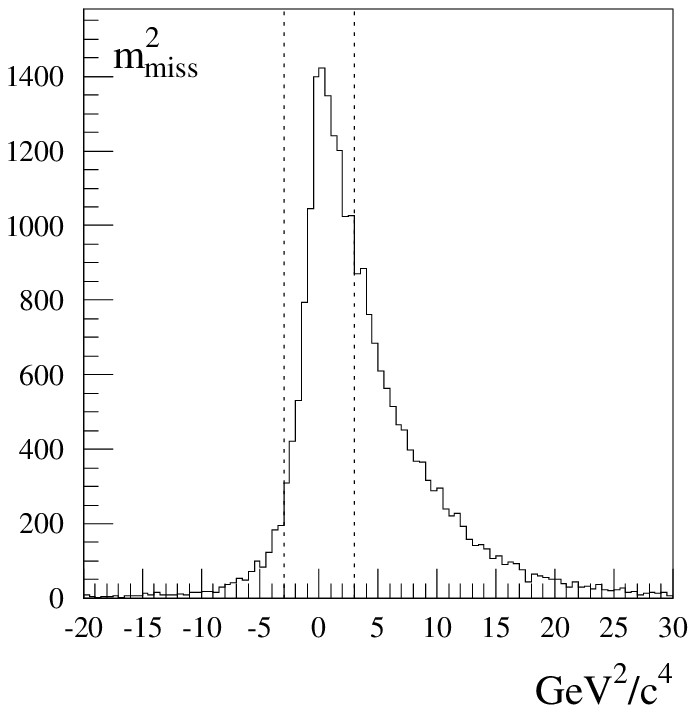}
\end{center}
\caption{Missing mass resolution in (left) a full reconstruction
analysis and (right) a classical neutrino reconstruction analysis.}
\label{fig:mm2comp}
\end{figure}

Figure~\ref{fig:vub_err_excl} shows the expected improvement of
the experimental error in $|V_{ub}|$ as a function of the integrated 
luminosity $L$, for the whole $q^2$ and the high $q^2$ regions.
Here, the errors are estimated by extrapolating the present Belle 
analysis using  semileptonic tagging, and are compared 
with the classical neutrino reconstruction.
One can see that the classical neutrino reconstruction will soon hit
the systematic limit.
At a few times 100 fb$^{-1}$, the tagging analysis will provide more 
precise results.
The experimental precisions in $|V_{ub}|$ expected at 5 and 50 ab$^{-1}$ 
are $2.3$\% and $1.9$\%, respectively, for the whole $q^2$ region,
and $2.9$\% and $2.1$\%, respectively, for the $q^2 > 16$GeV$^2$ 
region.

\begin{figure}[tbp]
\begin{center}
\includegraphics[width=0.49\textwidth,clip]{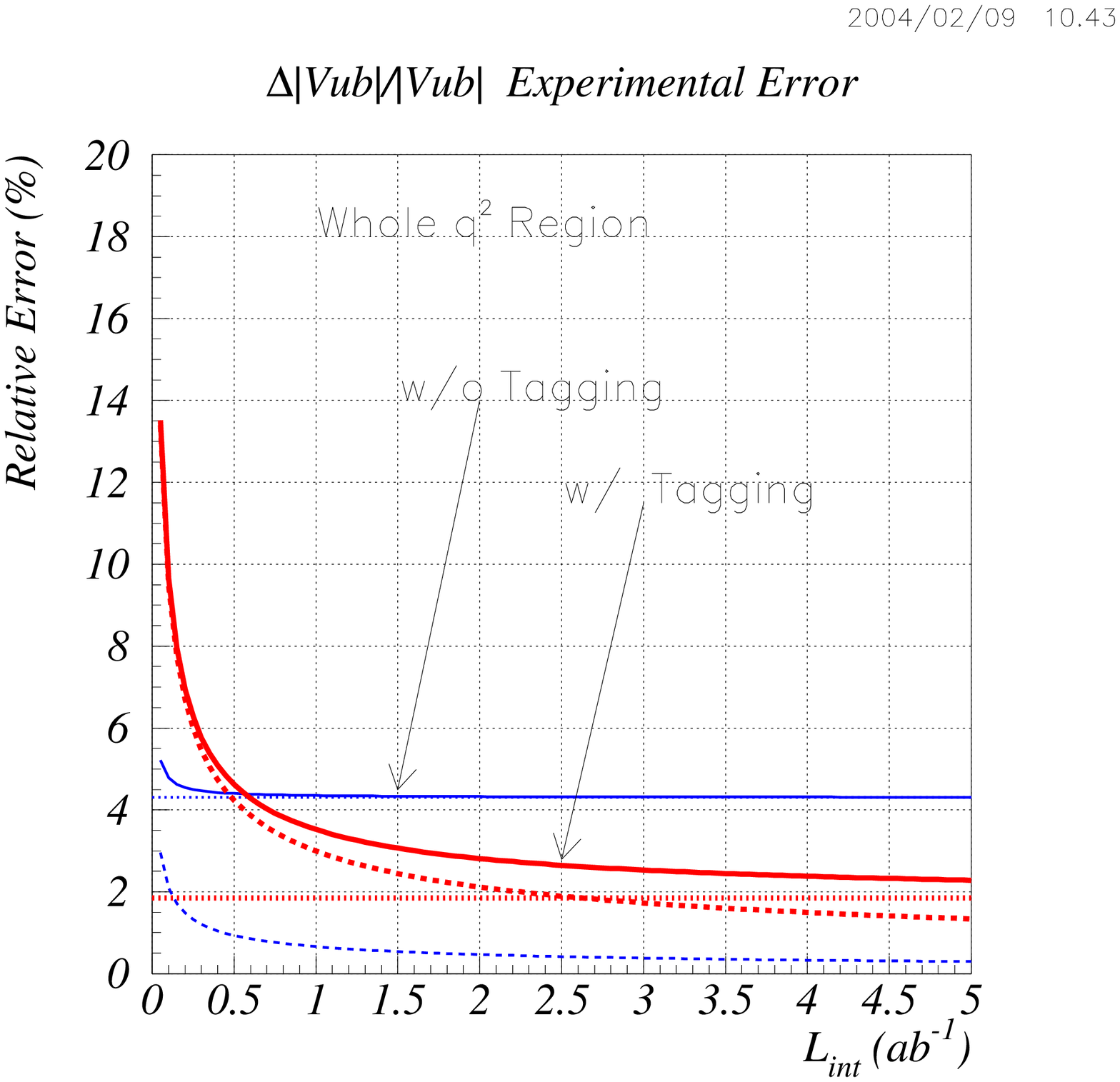}
\includegraphics[width=0.49\textwidth,clip]{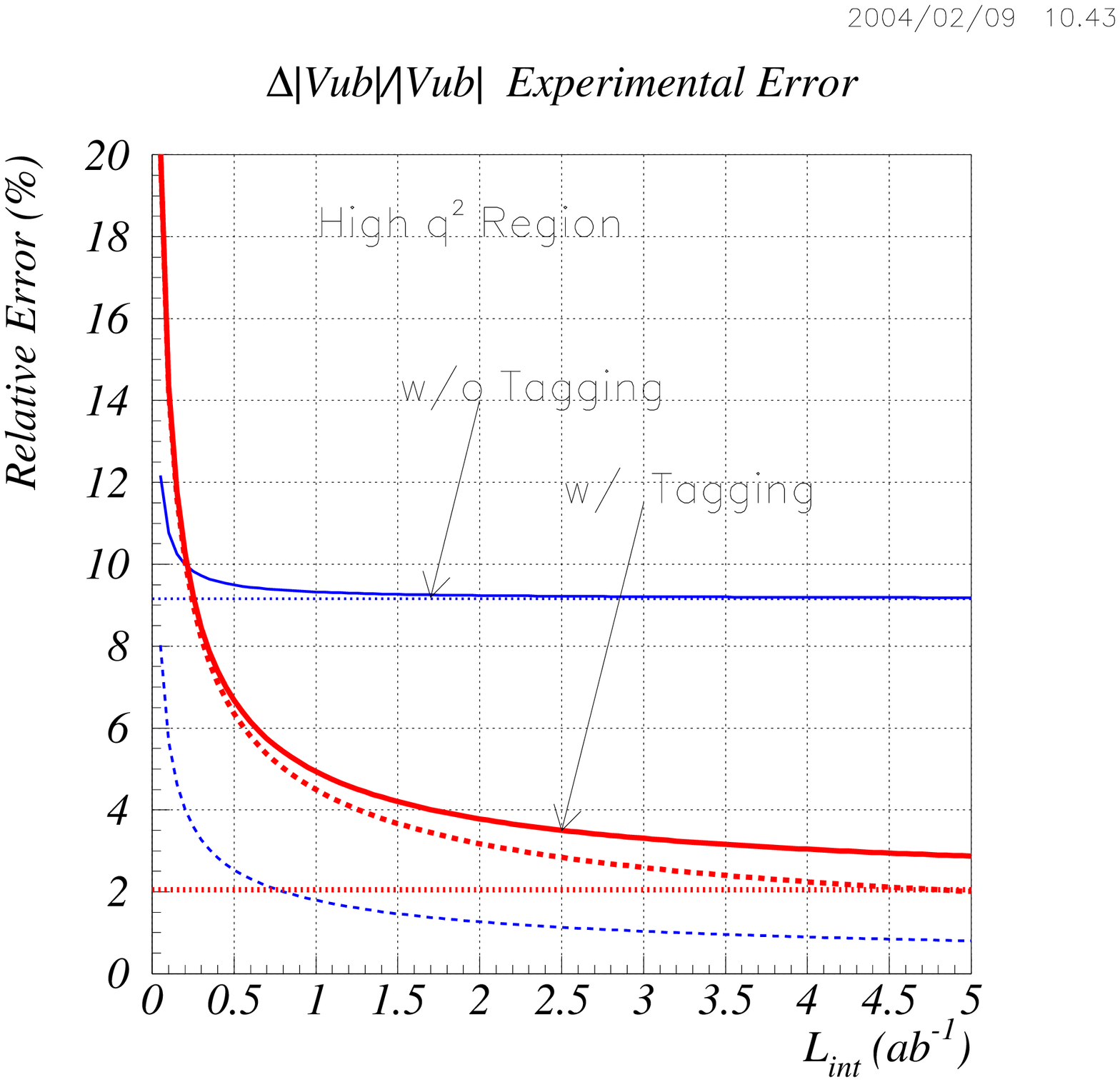}
\end{center}
\caption{Expected improvement of the experimental error in $|V_{ub}|$ 
as a function of the integrated luminosity $L$, (left) for the whole 
$q^2$ and (right) the high $q^2$ regions.}
\label{fig:vub_err_excl}
\end{figure}





%
%
%

\clearpage \newpage
\def\TMG{\tau^-\rightarrow\mu^-\gamma}
\def\TEG{\tau^-\rightarrow e^-\gamma}
\def\TME{\tau^-\rightarrow\mu^-\eta}
\def\TMEetc{\tau^-\rightarrow\mu^-(e^-)\eta, \mu^-(e^-)\eta', \mu^-(e^-)\pi^0}
\def\TMMM{\tau^-\rightarrow \mu^-\mu^+\mu^-}
\def\TLLL{\tau^-\rightarrow l^-l^+l^-}
\def\TmueG{\tau^-\rightarrow\mu^-{\rm (e^-)}\gamma}
\def\Tarrow{\tau\^\rightarrow}
\def\TT{\tau^+\tau^-}
\def\GT{\tau}
\def\GG{\gamma}
\def\GM{\mu}
\def\MG{\mu\gamma}
\def\MM{\mu\mu}
\def\BB{\rm{B\overline{B}}}
\def\MnotM{\GM^{not}\GM}
\def\Mnot{\GM^{not}}
\def\Minv{M_{inv}}
\def\DelE{\Delta E}

\section{Tau decays}

\subsection{Introduction}
The $\tau$ lepton is the only charged lepton that is heavy
enough to decay hadronically. 
Its variety of pure-leptonic and semi-leptonic decay modes makes
it possible to study various physics issues. 
With an anticipated luminosity of 
$5\times 10^{35}$ cm$^{-2}$sec$^{-1}$, SuperKEKB is expected to deliver 
a sample of $1.5\times 10^{10}$ tau pairs in three years
of data taking.
This huge data sample allows us to attain a single
event sensitivity of $3\times 10^{-10}$ for branching
fractions for a 10\% detection efficiency
with no backgrounds.
Research making the best use of such a large
data sample can be divided into two categories:
the sensitivity frontier and the precision frontier. 
The sensitivity frontier involves searches for new physics phenomena 
in rare or forbidden decays, while on the precision frontier
one searches for a small inconsistency with the SM in a high
precision measurement.
Among many possible subjects at the sensitivity frontier, 
here we discuss the search for
Lepton Flavor Violation (LFV) phenomena at SuperKEKB.

A theoretical overview of lepton-flavor-violating process
beyond the Standard Model is given in
Section~\ref{sec:Lepton_flavor_violation}.

\subsection{Present experimental status}

\begin{figure}[tbp]
  \begin{center}
    \includegraphics*[width=0.8\textwidth,clip]{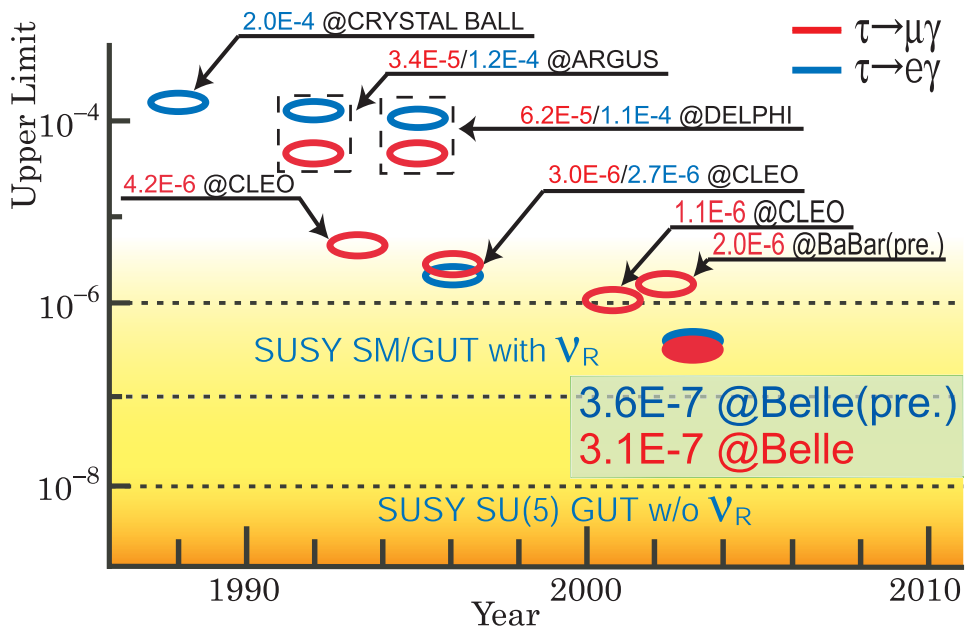}
  \end{center}
  \caption{Experimental status of the LFV search.}
  \label{fig:Ffive}
\end{figure}

So far, tau physics has been carried out mostly at 
electron-positron collider facilities. Samples of 
$e^+e^-\rightarrow\tau^+\tau^-$ reactions are selected based on
characteristic properties, such as low multiplicity with a
well collimated back-to-back jet-like pattern and 
missing momentum (energy) due to missing neutrinos.
While these requirements will remove $B\bar{B}$ and a large
portion of continuum reactions, an appreciable amount of
Bhabha, muon-pair and two-photon processes remain and become
severe backgrounds.
Since $\tau$ decays always include neutrinos, the $\tau$ cannot be
exclusively reconstructed. Therefore background
contamination cannot be totally avoided. 
The LFV processes discussed here are exclusive decays and
much higher sensitivity than for other
decay modes could be attained.  

Belle has so far analyzed 86 fb$^{-1}$ of data for
$\tau\rightarrow\mu^-(e^-)\gamma$, 
83 fb$^{-1}$ of data for $\tau^-\rightarrow\mu^-(e^-)h^0$ 
 (where $h^{0}$ denotes either $h^{0}=\eta, \eta^{\prime}$ or $\pi^{0}$),
and 87 fb$^{-1}$ of data for $\tau^-\rightarrow l^-l^+l^-$. 
The current experimental status is summarized
in Table~  \ref{tab:Tone} and 
the history of  $\tau^{-}\rightarrow\mu^-(e^-)\gamma$
searches 
is shown in
Figure~\ref{fig:Ffive}. 
It is seen that the older data, mostly collected by CLEO,
are no longer competitive with the best limits from the Belle
collaboration. 
Upper limits for branching fractions
go down to the 10$^{-7}$ level and
the searches are approaching regions sensitive to
new physics.   

In this section, we discuss the experience
in $\tau$ physics research that we have acquired
during the  analysis of the current Belle data.

While the signal side $\tau$ is exclusively reconstructed for
every LFV process, the tag side is required to be composed
of a single charged track and any number of photons with
neutrinos. 
The mode $\tau^-\rightarrow\mu^-(e^-)\gamma$ has the fewest
constraints, so that a certain amount
of background inevitably remains.
The $\tau^-\rightarrow\mu^-(e^-)h$ mode includes extra constraints
that provide additional
background rejection power, although there is some decrease of the detection
efficiency due to the higher multiplicity. 
Together with the kinematic constraints for the signal,
particle identification, especially for muons and electrons, 
play an essential role for event selection. 
The Belle detector provides a $\mu$-id efficiency of 90\% and 
an $e$-id efficiency of 97-98\%: the event selection power is 
strong and then background contamination is little at the
electron accompanying processes. 
The contamination due to the inefficiency of $\mu$-id
is 28\% for $\tau^- \rightarrow\mu^-\gamma$, while the rate due
to the inefficiency of $e$-id is 6\% in 
$\tau^{-}\rightarrow e^{-}\gamma$. 
  
We have introduced a new requirement on the relation between
the missing momentum and the missing mass-squared in
$\tau^-\rightarrow\mu^-(e^-)\gamma$, in addition to the
conventional requirements. 
It is quite effective; 98\% of generic tau-pairs and 80-90\%
of  radiative Bhabhas, muon-pairs and continuum are
removed, while 76\% of the signal is retained 
(See Figure~\ref{fig:Fsix}). 

\begin{figure}[tbp]
  \begin{center}
    \includegraphics[width=0.49\textwidth,clip]{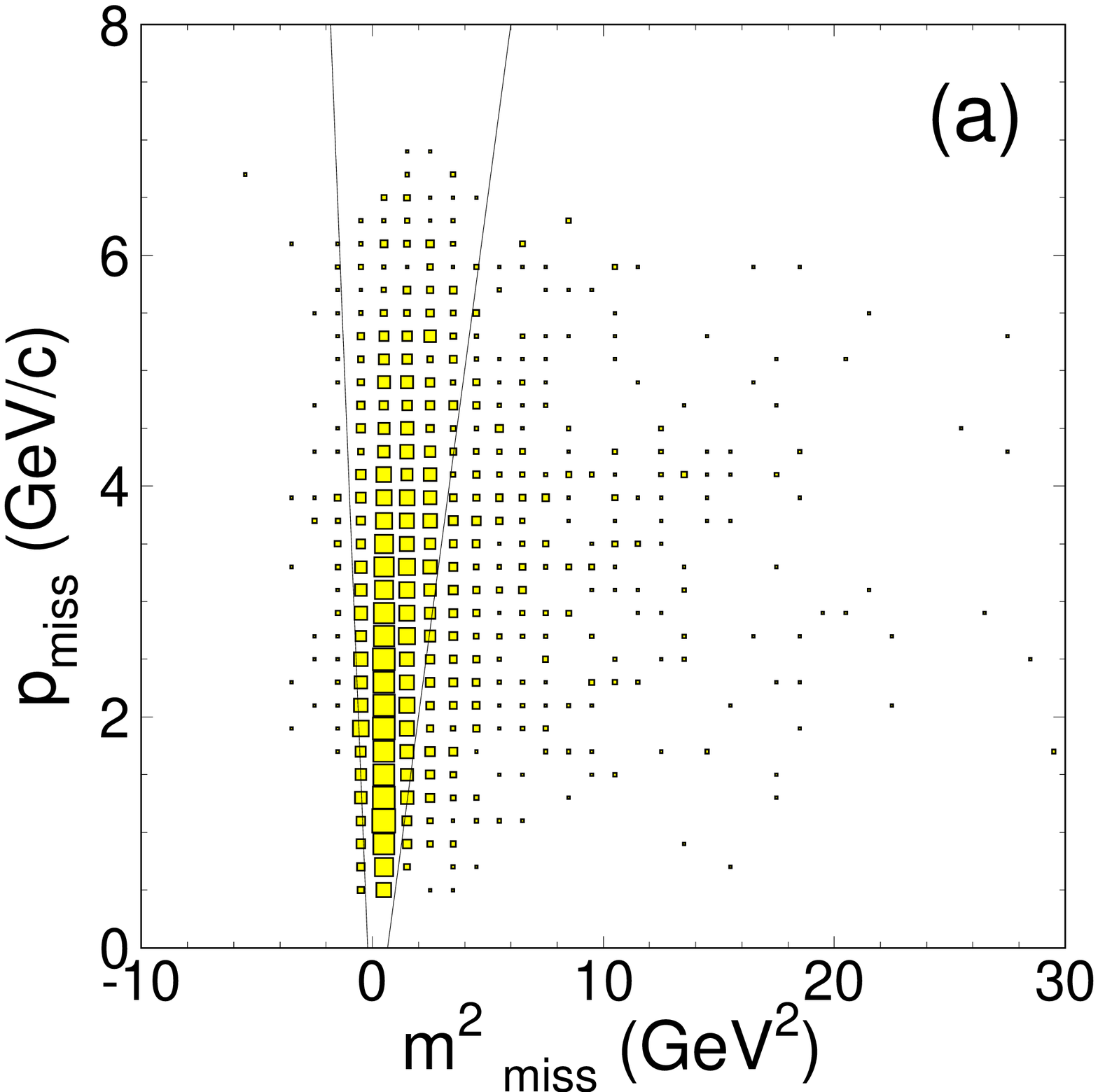}
    \includegraphics[width=0.49\textwidth,clip]{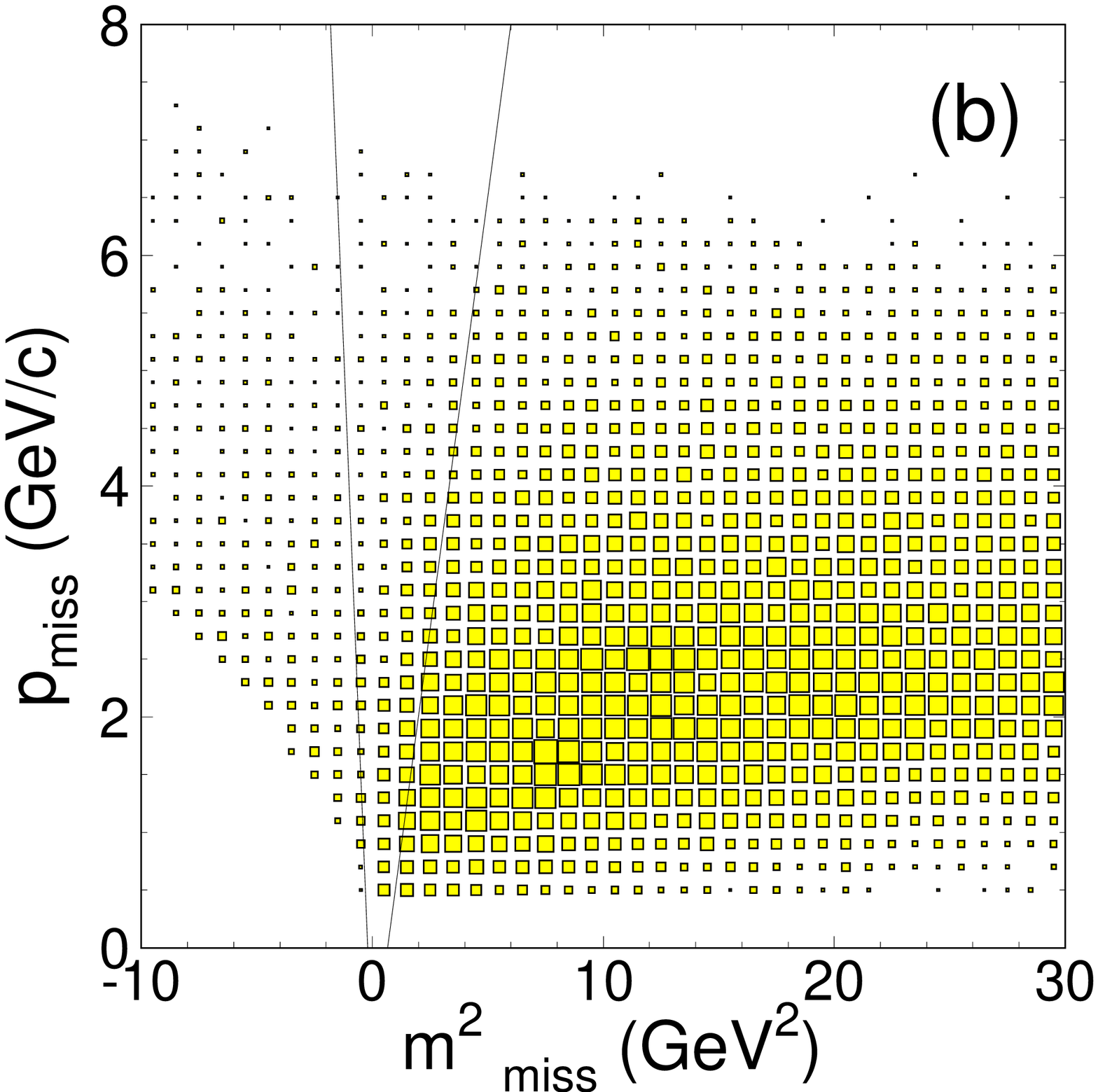}
  \end{center}
  \caption{
    A newly introduced selection regarding 
    the missing quantities,
    $p_{\mathrm{miss}}$~vs~$m_{\mathrm{miss}}^2$ for
    (a) signal-MC events, and (b) for
    generic tau-pair MC.  
    The selected region is the area between the lines. 
  }
  \label{fig:Fsix}
\end{figure}

\begin{figure}[tbp]
  \begin{center}
    \includegraphics[width=0.49\textwidth]{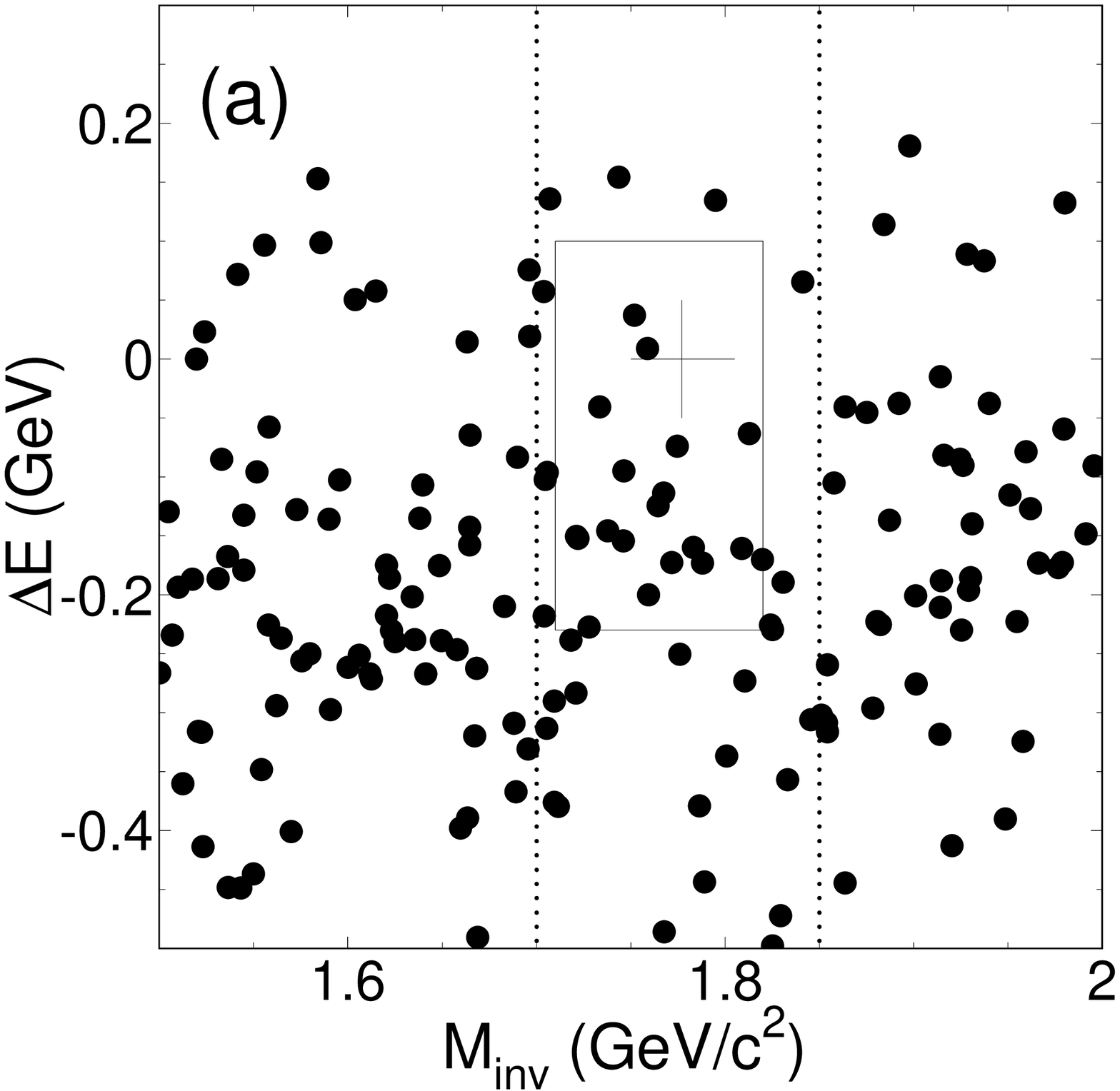}
    \includegraphics[width=0.49\textwidth]{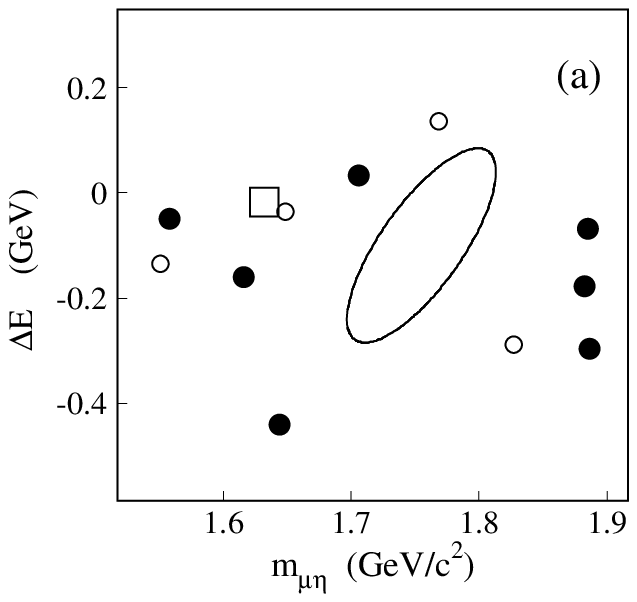}
  \end{center}
  \caption{
    Distributions of events after all selection requirements. 
    (a) $\tau\rightarrow\mu(e)\gamma$ with 86 fb$^{-1}$, 
    (b) $\tau\rightarrow\mu\eta$ with 83 fb$^{-1}$. 
  }
  \label{fig:Fseven}
\end{figure}

Within the data analyzed, no signal candidates are
found in the signal regions for
$\tau^{-}\rightarrow\mu^-(e^-)h^0$ and $\tau^{-}\rightarrow l^-l^+l^-$.
On the other hand, $\tau^{-}\rightarrow\mu^{-}(e^-)\gamma$ suffers
from backgrounds as seen in Figure~\ref{fig:Fseven} (a). 
The principal background remaining after the selections for
$\tau^-\rightarrow\mu^-(e^-)\gamma$ originates from $\tau^+\tau^-$.
In particular, the radiative tau-pair process 
($e^-e^{-} \rightarrow\tau^{+}\tau^{-}\gamma$) dominates. 
One of $\tau$'s decays semi-leptonically from which the lepton and the
radiated photon composes a tau candidate, while the other
tau decays leaving one charged track in the detector with a different
lepton flavor.
Multi-photon radiative muon-pair (and Bhabha events) 
yield the second
largest background: one of leptons and a
radiated photon form a tau candidate, and the other lepton
is mis-identified.
These backgrounds cannot be discriminated from the true
signal. 

\begin{figure}[tbp]
  \begin{center}
    \includegraphics[angle=90,width=0.8\textwidth,clip]{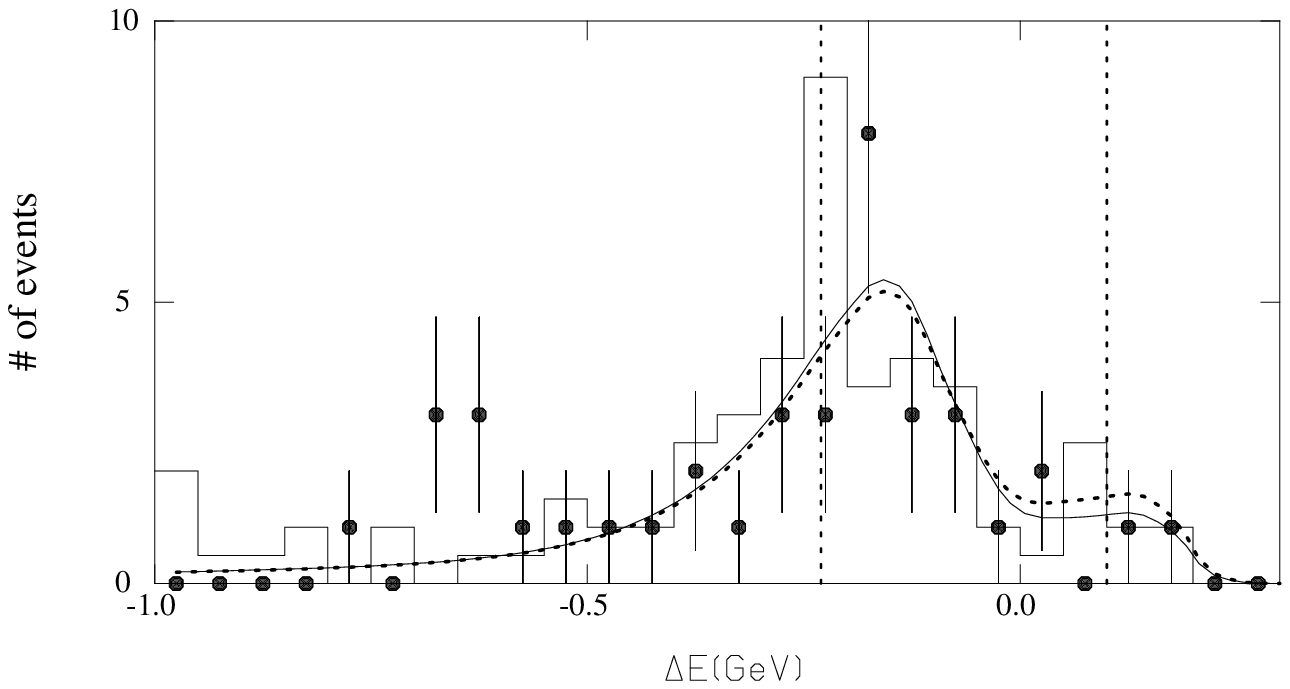}
  \end{center}
  \caption{Expected background distribution for $\TMG$.}
  \label{fig:Feight}
\end{figure}

For 
$\tau^-\rightarrow\mu^-(e^-)\eta$, $\mu^-(e^-)\eta'$,
$\mu^-(e^-)\pi^0$ and $\TLLL$, we evaluate
the expected backgrounds in the signal region from the
sidebands and extract the number of
signal candidates. 
No candidate is found. 
The upper limits are calculated according to a Baysian
approach. 
For $\TmueG$ the background distribution was
intensively studied using actual data and MC simulations, as
is seen in Figure~\ref{fig:Feight}. A thorough understanding
of the background's origin and properties was obtained.
Finally, the number of signal events was obtained by means of an
unbinned extended maximum likelihood method. 


\begin{table}[tbp]
\begin{center}
 \begin{tabular}{l  c c c c } \\
 \hline \hline
  Mode        &     Limit & Luminosity& Refs.& Limit in PDG2000    \\
              &           &  \quad $({\rm fb^{-1}})$  & \\
\hline
\hline 
$\tau^{-} \rightarrow\mu^{-}\gamma $ &  $<3.1\times 10^{-7}$ & 86 &
                      \cite{Abe:2003sx}&  $<11\times 10^{-7}$  \\
$\tau^{-} \rightarrow e^{-}\gamma $ &  $<3.6\times 10^{-7}$ &86  &
                           &  $<27\times 10^{-7}$  \\
$\tau^{-} \rightarrow \mu^{-}\eta $ &  $<3.4\times 10^{-7}$ &83 &
                      \cite{tau:mueta}&  $<96\times 10^{-7}$ 
                         \\
$\tau^{-} \rightarrow  e^{-}\eta $ &  $<6.9\times 10^{-7}$ &83 &
                      &  $<82\times 10^{-7}$  \\
$\tau^{-} \rightarrow  e^{-} e^{+} e^{-}  $ &  $<3.5\times 10^{-7}$ &87
                     & \cite{Yusa:2002ff}&  $<29\times 10^{-7}$ 
                       \\
$\tau^{-} \rightarrow  e^{-} e^{+} \mu^{-}  $ &  $<1.9\times 10^{-7}$ &87 
                     & \cite{Yusa:2002ff}&  $<17\times 10^{-7}$ 
                     \\
$\tau^{-} \rightarrow  e^{-} \mu^{+} e^{-}  $ &  $<1.9\times 10^{-7}$ &87 
                    & \cite{Yusa:2002ff}&   $<15\times 10^{-7}$ \\
$\tau^{-} \rightarrow  e^{-} \mu^{+} \mu^{-}  $ &  $<2.0\times 10^{-7}$ &87 
                   & \cite{Yusa:2002ff}&    $<18\times 10^{-7}$ \\
$\tau^{-} \rightarrow  \mu^{-} e^{+} \mu^{-}  $ &  $<2.0\times 10^{-7}$ &87 
                   & \cite{Yusa:2002ff}&    $<15\times 10^{-7}$ \\
$\tau^{-} \rightarrow  \mu^{-} \mu^{+} \mu^{-}  $ &  $<2.0\times 10^{-7}$ 
                      &87 
                     & \cite{Yusa:2002ff}&  $<19\times 10^{-7}$ \\
\hline
\end{tabular}
\end{center} 
 \caption{
     Limits on lepton-flavor-violating decay modes (90\% confidence level)
    obtained so far from Belle data. 
    The last column shows the previous limits from the PDG 2000 compilation. 
  }
  \label{tab:Tone}
\end{table}

\subsection{Achievable sensitivity at SuperKEKB and physics reaches}
Figures \ref{fig:Fnine} shows the sensitivities anticipated
at SuperKEKB. 
The sensitivities shown by the triangular symbols are
 the expected sensitivities obtained by assuming 
some signal-to-background conditions and applying 
the unbinned extended maximum likelihood method at SuperKEKB.
 The lower-solid line are obtained by assuming no candidate events 
in the signal-region.  

The former case corresponds to  $\TmueG$.
In the latter case, the upper limit decreases inversely 
proportional to the total luminosity $Br \propto 1/N_{\TT}$. 
The $\tau^-\rightarrow\mu^-(e^-)\eta$, $\mu^-(e^-)\eta'$,
$\mu^-(e^-)\pi^0$ and $\TLLL$ modes would follow a $\propto 1/N_{\TT}$ 
behavior for a short time as up-to a luminosity of a few 100 fb$^{-1}$ and 
then gradually change to a $\propto 1/\sqrt{N_{\TT}}$ dependence
as candidates begin to appear. 
Therefore, if the current signal-to-background condition is still maintained, 
the ultimate goal at  5,000 fb$^{-1}$ could be a branching fraction
sensitivity of several $\times 10^{-9}$ for
$\tau^-\rightarrow\mu^-(e^-)\eta$, $\mu^-(e^-)\eta'$,
$\mu^-(e^-)\pi^0$ and $\TLLL$.  
The $\TmueG$ mode could be with a branching fraction sensitivity of
a few $\times 10^{-8}$. 

\begin{figure}[tbp]
  \begin{center}
    \includegraphics[width=0.8\textwidth,clip]{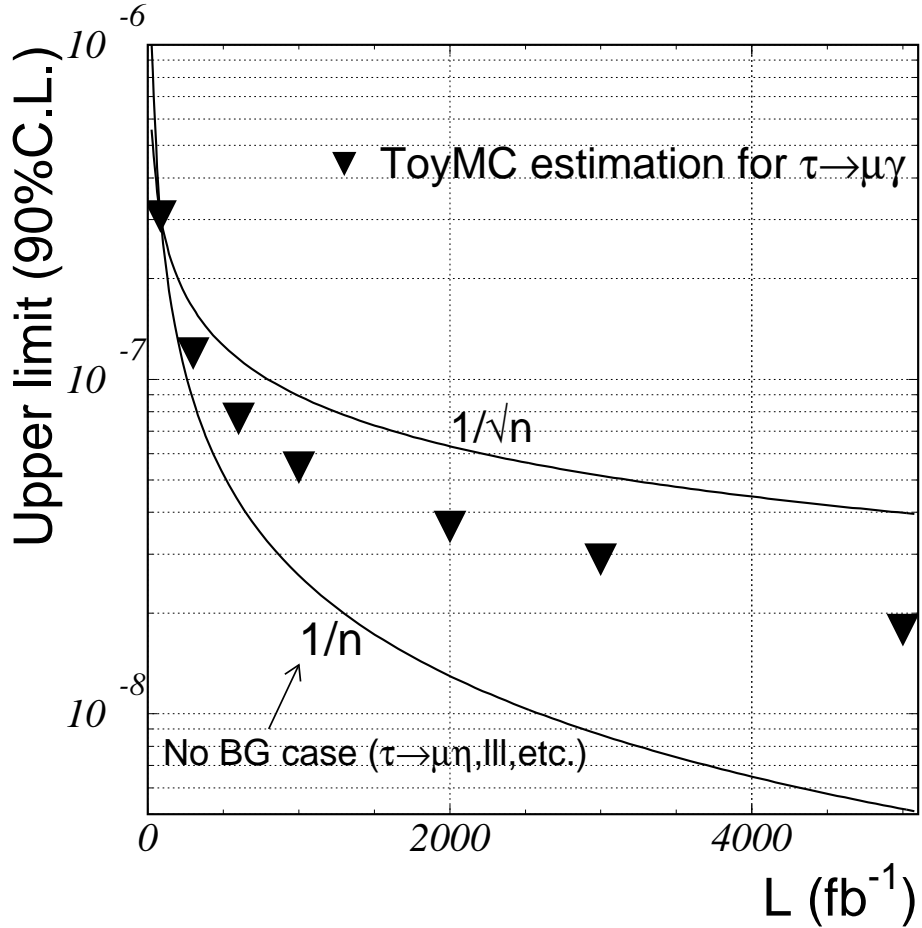}
  \end{center}
  \caption{Achievable upper limits for LFV decays at SuperKEKB.}
  \label{fig:Fnine}
\end{figure}

Can we further improve the sensitivity? 
Experimental sensitivity is in general determined by three 
key elements: statistics, resolutions, and a signal-to-background 
ratio; and its experimental reliability depends on how well 
the systematic uncertainty of these aspects is controlled. 
We discuss here possible measures for 
improvements, based on experience from the Belle $\TmueG$ analysis. 

\paragraph{Statistics}
Besides accumulating higher luminosity, it is important to
the detection efficiency including trigger performance. 
However, at a SuperKEKB,
the selection criteria will become  tighter and 
efficiencies will be lower than the current sizes. 
There is no way to increase the detection efficiency. 

\paragraph{Resolution}
Improvements on the momentum and energy resolution reduce
the size of the signal region. The signal-to-background ratio
and the resulting sensitivity are enhanced  because 
the background distributes uniformly for a narrow 
$\DelE$ vs. $\Minv$ region. 
As is seen in Figure~\ref{fig:Ften}, energy leakage in the  
calorimeter  and  initial state radiation yield a long low-energy 
tail, so that our signal region is defined to be quite large 
to include a sufficient portion of the signal, say, more than 90\%. 
While the effect of initial state radiation cannot be controlled, 
the effect of the calorimeter leakage can be improved,
for instance, by using a crystal with 
a longer radiation length and containing less lateral energy leakage 
in the analysis procedure. 
 
Improvement of particle identification ability is essential 
for all decay modes. 
The efficiency of $e$-id is already 98\% so 
that further improvement
would not be practical. 
On the other hand, the $\mu$-id efficiency is now 90\% so that 
the efficiency increase may reduce
background contamination, for example, 
$(\mu\gamma)$+not$-\mu$ events in the $\TMG$ search, and 
the sensitivity will increase. 
A KLM counter with finer segmentation in the radial direction might 
provide a better muon identification.
K/$\pi$ separation is also important in searches for decays 
into 3 hadrons. 
The Time-Of-Propagation counter being studied by the Nagoya group could
be a very good candidate for improvement here.

\begin{figure}[tbp]
  \begin{center}
    \includegraphics[width=0.6\textwidth,clip]{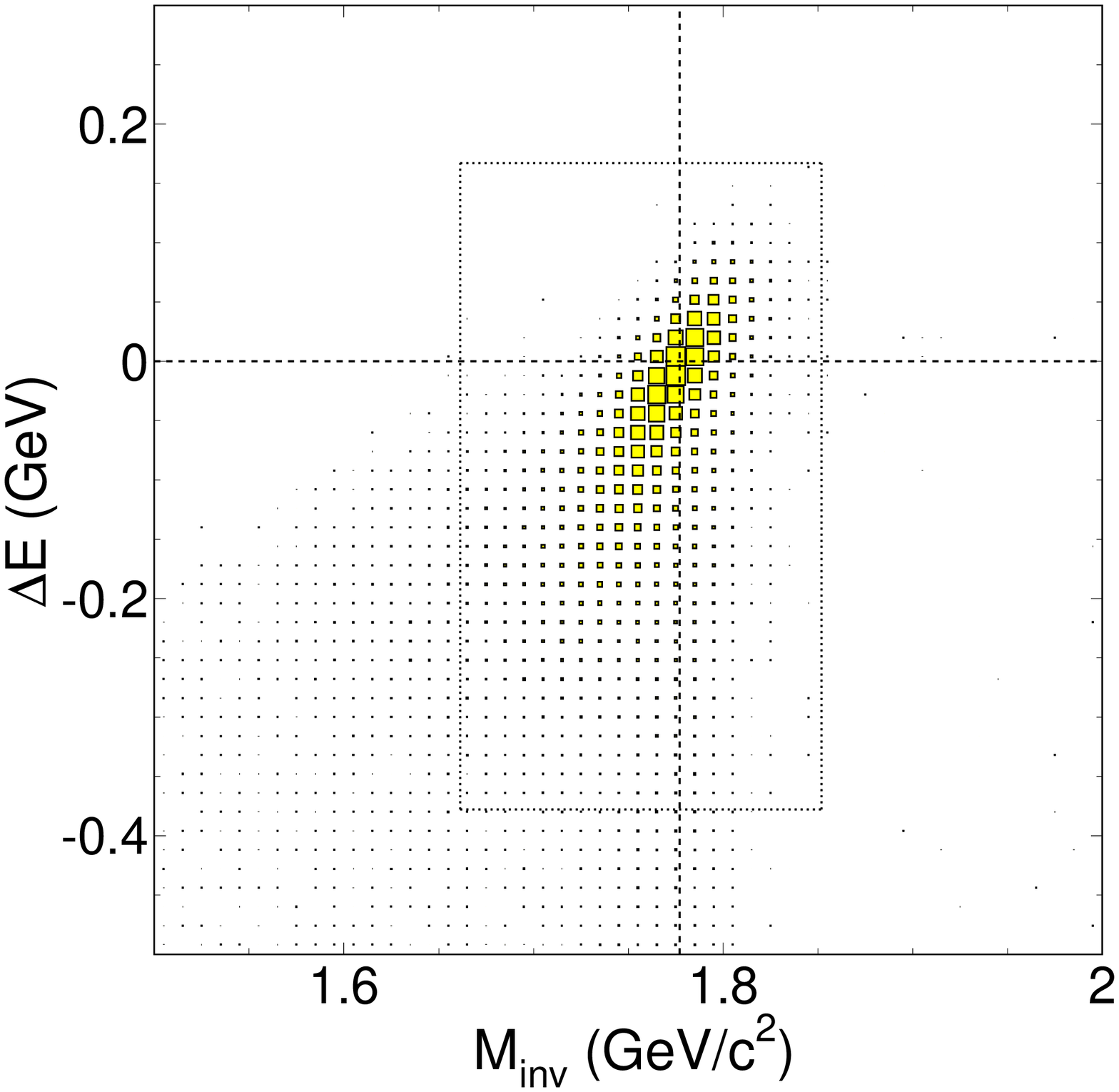}
  \end{center}
 \caption{ $\DelE$ vs. $\Minv$ distribution for
$\TMG$ decay.}
  \label{fig:Ften}
\end{figure}

\paragraph{Signal-to-background}
A newly introduced requirement for the missing quantities, 
$p_{miss}-M_{miss}^2$, plays an important rule for background 
as seen in Figure~\ref{fig:Fsix}.
It makes it possible to achieve a sensitivity higher than
that of the CLEO experiment.  
We also apply a new selection for $\TEG$, using the
opening angle between the tagged track and missing particle 
direction; this criterion quite effectively removes the radiative 
Bhabha background. 
We have to create this kind of new criteria for future analysis. 

\paragraph{Physics Reach}
\begin{itemize}
\item Figure~\ref{fig:ma-tanb} and \ref{fig:ellis-ma-tanb} show a possible exclusion 
  region in the $m_{A}-\tan\beta$ parameter space from 
 $\tau^{-}\rightarrow\mu^{-}\eta$.
  The region allowed by the current upper bound
  from the Belle experiment($<3.4\times 10^{-7}$) and the one
  from non-observation with the 1000~${\rm fb}^{-1}$ are shown. 
  In Fig.~\ref{fig:ma-tanb}, the boundary is evaluated by multiplying a factor 8.4 to the formula 
  given by (\ref{eq:tauetamu}), \textit{i.e.}
  \begin{equation}
    \Br(\tau^-\rightarrow\mu^{-}\eta)= 
    3.2 \times 10^{-6}
    \times |\delta^{L}_{\tau\mu}|^{2}\times
    \left( \frac{\tan\beta}{60}\right)^{6}\times \left( \frac{M_A}{100 GeV}
    \right)^{-4},
\label{eq:ma-tanb-eta}
  \end{equation}
  where $|\delta^{L}_{\tau\mu}| = 1$ is assumed
  for the evaluation of the boundary.
  While in Fig.\ref{fig:ellis-ma-tanb},
 the same formula is used  but 
 the Yukawa coupling and heavy neutrino mass in (3.42) are
 assumed to be $(Y^{\dagger}Y)_{32,33}=1$ and $M_{N}=10^{14}$ GeV
instead of the assumption in $|\delta^{L}_{\tau\mu}|$.
This is the  assumtion  used  by  A.~Dedes, J.~Ellis and M.~Raidal 
in their paper( PRL B549,159(2002)).

Two figures indicate how the branching fraction are sensitive to the
 value of parameters in the model.
All though the allowed region depend on the model parameters,
future experiments with 5000-50,000/fb can cover huge parameter space
in the model.
\end{itemize}

\begin{figure}[htbp!]
  \begin{center}
    \includegraphics[width=0.60\textwidth,clip]{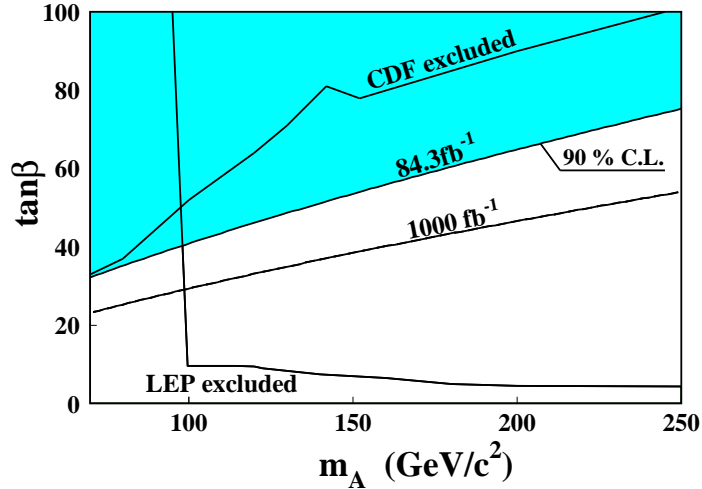} \caption{ 
Physics reach in $m_{A}- \tan\beta$ parameter space at 1000${\rm fb^{-1}}$ 
for the $\tau^- \rightarrow \mu^{-}\eta$ decay,
together with the regions excluded by direct searches at LEP and
in Tevatron experiments \cite{Hagiwara:fs,Affolder:2000rg}.  
Here, $m_{A}$ is the pseudo-scalar Higgs mass in the Higgs mediated model.
The boundary is based on the formula Eq.(\ref{eq:ma-tanb-eta}).
 }
  \label{fig:ma-tanb}
\end{center}
\end{figure}
\begin{figure}[htbp!]
  \begin{center}
    \includegraphics[width=0.60\textwidth,clip]{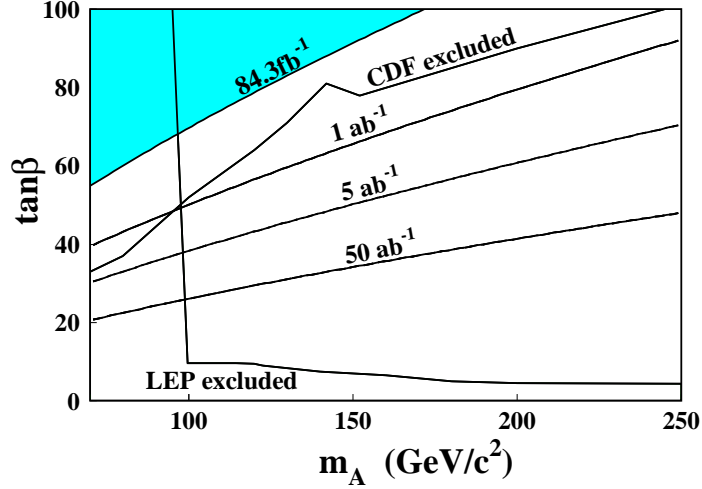} \caption{ 
Physics reach in $m_{A}- \tan\beta$ parameter space at 1000${\rm fb^{-1}}$ 
for the $\tau^- \rightarrow \mu^{-}\eta$ decay, in the Higgs mediated model.
 the Yukawa coupling and heavy neutrino mass in (3.42) are
 assumed to be $(Y^{\dagger}Y)_{32,33}=1$ and $M_{N}=10^{14}$ GeV.
 }
  \label{fig:ellis-ma-tanb}
\end{center}
\end{figure}

\begin{itemize}
\item Figure~\ref{fig:msusy-tanb} and \ref{fig:msusy-tanb-ellis} show
 a possible exclusion
  region in $m_{SUSY}-\tan\beta$ parameter space 
  for   $\tau^{-}\rightarrow\mu^{-}\gamma$.
   The excluded regions are from Belle's current upper-bound
  ($<3.1\times 10^{-7}$)  and from non-observation 
  at 1000~${\rm fb}^{-1}$.
In Fig.~\ref{fig:msusy-tanb}, 
  we use 
 we use the approximate formula for $Br(\tau^{-}\rightarrow\mu^{-}\gamma)$ 
   given as
  \begin{equation}
    \Br(\tau^-\rightarrow\mu^{-}\gamma)= 
    3.0 \times 10^{-6} \times
    \left(\frac{\tan\beta}{60}\right)^{2}
    \times
    \left(\frac{M_{SUSY}}{1 TeV}\right)^{-4}.
  \end{equation}
\end{itemize}
  The formula is obtained  from (\ref{eq:AtaulR})--(\ref{eq:AmueL})
   by taking that
  $|\delta^{R}_{\tau\mu}|=1$ and $|\delta^{L}_{\tau\mu}|=1$.

  While in Fig.~\ref{fig:msusy-tanb-ellis}, 
 the same formula is used but
 the Yukawa coupling and heavy neutrino mass in (3.42) are
 assumed to be $(Y^{\dagger}Y)_{32,33}=1$ and $M_{N}=10^{14}$ GeV.

\begin{figure}[!htbp]
\begin{center}
    \includegraphics[width=0.55\textwidth,clip]{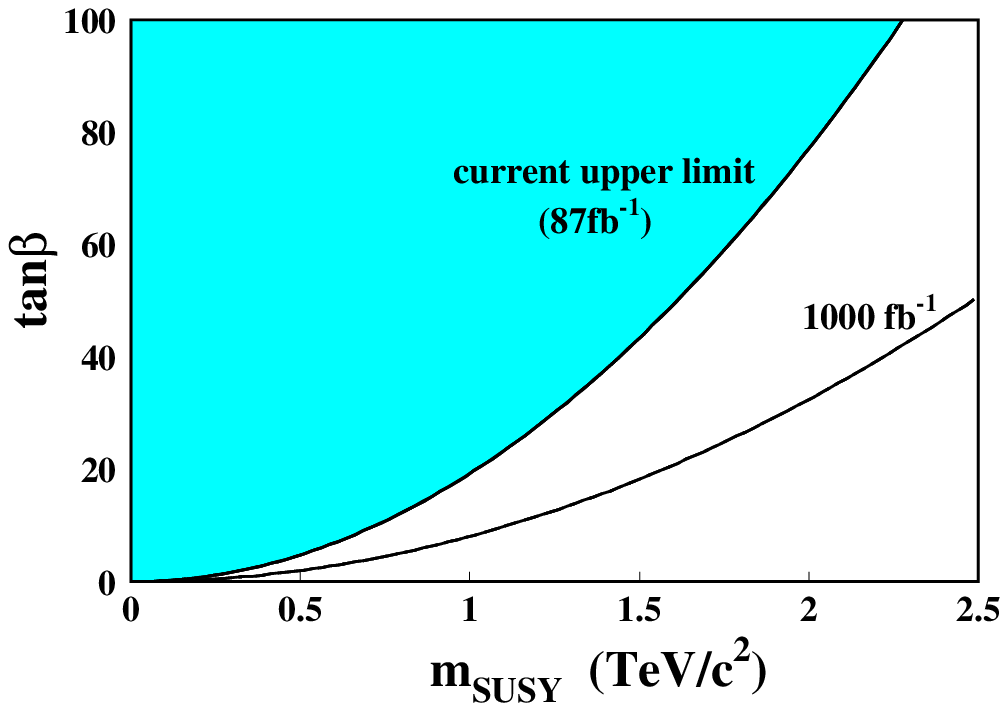}
  \caption{
    Physics reach in $m_{SUSY}- \tan\beta$ parameter space
    from Belle's current upper-bound and at 
    1000~${\rm fb^{-1}}$ luminosity for
    $\tau^{-}\rightarrow\mu^{-}\gamma$. 
    The branching fraction is obtained by taking 
    $|\delta^{R}_{\tau\mu}|=|\delta^{L}_{\tau\mu}|=1$ in 
    (\ref{eq:AtaulR})--(\ref{eq:AtaulL}).
  }
\label{fig:msusy-tanb}
\end{center}
\end{figure}
\begin{figure}[!htbp]
\begin{center}
    \includegraphics[width=0.55\textwidth,clip]{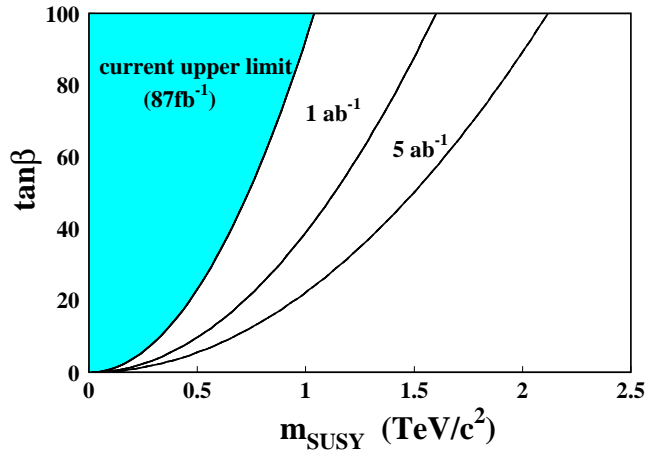}
  \caption{
    Physics reach in $m_{SUSY}- \tan\beta$ parameter space
    from Belle's current upper-bound and at 
    1000~${\rm fb^{-1}}$ luminosity for
    $\tau^{-}\rightarrow\mu^{-}\gamma$. 
    The  branching fraction is obtained by taking 
    $(Y^{\dagger}Y)_{32,33}=1$ and $M_{N}=10^{14}$ GeV in (3.42). 
  }
\label{fig:msusy-tanb-ellis}
\end{center}
\end{figure}

\subsection{Summary}
Many mechanisms for LFV decays are possible and 
a wide range of the relevant parameter space is allowed 
at present. 
Accordingly, many different models have been proposed as discussed
in Section ~\ref{sec:Lepton_flavor_violation}. 
Some predict relatively large branching fractions while 
others give extremely small values. 
Systematic and extensive investigation of various $\tau$ decay modes 
would provide a powerful means to select models, restrict their 
parameter space, and  possibly discover new phenomena 
beyond the Standard Model. 
SuperKEKB is the facility best suited to carry out this frontier research. 


\clearpage \newpage
\def\3{\ss}
\def\sinth{sin$^2\Theta_W$~}

\section{Diversity of physics at Super-KEKB--other possibilities}

\subsection{Charm physics}
At a $B$ factory, a large number of charm mesons are
produced from the $q\bar{q}$ continuum and also from decay
products of $B$ mesons. 
For example, with $11.1\,\mathrm{fb}^{-1}$
reconstructs $10^5$ $D^0$ ($\bar{D^0}$), $8\times 10^3$ $D^\pm$
and $6\times 10^3$ $D_s^\pm$ mesons in low multiplicity decay modes.
We can expect data samples a hundred times larger with a luminosity of
$10^{35}\, \mathrm{cm}^{-2} \mathrm{s}^{-1}$.

Due to the effectiveness of the GIM mechanism,
flavor-changing neutral current (FCNC) decays,
$D^0-\bar{D}^0$ mixing and $CP$ violation are small in the charm
sector. This is in sharp contrast with $K$ and $B$ 
FCNC processes, which are enhanced by the presence of top
quarks in loops. In many cases, extensions of the Standard Model (SM)
upset this suppression and give contributions sometimes orders
of magnitude larger than the SM. As a result, rare charm processes
are an excellent place to look for new physics.

The strength of $D^0-\bar{D}^0$ mixing is characterized by two
parameters $x=\Delta M/\Gamma$ and $y=\Delta \Gamma/2\Gamma$.
According to the conventional expectation of the SM, $x,y\leq 10^{-3}$.
However, in a recent treatment by 
Falk \textit{et al.}, the possibility of $y$ (and perhaps
$x$) $\sim 10^{-2}$ within the SM is raised~\cite{Falk:2001hx}. 
The current experimental limits are at the level of a few
times $10^{-2}$. 

Experimental searches for $D^0-\bar{D^0}$ mixing usually involve
hadronic decay modes such as $D^0\to K^+\pi^-$. For such modes,
there are contributions from both mixing and doubly Cabibbo
suppressed decays (DCSD), which can be distinguished by their time
dependences. In the $CP$ conserving limit, the rate for wrong sign
decays is $$r_{WS}(t) = [R_D +\sqrt{R_D} y^{\prime} t + 1/4
(x^{\prime 2}+y^{\prime 2}) t^2] e^{-t},$$
where $R_D$ is the DCSD rate, and  
$y^{\prime} = y
\cos\delta-x\sin\delta$ and $x^{\prime}= x \cos\delta + y \sin\delta$
are the mixing parameters $y$ and $x$ rotated by 
$\delta$, the relative strong phase between $D^0\to K^+\pi^-$
and $\bar{D}^0\to K^+\pi^-$.
In the absence of interference, mixing has a $t^2 e^{-t}$
dependence which peaks at 2 $D^0$ lifetimes, whereas DCSD follows
the usual $e^{-t}$ dependence. The interference term is
proportional to $t e^{-t}$ and dominates the sensitivity to mixing,  
since $(x^{\prime 2}+y^{\prime 2}) = (x^2 + y^2) \ll R_D$. 
Since the measurement of $y'$ and $x'$ requires that these three terms
be distinguished from each other, decay-time resolution is crucial:
improved vertexing at the Belle upgrade, together with the very large
$D^0$ samples available at $10^{35}\, \mathrm{cm}^{-2} \mathrm{s}^{-1}$,
will lead to an improvement in sensitivity over previous
experiments~\cite{Godang:1999yd} and the existing $B$-factories.

Interpretation of $D^0\to K^+\pi^-$ and other hadronic-decay mixing 
analyses is complicated by the strong phase difference $\delta$,
which may be large~\cite{Blaylock:1995ay,Bergmann:2000id}: 
it is important to obtain constraints on this quantity.
At a tau-charm facility, $\delta$ can be determined by using quantum
correlations with two fully reconstructed $D$
decays~\cite{Gronau:2001nr}. 
At a high luminosity $B$ factory, $\delta$ can be determined
by measuring related DCSD modes, including modes with $K_L$
mesons~\cite{Golowich:2001hb}. 

If $CP$ is violated in the $D$ system, then additional $D^0-\bar{D^0}$
mixing signals may be seen. $CP$ violation in the interference of 
$D^0$ decays with and without $D^0-\bar{D^0}$ mixing is parameterized
by the phase $\phi_D=\arg(q/p)$: 
the SM expectation is ${\cal O}(10^{-3}-10^{-2})$,
whereas in new physics scenarios it can be ${\cal O}(1)$.
This is in contrast with direct $CP$ violation, which occurs
in Cabibbo suppressed decays such as $D\to \rho\pi$ at the $10^{-3}$ level
in the SM: new physics scenarios are unlikely to change this expectation.

A time dependent asymmetry 
\begin{equation}
  \Gamma(D^0(t) - \bar{D^0}(t)) \propto x \sin\phi_D \Gamma t e^{-\Gamma t}
\end{equation}
may be measured by comparing $D^0\to K^+\pi^-$ and $\bar{D^0}\to K^-\pi^+$
decays~\cite{Wolfenstein:1995kv},
and would (unlike $CP$-conserving mixing) be a clear signal of new physics.
The corresponding asymmetry between $D^0$ and $\bar{D^0}$ decay rates to $K^+K^-$,
where the $D^0$ flavor is tagged by the pion from $D^{*+}\to D^0\pi^+$,
allows an especially clean measurement since the final state is identical in both
cases, and is only singly Cabibbo suppressed.
The analysis of this mode is similar to that used for 
time dependent $CP$ violation in $B$ decays.

There are several classes of rare $D$ decays where the large data samples
available at high luminosity will allow improved measurements. Two body decay
modes such as $D^0\to \gamma \gamma, \mu^+\mu^-$ and $\mu^{\pm}e^{\mp}$
are strongly suppressed in the Standard Model:
expectations are $10^{-8}$ for $D^0\to
\gamma\gamma$, $10^{-13}$ for $D^0\to\mu^+\mu^-$ and 0 for $D^0\to
\mu e$. In new physics scenarios, the rates can be orders of
magnitude larger~\cite{Burdman:2001tf}. 
For example, in both R-parity violating
and leptoquark models, the branching fraction for $D^0\to \mu^+\mu^-$
can be as large as $3\times 10^{-6}$ while that
for $D^0\to \mu^{\pm}e^{\mp}$ could be
$5\times 10^{-7}$. The current experimental bounds
for $D^0\to \mu^+\mu^-$ and $D^0\to \mu^{\pm} e^{\mp}$ are
$3.3\times 10^{-6}$ and $8.1\times 10^{-6}$, respectively.
For 3-body final states such as $\rho \ell^+\ell^-$, where SM expectations
are similar, orders of magnitude enhancements are expected at low $\ell^+\ell^-$
invariant masses in some new physics models.
In the case of radiative decays such as $D^0\to K^*\gamma, \rho\gamma$,
which are long-distance dominated, measurement at SuperKEKB could constrain
long-distance effects in the corresponding modes in the $B$ sector.

By the time SuperKEKB begins taking data, the tau-charm facility
at Cornell will also be operating. Although the design luminosity
is relatively low ($5\times 10^{32}\, \mathrm{cm}^2 \mathrm{s}^{-1}$),
correlated $D$ meson pairs
are produced at threshold from the $\psi''$ resonance. For
measurements where kinematic constraints from production at
threshold are essential, such as $f_D$ and $D$ absolute branching
fractions, the Cornell facility will remain competitive;
and a sensitivity to $D^0 - \bar{D^0}$ mixing at the $10^{-4}$ level
is claimed~\cite{Gronau:2001nr}.
SuperKEKB will have the advantage of precision vertexing for 
measurement of time-dependent decay distributions---especially important
if $CP$ violation is associated with mixing---and very large $D$ meson samples.
Other
facilities in the world such as ATLAS/CMS/CDF/D0 cannot do charm
physics. LHCb and BTeV may record large charm data samples if they
modify their trigger configurations, which are optimized for $B$
physics. However, these experiments cannot efficiently reconstruct
final states with neutrals or $K_L$ mesons.

\subsection{Electroweak physics}
The standard-electroweak model has two fundamental
parameters, which are directly related to measurable
quantities: the parameter 
$\rho = M_W^2/(M_Z^2\sin^2\Theta_W)$, 
which is unity in the Standard Model, and the Weinberg angle
$\Theta_W$, which determines the relative contributions of
electromagnetic and weak forces. 
Both parameters have been measured at the e$^+$e$^-$
colliders PETRA and TRISTAN before the high precision
measurements at the $Z$-pole became available from LEP. 

Recently a growing interest is being observed to revisit
this type of physics and repeat \sinth and $\rho$
measurements with high precision. 
There are two aspects to this renewed interest in a
precision determination of the fundamental electro-weak
parameters. 
One is related to a measurement of the NuTeV collaboration
at Fermilab \cite{Zeller:2001hh}, who observed values of
$\rho$ and \sinth, which are not in agreement with the
Standard Model. 
The second motivation is that the scale dependence of gauge
couplings has been observed for the strong coupling constant
$\alpha_s$ and the electromagnetic coupling $\alpha$,
however, not yet for the coupling constant of the weak
isospin group SU(2).

The quantity \sinth is related to the coupling parameters of
two gauge groups, U(1) for the electromagnetic part and
SU(2) for the weak isospin part of the standard electro weak
model. 
Both couplings have a different scale dependence resulting
in a scale dependence of a more complicated nature for
\sinth \cite{Ramsey-Musolf:1999qk} as shown in
Figure~\ref{fig:sin2theta}.  
To test the scale dependence, data below the $Z$-pole with
an accuracy of a few 10$^{-4}$ will serve the purpose.

\begin{figure}[tbp]
\begin{center}
   \includegraphics[width=0.8\textwidth,clip]{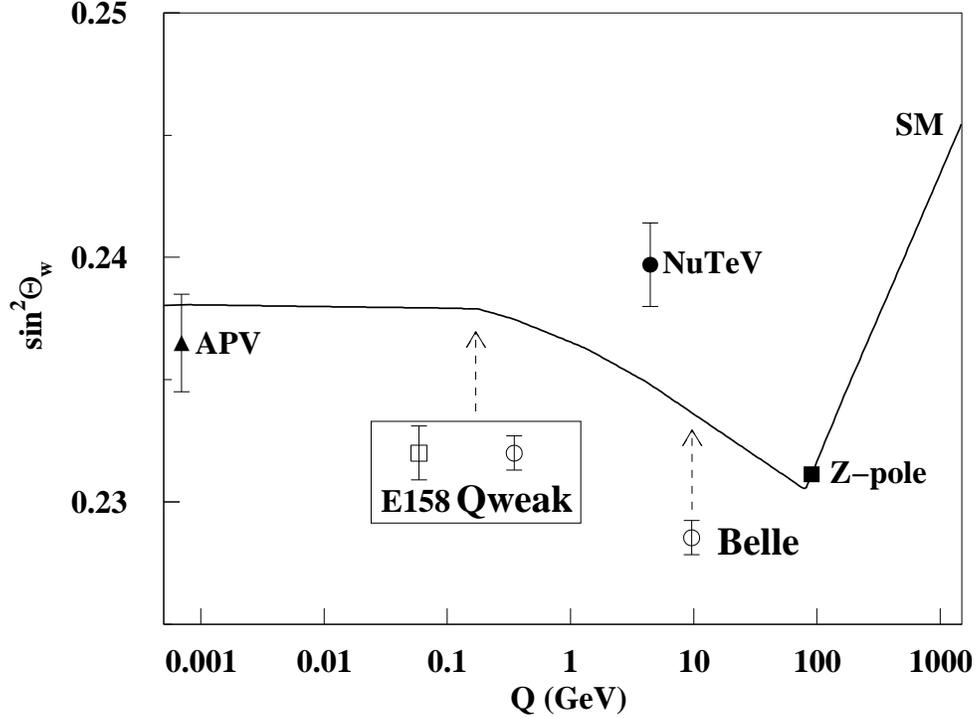}
\end{center}
\caption{
  Scale dependence of \sinth. 
  The full symbols show the current situation, while the
  open symbols with error bars for the proposed experiments
  QWEAK, SLAC E-158 and Belle are placed at the correct cm
  energy with arbitrarily chosen vertical positions. The
  previous measurements are determinations from atomic
  parity violation (AVP) \cite{Wood:zq}, 
  deep inelastic neutrino scattering (NuTeV), 
  and from $Z$-pole asymmetries (LEP/SLC).
}
\label{fig:sin2theta}
\end{figure}

Two experiments are being proposed to measure \sinth at
center of mass energies below 1~GeV, QWEAK at the Jefferson
Lab in Virginia and one at SLAC, E-158. 
The QWEAK experiment will deduce \sinth from elastic
scattering of polarised electrons off protons, while the
SLAC experiment measures M{\o}ller scattering of polarised
electrons. 

The NuTeV Detector at Fermilab consists of an 18~m long,
690~ton active steel-scintillator target with drift chambers
as tracking devices followed by an iron-toroid
spectrometer. 
High purity $\nu_{\mu}$ and $\overline{\nu_{\mu}}$ beams
resulting from interactions of 800~GeV protons in a BeO
target can be directed onto the detector. 

In principle \sinth can be derived from a measurement of the
ratio between neutral current (NC) and charged current
interactions (CC) in a nuclear target for just one neutrino
species. 
This approach is, however, subject to large QCD
corrections. The corrections can be minimized by a
determination of the ratio $R^-$ with 
\begin{equation}
R^- = \frac{\sigma(\nu_{\mu} N \rightarrow \nu_{\mu} X) - 
       \sigma(\overline{\nu_{\mu}} N \rightarrow \overline{\nu_{\mu}} X)}
       {\sigma(\nu_{\mu} N \rightarrow \mu^- X) - 
        \sigma(\overline{\nu_{\mu}} N \rightarrow \mu^+ X)}.
\end{equation}
NuTeV obtains with this method 
$\sin^2\Theta_W$ = 0.2277 $\pm$ 0.0013 $\pm$ 0.0009,
a value which is 3$\sigma$ above the value expected for the
standard electroweak model. 
From a two parameter fit to $\rho$ and \sinth they conclude
that one of the quantities, but not both of them can be made
to agree with the Standard Model value. 
Their result is sensitive to new physics in the W and Z
sector and suggests a smaller left-handed NC coupling to
light quarks than expected.  
But this result also depends on hadronic corrections and
nuclear structure functions, which is the main criticism
with respect to an interpretation in terms of deviations
from the Standard Model.

In the electro-weak process $e^+e^- \to \mu^+ \mu^-$
the values for \sinth and $\rho$ are derived from a fit to
the angular distribution ($\Theta^*$) of $\mu$ pairs with
respect to the axis of the incoming positron in the $e^+e^-$
center of mass system \cite{Marshall:1988bk}. 

\begin{equation}
\frac{d\sigma}{d\Omega} = \frac{\alpha^2}{4 s} (C_1(1+\cos^2\Theta^*)
+ C_2 \cos\Theta^*)
\end{equation}
with the following definitions:   
\begin{eqnarray}
  C_1 & = & 1 + 2v_e v_{\mu} \chi + (v_e^2 + a_e^2)
  (v_{\mu}^2 + a_{\mu}^2) \chi^2,
  \\
  C_2 & = & -4a_ea_{\mu} \chi + 8v_e a_e a_{\mu} \chi^2,
  \\
  v_{e,\mu} & = & -1 + 4\sin^2\Theta_W,\;\;\;\;\;
  a_{e,\mu} = -1.
\end{eqnarray}
The quantity $\chi$ may be written in two different ways, as
a function of \sinth or as a function of $\rho$
\begin{equation}
  \chi = \frac{1}{16 \sin^2\Theta_W \cos^2\Theta_W} \frac{s}{(s - M_Z^2)}
\end{equation}
or
\begin{equation}
  \chi = \frac{\rho\,G_F M_Z^2}{8 \pi \alpha \sqrt{2}} \frac{s}{(s - M_Z^2)}
\end{equation}
where $s$ is the square of the center of mass energy.

For the purpose of estimating statistical errors and
significance levels it is sufficient to consider the
forward-backward charge asymmetry integrated over all
angles
$A_{FB} = (3 C_2)/(8 C_1) = 1.5\chi$, while in an actual
experiment, one would fit the angular distribution. 

A guideline for estimating the significance of a measurement at 
$\sqrt{s}$ = 10~GeV would be the capability of
distinguishing the value of \sinth at the $Z$-pole from a
value it would assume if it were running according to the  
predictions of the Standard Model.
With the $Z$-pole value of \sinth = 0.232 the
forward-backward asymmetry assumes 
a value of $A_{FB}(\sqrt{s}=10~GeV) = 6.34 \times 10^{-3}$. 
For \sinth running according to the Standard Model, the
asymmetry is $A_{FB}(\sqrt{s}=10~GeV) = 6.38 \times 10^{-3}$. 
The charge asymmetries for a running and a fixed
coupling constant thus differ by 
$\delta(A_{FB}) = 4 \times 10^{-5}$.

At 10~GeV center of mass energy, the statistical error on
\sinth is 30 times larger than the error on the
forward-backward asymmetry: 
$\sigma(\sin^2\Theta_W) = 30~\sigma(A_{FB})$.
With a statistical error of 
$\sigma(A_{FB}) = \pm 1\times10^{-5}$ on the charge
asymmetry the corresponding error on \sinth is 
$\sigma(\sin^2\Theta_W) = 3 \times 10^{-4}$.

The number of events required to achieve this accuracy is
certainly smaller than 10$^{10}$ events, because when
fitting an angular distribution much more efficient use is
being made of the experimental data and, thus for an
accumulated luminosity of 1~ab$^{-1}$ with 10$^9$ events a
statistically significant measurement can be made, which is
compatible with that of the two dedicated experiments as may
be inferred from Figure~\ref{fig:sin2theta}. 

The $\mu$-pairs from the decay of the $\Upsilon(4S)$ will
have a different asymmetry from those of the continuum, as
with $\Upsilon$ as an intermediate state the Z-boson couples
to $b$-quarks and the relative weight between $\gamma$ and
$Z$-exchange is altered with respect to the continuum. 
The branching fraction of 
$BR(\Upsilon(4S)\to\mu^+\mu^-) = 3 \times10^{-5}$ is,
however, so small that corrections to the angular dependence
will become less important and they can be calculated with
sufficient accuracy, as all couplings and weak charges are
known. 
The experiment is thus a continuum experiment.

Radiative corrections are small and can be calculated with
sufficient accuracy. 
Also bin to bin migrations are expected to be small. 
A potential source of systematic errors are radiative
returns of the $\Upsilon(1S)$ resonance. 
These events can be removed by a mass cut on the invariant
two lepton mass.

The angular distribution of electron pairs is known with
high accuracy and therefore electron pairs can be used to
check the charge symmetry of the detector and other
systematic errors like those, which may be related to the
fact that the muon momenta in the forward and backward
hemispheres are different due to the asymmetric energies of
the colliding electron and positron beams.

Muon identification need not be very restrictive, because
there are only very few reactions, which could fake muon
pairs, if collinearity of the two tracks is requested. 
Tau pairs will be rejected by collinearity cuts and
invariant mass cuts, the cross section for pion pairs is too
small to present a serious background and electron pairs are
easily identified by their unique signature in the CsI
crystals. 
As a first guess, a loosely identified muon track would be
sufficient for one track, while the second collinear
particle need not be identified as a muon, it should be
compatible with a muon and incompatible with an electron. 

The presence of SU(2) breaking forces could result in a
different phenomenological pattern for reactions, where
quarks are present or only leptons are involved. 
Leptoquarks as an example could alter the result of a \sinth
measurement in neutrino-nucleon scattering without a
measurable impact on the charge asymmetry of muon pairs in
$e^+e^-$ annihilation. 
Therefore one could imagine to face a situation, where
purely leptonic experiments do agree with the Standard
Model, while reactions involving nucleons don't. In this
case one could extend the program at SuperKEKB to a
determination of the charge asymmetry of jets. 
This is a very ambitious measurement, which needs extensive
Monte Carlo studies beforehand to investigate the
feasibility of such an experiment.
  
In summary, SuperKEKB will be capable of performing a statistically
significant measurement of the Weinberg angle \sinth in
order to prove the running of the U(1)/SU(2) couplings of
the standard electroweak model and set limits on new weak isospin 
breaking interactions. 
In order to achieve this, data corresponding to about
1~ab$^{-1}$ are needed. 


\subsection{Charmonium physics}
\newcommand{\leplep}{\ell^{+}\ell^{-}}
\newcommand{\jp}{J/\psi}
\newcommand{\pipi}{\pi^{+}\pi^{-}}
\newcommand{\DE}{\Delta E}

In the history of physics,  many new and important insights
have derived from detailed studies of ``well understood'' 
systems: precise measurements of the motion
of planets in the solar system led to the discovery of
an anomalous precession of the perihelion of Mercury's orbit,
which provided an important impetus for General
relativity; high resolution measurements of atomic spectra
were the key to the discovery of fermion spin.

In hadronic physics, the most ``well understood'' systems are 
the quarkonium mesons, \textit{i.e,}  $c\bar{c}$ or
$b\bar{b}$ mesons.  
Here, because the quarks are massive, they
are nearly non-relativistic and ordinary quantum mechanics 
is applicable.  
Moreover, lattice calculations are particularly
well suited to heavy quark systems.  
As these improve we can expect reliable first-principle
calculations of quarkonium properties with good precision.  

In $B$ meson decays, the $b\to c\bar{c} s$ subprocess is CKM-favored 
and, thus, final states containing charmonium particles are common. 
A super-$B$ factory would provide {\em superb} opportunities
for precision, high sensitivity measurements of the charmonium
system.  

Even right at the $\Upsilon(4S)$ peak, the cross-section for
the continuum production of $c\bar{c}$ quark pairs is higher 
than that for $b\bar{b}$ pairs.  Thus, a ``super-$B$'' factory
is also a ``super-charm'' factory that will support a variety
of interesting studies of charmed particle and charmonium
physics.

\subsubsection{New results on charmonium from Belle/KEKB}

There are still a few undiscovered charmonium states 
that are predicted to have
masses below the relevant threshold for open charm
production and are, thus, expected to be narrow.  
These include the $n=1$ singlet~P state, the $h_c$, and
possibly the $n=1$ singlet and triplet spin-2~D states, \textit{i.e.}
the $1^1{\rm D}_{c2}$ and $1^3{\rm D}_{c2}$.
The discovery of these states and the measurements
of their properties are important for a number of reasons.
First of all, measurements of the masses of these states
will pin down unknown parameters of the charmonium model,
such as the strength of the fine and hyperfine terms in the 
inter-quark potential.
Second, the properties of these states are highly
constrained by theory. 
Measured variances from theoretical predictions
could indicate new and unexpected phenomena.
In Belle,  modest efforts of studying 
charmonium production in $B$ decays and 
continuum $e^+e^-$ processes have
produced interesting examples of both 
of these cases.

\begin{description}
\item[Discovery of $\eta_{c}'$:]
In 2002, with a 
42~fb$^{-1}$ data sample, Belle discovered the $\eta_{c}'$
via its  $K_S K\pi$ decay mode in exclusive $B^-\to K^- K_S K\pi$
decays~\cite{Choi:2002na} (see Figure~\ref{fig:etacprime}).  
This observation was subsequently confirmed by other Belle
measurements~\cite{Abe:2002rb,Abe:2003jac} as well as
BaBar~\cite{Aubert:2003pt} and CLEO~\cite{Asner:2003wv}.
Although nobody ever doubted the 
existence of the $\eta_{c}'$, it had evaded detection for nearly 
thirty years.  A ``candidate''  $\eta_{c}'$, reported by
the Crystal Ball in 1982~\cite{Edwards:1981mq}, indicated an 
$\psi'-\eta_{c}'$ mass splitting ($92\pm 7$~MeV)
that is above the range 
of theoretical expectations ($43 \sim 75$~MeV)
\cite{Buchmuller:1980su,Lahde:2001zd}. 
The Belle measurement of the $\eta_{c}'$ mass indicates that
mass splitting ($32\pm 10$~MeV) is, in fact, at the lower end of 
the  theoretically preferred range and has helped
pin down the strength of the 
hyperfine splitting terms in the charmonium potential.

\begin{figure}[tbp]
  \begin{center}
    \includegraphics[angle=0,width=0.8\textwidth]{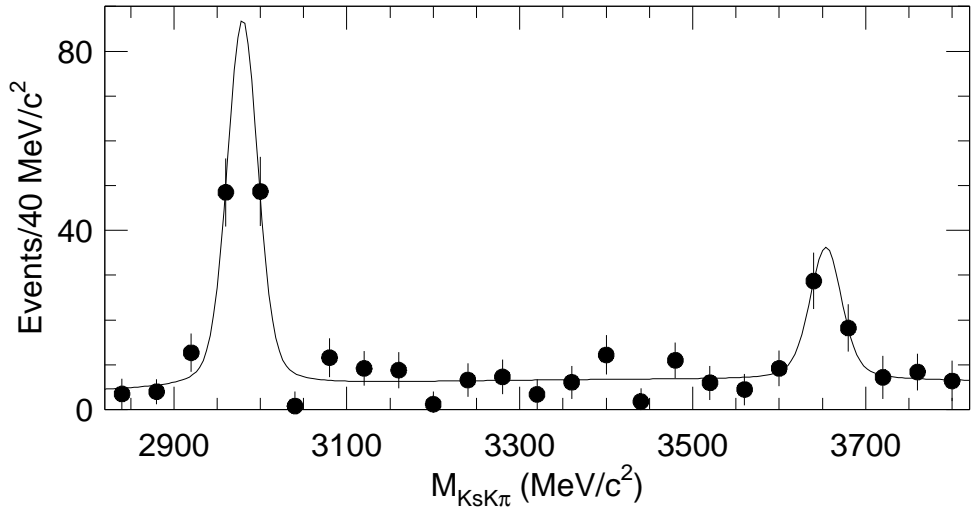}
    \caption{
      The $K_S K\pi$ invariant mass distribution
      for $B^-\to K^- K_S K\pi$ decays.  The large peak near 2980~MeV
      is the $\eta_c$; the smaller peak near 3650~MeV is attributed
      to the $\eta_{c}'$. 
    }
    \label{fig:etacprime}    
  \end{center}
\end{figure}

\item[Discovery of $X(3872)$:]
In 2003, with a nearly four-times larger data sample, Belle
discovered a narrow charmonium-like $\pi^+\pi^-\jp$ state
with a mass of 3872~MeV in exclusive $B\to K \pi\pi\jp$ 
decays~\cite{Choi:2003ue} (see Figure~\ref{fig:x3872}). 
This state, called the $X(3872)$,  was
originally considered to be the $^3D_{c2}$, however a
closer examination of its properties indicated problems
with this interpretation:
the 3872~MeV mass is substantially above model expectations of
$\sim 3815$~MeV;
the decay rate to $\gamma\chi_{c1}$ is too small;
the shape of the $M(\pi\pi)$ distribution is too
peaked at high $\pi\pi$ masses; and the 
inferred exclusive branching ratio for
$B\to K (^3D_{c2})$ is too large.
As a result, a number of theorists have speculated 
that this particle may not be a $c\bar{c}$ charmonium
state, but, instead, a new type of four-quark 
meson~\cite{Voloshin:ap,DeRujula:1976qd,Tornqvist:ng}.
Either the standard
charmonium model has to be modified, or the
$X(3872)$ is a first example of an altogether
new type of particle.  More data will help sort
this out.

\begin{figure}[tbp]
  \begin{center}
    \includegraphics[angle=0,width=0.85\textwidth]{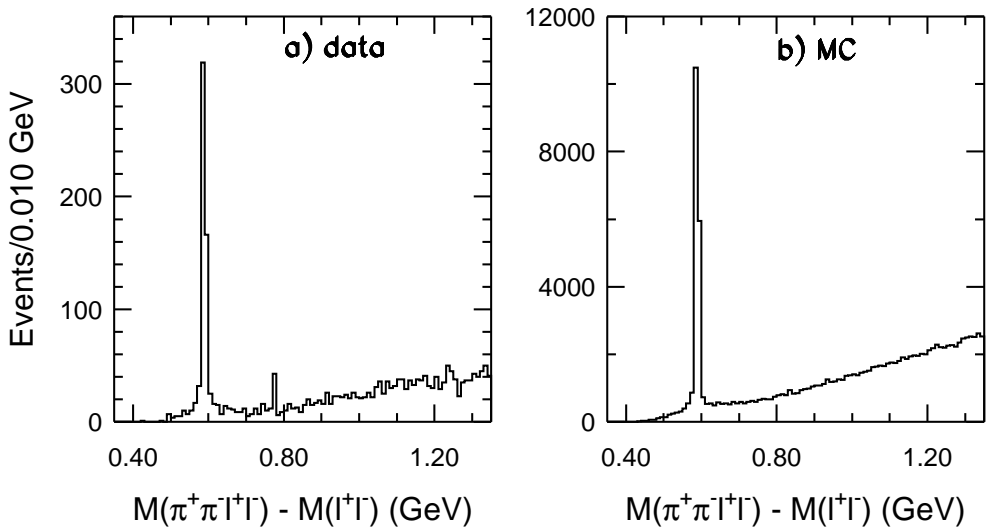}
  \end{center}
  \caption{ 
    Distribution of $M(\pipi\leplep)-M(\leplep)$ for
    selected events in the $\DE$-$\Mbc$ signal region for
    {\bf (a)} Belle data and  {\bf (b)} generic $\bbar$ MC
    events.
  }
  \label{fig:x3872}    
\end{figure}

\item[Discovery of the large rate for the continuum process 
  $e^+e^-\to c\bar{c}c\bar{c}$:]
A major problem for QCD are calculations of the production of physical hadrons.
This is because the long-range, low-$q^2$ processes that are involved,
can not be dealt with perturbatively.  Charmonium is an exception. Since
charmonium is formed at relatively short distances, perturbation theory
should be applicable.  An elegant effective theory called Non-relativistic
QCD (NRQCD) has been developed to deal with various aspects of charmonium
production~\cite{Kiselev:1994pu,Bodwin:2002kk}. 
In Belle,  studies of $J/\psi$ production in
continuum $e^+e^-$ annihilation processes were
started as a way to test the predictions of NRQCD.
This led to the remarkable, and totally unexpected discovery  
that continuum-produced $J/\psi$s are almost always 
accompanied by another $c\bar{c}$ system; 
As can be seen in Figure~\ref{fig:double_ccbar},
the recoil system appears to be totally saturated by
either other charmonium states
such as $\eta_c$, $\chi_{c0}$ or $\eta_{c}'$, or
by pairs of charmed particles~\cite{Abe:2002rb,Abe:2003jac}.
There is no evidence in the spectrum for low recoil
masses that NRQCD predicts should dominate; 
the measured cross-sections 
for exclusive and inclusive $e^+e^-\to c\bar{c}c\bar{c}$ processes
are an order-of-magnitude larger than
NRQCD predictions.   Theorists have not been able
to modify the model to accommodate the Belle experimental
results.  NRQCD specialist Bodwin has said that explaining
these results ``either requires the invention of charmonium 
production mechanisms within the Standard Model that have 
not yet been recognized, or physics beyond the Standard 
Model''~\cite{Bodwin:2003kj}.

\begin{figure}[tbp]
  \begin{center}
    \includegraphics[angle=0,width=0.9\textwidth]{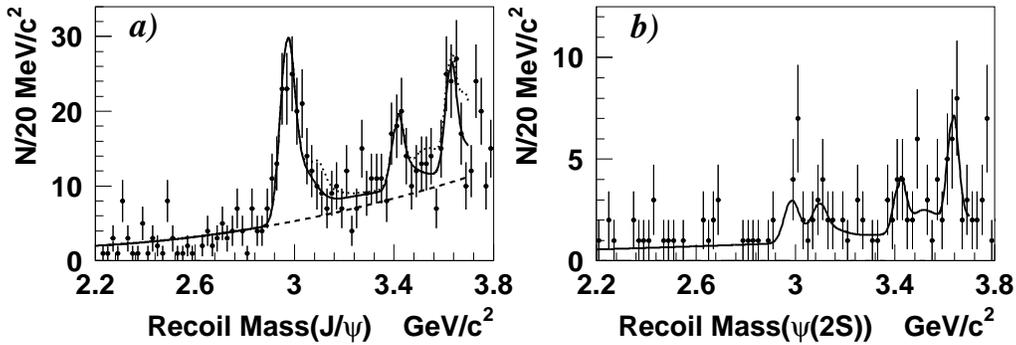}
  \end{center}
  \caption{ 
    Distribution of recoil masses from inclusive {\bf (a)} $J/\psi$
    and {\bf (b)} $\psi (2S)$ mesons produced in continuum $e^+e^-$
    annihilation processes near $\sqrt{s}=10.6$~GeV.  
    The three peaks in the $J/\psi$ recoil mass spectrum are
    the $\eta_c$, the $\chi_{c0}$ and the $\eta_c'$.  
    There is no evidence for events with recoil masses below 
    the $\eta_c$; there the observed level of events  is
    consistent with background. 
  }
  \label{fig:double_ccbar}
\end{figure}
\end{description}

\subsubsection{Strategies for a Super-B factory}
These examples demonstrate that when
data samples containing a significantly
larger number of $B$ mesons becomes available,
we can  reasonably expect that, as a bonus, 
charmonium studies might produce important
new discoveries. 
As the $X(3872)$ and continuum $e^+e^-\to c\bar{c}c\bar{c}$
discoveries demonstrate, these could provide new 
insights into hadronic physics.   

The $e^+e^-$ environment is well suited for these
investigations.  The $X(3872)\to \pi^+\pi^-\jp$
has an especially distinct signature that makes
it observable at hadron  colliders~\cite{Acosta:2003zx}.
However, understanding the nature of the particle
requires sensitivity to other, less distinct 
modes such as $\gamma\chi_{c1}$, and also 
$\pi^0\pi^0 J/\psi$, $\pi^+\pi^-\pi^0 J/\psi$
and $D^0\bar{D^0}\pi^0$~\cite{Voloshin:2003nt,Swanson:2003tb};
these are only accessible in the $e^+e^-$ environment.
Likewise, the recoil-mass technique that was used to
discover $e^+e^-\to c\bar{c}c\bar{c}$ is also
only possible in the $e^+e^-$ environment.

\begin{description}
\item[Search for the $h_{c}$:]
The $h_{c}$ has proven to be the most elusive of the
low-lying charmonium states.  A signal for an $h_c$
candidate in E760 at Fermilab~\cite{Armstrong:1992ae} was 
not reproduced in the subsequent experiment E835~\cite{e835_hc}.
The $h_c$ is expected to decay predominantly into final states that include
an $\eta_c$; these are experimentally not very distinct.
Searches in $B$ meson decays are also hampered by the
fact that the exclusive decay process $B\to K h_c$, which 
violates factorization, is expected to be small.
Using a technique based on a suggestion by Suzuki~\cite{Suzuki:2002sq},
Belle is looking for the $h_c$ via the $B\to K h_c$:
$h_c\to\gamma\eta_c$ decay chain.  Here the
hope is that the large expected decay branching ratio
for $h_c\to\gamma\eta-c$ ($\sim 60\%$) might compensate for the small,
factorization-suppressed rate for $B\to K h_c$.

So far, no signal is seen, but the branching ratio limit
(of $\sim 1\times 10^{-4}$) is not very restrictive.  This is
because of the low efficiency for reconstructing the $\eta_c$
in low-background channels.
Future searches at higher luminosity will improve this
limit, but only by a factor that goes
as the square-root of the increase in luminosity.

The process process $e^+e^-\to \eta_c h_c$ is allowed by
$C$-parity conservation.
Thus, another approach for $h_c$ searches would be to use 
experimentally distinct $\eta_c$ decay final states, such as 
$p\bar{p}$ and $4K$ modes, to study states recoiling 
against continuum $\eta_c$s in
$e^+e^-\to c\bar{c}c\bar{c}$ processes.  Since the useful
$\eta_c$ modes have branching fractions ($\sim 10^{-3}$)
that are $\sim 10^{-2}$ smaller than those for the $J/\psi$
this method will require a large data sample.  If we assume that
the $\sigma(e^+e^-\to\eta_c h_c)\simeq \sigma(e^+e^-\to J/\psi 
\chi_{c0})$, about 10~ab$^{-1}$ of data would be needed to
uncover the $h_c$ by this method.

\item[Search for the $^3D_{c2}$:]
Since it looks more-and-more unlikely that the $X(3872)$ is the 
$^3D_{c2}$, the discovery of the $^3D_{c2}$ remains an
open experimental question.  Since the exclusive decay
$B\to K ^3D_{c2}$ is expected to by strongly suppressed
by factorization, here also searches in exclusive decays
are not promising.  Since the $\gamma\chi_{c1}$ and 
$\pi^+\pi^- J/\psi$ modes  are expected to be strong and
are experimentally distinct, inclusive searches are
feasible.  (Belle has reported a strong signal for
the inclusive process $B\to \chi_{c2} X$~\cite{Abe:2002wp}, 
even though no evidence has been seen for it in exclusive
two-body decays.)
\end{description}

\subsubsection{Summary}
A super-$B$ factory will provide opportunities for
unprecedented studies of the properties of charmonium 
systems and to search for the missing charmonium states.
These measurements will provide stringent tests of hadronic
models where they are supposed to be most reliable.

Taking  history as our guide, we are confident that new and 
surprising discoveries will be made.

\subsection{Physics potential at $\Upsilon(5S)$}
\def\bs{$B^{}_s$}
\def\bd{$B^{}_d$}
\def\bsbar{$\overline{B}^{}_s$}
\def\cp{$CP$}
\def\ra{\!\rightarrow\!}

\newcommand{\GeVsq}{\mbox{${\rm GeV}^2$}}

The physics potential of the Super KEKB collider
can be extended significantly by running at the energy 
of the $\Upsilon$(10860) resonance, usually denoted as 
the $\Upsilon$(5S). The advantage is that at this center-of-mass energy
it is possible to produce pairs of \bs/\bsbar\ mesons, which
are kinematically forbidden when running at the $\Upsilon(4S)$ resonance.
This would provide an opportunity to study \bs\ decays in
a relatively low background environment as compared to that
at a hadron collider. In order to operate the KEKB accelerator 
in the $\Upsilon$(5S) mass range, the electron and positron 
beam energies would need to be increased by 2.7$\%$ relative
to their energies at the $\Upsilon$(4S) resonance;
this increase would result in the same Lorentz
boost factor of $\beta\gamma=0.425$.

The production ratio of \bs\ mesons at the $\Upsilon$(5S) resonance
to \bd\ mesons at the $\Upsilon$(4S) resonance is conservatively
estimated to be 1/10 
\cite{Besson:1984bd,Lovelock:nb,Lee-Franzini:gy}.
Thus for an integrated luminosity of 1~ab$^{-1}$,
which is equivalent to about 10$^9$ 
$\Upsilon$(4S)$\to B \bar{B}$ decays,
a total of 10$^8$ \bs\ decays
would be recorded. A reconstruction
efficiency of 10\% would therefore allow one to measure
branching fractions down to the level of~$10^{-7}$.

The $\Upsilon$(5S) can decay to $B\bar{B}$,
$B\bar{B^*}$, $B^*\bar{B^*}$, $B_s\bar{B_s}$, $B_s\bar{B_s^*}$ or
$B_s^*\bar{B_s^*}$ final states. The excited $B$ mesons
decay to ground states via $B^* \to B \gamma$ and
$B_s^* \to B_s \gamma$ radiative decays.
A detailed Monte Carlo simulation shows that, 
using the full reconstruction technique, \bs\ signals 
for these states are well-identified using the variables
$E_B^*-E_{beam}$ and $P_B^*$,
where $E_B^*$ and $P_B^*$ are the reconstructed energy and 
momentum of the \bs\ candidate in the $e^+e^-$ center-of-mass frame,
and $E_{beam}$ is the beam energy in this frame. The number of low 
energy photons from $B_s^* \to B_s \gamma$ decays can also be 
determined, providing a measurement of the total numbers of 
$B_s\bar{B_s^*}$ and $B_s^*\bar{B_s^*}$ final states produced.
Knowing this total yield would allow one to
measure absolute (rather than relative) branching
fractions, which are very difficult to measure at 
a hadron machine.

Many $B_s$ decay modes can be observed and
studied at an $e^+ e^-$ Super $B$ Factory. The strange ``partners''
of topical \bd\ decays can be reconstructed, such as Cabibbo-favored
$B_s \to D_s^- \pi^+$ and $B_s \to D_s^- D_s^+$ decays, color-suppressed
$B_s \to D K$ and $B_s \to J/\Psi \phi$ decays, Cabibbo-suppressed
$B_s \to D_s^- K^+$ and $B_s \to J/\Psi K^0$ decays, the electroweak
penguin decay $B_s \to \phi \gamma$, and the $b$ to $u$ transition
$B_s \to K \pi$. Final states containing $\pi^0$'s, which
are difficult to separate from background in a hadron collider 
environment, would be well-identified at an $e^+e^-$ machine.
In addition, $\Upsilon$(5S) decays are well-suited for 
studying large multiplicity \bs\ decays due to the lower
particle momenta, the $\sim$\,100\% trigger efficiency,
and the excellent $\pi/K$ discrimination.
Inclusive $B_s$ branching fractions can also be measured,
in particular the inclusive leptonic branching fraction.
Such inclusive measurements are easier to compare with
theoretical predictions.
Partial-reconstruction techniques can also be used, for example, 
to measure the exclusive decay $B_s \to D_s^+ l^- \nu$.
The numbers of \bs\ decays reconstructed in several topical decay
modes are listed in Table~\ref{tab:bfr}, along with the corresponding
numbers of events (when available) for the LHCB~\cite{Ball:2000ba,lhcb} and 
BTeV~\cite{Moroni:2003ms} hadron collider experiments. The event yields 
listed correspond to one year of running; the corresponding 
luminosity at the Super $B$ Factory would be 4~ab$^{-1}$.

\renewcommand{\arraystretch}{1.1}
\begin{table}[t!]
\vspace{-0.2cm}
\begin{center}
\begin{tabular}{|l|c|c|c|c|c|}
\hline
 & Branching  &  & Super $B$ & LHCB & BTeV \\
Decay & fraction & Efficiency & (4\,ab$^{-1}$) & (per year) & (per year) \\ \hline
$D_s^+ \pi^-$ & 5 $\cdot$ 10$^{-3}$ & 2 $\cdot$ 10$^{-2}$ & 40000 & 80000 & 120000 \\
$J/\Psi \phi$ & 1 $\cdot$ 10$^{-3}$ & 1.5 $\cdot$ 10$^{-2}$ & 6000 & 120000 & $\sim$ 90000 \\
$D_s^{(*)+} D_s^{(*)-}$ & 5 $\cdot$ 10$^{-2}$ & 5 $\cdot$ 10$^{-4}$ & 10000 &  &  \\ \hline
\end{tabular}
\caption{The number of fully-reconstructed $B_s$ mesons expected from
$\Upsilon$(5S) decays with $L_{int}=4$~ab$^{-1}$, and by the LHCB and 
BTeV hadron collider experiments per year of running.
\label{tab:bfr}}
\end{center}
\vspace{-0.1cm}
\end{table}

An important measurement in \bs\ physics is
that of the lifetime or decay width difference $\Delta \Gamma$
between the two mass eigenstates of the \bs-\bsbar\ system
\cite{Beneke:1996gn,Hartkorn:ga,Barate:2000kd}.
The ratio of the difference to the mean value is theoretically
predicted to be $\Delta \Gamma / \Gamma \approx 15\%$,
where $\Delta \Gamma\equiv \Gamma_{light} - \Gamma_{heavy}$
\cite{Beneke:1998sy,Ciuchini:2003ww,%
  Hashimoto:1999yh,Hashimoto:2000eh,Becirevic:2000sj,Flynn:2000hx,%
  Aoki:2002bh,Beneke:2000cu}.
Such a difference could be measured at an 
asymmetric $e^+ e^-$ collider. Assuming \cp\ conservation,
the mass eigenstates are \cp\ eigenstates, and one could 
compare the lifetime distributions of a \bs/\bsbar\ decaying 
to \cp-even\ and \cp-odd final states. In practice, one would 
plot the time difference $\Delta t=t^{}_{heavy}-t^{}_{light}$
between the decay vertices for $\Upsilon(5S)\ra B^{}_s\overline{B}^{}_s$
decays, in which one \bs\ decays to a \cp-even final state
and the other decays to a \cp-odd final state. This distribution
is proportional to $e^{-(\Gamma^{}_{heavy}\,\Delta t)}$ for $\Delta t>0$ and 
$e^{+(\Gamma^{}_{light}\,\Delta t)}$ for $\Delta t<0$; thus, fitting to this
distribution yields both $\Gamma^{}_{heavy}$ and $\Gamma^{}_{light}$.
For $\Upsilon(5S)\ra B^{}_s\overline{B}^*_s$ decays,
both \bs\ mesons decay to \cp-even or \cp-odd final
states (i.e., no mixture of \cp-even and \cp-odd), so
the $\Delta t$ distribution is symmetrical and would
yield an independent measurement of $\Gamma^{}_{heavy}$
or $\Gamma^{}_{light}$. Some \cp-definite final states 
with measurable branching fractions are $D_s^{(*)+} D_s^{(*)-}$, 
$K^+ K^-$, and $\phi \phi$. Using the above methods, it is 
estimated that an accuracy for $\Delta\Gamma/\Gamma$ of 
$\sim$\,2\% would be obtained with an integrated luminosity 
of 4~ab$^{-1}$. 

If these $\Delta t$ distributions show deviations
from pure exponential behavior, that would indicate
\cp\ violation. In particular, if the mass eigenstates
were not \cp\ eigenstates, then the $\Delta t$
distribution (i.e., for $\Delta t>0$ or $\Delta t <0$) 
would be the sum of two exponentials. Thus, looking for a
deviation from a single exponential allows one to search 
for \cp\ violation in the \bs-\bsbar\ system. 
Some theoretical models predict
a very large \bs-\bsbar\ mass difference $\Delta m^{}_s$
(i.e., $\Delta m^{}_s> 100$~ps$^{-1}$); in this case 
it would be very difficult to observe \bs-\bsbar\ mixing 
at a hadron collider, and consequently, \cp\ violation due
to interference between mixed and unmixed decay amplitudes.
Thus, an $e^+e^-$ machine
(and the quantum correlations inherent in production via the 
$\Upsilon(5S)$) may provide the best opportunity to 
observe \cp-violation in this system.

The time-integrated $CP$ asymmetry 
$A\equiv (N^{}_{\overline{B}^{}_s\rightarrow f}-N^{}_{B^{}_s\rightarrow\bar{f}})/
(N^{}_{\overline{B}^{}_s\rightarrow f}+N^{}_{B^{}_s\rightarrow\bar{f}})$
would allow one to measure {\it direct\/} \cp\ violation in \bs\ decays.
A good candidate for this measurement is $B_s \to K^- \pi^+$:
assuming a branching fraction of 10$^{-5}$ and an asymmetry of 
$-6$\%~\cite{Sun:2002rn}, one would observe a $3\sigma$ effect with
4~ab$^{-1}$ of data.
The penguin diagram causing the direct \cp-violation 
is very sensitive to nonstandard contributions (some of which can
give asymmetries much larger than 6\%), and measuring a direct \cp\ 
asymmetry could probe physics beyond the Standard Model.

\clearpage \newpage
\def\fcp{{\rm f}_{CP}}
\def\Nrec{N_{\rm rec}}
\def\Nev{N_{\rm ev}}
\def\sinbb{{\sin2\phi_1}}
\def\piz{\pi^0}
\def\pip{\pi^+} 
\def\pim{\pi^-}
\def\kz{K^0}
\def\kp{K^+}
\def\km{K^-}
\def\ks{K_S^0}
\def\bm{{B^-}}
\def\jpsi{J/\psi}
\def\kstarz{K^{*0}}
\section{Summary}
\label{sec:sss/summary}
Table~\ref{tbl:sensitivity_summary} summarizes
the sensitivities for some of key observables
described in the previous sections.
As a comparison, we also list expected sensitivities at
LHCb whenever available.
It is seen that most of key observables 
are accessible only at the $e^+e^-$ $B$ factories. 
The advantage of the clean environment at SuperKEKB is thus clear.
Note that the $B$ physics program at hadron colliders has
its own unique measurements that are not accessible
at $e^+e^-$ $B$ factories. Examples include
rare $B_s$ decays such as $B_s \to \mu^+\mu^-$.
Thus $B$ physics programs at hadron colliders
also help scrutinize the rich phenomenologies of
$B$ meson decays.
\begin{table}
\small
\begin{center}
\begin{tabular}{lrrrr}
\hline
Observable     &Belle 2003     &\multicolumn{2}{c}{SuperKEKB}& LHCb \\
               &(0.14ab$^{-1}$)&(5 ab$^{-1}$)&(50 ab$^{-1})$&(0.002ab$^{-1}$)\\
\hline 
$\deltasphiks$ & 0.51          & 0.079     & 0.031 & 0.2~\cite{bib:LHCbPhiKs} \\
$\deltaskkks$  & $^{+0.32}_{-0.26}$ & 0.056  & 0.026         &  \\
$\deltasetapks$& 0.27          & 0.049       & 0.024         & $\times$ \\
$\deltasksksks$& NA            & 0.14        & 0.04          & $\times$ \\
$\deltaspizks$ & NA            & 0.10        & 0.03          & $\times$ \\
$\sin2\chi$ ($B_s \to J/\psi\phi$)&$\times$&$\times$&$\times$& 0.058\\
\hline
$\calsksgamma$ & NA            & 0.14        & 0.04          & $\times$ \\
$\Br(\BtoXsgamma)$ & 26\% (5.8 fb$^{-1}$) & 5\%  & 5\%        & $\times$ \\
$\Acp(\BtoXsgamma)$ & 0.064    & 0.011       & 5$\times$10$^{-3}$ &$\times$\\
$C_9$    from $\AFB(\BtoKstarll)$ & NA & 32\% & 10\% & \\
$C_{10}$ from $\AFB(\BtoKstarll)$ & NA & 44\% & 14\% & \\
${\cal B}(B_s \rightarrow \mu^+\mu^-)$ & $\times$ & $\times$&$\times$&4$\sigma$ (3 years)~\cite{bib:LHCbBs2mumu}\\
\hline
${\cal B}(\bp\to\kp\nu\nu)$ & NA &           & 5.1$\sigma$   & $\times$ \\
${\cal B}(\bp\to D \tau \nu)$ & NA & 12.7$\sigma$ & 40.3$\sigma$ & $\times$ \\
${\cal B}(\bz\to D \tau \nu)$ & NA & 3.5$\sigma$ & 11.0$\sigma$  & $\times$ \\
\hline
$\sinbb$       & 0.06          & 0.019       & 0.014         & 0.022\\
$\phi_2$ ($\pi\pi$ isospin)& NA& 3.9$^\circ$ & 1.2$^\circ$   & $\times$ \\
$\phi_2$ ($\rho\pi$) & NA      & 2.9$^\circ$ & 0.9$^\circ$   & $\times$ \\
$\phi_3$ ($DK^{(*)}$)&20$^\circ$ & 4$^\circ$ & 1.2$^\circ$   & 8$^\circ$\\
$\phi_3$ ($B_s \rightarrow KK$) & $\times$ & $\times$ & $\times$ & 5$^\circ$\\ 
$\phi_3$ ($B_s \rightarrow D_sK$) & $\times$ &$\times$ &$\times$ &14$^\circ$\\ 
$|V_{ub}|$ (inclusive) & 16\%   & 5.8\%       & 4.4\%         & $\times$\\
\hline
${\cal B}(\tau\to\mu\gamma)$ 
               &$< 3.1\times 10^{-7}$ & $< 1.8 \times 10^{-8}$ &  &\\
${\cal B}(\tau\to \mu(e) \eta)$ 
               & $< 3.4(6.9)\times 10^{-7}$   & $< 5 \times 10^{-9}$ &  &\\
${\cal B}(\tau \to \ell\ell\ell)$ 
               & $<$ 1.4-3.1$\times 10^{-7}$ & $< 5 \times 10^{-9}$ &  &\\
\hline
\end{tabular}
\end{center}
\caption{Summary of sensitivity studies. Values for LHCb are taken from
\cite{bib:LHCb} unless otherwise stated.}
\label{tbl:sensitivity_summary}
\end{table}



\def\calsksgamma{\cals_{K^{*0}\gamma}}
\chapter{Study of New Physics Scenarios at SuperKEKB}

\section{New physics case study}
\label{sec:new_physics_case_study}
As discussed in the previous sections, there are a number of
processes in which one may find observable effects of
physics beyond the Standard Model (BSM).
In particular, processes induced by loop diagrams can be
sensitive to new particles at the TeV energy scale, as
the Standard Model contribution is relatively suppressed.
Such processes include $\mathcal{S}_{\phi \ks}$,
$A_{CP}^{\mathrm{dir}}(b\to s\gamma)$, 
$A_{CP}^{\mathrm{mix}}(b\to s\gamma)$, 
$Br(b\to s\ell^+\ell^-)$, $A_{FB}(b\to s\ell^+\ell^-)$, and
$Br(\tau\to\mu\gamma)$.
In this section we consider some specific new physics models
in the context of supersymmetry (SUSY) and investigate how their
effects could be seen in these processes. 

The hierarchy problem of the Higgs mass suggests that the
physics beyond the Standard Model most likely exists at the
TeV energy scale, and hence the LHC has a good chance of
discovering some new particles. 
However, the flavor structure of the BSM will still remain to be
investigated, as hadron collider experiments are 
largely insensitive to flavor-violating processes.
In SUSY models, for example, flavor physics is the
key to investigating the mechanism of SUSY breaking, which
is expected to lie at a higher energy scale.

As discussed in Section~\ref{sec:Supersymmetric_models},
even in the minimal supersymmetric extension of the SM
(MSSM), most of the parameter space is already excluded by
the present FCNC constraints, and hence one has to consider
some structure in the soft SUSY breaking terms such that the 
FCNC processes are naturally suppressed.
In this study we consider the following three models
according to refs.~\cite{Goto:2002xt,Goto:2003iu}.
\begin{itemize}
\item \textit{Minimal supergravity model (mSUGRA)}\\
  In this model one assumes that supersymmetry is broken
  in some invisible (hidden) sector and its effect is
  mediated to the visible (observable) world only through the
  gravitational interaction.
  Since gravity is insensitive to flavor, the induced
  soft breaking terms are flavor blind at the scale where they
  are generated.
  Namely, all sfermions have degenerate mass $m_0^2$;
  trilinear couplings of scalars are proportional to Yukawa
  couplings; all gauginos are degenerate with mass $M_{1/2}$.
  Even though the soft breaking terms are flavor blind, they
  could induce flavor violation as they evolve from the SUSY
  breaking scale 
  (assumed here to be $M_X\simeq 2\times 10^{16}$~GeV for
  simplicity) to the electroweak scale by the
  renormalization group equations, since the Yukawa
  couplings are flavor dependent.
  In general, FCNC effects are small for this model.
\item \textit{$SU(5)$ SUSY GUT with right-handed neutrinos}\\
  Grand Unification (GUT) is one of the motivations to
  consider supersymmetric models, as they naturally lead
  to the unification of the gauge couplings.
  In GUT models quarks and leptons belong to the
  same multiplet at the GUT scale.
  Therefore the flavor mixings of the quark
  and lepton sectors are related to each other.
  In particular, large neutrino mixing can induce large 
  squark mixing for right-handed down-type squarks
  \cite{Baek:2000sj,Moroi:2000mr,Moroi:2000tk,Akama:2001em,%
    Baek:2001kh,Chang:2002mq}.
  In this class of models we consider two cases where the
  masses of the right-handed neutrinos are \textit{degenerate}
  or \textit{non-degenerate}.
  Because of the large neutrino mixing, the bulk of the
  parameter space is excluded in the degenerate case due to
  the $\Br(\mu\to e\gamma)$ constraint, and as a
  result the effect on other FCNC processes is limited.
  On the other hand, for the non-degenerate case one can
  expect larger FCNC effects.
\item \textit{$U(2)$ flavor symmetry}\\
  The family structure of the quarks and leptons could be
  explained by some flavor symmetry among generations.
  Although $U(3)$ is a natural candidate for such a symmetry, it
  is badly broken by the top quark Yukawa coupling.
  Therefore, here we consider a $U(2)$ flavor symmetry among
  the first two generations in the context of SUSY.
  We assume some breaking structure of the $U(2)$ symmetry for
  the Yukawa couplings, squark mass matrices, and the scalar
  trilinear couplings.
\end{itemize}

These models still contain many parameters.
We randomly choose a number of points from the
multi-dimensional parameter space, while fixing the gluino
mass at 600~GeV/c$^2$ and $\tan\beta=30$.
Among these points, those that are already excluded from
existing mass bounds and FCNC constraints, such as
$\epsilon_K$, $b\to s\gamma$ branching fraction, and 
$\mu\to e\gamma$ branching ratio, are eliminated. 
The results for several quantities are then shown in
the following scatter plots.

\begin{figure}[tbp]
\begin{center}
\begin{minipage}[t]{9cm}
  (a)\\
  \includegraphics[width=1.0\textwidth,clip]{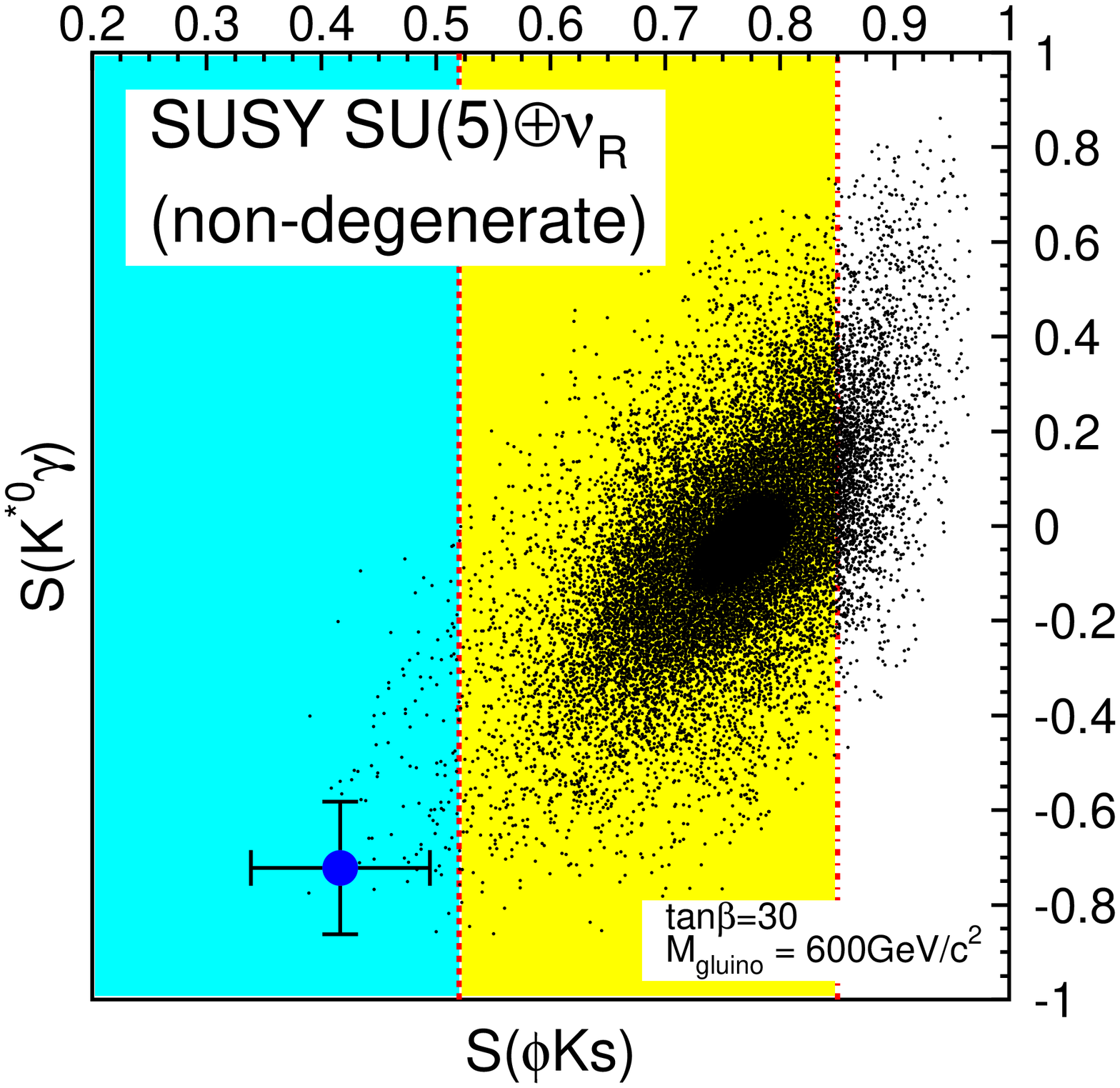}
\end{minipage}
\begin{minipage}[t]{5cm}
  (b)\\
  \includegraphics[width=1.0\textwidth,clip]{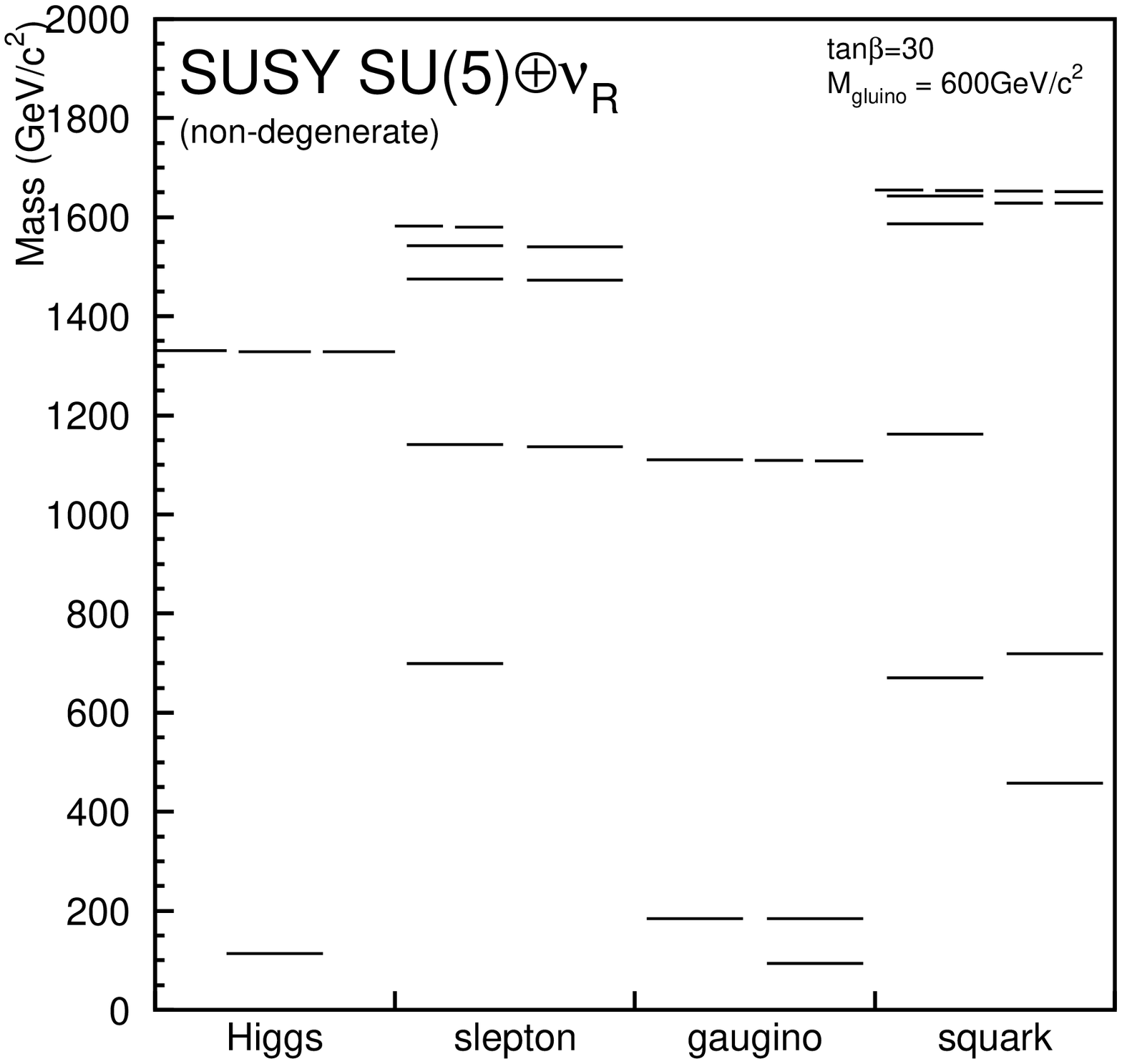}
\end{minipage}
\begin{minipage}[t]{9cm}
  (c)\\
  \includegraphics[width=1.0\textwidth,clip]{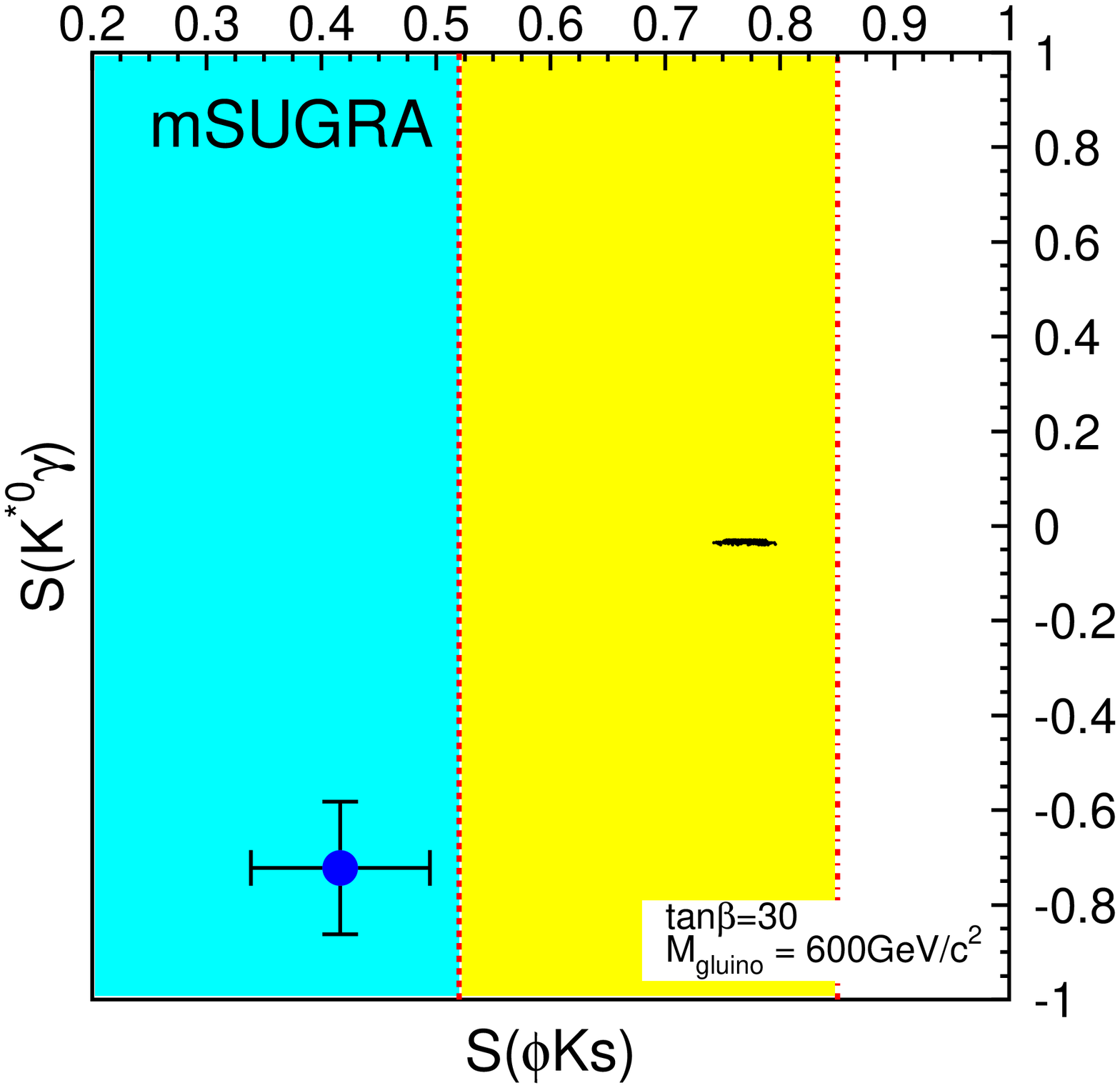}
\end{minipage}
\begin{minipage}[t]{5cm}
  (d)\\
  \includegraphics[width=1.0\textwidth,clip]{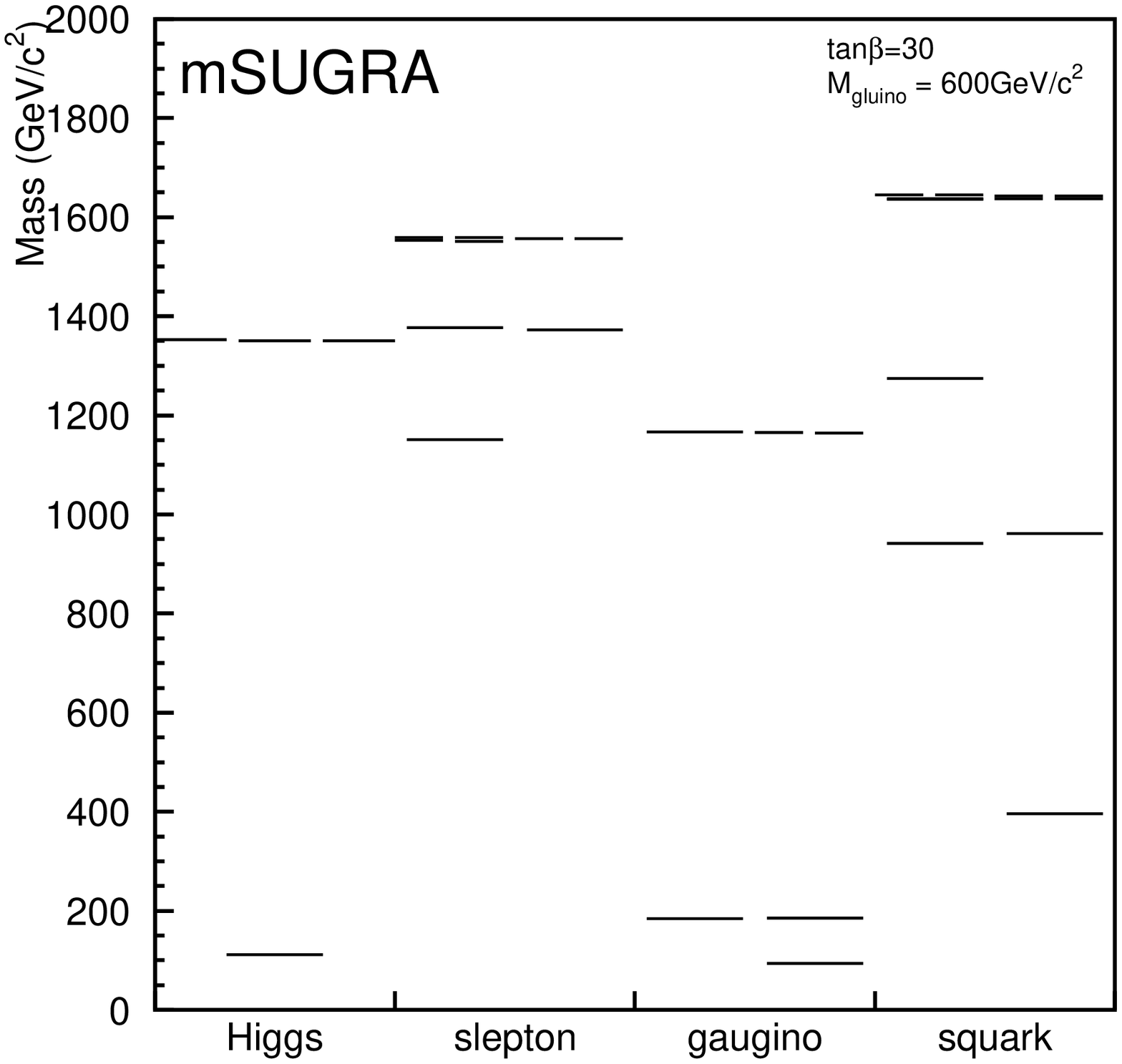}
\end{minipage}
\end{center}
\caption{
  $\calsksgamma$ and $\calsphiks$ for various parameters in
  (a) SUSY $SU(5)$ with right-handed neutrinos
  (the non-degenerate case), and (c) mSUGRA.  
  Circles with error bars indicate
  an expected result from a certain parameter set in the $SU(5)$ SUSY GUT,
  where the errors are obtained at 5~ab$^{-1}$.
  A present experimental bound at 2$\sigma$ (3$\sigma$) level
  is also shown by the dashed (dot-dashed) vertical line.
  Associated small figures show examples of
  mass spectra of SUSY particles for (b) the SUSY $SU(5)$
  and (d) mSUGRA.
  }
  \label{fig:sbsg_vs_sphiks}
\end{figure}

First of all, the expectations from these models for two
interesting FCNC processes 
$\mathcal{S}_{\phi \ks}$ and $\mathcal{S}_{b\to s\gamma}$ 
are shown in Fig.~\ref{fig:sbsg_vs_sphiks}, together with
the expected sensitivity at SuperKEKB.
Each dot in the plots shows a randomly chosen parameter
point of the model.
One can observe a significant deviation from the SM for the
$SU(5)$ SUSY GUT with a right-handed neutrino (non-degenerate
case) while the mSUGRA model gives comparatively small deviations
from the SM for these FCNC processes.
We emphasize that these different models could have almost
an identical mass spectrum of SUSY particles as shown on the
bottom panels in the figure.
This means that the flavor structure of BSM models
is barely probed by measurements at high energy collider experiments. 

\begin{figure}[tbp]
\begin{center}
\includegraphics[width=1.0\textwidth,clip]{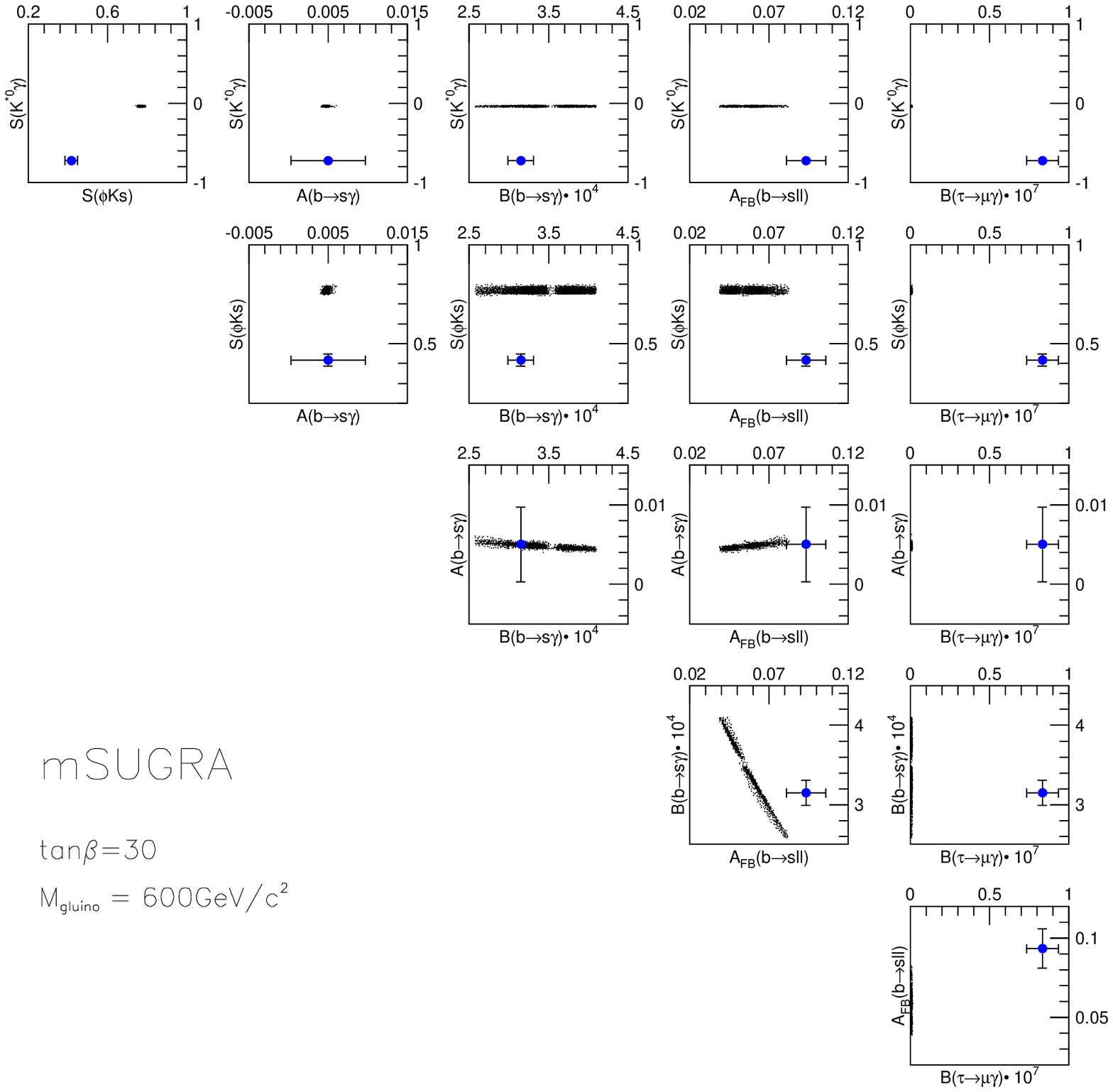}
\end{center}
\caption{Correlations between key observables in the case of mSUGRA.
         Dots show the possible range in mSUGRA.
         The circles correspond to a certain parameter set of
         non-degenerate SUSY SU(5) GUT model with $\nu_R$.
         Expected errors with an integrated luminosity of 50 ab$^{-1}$
         are shown by bars associated with the circles.}
\label{fig:allcorr_msg}
\end{figure}

\begin{figure}[tbp]
\begin{center}
\includegraphics[width=1.0\textwidth,clip]{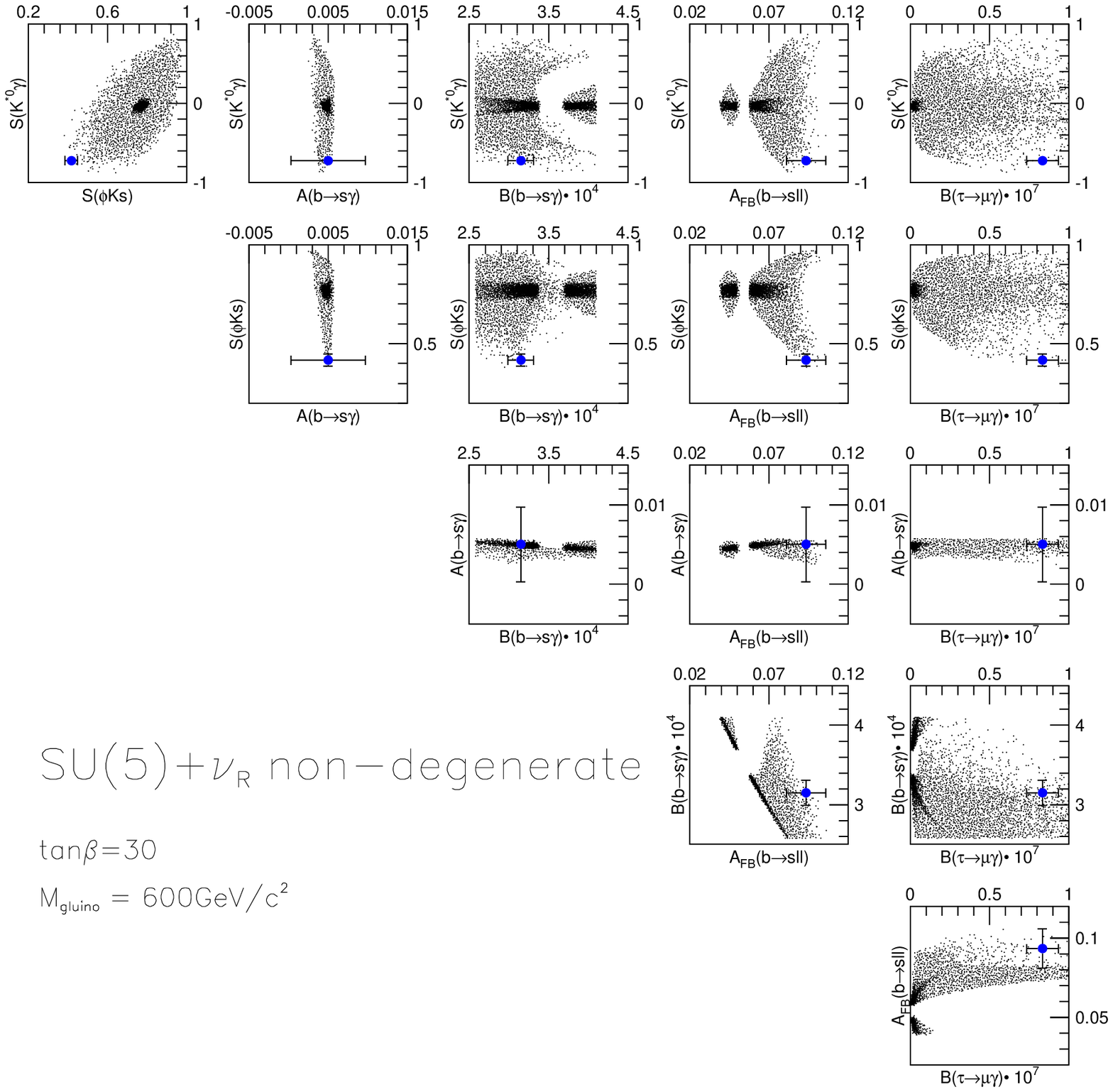}
\end{center}
\caption{Correlations between key observables in the case of 
         the non-degenerate SUSY SU(5) GUT model with $\nu_R$.
         Dots show the possible range in the model.
         The circles correspond to a certain parameter set in this model.
         Expected errors with an integrated luminosity of 50 ab$^{-1}$
         are shown by bars associated with the circles.}
\label{fig:allcorr_nondeg}
\end{figure}

\begin{figure}[tbp]
\begin{center}
\includegraphics[width=1.0\textwidth,clip]{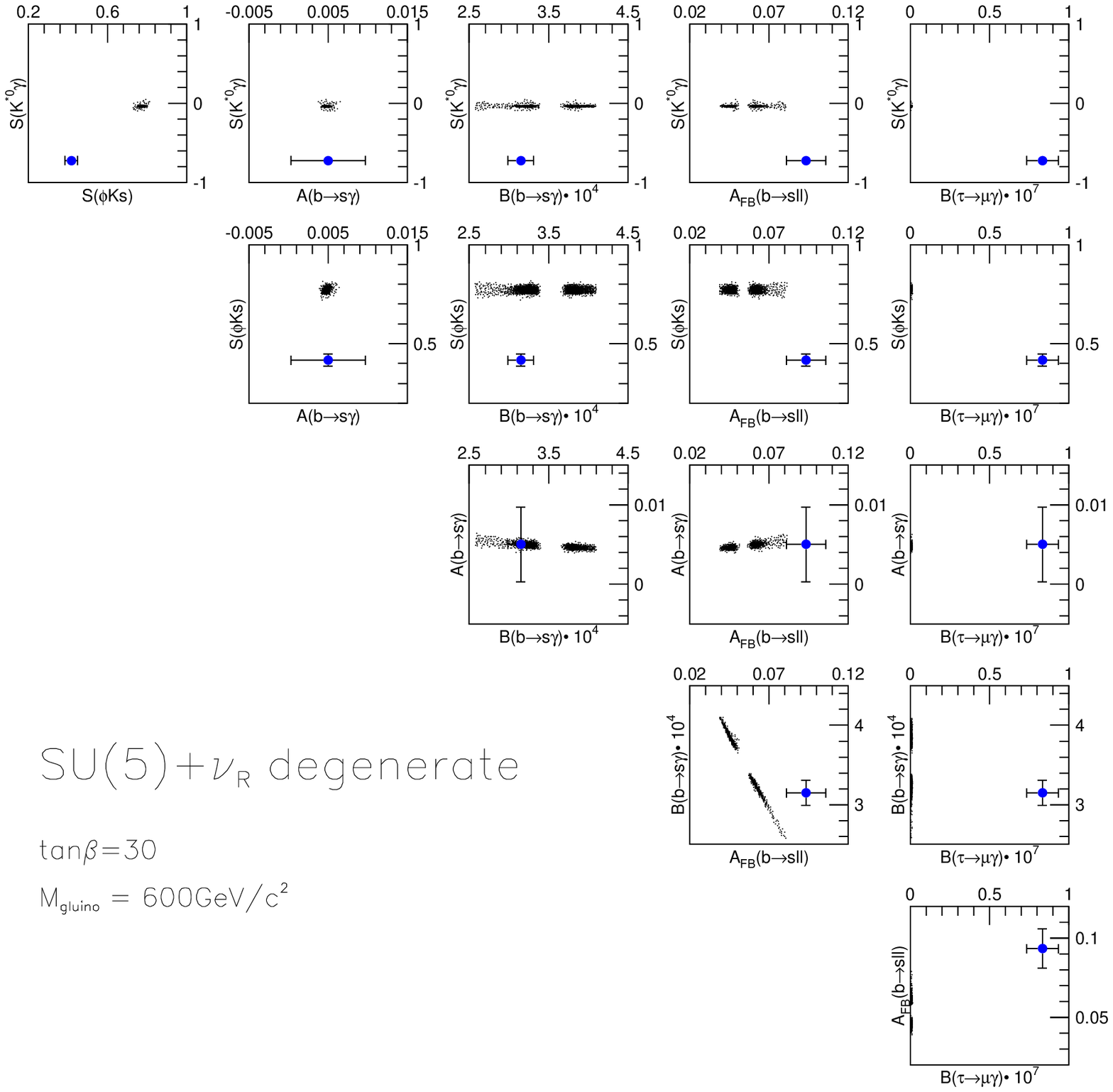}
\end{center}
\caption{Correlations between key observables in the case of 
         the degenerate SUSY SU(5) GUT model.
         Dots show the possible range in the model.
         The circles correspond to a certain parameter set of the
         non-degenerate SUSY SU(5) GUT model with $\nu_R$.
         Expected errors with an integrated luminosity of 50 ab$^{-1}$
         are shown by bars associated with the circles.}
\label{fig:allcorr_deg}
\end{figure}

\begin{figure}[tbp]
\begin{center}
\includegraphics[width=1.0\textwidth,clip]{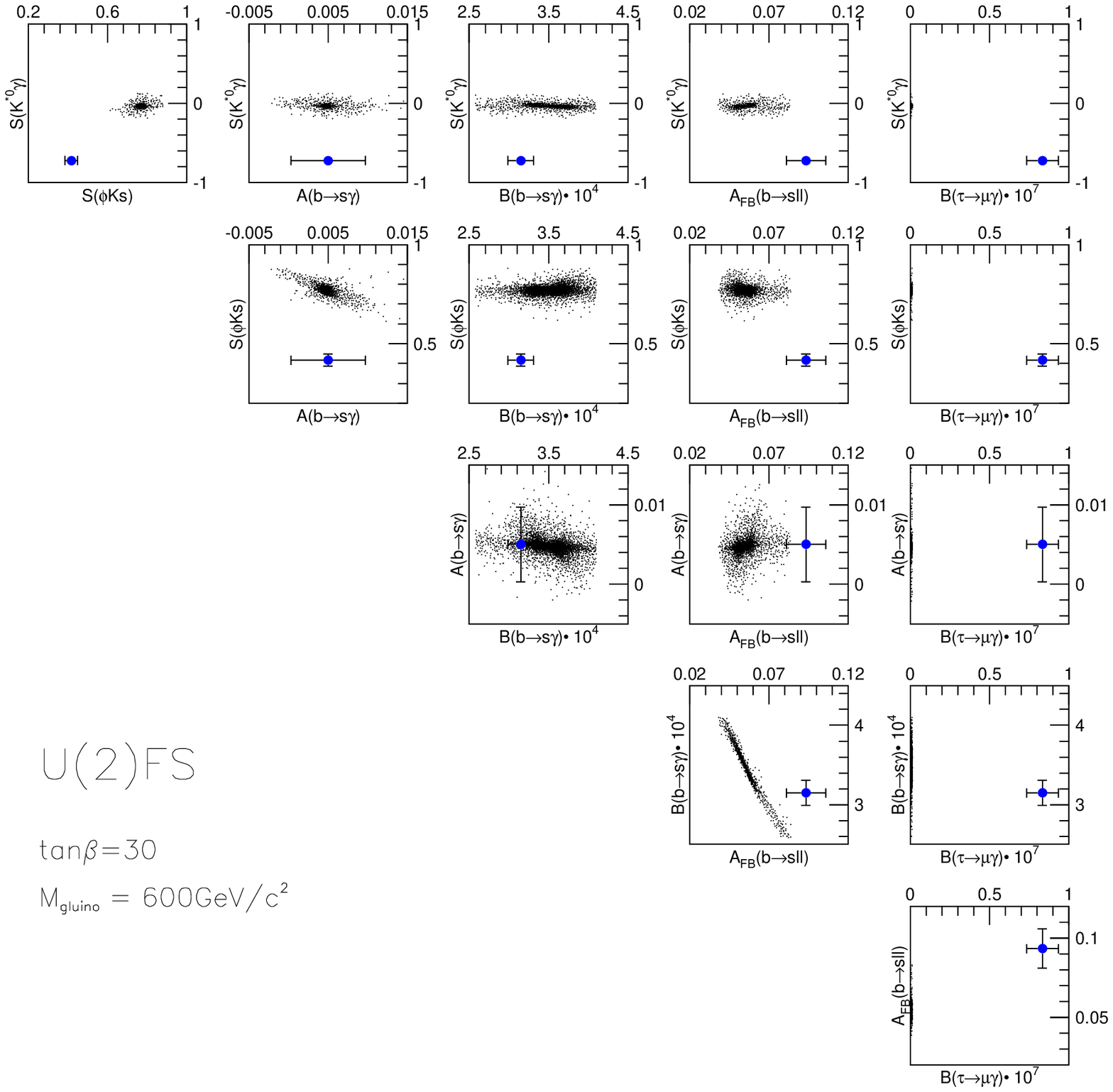}
\end{center}
\caption{Correlations between key observables in the case of 
         the U(2) flavor symmetry model.
         Dots show the possible range in the model.
         The circles correspond to a certain parameter set of the
         non-degenerate SUSY SU(5) GUT model with $\nu_R$.
         Expected errors with an integrated luminosity of 50 ab$^{-1}$
         are shown by bars associated with the circles.}
\label{fig:allcorr_u2}
\end{figure}

Correlations for other interesting FCNC processes are shown in
Figs.~\ref{fig:allcorr_msg}--\ref{fig:allcorr_u2} for the
above four SUSY models. Also shown in the figures are
expected errors  at 50 ab$^{-1}$. 
With these additional correlations, we can further
narrow down possible scenarios; 
for instance, 
large direct $CP$ violation in $b \to s\gamma$ would imply that
the U(2) model was favored. On the other hand,
observation of $\tau\to\mu\gamma$ decays would strongly
suggest the non-degenerate SU(5) model.
Even if the deviation from the mSUGRA in
$\mathcal{S}_{\phi \ks}$ and $\mathcal{S}_{b\to s\gamma}$ 
turns out to be too small to distinguish among
different scenarios, there are many other possibilities
that can be pursued at SuperKEKB. 

\section{Model independent approaches}
At SuperKEKB, there are a number of processes
from which we can measure the fundamental parameters of the
Standard Model (SM), 
\textit{i.e.} quark mixing angles and phases.
In this section we discuss how these measurements at
SuperKEKB may be used to gain insight into the flavor
structure of new physics without assuming some particular
model. 

Fitting the CKM parameters using the available data is a
common practice.
One draws the constraints obtained from several measurements
in the parameter space of $(\bar{\rho},\bar{\eta})$, which
are the least known parameters in the usual parametrization of the
CKM matrix. 
If one found an inconsistency among the different input quantities,
it would indicate the existence of a new physics effect for some
quantities.
We investigate how SuperKEKB can contribute to narrowing
down the allowed region from each measurement.

Once deviations from the SM are established, the next
question is which quantities are affected by the new physics
and how. 
To consider such questions, it is natural to assume that
tree-level processes are not affected much by the new
physics.
If the Standard Model is a low energy effective theory of
some more fundamental theory, the new physics effects will
manifest themselves as non-renormalizable higher dimensional operators.
In tree-level processes such operators are relatively
suppressed, while in loop-induced FCNC processes the
higher dimensional operators could compete with the loop
effects of the Standard Model.

Therefore, we may proceed in three steps.
Firstly, we obtain a constraint on 
$(\bar{\rho},\bar{\eta})$ using only tree-level
processes, \textit{i.e.} $|V_{ub}|$ from the semileptonic
decay $b\rightarrow u\ell\bar{\nu}$ and $\phi_3$ from the $DK$
asymmetry.
The result can be considered as the SM value of
$(\bar{\rho},\bar{\eta})$ even in the presence of the new
physics.
Secondly, the allowed region for $(\bar{\rho},\bar{\eta})$ is compared
with the determination through $B^0-\bar{B}^0$ mixing,
\textit{i.e.} $\Delta m_d$ and $\sin 2\phi_1$ from
$B\rightarrow J/\psi \ks$.
If we assume that the decay amplitude for
$B\rightarrow J/\psi \ks$ is dominated by the tree-level
contribution, this region of $(\bar{\rho},\bar{\eta})$ contains
loop effects from $\Delta B=2$ processes.
Finally, we consider a parametrization of new
physics contributions to $B^0-\bar{B}^0$ mixing with the amplitude 
$M_{12}$ expressed as 
$M_{12}=M_{12}^{\rm SM}+M_{12}^{\rm new}$, and obtain a
constraint on $M_{12}^{\rm new}$
\cite{Soares:1992xi,Goto:1995hj,Cahn:1999gx}.
Such a study of the CKM parameter fit is described in the next
section.

The separation of the Standard Model and new physics
contributions can also be formulated for radiative and
semileptonic $B$ decays: $b\rightarrow s\gamma$ and
$b\rightarrow s\ell^+\ell^-$.
As discussed in Section~\ref{sec:new_physics/btos}, in the 
language of the effective Hamiltonian the new physics
contribution can be expressed in terms of Wilson coefficients.

A similar strategy can be applied to hadronic decay
amplitudes. 
For instance, the difference between the values
of $\sin 2\phi_1$ measured by
$J/\psi \ks$ and $\phi \ks$ implies some new physics
effect in the $b\rightarrow s\bar{s}s$ penguin process.
One can parametrize the $b\rightarrow s\bar{s}s$ amplitude
in terms of SM and new physics amplitudes.
In this case, however, the calculation of the decay
amplitudes from the fundamental theory is much more
difficult and the utility of the parametrization is limited.

\section{The CKM fit}
\label{sec:The_CKM_fit}
We perform a global fit to the CKM parameters
$(\bar{\rho},\bar{\eta})$ using the CKMfitter package
\cite{Hocker:2001xe}\footnote{
  We use the CKMFitter package with the $R$-fit option. 
  The parameter values are the default values used in the
  package if not mentioned explicitly.
}.
If no new physics effect exists, all the measurements should 
agree with each other in the $(\bar\rho,\bar\eta)$ plane and
can be combined to obtain the constraint on them.
On the other hand, the effect of new physics may cause 
discrepancies among these measurements.

The strategy is to look for the new physics effects from only
one source at a time, in order to quantitatively identify
the new physics contribution.
Assuming that the tree-level processes are insensitive to
new physics effects, we determine the SM value of
$(\bar\rho,\bar\eta)$ using $|V_{ub}|$ from $B\to X_u
\ell\nu$ decays and $\phi_3$ from 
$B^\pm \to D^0 [\to K^0_S \pi^+\pi^-] K^\pm$
\footnote{
  There is a possible effect of new physics on the Dalitz
  distributions through $D^0-\bar{D}^0$ mixing or $CP$
  violation in $D^0$.
  However, we can check and measure the effect from the data
  using soft-pion tagged $D^0$ decays ($D^{*+}\to D^0 \pi^+_s$)
  \cite{Asner:2003uz}, and verify whether its effect to the
  $\phi_3$ measurement is small enough. 
}.
We then examine the $b\to d$ box diagram, \textit{i.e.}
$B^0-\bar{B}^0$ oscillations, for which the available
measurements are 
$\sin 2\phi_1$ from $B^0\rightarrow c\bar{c}K^0$
and $\Delta m_d$. 

\begin{table}[tbp]
  \begin{center}
    \begin{tabular}{|c|c|ccc|}
      \hline
      parameters & central values & & errors & \\
      & & 0.5~ab$^{-1}$ & 5~ab$^{-1}$ & 50~ab$^{-1}$ \\
      \hline\hline
      $\sin 2\phi_1$ & 0.753 & 
      $\pm$4.4\% & $\pm$2.5\% & $\pm$1.9\%
      \\
      $\phi_3$ (Dalitz) & 65$^\circ$ & 
      $\pm$23\% & $\pm$7.2\% & $\pm$2.3\%
      \\
      $|V_{ub}|$ & 3.79$\times 10^{-3}$ & 
      $\pm 6.9\%\pm 8\%$ & $\pm 3.7\%\pm 4.5\%$ & $\pm 3.2\%\pm 3\%$
      \\ 
      $\Delta m_d$ & 0.496 ps$^{-1}$ & 
      $\pm$0.8\% & $\pm$0.5\% & $\pm$0.4\%
      \\
      $f_{B_d}\sqrt{B_{B_d}}$ & 0.230 GeV & 
      $\pm 0.011\pm 0.026$ & $\pm 0.011\pm 0.026$ & $\pm 0.005\pm 0.015$
      \\
      \hline
      $\phi_2$ & 91$^\circ$ &
      $\pm 9^\circ$ & $\pm 2.9^\circ$ & $\pm 0.9^\circ$ 
      \\
      \hline
    \end{tabular}
    \caption{
      Summary of inputs to in the CKM-fit. 
      If two errors are given, the second error is treated as flat 
      in the fit \cite{Hocker:2001xe}.
      Otherwise, errors are taken to be Gaussian.
      $|V_{cb}|$ is fixed at $4.04 \times 10^{-3}$ in the fit.
      Central values correspond to 
      $\bar\rho = 0.176$ and $\bar\eta = 0.376$.
    }
    \label{tab:ckmfit_input}
  \end{center}
\end{table}

The input parameters used in the fit are listed in
Table~\ref{tab:ckmfit_input}, where the errors are based on the
sensitivity studies in the previous sections.
The errors for $\Delta m_d$ are extrapolated from the
current measurements with an irreducible systematic error of 
0.4\% \cite{Ronga:2003bm}. 
The current value of $f_{B_d}\sqrt{B_{B_d}}$ is from a
recent unquenched lattice calculation 
$215(11)(^{+\ 0}_{-23})(15)$~MeV \cite{Aoki:2003xb},
where the first error is statistical, the second is the
uncertainty from the chiral extrapolation and the third is
due to other systematic errors. 
In Table~\ref{tab:ckmfit_input} the two systematic errors are
combined and the error is symmetrized.
To improve the accuracy of the lattice calculation,
simulations at significantly smaller sea quark masses will
be required.
With the staggered fermion formulation for sea quarks this
is possible, albeit with additional complications, by
introducing more flavors (or tastes) of fictitious quarks
\cite{Davies:2003ik}.
Within several years it will also become possible to perform
unquenched simulations using lattice fermions with an exact
chiral symmetry, such as overlap or domain-wall
fermions. 
The reduction of errors down to a level of several percent is
feasible within the time scale of the construction
of SuperKEKB.

The constraint on $(\bar{\rho},\bar{\eta})$ using all these 
measurements for the SM case is shown in
Figure~\ref{fig:ckmfit_sm}.
The plots are shown for (a) 500~fb$^{-1}$, (b) 5~ab$^{-1}$,
and (c) 50~ab$^{-1}$.
Drastic improvements are expected with the statistical power
of the SuperKEKB.
The error in the determination of $(\rho,\eta)$ will be
eventually reduced to a $\pm 0.03$ level with 50~ab$^{-1}$.

\begin{figure}[tbp]
  \begin{flushleft}
    \begin{minipage}[t]{7cm}
      (a)\\
      \includegraphics[width=1.0\textwidth]{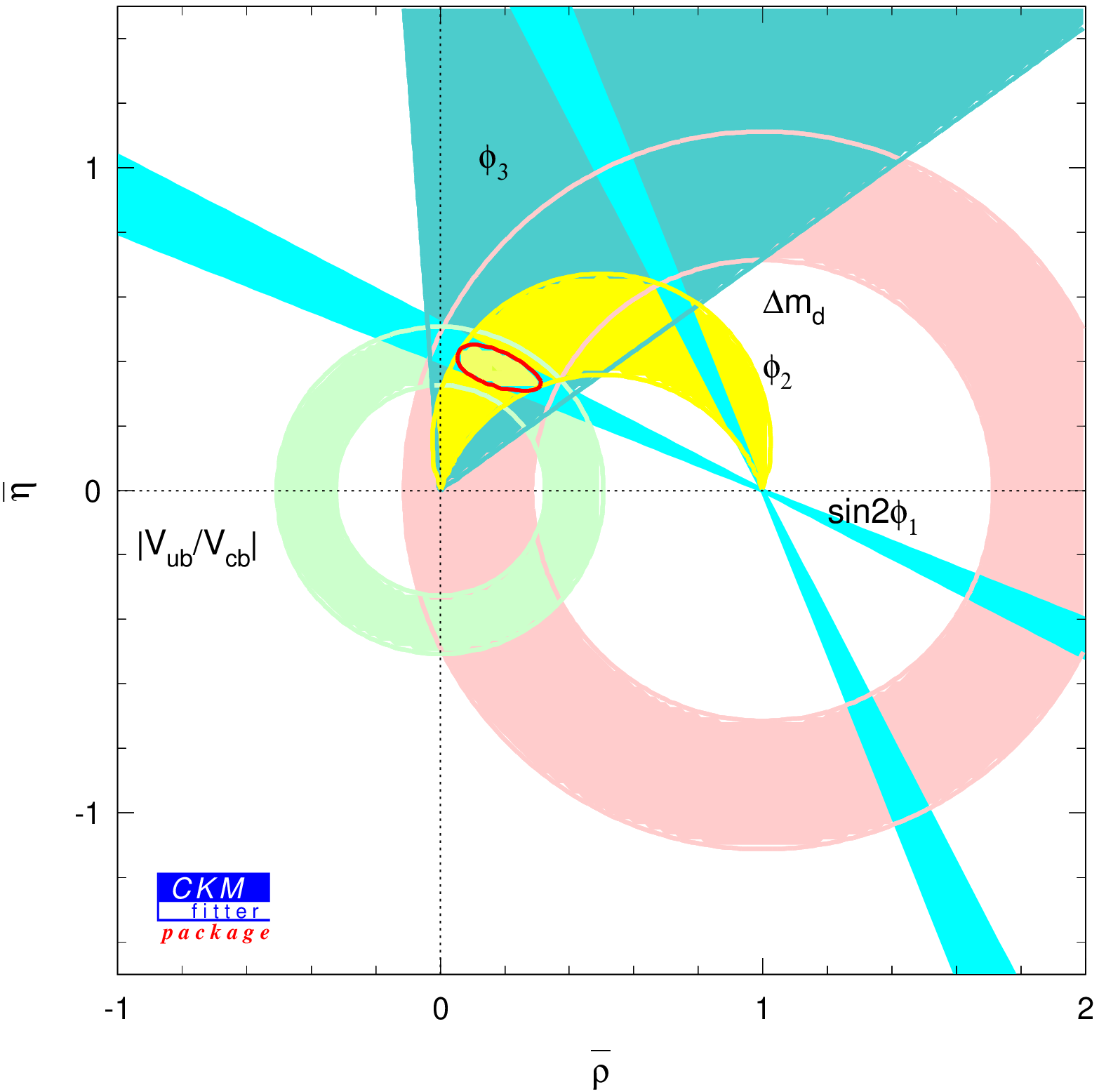}
    \end{minipage}
    \begin{minipage}[t]{7cm}
      (b)\\
      \includegraphics[width=1.0\textwidth]{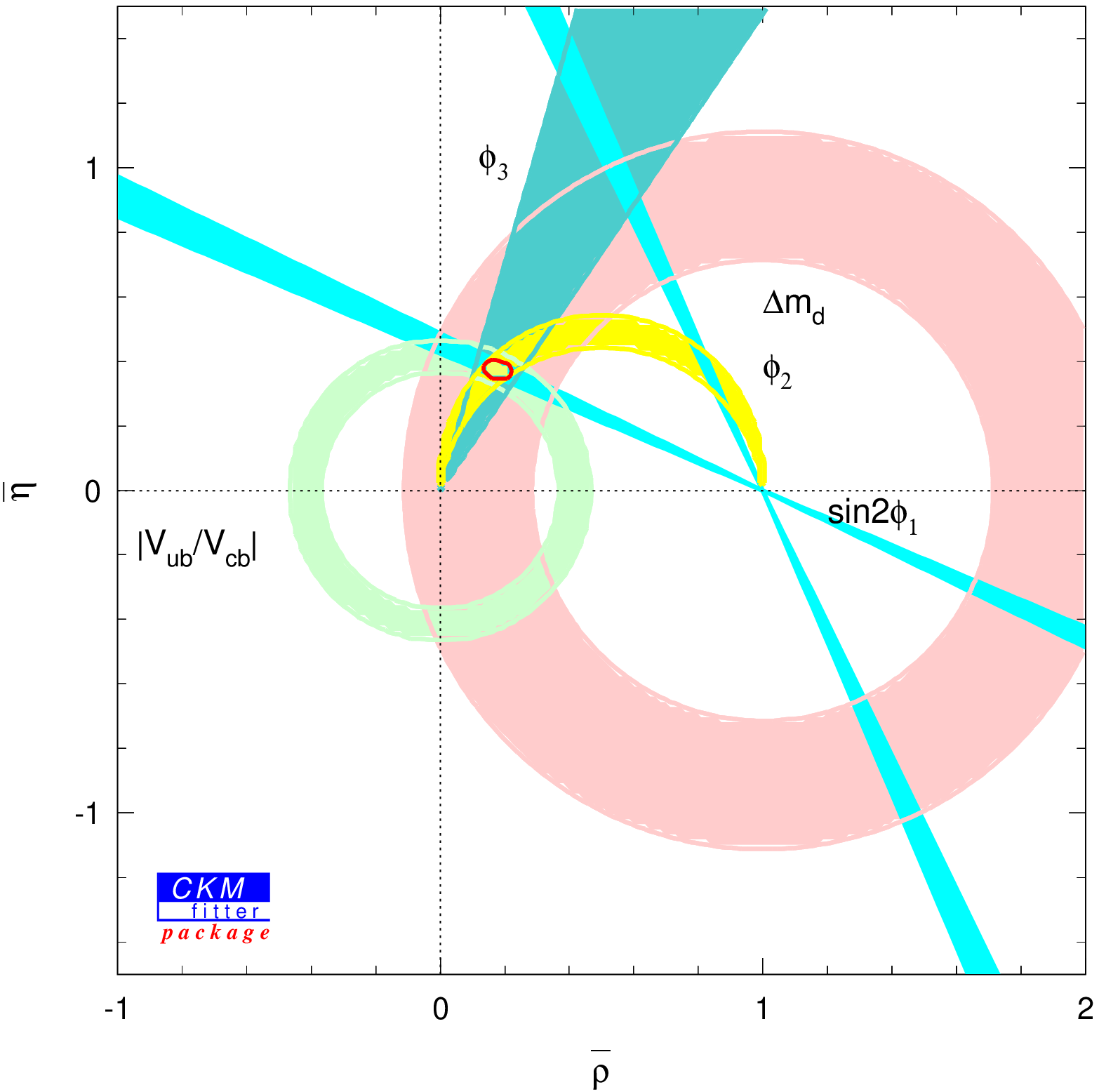}
    \end{minipage}
    
    \begin{minipage}[t]{7cm}
      (c)\\
      \includegraphics[width=1.0\textwidth]{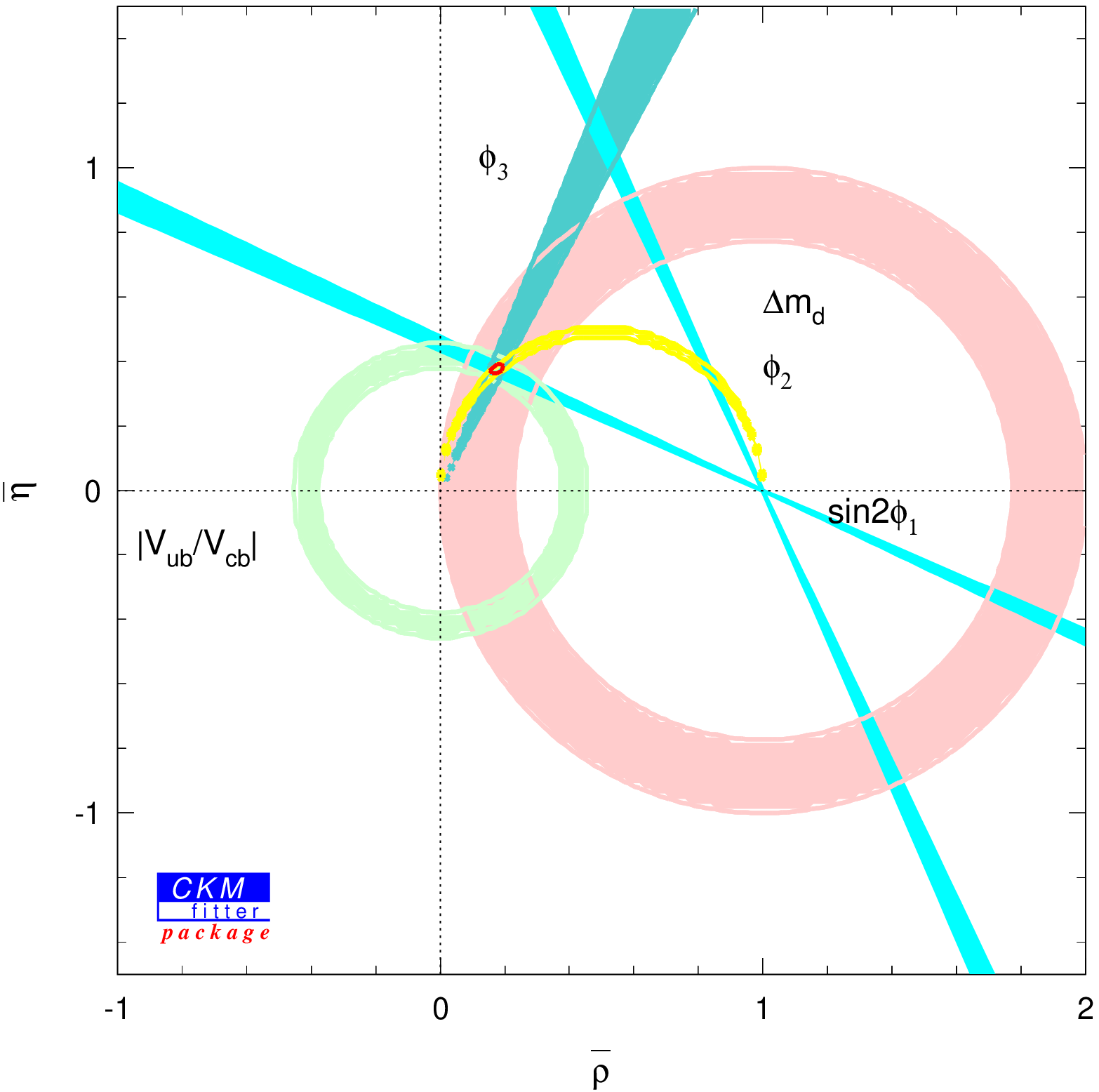}
    \end{minipage}
  \end{flushleft}
  \caption{
    Constraints on the CKM unitarity triangle at
    (a) 0.5~ab$^{-1}$, (b) 5~ab$^{-1}$, and (c) 50~ab$^{-1}$.
  }
  \label{fig:ckmfit_sm}
\end{figure}

Among those input parameters the measurement of the angle
$\phi_3$ is not yet precise at present.
Therefore, despite the good agreement of
$(\bar{\rho},\bar{\eta})$ from different measurements, there
is a possibility that $\phi_3$ does not agree with the
current world average obtained with the CKM fit.
It means that the SM value of $(\bar{\rho},\bar{\eta})$
could still be significantly different from the current
central values, if other loop-induced processes contain
large new physics effects.
To see what could be observed at SuperKEKB, here we
show a plot assuming that the central value of the $\phi_3$ 
measurement is 120$^\circ$ instead of the current average
65$^\circ$ (Figure~\ref{fig:ckmfit_phi3=120}).
The CKM fits are independently done for tree-level processes
($|V_{ub}|$ and $\phi_3$) and for $b\to d$ mixing processes
($\sin 2\phi_1$, $\Delta m_d$ and $\phi_2$). 
Clear separation of these two regions is achieved with the
integrated luminosity of 5 ab$^{-1}$, while
the deviation is not significant at 0.5 ab$^{-1}$
that can be achieved at the present KEKB/Belle.

Measurements of $\phi_2$ using $B^0\to\pi^+\pi^-$ or
$B^0\to\rho\pi$ decay modes are contaminated by the $b\to d$
penguin diagram, which can be eliminated using the isospin
relations within the Standard Model.
However, the spin and flavor structure of new physics
contribution to the $b\to d$ box and penguin amplitudes is
model-dependent, and the isospin relations could be violated
for some class of new physics models.
Here, for simplicity we assume that the ``penguin trapping''
technique based on the isospin relation is not affected by
the new physics contribution in the penguin-loop, which is
satisfied if the new physics diagrams only contribute to
$\Delta I = 1/2$ amplitudes.  
In this case, the sum of the SM penguin and new physics
penguin amplitudes is treated as a penguin amplitude to be
trapped out. 

\begin{figure}[tbp]
  \begin{flushleft}
    \begin{minipage}[t]{7cm}
      (a)\\
      \includegraphics[width=1.0\textwidth]{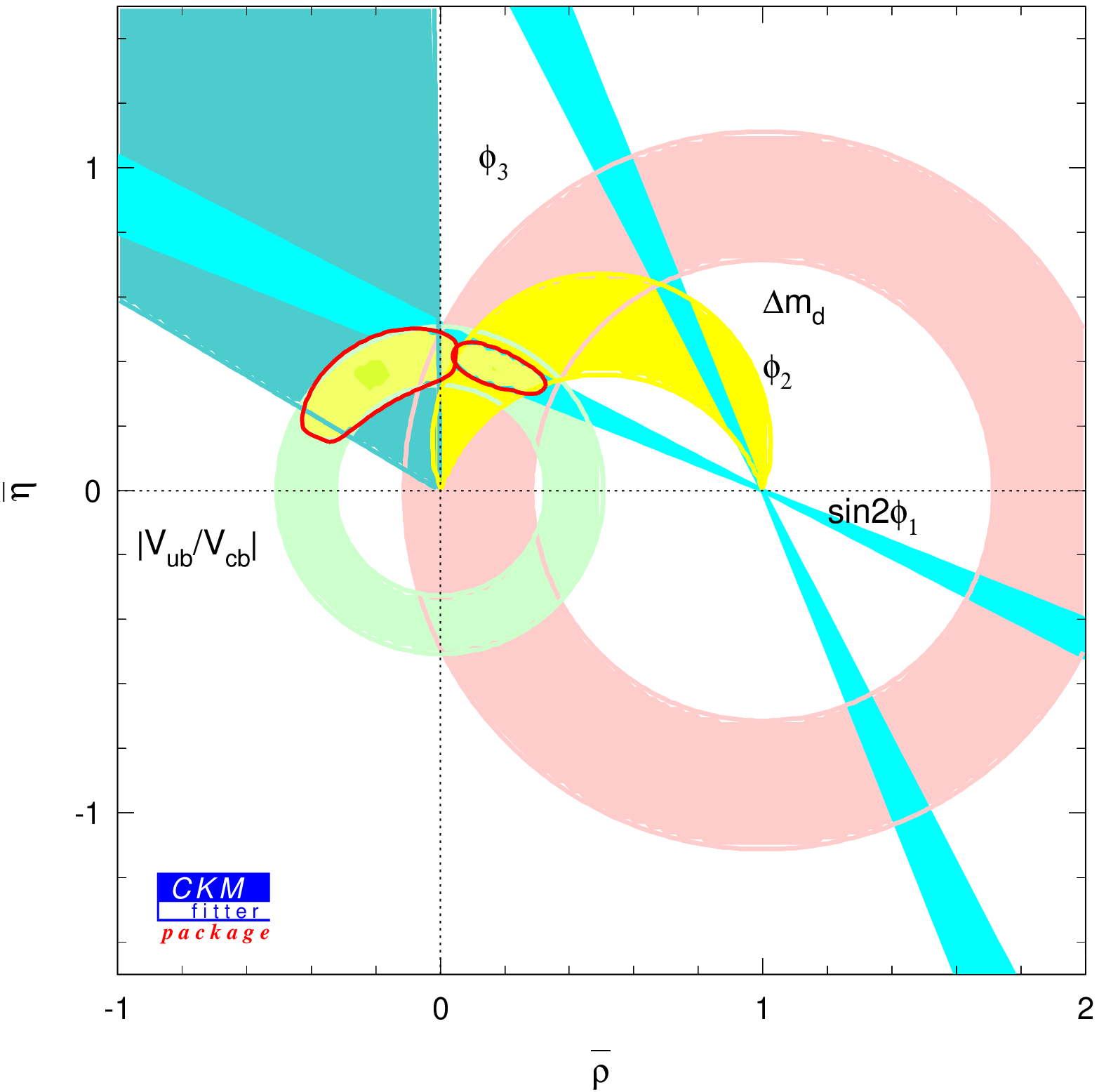}
    \end{minipage}
    \begin{minipage}[t]{7cm}
      (b)\\
      \includegraphics[width=1.0\textwidth]{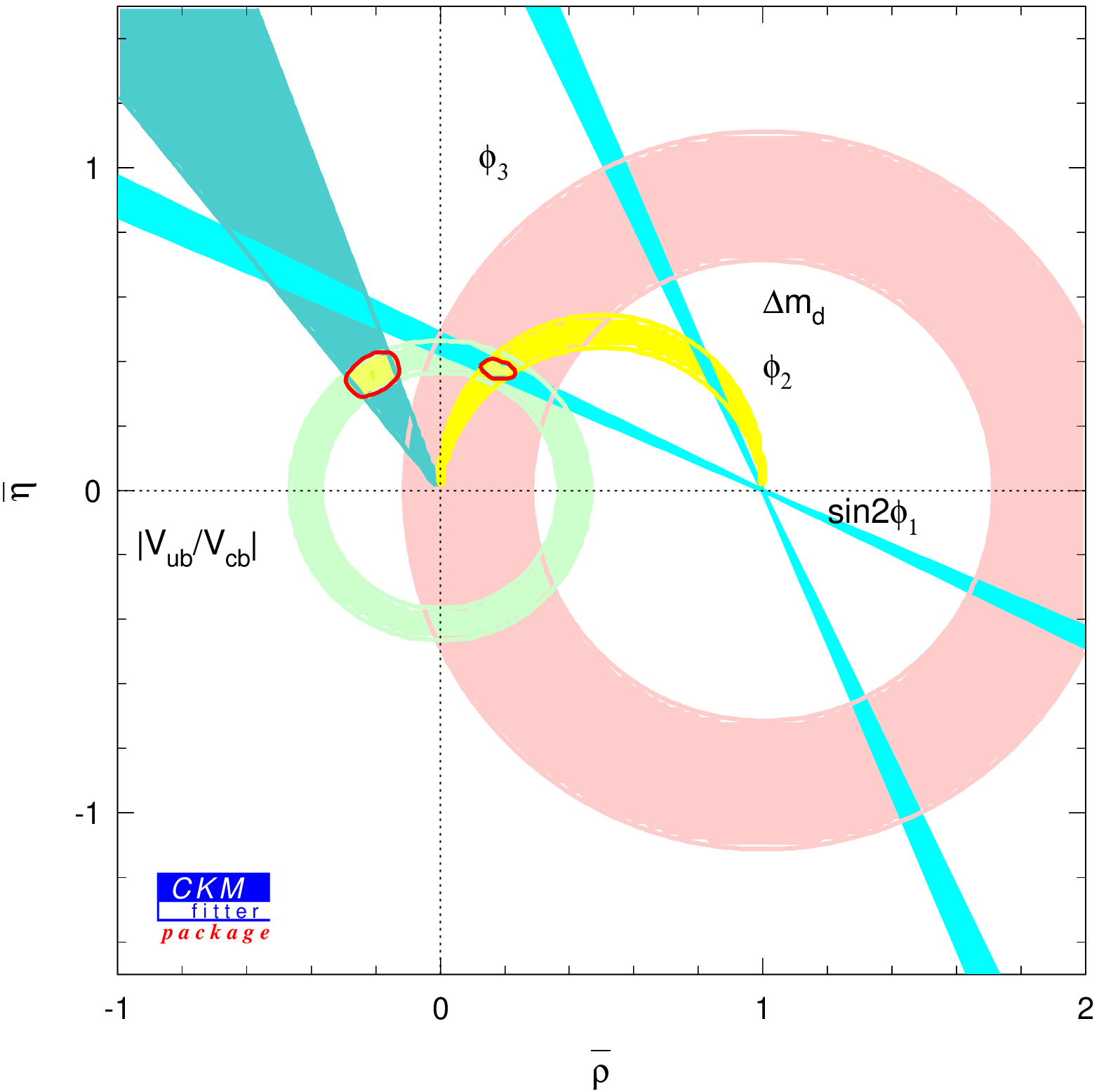}
    \end{minipage}
    
    \begin{minipage}[t]{7cm}
      (c)\\
      \includegraphics[width=1.0\textwidth]{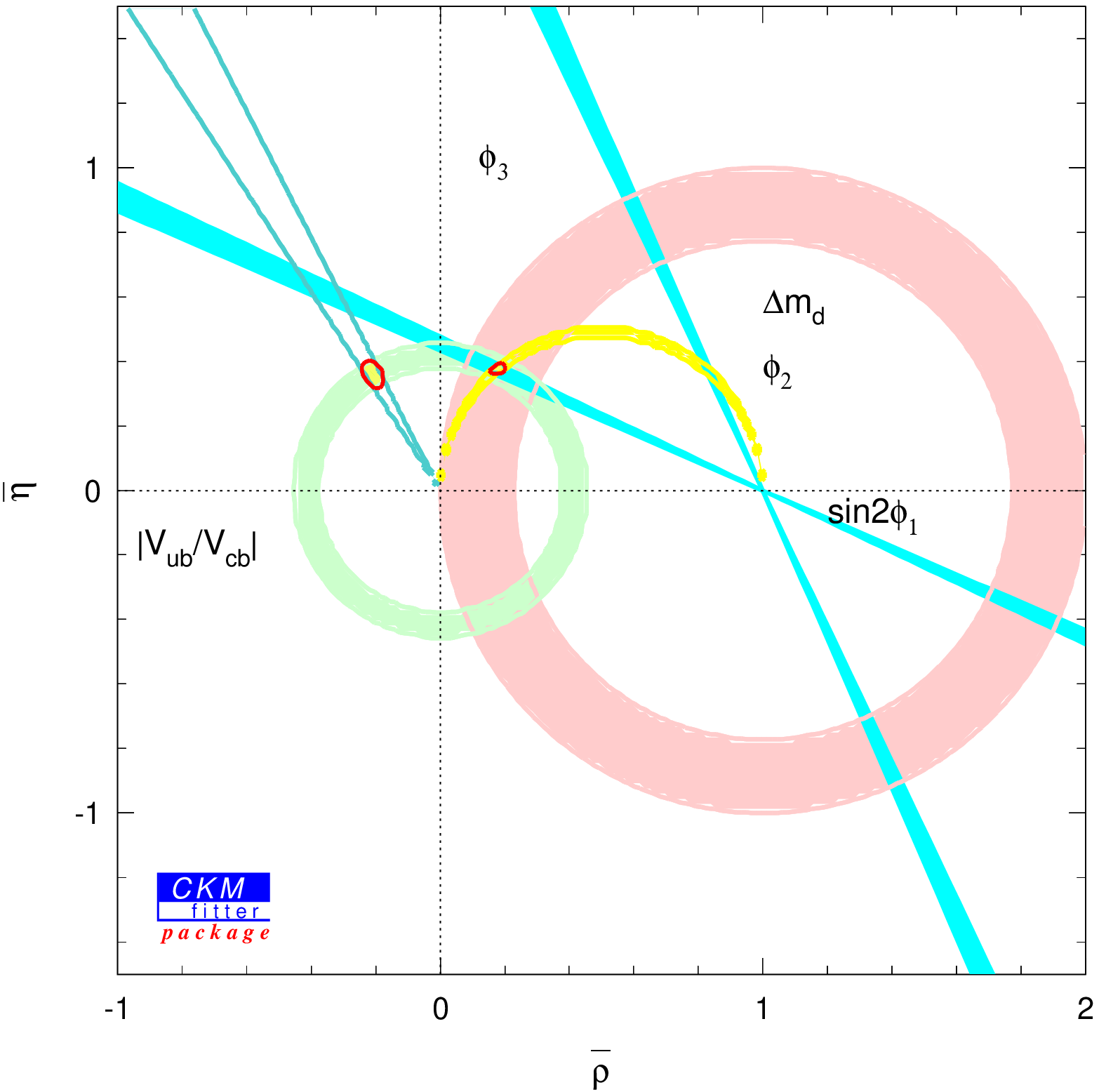}
    \end{minipage}
  \end{flushleft}
  \caption{
    Constraints on the CKM unitarity triangle assuming
    $\phi_3=120^\circ$ at
    (a) 0.5~ab$^{-1}$, (b) 5~ab$^{-1}$, and (c) 50~ab$^{-1}$.
    Independent constraints are shown for tree-level
    processes ($|V_{ub}|$ and $\phi_3$) and for $b\to d$
    mixing processes ($\sin 2\phi_1$, $\Delta m_d$ and $\phi_2$).
  }
  \label{fig:ckmfit_phi3=120}
\end{figure}

The effect of new physics on the $b\to d$ box diagram
can be expressed as
\begin{equation}
  M_{12} = M_{12}^{\mathrm{SM}} + M_{12}^{\mathrm{new}},
\end{equation}
where $M_{12}^{\mathrm{SM}}$ ($M_{12}^{\mathrm{new}}$) is the
contribution of the amplitude to the $b\to d$ box diagram
from the SM (new physics particles)
\cite{Soares:1992xi,Goto:1995hj,Cahn:1999gx}.
The fit result for the constraints in the 
$(\mathrm{Re}~M_{12}^{\mathrm{new}},\mathrm{Im}~M_{12}^{\mathrm{new}})$
plane is shown in Figure~\ref{fig:M12new}.

For the SUSY models discussed in
Section~\ref{sec:new_physics_case_study},
the $M_{12}^{\mathrm{new}}$ is tiny and the details of the
models could not be distinguished from the fit on the
$(\mathrm{Re}~M_{12}^{\mathrm{new}},\mathrm{Im}~M_{12}^{\mathrm{new}})$
plane.
However, it implies that if one finds a significant
deviation of 
$(\mathrm{Re}~M_{12}^{\mathrm{new}},\mathrm{Im}~M_{12}^{\mathrm{new}})$
from the origin, those specific SUSY models are excluded.
Thus, this analysis provides another method to probe
different kinds of new physics models that are sensitive to the
$b\to d$ transitions.

\begin{figure}[tbp]
  \centering
  \includegraphics[width=8cm]{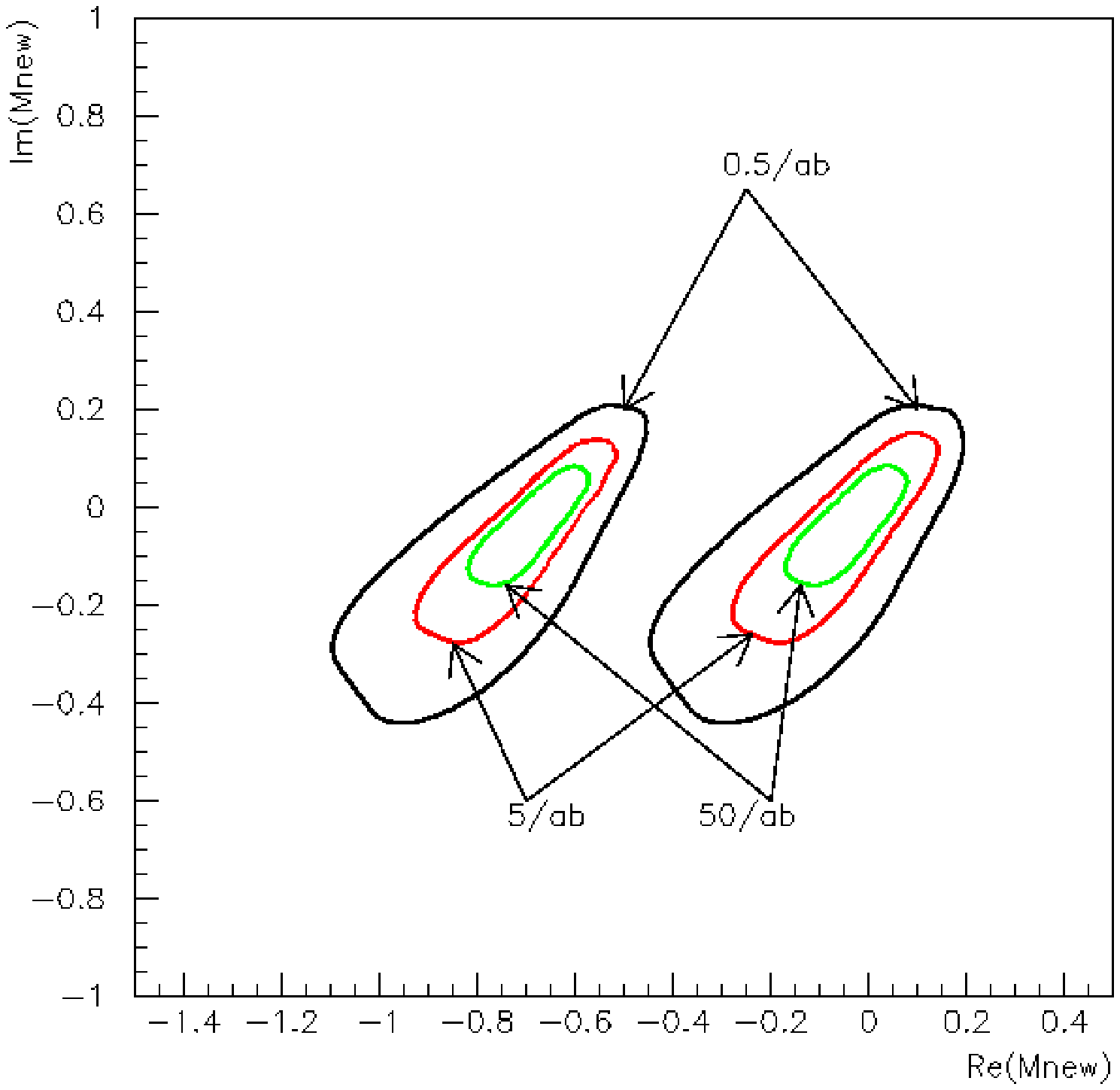} 
  \caption{
    Constraints on the new physics contribution to the 
    $b\to d$ box amplitude $M_{12}^{\rm new}$.
    Contours correspond to 0.5~${\rm ab}^{-1}$ (black),
    5~${\rm ab}^{-1}$ (red) and 50~${\rm ab}^{-1}$ (green).
  }
  \label{fig:M12new}
\end{figure}

The method we described above can be extended to the 
$b\to s$ box diagram 
($\Delta m_s$) and time dependent $CP$ violation using 
$B^0_s \to J\psi \phi$ \textit{etc.},
$s\to d$ box diagram ($\epsilon_K$), and 
$b\to d$ penguin diagram ($b \to d \gamma$ \textit{etc.}).
The exploration of new physics effects in flavor physics
will be most powerful from a combination of
these many different directions.



\newpage
%
\chapter{Summary}
%

The grand challenge of elementary particle physics is
to identify the fundamental elements of Nature and
uncover the ultimate theory of their 
creation, interactions and annihilation.
To this end, all elementary particles and
the forces between them should be described in a unified picture.
Such unification naturally requires 
a deep understanding of physical laws at a very high-energy
scale; for instance
the unification of electroweak and strong forces
is expected to occur at around 10$^{16}$ GeV, 
which is often called the GUT scale.

It is unlikely that the GUT scale will be realized
at any accelerator-based experiment even in the distant future.
However, there are a few very promising ways to promote
our grand challenge.
One such approach is to elucidate the nature of
quantum loop effects by producing as many particles
as possible.
This provides the rationale to pursue the luminosity frontier. 

There is no doubt that past experiments at 
the luminosity (or intensity) frontier of the age
have yielded epoch-making results.
%
This good tradition has been followed by 
Belle and BaBar, experiments at energy-asymmetric
$e^+e^-$ $B$ factories KEKB and PEPII,
which have observed $CP$ violation in the neutral 
$B$ meson system.
The result is in good agreement with the
constraints from the Kobayashi-Maskawa (KM) model of
$CP$ violation. We are now confident that
the KM phase is the dominant source of $CP$ violation.
In 2003 the KEKB collider achieved
its design luminosity of
1$\times$10$^{34}$ cm$^{-2}$s$^{-1}$. 
The Belle experiment will accumulate
an integrated luminosity of 500~fb$^{-1}$
within a few years. This will suffice to determine
the Unitarity Triangle with a precision of ${\cal O}(10)$\%.
Various other quantities in $B$ meson decays will also
become accessible. In particular, the first observation of
direct $CP$ violation in charmless $B$ decays is anticipated.

Over the past thirty years,
the success of the Standard Model, which incorporates
the KM mechanism, has become increasingly firm.
This strongly indicates that the Standard Model is {\it the}
effective low-energy description of Nature.
Yet there are several reasons to believe that
physics beyond the Standard Model should exist.
One of the most outstanding problems is the
quadratically divergent radiative correction 
to the Higgs mass, which requires a fine tuning of
the bare Higgs mass unless the cutoff scale is $\cal O$(1) TeV.
This suggests that the new physics lies
at the energy scale of ${\cal O}(1)$ TeV.
There is a good chance that LHC will
discover new elementary particles such as supersymmetric
(SUSY) particles.
With this vision in mind, we raise an important question
``what should be a role of the luminosity frontier
in the LHC era ?''

To answer the question, we note that
the flavor sector of the Standard Model is quite successful 
in spite of the problem in the Higgs sector.
This is connected to the observation that
Flavor-Changing-Neutral-Currents (FCNCs) 
are highly suppressed.
Indeed, if one considers a general new physics model
without any mechanism to suppress FCNC processes, 
present experimental results on $B$ physics imply that
the new physics energy scale should be larger than ${\cal O}(10^3)$ TeV.
This apparent mismatch is called 
{\it the new physics flavor problem}.
To overcome the problem, any new physics at the TeV scale
should have a mechanism to suppress FCNC processes,
which often results in a distinctive flavor structure at low energy.
Therefore, the indispensable roles of the luminosity frontier are
to observe deviations from the Standard Model in flavor physics, and
more importantly, to distinguish between different new physics models
by a close examination of the flavor structure.
Comprehensive studies of $B$ meson decays in the clean
$e^+e^-$ environment provide the ideal solution for this purpose,
which is not possible at LHC nor even at the future
linear collider.

These provide the primary motivation for SuperKEKB,
a major upgrade of KEKB.
Its design luminosity is
$5 \times 10^{35}~{\rm cm}^{-2}{\rm s}^{-1}$, which is
50 times as large as the peak luminosity achieved by
KEKB. 
Various FCNC processes,
such as the radiative decay $b\rightarrow s\gamma$,
the semileptonic decay $b\rightarrow sl^+l^-$, and
the hadronic decays $b\rightarrow s\bar{q}q$ and
$b\rightarrow d\bar{q}q$, can be studied with
unprecedented precision.
All of these processes are suppressed in the Standard Model
by the GIM mechanism, and therefore the effect of new
physics is relatively enhanced.
New observables that are currently out of reach
will also become accessible.
In addition to $B$ meson decays, FCNC processes in
$\tau$ and charm decays will also be studied at SuperKEKB.


The Belle detector will be upgraded to take full
advantage of the high luminosity of SuperKEKB.
In spite of harsh beam backgrounds, 
the detector performance will be at least as good as the present
Belle detector and improvements in several aspects are envisaged.
Table~\ref{exe:tbl:sensitivity_summary} summarizes
the physics reach at SuperKEKB. As a reference, measurements
expected at LHCb are also listed.
One of the big advantages of SuperKEKB is the capability
to reconstruct rare decays that have $\gamma$'s, $\pi^0$'s,
$\kl$'s or
neutrinos in the final states. 
There are several key observables in Table~\ref{exe:tbl:sensitivity_summary}
that require this capability.
Also important are time-dependent $CP$ asymmetry measurements
using only a $\ks$ and a constraint from the interaction point to
determine the $B$ decay vertices.
Examples include $\bz\to K^{*0}(\to \ks\pi^0)\gamma$, 
$\pi^0\ks$ and $\ks\ks\ks$.
These fundamental measurements cannot be carried out at hadron colliders.

Figure~\ref{exe:fig:newCPV}(a) shows a comparison 
between time-dependent $CP$ asymmetries in $\bz\to\jpsi\ks$,
which is dominated by the $\btoccs$ tree process,
and $\bz\to\phi\ks$, which is governed by
the $\btosss$ FCNC (penguin) process.
It demonstrates how well a possible new $CP$-violating phase 
can be measured. Such a new source of $CP$ violation
may revolutionize the understanding of
the origin of the matter-dominated Universe, which is
one of the major unresolved issues in cosmology.
Figure~\ref{exe:fig:newCPV}(b) shows
correlations between time-dependent $CP$ asymmetries in 
$\bztoksgamma$ and $\bztophiks$ decays
in two representative new physics
models with different SUSY breaking scenarios;
the SU(5) SUSY GUT with right-handed neutrinos and
the minimal supergravity model.
The two can be clearly distinguished.
This demonstrates that SuperKEKB is
sensitive to a quantum phase even at the GUT scale.
Note that these two models may have rather similar
mass spectra. It will therefore be very difficult to distinguish
one from the other at LHC.
If SUSY particles are discovered at LHC,
the origin of SUSY breaking will become one of the
primary themes in elementary particle physics.
SuperKEKB will play a leading role in such studies.

%
%
We emphasize that the example above is just one of several
useful correlations that can be measured only at SuperKEKB.
The true value of SuperKEKB is a capability to observe
the pattern as a whole, which allows us to
differentiate a variety of new physics scenarios.
It is so to speak
``{\it DNA identification of new physics}'',
in that each measurement does not
yield a basic physical parameter of the new physics
but provides an essential piece of the overall flavor structure.
This strategy works better when we accumulate more data.
Thus the target annual integrated luminosity
of 5 ab$^{-1}$ is not a luxury but necessity, and
stable long-term operation of SuperKEKB is necessary
to meet the requirements.


Determination of the Unitarity Triangle will also be pushed
forward and will be incorporated in the global pattern
mentioned above.
This can be done at SuperKEKB using redundant measurements of
all three angles and all three sides of the Unitarity Triangle. 
In particular, $\phi_2$ measurements and $V_{ub}$ measurements
require the reconstruction of $\pi^0$ mesons and neutrinos
and are thus unique to a Super $B$-Factory.
An inconsistency among these measurements
implies new physics. 
Figure~\ref{exe:fig:ckmfit_np}(a) and (b) show the expected
constraints at 50 ab$^{-1}$ in two cases
that are both allowed with the present experimental constraints.
The clear difference between two figures demonstrates the
power of the ultimate precision of
${\cal O}(1)$\%, which will be obtained at SuperKEKB.

We thus conclude that the physics case at SuperKEKB 
is compelling. It will be the
place to elucidate {\it the new physics flavor problem}
in the LHC era. The physics program at SuperKEKB
is not only complementary to the next-generation energy 
frontier programme, but is an essential element of
the grand challenge in elementary particle physics.

%
%
%
%
%

%
%
\begin{table}
\small
\begin{center}
\begin{tabular}{|l|rr|r|}
\hline
Observable     &\multicolumn{2}{|c|}{SuperKEKB}& LHCb \\
               &(5 ab$^{-1}$)&(50 ab$^{-1})$&(0.002ab$^{-1}$)\\
\hline 
$\deltasphiks$ & 0.079       & 0.031         & 0.2 \\
$\deltaskkks$  & 0.056  & 0.026              &  \\
$\deltasetapks$& 0.049       & 0.024         & $\times$ \\
$\deltasksksks$& 0.14        & 0.04          & $\times$ \\
$\deltaspizks$ & 0.10        & 0.03          & $\times$ \\
$\sin2\chi$ ($B_s \to J/\psi\phi$)&$\times$&$\times$& 0.058\\
\hline
$\calsksgamma$ & 0.14        & 0.04          & $\times$ \\
$\Br(\BtoXsgamma)$ & 5\%         & 5\%           & $\times$ \\
$\Acp(\BtoXsgamma)$ & 0.011       & 5$\times$10$^{-3}$ &$\times$\\
$C_9$    from $\AFB(\BtoKstarll)$ & 32\% & 10\% & \\
$C_{10}$ from $\AFB(\BtoKstarll)$ & 44\% & 14\% & \\
${\cal B}(B_s \rightarrow \mu^+\mu^-)$ & $\times$ & $\times$ &$4\sigma$ (3 years)\\
\hline
${\cal B}(\bp\to\kp\nu\nu)$ &           & 5.1$\sigma$   & $\times$ \\
${\cal B}(\bp\to D \tau \nu)$ & 8\% & 2.5\% & $\times$ \\
${\cal B}(\bz\to D \tau \nu)$ & 3.5$\sigma$ & 9\% & $\times$ \\
\hline
$\sinbb$       & 0.019          & 0.014         & 0.022\\
$\phi_2$ ($\pi\pi$ isospin)& 3.9$^\circ$ & 1.2$^\circ$   & $\times$ \\
$\phi_2$ ($\rho\pi$) & 2.9$^\circ$ & 0.9$^\circ$   & $\times$ \\
$\phi_3$ ($DK^{(*)}$)& 4$^\circ$  & 1.2$^\circ$    & 8$^\circ$\\
$\phi_3$ ($B_s \rightarrow KK$) & $\times$ & $\times$ & 5$^\circ$\\ 
$\phi_3$ ($B_s \rightarrow D_sK$) &$\times$ &$\times$ &14$^\circ$\\ 
$|V_{ub}|$ (inclusive) & 5.8\%       & 4.4\%         & $\times$\\
\hline
${\cal B}(\tau\to\mu\gamma)$ 
               & $< 1.8 \times 10^{-8}$ & &\\
\hline
\end{tabular}
\end{center}
\caption{Summary of the physics reach at SuperKEKB.
Expected errors for the key observables are listed
for an integrated luminosity of
5 ab$^{-1}$, which corresponds to
one year of operation, and with 50 ab$^{-1}$.
$\Delta\cals_f$ is defined by
$\Delta\cals_f \equiv (-\xi_f)\cals_f - \cals_{J/\psi\ks}$,
where $\xi_f$ is the $CP$ eigenvalue of the final state $f$.
For comparison,
expected sensitivities at LHCb with one year of operation
are also listed if available.
The $\times$ marks indicate measurements that
are very difficult or impossible. 
} 
\label{exe:tbl:sensitivity_summary}
\end{table}
%
\begin{figure}[tbp]
\begin{center}
\begin{minipage}[t]{7cm}
(a)\\
\includegraphics[width=1.0\textwidth]{sensitivity_sss/asym_phiks_5ab-1.eps}
\end{minipage}
\begin{minipage}[t]{7cm}
(b)\\
\includegraphics[width=1.0\textwidth,clip]{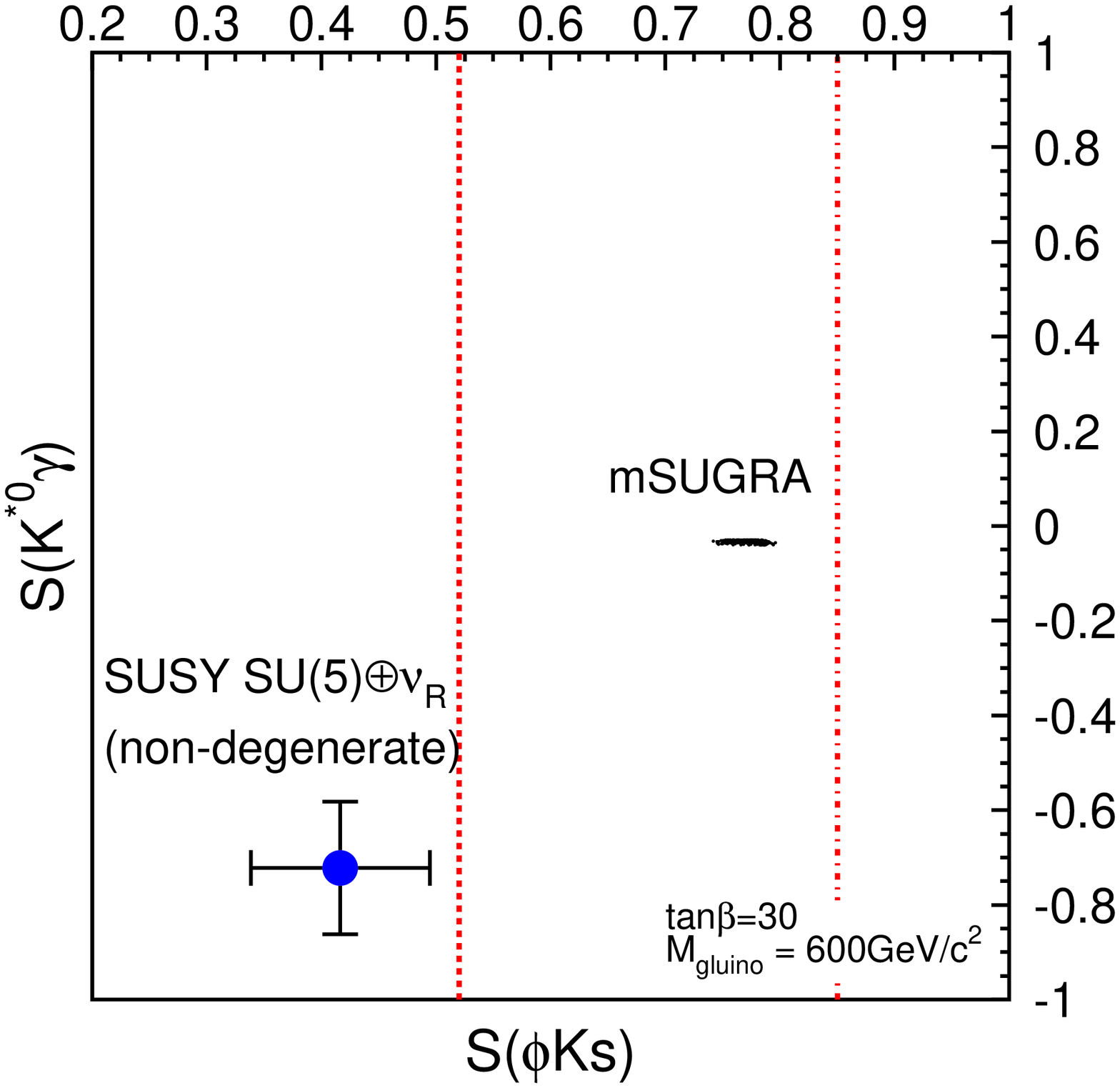}
\end{minipage}
\end{center}
\caption{
  (a) Time-dependent $CP$ asymmetries in $\bztophiks$ and 
      $\bz\to\jpsi\ks$ decays expected with one year of operation at SuperKEKB
      (5 ab$^{-1}$). The world average values (as of August 2003)
      for the modes governed by the $b \to s$ transition are used 
      as the input values of $\calsphiks = +0.24$ and $\calaphiks = +0.07$.
  (b) A correlation between time-dependent $CP$ asymmetries in 
      $\bztoksgamma$ and $\bztophiks$. The dots show the range
      in the minimal supergravity model (mSUGRA). The circle corresponds to
      a possible point of supersymmetric 
      SU(5) GUT model with right-handed neutrinos.
      Error bars associated
      with the circle indicate expected errors
      with one year of operation at SuperKEKB. 
      The present experimental bound of $\calsphiks < +0.52$ 
      ($\calsphiks < +0.85$)
      at the 2$\sigma$ (3$\sigma$) level
      is also shown by the dashed (dot-dashed) vertical line.
}
\label{exe:fig:newCPV}
\end{figure}
%
\begin{figure}[tbp]
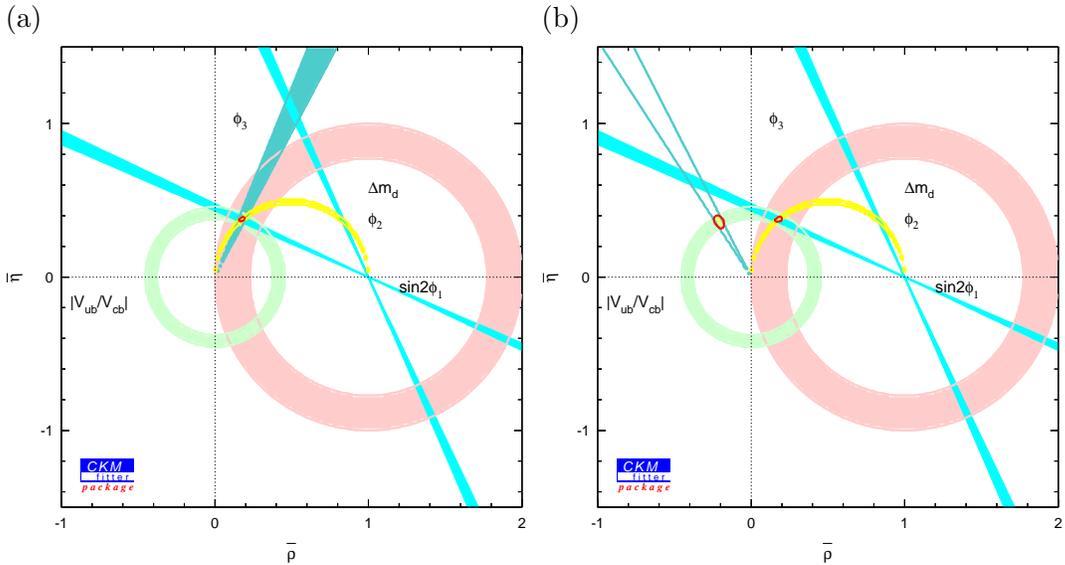

  \begin{center}
  \begin{minipage}[t]{7cm}
   (a)\\
   \includegraphics[width=1.0\textwidth]{figures/rhoeta_sm_phi2_50000.eps}
   \end{minipage}
   \begin{minipage}[t]{7cm}
   (b)\\
   \includegraphics[width=1.0\textwidth]{figures/rhoeta_phi3=120_50000.eps}
   \end{minipage}
  \end{center}
  \caption{
Constraints on the CKM unitarity triangle at 50~ab$^{-1}$
(a) in the case consistent with the SM, and (b) with
a large deviation from the SM. Both cases are within present
experimental constraints.
  }
  \label{exe:fig:ckmfit_np}
\end{figure}

%
%
%
%
%
%
%


\end{document}